\documentclass[12pt,oneside]{book}
\usepackage[margin=0.9in]{geometry}
\usepackage[toc,page]{appendix}
\usepackage{graphicx}
\usepackage[square,numbers]{natbib}
\usepackage{caption}
\usepackage{pdfpages}
\usepackage[shortlabels]{enumitem}
\usepackage{mathtools}
\usepackage{fancyhdr}
\usepackage{nicefrac}
\usepackage{bbm}
\input{Qcircuit}
\usepackage{upgreek}
\newcommand{\changefont}{%
    \fontsize{9}{11}\selectfont
}
 \usepackage{makecell}
 \usepackage{placeins}
  \usepackage{afterpage}
\usepackage{lettrine}
\usepackage[T1]{fontenc}
\usepackage{lmodern}

\usepackage{titlesec} 
\titleformat{\chapter}[display] %shape
      {\bfseries\Large}
  {\filleft\MakeUppercase{\chaptertitlename} \Huge\thechapter}
  {4ex}
  {\titlerule
   \vspace{1ex}%
   \filright}
  [\vspace{2ex}%
\titlerule]

\usepackage{mathrsfs}

\usepackage{graphics}
\usepackage{amsfonts,amssymb,amsmath}            % for math symbols.
\usepackage{ifthen}
\usepackage{amsthm, thm-restate}
\usepackage{dsfont}
\usepackage{float}
\usepackage{MnSymbol}

\usepackage[caption=false]{subfig}
\PassOptionsToPackage{hyphens}{url}
\usepackage[hyperfootnotes=false]{hyperref}
\usepackage{xcolor}
\definecolor{mylinkcolor}{rgb}{0,0,0.7} % set link color here as red,green,blue.
\hypersetup{unicode=true,%
  bookmarksnumbered=false,bookmarksopen=false,bookmarksopenlevel=1, %
  breaklinks=true,pdfborder={0 0 0},colorlinks=true}%
\hypersetup{%
  anchorcolor=mylinkcolor,citecolor=mylinkcolor, %
  filecolor=mylinkcolor,linkcolor=mylinkcolor, %
  menucolor=mylinkcolor,runcolor=mylinkcolor, %
  urlcolor=mylinkcolor} 
\usepackage{tikz}
\usetikzlibrary{arrows}
\usetikzlibrary{arrows.meta}
\usepackage{rotating, xcolor}
\usetikzlibrary{shapes}
\usetikzlibrary{hobby}
\usetikzlibrary{decorations.markings}
\usetikzlibrary{arrows.meta}
\usetikzlibrary{snakes} 
\usepackage{tikz-cd}
\tikzset{degil/.style={
            decoration={markings,
            mark= at position 0.5 with {
                  \node[transform shape] (tempnode) {$\backslash$};
                  }
              },
              postaction={decorate}
}
}

\newcommand\longrsquigarrow{
\begin{tikzpicture}
\draw [decorate, decoration={zigzag, segment length=+6pt, amplitude=+.95pt,post length=+2pt}, arrows={-Classical TikZ Rightarrow}]  (0,0.1) -- (0.6,0.1); \draw[draw=none] (0,0)--(0.6,0);
\end{tikzpicture}
}

\newcommand\longlsquigarrow{
\begin{tikzpicture}
\draw [decorate, decoration={zigzag, segment length=+6pt, amplitude=+.95pt,post length=+2pt}, arrows={-Classical TikZ Rightarrow},rotate around={180:(0.3,0.1)}]  (0,0.1) -- (0.6,0.1); \draw[draw=none] (0,0)--(0.6,0);
\end{tikzpicture}
}

\makeatletter
\newcommand*{\da@rightarrow}{\mathchar"0\hexnumber@\symAMSa 4B }
\newcommand*{\da@leftarrow}{\mathchar"0\hexnumber@\symAMSa 4C }
\newcommand*{\xdashrightarrow}[2][]{%
  \mathrel{%
    \mathpalette{\da@xarrow{#1}{#2}{}\da@rightarrow{\,}{}}{}%
  }%
}
\newcommand{\xdashleftarrow}[2][]{%
  \mathrel{%
    \mathpalette{\da@xarrow{#1}{#2}\da@leftarrow{}{}{\,}}{}%
  }%
}
\newcommand*{\da@xarrow}[7]{%
  \sbox0{$\ifx#7\scriptstyle\scriptscriptstyle\else\scriptstyle\fi#5#1#6\m@th$}%
  \sbox2{$\ifx#7\scriptstyle\scriptscriptstyle\else\scriptstyle\fi#5#2#6\m@th$}%
  \sbox4{$#7\dabar@\m@th$}%
  \dimen@=\wd0 %
  \ifdim\wd2 >\dimen@
    \dimen@=\wd2 %   
  \fi
  \count@=2 %
  \def\da@bars{\dabar@\dabar@}%
  \@whiledim\count@\wd4<\dimen@\do{%
    \advance\count@\@ne
    \expandafter\def\expandafter\da@bars\expandafter{%
      \da@bars
      \dabar@ 
    }%
  }%  
  \mathrel{#3}%
  \mathrel{%   
    \mathop{\da@bars}\limits
    \ifx\\#1\\%
    \else
      _{\copy0}%
    \fi
    \ifx\\#2\\%
    \else
      ^{\copy2}%
    \fi
  }%   
  \mathrel{#4}%
}
\makeatother

 \newcommand\mlnode[1]{\fbox{\begin{tabular}{@{}c@{}}#1\end{tabular}}}

\usepackage[nodayofweek]{datetime}
\newcommand{\comment}[1]{}
\newcommand{\ket}[1]{| #1 \rangle}
\newcommand{\bra}[1]{\langle #1 |}
\newcommand{\braket}[2]{\langle #1|#2\rangle}

\newcommand{\proj}[1]{|#1\rangle\!\langle#1|}
\newcommand{\id}{\mathds{1}}
\newcommand{\tr}{{\rm tr}}
\newcommand{\cA}{\mathcal{A}}
\newcommand{\cB}{\mathcal{B}}
\newcommand{\cC}{\mathcal{C}}

\newcommand{\cF}{\mathcal{F}}
\newcommand{\cG}{\mathcal{G}}
\newcommand{\cH}{\mathscr{H}}

\newcommand{\cP}{\mathcal{P}}

\newcommand{\cS}{\mathcal{S}}
\newcommand{\cT}{\mathcal{T}}
\newcommand{\cX}{\mathcal{X}}
\newcommand{\cY}{\mathcal{Y}}
\newcommand{\cZ}{\mathcal{Z}}

\newcommand{\rC}{\mathrm{C}}

\newcommand{\Conv}{\mathrm{Conv}}
\newcommand{\Cone}{\mathrm{Cone}}
\newcommand{\Aff}{\mathrm{Aff}}
\newcommand{\chsh}{\mathrm{CHSH}}
\newcommand{\ot}{\otimes}

\newcommand{\PR}{\mathrm{PR}}
\newcommand{\Cl}{\mathrm{C}}
\newcommand{\BC}{\mathrm{BC}}
\newcommand{\QM}{\mathrm{QM}}
\newcommand{\CG}{\mathrm{CG}}
\newcommand{\RL}{\mathrm{R}}
\newcommand{\E}{\mathrm{E}}
\newcommand{\iso}{\mathrm{iso}}
\newcommand{\mix}{\mathrm{mix}}
\newcommand{\noise}{\mathrm{noise}}

\newcommand{\NL}{\mathrm{NL}}
\newcommand{\Loc}{\mathrm{L}}
\newcommand{\indep}{\upmodels}

\DeclareMathOperator{\Tr}{Tr}

\theoremstyle{plain}
\newtheorem{theorem}{Theorem}[section]
\newtheorem{conjecture}{Conjecture}[section]
\newtheorem{lemma}{Lemma}[section]
\newtheorem{corollary}{Corollary}[section]
\newtheorem{proposition}{Proposition}[section]
\newtheorem{example}{Example}[section]
\newtheorem*{assumption}{Assumption}
\theoremstyle{definition}
\newtheorem{definition}{Definition}[section]
\newtheorem{remark}{Remark}[section]
\declaretheorem[]{axiom}

\usepackage{cleveref}

\newcommand{\footnoteremember}[2]{\footnote{\label{#1}#2}\newcounter{#1}\setcounter{#1}{\value{footnote}}}
\newcommand{\footnoterecall}[1]{\hyperref[#1]{\footnotemark[\value{#1}]}}

\begin{document}

\captionsetup[figure]{margin=1.1cm,font=small,labelfont={bf},name={Figure},labelsep=colon}
\captionsetup[table]{font=small,labelfont={bf},name={Table},labelsep=colon}
%\captionsetup[table]{margin=1.5cm,font=small,labelfont={bf},name={Table},labelsep=colon,textfont={it}}

\frontmatter

\begin{titlepage}
	\centering
	{\Large\bfseries Approaches to causality and multi-agent paradoxes in non-classical theories\par}
	\vspace{3cm}
	{\scshape By\\\vspace{0.5cm}
	Vilasini Venkatesh\par}
	\vspace{4cm}
	{\scshape Doctor of Philosophy\par}
	\vspace{2cm}
	{\scshape University of York \par}
%	\vspace{0.1cm}
	{\scshape Mathematics \par}
	\vspace{2cm}
	{\scshape February 2021\par}	
	\vfill
\end{titlepage}
\clearpage
\begin{center}
    \thispagestyle{empty}
    \vspace*{\fill}
 \Large{\textbf{\textit{Dedication}}}\\\vspace{2mm}\textit{\large{To my mother, Sujatha}}
    \vspace*{\fill}
\end{center}
\clearpage

\newpage
\chapter*{\Large \center Abstract}
\addcontentsline{toc}{chapter}{Abstract}
Causality and logic are both fundamental to our understanding of the universe, but our intuitions about these are challenged by quantum phenomena. This thesis reports progress in the analysis of causality and multi-agent logical paradoxes in quantum and post-quantum theories. Both these research areas are highly relevant for the development of quantum technologies such as quantum cryptography and computing. 

Part I of this thesis focuses on causality. Firstly, we develop techniques for using generalised entropies
to analyse distinctions between classical and non-classical causal structures. We derive new properties of classical and quantum Tsallis entropies of systems that follow from the relevant causal structure, and apply these to obtain new necessary constraints for classicality in the Triangle causal structure. Supplementing the method with the post-selection technique, we provide evidence that Shannon and Tsallis entropic constraints are insufficient for detecting non-classicality in Bell scenarios with non-binary outcomes. This points to the need for better methods of characterising correlations in non-classical causal structures. Secondly, we investigate the relationships between causality and space-time by developing a framework for modelling cyclic and fine-tuned influences in non-classical theories. We derive necessary and sufficient conditions for such causal models to be compatible with a space-time structure and for ruling out operationally detectable causal loops. In particular, this provides an operational framework for analysing post-quantum theories admitting jamming non-local correlations.
  
In Part II of this thesis, we investigate multi-agent logical paradoxes, of which the quantum Frauchiger-Renner paradox has been the only example. We develop a framework for analysing such paradoxes in arbitrary physical theories. Applying this to box world, a post-quantum theory, we derive a stronger paradox that does not rely on post-selection. Our results reveal that reversible, unitary evolution of agents' memories is not necessary for deriving multi-agent logical paradoxes, and suggest that certain forms of contextuality might be.

\definecolor{mylinkcolor}{rgb}{0,0,0}
\newpage
\addcontentsline{toc}{chapter}{Contents}
\setcounter{tocdepth}{1}
\tableofcontents

\newpage
\addcontentsline{toc}{chapter}{List of Tables}
\listoftables

\newpage
\addcontentsline{toc}{chapter}{List of Figures}
\listoffigures

%\newpage
%\chapter*{\Large \center Preface (or Introduction) (if any)}
%\addcontentsline{toc}{chapter}{Preface}

%An introduction to the thesis as a whole - not to be confused with Chapter 1.

\newpage
\chapter*{\Large \center Acknowledgements}
\addcontentsline{toc}{chapter}{Acknowledgements}
First and foremost, I am grateful to my supervisor, Roger Colbeck for his gentle guidance
and deep insights without which this thesis would not be possible. I have greatly benefited both
from his expertise and from his encouragement to develop my own research ideas and to pursue
external collaborations. His ability to express complex concepts in the most effective verbal
and mathematical form have inspired me to strive for perfection in my scientific communication.

I wish to thank the Department of Mathematics for my PhD fellowship and all the timely administrative support. I extend my gratitude to other members of our department, in particular Tony Sudbery and Matt Pusey, for valuable discussions that have enriched my knowledge. I am thankful to Mirjam for several stimulating discussions on causal structures, for sharing snippets of her Mathematica code, and to her and Vicky for their help during my initial days in York. I deeply appreciate the experience I have had with all my peers, in particular Peiyun, Giorgos, Rutvij, Vincenzo, Vasilis, Max and Ali--- the enjoyable conversations, moral support and spontaneous music-making sessions have made my PhD experience very memorable. I express my deep regards to my collaborators at ETH Z\"urich. Firstly to L\'idia del Rio for the numerous exciting exchanges about quantum foundations, career and life advice, and much more. Thanks are also due to Nuriya Nurgalieva, especially for the pretty diagrams on our joint paper. I am grateful to Renato Renner, whose profound ideas and clarity of thought have always inspired me, and to him and Christopher Portmann for truly insightful interactions that have undoubtedly benefited my work. Further, I thank my examiners, Matt Pusey and Stefan Wolf for their useful comments that have enhanced the final version of this thesis.

The unwavering support of my family and friends have been vital to the completion of this thesis. My aunt, Indumathi for several enlightening discussions on abstract mathematical concepts since high school days. Meghana and Avni for being ever-prepared to endure my excited monologues about quantum physics, and being the main contributors to my knowledge of fascinating scientific facts outside of physics. Nemanja for his invaluable support with all things tech and for being a patient audience to several practice talks. Lastly but importantly, my mother, Sujatha for inspiring my academic pursuits at a young age by impressing upon me that knowledge should be sought not merely as a means to an end, but as an end in itself.

\newpage
\chapter*{\Large \center Declarations}
\addcontentsline{toc}{chapter}{Declarations}

\begin{quote}
I declare that this thesis is a presentation of original work and I am the sole author. This work has been carried out under the supervision of Prof. Roger Colbeck and has not previously been presented for an award at this, or any other University. Chapters~\ref{chapter: Tsallispaper} and \ref{chapter: mixingpaper} are primarily based on publications [1] and [2] listed below which were carried out with Prof. Colbeck, while Chapter~\ref{chapter: PRdoxespaper} is based on the publication [3] which is joint work with collaborators from ETH Z\"{u}rich, Switzerland. Chapter~\ref{chapter: jammingpaper}  is based on yet unpublished work [4] carried out with Prof. Colbeck. Further, the Mathematica package [5] was developed for use in publications [1] and [2]. Additional research work carried out or published during the doctoral studies in related topics, but not reported in this thesis are given in [6] and [7]. All other sources are acknowledged and listed in the bibliography.
\end{quote}

\subsection*{List of publications included in this thesis}\renewcommand*\labelenumi{[\theenumi]}
\begin{enumerate}
    \item Vilasini, V. and Colbeck, R.  \emph{Analyzing causal structures using Tsallis entropies.} Physical Review A 100, 062108 (2019).
    \item Vilasini,  V.  and  Colbeck,  R.   \emph{Limitations of entropic inequalities for detecting nonclassicality in the postselected Bell causal structure}. Physical Review Research 2, 033096 (2020). 
    \item Vilasini, V., Nurgalieva, N. and del Rio, L.  \emph{Multi-agent paradoxes beyond quantum theory.} New Journal of Physics 21, 113028 (2019).
  
\end{enumerate}
\subsection*{In preparation and code}
\begin{enumerate}
\setcounter{enumi}{3}
   \item Vilasini,  V.  and  Colbeck,  R. \emph{Cyclic and fine-tuned causal models and compatibility with relativistic principles.} In preparation (2020).
        \item Colbeck,  R. and Vilasini,  V.  \emph{LPAssumptions (A Mathematica package for solving linear programming problems with free parameters, subject to assumptions on these parameters)} (2019). \url{https://github.com/rogercolbeck/LPAssumptions}.
\end{enumerate}
\subsection*{Additional research work not included in this thesis}
\begin{enumerate}
\setcounter{enumi}{5}
\item  Vilasini,  V.,  Portmann,  C.  and  del  Rio,  L.   \emph{Composable  security  in  relativistic  quantum  cryptography.} New Journal of Physics 21, 043057 (2019).  
\item Vilasini,  V.,  del  Rio,  L. and Renner, R. \emph{Causality in definite and indefinite space-times.} In preparation (2020). Extended abstract at: \url{https://wdi.centralesupelec.fr/users/valiron/qplmfps/papers/qs01t3.pdf}
\end{enumerate}
\renewcommand*\labelenumi{\theenumi.}

\definecolor{mylinkcolor}{rgb}{0,0,0.7}
\mainmatter

\chapter{Introduction}
\section{Preface}
\lettrine[nindent=0em, slope=-.5em,lines=2]{Q}{uantum} theory makes several predictions that fly in the face of common intuition and have puzzled its very founders--- energy appearing only in discrete quanta, particles exhibiting wave-like properties such as interference and superposition, entanglement of distant particles, the list goes on. Nevertheless, quantum theory has passed several decades of stringent experimental tests, establishing itself as one of the most successful theories of nature currently available.  Why does nature seem to comply with the predictions of quantum theory? What are the physical principles that uniquely define quantum theory? These questions continue to be actively researched, even after a century since the inception of the theory. The process has repeatedly revealed the inter-dependent and mutually reinforcing relationship between technological progress and advancements in fundamental scientific research--- an understanding of fundamental  principles enables us to harness their potential for technologies, which in turn allow us to probe nature at greater detail. Revolutionary technologies such as the laser, nuclear energy, positron emission tomography (i.e., PET medical imaging) and quantum computing would not exist if not for fundamental explorations into the microscopic regime, fed by human curiosity to comprehend the ways of nature. On the other hand, breakthroughs in high-precision measurements are enabling the observation of quantum effects at newer physical regimes, making it imperative to model the implications of extending quantum theory to larger and more complex systems. 
%Therefore continued research into these foundational questions furthers us on the path to a more complete understanding of the physical universe and also offers tremendous potential for future technologies.

In the modern era of physics, the field of quantum information theory has led to great progress in understanding the foundational aspects of the quantum regime, while also bringing to attention the increased information-processing capabilities of quantum theory. A lot of this progress traces back to a seminal theorem by John Bell in 1964 \cite{Bell} proving the incompatibility of quantum predictions with certain classical models of the physical world. The remarkable fact is that this incompatibility can in principle be witnessed at the level of empirical data such the settings of knobs and values of pointers on measurement devices, by testing the strength of correlations between the data obtained by non-communicating parties. This has made possible cryptographic protocols such as key distribution whose security is guaranteed by the laws of physics \cite{Ekert91,MayersYao,BHK}, as opposed to weighing an adversary's computing power against the difficulty of solving a mathematical problem. It has also enabled the generation and certification of true randomness \cite{Pironio2009, CK, Acin2016}, leading to the development of quantum random number generators that have now become commercially available. In the field of computing, Shor's quantum algorithm \cite{Shor1994} enables prime factorisation in polynomial time, offering a significant speedup over existing classical algorithms, which has major implications for the security of modern day cryptosystems.\footnote{The RSA encryption scheme is based on the complexity of this problem.} This has attracted huge investments from governments as well as companies such as Google, IBM and Microsoft, that are making steady progress towards the development of scalable quantum computers as well as quantum cryptographic primitives. Furthermore, quantum advantages in several communication tasks have also been reported \cite{Vitter1999, Buhrman2010}, and there are global efforts towards establishing quantum communication in space through satellite networks and also developing the quantum internet. 

These technological feats are backed by theoretical research into quantum correlations, quantum networks, relativistic quantum information and quantum circuits, among others. Many of these areas fall in the ambit of the study of causality and causal structures in non-classical theories, of which there are several approaches. For example, reformulating Bell's theorem in terms of the different constraints implied by causal structures on correlations in classical and non-classical theories has revealed possible ways to generalise it to more complex scenarios involving multiple parties and sources of correlations \cite{Wood2015, Chaves2015}. Analysing causal structures beyond the standard CHSH Bell scenario \cite{Clauser1974} are important for building quantum networks for communication and distributed computing. Furthermore, in protocols involving multiple agents distributed over space-time, relativistic principles such as the finite speed of signalling, also play a role in restricting the information processing possibilities and consequently, in the security of relativistic cryptographic protocols \cite{RogerThesis,Vilasini_crypto}. The study of causation in quantum theory has also extended the analysis of quantum properties such as superposition and entanglement from the spatial to the temporal domain, and generalisations of quantum circuits have been introduced  for modelling spatio-temporal correlations \cite{Chiribella2013, Oreshkov2012, Portmann2017}. Such scenarios have not only been physically implemented \cite{Megidish2013, Procopio2015, Rubino2017}, they have also been shown to offer additional information-theoretic advantages in numerous information-processing tasks \cite{Brukner2004, Araujo2014b, Guerin2016, Chiribella2018}. These examples can also be used to simulate thought experiments where the space-time structure itself exhibits quantum properties, arising from quantum gravitational effects \cite{Zych2019}. Therefore a more complete understanding of causation and its relation to space-time structure in quantum theory is essential for harnessing its full information-processing potential and is likely to aid the long sought after unification of quantum theory with general relativity.

The ability to create and manipulate quantum superpositions of larger composite systems is an important consequence of the advancements in quantum networks and computing technologies. While we don't observe quantum effects at macroscopic scales, the possibility of such an observation under careful experimental conditions is not forbidden by any known physical principle. Several theoretical and experimental research groups around the globe have been invested in probing the quantum to classical transition, whether there is a fundamental scale at which such a transition should occur remains a burning open question. Recent work has revealed that extending quantum theory to more complex systems can lead to a conflict with simple and commonly employed principles of logical reasoning \cite{Frauchiger2018, Nurgalieva2018}. Continued research into this domain would hence enable a better characterisation of the structure of logic that apply to larger scale quantum systems, such as quantum computers.  It would also generate insights into whether the principles of quantum theory are universally applicable, or whether they are merely a very good approximation for the microscopic realm offered by a more fundamental theory that we are yet to discover.

This thesis concerns itself with the two areas of quantum information research enunciated above, namely the study of causality and of multi-agent logical paradoxes that can arise when quantum theory is extended to more complex systems. Our analysis of these topics also extends to post-quantum theories, since it is often the case that stepping beyond quantum theory to a more general setting informs an understanding of what makes quantum theory special among the plethora of other physical theories. We summarise the main contributions of this thesis below, and its relevance for the broader open questions described here.

\section{Synopsis}
This thesis divided into two parts and consists of a total of nine chapters. The first part focuses on causality while the second on multi-agent paradoxes in non-classical theories. Chapters~\ref{chapter: Tsallispaper}, \ref{chapter: mixingpaper}, \ref{chapter: jammingpaper} and~\ref{chapter: PRdoxespaper} contain the majority of the original contributions of this thesis which are based on published \cite{Vilasini2019, Vilasini2020, Vilasini_PRdoxes} as well as unpublished work\footnote{Vilasini,  V.  and  Colbeck,  R. \emph{Cyclic and fine-tuned causal models and compatibility with relativistic principles.} In preparation (2020).}. Leaving out the present chapter, of which the current section is the last, we summarise the main contents and/or contributions of the remaining eight chapters below.

\paragraph{Chapter 2: Preliminaries} Here we introduce the background concepts and mathematical tools that will be important for the rest of this thesis. An expert reader may choose to skip over parts or the whole of this chapter. We first cover the basics of polyhedral geometry and computation, followed by an overview of information-theoretic entropic measures and their properties. Both these topics are central to the results of Chapters~\ref{chapter: Tsallispaper} and~\ref{chapter: mixingpaper}. We then discuss a class of post-quantum theories, namely generalised probabilistic theories, before moving on to the preliminary concepts about causal structures in classical, quantum and generalised probabilistic theories. The former will be relevant for both parts of the thesis while the latter will predominantly be employed in Part I. We conclude this chapter with a review of epistemic modal logic, a concept that will feature in Part II of the thesis.

\subsection*{Part I: Approaches to causality in non-classical theories}
Causality is a fundamental part of our perception of the universe. A precise understanding of cause-effect relationships is integral to the scientific method, having found applications in drug trials, economic predictions, machine learning as well as biological and physical systems. Quantum mechanics, a theory that has been immensely successful in explaining microscopic physics, strongly challenges our classical intuitions about causation, deeming classical models inadequate for describing quantum phenomena. Apart from this fundamental implication, a plethora of information-theoretic applications arise from the study of causal structures, that benefit from a quantum over classical advantage. Extending this study to more general, post-quantum theories facilitates this understanding by highlighting the aspects of causality that might be special to quantum theory. Part I of this thesis is dedicated to the analysis of causal structures, both acyclic and cyclic, and using both entropic and non-entropic techniques. 

%following an overview of existing techniques in Chapter~\ref{chap: techniques}, we develop several new techniques for analysing causal structures in quantum and post-quantum theories. We first consider the problem of certifying the non-classicality of a causal structure from the observed correlations in Chapters~\ref{chapter: Tsallispaper} and~\ref{chapter: mixingpaper}, combining both entropic and non-entropic techniques and employing a number of different entropy measures. We then construct a framework for analysing cyclic causal structures in non-classical theories, and characterise how they can be embedded in an underlying space-time in Chapter~\ref{chapter: jammingpaper}. 

\paragraph{Chapter 3: Overview of techniques for analysing causal structures} Here, we provide a more specific overview of the techniques for certifying the non-classicality of causal structures, that will be employed in Chapters~\ref{chapter: Tsallispaper} and~\ref{chapter: mixingpaper}. We first discuss the techniques employed when the observed correlations in a causal structure are represented in probability space, focusing on the geometry of the correlation sets in the bipartite Bell causal structure. Following this, we present an overview of methods to certify non-classicality in entropy space. This includes a summary of the entropy vector method and the post-selection technique which is often employed with this method. This chapter is based on the introductory sections of our published papers \cite{Vilasini2019, Vilasini2020}.

\paragraph{Chapter 4: Entropic analysis of causal structures without post-selection} In this chapter, we develop new techniques for using generalised entropy measures to analyse causal structures, particularly Tsallis entropies. Employing the Shannon entropy for this purpose is known to have certain limitations \cite{Weilenmann16, Weilenmann2018} and the use of generalised entropies have not been previously considered. Our first contribution is a new set of constraints on the Tsallis entropies that are implied by (conditional) independence between classical random variables, which we  apply to causal structures.\footnote{Additionally, we also generalise these constraints to quantum Tsallis entropies in certain cases, under a suitable notion of quantum conditional independence. But this is not central to the remaining results.} Finding that the standard entropic method becomes computationally intractable even for small causal structures, we propose a way to circumvent this issue and obtain a partial solution, namely a new set of Tsallis entropic necessary constraints for classicality in the Triangle causal structure. Tsallis entropies have found important applications in other areas of information theory and in the field of non-extensive statistical physics, hence the properties of Tsallis entropies derived here can be useful even beyond the analysis of causal structures. Further, we also find that R\'enyi entropies pose significant limitations for certifying the non-classicality of causal structures. This chapter is based on our published work \cite{Vilasini2019}.

\paragraph{Chapter 5: Entropic analysis of causal structures with post-selection} We investigate whether the entropic method combined with the post-selection technique allows non-classicality to be detected whenever it is present. We focus on the bipartite Bell causal structure, where the answer is known to be positive for the case of binary valued measurement outcomes \cite{Chaves2012}. Developing an analogue of the technique used in \cite{Chaves2012} to the scenario where there are three outcomes and two inputs per party, we identify two families of non-classical distributions here, namely those whose non-classicality can/cannot be detected through this technique. We then investigate extensions of the technique by allowing the observed correlations to be post-processed according to a natural class of non-classicality non-generating operations prior to testing them entropically. We provide numerical as well as analytical evidence that indicate that even under a natural class of post-processings, entropic inequalities for either the Shannon or Tsallis entropies are generally not sufficient to detect non-classicality in the Bell causal structure. Our work provides insights into some of the advantages and the limitations of the entropic technique, and is the first to report drawbacks of the technique in post-selected causal structures. This chapter is based almost entirely on published material \cite{Vilasini2020}. 
%In addition, we also characterise the vertex representation of the polytope of non-signaling distributions in the Bell scenario with two inputs and three outputs per party, that satisfy all the CHSH-type inequalities, a result which could be of independent interest.

\paragraph{Chapter 6: Cyclic and fine-tuned causal models and compatibility with relativistic principles} In this chapter, we investigate the relationships between causation and space-time structure. We develop an operational framework for modelling cyclic and fine-tuned causal influences in non-classical theories, and characterise their compatibility with an underlying space-time structure. Distinguishing between causal loops based on the ability to operationally detect them, we derive necessary and sufficient conditions for ruling out operationally detectable loops in these causal models, and for the model to not signal superluminally with respect to a space-time structure. This provides a mathematical framework for analysing generalisations of the so-called ``jamming'' scenario \cite{Grunhaus1996, Horodecki2019} whereby a party superluminally influences correlations between other space-like separated parties. We prove, through an explicit protocol that these scenarios can lead to superluminal signalling (without additional assumptions), contrary to the original claim that they are compatible with relativistic principles. Furthermore, we analyse a claim made in \cite{Horodecki2019} that a set of no-signaling conditions are necessary and sufficient for ruling out causal loops in multipartite Bell scenarios. Our results also identify missing assumptions in these claims and generalise them to arbitrary causal structures.  Finally, we present ideas for building a more complete framework for causally modelling these general scenarios, which is likely to be of broader relevance. This work is yet unpublished.

\subsection*{Part II: Multi-agent paradoxes in non-classical theories}
%In everyday life and in research alike, we often make common-sensical deductions such as ``if I know a and I know that a implies b, then I know b'', irrespective of the nature of a and b. At several instances, we must also reason about the knowledge of others to reach certain conclusions about our world, this involves identifying a trustworthy source S of information and making inferences such as ``I know that S knows a, and I trust the methods by which S arrived at a, therefore I know a''. 
Logical deduction is another rudimentary aspect of our understanding of the world around us, which we begin to employ since early childhood. In recent years, it has been shown that quantum theory, when extended to the level of the observer, challenges some of the commonly employed rules of logical reasoning. More precisely, when observers model each other's memories (where they store their measurement outcomes) as quantum systems and use simple logical principles to reason about each other's knowledge, they can arrive at a deterministic contradiction \cite{Frauchiger2018, Nurgalieva2018}. This multi-agent logical paradox proposed by Frauchiger and Renner has generated a surge of scientific interest in understanding the role of observers as quantum systems, and the objectivity of their physical observations. An observer here need not necessarily be a conscious entity, but one capable of implementing quantum measurements and performing simple deductions based on the outcomes--- something that a relatively small quantum computer can be programmed to do (in principle). Experimental efforts towards the developing scalable quantum computers could make possible a physical implementation of such scenarios in the near future, making it imperative to develop a better theoretical understanding thereof. Part II of this thesis focuses on multi-agent paradoxes, in and beyond quantum theory, for we must step outside the quantum formalism to better comprehend its peculiarities. 
%In Chapter~\ref{chapter: FR}, we present a simplified version of the Frauchiger-Renner thought experiment and paradox. In Chapter~\ref{chapter: PRdoxespaper} we develop a framework to study multi-agent paradoxes beyond quantum theory, motivated by the question: Which physical theories lead to a contradiction between simple reasoning principles and modelling observers' memories as physical systems?

\paragraph{Chapter 7: Multi-agent paradoxes in quantum theory} In this chapter, we present an entanglement-based version of the Frauchiger-Renner thought experiment \cite{Frauchiger2018} and explain its seemingly paradoxical result. We briefly analyse the relationship of this to other results with similar implications for the objectivity of agents' observations. We conclude with a discussion about the implications of this result for the interpretations of quantum theory.

\paragraph{Chapter 8: Multi-agent paradoxes beyond quantum theory} We generalize assumptions involved of the Frauchiger-Renner result so that they can be applied to arbitrary physical theories. This provides an operational framework for modelling agents as systems of a general physical theory, which we apply for modelling how observers' memories may evolve in box world, a particular post-quantum, probabilistic theory. We use this to find a deterministic contradiction in the case where agents share a PR box, a bipartite box world system. Our version of the paradox in box world is stronger than the quantum one of Frauchiger and Renner, in the sense that it does not rely on post-selection. It also reveals that reversibility of the memory update, akin to quantum unitarity is not necessary for witnessing such paradoxes, suggesting that certain forms of contextuality are to be held responsible. Obtaining an inconsistency in the framework of generalised probabilistic theories broadens the landscape of theories which are affected by the application of classical rules of reasoning to physical agents, and enables a deeper understanding of the features of quantum theory that lead to an incompatibility with classical logical structures. This chapter is based on our published paper \cite{Vilasini_PRdoxes}.

\paragraph{Chapter 9: Conclusions and oulook} We present concluding remarks on the contributions made in both parts of this thesis, and their potential scope for addressing immediate as well as broader open problems in the foundations of physics and information theory.

\chapter{Preliminaries}
\lettrine[nindent=0em, slope=-.5em,lines=2]{I}{n}  this chapter, we review the necessary background and main mathematical tools that will be employed in this thesis. We begin with an overview of convex geometry, optimization and polyhedral computations in Section~\ref{sec: convex}. In Section~\ref{sec: entropies}, we present the information-theoretic entropy measures that are relevant to this thesis, namely Shannon, von-Neumann, Tsallis and R\'enyi entropies, and outline some of their useful properties. We then review the framework of generalised probabilistic theories (GPTs) in Section~\ref{sec: GPT}, mainly following the work of Barrett \cite{Barrett07}. Section~\ref{sec: causalstr} summarises the basics of classical and non-classical causal models which will be primarily based on the classical Bayesian networks approach of Judea Pearl \cite{Pearl2009} and the framework of generalised causal structures developed by Henson, Lal and Pusey \cite{Henson2014} that apply to quantum and GPT causal structures. These concepts form the backbone of Part I of this thesis which concerns causality in classical and non-classical theories. In Section~\ref{sec: logic} we summarise the main structures and axioms underlying the modal logic framework. This, along with the framework for GPTs mentioned above form the main prerequisites for Part II of this thesis, which concerns multi-agent logical paradoxes in quantum and GPT settings. We first present a brief overview of notational conventions used in this thesis.

\section{Overview of notational conventions}

Throughout this thesis (unless specified otherwise), we will employ the following notational conventions.

\paragraph{Random variables,  probabilities, entropies: } We will use capital letters (typically from the second half of the English alphabets) to denote random variables or RVs for short e.g., $X$, $Y$, $Z$. We will also use the same label to denote the set of possible values taken by the random variable, the corresponding small letter to denote a particular value of the random variable and $|.|$ to denote cardinality of sets. For example, the random variable $X$ takes values $x\in X$, and there are $|X|$ possible values that $X$ can take. We will only consider finite and discrete random variables throughout this thesis. For a probability distribution over a set of random variables $\{X_1,...,X_n\}$, we will use $P_{X_1,...,X_n}\in \mathscr{P}_n$, where $\mathscr{P}_n$ denotes the set of all probability distributions over $n$ discrete random variables. Whenever it is necessary to mention the specific values of the random variables, we will abbreviate $P_{X_1,...,X_n}(X_1=x_1,...,X_n=x_n)$ to $P(X_1=x_1,...,X_n=x_n)$ or simply $P(x_1,...,x_n)$. When an equation contains only distributions labelled by variables denoted in capital letters, it should be interpreted as being satisfied for every value of those variables (denoted by the corresponding small letter). For example $P_{XY}=P_XP_Y$ should be read as $P(xy)=P(x)P(y)$ $\forall x\in X$, $y\in Y$, where $P(xy)$ is short for $P_{XY}(X=x,Y=y)$. We will use the short form $XY$ or sometimes $X,Y$ to denote the union of two variables/sets of variables. Further, we will use $H(XY...)$ (or sometimes, $H(X,Y,...)$) to denote joint entropies of sets of random variables $X$, $Y$, .... 
\paragraph{Quantum formalism: } Given a Hilbert space $\mathscr{H}$, we will use  $\mathscr{P}(\mathscr{H})$ to represent the set of positive, semi-definite operators on $\mathscr{H}$, and $\mathscr{S}(\mathscr{H})$ to denote the set of positive semi-definite and trace one operators on $\mathscr{H}$. Quantum states will, in general be represented by density operators, i.e., $\rho\in \mathscr{S}(\mathscr{H})$. Hermitian conjugate will be denoted using the usual dagger notation, $\rho^{\dagger}$ and matrix transpose by the subscript $T$, $\rho^T$. Rank 1 density operators i.e., pure states will be denoted as elements of the Hilbert space $\mathscr{H}$, in the conventional bra-ket notation $\ket{\psi}\in \mathscr{H}$ and $\ket{\psi}^{\dagger}=\bra{\psi}\in\mathscr{H}^*$, where $\mathscr{H}^*$ is the dual space of $\mathscr{H}$. Quantum channels or transformations on quantum states correspond to \emph{completely positive trace-preserving} (CPTP) maps $\mathscr{E}:\mathscr{S}(\mathscr{H}_A)\mapsto \mathscr{S}(\mathscr{H}_B)$ that map an input state $\rho_A\in \mathscr{S}(\mathscr{H}_A)$ to an output state $\rho_B=\mathscr{E}(\rho_A)\in \mathscr{S}(\mathscr{H}_B)$. Quantum measurements will be described by \emph{positive operator valued measures} (POVMs): a POVM is a set $\{E^x\}_{x\in X}$ with $E^x\in \mathscr{P}(\mathscr{H})$, labelled by the classical values $x$, and summing to identity $\sum_{x\in X}E^x=\mathds{1}$. We will often abbreviate $\{E^x\}_{x\in X}$ to $\{E^X\}_X$. For the purposes of this thesis, we will only consider a discrete and finite set of measurement outcomes, so $X$ corresponds to a discrete and finite random variable taking values $x\in X$.

\paragraph{Causal structures and space-time diagrams: } Everywhere in the thesis, except for Chapter~\ref{chapter: jammingpaper}, we will use the regular arrows $\longrightarrow$ to denote causal influence (from cause to effect). In Chapter~\ref{chapter: jammingpaper}, we will use $\longrsquigarrow$ for the same purpose because we would like to categorise these into solid arrows $\longrightarrow$ and dashed arrows $\xdashrightarrow{}$ based on how these causal influences are detected operationally. A vast majority of the causal structures appearing in the rest of the thesis correspond to the $\longrightarrow$ case, which justifies this notation. All space-time diagrams will have space along the x-axis and time along the y-axis.
 
\paragraph{Logical operators and others: } We will use the standard logical operators i.e., $\neg$ for ‘not’, $\land$ for ‘and’, $\lor$ for ‘or’, $\Rightarrow$ for ‘if...then’, and $\Leftrightarrow$ for ‘equivalent.’ Further, $\oplus$ will be used to denote modulo-2 addition of classical bits i.e., the bitwise OR. Logarithms will use base $e$ i.e., we will consider natural logarithms and denote it by $\ln$.

\section{Convex geometry and polyhedral computations}
\label{sec: convex}
 The lecture notes \cite{Fukuda2016} provide a thorough introduction to polyhedral computation, while \cite{Boyd2004, Grotschel2012, Martin1999} are good resources for convex geometry and their application in optimisation problems. In this section, we give an overview of the main aspects of these topics that will be relevant to the current thesis and will focus our attention on the vector spaces $V=\mathbb{R}^n$ over the ordered field of reals $\mathbb{R}$.

\subsection{The geometry of convex polyhedra}
The following special types of linear combinations are central to the concepts of polyhedral theory and convex optimization.

\begin{definition}[Affine, conic and convex combinations]
Consider a set of points $x_1,...,x_m\in V$ and their linear combination $\sum_{i=1}^m\alpha_ix_i$ with real coefficients $\alpha_i$. Then,
\begin{enumerate}
    \item $\sum_{i=1}^m\alpha_ix_i$ is an \emph{affine combination} if $\sum_{i=1}^m\alpha_i=1$.
      \item $\sum_{i=1}^m\alpha_ix_i$ is a \emph{conic combination} if $\alpha_i\geq 0$ $\forall i$.
        \item $\sum_{i=1}^m\alpha_ix_i$ is a \emph{convex combination} if it is both affine and conic i.e.,  $\sum_{i=1}^m\alpha_i=1$ and $\alpha_i\geq 0$ $\forall i$.
\end{enumerate}
\end{definition}

The affine, conic and convex hull of a set of points are then naturally defined with respect to the corresponding type of linear combination.

\begin{definition}[Affine, conic and convex hull]
The affine hull of a set of points $S\subseteq V$, is the set of all affine combinations of
points in $S$ and is denoted as $\Aff(S)$. Similarly the conic and convex hull of $S$ correspond to the sets of all conic and convex combinations of $S$ and are denoted by $\Cone(S)$ and $\Conv(S)$ respectively.
\end{definition}

Analogous to the concept of linear independence, a set of points $S$ is said to be \emph{affinely independent} if and only if none of the points in $S$ can be expressed as an affine combination of the remaining points in $S$. A set $S\subseteq V$ that is closed under affine combinations is called an \emph{affine subspace} of $V=\mathbb{R}^n$. 

We now proceed to discuss the central objects of polyhedral theory namely, convex sets, convex cones, polyhedra and polytopes, based on the concepts defined above. 
\begin{definition}[Convex set]
\label{definition: convset}
A subset $\mathcal{S}\subseteq V$ is said to be \emph{convex} if for any two points $x_1, x_2 \in \mathcal{S}$, the line segment $[x_1,x_2]:=\{x:x_1+(1-\alpha)x_2,0\leq \alpha \leq 1$\} is contained in $\mathcal{S}$.
\end{definition}

%\begin{definition}[Convex hull]
%\label{definition: convhull}
%The \emph{convex hull} of a set of points $S=\{x_1,...,x_m\}\subseteq V$ is defined as
%\begin{equation}
%    \text{conv}(S):=\{\sum_{i=1}^m\alpha_ix_i:\alpha_i\geq 0, \forall i \text{ and } \sum_{i=1}^m\alpha_i\}
%\end{equation}
%\end{definition}
 By applying Definition~\ref{definition: convset} inductively to the set $S$, it can be shown that a set is convex if and only if it contains every convex combination of its points. Convex sets of particular interest to us are \emph{convex cones}, and a further subset of those, \emph{polyhedral cones}.

\begin{definition}[Convex cone]
A set $\cC\subseteq V$ is called a \emph{cone} if for every $x\in \cC$, and $\alpha\geq 0$, $\alpha x\in \cC$. $\cC$ is a \emph{convex cone} if it is convex and a cone i.e., for any $x_1, x_2\in\cC$, and $\alpha_1, \alpha_2\geq 0$,
\begin{equation}
    \alpha_1x_1+\alpha_2x_2\in \cC
\end{equation}
\end{definition}
Note that a set is a convex cone if and only if it contains every conic combination of its points. 

\begin{definition}[Polyhedral cone]
A set $\mathcal{P}\subseteq V$ is said to be a \emph{polyhedral cone} if it is a \emph{polyhedron} and a \emph{cone}, where a polyhedron $\mathcal{P}$ is a subset of $V$ that can be expressed as the solution set of a finite number of linear inequalities,\footnote{Note that the equalities $Cx=d$ in Equation~\eqref{eq: polycone} can be equivalently written in terms of inequalities as $Cx\geq d$ and $Cx\leq d$, but we will often write these separately (as is common in the convex optimization literature).}
\begin{equation}
\label{eq: polycone}
  \mathcal{P}=\{x\in V:Ax\leq b \text{ and } Cx=d\}, \quad A\in \mathbb{R}^{m\times n}, b\in \mathbb{R}^m, C\in \mathbb{R}^{k\times n}, d\in \mathbb{R}^k.
\end{equation}
\end{definition}

By construction, a polyhedron corresponds to an intersection of \emph{half-spaces} $\{x\in V|Ax\leq b\}$ and \emph{hyperplanes} $\{x\in V|Cx= d\}$, and is convex. This is known as the $\mathcal{H}$-representation (where $\mathcal{H}$ stands for half-space) of a polyhedron, and bounded polyhedra are called \emph{polytopes}.\footnote{We will adopt these definitions in the rest of thesis, noting that in the literature, the opposite convention for defining polyhedra and polytopes is sometimes used.} Since a cone, by definition contains the origin cone, $\mathbf{0}\in V=\mathbb{R}^n$, polyhedral cones can be expressed in the more concise form $\mathcal{P}=\{x\in V:Ax\leq 0\}$. Polyhedra (and hence polytopes) can also be represented through a $\mathcal{V}$-representation (where $\mathcal{V}$ stands for vertex)\footnote{Strictly speaking, unbounded polyhedra have extremal rays rather than extremal points or vertices (defined later in the section). Nevertheless, the representations in terms of extremal rays (in the unbounded case) as well as in terms of vertices (bounded case) are known as the $\mathcal{V}$-representation. } which follows from an important result in polyhedral theory, the Minkowski-Weyl Theorem.

\begin{theorem}[Minkowski-Weyl Theorem]
\label{theorem: minkweyl}
For $\mathcal{P}\subseteq V$, the following statements are equivalent
\begin{enumerate}
    \item $\mathcal{P}$ is a polyhedron,
    \item $\mathcal{P}$ is finitely generated, i.e., there exist finite sets $S,T \subset V$ such that
    \begin{equation}
        \mathcal{P}=\Conv(S)+\Cone(T), 
        \end{equation}
        where addition of sets is defined with respect to the Minkowski sum, i.e., $S_1+S_2:=\{s_1+s_2:s_1\in S_1, s_2\in S_2\}$ for $S_1,S_2\subseteq V$.
\end{enumerate}
\end{theorem}
 The second statement defines the $\mathcal{V}$-representation of $\mathcal{P}$. In the case of a polyhedral cone given by $\mathcal{P}=\{x\in V:Ax\leq 0\}$ in the $\mathcal{H}$-representation, $S\subset V$ can be taken to be the empty set in the corresponding $\mathcal{V}$-representation. On the other hand, a polytope (in its $\mathcal{V}$-representation) can be written as the convex hull of a finite set of points, and $T$ can be taken to be the empty set in this case. A \emph{simplex} is a polytope that can be written as the convex hull of a finite set of \emph{affinely independent} points. 
 
The geometry of a polyhedron $\mathcal{P}\subseteq V$ is characterised by its \emph{faces} which are defined in terms of valid linear inequalities of $\mathcal{P}$. A linear inequality is valid for $\mathcal{P}$ if it holds for all $x\in \mathcal{P}$.

\begin{definition}[Faces of a polyhedron]
For a polyhedron $\mathcal{P}\subseteq V$, a subset $\mathcal{F}\subseteq \mathcal{P}$ is called a is called a face of $\mathcal{P}$ if it is represented as
\begin{equation}
    \mathcal{F}=\mathcal{P}\cap \{x:c^Tx=d\},
\end{equation}
for some valid inequality $c^T x \leq d$.
\end{definition}

All faces of a polyhedron are by construction, polyhedrons themselves. Faces of dimensions 0, 1 and $dim(\mathcal{P})-1$ are called \emph{vertices}, \emph{edges} and \emph{facets} respectively. It can be shown that vertices of a polyhedron are equivalent to its \emph{extreme points} which are points in $\mathcal{P}$ that cannot be written as a convex combination of other points in $\mathcal{P}$. Hence will use these terms interchangeably. Further, when an edge of a polyhedron is unbounded, it can either be a line (unbounded in both directions or half-line (starting from a vertex and unbounded in one direction). In the latter case, the edge is called an \emph{extremal ray}.\footnote{Extremal rays can also be defined as a subset $R$ of $\mathcal{P}$ that cannot be expressed as a (non-trivial) conic combination of points not belonging to $R$.} In the $\mathcal{V}$-representation of a polyhedron given by statement 2 of Theorem~\ref{theorem: minkweyl}, it is often convenient to take the sets $S$ and $T$ to correspond to the set of \emph{extremal points} and \emph{extremal rays} respectively.

\subsection{Projections of polyhedra: Fourier-Motzkin Elimination}
\label{ssec: FME}

Fourier-Motzkin Elimination is a method for projecting higher dimensional polyhedra to lower dimensional ones through variable elimination, and forms an important part of the computational methods employed in this thesis. Here, we describe the mathematical concepts underpinning this algorithm. For this, we start with linear transformations of which the transformations of interest, projections are a subset. 
\begin{definition}[Linear transformation]
A \emph{linear transformation} is a map $f:\mathbb{R}^n\to \mathbb{R}^m$ that acts as $f: x\mapsto Ax$, where $A\in \mathbb{R}^{m\times n}$.
\end{definition}

It can be immediately shown that linear transformations preserve convexity (since convex combinations are particular cases of linear combinations). Additionally, linear transformation  also map convex cones to convex cones and polyhedral cones to polyhedral cones. Projections $\pi$ are linear transformations that are idempotent i.e., $\pi.\pi=\pi$. In particular, we will consider orthogonal projections which are projections on a Hilbert space that satisfy $\langle \pi(u),v\rangle=\langle u,\pi(v)\rangle$ for the Hilbert space inner-product $\langle,\rangle$ between any vectors $u$ and $v$ in the space. In our case this Hilbert space is simply $\mathbb{R}^n$ and $\langle,\rangle$ is the corresponding dot product, and in the remainder of the thesis, projections must be taken to mean orthogonal projections.

For a polyhedron $\mathcal{P}\subseteq V$ expressed as $\mathcal{P}=\{(x,y)\in \mathbb{R}^{n_1}\times \mathbb{R}^{n_2}: Ax+By\leq b\}$, with $n_1+n_2=n$, the projection of $\mathcal{P}$ into the subspace of the $y$ variables $\pi_y(\mathcal{P}):V\mapsto V$ is defined as
\begin{equation}
\label{eq: projmain}
 \pi_y(\mathcal{P}):=\{(\mathbf{0},y):y\in \mathbb{R}^{n_2}, \text{ and } \exists x\in \mathbb{R}^{n_1} \text{ such that } (x,y)\in \mathcal{P}\}.  
\end{equation}

$\pi_y$ is a linear transformation that can be represented by a $n\times n$ matrix with an $n_2\times n_2$ identity matrix as its second block, and zeroes everywhere else.\footnote{Here, the projection $\pi$ is defined as an endomorphism on $V$ i.e., a map from $V$ to itself. However, it can be equivalently seen as a map from $V$ to a lower dimensional space by dropping additional zeroes, i.e., replacing $(\mathbf{0},y)\in \mathbb{R}^n$ with $y\in \mathbb{R}^{n_2}$ in the Equation~\eqref{eq: projmain}.} In the $\mathcal{V}$ representation, a projection $\pi$ transforms the convex/conic hull of a set of points $\{x_1,...,x_m\}$ to the convex/conic hull of the points $\{\pi(x_1),...,\pi(x_m)\}$. Given a polyhedron $\mathcal{P}$ in the $\mathcal{H}$-representation $\{x\in V: Ax\leq b\}$ as input, the Fourier-Motzkin Elimination (FME) procedure outputs the $\mathcal{H}$-representation, or the inequalities describing the projected polyhedron $\pi_y(\mathcal{P})$. The projection $\pi_y$ is implemented through FME by eliminating the first $n_1$ variables in the system of inequalities defining $\mathcal{P}$. Taking $x$ to represent the $n$-dimensional vector with components $x_i$, the following steps detail the Fourier-Motzkin Elimination procedure for eliminating the variable $x_1$. This procedure can then be iterated $n_1$ times to eliminate all required variables.
\begin{enumerate}
    \item The inequalities $Ax=b$ are partitioned into three sets $I_+$, $I_-$ and $I_0$. $I_+$ denotes the set of all inequalities in $Ax\leq b$ where the variable $x_1$ has a strictly positive coefficient, i.e., all inequalities where $A_{i1}>0$. Similarly, $I_-$ is the set of all inequalities where $x_1$ has a strictly negative coefficient, and $I_0$ is the set of all inequalities where $x_1$ does not appear i.e., its coefficient is 0.
    \item If $I_+$ is empty, then the inequalities in $I_-$ are ignored, and vice-versa. If both $I_+$ and $I_-$ are non-empty, then the following steps are undertaken. Every inequality $A_ix\leq b_i$ in $I_+$ is rearranged and expressed as
    \begin{equation}
    \label{eq: FME1}
     x_1\leq \frac{1}{A_{i1}}(b_i-\sum_{k\neq 1}A_{ik}x_k):=f_i(x_2,...,x_n).  
    \end{equation}
    There are $|I_+|$ such inequalities. Similarly, every inequality $A_jx\leq b_j$ in $I_-$ is expressed as (noting that $A_{j1}<0$ in this case)
     \begin{equation}
     \label{eq: FME2}
     x_1\geq \frac{1}{A_{j1}}(b_j-\sum_{l\neq 1}A_{jl}x_l):=g_j(x_2,...,x_n),  
    \end{equation}
    and there are $|I_-|$ inequalities of this type. Equations~\eqref{eq: FME1} and \eqref{eq: FME2} are then combined to give the following set of $|I_+|.|I_-|$ inequalities that are independent of the variable $x_1$,
    \begin{equation}
        \label{eq: FME3}
        g_j(x_2,...,x_n)\leq f_i(x_2,...,x_n).
    \end{equation}
    \item The union of the $|I_+|.|I_-|$ inequalities of Equation~\eqref{eq: FME3} with those of $I_0$ characterise the final output set of inequalities corresponding to the projection into the space of the $n-1$ variables $x_2,...,x_n$.
\end{enumerate}
Note that the final set of inequalities obtained in the FME procedure is often not minimal, in the sense that it can contain several redundancies. This can however be checked efficiently through a linear program, and a characterisation of the projected polyhedron in terms of fewer, non-redundant inequalities can be obtained (as described in Section~\ref{sssec: LP} and Figure~\ref{fig:simplexalgo}).

\begin{figure}[t!]
	\centering
	\subfloat[]{\includegraphics[width=.4\textwidth]{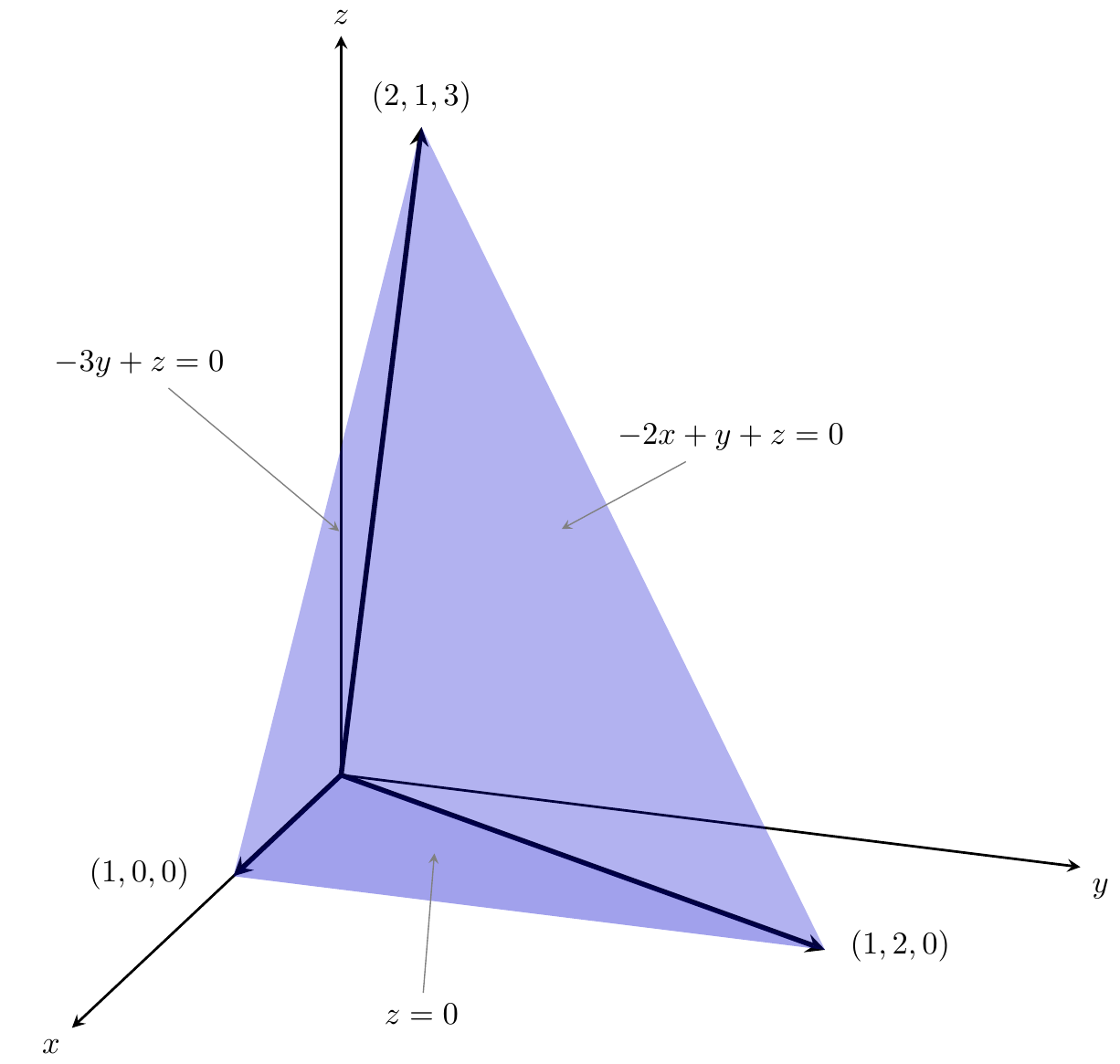}\label{fig: proja}}\quad
\subfloat[]{\includegraphics[width=.3\textwidth]{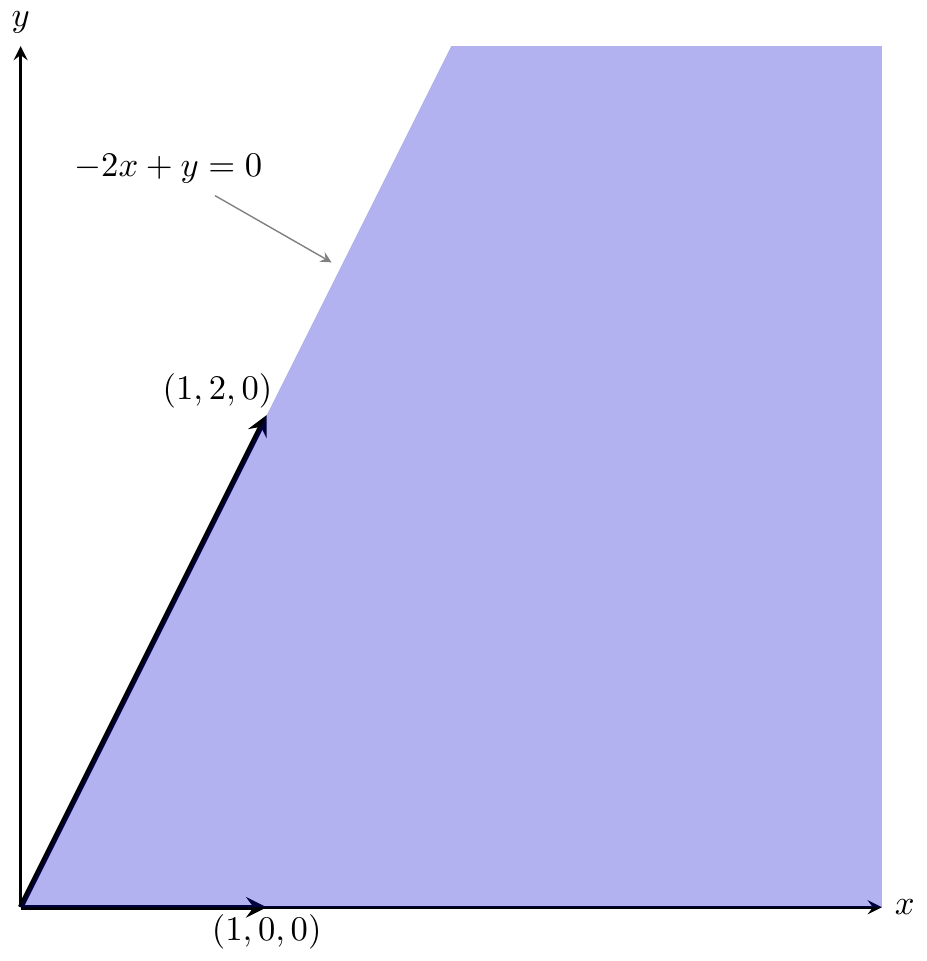}\label{fig: projb}}
\\
\subfloat[]{\includegraphics[width=.3\textwidth]{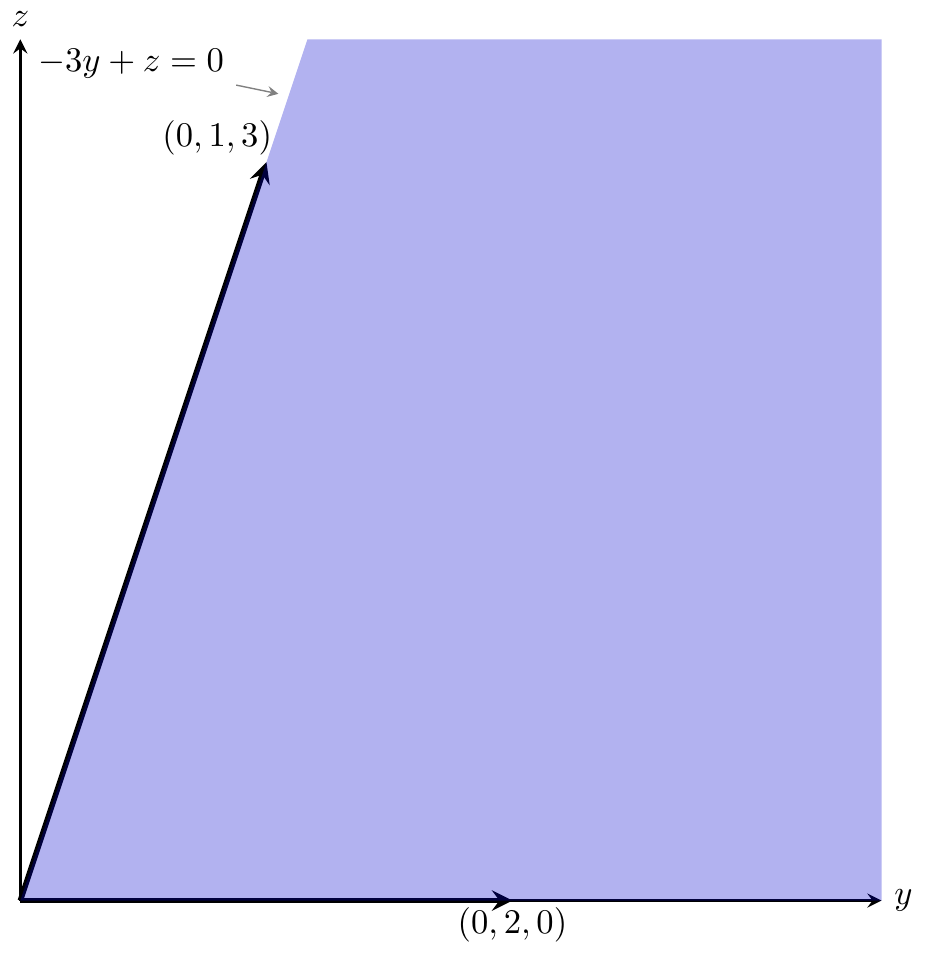}\label{fig: projc}}\qquad\qquad\quad
\subfloat[]{\includegraphics[width=.3\textwidth]{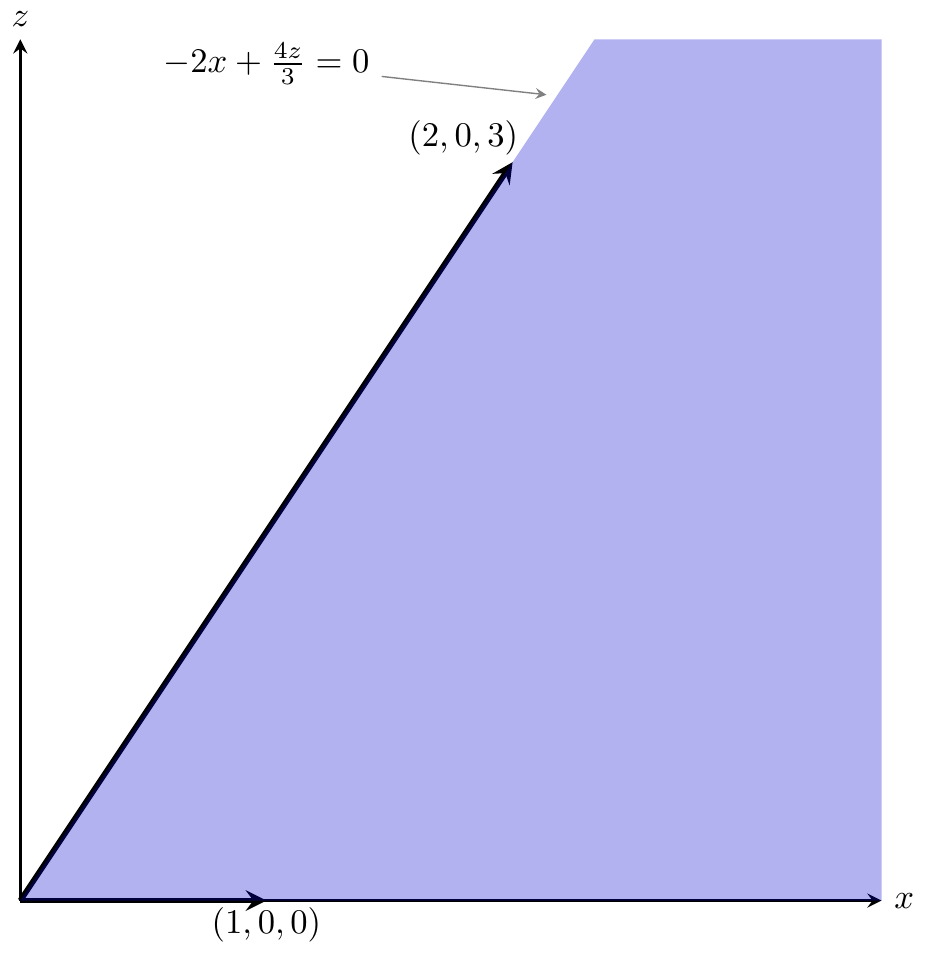}\label{fig: projd}}
	\caption[Projections of a polyhedral cone]{\textbf{Projections of a polyhedral cone (Example~\ref{example: FME}): }(a) illustrates a polyhedral cone ($\mathcal{P}$), with 3 faces ($\mathcal{H}$-representation) and 3 extremal rays ($\mathcal{V}$-representation) as indicated. Projections of $\mathcal{P}$ on to the $xy$, $yz$ and $xz$-planes are also polehedra, as shown in (b), (c) and (d) respectively. In all four subfigures, the corresponding polyhedron (blue region) is unbounded, and extends to infinity along the direction of the extremal rays indicated.}
	\label{fig: projection}
	\end{figure}

\begin{example}
\label{example: FME}
Consider the polyhedral cone $\mathcal{P}\in V$ of Figure~\ref{fig: proja} which, in the $\mathcal{H}$-representation is given by
\begin{equation}
\label{eq: egFME1}
    \mathcal{P}=\left\{\left. \begin{pmatrix}
    x \\ y \\ z 
  \end{pmatrix}
 \in \mathbb{R}^3\right\vert \begin{pmatrix}
    -2&1&1 \\ 0&-3&1 \\ 0&0&-1 
  \end{pmatrix}\cdot \begin{pmatrix}
    x \\ y \\ z 
  \end{pmatrix}\leq\begin{pmatrix}
    0 \\ 0 \\ 0 
  \end{pmatrix} \right\}
\end{equation}
The same polyhedron in expressed in the $\mathcal{V}$-representation as
\begin{equation}
    \mathcal{P}=\Cone\left(\begin{pmatrix}
    1 \\ 0 \\ 0 
  \end{pmatrix},\begin{pmatrix}
    1 \\ 2 \\ 0 
  \end{pmatrix},\begin{pmatrix}
    2 \\ 1 \\ 3 
  \end{pmatrix}\right)
\end{equation}
The projection matrices in the 3 coordinate planes are,
\begin{equation}
    \pi_{xy}=\begin{pmatrix}
    1&0&0 \\ 0&1&0 \\ 0 &0&0
  \end{pmatrix},\quad  \pi_{yz}=\begin{pmatrix}
    0&0&0 \\ 0&1&0 \\ 0 &0&1
  \end{pmatrix},\quad
    \pi_{xz}=\begin{pmatrix}
    1&0&0 \\ 0&0&0 \\ 0 &0&1
  \end{pmatrix},\quad
\end{equation}
Then the projections of $\mathcal{P}$ onto each of the coordinate planes correspond to the polyhedra $\mathcal{P}_{xy}$ (Figure~\ref{fig: projb}), $\mathcal{P}_{yz}$ (Figure~\ref{fig: projc}) and $\mathcal{P}_{xz}$ (Figure~\ref{fig: projd}) expressed in the $\mathcal{V}$-representation as
\begin{equation}
\label{eq: egFME2}
      \mathcal{P}_{xy}=\Cone\left(\begin{pmatrix}
    1 \\ 0 \\ 0 
  \end{pmatrix},\begin{pmatrix}
    1 \\ 2 \\ 0 
  \end{pmatrix},\begin{pmatrix}
    2 \\ 1 \\ 0
  \end{pmatrix}\right),\quad    \mathcal{P}_{yz}=\Cone\left(\begin{pmatrix}
    0 \\ 0 \\ 0 
  \end{pmatrix},\begin{pmatrix}
    0 \\ 2 \\ 0 
  \end{pmatrix},\begin{pmatrix}
    0 \\ 1 \\ 3 
  \end{pmatrix}\right)\quad     \mathcal{P}_{xz}=\Cone\left(\begin{pmatrix}
    1 \\ 0 \\ 0 
  \end{pmatrix},\begin{pmatrix}
    1 \\ 0 \\ 0 
  \end{pmatrix},\begin{pmatrix}
    2 \\ 0 \\ 3 
  \end{pmatrix}\right)
\end{equation}
The same projections can be obtained by performing Fourier-Motzkin Elimination on the $\mathcal{H}$-representation of $\mathcal{P}$ (Equation~\eqref{eq: egFME1}) as shown below.
\begin{enumerate}
    \item For the projections $\pi_{xy}$, $\pi_{yz}$, $\pi_{xz}$, we wish to eliminate the variables $z$, $x$ and $y$ respectively. Denoting the variable to be eliminated by a superscript, we have
    \begin{center}
             \begin{tabular}{ccc}
          $ I^z_+=\{-2x+y+z\leq 0, -3y+z\leq 0\}$,   & $I^z_-=\{-z\leq 0\}$, & $I^z_0=\emptyset$\\
       $I^x_+=\emptyset$,& $I^x_-=\{-2x+y+z\leq 0\}$,& $I^x_0=\{-3y+z\leq 0,-z\leq 0\}$\\
         $I^y_+=\{-2x+y+z\leq 0\}$,& $I^y_-=\{-3y+z\leq 0\}$,& $I^y_0=\{-z\leq 0\}$
        \end{tabular}   
    \end{center}
\item Since $I_+^x$ is empty, we can ignore $I_-^x$ and only need $I_0^x$ to describe $\mathcal{P}_{yz}$. $I^{y}_{+}$, $I^{y}_{-}$, $I^{z}_{+}$ and $I^{z}_{-}$ are all non-empty, so we rearrange the corresponding inequalities as $I^z_+=\{z\leq 2x-y,z\leq 3y\}$, $I^z_-=\{z\geq 0\}$ and $I^y_+=\{y\leq 2x-z\}$, $I^y_-=\{y\geq \frac{z}{3}\}$. Combining the inequalities in each case we have $2x-y\geq 0,3y\geq 0$ for the projection $\pi_{xy}$, $3y-z\geq 0$, $z\geq 0$ for the projection $\pi_{yz}$ and $2x-4\frac{z}{3}\geq 0$, $z\geq 0$ for the projection $\pi_{xz}$.
\item We obtain the $\mathcal{H}$-representation of the projected polytopes as
\begin{equation}
\begin{split}
      \mathcal{P}_{xy}&=\left\{\left. \begin{pmatrix}
    x \\ y  
  \end{pmatrix}
 \in \mathbb{R}^2\right\vert \begin{pmatrix}
    -2&1 \\ 0&-3 
  \end{pmatrix}\cdot \begin{pmatrix}
    x \\ y  
  \end{pmatrix}\leq\begin{pmatrix}
    0 \\ 0  
  \end{pmatrix} \right\}\\
        \mathcal{P}_{yz}&=\left\{\left. \begin{pmatrix}
    y \\ z  
  \end{pmatrix}
 \in \mathbb{R}^2\right\vert \begin{pmatrix}
    -3&1 \\ 0&-1 
  \end{pmatrix}\cdot \begin{pmatrix}
    y \\ z 
  \end{pmatrix}\leq\begin{pmatrix}
    0 \\ 0  
  \end{pmatrix} \right\}\\
   \mathcal{P}_{xz}&=\left\{\left. \begin{pmatrix}
    x \\ z  
  \end{pmatrix}
 \in \mathbb{R}^2\right\vert \begin{pmatrix}
    -2&\frac{4}{3} \\ 0&-1 
  \end{pmatrix}\cdot \begin{pmatrix}
    x \\ z 
  \end{pmatrix}\leq\begin{pmatrix}
    0 \\ 0  
  \end{pmatrix} \right\}
  \end{split}
\end{equation}

One can then see that the $\mathcal{V}$-representation of the polyhedra are indeed those found in Equation~\eqref{eq: egFME2}, as illustrated in Figure~\ref{fig: projection}.
\end{enumerate}
\end{example}

\paragraph{Algorithmic complexity of FME: } The FME algorithm can be computationally costly due to its high algorithmic complexity, which is, in the worst case double exponential. For an initial system of $n_i$ inequalities, a maximum of $(\frac{n_i}{2})^2$ inequalities can be obtained after one elimination step.\footnote{This corresponds to the case where $|I_+|=|I_-|=\frac{n_i}{2}$ and $|I_0|=0$ for the variable being eliminated.} After $k$ elimination steps, the algorithm can produce a maximum of $4(\frac{n_i}{4})^{2^k}$ inequalities, which scales double-exponentially in the number of iterations $k$. This makes the algorithm quite inefficient in general. However, the method remains useful in several cases for which the scaling is far from the worst-case scaling. It can also be marginally improved by removing some of the redundancies using the so-called Chernikov rules \cite{Chernikov1960,Chernikov1965}.  For the work presented in this thesis, the algorithm was mainly implemented on the {\sc porta} software \cite{porta} that takes into account some of the Chernikov rules. 

\subsection{Convex optimization}

\subsubsection{Linear programs}
\label{sssec: LP}
A \emph{linear program} (LP) consists of optimising a linear function subject to linear equality/inequality constraints. Every linear program can be expressed as a \emph{primal} and a \emph{dual} problem. For $x,y\in \mathbb{R}^n$, $A\in \mathbb{R}^{m\times n}$, $c\in \mathbb{R}^n$ and $b\in \mathbb{R}^m$ we have,

\begin{minipage}{0.5\textwidth}
\begin{center}
 \textbf{Primal program}
\begin{align*}
\begin{split}
      \text{Minimise }\qquad &c^Tx\\
      \text{Subject to }\qquad &Ax\geq b, x\geq 0
\end{split}
\end{align*}   
\end{center}
\end{minipage}\begin{minipage}{0.5\textwidth}
 \begin{center}
     \textbf{Dual program}
\begin{align*}
\begin{split}
      \text{Maximise } \qquad &b^Ty\\
      \text{Subject to } \qquad &A^Ty\leq c, y\geq 0
\end{split}
\end{align*}  
 \end{center}
 \end{minipage} 
 \par
A vector satisfying all the constraints of the primal/dual program is said to be a \emph{feasible solution} of that program. Hence the sets of feasible solutions for the primal and the dual programs are $\{x\in V:Ax\geq b, x\geq 0\}$ and  $\{y\in V:A^Ty\leq c, y\geq 0\}$ respectively. An \emph{optimal solution} for the primal (/dual) program is a feasible solution of that program that attains the smallest (/largest) value of the objective function $c^Tx$ (/$b^Ty$). A linear program is called \emph{feasible} if the corresponding feasible set of solutions is non-empty and it is said to be \emph{unbounded} if the objective function is not bounded above in the case of maximization or not bounded below in the case of minimization problems. Note that the geometry of the feasible region of a linear program is a polyhedron, and hence convex. In the rest of this thesis, we will refer to this as the \emph{standard form} of the LP\footnote{This might differ from other texts but we choose this convention for convenience, since this is also the format used by Mathematica, the main computational software used in this thesis.}. 

\paragraph{Duality: } Using the definitions of the primal and dual programs, we can see that $c^Tx\geq (A^Ty)^Tx=y^TAx\geq y^Tb=b^Ty$. In other words, for any feasible solutions $x$ and $y$ of the primal (minimization) problem and the dual (maximization) problem, $c^Tx\geq b^Ty$ holds. This property is called \emph{weak duality}. Another property which is not immediately evident, but nevertheless true is that of \emph{strong duality}. It states that if the primal and dual problems are feasible, then there exist a pair of feasible solutions $x^*$ and $y^*$ of these problems such that $c^Tx^*=b^Ty^*$. By weak duality, these solutions will be optimal. 

\paragraph{Removing redundant constraints using LPs: } Linear programs can be efficiently used to check whether a given system of linear inequalities contains any redundant inequalities that can be dropped without affecting the solution set.  In general, to check whether the $i^{th}$ inequality in $Ax\geq b$ is redundant, one can minimise $A_ix$ subject to the remaining inequalities. If the optimal feasible solution is strictly smaller than $b_i$, then the inequality $A_ix\geq b_i$ is not redundant. Otherwise, the inequality is redundant and can be ignored. An example is illustrated in Figure~\ref{fig:simplexalgo}. Iterating this procedure for all inequalities in the system, we can obtain a minimal set of non-redundant inequalities that are equivalent to the original system. For the work presented in the current thesis, this method was used to remove redundant inequalities in the output of the Fourier-Motzkin elimination procedure. Reducing inequalities at the output of each iteration also helps in speeding up the computational procedure of FME. This was primarily implemented on Mathematica using the inbuilt {\sf LinearProgramming} function. 

The {\sf LinearProgramming} function in Mathematica can only handle linear programs where the objective vector $c$, constraint matrix $A$ and the constraint vector $b$ do not contain unspecified variables. For example, we might wish to optimize $x_1+ax_2$, where $a$ is an unspecified constant in the range $0\leq a\leq 1$, returning the solution for all values of $a$ in the range. For solving linear programs involving such additional unknown variables (other than those being optimised over), subject to assumptions on them, we developed a Mathematica Package, {\sc LPAssumptions} \cite{LPAssumptions} that was used to produce some of the main computational results of Chapters~\ref{chapter: Tsallispaper} and \ref{chapter: mixingpaper}. Our package is based on the two-phase simplex method for solving linear programming problems. 

\subsubsection{The Simplex method for solving linear programs}
\label{ssec:simplexalgo}
The simplex algorithm is widely used for solving linear programming problems and was originally proposed by George Dantzig \cite{Dantzig1963}. The algorithm is fairly involved, consisting of several steps. Our Mathematica Package \cite{LPAssumptions} (see Appendix~\ref{appendix: LPA} for details) implements this algorithm, however the full details are not required for understanding the numerical results produced using this package. Hence, we provide only a brief overview of the main mathematical concepts underpinning this algorithm and refer the reader to \cite{Dantzig1990, Martin1999, Grotschel2012} for further details of the algorithm, and other methods for solving linear programs.

As we have previously noted, the feasible region of a linear program is a polyhedron, and by Theorem~\ref{theorem: minkweyl}, every polyhedron can be represented as a Minkowski sum of two sets: the former a convex hull of a finite set of vertices, and the latter a conic hull of a finite set of extremal rays. An extreme point or vertex of the polyhedron defining the feasible region of a linear program is called a \emph{basic feasible solution} (BFS). The following theorem (Theorem 3.3 of \cite{Katta1983}) is at the core of the simplex method.

\begin{theorem}[\cite{Katta1983} Theorem 3.3]
\label{theorem:simplex1}
Let $\mathcal{P} \subset V$ be polyhedron with at least one extreme point. Then every linear program
with the feasible solution set  $\{x\in \mathcal{P} \}$, is either unbounded or attains its optimal
value at an extreme point of $\mathcal{P} $.
\end{theorem}
 
Further, if a given extreme point $x_e$ of $\mathcal{P}$ is not an optimal solution, then it can be shown that there exists an edge of $\mathcal{P}$ containing $x_e$ such that the objective function strictly improves (i.e., increases in case of a maximization LP and decreases in case of a minimization LP) as we move away from $x_e$ along that edge (Section 3.8 of \cite{Katta1983}). Finite edges connect extreme points to extreme points and moving along this direction will lead to a new extremal point $x'_e$ with a better value of the objective function. If the identified edge emanating from $x_e$ is unbounded, then the linear program has no finite solution. The simplex method is based on this intuition--- starting with an initial vertex of the feasible set $\mathcal{P}$ i.e., an initial basic feasible solution, one moves along edges of $\mathcal{P}$ to other vertices that improve the objective function value, until an optimal vertex is reached or it is revealed that the problem is unbounded.

\begin{figure}[t!]
    \centering
    \includegraphics[scale=1.5]{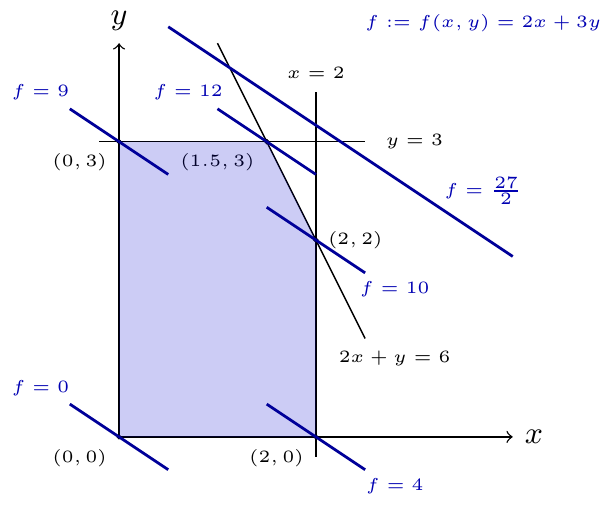}
    \caption[Intuition behind the Simplex algorithm]{\textbf{Intuition behind the Simplex algorithm: } This figure illustrates the simple linear program of Example~\ref{example:simplexalgo}, where the feasible region corresponds to a polytope (shaded in blue). The thick blue lines correspond to different level curves $f(x,y)=c$ of the objective function $f$. The maximum value of $c$ for which $(x,y)$ belongs to the feasible region is our required solution. In the present example, the maximum value of $f=12$ is attained at the vertex $(1.5,3)$ (in accordance with Theorem~\ref{theorem:simplex1}).  Starting at any other vertex, there is always an edge (in this 2D case, edges are also the facets) along which the value of the $f$ increased, and following the path specified by such an edge at each subsequent vertex, we can always reach the optimal vertex. This figure also illustrates how redundant inequalities can be removed using linear programming (explained in Section~\ref{sssec: LP}). Suppose that we have the 5 inequalities of Example~\ref{example:simplexalgo}, expressed as in the standard form of an LP, along with a sixth inequality $-2x-3y\geq -\frac{27}{2}$. From the figure, it is evident that the sixth inequality is not required to characterise the blue polytope, and can be dropped. This is checked by minimising $-2x-3y$ subject to the other inequalities, which is the LP of Example~\ref{example:simplexalgo}. Since the minimum value $-12\geq -\frac{27}{2}$, we conclude that this inequality is redundant. On the other hand, $-2x-3y\geq -9$  along with the other 5 inequalities would define a smaller polytope and would hence not be redundant.}
    \label{fig:simplexalgo}
\end{figure}

\begin{example}[Intuition behind the Simplex algorithm]
\label{example:simplexalgo}
Consider the following simple linear program (left), expressed in the standard form (right).

\begin{minipage}{0.5\textwidth}
\begin{align*}
\begin{split}
      \text{Maximise }\qquad &2x+3y\\
      \text{Subject to }\qquad &2x+y\leq 6,\\ &x\leq 2,\\&y\leq 3,\\
      &x,y\geq 0
\end{split}
\end{align*}
\end{minipage}
\begin{minipage}{0.5\textwidth}
\begin{align*}
\begin{split}
      \text{Minimise }\qquad &(-2,-3)^T\cdot (x,y)\\
      \text{Subject to }\qquad &\begin{pmatrix}
      -2 & -1\\
      -1 & 0\\
      0& -1\\
      \end{pmatrix}\cdot
      \begin{pmatrix}
      x\\
      y
      \end{pmatrix}\geq 
        \begin{pmatrix}
      -6\\
      -2\\
      -3
      \end{pmatrix},\\
      &x,y\geq 0
\end{split}
\end{align*}
\end{minipage}

 As shown in Figure~\ref{fig:simplexalgo}, the feasible region is a polytope (bounded polyhedron) in 2 dimensions with 5 vertices, $\mathcal{V}=\{(0,0),(0,3),(\frac{3}{2},3),(2,2),(2,0)\}$ and 5 facets $\mathcal{F}=\{x=0,y=0,y=3,2x+y=6,x=2\}$. From the figure, it is evident that the maximum must be attained at a vertex since the objective function $f(x,y)=2x+3y$ is linear. Calculating the value of the objective function $f$ at all 5 vertices, we immediately see that the maximum value $f=12$ is attained at the vertex $(\frac{3}{2},3)$. That this is indeed the maximum value of the constrained optimization problem, and that it is attained at the said point, can also be verified by using the method of Lagrange multipliers, for example.
\end{example}

\paragraph{Phases: } The simplex algorithm proceeds in two phases. Phase I concerns the problem of finding an initial \emph{basic feasible solution}. This is done by formulating a new linear program $\overline{LP}$ based on the original $LP$ for which an initial BFS $\overline{x}$ is readily found. The construction of $\overline{LP}$ is such that applying the simplex method to $\overline{x}$ either yields an optimum $x$ that would be an initial BFS for the original problem $LP$, or reveals that $LP$ is infeasible. If feasible, Phase II uses the initial BFS $x$ obtained in Phase I to find an optimum solution (if a finite solution exists) of the original problem $LP$ through the simplex method.

\paragraph{Column geometry and efficiency: }  
It can be shown that a linear programming problem with $m$ constraints and $n$ variables has at most $\binom{m}{n}=\frac{m!}{n!(m-n)!}$ basic feasible solutions when $m\geq n$.
Hence a brute-force method for finding the optimum solution would be quite inefficient in general. The merit of the simplex method is derived from its ability to find the optimum much more efficiently (relative to a brute-force method) by exploiting the geometry of convex sets. This consists of first expressing the original LP involving inequality constraints in an equivalent form involving only equality constraints, by introducing certain additional \emph{slack variables}. Once the LP is bought to the desired equivalent form, the geometry defined by the column vectors of the new constraint matrix provide a way to systematically compute a better BFS at each step or to decide whether the LP is unbounded.\footnote{Here, the columns of the constraint matrix provide a basis, an appropriate linear combination of which gives the required BFS. Geometrically, the basis vectors in each iteration correspond to the vertices of a simplex and the associated BFS lies in this simplex i.e., belongs to the convex hull of the vertices represented by those basis vectors. The geometry of the simplex provides a minimal, affinely independent basis for representing the basic solution at each step.} The improved BFS in every iteration is found by computing its so-called \emph{reduced cost} that indicates how much the objective function value will be improved by moving to that BFS. Hence the method manages to reach the optimum solution (if it exists) in much fewer iterations on an average, as compared to brute-force and other graph-traversal methods \cite{Dantzig1990, Martin1999}. It was originally found that for many linear programs admitting an optimal solution, the simplex method found it in $O(n)$ iterations \cite{Dantzig1963, Dantzig1990}, which was the main reason for the success of the algorithm. However, over the years, a number of examples have surfaced where the number of iterations of the simplex method scales exponentially in the number of variables. In fact, determining the number of iterations needed for solving a given linear program, is known to be an NP-hard problem. Nevertheless, the simplex method continues to be useful in several practical cases (which includes the work presented in this thesis).

\subsection{Polyhedral representation conversion}

The task of converting from the $\mathcal{H}$ to the $\mathcal{V}$ representation of a convex polyhedron is called vertex enumeration while the reverse is called as facet enumeration. Representation conversion problems are in general known to be computationally hard--- there is no known vertex/facet enumeration algorithm that is polynomial in the input and output size and in the dimension of a polyhedron, and in the case of unbounded polyhedra, the problem is known to be NP hard \cite{Khachiyan2008}. In general, it is difficult to characterise the size of the output of such problems given the size of the input. For example, a hypercube in $n$-dimensions has $2n$ facets and $2^n$ vertices. The output of the vertex enumeration in this case would scale exponentially in the input while the output of the facet enumeration would be quite small relative to the input size. The computational complexity of vertex/facet enumeration problems continues to be a subject of current research and several algorithms have been proposed for the same.

The most basic algorithm for facet enumeration involves a direct application of Fourier-Motzkin Elimination, which is not always very efficient (as seen in Section~\ref{ssec: FME}). There are also better methods, such as the double description method, that use similar steps as FME, while additionally exploiting properties of polyhedral representations (such as the Minkowski-Weyl Theorem~\ref{theorem: minkweyl}) to achieve better efficiency. Broadly, there are two main classes of polyhedral representation conversion algorithms, namely incremental and graph traversal algorithms. The double description method mentioned above corresponds to an incremental algorithm. In the case of polytopes, one method for implementing the latter is through reversing the procedure of the simplex method (discussed in Section~\ref{ssec:simplexalgo}), whereby starting at a vertex that optimises a suitably chosen objective function, all other vertices are systematically enumerated until every vertex has been visited at least once \cite{Fukuda2016}.

\begin{figure}[t!]
    \centering
    \includegraphics[scale=0.4]{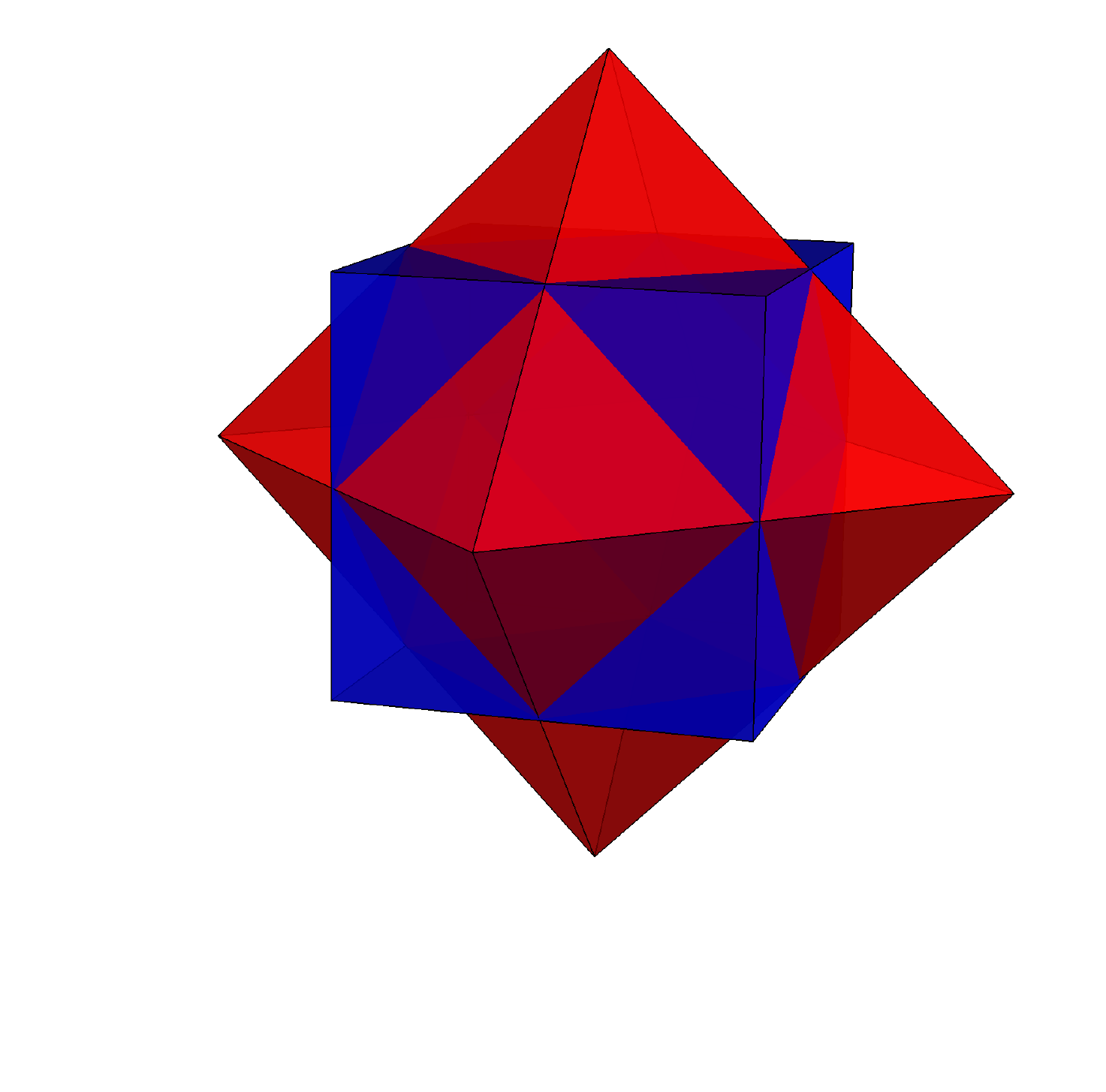}
    \caption[Dual of a polytope]{\textbf{A polytope (blue cube) and its dual (red octahedron): } The dual of a polyhedron has the elegant property that there is a one-to-one correspondence between vertices of the original polyhedron and facets of the dual, and between facets of the original and vertices of the dual. Using this concept, any algorithm for facet enumeration also yields an algorithm for vertex enumeration. The notion of a polar dual that is required for this inter-conversion is defined formally in the main text. Note that this figure is only for the purpose of illustrating the concept of a dual, and does not exactly correspond to the polar dual.}
    \label{fig:dualpolytope}
\end{figure}

Softwares such as {\sc porta} \cite{porta}, on which the main computational work of this thesis has been carried out,  employ FME-based incremental algorithms. We present an overview of a simple FME-based approach for enumerating facets of polyhedral cones, without going into the details of more efficient incremental or graph-traversal algorithms. The concept of a \emph{polar dual} of a polyhedron, which essentially interchanges the role of the vertices and faces of a polyhedron  (see Figure~\ref{fig:dualpolytope}) is then used such that we also get vertex enumeration for free, given an algorithm for facet enumeration. 

\paragraph{Facet enumeration: }
Consider a polyhedron $\mathcal{P}\subseteq V$ expressed as $\mathcal{P}=\Conv(x_1,...,x_k)+\Cone(y_1,...,y_l)$ in the $\mathcal{V}$-representation. To convert to the $\mathcal{H}$-representation of $\mathcal{P}$,
\begin{enumerate}
    \item Express the convex and conic hulls in terms of inequalities as follows
    \begin{align}
        \begin{split}
            x&=\alpha_1x_1+...+\alpha_kx_k+\lambda_1y_1+...+\lambda_ly_l,\\
       1&=\alpha_1+...+\alpha_k,\\
         &\alpha_1,...,\alpha_k,\lambda_1,...,\lambda_l\geq 0   
        \end{split}
    \end{align}
    \item The variables in the above system of equations are $x,\alpha_1,...,\alpha_k,\lambda_1,...,\lambda_l$. Project out the variables $\alpha_1,...,\alpha_k,\lambda_1,...,\lambda_l$ using Fourier-Mozkin Elimination (See Section~\ref{ssec: FME}) to obtain linear inequalities only involving components of $x$, which corresponds to the required $\mathcal{H}$-representation.
\end{enumerate}

In order to convert from the $\mathcal{H}$ to $\mathcal{V}$-representation, we require the concept of the polar or polar dual of a polyhedron. 

\begin{definition}[Polar of a polyhedron]
The polar $\mathcal{P}^{\Delta}$ of a polyhedron $\mathcal{P}\subseteq V$ is defined as
\begin{equation}
    \mathcal{P}^{\Delta}=\{c\in V :c^Tx\leq 1,\forall x \in \mathcal{P}\}.
\end{equation}
\end{definition}
 The following theorem (Theorem 9.1 of  \cite{Schrijver1986}) elucidates some elegant properties of the polar $\mathcal{P}^{\Delta}$ and relates its $\mathcal{H}$/$\mathcal{V}$ representations to those of the original polytope $\mathcal{P}$. 

 \begin{theorem}[Properties of the polar]
 \label{theorem: polar}
 Let $\mathcal{P}\subseteq V$ be a polyhedron that contains the origin. Then,
 \begin{enumerate}
     \item $\mathcal{P}^{\Delta}$ is a polyhedron and $\mathcal{P}^{\Delta\Delta}=\mathcal{P}$,
     \item If $\mathcal{P} = \Conv(x_1,...,x_k)+\Cone(y_1,...,y_l)\subseteq V$, then $\mathcal{P}^{\Delta}=\{u\in V :x_i^Tu\leq 1,\text{ } \forall i=1,...,k$ and $y_j^Tu\leq 0, \text{ }\forall j=1,...,l\}$,
     \item If $\mathcal{P} = \{u\in V :x_i^Tu\leq 1,\text{ } \forall i=1,...,k, \text{ and } y_j^Tu\leq 0,\text{ }\forall j=1,...,l\}$, then $\mathcal{P}^{\Delta} = \Conv(x_1,...,x_k)+\Cone(y_1,...,y_l)$.
 \end{enumerate}
 \end{theorem}

 \paragraph{Vertex enumeration: }
 Consider a polyhedron $\mathcal{P}\subseteq V$ that contains the origin in its interior. If a polyhedron does not contain the origin, then by an appropriate coordinate transformation, it can be transformed into one that does. This allows $\mathcal{P}$ to be expressed in the $\mathcal{H}$-representation as $\mathcal{P}=\{x\in V:Ax\leq b\}$, with $b_i\geq 0$ $\forall i$. To convert to the $\mathcal{V}$-representation of $\mathcal{P}$,
 \begin{enumerate}
  \item For all inequalities $A_ix\leq b_i$ such that $b_i> 0$, divide throughout by $b_i$ to express in the form $\tilde{A}_ix\leq 1$. Without loss of generality, let these be the first $k$ inequalities. Then $\mathcal{P}=\{x\in V: \tilde{A}_ix\leq 1, \text{} \forall i=1,...,k \text{ and } A_ix\leq 0,\text{} \forall i=k+1,...,m \}$, where $m$ is the total number of inequalities.
     \item Use the third statement of Theorem~\ref{theorem: polar} to find the $\mathcal{V}$-representation of the polar $\mathcal{P}^{\Delta}$ as $\mathcal{P}^{\Delta}=\Conv(\tilde{A}_{1}^T,...,\tilde{A}_k^T)+\Cone(A_{k+1}^T,...,A_m^T)$. 
     \item Convert $\mathcal{P}^{\Delta}$ to its $\mathcal{H}$-representation using facet enumeration.
     \item Use the first an third statements of Theorem~\ref{theorem: polar} to find the $\mathcal{V}$-representation of $\mathcal{P}=(\mathcal{P}^{\Delta})^{\Delta}$ using the $\mathcal{H}$-representation of $\mathcal{P}^{\Delta}$ obtained in the previous step.
 \end{enumerate}

\section{Information-theoretic entropy measures and their properties}
\label{sec: entropies}

Entropies have played a crucial role in the study of thermodynamic phenomena and statistical mechanics since the 19th century \cite{Clausius1850, Boltzmann1866, Gibbs1902}, and in 1932, John von Neumann generalised the concept of entropies to quantum systems \cite{vonNeumann1932}. An information-theoretic understanding of entropy on the other hand, was first developed by Claude Shannon in his seminal 1948 paper \cite{Shannon1948}, where he introduced the Shannon entropy as a measure of the information content of a message and used it to characterise the information capacity of communication channels. Interestingly, the von Neumann entropy, even though chronologically prior, turns out to be a quantum generalisation of the Shannon entropy.\footnote{In fact, Shannon had corresponded with von Neumann, who is said to have suggested that Shannon call his new uncertainty measure ``entropy'', due to its similarities with prior notions of thermodynamic and statistical entropies.}
Since then, entropies have become an integral part of both classical and quantum information theory. Several other information-theoretic entropy measures have also been proposed (such as R\'enyi \cite{Renyi1961}, Tsallis \cite{Tsallis1988}, min and max entropies \cite{Renner2006}), and serve as useful mathematical tools for tackling a plethora of information-theoretic problems such as data compression \cite{Shannon1948, Preskill2016}, randomness extraction \cite{Bennett1995, Konig2005}, key distribution \cite{Renner2006, Arnon-Friedman2018} as well as causal structures \cite{Chaves2012, FritzChaves2013, Wajs15, Weilenmann2018}. Here, we provide a brief overview of some of the classical and quantum information-theoretic entropy measures and their properties that are relevant for the work presented in this thesis.

\subsection{Shannon and von Neumann entropies}
\label{ssec: ShanEnt}
\begin{definition}[Shannon entropy]
Given a random variable $X$ distributed according to the discrete probability
distribution $P_X$, the \emph{Shannon entropy} of $X$ is given by\footnote{Note that it is common to
  take logarithms in base 2 and measure entropy in \emph{bits}; here
  we use base $e$ corresponding to measuring entropy in \emph{nats}.} 
\begin{equation}
    H(X)=-\sum_x P(x)\ln P(x).
\end{equation}
\end{definition}

The Shannon entropy $H(X)$ is a continuous and strictly concave function of the probability distribution over $X$ i.e., for two probability distributions $P$ and $Q$ over $X$ and $\lambda\in [0,1]$, denoting the dependence on the distribution by a subscript,
\begin{equation}
    H(X)_{\lambda P+(1-\lambda)Q}\geq \lambda H(X)_P+(1-\lambda)H(X)_{Q}.
\end{equation}

\begin{definition}[Conditional Shannon entropy]
Given two random variables $X$ and $Y$, distributed according to
$P_{XY}$, the \emph{conditional Shannon entropy} is defined by
\begin{equation}
\label{eq: shancond}
    H(X|Y)=-\sum_{x,y}P(xy)\ln \frac{P(xy)}{P(y)}
\end{equation}
\end{definition}

It is also useful to define a quantity called the \emph{mutual information}, which, as the name suggests, quantifies the amount of information that two random variables carry about each other. The conditional version of this quantity quantifies how much information two variables share about each other, given the value of a third variable. 

\begin{definition}[Shannon mutual information]
\label{definition: ShanMI}
Given 3 random variables, $X$, $Y$ and $Z$, the \emph{Shannon mutual information} and \emph{Shannon conditional mutual information} between $X$ and $Y$, and $X$ and $Y$ conditioned on $Z$ are respectively given by 
\begin{equation}
\label{eq: ShanMI}
\begin{split}
    I(X:Y)&=H(X)-H(X|Y),\\
    I(X:Y|Z)&=H(X|Z)-H(X|YZ).
  \end{split}  
\end{equation}
\end{definition}

\paragraph{Properties of Shannon entropy:} Some useful information-theoretic properties satisfied by the Shannon entropy and associated mutual information are discussed below. Some of these properties can be expressed in different, but equivalent forms. We will distinguish between these because equivalences that apply in the case of the Shannon entropy may not necessarily apply for other entropies. Whenever we have an entropic equation/inequality where no conditional entropies are explicitly involved (such as the left most ones in all of the properties below), we will refer to it as the \emph{unconditional form} of that property, otherwise it will be called the \emph{conditional form}. According to the defining of mutual information (Equation~\eqref{eq: ShanMI}) adopted in this thesis, the forms of properties expressed in terms of the mutual informations qualify as conditional forms as well. In the following, $R$, $S$ and $T$ are three disjoint sets of random variables.

\begin{enumerate}
\item \textbf{Non-negativity and upper bound (UB): } The Shannon entropy is positive i.e., $H(R)\geq 0$ and upper bounded as $H(R)\leq \ln|R|$, where $|.|$ denotes cardinality. Equality for the upper bound is achieved if and only
    if $P_R(r)=1/|R|$ for all possible values $r\in R$ (i.e., if the distribution on $R$ is
    uniform).
\item \textbf{Chain rule (CR): } For disjoint sets $R_1$,...,$R_n$, $S$ of random variables,  
    \begin{equation}
        \label{eq: chainrule}
           H(R_1,R_2,\ldots,R_n|S)=\sum_{i=1}^n H(R_i|R_{i-1},\ldots,R_1,S),
    \end{equation}
which in particular implies the two useful identities $H(R|S)=H(RS)-H(S)$ and $H(R|ST)=H(RS|T)-H(S|T)$.
\item \textbf{Additivity (A): } If $R$ and $S$ are independent i.e., if $P_{RS}=P_RP_S$, then
\begin{equation}
\label{eq: additivity}
       H(RS)=H(R)+H(S)\quad\xLeftrightarrow[]{CR}  \quad H(R|S)=H(R)
\end{equation}
    \item \textbf{Monotonicity (M): } From the definition~\eqref{eq: shancond} of the conditional Shannon entropy, it follows that 
    \begin{equation}
    \label{eq: monotonicity}
             H(S)\leq H(RS)
             \quad\xLeftrightarrow[]{CR}  \quad  H(R|S)\geq 0
    \end{equation}
    \item \textbf{Subadditivity (SA): }
      \begin{equation}
    \label{eq: SA}
          H(R)+H(S)\geq H(RS) \quad\xLeftrightarrow[]{CR}  \quad  H(R)\geq H(R|S) \quad\xLeftrightarrow[]{\eqref{eq: ShanMI}}  \quad  I(R:S)\geq 0 
    \end{equation}
  
    \item \textbf{Strong subadditivity (SSA): } 
          \begin{equation}
    \label{eq: SSA}
         H(RT)+H(ST)\geq H(RST)+H(T)
         \quad\xLeftrightarrow[]{CR}  \quad   H(R|T)\geq H(R|ST) \quad\xLeftrightarrow[]{\eqref{eq: ShanMI}}  \quad I(R:S|T)\geq 0 
    \end{equation}
\end{enumerate}
Note that SSA implies SA when $T$ is the empty set, but SA does not imply SSA. The unconditional form of SSA is often called \emph{submodularity}. The conditional forms of SA and SSA (often referred to as \emph{data-processing} inequalities) tell us that entropy cannot increase when conditioning on more random variables, which is a crucial information-theoretic property of the entropy. The conditional and unconditional forms of the properties given above are equivalent for entropic measures that satisfy the chain rule (CR) in the form given in Equation~\eqref{eq: chainrule}. However for entropies that do not satisfy this chain rule, the two forms will be inequivalent and we will explicitly specify which form is being referred to in such cases. In the rest of this thesis, when we say that an entropy satisfies the properties given by Equations~\eqref{eq: chainrule}-\eqref{eq: SSA}, this should be understood as--- the corresponding property holds when the Shannon entropy $H()$ in these equations is replaced by the entropy measure under consideration.

The von Neumann entropy generalises the Shannon entropy to quantum states (represented by density operators), and is defined as follows.
\begin{definition}[von Neumann entropy]
\label{definition: vNent}
The \emph{von Neumann entropy} of a density operator $\rho_A\in \mathscr{S}(\mathscr{H}_A)$ is defined as
\begin{equation}
    H(A)_{\rho}:=-\tr(\rho\ln(\rho)).
\end{equation}
\end{definition}
It follows that the von Neumann entropy is zero for pure quantum states.

We will often drop the subscript $\rho$ in the von Neumann entropy even though this will lead to the same notation as the Shannon case. However, this is justified since classical probability distributions can be equivalently encoded using diagonal density matrices (with the distribution along the diagonal), in which case we indeed recover the Shannon entropy. We will often use $X$, $Y$, $Z$ etc. to denote classical random variables and $A$, $B$, $C$ etc. to denote labels of quantum subsystems, to distinguish between the two entropies. With this, the conditional von Neumann entropy is simply defined through the chain rule, i.e.,
\begin{equation}
\label{eq: condvN}
    H(A|B)=H(AB)-H(B).
\end{equation}
The von Neumann mutual information and its conditional version are defined exactly as in the Shannon case, i.e., through Equation~\eqref{eq: ShanMI} with $H()$ denoting the von Neumann entropy instead.

\paragraph{Properties of von Neumann entropy:} The von Neumann entropy is nonnegative and satisfies additivity (Equation~\eqref{eq: additivity}), chain rule (Equation~\eqref{eq: chainrule}), subadditivity (Equations~\eqref{eq: SA}) and strong subadditivity\footnote{The strong subadditivity of the von Neumann entropy is an important theorem in quantum information theory, proven by Lieb and Ruskai \cite{Lieb1973}. The original proof (which is also the one presented in standard textbooks \cite{NielsenChuang}) is particularly known for its complexity and a number of simpler proofs have been proposed since then (see for example \cite{Renner2006}).} (Equations~\eqref{eq: SSA}). Further, it also admits a corresponding upper bound, $H(A)\leq \ln(d_A)$ where $d_A$ denotes the dimension of the Hilbert space $\mathscr{H}_A$, with equality if an only if $\rho_A$ is the maximally mixed state $\frac{\mathds{1}}{d_A}$. Note however that the von Neumann entropy \emph{does not} satisfy monotonicity (Equation~\eqref{eq: monotonicity}) i.e., the conditional entropy is not positive. This can be seen by taking $\rho_{AB}$ to be an entangled pure state in Equation~\eqref{eq: condvN}, such that $H(A),H(B)>0$ even though $H(AB)=0$. Instead, the von Neumann entropy satisfies a weaker property (\emph{weak monotonicity}) that states that for any tripartite state $\rho_{ABC}$, the following holds.
\begin{equation}
    \label{eq: weakmono}
    H(A|B)_{\rho}\geq -H(A|C)_{\rho}
\end{equation}

\subsection{Tsallis entropies}
\label{ssec: tsalprop}

\begin{definition}[(Classical) Tsallis entropies]
\label{definition: Tsallis}
Given a random variable $X$ distributed according to the discrete probability distribution $P_X$, the order $q$ Tsallis entropy of $X$ for a non-negative real parameter $q$ is defined as~\cite{Tsallis1988}
\begin{equation}
\label{eq: tsallis1}
 S_q(X)= 
 \begin{cases}
 -\sum_{\{x\in X:P_X(x)>0\}}P(x)^q\ln_q P(x) & \text{if } q\neq 1\\
 H(X) & \text{if } q=1
 \end{cases}
\end{equation}
where $\ln_q P(x)=\frac{P(x)^{1-q}-1}{1-q}$.
\end{definition}

The $q$-logarithm function converges to the natural logarithm
in the limit $q \rightarrow 1$ so that
$\lim\limits_{q\rightarrow 1}S_q(X)=H(X)$ and the function is
continuous in $q$. For brevity, we will henceforth write $\sum_x$
instead of $\sum_{\{x:P(x)>0\}}$, keeping it implicit that probability
zero events do not contribute to the sum.\footnote{Note that this
  means the Tsallis entropy for $q<0$ is not robust in the sense that
  small changes in the probability distribution can lead to large
  changes in the Tsallis entropy.} An equivalent form of Equation~\eqref{eq: tsallis1} is the following.
  \begin{equation}
      \label{eq: tsallis2}
       S_q(X)= 
 \begin{cases}
\frac{1}{1-q}\big(\sum_{x\in X} P(x)^q-1\big) & \text{if } q\neq 1\\
 H(X) & \text{if } q=1
 \end{cases}
  \end{equation}
A number of non-equivalent ways of defining the conditional Tsallis entropy have been proposed in the literature \cite{Furuichi04, ABE2001157}. The definitions of \cite{ABE2001157} does not satisfy the chain rule that is satisfied by the conditional Shannon entropy (Equation~\eqref{eq: chainrule}), while the definition of \cite{Furuichi04} does. Hence we will stick to the latter definition throughout this thesis, which is as follows.

\begin{definition}[Conditional Tsallis entropies]
Given two random variables $X$ and $Y$, distributed according to $P_{XY}$, the order $q$ conditional Tsallis entropy for $q\geq 0$ is defined by
\begin{equation}
    \label{eq: tsalliscond}
    S_q(X|Y):=  \begin{cases}
 -\sum_{x,y} P(xy)^q \ln_q P(x|y)& \text{if } q\neq 1\\
 H(X|Y) & \text{if } q=1
 \end{cases}
\end{equation}
\end{definition}
Note that $S_q(X|Y)$ converges to the Shannon conditional entropy $H(X|Y)$ in the limit $q\rightarrow 1$. The unconditional and conditional Tsallis mutual informations are defined analogously to Equation~\eqref{eq: ShanMI} for the Shannon case with $I$ replaced by $I_q$ and $H$ replaced by $S_q$.
\begin{equation}
\label{eq: tsalmi1}
    \begin{split}
       &I_q(X:Y)=S_q(X)-S_q(X|Y),\\
      &I_q(X:Y|Z)=S_q(X|Z)-S_q(X|YZ).
    \end{split}
\end{equation}

\paragraph{Properties of Tsallis entropies:} Tsallis entropies are non-negative and satisfy the unconditional and conditional forms of monotonicity (Equation~\eqref{eq: monotonicity}) and the chain rule (Equation~\eqref{eq: chainrule}) for all $q\geq 0$. They also satisfy both the forms of subadditivity (Equations~\eqref{eq: SA}) and strong subadditivity (Equations~\eqref{eq: SSA}) for all $q\geq 1$. For $q\geq 0$, they admit an analogous upper-bound as the Shannon case i.e., $S_q(X)\leq\ln_q|X|$, where for $q>0$, equality is achieved if and only
    if $P_X(x)=1/|X|$ for all $x$ (i.e., if the distribution on $X$ is
    uniform). However, Tsallis entropies are not additive (Equation~\eqref{eq: additivity}) in general and instead satisfy a weaker condition known as \emph{pseudo-additivity} \cite{Curado1991}--- for two independent random
      variables $X$ and $Y$ i.e., $P_{XY}=P_XP_Y$, and for all $q$, the Tsallis entropies satisfy
    \begin{equation}
        \label{eq: pseudoadd}
        S_q(XY)=S_q(X)+S_q(Y)+(1-q)S_q(X)S_q(Y).
    \end{equation}
    Note that in the Shannon case ($q=1$), we recover additivity for
    independent random variables. For $q<1$, strong subadditivity (Equation~\eqref{eq: SSA}) does not hold in
general~\cite{Furuichi04}. The results presented in this thesis rely on this property and hence we will restrict to the $q\geq
1$ case in the remainder of this thesis. The quantum generalisation of the (classical) Tsallis entropy is defined as follows. Note that in the rest of this thesis, ``Tsallis entropy'' will stand for the ``classical Tsallis entropy'' defined in Definition~\ref{definition: Tsallis}. In the quantum case, we will explicitly use ``quantum Tsallis entropy''.

\begin{definition}[Quantum Tsallis entropies]
\label{definition: qTsallis}
The order $q$ \emph{quantum Tsallis entropy} of a density operator $\rho_A\in\mathscr{S}(\mathscr{H}_A)$, for a non-negative real parameter $q$ is defined as~\cite{Tsallis1988}
\begin{equation}
\label{eq: qtsallis1}
 S_q(A)= 
 \begin{cases}
 \frac{1}{1-q}\big(\tr(\rho^q)-1\big) & \text{if } q\neq 1\\
 H(A) & \text{if } q=1,
 \end{cases}
\end{equation}
where $H(A)$ denotes the von Neumann entropy (Definition~\ref{definition: vNent}) of $\rho_A$.
\end{definition}

Quantum Tsallis entropies can be seen as a generalisation of the von Neumann entropy since they converge to $H(A)$ in the limit of $q$ going to 1. The conditional quantum Tsallis entropy for a density operator $\rho_{AB}\in\mathscr{S}(\mathscr{H}_A\otimes\mathscr{H}_B)$ can be simply defined as
\begin{equation}
\label{eq: condqtsal}
    S_q(A|B)=S_q(AB)-S_q(B)
\end{equation} (analogous to the von Neumann case, Equation~\eqref{eq: condvN}).

\paragraph{Properties of quantum Tsallis entropies:} Quantum Tsallis entropies are non-negative and satisfy the chain rule (Equation~\eqref{eq: chainrule}) by definition of the conditional entropy. They also satisfy pseudo-additivity (Equation~\eqref{eq: pseudoadd}) $\forall q\geq 0$, both forms of subadditivity (Equation~\eqref{eq: SA})  $\forall q\geq 1$ and are upper-bounded as $S_q(A)\leq \ln_q (d_A)$, where the bound is saturated for $q>0$ if and only if $\rho_A$ is the maximally mixed state. However, they no not satisfy monotonicity and strong-subadditivity in general. The former is evident since the von Neumann entropy, a special case of quantum Tsallis entropies, also does not satisfy this property. The latter was noted in \cite{Petz2014} where sufficient conditions for the strong subadditivity of quantum Tsallis entropies expressed in its unconditional form, were also analysed.
\setlength{\tabcolsep}{0pt}
\begin{table}[t!]
    \centering
\begin{tabular}{|c|ccccc|}
\hline
     & \makecell{A\\ Eq.~\eqref{eq: additivity}} & \makecell{CR\\ Eq.~\eqref{eq: chainrule}} & \makecell{M\\ Eq.~\eqref{eq: monotonicity}} & \makecell{SA\\ Eq.~\eqref{eq: SA}} &\makecell{SSA\\ Eq.~\eqref{eq: SSA}}\\
     \hline
    Shannon  & $\checkmark$ \cite{Shannon1948} &$\checkmark$ \cite{Shannon1948} &$\checkmark$ \cite{Shannon1948} &$\checkmark$ \cite{Shannon1948} &$\checkmark$ \cite{Shannon1948}\\
    Von Neumann  &$\checkmark$ \cite{vonNeumann1932} &\makecell{$\checkmark$ \\ \small{\emph{(By definition)}}} &\makecell{$\times$ \cite{Cerf1997} \\ \small{\emph{(Weak Mono. \cite{Lieb1973}) }}}&$\checkmark$\cite{Araki1970} &$\checkmark$ \cite{Lieb1973}\\
    Tsallis  & \makecell{$\times$ \cite{Tsallis1988}\\ \small{\emph{(Pseudo-add.)}}}& $\checkmark$ \cite{Furuichi04} &$\checkmark$ \cite{Daroczy1970}&\makecell{$\checkmark$ \cite{Furuichi04}\\ \small{\emph{($q\geq 1$)}}}&\makecell{$\checkmark$ \cite{Furuichi04}\\ \small{\emph{($q\geq 1$)}}}\\
    Q. Tsallis  &\makecell{$\times$ \cite{Tsallis1988} \\ \small{\emph{(Pseudo-add.)}}}&\makecell{$\checkmark$\\ \small{\emph{(By definition)}}}&$\times$ \cite{Cerf1997}&\makecell{$\checkmark$ \cite{Audenaert2007}\\ \small{\emph{($q\geq 1$)}}}&$\times$ \cite{Petz2014}\\
    R\'enyi  &$\checkmark$ \cite{Renyi1961}&\makecell{$\times$\cite{Iwamoto2013}\\ \small{\emph{(Alternate CR \cite{Muller-Lennert2013})}}}&$\checkmark$\cite{Linden2013}&\makecell{$\checkmark$\\ \small{\emph{(C. form \cite{Iwamoto2013})}}}&\makecell{$\checkmark$\\ \small{\emph{(C. form \cite{Iwamoto2013})}}}\\
    Min  & $\checkmark$ \cite{Renner2006}& $\times$ \cite{Renner2006}&$\times$ \cite{Renner2006}&\makecell{$\checkmark$\\ \small{\emph{(C. form \cite{Renner2006})}}}&\makecell{$\checkmark$\\ \small{\emph{(C. form \cite{Renner2006})}}}\\
   Max  & $\checkmark$ \cite{Renner2006}& $\times$ \cite{Renner2006}&$\times$ \cite{Renner2006}&\makecell{$\checkmark$\\ \small{\emph{(C. form \cite{Renner2006})}}}&\makecell{$\checkmark$\\ \small{\emph{(C. form \cite{Renner2006})}}}\\
   \hline
\end{tabular}
    \caption[Information-theoretic properties of classical and quantum entropies]{\textbf{Information-theoretic properties of generalised entropies: } This table summarises the main properties of the entropy measures discussed in the main text along with those of the min and max entropies. The acronyms A, CR, M, SA, SSA and C. form denote additivity, chain rule, monotonicity, subadditivity, strong subadditivity and conditional form respectively. R\'enyi, Min and Max entropies satisfy SA and SSA only in the conditional form while the remaining entropies satisfy them (wherever indicated so) in both conditional and unconditional forms. For the von Neumann and quantum Tsallis entropies, the chain rule follows by definition (Equations~\eqref{eq: condvN} and \eqref{eq: condqtsal}). Monotonicity fails for the von Neumann entropy and hence the quantum Tsallis entropy which is a generalisation of it for $q\neq 1$. Weak Monotonicity~\eqref{eq: weakmono} holds in the von Neumann case but an analogue for the more general $q\geq 0$ quantum Tsallis entropies is not known (to the best of our knowledge). Pseudo-additivity (Equation~\eqref{eq: pseudoadd}) for quantum Tsallis entropies follows in the same manner as it does for the classical case shown in \cite{Tsallis1988}. The table only lists the properties of the classical R\'enyi entropy, since the quantum version is not relevant for the work presented in this thesis. Multiple definitions of quantum R\'enyi entropies have been proposed \cite{Tomamichel2009, Muller-Lennert2013}, and these properties may differ in each case. For example SSA in conditional form is only known to hold for $\alpha\in [0,2]$ for the quantum R\'enyi entropy of \cite{Tomamichel2009}. The R\'enyi entropies do not satisfy the chain rule of Equation~\eqref{eq: chainrule}, but they satisfy a weaker, cardinality-dependent chain rule of Equation~\eqref{eq: RenyiCR}. The min and max entropies (which are originally defined as quantum entropies) are also included in this table for comparison, their definition and further properties can be found in \cite{Renner2006}.}
    \label{tab:ent_prop}
\end{table}

\subsection{R\'enyi entropies}
\label{ssec: renyi}
Here we define only the classical versions of R\'enyi entropies, as the quantum version will not be used in the remainder of this thesis. 

\begin{definition}[(Classical) R\'enyi entropies]
Given a random variable $X$ distributed according to the discrete probability distribution $P_X$, the order $\alpha$ R\'enyi entropy of $X$ for a non-negative real parameter $\alpha$ is defined as
\begin{equation}
\label{eq: renyi1}
 R_{\alpha}(X)= 
 \begin{cases}
 \frac{1}{1-\alpha}\ln\Big(\sum_{x\in X}P(x)^{\alpha}\Big) &\text{if } \alpha\neq 1\\
 H(X) & \text{if } \alpha=1
 \end{cases}
\end{equation}
\end{definition}

There are several inequivalent definitions of the conditional R\'enyi entropy proposed in the literature \cite{Iwamoto2013, Fehr2014}, in this thesis, we will consider the following definition.

\begin{definition}[Conditional R\'enyi entropy]
\label{definition: condRenyi}
Given two random variables $X$ and $Y$, distributed according to $P_{XY}$, the order $\alpha$ conditional R\'enyi entropy for $\alpha\geq 0$ is defined by
\begin{equation}
\label{eq: renyi2}
 R_{\alpha}(X|Y)= 
 \begin{cases}
\frac{1}{1-\alpha}\log(\sum_{xy}P(xy)^\alpha P(y)^{1-\alpha}) &\text{if } \alpha\neq 1\\
 H(X|Y) & \text{if } \alpha=1
 \end{cases}
\end{equation}
\begin{equation}
\end{equation}
\end{definition}

Both the unconditional and conditional R\'enyi entropies converge to the corresponding Shannon entropy as $\alpha$ tends to 1.

\paragraph{Properties of R\'enyi entropies: } The R\'enyi entropies are nonnegative and upper-bounded as $R_{\alpha}(X)\leq \ln|X|$ for all $\alpha\geq 0$, with equality (for $\alpha>0$) if and only if $P_X$ is the uniform distribution over $X$ \cite{Renyi1961}. R\'enyi entropies do not satisfy the chain rule (Equation~\eqref{eq: chainrule}) in general for $\alpha\neq 1$. However, an alternate, dimension dependent chain rule has been proposed for quantum R\'enyi entropies in \cite{Muller-Lennert2013}, which therefore also hold for classical R\'enyi entropies (where dimension would correspond to the cardinality of the variables). This alternate chain rule is given as
\begin{equation}
    \label{eq: RenyiCR}
     H_\alpha(X|YZ)\geq H_\alpha(XZ|Y)-\ln|Z|.
\end{equation}
Further, R\'enyi entropies also satisfy monotonicity (Equation~\eqref{eq: monotonicity}) in both the conditional and unconditional forms i.e., $H_\alpha(X)\leq H_\alpha (XY)$ and $H_\alpha(X|Y)\geq 0$, even though these forms are not equivalent due to failure of the chain rule (in contrast to the Shannon case). Classical R\'enyi entropies have been shown to satisfy subadditivity (Equation~\eqref{eq: SA}) as well as strong subadditivity (Equation~\eqref{eq: SSA}), both only in the conditional forms. They fail to satisfy the (inequivalent) unconditional form of these properties for $\alpha\neq 1$.

Table~\ref{tab:ent_prop} summarises some of the important information-theoretic properties of the entropies discussed so far, as well as those of the min and max entropies proposed in \cite{Renner2006} for comparison. We will not define or discuss these and other entropy measures here as they are not relevant for the work presented in this thesis.

\section{Generalised probabilistic theories}
\label{sec: GPT}

Quantum theory has worked and continues to work remarkably well to explain and predict empirical observations. However there is no consensus on a set of natural physical principles that single out quantum theory from a plethora of other possible theories that are also compatible with relativistic principles (such as finite signalling speed) but produce stronger-than-quantum correlations \cite{Popescu1994}. Generalised probabilistic theories (GPTs), of which classical and quantum theories can be seen as particular members, were developed out the motivation to derive the mathematical formalism of quantum theory from fundamental physical principles or axioms, and to probe deeper, theory-independent connections between information processing and physical principles \cite{Segal1947, Ludwig1964, Ludwig1967, Ludwig1968, Hardy01, Barrett07, Pawlowski2009}. Several properties such as (Bell) non-locality, contextuality, entanglement, non-unique decomposition of a mixed state into pure states, monogamy of correlations, information-theoretically secure cryptography and no-cloning have been shown to extend beyond quantum theory \cite{Popescu1994, Hardy01, BHK, Masanes2006, Barrett07}. However, teleportation and entanglement swapping are not always possible in non-classical GPTs despite the existence of entangled states \cite{Short2006, Barrett07, Skrzypczyk2008, Gross2010, Weilenmann2020}.\footnote{The intuition here is that there is a trade-off between the state space and the effect space (the dual of the former). A larger state space implies that the set of effects leading to valid probabilities would be smaller, leading to a smaller effects space. Hence, there exist GPTs with entangled states but no analogue of entangling measurements, both of which are needed for teleportation-like protocols.} Further, from a foundational point of view, it is of interest to understand whether interpretational issues are peculiarities of quantum theory. The contextuality/non-locality of GPTs suggests that similar interpretational problems could arise here. We will discuss this aspect in more detail in Chapter~\ref{chapter: PRdoxespaper} where we derive a Wigner's friend type paradox in \emph{box world} (a particular GPT). 

In the present chapter, we provide an overview of the framework for information processing in GPTs proposed by Barrett in \cite{Barrett07}. In the remainder of this thesis, we will use the more colloquial term, ``box world'' to denote the set of theories that Barrett originally calls \emph{Generalised no-signaling Theories}. This theory allows arbitrary correlations between measurements on separated systems, as long as they are non-signaling i.e., choice of measurement on one subsystem does not affect the measurement outcome on the other. Here separation between systems is expressed in terms of a tensor product structure (that allows us to identify subsystems), and without reference to space-time locations/space-like separation\footnote{Of course this leads to systems and correlations that are non-signaling also with respect to a space-time structure once this is introduced into the picture.}. Not all GPTs need to have a tensor product structure for representing composite systems, but will focus only on those that do, for the purposes of this thesis. The following is based on the review sections of our paper \cite{Vilasini_PRdoxes}.
 
 %In Section~\ref{sssec: gen_causalstr}, we discussed the operation-probabilistic framework of \cite{Chiribella2010} used to describe the generalised causal structures of \cite{Henson2014}.  This can be seen as providing an operational semantic to GPTs in terms of generalised circuits (or diagrams) made up of composable elements (preparations, transformations, observations). Alternatively, Barrett's framework for GPTs can be seen as a description of the states, transformations and measurements underlying the generalised circuits.

\subsection{States and transformations}
\label{ssec: statesop}

\begin{figure}[t]
    \centering
\hspace{-0.5cm}\subfloat[\textbf{G-bit.} A gbit is a function with binary input and output, characterized by the  probability vector $\vec P_{gbit}$, also called the state vector. ]{
    \vspace{1cm}
    \qquad\quad\includegraphics[scale=0.3]{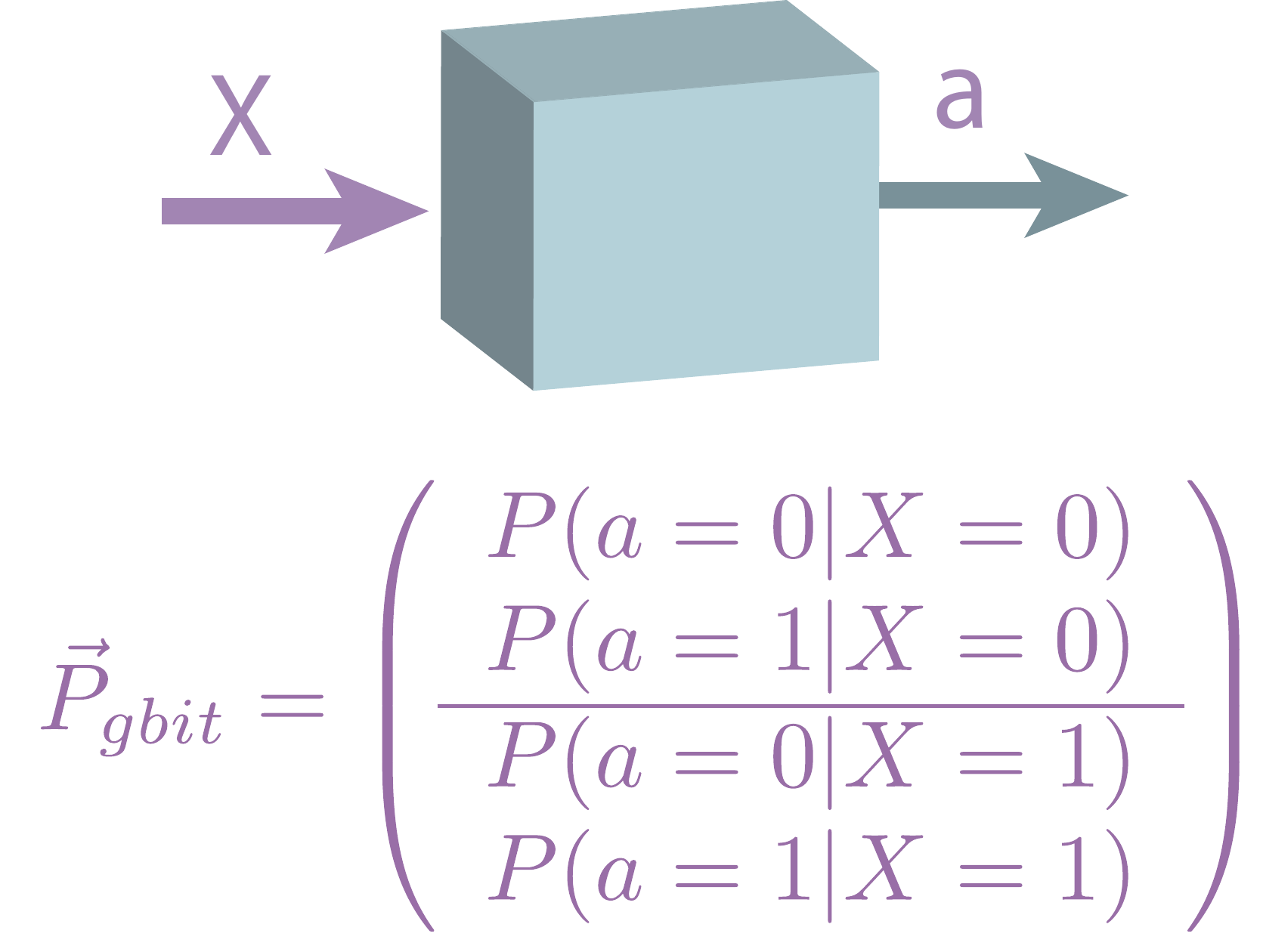}\qquad\quad
    \label{fig:gbit}
}\hspace{-1cm}\subfloat[\textbf{PR box.} The PR box has two binary inputs $X, Y$ and two binary outputs $a,b$, satisfying $XY = a\oplus b$, and otherwise uniformly random (state vector on the right). Usually it is applied in the context of two space-like separated agents, each providing one of the inputs and obtaining the respective output. The box is non-signaling, and maximally violates the CHSH inequality~\cite{Popescu1994}.]{ \centering
    \qquad\includegraphics[scale=0.3]{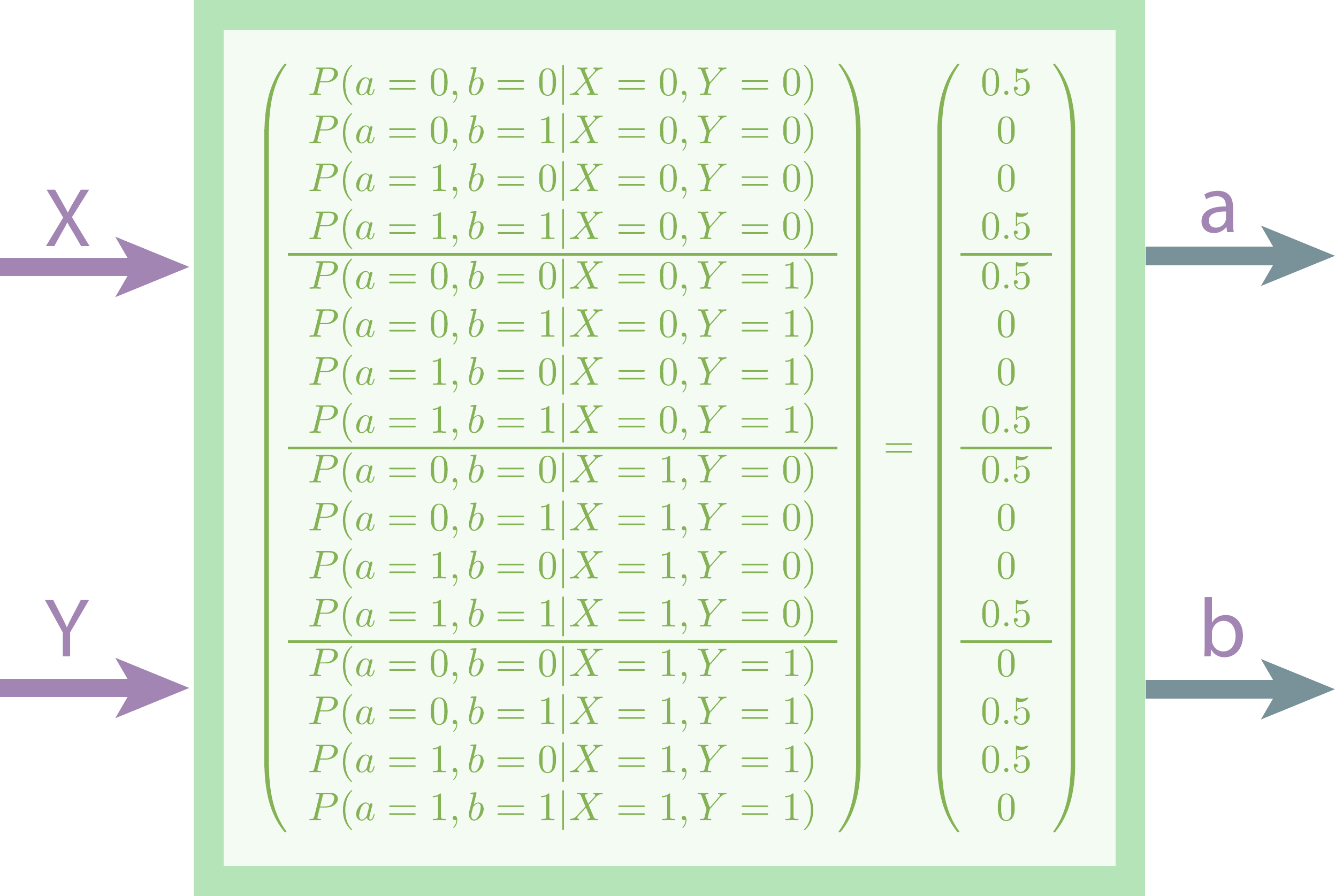}\qquad
    \label{fig:PRbox}
}

\caption[Individual and composite systems in box world]{{\bf Boxes in Generalized Probabilistic Theories.} The modular objects of GPTs are input/output functions depicted as boxes and characterized by probability vectors. Each function (or box) can be evaluated once, and it may or not correspond to a physical system being probed; even if it is, nothing is assumed about the post-evaluation state of the system (unlike quantum theory, which specifies the post-measurement state of a system given its initial state and the measurement device).\footnotemark}
\label{fig:GTP_states}
\end{figure}

\paragraph{Individual states.} The so-called generalised bit or \emph{gbit} is a system completely characterized by two binary measurements which can be performed on it~\cite{Barrett07}. Such sets of measurements that completely characterise the state of a system are known as \emph{fiducial measurements}. The state of a gbit is thus fully specified by the vector
\begin{equation}
\label{eq: gbit}
    \vec{P}_{gbit}=\left(
\begin{array}{c}
 P(a=0|X=0)\\
 P(a=1|X=0)\\
\hline
 P(a=0|X=1)\\
 P(a=1|X=1)\\
\end{array}
\right),
\end{equation}
where $X=0$ and $X=1$ represent the two choices of measurements and $a \in \{0,1\}$ are the possible outcomes (Figure~\ref{fig:gbit}).
Analogously, a classical bit is a system characterized by a single binary fiducial measurement, 
\begin{equation}
\label{eq: bit}
    \vec{P}_{bit}=\left(
\begin{array}{c}
 P(a=0|X=0)\\
 P(a=1|X=0)
\end{array}
\right),
\end{equation}

and, in quantum theory, a qubit is characterized by three fiducial measurements (corresponding, for example, to three directions $X$, $Y$ and $Z$ in the Bloch sphere), 
\begin{equation}
\label{eq: qubit}
    \vec{P}_{qubit}=\left(
\begin{array}{c}
 P(a=0|X=0)\\
 P(a=1|X=0)\\
\hline
 P(a=0|X=1)\\
 P(a=1|X=1)\\
 \hline
 P(a=0|X=2)\\
 P(a=1|X=2)\\
\end{array}
\right).
\end{equation}

\footnotetext{This figure is taken from our paper \cite{Vilasini_PRdoxes} which is joint work with Nuriya Nurgalieva and L\'idia del Rio, and was made by L\'idia.}

For normalized states, we have $|\vec{P}|=\sum_i P(a=i|X=j)=1,   \forall\, j$.
The set of possible states of a gbit is convex, with the extreme points
\begin{align}
\label{eq: gbitpure}
   \vec{P}_{00}=\left(
\begin{array}{c}
1\\
0\\
\hline
1\\
0\\
\end{array}
\right), \qquad
 \vec{P}_{01}=\left(
\begin{array}{c}
1\\
0\\
\hline
0\\
1\\
\end{array}
\right), 
\quad
   \vec{P}_{10}=\left(
\begin{array}{c}
0\\
1\\
\hline
1\\
0\\
\end{array}
\right),
\qquad
 \vec{P}_{11}=\left(
\begin{array}{c}
0\\
1\\
\hline
0\\
1\\
\end{array}
\right).
\end{align}
These correspond to pure states, and the state space of a gbit is a polytope (due to the finite number of extreme points). In the qubit case, the extreme points correspond to all the (infinitely many) pure states on the surface of the Bloch sphere, some of these are
\begin{equation}
    \vec{P}_{\ket+}=\left(
\begin{array}{c}
 1\\
 0\\
\hline
 \nicefrac12 \\
\nicefrac12 \\
 \hline
 \nicefrac12 \\
 \nicefrac12 \\
\end{array}
\right), 
\qquad
    \vec{P}_{\ket-}=\left(
\begin{array}{c}
 0\\
 1\\
\hline
 \nicefrac12 \\
\nicefrac12 \\
 \hline
 \nicefrac12 \\
 \nicefrac12 \\
\end{array}
\right), 
\qquad
    \vec{P}_{\ket0}=\left(
\begin{array}{c}
 \nicefrac12 \\
 \nicefrac12 \\
\hline
 \nicefrac12 \\
 \nicefrac12 \\
 \hline
 1\\
 0\\
\end{array}
\right)\qquad
    \vec{P}_{\ket1}=\left(
\begin{array}{c}
 \nicefrac12 \\
 \nicefrac12 \\
\hline
 \nicefrac12 \\
 \nicefrac12 \\
 \hline
 0\\
 1\\
\end{array}
\right).
\end{equation}
Note that in box world, pure gbits are deterministic for both alternative measurements, whereas in quantum theory at most one fiducial measurement can be deterministic for each pure qubit, as reflected by uncertainty relations. 
We denote the set of allowed states of a system $A$ by $\mathscr{S}^A$.

\paragraph{Composite states.} The state of a bipartite system $AB$, denoted by $\vec{P}^{AB} \in \mathscr{S}^{AB}$ can be written in the form $\vec{P}^{AB}=\sum_i r_i\ \vec{P}^A_i\otimes \vec{P}^B_i$ where $r_i$ are real coefficients\footnote{Note that it is not necessary that the coefficients $r_i$ be positive and sum to one. If this is the case, then the composite state would be separable and hence local, otherwise, the state is entangled \cite{Barrett07}.} and $\vec{P}^A_i\in \mathscr{S}^A$, $\vec{P}^B_i\in \mathscr{S}^B$ can be taken to be pure and normalised states of the individual systems $A$ and $B$~\cite{Barrett07}. Thus, a general 2-gbit state $\vec{P}^{AB}_2$ can be written as in Figure~\ref{fig:PRbox} (left), where $X, Y\in \{0,1\}$ are the two fiducial measurements on the first and second gbit and $a,b \in \{0,1\}$ are the corresponding measurement outcomes. The PR box $\vec{P}_{PR}$, on the right, is an example of such a 2 gbit state that is valid in box world, which satisfies the condition $a\oplus b = x y$~\cite{Popescu1994}.

\begin{figure}[t]
    \centering
\begin{tikzpicture}
\draw[arrows={-stealth}] (2.2,0)--(2.2,-1); \draw[arrows={-stealth}] (6.2,0)--(6.2,-1);
\node[align=center] at (2.5,-0.5) {$X$}; \node[align=center] at (6.5,-0.5) {$Y$};
\draw (1.6,-2) rectangle node[align=center]{$\mathbf{A}$} (2.8,-1);
\draw (5.6,-2) rectangle node[align=center]{$\mathbf{B}$} (6.8,-1); 
\draw (5.3,0) rectangle node[align=center]{$Y=f_1(a)$} (7.1,1);\draw[dashed] (2.8,-1.5)--(5.6,-1.5); \draw[arrows={-stealth}] (6.2,-2)--(6.2,-3); \node[align=center] at (6.5,-2.5) {$b$}; \draw (5.6,-4.2) rectangle node{$o=f_2(a,b)$} (8.6,-3);
 \draw[arrows={-stealth}] (7.7,-2)--(7.7,-3);   \node[align=center] at (8,-2.5) {$a$};  
\draw[arrows={-stealth}] (7.1,-4.2)--(7.1,-5.2);  \node[align=center] at (7.1,-5.5) {$o$}; \draw (2.2,-2)--(2.6,-2.5); \draw (2.6,-2.5) to [out=315,in=135] (5.8,1.5); \draw[arrows={-stealth}] (5.8,1.5)--(6.2,1);
\node[align=center] at (2.2,-2.4) {$a$}; 
%\draw plot [smooth] coordinates {(7.7,1) (7.7,0.7)};

\draw (7.7,1)-- (7.7,-2); \draw (5.8,1.5)to [out=45,in=135] (7.7,1);
\end{tikzpicture}
    \caption[Bipartite measurements in box world]{{\bf Bipartite measurements in box world.} Any bipartite measurement on a 2-gbit box world system can be decomposed into a procedure (or convex combinations thereof) of the following form. Alice first performs a measurement $X$ on one of the gbits (labelled $A$), and forwards the outcome $a$ to Bob. Bob then performs a measurement $Y=f_1(a)$, which may depend on $a$, on the other gbit (labelled $B$), obtaining the outcome $b$. The final measurement outcome $o$ of the joint measurement can be computed by Bob as a function $f_2(a,b)$. All allowed bipartite measurements are convex combinations of this type of classical wirings~\cite{Barrett07}.}
    \label{fig: bipartitemeas}
\end{figure}
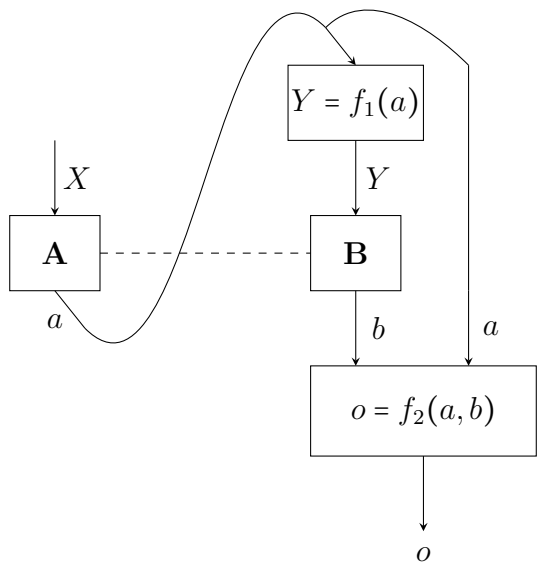

\paragraph{State transformations.} Valid operations are represented as matrices that transform valid state vectors to valid state vectors. In addition, we only have access to the (single-shot) input/output behaviour of systems, so in practice all valid operations in box world take the form of classical wirings between boxes, which correspond to pre- and post-processing of input and output values, and convex combinations thereof~\cite{Barrett07}.  
For example, 
 bipartite joint measurements on a 2-gbit system can be decomposed into convex combinations of classical ``wirings'', as shown in  Figure~\ref{fig: bipartitemeas}.
In contrast, quantum theory allows for a richer structure of bipartite measurements by allowing for entangling measurements (e.g.\ in the Bell basis), which cannot be decomposed into classical wirings. 
  Bipartite transformations on multi-gbit systems turn out to be  classical wirings as well \cite{Barrett07}. Reversible operations in particular  consist only of trivial wirings: local operations and permutations of systems \cite{Gross2010}. 
 One cannot perform entangling operations such as a coherent copy (the quantum CNOT gate) \cite{Barrett07, Short2006, Gross2010}.
 
 \subsection{Measurements: observing outcomes}
 \label{ssec: boxworldoutcomes}
In quantum theory, systems are described by states that live in a Hilbert space, measurements and transformations on these states are represented by CPTP maps and the Born rule specifies how to obtain the probabilities of possible measurement outcomes give these states and measurements. In more general theories, there is no reason to assume Hilbert spaces or CPTP maps. In fact such a description of the state space and operations may not even be available, systems may be described as black boxes taking in classical inputs (choice of measurements) and giving classical outputs (measurement outcomes). What we can demand is that the theory provides a way for agents to predict the probabilities of obtaining various outputs based on their input choice and some operational description of the box. We have briefly reviewed states and transformations in Barrett's framework for GPTs \cite{Barrett07}, we now discuss how measurements and associated probabilities are characterised here.  

%Later, Gross \emph{et al.}\ found restrictions on the reversible dynamics of maximally non-local GPTs \cite{Gross2010} showing that all reversible operations on box-world are trivial i.e., they map product states to product states and cannot correlate initially uncorrelated systems. In accordance with this, our memory update procedure that maps the initial product state $\vec{P}_{in}^{SM}$ (Equation~\eqref{eq: PSMin}) to the final correlated state of the system and memory $ \vec{P}_{fin}^{SM}$ (Equation~\eqref{eq: PSMfin} or equivalently Equation~\eqref{eq: finshort}) is an irreversible transformation in contrast to the quantum case where the corresponding transformation is a unitary and hence reversible. 

Consider a GPT, $\mathbb{T}$. Denoting the set of all allowed states of a system in $\mathbb{T}$ by $\mathscr{S}$, any valid transformation on a normalised GPT state $\vec{P}\in \mathscr{S}$ maps it to another normalised GPT state in $\mathscr{S}$. The set of allowed normalized states is assumed to be closed and convex\footnote{The motivation for this assumption is that if it is possible to prepare two states $P_1$ and $P_2$, then it should be possible to prepare a probabilistic mixture of the two (for example, by tossing a coin).} and transformations on these states, assumed to be linear. Each valid transformation is then represented by a matrix $M$ such that $\vec{P}\rightarrow M.\vec{P}$ under this transformation and $M.\vec{P}\in \mathscr{S}$ \cite{Barrett07}. Further, operations that result in different possible outcomes can be associated with a set of transformations, one for each outcome. These also give an operational meaning to unnormalised states where $|\vec{P}|=\sum_i P(a=i|X=j)=c \quad \forall j, c\in [0,1]$ (i.e., the norm is independent of the value of $j$). Such an operation $M$ on a normalised initial state $\vec{P}$ can be associated with a set of matrices $\{M_i\}$ such that the unnormalised state corresponding to the $i^{th}$ outcome is $M_i.\vec{P}$. Then the probability of obtaining this outcome is simply the norm of this unnormalised state, $|M_i.\vec{P}|$ and the corresponding normalized final state is $M_i.\vec{P}/|M_i.\vec{P}|$. A set $\{M_i\}$ represents a valid operation if the following hold \cite{Barrett07}.
\begin{subequations}
\begin{equation}
\label{eq: op1}
    0 \leq |M_i.\vec{P}| \leq 1 \qquad \forall i, \vec{P} \in \mathscr{S}
\end{equation}
\begin{equation}
\label{eq: op2}
    \sum_i |M_i.\vec{P}|=1 \qquad \forall \vec{P}\in \mathscr{S}
\end{equation}
\begin{equation}
\label{eq: op3}
    M_i.\vec{P} \in \mathscr{S} \qquad \forall i, \vec{P}\in \mathscr{S}
\end{equation}
\end{subequations}

 This is the analogue of quantum Born rule for GPTs. Box world is a GPT where the state space $\mathscr{S}$ consists of all normalized states $\vec{P}$ whose entries are valid probabilities (i.e., $\in [0,1]$) and satisfy the \emph{no-signaling} constraints i.e., for a $N$-partite state $\vec{P}$, the marginal term $$\sum_{a_i}P(a_1,..,a_i,..,a_N|X_1,..,X_i,..,X_N)$$ is independent of the setting $X_i$ for all $i\in \{1,...,N\}$\footnote{This is in the spirit of relativistic causality since one would certainly expect that the input of one party does not affect the output of others when the are all space-like separated from each other.}
 
 When the GPT $\mathbb{T}$ is box world, the conditions of Equations~\ref{eq: op1}-\ref{eq: op3} result in the characterization of measurements and transformations in the theory in terms of classical circuits or \emph{wirings} as shown in \cite{Barrett07}. We summarise the results of \cite{Barrett07} characterising allowed transformations and measurements in box world and will only consider normalization-preserving transformations.
 
 %It suffices for the purpose of this paper to take that characterisation as the common knowledge of agents in the theory. In the original quantum paradox \cite{Frauchiger2018}, the Born rule is taken as common knowledge and here, the common knowledge consists of characterisations that follow from the box world analogue of the born rule (Equations~\ref{eq: op1}-\ref{eq: op3}).
\begin{itemize}
    \item \textbf{Transformations: } \begin{itemize}
        \item \emph{Single system:} All transformations on single box world systems are relabellings of fiducial measurements or outcomes or a convex combination thereof.
        \item \emph{Bipartite system:} Let $X$ and $Y$ be fiducial measurements performed on the transformed bipartite system with corresponding outcomes $a$ and $b$, then all transformations of 2-gbit systems can be decomposed into convex combinations of classical circuits of the following form: A fiducial measurement $X'=f_1(X,Y)$ is performed on the initial state of the first gbit resulting in the outcome $a'$ followed by a fiducial measurement $Y'=f_2(X,Y,a')$ on the initial state of the second gbit resulting in the outcome $b'$. The final outcomes are given as $(a,b)=f_3(X,Y,a',b')$, where $f_1$, $f_2$ and $f_3$ are arbitrary functions.
    \end{itemize}
     \item \textbf{Measurements: } \begin{itemize}
        \item \emph{Single system:} All measurements on single box world systems are either fiducial measurements with outcomes relabelled or convex combinations of such.
        \item \emph{Bipartite system:} All bipartite measurements on 2-gbit systems can be decomposed into convex combinations of classical circuits of the following form (Figure~\ref{fig: bipartitemeas}): A fiducial measurement $X$ is performed on the initial state of the first gbit resulting in the outcome $a'$ followed by a fiducial measurement $Y=f(a')$ on the second gbit resulting in the outcome $b'$. The final outcome is $a=f'(a',b')$, where $f$ and $f'$ are arbitrary functions.
    \end{itemize}
\end{itemize}

% \begin{remark}
% Note that an agent Alice who measures a box world system only sees a classical final state, which corresponds the classical measurement outcome, since the box is a single-shot input/output function. Alice could use Equations~\ref{eq: op1}-\ref{eq: op3} to calculate the probabilities of obtaining different outcomes given the measurement she performs and prepare a new box (a new input/output function) depending on the measurement and outcome she just obtained (and has stored in her memory), as in Figure~\ref{fig:measurement_Alice}. An outside agent who does not know Alice's measurement outcome would see correlations between Alice's system and memory and would describe the measurement by an irreversible transformation, more specifically  a classical wiring between Alice's system and memory as shown in the following section.
% \end{remark}

\section{Causal structures and causal models}
\label{sec: causalstr}

Cause-effect relationships between systems constrain the possible correlations that can be observed between them. These relationships are naturally represented using \emph{causal structures} or directed graphs. Given a causal structure, a \emph{causal model}  provides a set of mathematical rules to define the possible correlations that are compatible with the causal structure. Additionally, it can also provide rules for how the correlations compatible with a causal structure change under external manipulations of the systems involved. There are different approaches for mathematically formalising these concepts in general physical theories, depending on what the nodes and edges of the causal graph correspond to, the compatibility condition required by the causal model and how external manipulations are described. For classical causal models, a common approach is that of Bayesian networks, pioneered by Judea Pearl \cite{Pearl2009}. This has been generalised to quantum and more general, post-quantum causal structures in \cite{Henson2014}. We will mainly adopt this approach in the thesis, and will briefly discuss other approaches for quantum causal modelling at the end of Section~\ref{ssec: caus_other}. 

%Causal structures typically have a set of systems that are observed and the remaining which are unobserved and hence inaccessible. The nature of the unobserved systems, classical, quantum or post-quantum defines the nature of the causal structure (and the corresponding causal model). In general classical and non-classical causal structures impose different constraints on the possible correlations over the observed systems. Certifying this difference has been a subject of active research in quantum information theory for several years, and is also known to be a difficult (NP-hard) problem in general \cite{Pitowski89}. The main results of Chapters~\ref{chapter: Tsallispaper} and \ref{chapter: mixingpaper} are related to this problem of certifying non-classicality of causal structures based on the observed correlations produced, which we analyse for specific causal structures using entropies as a correlation measure. The results of Chapter~\ref{chapter: jammingpaper}  are related to generalising standard causal models to scenarios that include cyclic causal relationships (such as causal loops) and certain hidden causal influences, and understanding when such causal models are compatible with an underlying space-time structure. We now review the conceptual and mathematical preliminaries related to classical and non-classical (quantum/post-quantum) causal models required for these chapters. 

\paragraph{General definitions and terminology:} 
Throughout this thesis, we will only consider discrete and finite random variables, and corresponding causal structures with finite number of systems/nodes. More formally, we have the following definitions.

\begin{definition}[Causal structure]
A causal structure is a directed
graph $\cG$ over a set of nodes, some of which are labelled observed
and the rest unobserved. Each observed node $X$ corresponds to a classical random variable\footnote{Observed variables may represent classical inputs (such as measurement settings) or outputs (such as measurement outcomes) of an experiment.} of the same name, while unobserved nodes can correspond to  classical, quantum or post-quantum systems. For two (observed or unobserved) nodes $X$ and $Y$ of $\cG$, $X$  is said to be a \emph{cause} of a $Y$ if and only if there exists a directed path from $X$ to $Y$ in $\cG$.
\end{definition}

The majority of the literature on causal structures as well as Chapters~\ref{chapter: Tsallispaper} and \ref{chapter: mixingpaper} concern the case of acyclic causal structures represented by directed acyclic graphs (DAGs). In Chapter~\ref{chapter: jammingpaper} , we attempt to model cyclic causal structures as well, but an understanding of the acyclic case is a pre-requisite for this as well. Hence, we will focus on the acyclic case in this review.

 Note that by the above definition of \emph{cause}, $X$ is said to be a cause of itself if and only if there is a directed path from $X$ to itself in $\cG$, which is not possible if the graph is acyclic.  We will use \emph{family relations} to describe causal relationships between nodes of a directed graph--- for $i\neq j$, if $X_i$ is a direct or indirect cause of $X_j$, then $X_i$ is said to be an \emph{ancestor} of $X_j$ or equivalently, $X_j$ a \emph{descendant} of $X_i$. We will use either explicitly as $\text{anc}(X_i)$ or more compactly as $X^{\downarrow}$ to denote the set of all ancestors of $X_i$ in $\cG$ and similarly, $\text{desc}(X_i)$ or $X_i^{\uparrow}$ for the set of all its descendants. If $X_i$ is a direct cause of $X_j$, then $X_i$ is said to be a \emph{parent} of $X_j$ or equivalently $X_j$, a \emph{child} of $X_i$. We will use $\text{par}(X_i)$ or $X_i^{\downarrow_1}$ and $\text{child}(X_i)$ or $X_i^{\uparrow_1}$ to denote the sets of all parents and children respectively, of $X_i$ in $\cG$. If a node $X_i$ has no parents in $\mathcal{G}$, then it is said to be \emph{exogenous} with respect to that graph, and is called \emph{endogenous} otherwise. 
 
\begin{figure}[t!]
	\centering
	\subfloat[]{\includegraphics[scale=1.2]{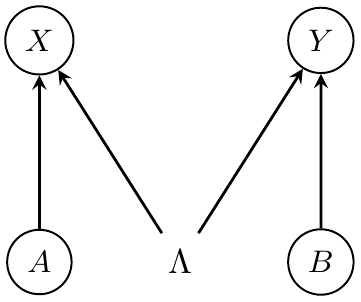}
	\label{fig: Bell}}\qquad\qquad\qquad\qquad\qquad
\subfloat[]{\includegraphics[scale=1.2]{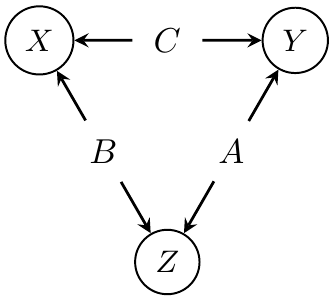}
	\label{fig: Triangle}}
\caption[The bipartite Bell and Triangle causal structures]{\textbf{The bipartite Bell and Triangle causal structures:} Two widely studied causal structures that are known to support a gap between correlations realisable through classical vs non-classical unobserved causes. Here observed nodes are circled while the uncircled nodes are unobserved. \textbf{(a)} The bipartite Bell causal structure: This corresponds to an information processing protocol involving two parties Alice and Bob where the observed nodes $A$ and $B$ represent the random variables corresponding their choice of input (measurement setting) while $X$ and $Y$ represent the random variables corresponding to their outputs (measurement outcome). The outcomes $X$ and $Y$ are obtained by measuring the respective halves of a shared system $\Lambda$, which may be classical, quantum or post-quantum. \textbf{(b)} The Triangle causal structure: Here, the three observed nodes $X$, $Y$ and $Z$ have unobserved, pairwise common causes $A$, $B$ and $C$, but no joint common cause. This corresponds to a information processing protocol involving three parties who produce correlated data ($X$, $Y$ and $Z$) by pre-sharing information pairwise, and never having interacted as a group.}
\label{fig: causalstruc}
\end{figure}

 Figure~\ref{fig: causalstruc} illustrates two causal structures that will be considered extensively Part I of this thesis, particularly in Chapters~\ref{chapter: Tsallispaper} and~\ref{chapter: mixingpaper}. These are the biparite Bell and Triangle causal structures, which we will denote by $\cG_B$ and $\cG_T$ respectively. Both the Bell \cite{Bell, CHSH69, Wood2015, Kaszlikowski2002, CollinsGisin04, Pitowski89, Masanes2002, Bancal2010, Cope2019, BraunsteinCaves88, Weilenmann16} and Triangle \cite{Fritz2012, Chaves2014, Chaves2015, Branciard2012, Weilenmann2017,Renou2019,Kraft2020} causal structures have been studied extensively and continue to be a subject of ongoing research, by virtue of being relatively small causal structures that are known to support non-classical correlations. We first outline the preliminaries of classical causal models \cite{Pearl2009} required for this thesis, and then discuss the generalisation to quantum and post-quantum causal structures, mainly following the approach of \cite{Henson2014}.
 We now enunciate the following broad definition of a causal model. 
 \begin{definition}[General definition of a causal model]
A causal model consists of a causal structure $\cG$, and a joint distribution $P_{X_1,...,X_n}$ over the set $\{X_1,...,X_n\}$ of observed nodes of $\cG$ that is compatible with the causal structure according to a \emph{compatibility condition}. 
\end{definition}

%Causal structure figures

\subsection{Classical causal models}
\label{ssec: classCM}
The section is primarily based on the framework of
classical causal models developed by Judea Pearl~\cite{Pearl2009}.
\subsubsection{Directed graphs and compatible probability distributions}
\begin{definition}[(Classical) causal structures]
\label{definition: Ccausalstructure}
 A causal structure is called classical and denoted $\cG^C$ if all its unobserved nodes correspond to classical random variables (denoted by the same name as the associated node).
\end{definition}

Hence for classical causal structures, all nodes (observed and unobserved) correspond to classical random variables and we will use the same label to denote the node of the DAG as well as the random variable it represents. The framework of classical causal models~\cite{Pearl2009} then specifies when a joint probability distribution over all nodes is said to be compatible with the causal structure.

\begin{definition}[Compatibility of distribution with $\cG^C$]
\label{definition: CDAGcompat}
A distribution $P_{X_1,\ldots,X_n}\in \mathscr{P}_n$ over the random variables $\{X_1,\ldots,X_n\}$ is said to be \emph{compatible} with a classical causal structure $\mathcal{G}^\rC$ (with these variables as nodes) if it satisfies the \emph{causal Markov condition} i.e., the joint distribution decomposes as
\begin{equation}
\label{eq: markov}
    P_{X_1,\ldots,X_n}=\prod\limits_{i=1}^{n} P_{X_i|X_i^{\downarrow_1}},
\end{equation}
where $X_i^{\downarrow_1}$ denotes the set of all parent nodes of the
node $X_i$ in the DAG $\mathcal{G}^\rC$. 
\end{definition}
A crucial concept relating to the joint distribution of a causal structure is \emph{conditional independence} defined as follows.

\begin{definition}[Conditional independence]
Given pairwise disjoint subsets of random variables $X$, $Y$ and $Z$ (where $X$ and $Y$ are non-empty) and a joint distribution $P_{XYZ}$ over them, $X$ and $Y$ are said to be \emph{conditionally independent} given $Z$, denoted $X\indep Y|Z$ if
\begin{equation}
    \label{eq: condindep}
    P_{XY|Z}=P_{X|Z}P_{Y|Z}.
\end{equation}
\end{definition}
The relationship between the \emph{Markov compatibility} condition of Definition~\ref{definition: CDAGcompat} and conditional independence as defined above are given by the following Theorem (Theorem 1.2.7 in \cite{Pearl2009}). 

\begin{theorem}[Pearl 2009]
\label{theorem: localMarkov}
A distribution $P\in \mathscr{P}_n$ is compatible with a classical causal structure $\cG^C$ over a node set $\{X_1,...,X_n\}$ if and only if every node $X_i$ is conditionally independent from its non-descendants, denoted
$X_i^{\nuparrow}$ given its parents $X_i^{\downarrow_1}$ in
$\mathcal{G}^C$ i.e., $\forall i\in \{1,\ldots,n\}$ $X_i \indep X_i^{\nuparrow}|X_i^{\downarrow_1}$, or explicitly
\begin{equation}
\label{eq: localMarkov} 
  P_{X_i
  X_i^{\nuparrow}|X_i^{\downarrow_1}}=P_{X_i|X_i^{\downarrow_1}}P_{X_i^{\nuparrow}|X_i^{\downarrow_1}}  
\end{equation}
\end{theorem}
The equivalent compatibility condition in terms of conditional independence provided by Theorem~\ref{theorem: localMarkov} is often referred to as \emph{local Markov} or \emph{parental Markov} condition in the literature. It is possible to read off the conditional independences satisfied by a Markov compatible distribution from the corresponding causal graph, rules for doing so were independently developed by Geiger~\cite{Geiger1987}
and Verma and Pearl~\cite{Verma1988}. The following graph theoretic notions are required to understand how this can be done.

%\begin{definition}[Paths and colliders]
%Given two distinct nodes $X$ and $Y$ of a DAG $\cG$, a \emph{path} between $X$ and $Y$ in $\cG$ is a sequence of nodes $N_0, N_1,...,N_{k-1},N_k$ with $X$ and $Y$ as end points such that any pair of adjacent nodes in the sequence are connected by an edge in $\cG$. A \emph{directed path} from $X$ to $Y$ is a path between $X$ and $Y$ with $N_0=X$ and $N_k=Y$, where all edges in the path are of the form $N_i\rightarrow N_{i+1}$ i.e., we can get from $X$ to $Y$ in the graph by following the direction specified by the edges in the path. A \emp{collider} is a node $W$ such that there exists a 3-node path $A\rightarrow W\leftarrow B$ for distinct nodes $A$ and $B$ in $\cG$. If no such path with $W$ as a central node exists in $\cG$, then $W$ is called a \emph{non-collider}.
%\end{definition}
%\begin{definition}[Collider and common cause]
%A node $W$ of a causal structure $\cG$ is called a \emph{collider} if there exist two distinct nodes $A$ and $B$ such that $A\longrightarrow W \leftarrow B$ is a sub-graph of $\cG$, otherwise it is called a \emph{non-collider}. It is called the \emph{common cause} of two nodes $A$ and $B$ if $A\leftarrow W\longrightarrow B$ is a sub-graph of $\cG$.
%\end{definition}
%We will often use the terms \emph{collider/non-collider} to refer to the node $W$ as well the associated subgraph involving the additional nodes $A$ and $B$ with $W$ as the central node. 

\begin{definition}[Blocked paths]
Let $\mathcal{G}$ be a DAG in which $X$ and $Y$ are distinct nodes and $Z$ be a
set of nodes not containing $X$ or $Y$.  A path from $X$ to $Y$ is
said to be \emph{blocked} by $Z$ if it contains either $A\longrightarrow W\longrightarrow B$
with $W\in Z$, $A\longleftarrow W\longrightarrow B$
with $W\in Z$ or $A\longrightarrow W\longleftarrow B$ such that neither $W$ nor any descendant of $W$ belongs to $Z$.
\end{definition}

In the following, three node subgraphs of the form $A\longrightarrow W\longleftarrow B$ will be known as \emph{colliders} where $W$ is referred to as the collider node. Other three node subgraphs will be referred to as \emph{non-colliders}, in particular $A\longleftarrow W\longrightarrow B$ corresponds to the \emph{common cause subgraph} where the node $W$ is a common cause of $A$ and $B$.

\begin{definition}[d-separation]
\label{definition:dsep}
Let $\mathcal{G}$ be a DAG in which $X$, $Y$ and $Z$ are disjoint
sets of nodes.  $X$ and $Y$ are \emph{d-separated} by $Z$ in
$\mathcal{G}$, denoted as $(X\perp^d Y|Z)_{\cG}$ (or simply $X\perp^d Y|Z$ if $\cG$ is obvious from context) if every path from a variable in $X$ to a variable in
$Y$ is \emph{blocked} by $Z$, otherwise, $X$ is said to be \emph{d-connected} with $Y$ given $Z$.
\end{definition}

The concept of \emph{d-separation} was developed independently in \cite{Geiger1987} and \cite{Verma1988}. It follows from the above definition that d-separation is a symmetric relation in the first two arguments (as is conditional independence) i.e., $X\perp^d Y|Z$ is equivalent to $Y\perp^dX|Z$. When $Z$ is the empty set $\emptyset$, we will simply denote this as $X\perp^d Y$, and similarly for conditional independences (which will be called \emph{independences} when the conditioning set $Z$ is empty). For arbitrary disjoint subsets $X$, $Y$ and $Z$ of the nodes,
d-separation can be used to determine whether $X$ and $Y$ are conditionally
independent given $Z$ for a distribution compatible with the causal structure (according to Definition~\ref{definition: CDAGcompat}). The following result which is Theorem 1.2.5 of \cite{Pearl2009}, and was originally presented in \cite{Verma1988} illustrates this. 

\begin{theorem}[Verma and Pearl 1988]
\label{theorem: dsepcomplete}
Let $\cG^C$ be a classical causal structure and let $X$, $Y$ and $Z$
be pairwise disjoint subsets of nodes in $\cG^C$. If a probability distribution $P$ is compatible with $\cG$ (according to Definition~\ref{definition: CDAGcompat}), then the d-separation $X\perp^d Y|Z$ implies the conditional independence $X\indep Y|Z$. Conversely, if all distributions $P$ compatible with $\cG^C$ satisfy the conditional independence $X\indep Y|Z$, then the corresponding d-separation relation $X\perp^d Y|Z$ holds. 
%In other words, for all distributions $p$ compatible with $\cG$,
%\begin{equation}
 %   \label{eq: dsepcomplete}
%    X\perp^d Y|Z \Leftrightarrow X\indep Y|Z.
%\end{equation}
\end{theorem}
The forward direction of the above Theorem is referred to as the \emph{soundness of d-separation} and the reverse direction as the \emph{completeness of d-separation} for determining conditional independences. Note that for the latter, it is important that \emph{all} compatible distributions satisfy the conditional independence. It is possible for a distribution $P$ that is compatible with a graph $\cG$ to satisfy a conditional independence $X\indep Y|Z$ even when the d-separation relation $X\perp^d Y|Z$ does not hold in $\cG$. A causal model defined by such a distribution $P$ and causal structure $\cG$ is said to be \emph{unfaithful} or \emph{fine-tuned}. Therefore for a \emph{faithful} causal model, the only conditional independence relations are those that can be read off from the d-separation relations in the associated causal graph. For a faithful causal model, it can be shown that
all the conditional independences in the distribution are implied by the $n$ conditional independence constraints (one for each node) given by the \emph{parental/local Markov condition} (Equation~\eqref{eq: localMarkov}), and can be derived from these
constraints and standard probability calculus based on Bayes' rule. In general, not all of these $n$ constraints may be required. The following example (taken from \cite{Scheines1997}) illustrates a compatible distribution that is unfaithful with respect to its causal graph.

\begin{figure}[t!]
    \centering
 \begin{tikzpicture}
  \node[shape=circle,draw=black] (S) at (0,0) {$S$};  \node[shape=circle,draw=black] (E) at (-2,2) {$E$};  \node[shape=circle,draw=black] (H) at (2,2) {$H$}; 
  \draw [thick, arrows=-stealth] (S) -- (E) node[midway,above] {$+$};   \draw [thick, arrows=-stealth] (S) -- (H) node[midway,above] {$-$};   \draw [thick, arrows=-stealth] (E) -- (H) node[midway,above] {$+$};
 \end{tikzpicture}
    \caption[Example of fine-tuned correlations compatible with a causal structure]{\textbf{Example of fine-tuned correlations compatible with a causal structure: } Even though $S$ is a cause of $H$, it is possible for the correlations $p_{SEH}$ to be fine-tuned such that the positive dependence $S\xrightarrow[]{+}E\xrightarrow[]{+}H$ exactly cancels out the negative dependence $S\xrightarrow[]{-}H$ resulting in $S$ and $H$ being uncorrelated (see Example~\ref{example: fine-tune1}}
    \label{fig:eg_unfaith}
\end{figure}
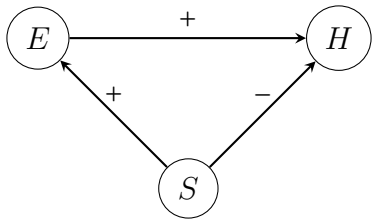

\begin{example}[A fine-tuned causal model]
\label{example: fine-tune1}
Consider the causal structure of Figure~\ref{fig:eg_unfaith} where $+$ or $-$ on an edge indicates whether the nodes connected by that edge are positively or negatively correlated as a result of that causal influence. Note that the Markov condition of Equation~\eqref{eq: markov} does not impose any constraints on joint distributions $P_{SEH}$ compatible with this causal structure, and neither does d-separation. Any probability distribution $P_{SEH}$ can be expressed as $P_{SEH}=P_SP_{E|S}P_{H|ES}$ using Bayes' rules, which is the same factorization obtained by applying the Markov condition to this graph. Further, $S$ influences $H$ through two causal mechanisms--- a direct, negative causal influence and an indirect, positive influence. It is therefore possible to have a compatible distribution $P_{SEH}$ in which these two causal mechanisms exactly cancel each other out such that $S$ and $H$ appear to be uncorrelated i.e., $P_{SH}=P_SP_H$, even though $S$ and $H$ are clearly d-connected in the graph.
\end{example}
%This can be seen more intuitively if we take the variables $S$, $E$ and $H$ to correspond to smoking, exercise and health respectively. Suppose smoking negatively impacts the health of an individual, but for some reason also causes them to exercise more, which in turn positively impacts their health. Then it is possible for the net result to be that the smoking tendencies and the health of the individual are uncorrelated, even though smoking has a causal influence on the individual's health. 
Note that such examples are \emph{fine-tuned} in the sense that the causal influences will have to conspire in a very specific way in order to just cancel out, and even slight deviations in the strength of the correlations will destroy the apparent independence. We will consider such examples in further detail (including explicit distributions) in Chapter~\ref{chapter: jammingpaper}. In the rest of this section, we will focus only on faithful correlations and causal models. We now provide an illustrative example for the concepts of d-separation and conditional independence.

\begin{figure}[t!]
    \centering
\begin{tikzpicture}
   \node[shape=circle,draw=black] (X) at (0,0) {$X$}; \node[shape=circle,draw=black] (R) at (2,0) {$R$}; \node[shape=circle,draw=black] (S) at (4,0) {$S$}; \node[shape=circle,draw=black] (T) at (6,0) {$T$}; \node[shape=circle,draw=black] (U) at (8,0) {$U$}; \node[shape=circle,draw=black] (V) at (10,0) {$V$}; \node[shape=circle,draw=black] (Y) at (12,0) {$Y$}; \node[shape=circle,draw=black] (W) at (6,-2) {$W$}; 
   \path [thick, arrows=-stealth] (X) edge (R); \path [thick, arrows=-stealth] (R) edge (S); \path [thick, arrows=-stealth] (S) edge (T); \path [thick, arrows=-stealth] (U) edge (T); \path [thick, arrows=-stealth] (T) edge (W); \path [thick, arrows=-stealth] (V) edge (U); \path [thick, arrows=-stealth] (V) edge (Y);
\end{tikzpicture}
    \caption[Example illustrating d-separation and conditional independence]{\textbf{Causal structure of Example~\ref{example: d-sep}}}
    \label{fig:eg_dsep}
\end{figure}
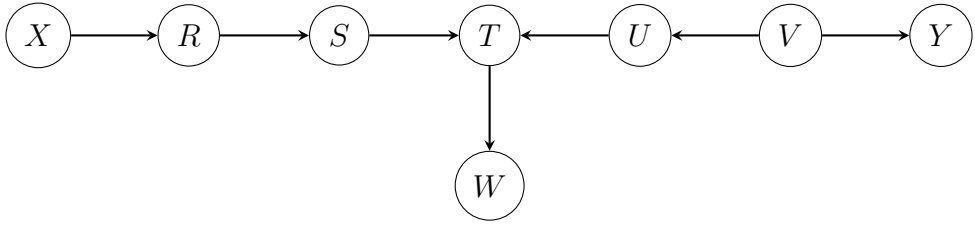

\begin{example}[d-separation and conditional independence]
\label{example: d-sep}
Consider the causal structure of Figure~\ref{fig:eg_dsep}, where all nodes are observed. Here, the node $T$ (and the corresponding sub-graph $S\longrightarrow T\longleftarrow U$) is a \emph{collider} (and all remaining nodes are \emph{non-colliders}), and $V$ is a common cause of $U$ and $Y$. We now consider d-separation relations involving different conditioning sets $Z$. If $Z$ is the empty set, $X\perp^d Y|Z$ (or simply $X\perp^d Y$) because the path between $X$ and $Y$ is blocked by the collider $T$ that does not belong to $Z=\emptyset$. If $\{T,W\}$ or a subset thereof is contained in $Z$, then $X$ and $Y$ are not d-separated (or equivalently, they are \emph{d-connected}) by $Z$ since the path contains the node $T$ that either belongs to $Z$ or has a descendant $W$ that belongs to $Z$. Further, $X$ and $T$ as well as $T$ and $Y$ are d-connected by the empty set $Z=\emptyset$. The former become d-separated when $Z$ contains either one or both of $R$ and $S$ since these block the directed path from $Z$ to $T$. Similarly, $T$ and $Y$ become d-separated by $Z$ if it contains one or both of $U$ and $V$. By the \emph{soundness of d-separation}, this implies the conditional independence relations $X\indep Y$, $X\indep T|\{R,S\}$, $T\indep Y|\{U,V\}$. By finding the set of non-descendants and parents for all 8 nodes, we have the following set of 8 conditional independences corresponding to Equation~\eqref{eq: localMarkov}. For a faithful causal model over the causal structure~\ref{fig:eg_dsep}, all conditional independence relations in every compatible distribution (including those mentioned previously) are implied by these 8 relations. 
\begin{equation}
\label{eq: eq_allCI}
\begin{split}
      X&\indep \{U,V,Y\},\\
      R&\indep \{U,V,Y\}|X,\\ 
      S&\indep \{U,V,Y\}|R,\\
      T&\indep \{X,R,V,Y\}|\{S,U\},\\ 
      W&\indep \{X,R,S,U,V,Y\}|T, \\
      U&\indep \{X,R,S,Y\}|V,\\
      V&\indep \{X,R,S\},\\
      Y&\indep \{X,R,S,T,W,U\}|V.
\end{split}
\end{equation}
\end{example}

\paragraph{d-separation and entropic causal constraints:}
An equivalent statement of Theorem~\ref{theorem: dsepcomplete} can be obtained by encoding conditional independences in terms of the entropies (rather than the probability distribution) over the nodes. For a classical causal structure $\cG$ with pairwise disjoint subsets of nodes $X$, $Y$ and $Z$, $X\indep Y|Z$ is equivalent to the entropic constraint $I(X:Y|Z)=0$, where $I(X:Y|Z)$ is the Shannon conditional mutual information (Definition~\ref{definition: ShanMI}). Then, $X$ and $Y$ are d-separated by $Z$ in $\cG$ if and only if
$I(X:Y|Z)=0$ for all distributions compatible with
$\cG$~\cite{Pearl2009}. The complete set of d-separation
conditions give all the conditional independence relations implied by
the DAG (by soundness of d-separation). In the case of Shannon entropy for a DAG with $n$ nodes
these are all implied by the $n$ constraints
\begin{equation}
\label{eq: shancausmain}
  I(X_i:X_i^{\nuparrow}|X_i^{\downarrow_1})=0 \quad \forall i\in\{1,\ldots,n\}. 
\end{equation}
In other words, a distribution over $n$ variables satisfies Equation~\eqref{eq: localMarkov} and hence
Equation~\eqref{eq: markov} (by Theorem~\ref{theorem: localMarkov}) if and only if it satisfies
Equation~\eqref{eq: shancausmain}.

\subsubsection{Interventions and counterfactuals}
\label{sssec: interventions}
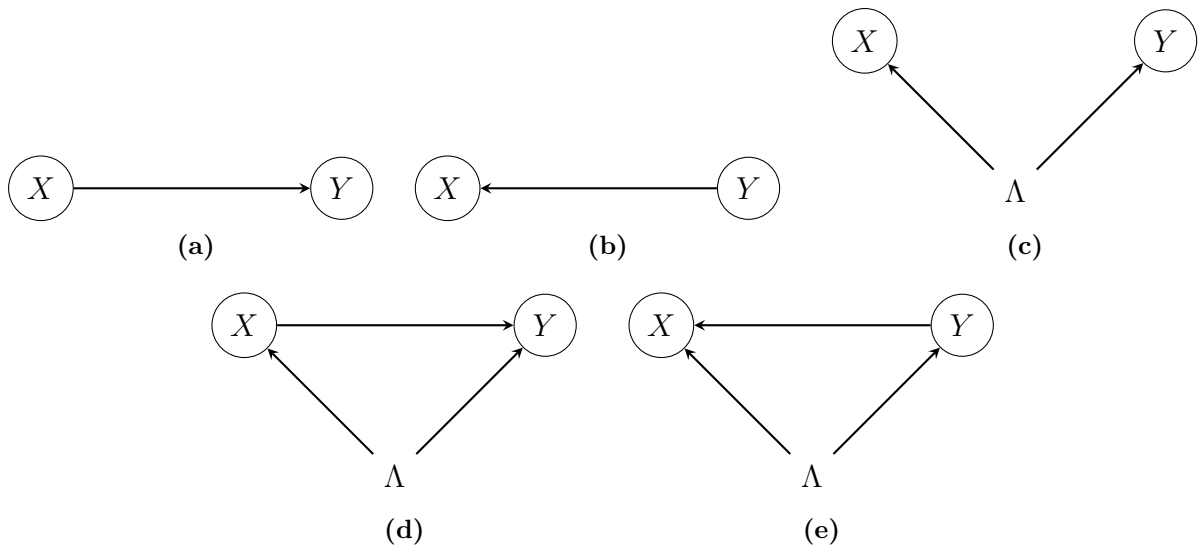
\begin{figure}[t!]
    \centering
\subfloat[]{\begin{tikzpicture}
\node[shape=circle,draw=black] (X) at (0,0) {$X$}; \node[shape=circle,draw=black] (Y) at (4,0) {$Y$}; \draw[thick, arrows=-stealth] (X)--(Y);
\end{tikzpicture}
	 \label{fig:2nodeCSa}}\quad
	\subfloat[]{\begin{tikzpicture}
\node[shape=circle,draw=black] (X) at (0,0) {$X$}; \node[shape=circle,draw=black] (Y) at (4,0) {$Y$}; \draw[thick, arrows=-stealth] (Y)--(X); 
\end{tikzpicture}
 \label{fig:2nodeCSb}	}\quad
	\subfloat[]{\begin{tikzpicture}
\node[shape=circle,draw=black] (X) at (0,0) {$X$}; \node[shape=circle,draw=black] (Y) at (4,0) {$Y$}; \node[shape=circle,draw=none] (L) at (2,-2) {$\Lambda$};\draw[thick, arrows=-stealth] (L)--(X); \draw[thick, arrows=-stealth] (L)--(Y); 
\end{tikzpicture}
 \label{fig:2nodeCSc}	}
	
		\subfloat[]{\begin{tikzpicture}
\node[shape=circle,draw=black] (X) at (0,0) {$X$}; \node[shape=circle,draw=black] (Y) at (4,0) {$Y$}; \node[shape=circle,draw=none] (L) at (2,-2) {$\Lambda$};\draw[thick, arrows=-stealth] (L)--(X); \draw[thick, arrows=-stealth] (L)--(Y); \draw[thick, arrows=-stealth] (X)--(Y);
\end{tikzpicture}
 \label{fig:2nodeCSd}	}\quad
	\subfloat[]{\begin{tikzpicture}
\node[shape=circle,draw=black] (X) at (0,0) {$X$}; \node[shape=circle,draw=black] (Y) at (4,0) {$Y$}; \node[shape=circle,draw=none] (L) at (2,-2) {$\Lambda$};\draw[thick, arrows=-stealth] (L)--(X); \draw[thick, arrows=-stealth] (L)--(Y); \draw[thick, arrows=-stealth] (Y)--(X);
\end{tikzpicture}
 \label{fig:2nodeCSe}	}
    \caption[Two node causal structures with same set of compatible distributions]{\textbf{Two node causal structures with same set of compatible distributions: } All of the above causal structures contain two observed nodes $X$ and $Y$. In all cases, neither the Markov compatibility condition~\eqref{eq: markov} nor d-separation imply any constraints on the marginal distribution $P_{XY}$ over the observed nodes. Therefore any 2-variable distribution $P_{XY}$ is compatible with all 5 of these graphs, and the causal structure cannot be inferred directly from these correlations. However, if we were given, or could deduce additional information regarding the results of active interventions on $X$ and $Y$, we would be able to rule out some of these explanations. For example, if forcibly setting $X$ to some value $x$ (without changing anything else), resulted in a change in the distribution over $Y$, we would say that $X$ is a cause of $Y$. In this case, the opposite direct cause explanations (b) and (e) as well as the purely common cause explanation (c) would  be ruled out, leaving us with (a) and (d) as possible explanations.}
    \label{fig:2nodeCS}
\end{figure}
Given a causal structure, we can apply Definition~\ref{definition: CDAGcompat} to obtain the set of all joint distributions compatible distributions with it. Causal inference concerns the reverse problem--- given a compatible distribution (and possibly some additional information), can we infer the causal structure that it is compatible with? It is not possible to infer an underlying causal structure from correlations alone: correlations are symmetric while causal relationships are directional. For example,  if two variables $X$ and $Y$ are correlated, Reichenbach's principle \cite{Reichenbach1956} asserts that either $X$ must be a cause of $Y$, $Y$ must be a cause of $X$, $X$ and $Y$ share a common cause or any combination thereof and any distribution $P_{XY}$ over these variables is compatible with all five of these causal structures illustrated in Figure~\ref{fig:2nodeCS}. Hence these causal explanations cannot be distinguished immediately using observed correlations alone, and additional information/deductions are required in order to infer the underlying causal structure.  Intuitively, we can argue that if actively intervening or ``doing'' something only to $X$ changed something about $Y$, then we definitely know that $X$ is a cause of $Y$.  This intuition is formalised in terms of interventions and do-conditionals \cite{Pearl2009}, which in addition to the observed correlations allow for the inference of the underlying causal structure in several scenarios that cannot be identified from the correlations alone. We will follow the augmented graph approach (Section 3.2.2 of \cite{Pearl2009}) to define interventions and do-conditionals and only consider the case of perfect or ideal interventions in this thesis. 

First, consider a causal model associated with a causal structure $\cG$ over a set $\{X_1,...,X_n\}$ of nodes, all of which are observed. External intervention on a node $X_i$ can be described using an augmented graph $\cG_{I_{X_i}}$ which is obtained from the original graph $\cG$ by adding a node $I_{X_i}$ and an edge $I_{X_i}\longrightarrow X_i$ (with everything else unchanged). The intervention variable $I_{X_i}$ can take values in the set $\{idle, \{\text{do}(x_i)\}_{x_i\in X_i}\}$, where $I_{X_i}=idle$ corresponds to the case where no intervention is performed (i.e., the situation described by the original causal model) and $I_{X_i}=\text{do}(x_i)$ forces $X_i$ to take the value $x_i$ by cutting off its dependence on all other parents. Hence the parents of the node $X_i$ in the original and augmented graphs are related as $\text{par}_{\cG_{I_{X_i}}}(X_i)=\text{par}_{\cG}(X_i)\cup I_{X_i}$. We will denote $\text{par}_{\cG_{I_{X_i}}}(X_i)$ simply as $\text{par}_{I}(X_i)$ and the $\text{par}_{\cG}(X_i)$ as $\text{par}(X_i)$, in short-hand.  The conditional probability of $X_i=x_i$ conditioned on its parents $\text{par}(X_i)$/$\text{par}_{I}(X_i)$ taking the value $p_{X_i}$/$p_{I_{X_i}}$ is given as.
\begin{equation}
\label{eq: intervention}
   P(x_i|p_{I_{X_i}}):= \begin{cases}
   P(x_i|p_{X_i}),\quad &\text{if } I_{X_i}=idle\\
   0,\quad &\text{if } I_{X_i}=\text{do}(x_i') \text{ and } x_i'\neq x_i \\
   1,\quad &\text{if } I_{X_i}=\text{do}(x_i') \text{ and } x_i'= x_i\\
    \end{cases}
\end{equation}

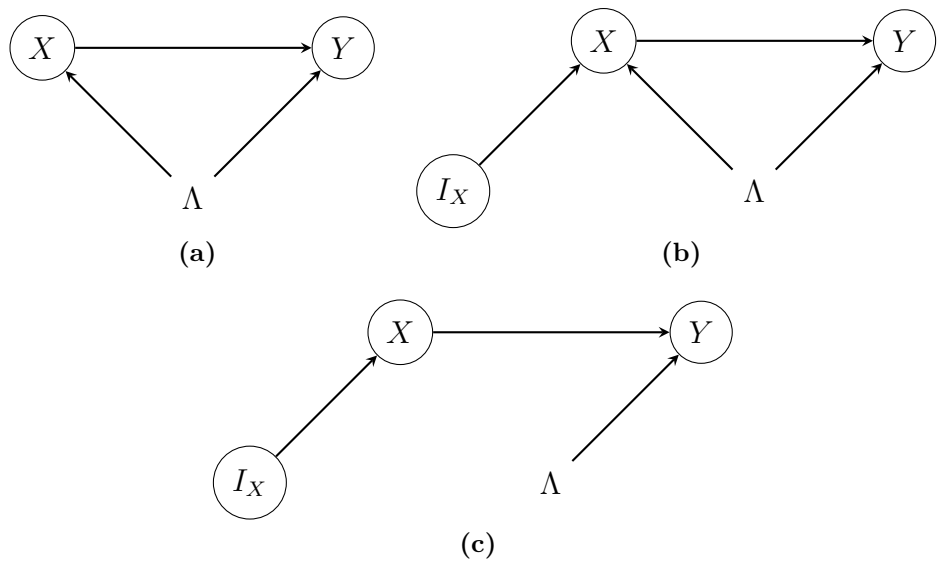
\begin{figure}[t!]
    \centering
		\subfloat[]{\begin{tikzpicture}
\node[shape=circle,draw=black] (X) at (0,0) {$X$}; \node[shape=circle,draw=black] (Y) at (4,0) {$Y$}; \node[shape=circle,draw=none] (L) at (2,-2) {$\Lambda$};\draw[thick, arrows=-stealth] (L)--(X); \draw[thick, arrows=-stealth] (L)--(Y); \draw[thick, arrows=-stealth] (X)--(Y);
\end{tikzpicture}
	}\quad	\subfloat[]{\begin{tikzpicture}
\node[shape=circle,draw=black] (X) at (0,0) {$X$}; \node[shape=circle,draw=black] (Y) at (4,0) {$Y$}; \node[shape=circle,draw=none] (L) at (2,-2) {$\Lambda$}; \node[shape=circle,draw=black] (I) at (-2,-2) {$I_X$};\draw[thick, arrows=-stealth] (L)--(X); \draw[thick, arrows=-stealth] (L)--(Y); \draw[thick, arrows=-stealth] (X)--(Y); \draw[thick, arrows=-stealth] (I)--(X);
\end{tikzpicture}
	}\quad
	\subfloat[]{\begin{tikzpicture}
\node[shape=circle,draw=black] (X) at (0,0) {$X$}; \node[shape=circle,draw=black] (Y) at (4,0) {$Y$}; \node[shape=circle,draw=none] (L) at (2,-2) {$\Lambda$}; \node[shape=circle,draw=black] (I) at (-2,-2) {$I_X$}; \draw[thick, arrows=-stealth] (L)--(Y); \draw[thick, arrows=-stealth] (X)--(Y); \draw[thick, arrows=-stealth] (I)--(X);
\end{tikzpicture}
	}
    \caption[Pre-intervention, augmented and post-intervention causal structures]{\textbf{Pre-intervention, augmented and post-intervention causal structures: } Taking the original, pre-intervention causal structure to be that of Figure~\ref{fig:2nodeCSd} (repeated here in (a) for completeness), parts (b) and (c) of this figure illustrate the augmented and post-intervention causal structures corresponding to an intervention on $X$ in the causal structure of (a). In the former, the variable $I_X$ can take values in the set $\{idle,\{\text{do}(x)\}_{x\in X}\}$ while in the latter, it can only take the values $\{\text{do}(x)\}_{x\in X}$ corresponding to an active intervention. Conditioned on $I_X=idle$, we effectively obtain the original causal structure (a) which corresponds to no intervention being performed, as specified by Equation~\eqref{eq: do}.}
    \label{fig:augCS}
\end{figure}

From this, we see that whenever $I_{X_i}\neq idle$, $X_i$ no longer depends on its original parents par$(X_i)$. Therefore, conditioned on $I_{X_i}\neq idle$, it is illustrative to consider a new graph which we denote by $\cG_{\text{do}(X_i)}$ that represents the post-intervention causal structure after a non-trivial intervention has been performed. The causal graph $\cG_{\text{do}(X_i)}$  is obtained by cutting off all incoming arrows to $X_i$ except the one from $I_{X_i}$ in the causal graph $\cG_{I_{X_i}}$, with everything else unchanged. An example of the graphs $\cG$, $\cG_{I_{X_i}}$ and $\cG_{\text{do}(X_i)}$ is given in Figure~\ref{fig:augCS}. Then the effect of an intervention setting $X_i=x_i'$, i.e., performing $\text{do}(x_i')$ is to transform the original probability distribution $P(X_1=x_1,...,X_n=x_n)$ into a new probability distribution $P(X_1=x_1,...,X_n=x_n|\text{do}(X_i=x_i))$ (the \emph{do-conditional}) given as
\begin{equation}
\label{eq: do}
  P(x_1,...,x_n|\text{do}(x_i)):=P_{\cG_{\text{do}(X_i)}}(x_1,...,x_n|I_{X_i}=\text{do}(x_i)),
\end{equation}
where $P(x_1,...,x_n,I_{X_i}=\text{do}(x_i))$ is compatible with the graph $\cG_{\text{do}(X_i)}$ and  Equation~\eqref{eq: intervention},
which implies that $P(x_1,...x_i',...,x_n|\text{do}(x_i))=0$ if $x_i'\neq x_i$. Applying the Markov compatibility condition (Definition~\ref{definition: CDAGcompat}) to the original, pre-intervention graph $\cG$ implies that all distributions compatible with this graph must satisfy $P_{X_1,...,X_n}=\Pi_{j=1}^nP_{X_j|\text{par}(X_j)}$. In the post-intervention graph $\cG_{\text{do}(X_i)}$, the only causal relations that change are those involving the intervened node $X_i$ where the set of parents of $X_i$ in $\cG_{\text{do}(X_i)}$ is the single element set $\{I_{X_i}\}$, and the parent sets of all other nodes remain the same as those of $\cG$. Then any distribution $P_{X_1,...,X_n,I_{X_i}}$ with $I_{X_i}$ taking values in $\{\text{do}(x_i)\}_{x_i\in X_i}$ (i.e., not idle), that is compatible with $\cG_{\text{do}(X_i)}$ satisfies $P_{X_1,...,X_n|I_{X_i}}=\Pi_{j=1,j\neq i}^nP_{X_j|\text{par}(X_j)}P_{X_i|I_{X_i}}$, for any distribution over $I_{X_i}$. Noting that $P_{X_i|I_{X_i}}$ is the deterministic distribution which is 1 whenever $X_i$ takes the same value set by $I_{X_i}$ and 0 otherwise, this gives the following transformation between the do-conditional (Equation~\eqref{eq: do}) and the original distribution,
\begin{equation}
    \label{eq: do_transf}
    P(x_1,...,x_n|\text{do}(x_i'))=\begin{cases}
    \frac{P(x_1,...,x_n)}{P(x_i|\text{par}(x_i))},\quad &\text{if } x_i'=x_i\\
    0,\quad &\text{if } x_i'\neq x_i.
    \end{cases}
\end{equation}

Importantly, this allows us to decide, given two nodes $X$ and $Y$, whether or not $X$ is a cause of $Y$--- if there exists a value $x\in X$ such that $P(y|\text{do}(x))\neq P(y)$ for some $y\in Y$, then $X$ is a cause of $Y$ (or equivalently, there is a directed path from $X$ to $Y$). Otherwise, we can conclude that $X$ is not a cause of $Y$ or equivalently that there is no directed path from $X$ to $Y$ (if the causal model is faithful)\footnote{Note that in unfaithful causal models, it is possible to have $P(y|\text{do}(x))= P(y)$ for all $x\in X$ and $y\in Y$ even when $X$ is a cause of $Y$, and checking for the violation of this condition only provides a sufficient test of causation but not a necessary one. This point is crucial for the results of Chapter~\ref{chapter: jammingpaper} , which concern unfaithful causal models.}. This formalises the intuition explained in Figure~\ref{fig:2nodeCS}, where checking whether an active intervention ($I_X\neq idle$) on $X$ changes the distribution over $Y$ allowed us to rule out 3 of 5 causal explanations. As we will see in the following paragraphs, checking whether the do-conditional $P(y|\text{do}(x))$ differs from the regular conditional $P(y|x)$, can allow us to pick out a unique causal structure in this case. 

\paragraph{Interventions on subsets of nodes: } This procedure naturally extends to interventions on subsets of the nodes. If a simultaneous intervention is performed on a subset $X$ of the nodes $\{X_1,...,X_n\}$, an intervention variable $I_{X_i}$ will be introduced for each element $X_i$ of the subset $X$, along with the corresponding edge $I_{X_i}\longrightarrow X_i$. Then conditioning on each of the $I_{X_i}$ for $X_i\in X$ being $idle$, the original graph is obtained and conditioning on $I_{X'}\neq idle$ for $X'\subseteq X$ (i.e., a ``do'' operation has been performed on every node in $X'$), the corresponding graph $\cG_{\text{do}(X')}$ is defined by cutting off all incoming edges (except $I_{X_i}\longrightarrow X_i$ for all $X_i\in X'$) in $\cG$. Then the do conditional $P(x_1,...,x_n|\text{do}(x'))$ and its relation to the original distribution $P(x_1,...,x_n)$ are defined analogously, using Equations~\eqref{eq: do} and~\eqref{eq: do_transf} for the graph $\cG_{\text{do}(X')}$.

The importance of the transformation~\eqref{eq: do_transf} is that it allows for \emph{counter-factual reasoning} by mathematically relating the post-intervention distribution to the pre-intervention distribution. It tells us that if all the nodes in a causal structure are observed, then we can deduce the result of any intervention (possibly on a subset of nodes) just by passively observing the correlations over the variables and using mathematical manipulations, \emph{without physically performing any interventions}. This ability to deduce the post-intervention distribution only from the observed distribution in a consistent manner is known as \emph{identifiability}. Identifiability of interventional effects is of crucial practical use, especially in situations where an actually intervention may be physically (or ethically) forbidden, for example testing the effect of certain drugs and treatments on humans. In causal structures that include unobserved nodes, this kind of counter-factual reasoning may not always be possible i.e., the effect of interventions may not be identifiable from the observed distributions, as explained below. 

\paragraph{Interventions in the presence of unobserved nodes: } For a causal structure containing unobserved nodes, interventions can only be performed on the observed nodes. For a classical causal structure $\cG^C$, the unobserved nodes are also random variables. Let $\{X_1,...,X_n\}$ and $\{\Lambda_1,...,\Lambda_m\}$ denote the set of observed and unobserved nodes of $\cG^C$ respectively. Any observed distribution $P_{X_1,...,X_n}$ compatible with $\cG$ can be obtained by applying the Markov condition~\eqref{eq: markov} to the hypothetical distribution over all nodes and summing over all possible values of the unobserved nodes i.e.,
\begin{equation}
    P_{X_1,...,X_n}=\sum_{\Lambda_1,...,\Lambda_m}\Pi_{i=1}^nP_{X_1|\text{par}(X_i)}\Pi_{j=1}^mP_{\Lambda_j|\text{par}(\Lambda_j)}.
\end{equation}
Considering interventions only on observed nodes, Equations~\eqref{eq: do} and ~\eqref{eq: do_transf} can be derived analogously using this compatibility condition. However there are two important distinctions to be noted. Firstly, the observed distribution $P_{X_1,...,X_n}$ is in general not Markov (and hence not compatible) with respect to the sub-graph involving only the observed nodes and edges connecting them. However the overall distribution over observed and unobserved nodes is Markov with respect to the complete DAG involving all these nodes if we assume that graph accounts for all the unobserved common causes and corresponds to a classical causal structure. Secondly, the post-intervention distribution may not be identifiable from the observed, pre-intervention distribution alone. This is because, an intervened node $X_i$ may contain unobserved parents in the original causal structure and Equation~\eqref{eq: do_transf} would then require division by a quantity that cannot be obtained from the observed distribution alone. More explicitly, the following example describes a non-identifiable causal structure involving an unobserved node.

\begin{example}[A non-identifiable causal structure]
\label{example:nonidentifiable}
Consider the causal structure of Figure~\ref{fig:2nodeCSd} with $X$, $Y$ observed and $\Lambda$ unobserved. The augmented and post-intervention causal structures corresponding to an intervention on $X$ are illustrated in Figure~\ref{fig:augCS}. Observed distributions compatible with the original causal structure are given as $P_{XY}=\sum_{\Lambda}P_{\Lambda XY}=\sum_{\Lambda}P_{\Lambda}P_{X|\Lambda}P_{Y|X\Lambda}$. Note that \emph{any} distribution $P_{XY}$ can be written in this form using Bayes' rule and marginalising over some variable $\Lambda$. The post-intervention distribution $P_{XY|I_X}$ (with $I_X\neq idle$) is compatible with the graph $\cG_{\text{do}(X)}$ and is given as $P_{XY|I_X}=\sum_{\Lambda}P_{\Lambda}P_{X|I_X}P_{Y|X\Lambda}$. Equation~\eqref{eq: do_transf} in this case corresponds to
\begin{equation}
    P(xy|\text{do}(x'))=\begin{cases}
   \sum_{\lambda\in\Lambda}\frac{P(\lambda xy)}{P(x|\lambda)}= \sum_{\lambda\in\Lambda}P(\lambda)P(y|x\lambda),\quad &\text{if } x'=x\\
   0,\quad &\text{if } x'\neq x.
    \end{cases}
\end{equation}
Note that the transformation involves division by a value corresponding to the distribution $P_{X|\Lambda}$ which cannot be determined from the observed distribution $P_{XY}$ alone. Hence the effect of the intervention is not identifiable for this causal structure. We will discuss identifiability of causal structures in further detail in Chapter~\ref{chapter: jammingpaper} .
\end{example}
In the above example, the do-conditional $P(y|\text{do}(x))$ is simply the non-zero expression in the above equation i.e., $P(y|\text{do}(x))=\sum_{\lambda\in\Lambda}P(\lambda)P(y|x\lambda)$. The regular conditional $P(y|x)$ is obtained using the Markov compatibility condition as $P(y|x)=\frac{\sum_{\lambda\in\Lambda}P(\lambda)P(x|\lambda)P(y|x\lambda)}{P(x)}$. As long as $P(x|\lambda)\neq P(x)$ for some values $x\in X, \lambda\in \Lambda$, the two conditionals will be different. Note that this must indeed hold i.e., $X$ and $\Lambda$ cannot be independent if the distribution is faithful, since they are d-connected in the causal structure. Therefore, for faithful causal models over the causal structure~\ref{fig:2nodeCSd}, there exist values $x\in X,y\in Y$ such that $P(y|\text{do}(x))\neq P(y|x)$. On the other hand, one can easily check that for the causal structure~\ref{fig:2nodeCSa}, $P(y|\text{do}(x))= P(y|x)$ always holds--- the intuition being that, in this case $X$ is a parentless (i.e., exogenous) node, hence the intervention does not change any of the causal relations that were present in the original graph. In general, interventions on an exogenous node $X$ does not change the distribution, and whether or not $X$ is a cause of $Y$ in this case can simply be decided by checking whether or not $X$ and $Y$ are correlated i.e., whether $P_{Y|X}=P_Y$. In conclusion, while the do-conditional $P(y|\text{do}(x))$ for the causal structure~\ref{fig:2nodeCSd} is not identifiable from the original observed distribution $P(xy)$, it can be determined experimentally by physically performing the intervention. Comparing this experimentally obtained estimate of the do-conditional with the original distribution will allow us to uniquely identify one among the 5 causal structures of Figure~\ref{fig:2nodeCS}, if we assume that the causal model is faithful.

\begin{remark}
Note that there is a slight but harmless abuse of notation in this section. The intervention variable $I_X$ takes different values in the augmented graph $\cG_{I_X}$ and the post-intervention graph $\cG_{\text{do}(X)}$. It takes values in $\{idle,\{\text{do}(x)\}_{x\in X}\}$ and $\{\text{do}(x)\}_{x\in X}$ in the two cases respectively. We nevertheless use the variable name to also denote the set of possible values it takes, but will be careful to specify which graph we are referring to whenever this happens.
\end{remark}

\subsection{Non-classical causal models}
\label{ssec: NCcausstr}
A typical laboratory experiment uses devices that take classical inputs (settings of knobs or buttons) and produce classical outputs (reading of a pointer or numbers on a screen). This classical, observed data may however be produced through quantum mechanisms. Bell \cite{Bell} famously showed that local measurements on a shared quantum system can lead to stronger correlations than those on any shared classical system. In recent years, Wood and Spekkens \cite{Wood2015} have reformulated Bell's Theorem in the language of Bayesian networks \cite{Pearl2009}, showing that quantum correlations cannot be explained by faithful classical causal models, and therefore the framework of \cite{Pearl2009} must be suitably generalised for defining quantum causal models\footnote{Unless one is willing to accept fine-tuned classical causal explanations that involve hidden influences propagating superluminally or retrocausally.}. More generally, one can also consider post-quantum systems such as non-local or non-signaling boxes \cite{Tsirelson1993, Popescu1994}, that produce stronger-than-quantum correlations. These can be seen as elements of a more general class of theories, generalised probabilistic theories (GPTs, see Section~\ref{sec: GPT}), of which quantum and classical theories emerge as particular cases. Hensen, Lal and Pusey (HLP) \cite{Henson2014} were the first to define a generalised causal model that allows unobserved nodes to be either classical, quantum or more generally, GPT. This enables a characterisation of the theory-independent aspects of causal models and allows us to derive bounds on the correlations realisable in causal structures depending on the nature of their unobserved nodes, which has foundational as well as practical applications.  
\subsubsection{Generalised causal structures and compatibility}
\label{sssec: gen_causalstr}
\begin{definition}[Generalised causal structure]
A \emph{generalised causal structure}, denoted $\cG^G$ is a causal structure in which every observed node is associated with a classical random variable and every unobserved node is associated with a system belonging to a generalised probabilistic theory. Generalised causal structures include quantum $\cG^Q$ and classical causal structures $\cG^C$. In the former case, each unobserved node is associated with a quantum system, and in the latter case, each unobserved node is associated with a classical random variable. 
\end{definition}

For a classical causal structure $\cG^C$, the Markov factorization condition~\eqref{eq: markov} provides a compatibility condition for a joint distribution with the causal structure. In case there are unobserved nodes, the observed distribution is simply obtained by taking a joint distribution over all the nodes that factorises according to the Markov condition, and then marginalising over the unobserved variables. If the unobserved systems are non-classical, this does not work since a joint distribution over all the nodes may no longer exist\footnote{Quantum and GPT systems do not co-exist with the outcomes of measurements performed on them, therefore one cannot in general assign a joint probability distribution over quantum/GPT nodes and their children.} and a more general compatibility condition that takes into account quantum and GPT mechanisms is required. 

\paragraph{Post-quantum/GPT causal structures: }The operational probabilistic framework of \cite{Chiribella2010} provides a method of representing operations and transformations on GPT systems using circuit-like diagrams. Elements of this generalised circuit are called \emph{tests}, which take as input a system and produce as output another system, and a classical outcome variable. Inputs and outputs of different tests are connected through \emph{wires} which carry the associated systems along them. If a test has a trivial input system, then it is called a \emph{preparation test} and if it has a trivial output system, it is called an \emph{observation test}. Any test from the trivial system to itself is a probability distribution, hence probabilities can be generated by composing tests using wires.  In the HLP framework \cite{Henson2014}, every node of a generalised causal structure is associated with a test. In the case of observed nodes, the output system of the test is simply a classical random variable and for unobserved nodes, this corresponds to a GPT system. Further, the systems described by the operational probabilistic framework \cite{Chiribella2010} are \emph{causal} in the sense that an outcome at an earlier time cannot depend on the choice of operation performed on the system at a later time. Then a distribution $P$ over the observed nodes of a generalised causal structure $\cG^G$ is said to be \emph{compatible} with $\cG^G$ if there exists a causal operational probabilistic theory, a test corresponding to each node and a system corresponding to each edge of $\cG$ in the theory that generate the observed distribution $P$. The formula for generating $P$ in this manner is analogous to the Markov condition~\eqref{eq: markov}, where, instead of a product of conditional distributions $P_{X_i|X_i^{\downarrow_1}}$, we have a product of tests that map the unobserved parents of a node to its unobserved children, conditioned on the observed parents. This is called the \emph{generalised Markov condition}, and the corresponding generalised causal structure is represented by a directed acyclic graph or a GDAG. The set of all distributions compatible with a generalised causal structure $\cG^G$ will be denoted by $\mathscr{P}(\cG^G)$. Note that by construction these are distributions over observed nodes only.

\paragraph{Quantum causal structures: } A GPT of particular interest is quantum theory and the causal structure framework of \cite{Henson2014} includes quantum causal structures as a special case. There are however several frameworks for describing quantum causal structures that are compatible with each other, but differ essentially on how nodes and edges of the causal structure are identified with quantum systems and transformations. We will mainly stick to the description that follows from the generalised causal structures of \cite{Henson2014}, to provide a unified picture, but use a language and notation similar to that of \cite{Weilenmann2017} for convenience. We will discuss the relation to other approaches at the end of this section. For a quantum causal structure $\cG^Q$, every exogenous node $A$ i.e., those without any incoming edges correspond to a preparation of a density matrix $\rho_A\in \mathscr{S}(\mathscr{H}_A)$. In the case of observed nodes, $\rho_A$ is classical. This can be seen as a preparation test, where the trivial input system is associated with the one dimensional Hilbert space $\mathscr{H}=\mathbb{C}$. Each directed edge corresponds to a quantum system and hence to a Hilbert space, which we will label by the starting and ending node of the edge. For example, if a node $A$ only has two children $Y$ and $Z$, then the Hilbert space of $A$ will be factorised as $\mathscr{H}_A=\mathscr{H}_{A_Y}\otimes\mathscr{H}_{A_Z}$, where the subsystem $\mathscr{H}_{A_Y}$ corresponds to the edge from $A$ to $Y$, and $\mathscr{H}_{A_Z}$ to the edge from $A$ to $Z$. If $A$ is a classical node, then the two factors $A_Y$ and $A_Z$ are taken to be copies of $A$, since classical information can be copied. 

As in the HLP framework, each node is labelled by its output state which is classical for observed nodes and quantum for unobserved nodes in $\cG^Q$. Every observed node $X$ represents the output statistics of a measurement performed on the incoming systems. If this node has unobserved parents then $X$ can be seen as the result of a quantum measurement (implemented through a POVM) on those incoming quantum states, where the choice of measurement can possibly depend on the classical value of other observed parents of $X$. If $X$ only has observed parents, then it can be seen as the output as a stochastic map on the incoming classical random variables. More generally, tests correspond to CPTP maps that map the joint state of all incoming edges to the joint state of all outgoing edges. Preparations, measurements (POVMs) and arbitrary classical stochastic maps can be seen as special cases of CPTP maps. A distribution, $P$ over the observed nodes of a causal structure $\cG^Q$ is \emph{compatible} with $\cG^Q$ if there exists a quantum state labelling each unobserved node (with subsystems for each unobserved edge) and CPTP maps, i.e., preparations or quantum transformations for each unobserved node as well as POVMs or stochastic maps for each observed node, that allow for the generation of $P$ in accordance with the Born rule. We will denote the set of all observed distributions compatible with a quantum causal structure $\cG^Q$ by $\mathscr{P}(\cG^Q)$, and for a classical causal structure $\cG^C$ by $\mathscr{P}(\cG^C)$.

\paragraph{Theory dependent and theory-independent aspects: } We already noted that in causal structures with non-classical nodes, a joint distribution over all nodes may no longer be defined. Further in classical causal structures, it is implicitly assumed that the complete information about a node is transmitted to each of its children. Then each node receives complete information about all its parents and performs a classical stochastic map to obtain the output. The stochastic nature of this map is taken to arise from lack of knowledge about additional ``error variables'', and once the parents and the error variables are given, the map corresponds to a deterministic function\footnote{Introducing suitable error variables and expressing the functional dependence of each node on its parents through structural equations is the basis of the structural causal model approach for classical causal structures. This is compatible with the Bayesian networks approach discussed in Section~\ref{ssec: classCM}, and can been seen as providing underlying classical mechanisms to the probability assignments. Both approaches are covered in Pearl's book \cite{Pearl2009}.}. These aspects are not problematic in the classical case since classical information can be copied. However arbitrary states of quantum and GPT systems cannot be copied due to the \emph{no-cloning theorem}. Hence, after a measurement on such a system, the measurement outcome and the initial state do not co-exist and one can no longer condition on unobserved systems in the same manner\footnote{The quantum causal models of \cite{Allen2017, Leifer2013} tackle this issue by introducing the notion of a conditional quantum state of an output conditioned on an earlier input. These are interpreted not as quantum states in the usual sense (since they correspond to systems at different times that do not coexist), but as a mathematical representations of the quantum channel transforming the input to the output system.}. Neither can complete information about the state of an unobserved node be transmitted to all its children. Further, arbitrary quantum or GPT maps cannot be reduced to deterministic functions by including additional information.
These theory-dependent differences mean that the sets of compatible correlations $\mathscr{P}(\cG^C)$, $\mathscr{P}(\cG^Q)$,  and $\mathscr{P}(\cG^G)$ for a causal structure $\cG$ need not coincide in general. Causal structures that support such a gap between the sets of classical and non-classical correlations are termed as \emph{interesting causal structures}, and \cite{Henson2014} identify certain necessary conditions for a causal structure to be interesting. To do so, they prove the following \emph{theory-independent} property of generalised causal structures, which is an analogue of Theorem~\ref{theorem: dsepcomplete}, showing that the d-separation condition applies to classical, quantum and post-quantum causal structures alike. Identifying such theory-independent aspects is also useful because for instance, there is no consensus on the representation of states and transformations in arbitrary GPTs, even though this exists for specific GPTs such as \emph{box world} (See Section~\ref{sec: GPT}).

\begin{theorem}[Hensen, Lal, Pusey 2014]
\label{theorem: gen_dsep}
Let $\cG^G$ be a generalised causal structure and let $X$, $Y$ and $Z$ be pairwise disjoint subsets of observed nodes in $\cG^G$ . If a probability distribution $P$ is compatible with $\cG^G$, then the d-separation $X\perp^dY|Z$ implies the conditional independence
$X\indep Y|Z$. Conversely, if for every distribution $P$ compatible with $\cG^G$, the conditional independence $X\indep Y|Z$ holds, then the d-separation relation $X\perp^d Y|Z$ also holds in $\cG^G$.
\end{theorem}

Using this, necessary conditions for a causal structure $\cG$ to be interesting are found in \cite{Henson2014} by first identifying sufficient conditions under which the only restrictions on $\mathscr{P}(\cG^C)$ are those implied by d-separation, and then using Theorem~\ref{theorem: gen_dsep} to conclude that such causal structures must be uninteresting i.e., $\mathscr{P}(\cG^C)=\mathscr{P}(\cG^Q)=\mathscr{P}(\cG^G)$. Simple examples of causal structures that satisfy this equivalence and are hence uninteresting, are those of Figures~\ref{fig:2nodeCSc}- \ref{fig:2nodeCSe}. In these causal structures, there are no restrictions implied by d-separation on the observed distribution $P_{XY}$ and hence arbitrary distributions are compatible with these causal structures irrespective of the nature of the unobserved node $\Lambda$. Examples of causal structures that do not satisfy this equivalence and are hence \emph{interesting} are the bipartite Bell ($\cG_B$) and Triangle ($\cG_T$) causal structures illustrated in Figure~\ref{fig: causalstruc}. In the former case, d-separation implies the independence of the inputs $A$ and $B$ as well as the no-signaling conditions ($X\indep B|A$ and $Y\indep A|B$) which are satisfied by all observed distributions $\mathscr{P}(\cG_B^C)$, $\mathscr{P}(\cG_B^Q)$ and $\mathscr{P}(\cG_B^G)$. However, $\mathscr{P}(\cG_B^C)$ satisfy additional constraints given by Bell inequalities, which are not satisfied by all distributions in $\mathscr{P}(\cG_B^Q)$ and $\mathscr{P}(\cG_B^G)$. In fact, there is also a gap between the sets $\mathscr{P}(\cG_B^Q)$ and $\mathscr{P}(\cG_B^G)$, the former satisfies Tsirelson's inequalities while the latter does not. Example~\ref{example: Bellsets} describes the sets $\mathscr{P}(\cG_B^C)$ and $\mathscr{P}(\cG_B^C)$ for the bipartite Bell causal structure $\cG_B$. That $\cG_T$ is also an interesting causal structure can be witnessed by suitably embedding the Bell scenario $\cG_B$ in the Triangle $\cG_T$ such that non-classicality of correlations in $\cG_T$ follows from that of $\cG_B$ \cite{Fritz2012} (this is explained in Section~\ref{ssec: qviol} of Chapter~\ref{chapter: Tsallispaper}). The simplest causal structure that is interesting i.e., supports a classical-quantum gap is the \emph{instrumental causal structure} of \cite{Pearl1995, Chaves2018, Himbeeck2019}, which is essentially the causal structure of Figure~\ref{fig:2nodeCSd} (which by itself is not interesting) but with an additional node and edge $Z\longrightarrow X$. 
%For causal structures with up to six nodes, there are 21 examples (and over 10000 adaptations thereof) that are known to support a gap between the classical and non-classical correlation sets \vilasini{cite}. The above mentioned examples are included in these.

\begin{example}[Sets of compatible correlations in the bipartite Bell causal structure $\cG_B$]
\label{example: Bellsets}
In the classical causal structure $\cG_B^C$, the set of compatible (observed) distribution is obtained by assuming a joint distribution $P_{\Lambda XYAB}\in \mathscr{P}_5$ over all nodes, that satisfies the Markov condition~\eqref{eq: markov} and  marginalising over the unobserved node $\Lambda$,
\begin{equation}
    \mathscr{P}(\cG_B^C):=\{P_{XYAB}\in\mathscr{P}_4|P_{XYAB}=\sum_\Lambda P_{\Lambda}P_AP_BP_{X|A\Lambda}P_{Y|B\Lambda}\}.
\end{equation}
If $\Lambda$ is a continuous random variable, the sum is replaced by an integral over $\Lambda$. This compatibility condition for the classical causal structure $\cG_B^C$ is in fact identical to the \emph{local-causality condition} used in the derivation of Bell inequalities (see \cite{Brunner2014} for a comprehensive review). In the quantum causal structure $\cG_B^Q$, the unobserved node $\Lambda$ corresponds to a quantum state $\rho_{\Lambda}\in\mathscr{S}(\cH_{\Lambda})=\mathscr{S}(\cH_{\Lambda_X}\otimes\cH_{\Lambda_Y})$, and the observed nodes $X$ and $Y$ are associated with the POVMs  $\{E^X_A\}_X$ and $\{E^Y_B\}_Y$, that act on the subsystems $\cH_{\Lambda_X}$ and $\cH_{\Lambda_Y}$, depending on the inputs $A$ and $B$ respectively to generate the output distribution.
\begin{equation}
    \mathscr{P}(\cG_B^Q):=\{P_{XYAB}\in\mathscr{P}_4|P_{XYAB}=\tr\Big((E^X_A\otimes E^Y_B) \rho_{\Lambda}\Big)P_AP_B\}.
\end{equation}
\end{example}

\subsubsection{Interventions and counterfactuals in non-classical causal structures}
\label{ssec: qinterventions}
Interventions in generalised causal structures can be modelled in the same manner as the classical case (Section~\ref{sssec: interventions}), where a non-trivial intervention on an observed node $X$ (or a subset of of nodes) introduces a new edge $I_X\longrightarrow X$ and cuts off all other incoming arrows to $X$. This means that we can only intervene upon classical variables and not quantum systems. Note however that we can model situations involving the manipulation of quantum/GPT systems by allowing the choice of preparation, transformation or measurement on the system to depend on a classical variable. Then an intervention on such a variable will influence the map performed on the quantum/GPT system. Equation~\eqref{eq: do} relates the conditional probabilities $P_{X_i|\text{par}(X_i)}$ for the pre and post intervention causal structures, but this can involve conditioning on possibly unobserved parents which is not possible in generalised causal structures. Equation~\eqref{eq: do} can however be generalised by specifying how the test (i.e., the causal mechanisms) corresponding to each node transforms under an intervention. In \cite{Henson2014}, each node is associated with a test that maps all joint state of the incoming (unobserved) subsystems to the joint of the outgoing (unobserved) subsystems, depending on the values of the observed parents. Under an intervention on the node $X$, the corresponding test would remain unchanged if $I_X=idle$, and would effectively correspond to a preparation test that prepares $X$ in a specific value $x\in X$ (deterministically) whenever $I_X=\text{do}(x)$, and traces out the systems associated with all other incoming edges.\footnote{Note that we can view $\text{do}(x)$ effectively as a preparation test even though the node $X$ has an input edge $I_X\longrightarrow X$, because a) $I_X$ is just a hypothetical variable that is introduced to provide an agency to the intervention in the augmented graph approach, but interventions can be equivalently described by an alternative approach that does not introduce any additional variables \cite{Pearl2009}, and b) The value of $X$ is deterministically correlated with $I_X$, such that if $\text{do}(x)$ is seen as a preparation on the exogenous variable $I_X$, this deterministically induces the preparation $X=x$ on $X$.} 

So far, the procedure is similar to the classical case, however, there is one crucial difference. Equation \eqref{eq: do_transf} which relates the pre and post intervention distributions and thereby provides a rule for counterfactual inference no longer holds for non-classical causal structures, since this equation is derived using the Markov compatibility condition~\eqref{eq: markov}, which does apply to this case. Since non-classical causal structures by construction include at least one unobserved node, one would expect that the observed post-intervention distribution may not always be determined by the observed pre-intervention distribution, as is the case also with classical causal structures involving unobserved nodes (see Example~\ref{example:nonidentifiable}). \cite{Pienaar2020} shows that it is impossible to derive any general inference rule relating the observed distributions in pre and post intervention quantum causal structures.\footnote{The causal model framework of \cite{Pienaar2020} is slightly different than what we have presented here, but nevertheless compatible with this picture. This is explained further in the next subsection.} This means that it may not always be possible to deduce the post-intervention distribution counterfactually using the pre-intervention distribution, instead one would have to physically perform the intervention (i.e., an additional experiment, after the original experiment that collected the pre-intervention statistics) to characterise this distribution and all the do-conditionals. However, \cite{Pienaar2019} and \cite{Pienaar2020} provide conditions for identifying quantum causal structures where an inference rule analogous to~\eqref{eq: do_transf} does exist and fully counterfactual inference is possible. This is still a field of ongoing research, we will discuss interventions in non-classical causal structures in further detail in Chapter~\ref{chapter: jammingpaper}.

\subsubsection{Other approaches for causality beyond the classical setting}
\label{ssec: caus_other}
The main approach to quantum and post-quantum causal structures adopted in this thesis, that we have discussed so far is based on the framework of \cite{Henson2014}. This approach retains d-separation and conditional independence criteria from the classical Bayesian networks approach \cite{Pearl2009} while generalising the Markov condition~\eqref{eq: markov}. There are a several other frameworks for quantum causal models that build on different aspects of Pearl's classical framework \cite{Pearl2009}. These frameworks are consistent with each other but can differ in scope due to how they formalise the notion of a causal model in quantum settings i.e., how the nodes and edges are represented, what compatibility conditions are used, and type of interventions/manipulations that are considered. 

\cite{Pienaar2015, Costa2016, Allen2017} have proposed generalisations of the Reichenbach principle \cite{Reichenbach1956} to the quantum setting. Among these, \cite{Pienaar2015} is based on a new graph separation condition for quantum causal structures called q-separation, while \cite{Costa2016, Allen2017} are based on formulating quantum networks using quantum channels and their Choi states. \cite{Allen2017} as well as an earlier work \cite{Leifer2013} construct quantum networks by representing quantum channels as conditional quantum states (analogous to conditional probabilities). The framework of \cite{Allen2017} will employed in one of our results, we briefly outline it here and present the relevant details in Chapter~\ref{chapter: Tsallispaper} where it will be used. In \cite{Allen2017}, nodes correspond to systems and edges to transformations, in contrast to the HLP framework \cite{Henson2014} where transformations (tests) occur at the nodes and the edges propagate systems. All nodes are taken to be quantum systems and a distinction between observed and unobserved nodes is not made. Then a conditional quantum state $\rho_{A|\text{par}(A)}$ for a node $A$ conditioned on its parents (analogous to conditional distributions $P_{X|\text{par}(X)}$) is defined through the Choi-Jamiolkowski representation of the quantum channel that maps the quantum systems corresponding to $\text{par}(A)$ to that corresponding to $A$. With this, a quantum Markov condition, completely analogous to the classical case~\eqref{eq: markov} is defined. This also allows for the definition of a joint state $\sigma_{A_1,...,A_n}$ over all the nodes $\{A_1,...,A_n\}$, even though the systems corresponding to these nodes may not coexist. This joint state $\sigma$ characterises the structure of the channels connecting the various nodes and not the quantum states of the systems. Interventions correspond to CPTP maps that act on quantum nodes and the causal model provides a prescription for obtaining the post-intervention distribution given the joint state $\sigma$ and the CPTP maps corresponding to the interventions, through a formula analogous to the Born rule. 

\interfootnotelinepenalty=10000
In another recent framework for quantum causal models \cite{Pienaar2019,Pienaar2020} (also noted in the previous subsection), all nodes are taken to be classical, while edges correspond to propagating quantum systems that are acted upon by transformations that depend on the values of the classical nodes.\footnote{For example, three sequential operations on a quantum state are represented by the causal structure $X\longrightarrow Y\longrightarrow Z$, where $X$ specifies the preparation of a quantum state $\rho_X$ that travels to the next node where a CPTP map $\mathscr{E}_Y$ labelled by $Y$ acts on $\rho_X$, and finally a POVM labelled by $Z$, $\{E^Z\}_Z$ acts on the output of the CPTP map $\mathscr{E}_Y(\rho_X)$.}  Further, frameworks such as \cite{Ried2015} have shown that quantum causal models can provide an information-theoretic advantage for causal inference by utilising quantum properties such as entanglement and superposition.
%\cite{Pienaar2020} also explicitly models a new kind of manipulation which they call an \emph{unmeasurement}, which corresponds to the act of physically moving a measurement device from a space-time location. This becomes relevant in quantum scenarios such as interference experiments where the presence/absence of a measuring device near one arm of the interferometer can affect the experiment even when a measurement is not actually performed.

The discussion so far has focussed on the assumption that a fixed causal structure (between classical/quantum or post-quantum systems) exists, even if it may be unknown. More general approaches to non-classical notions of causality have considered scenarios where not only the systems in a causal structure, but the causal structure itself is of a quantum nature \cite{Hardy2005, Oreshkov2012, Zych2019}. In such frameworks, quantum superpositions of the causal and temporal orders between events are typically considered by dropping the assumption of a fixed background space-time or causal structure, and the study of such indefinite causal structures have garnered much research interest in recent years and shown to provide additional information-theoretic advantages \cite{Chiribella2013, Araujo2014b, Araujo2015, Araujo2016, Baumeler2016, Guerin2016, Branciard2016, Oreshkov2016, Portmann2017, Chiribella2018, Oreshkov2019, Barrett2020}. Here, analogous to Bell inequalities, which certify the non-classicality of correlations in a fixed causal structure, causal inequalities have been proposed for certifying the non-classicality of the causal structure itself \cite{Oreshkov2012}. Unlike the former which have been violated in several experiments, violations of the latter are modelled in several theoretical frameworks \cite{Oreshkov2012, Guerin2018, Branciard2016, Barrett2020}, but whether causal inequalities can be physically violated remains a thriving open question. Further, scenarios where quantum systems are superposed/entangled over space as well as in time have been physically implemented. The so-called causal box framework \cite{Portmann2017} models information processing tasks where quantum states can be delocalised in states as well as time, in a manner compatible with our relativistic causality. This framework has applications for relativistic quantum cryptography as well as for foundational questions regarding causal inequalities, as we have shown in a previous work \cite{Vilasini_crypto}, and an ongoing work\footnote{Vilasini, V., del Rio, L. and Renner, R. Causality in definite and indefinite space-times. In preparation (2020).} (both of which are not included in this thesis). Finally, there also exist general frameworks for quantum causality that take into account gravitational effects, with the aim of developing an operational understanding of quantum gravity \cite{Hardy2005, Hardy2016, Zych2019, Castro-Ruiz2020}. 
\section{Modal logic}
\label{sec: logic}

In daily life and as part of the scientific method, we often use logic to reason about our observations and to predict the possibility of future observations. We do so by assigning truth values to different statements depending on what we observed and what we might already know. Modal logic is broadly speaking, a family of formal systems of logic that are associated with different modalities for assigning truth values to statements, such as ``it is necessary that'' (alethic), ``it is known that'' (epistemic), ``it ought to be the case that'' (deontic), or ``it has been the case that'' (temporal). In this thesis, we will focus on the branch of modal logic, namely \emph{epistemic modal logic} that is applied for reasoning about knowledge. This is particularly useful in scenarios (both classical and quantum) where  where multiple agents reason about each others' knowledge to make logical deductions. Examples include games like poker, \href{https://en.wikipedia.org/wiki/Hat_puzzle}{logical hat puzzles}\footnote{For an analysis of classical hat puzzles in the epistemic modal logic language, see \cite{Nurgalieva2018}.}, extended Wigner's friend experiments in quantum theory \cite{Frauchiger2018, Nurgalieva2018} (Chapter~\ref{chapter: FR}) and in post-quantum theories \cite{Vilasini_PRdoxes} (Chapter~\ref{chapter: PRdoxespaper}), among others \cite{Carnap1988, Fagin2004, Lewis1918, Lewis1959}.

Here we provide a brief summary including the main features of epistemic modal logic that are relevant for this thesis.  Modal logic applies to most classical multi-agent setups, and can be seen as a compact mathematical way to capture some of intuitive laws commonly used for reasoning in classical settings. We will first review the basics of the standard framework \cite{Kripke2007}, and then discuss the weaker versions of the axioms proposed recently in \cite{Nurgalieva2018} for modelling scenarios involving multiple quantum agents. For further reading and a more in depth understanding of the modal logic framework, we refer the reader to \cite{Carnap1988, Goldblatt2006, Fagin2004, Lewis1918, Lewis1959}.

%Statements expressing the knowledge of agents, such as “Malavika knows that Jimi Hendrix was a great guitarist,” are represented by associating a \emph{knowledge operator} $K_i$ to each agent $A_i$, $K_M\phi_1$. Here $K_i\phi$ stands for “Agent $i$ knows $\phi$.” By extension, more complex, chained statements such as “Bart doesn't know that Malavika knows that Jimi Hendrix is a great guitarist and Bart knows that the grass is green” are more compactly expressed as $(\neg K_B K_M\phi_1)\land K_B\phi_2$.

\subsection{Knowledge, possibilities and truth: Kripke structures}

 We will consider a finite set of agents $\{A_1,...,A_n\}$ who describe the world by means of primitive propositions $\phi_1,\phi_2,...$ belonging to a some set $\Phi$. These primitive propositions correspond to simple statements about the world, for example, $\phi_1=$ ``Alice is a person'' or $\phi_2=$ ``Alice has a secret key''. The framework of modal logic allows for compactly representing the knowledge that agents possess about these primitive statements as well more complex, chained statements about the knowledge of other agents, such as $\tilde{\phi}=$“Alice knows that Bob knows that Eve doesn't know the secret key $k$, and Alice further knows that $k=1$”. It also provides simple rules for combining such statements to perform logical deductions.

More formally, we consider a set $\Sigma$ of possible states (or alternatives, or \emph{worlds}\footnote{This is an entirely classical concept and is not to be confused with ``many worlds'' type interpretations of quantum theory.  \emph{World} here only represents potential alternative situations i.e. a classical list of possibilities, it does not mean that all these situations are actually realised.}) as introduced by \cite{Kripke2007} in the context of modal logic: for example, in a world $s_1\in \Sigma$ the key value is $k=1$ and Eve does not know it, and in another world $s_2\in \Sigma$ Eve could know that $k=0$. The truth value of a proposition $\phi$ is then assigned depending on the possible world in $\Sigma$, and can differ from one possible world to another. The setup we will employ for modelling multiple reasoning agents is associated the following structure.

\begin{definition}{\textbf{(Kripke structure)}}
A Kripke structure $M$ for $n$ agents over a set of statements $\Phi$ is a tuple $\langle \Sigma, \pi,\mathcal{K}_1,...,\mathcal{K}_n\rangle$ where $\Sigma$ is a non-empty set of states, or possible worlds, $\pi$ is an interpretation, and $\mathcal{K}_i$ is a binary relation on $\Sigma$.\\
The interpretation $\pi$ is a map $\pi:\Sigma\times\Phi\to\{\textbf{true},\textbf{false}\}$, which defines a truth value of a statement $\phi\in\Phi$ in a possible world $s\in\Sigma$.\\
$\mathcal{K}_i$ is a binary equivalence relation on a set of states $\Sigma$, where $(s,t)\in\mathcal{K}_i$ if agent $i$ considers world $t$ possible given his information in the world $s$. 
\end{definition}

The truth assignment  tells  us if the proposition $\phi\in\Phi$ is true or false in a possible world $s\in\Sigma$; for example, if $\phi=$ ``you are reading this PhD thesis'' and $s$ is the present world in which you have indeed chosen to read this thesis, then $\pi(s,\phi)=\textbf{true}$. On the other hand, if $\phi'=$``There is a mammoth in the snow'' then $\pi(s,\phi')=\textbf{false}$ in the current world where mammoths are (unfortunately) extinct, however in a world $s'$ corresponding to the ice age $\phi'$ can indeed be a true statement. Therefore the truth value of a statement in a given structure $M$ might vary from one possible world to another; we will denote that $\phi$ is true in world $s$ of a structure $M$ by $(M,s)\models\phi$, and $\models\phi$ will mean that $\phi$ is true in any world $s$ of a structure $M$.

Naturally, agents need not possess complete information about a possible world they are in, and may consider other alternative worlds possible; for example, if Bob doesn’t know whether Alice has a secret key, he can consider as possible both the world where Alice has such a key, and one where she doesn't. This situation is captured by binary relations $\mathcal{K}_i$, as formalised in the following definition.
\begin{definition}[Knowledge operators $K_i$]
 We say that agent $A_i$ ``knows'' $\phi$ in a world $s\in \Sigma$ i.e.,
 $$(M,s)\models K_i\phi$$
 if and only if for all possible worlds $t\in \Sigma$ such that $(s,t) \in \mathcal{K}_i$ (that is, all the worlds deemed possible by the agent, given their knowledge), it holds that $(M,t) \models \phi$.
\end{definition}

Knowledge operators combined with the standard logical operators allow us to compactly express complex statements, for instance, our earlier example $\tilde{\phi}=$``Alice knows that Bob knows that Eve doesn't know the secret key $k$, and Alice further knows that $k=1$'' becomes $$K_A\ [(K_B\ \neg K_E\  k ) \ \wedge \ k=1].$$

\subsection{Reasoning about other's knowledge: Axioms of knowledge}

The axioms of knowledge \cite{LogicStanford} provide certain rules for combining the statements produced by different agents and to compress them for deducing new statements. These axioms might appear to be common-sensical, but it is imperative to state them formally since not all of these axioms apply in the quantum world, that often surprises our daily common-sensical intuitions.
%We first state the the basic axioms of the standard, classical framework before discussing the modification proposed in \cite{Nurgalieva2018} for quantum multi-agent settings.

The distribution axiom allows agents combine statement which contain inferences:
\begin{axiom}[Distribution axiom.]
\label{axiom:distribution}
If an agent is aware of a proposition $\phi$ and that another proposition $\psi$ follows from $\phi$, then the agent can conclude that $\psi$ holds:
$$(M,s)\models(K_i\phi\wedge K_i(\phi\Rightarrow\psi))\Rightarrow (M,s)\models K_i\psi.$$
\end{axiom}

Knowledge generalization rule permits agents use commonly shared knowledge:
\begin{axiom}[Knowledge generalization rule.]
All agents know all the propositions that are  always valid in a structure:
\begin{center}
$(M,s)\models\phi \ \forall s\in \Sigma \Rightarrow \models K_i\phi \ \forall i.$
\end{center}
\end{axiom}
At first sight, this might appear to be a strong assumption. However, this depends on what $\Sigma$ is chosen to be for a particular purpose, and characterises what statements can be assumed to be the basic, common knowledge of all agents. For example, we would find it reasonable to expect that the propositions $\phi_1$=``objects fall when we drop them'', $\phi_2$=``guitars have strings'', $\phi_3$=``we cannot signal from future to past'' to be common knowledge, such that they hold in every $s\in \Sigma$ that we wish to consider. However, if we wish to consider hypothetical worlds where guitars are wind instruments or where we can signal to the past (as physicists working in foundations, including ourselves often do), we can simply include a world $s'\in \Sigma$ where the statements $\phi_2$ or $\phi_3$ are false, and hence do not correspond to the common knowledge of the agents. In fact, parts of Chapter~\ref{chapter: jammingpaper}, will indeed be set in such a hypothetical world where $\phi_3$ is not true. One can also treat a theory (such as quantum theory) or aspects of the theory (such as the Born rule) to be the common knowledge of the agents that we wish to consider, as we will see in Chapter~\ref{chapter: PRdoxespaper}. 

\begin{axiom}[Truth axiom]
\label{axiom:truth}
If an agent knows that a proposition is true then the proposition is true,
$$(M,s)\models K_i\phi \Rightarrow (M,s)\models \phi$$
\end{axiom}

In philosophy, the truth axiom is often considered as a candidate for a property that distinguishes knowledge from belief.

Positive and negative introspection axioms highlight the ability of agents to reflect upon their knowledge:
\begin{axiom}[Positive and negative introspection axioms.] 
Agents can perform  introspection regarding their knowledge:
\begin{center}
$(M,s)\models K_i\phi\Rightarrow (M,s)\models K_iK_i\phi$ (Positive Introspection),\\
$(M,s)\models \neg K_i\phi\Rightarrow (M,s)\models K_i\neg K_i\phi$ (Negative Introspection).
\end{center}
\end{axiom}

\paragraph{Weaker version: }
The truth axiom assigns objective reality to all statements that agents know which can be particularly problematic for quantum settings. Hence a weaker alternative to the truth axiom was proposed in \cite{Nurgalieva2018}. 

%We discuss trust structure below as this is more central to our results and will leave the discussion of context to a later point in the thesis where the quantum thought experiment of \cite{Frauchiger2018} has been appropriately presented.

The trust structure governs the way the information is passed on between agents:
\begin{definition}[Trust]
\label{def:trust}
    We say that an agent $i$ trusts an agent $j$ (and denote it by $j \leadsto i$ ) if and only if $$(M,s)\models K_i \ K_j \ \phi \implies K_i \ \phi,\qquad \forall \phi \in \Phi, s\in \Sigma$$
\end{definition}

Note that trust is neither a symmetric nor a transitive relation and hence does not define a pre-order on the agents. This is also in line with what we might expect common-sensically; I might trust my favourite news reporter (R) to deliver reliable news such that $K_IK_R\phi\Rightarrow K_I\phi$ would hold in all worlds $s$ relevant to me and for all statements $\phi$ relating to the news (which I could take to characterise the sets $\Sigma$ and $\Phi$) i.e., $R\leadsto I$. However R, who does not know me has no reason to trust what I say, even if it relates to the news hence $K_IK_R\phi\not\Rightarrow K_I\phi$ or equivalently $R\not\leadsto I$. Similarly, R might have a trusted source S for acquiring the news, and S in turn may acquire that news through a source T whom she trusts i.e., $S\leadsto R$ and $T\leadsto S$. While R would trust the same information when relayed indirectly through the agent S, R may not know of T's existence and hence not trust T if he approached R directly, hence $T \leadsto R$ does not necessarily hold. 

While replacing truth with the notion of trust may suffice in standard quantum settings, as we will see in Chapters~\ref{chapter: FR} and~\ref{chapter: PRdoxespaper}, it turns out to be incompatible with situations where reasoning agents are themselves modelled as physical systems of quantum or post-quantum theories \cite{Nurgalieva2018}.

\part{Approaches to causality in non-classical theories}
\begin{center}
    \thispagestyle{empty}
    \vspace*{\fill}
   \large{\textit{ Correlation doesn’t imply causation, but it does waggle its eyebrows suggestively and gesture furtively while mouthing ‘look over there.’
}}\\\vspace{3mm}
\raggedleft\large{\textit{  - Randall Munroe }}
    \vspace*{\fill}
\end{center}
\chapter{Overview of techniques for analysing causal structures}
\label{chap: techniques}
\lettrine[nindent=0em, slope=-.5em,lines=2]{I}{n} Section~\ref{sec: causalstr}, we provided an overview of the frameworks used for modelling causal structures in classical, quantum and post-quantum theory. We motivated the problem of certifying the non-classicality of causal structures, which has become a focal point of research interest in quantum information by virtue of having deep foundational as well as practical implications.
In this chapter, we outline some of the techniques used for this purpose. Section~\ref{sec: Probspace} and Section~\ref{sec:entvec} provide an overview of the main probabilistic and entropic techniques employed in Chapters~\ref{chapter: Tsallispaper} and~\ref{chapter: mixingpaper}, taking as example the bipartite Bell causal structure (Figure~\ref{fig: Bell}), which is a causal structure of particular interest in this thesis. These sections are largely based on the review sections of our published papers \cite{Vilasini2019} and \cite{Vilasini2020}. The main results of these papers are presented in Chapters~\ref{chapter: Tsallispaper} and~\ref{chapter: mixingpaper}. All of the above focus on the case of faithful and acyclic causal structures, as does the majority of literature in the field, and justifiably so, since our observations suggest that causal influences propagate in one direction i.e., from past to future. The present chapter will also focus only on the acyclic case. In Chapter~\ref{chapter: jammingpaper}, we develop a framework for modelling cyclic and fine-tuned causal structures in non-classical theories and will review the necessary preliminaries for cyclic causal structures there.

\section{Certifying non-classicality in probability space}
\label{sec: Probspace}
%\label{ssec: prob}
In the probability space characterisation of bipartite Bell causal structure $\cG_B$ (Figure~\ref{fig: Bell}), the observed correlations are represented through the conditional distribution $P_{XY|AB}$. Each point in the probability space corresponds to such a conditional distribution and it will be convenient to denote the cardinalities $|A|$, $|B|$, $|X|$ and $|Y|$ as $i_A$, $i_B$, $o_A$ and $o_B$ (representing the inputs and outputs of the parties Alice and Bob) respectively. Each Bell scenario is defined by the 4-tuple $(i_A,i_B,o_A,o_B)$. In a given scenario we will be interested in the sets of \emph{local} and \emph{non-signaling} distributions, which are the sets of conditional distributions $P_{XY|AB}$ corresponding to the sets $\mathscr{P}(\cG^C_B)$ and $\mathscr{P}(\cG^G_B)$ (Section~\ref{sssec: gen_causalstr}) of observed correlations $P_{XYAB}$. Following the notation of~\cite{Cirelson93}, we will express the distribution using a matrix, rather than a vector. For instance, in the case
where all the variables take values in $\{0,1\}$ and using $P(xy|ab)$
as an abbreviation for $P_{XY|AB}(xy|ab)$, this is done as
\begin{equation}
\label{eq: dist}
    P_{XY|AB}=
\begin{array}{ |c|c|} 
 \hline
 P(00|00) \quad P(01|00) & P(00|01) \quad P(01|01)\\ 
 P(10|00) \quad P(11|00) & P(10|01) \quad P(11|01)\\ 
 \hline
 P(00|10) \quad P(01|10) & P(00|11) \quad P(01|11)\\ 
 P(10|10) \quad P(11|10) & P(10|11) \quad P(11|11)\\ 
 \hline
\end{array}
\end{equation}
and the generalisation to larger alphabets is analogous (see,
e.g.,~\cite{Cirelson93}).  This format is convenient because it makes
it easy to check whether a distribution is non-signaling, i.e., to check
that $P_{X|AB}$ is independent of $B$ and that $P_{Y|AB}$ is
independent of $A$. Both the local and non-signaling sets form convex polytopes, that are highly symmetric. In particular, such polytopes are invariant under local relabellings and/or relabelling parties. By \emph{local relabellings} we mean combinations of relabelling the inputs (e.g., $A\mapsto A\oplus 1$) and outputs conditioned on the local input (e.g., $X\mapsto X\oplus\alpha A\oplus\beta$ where $\alpha, \beta \in \{0,1\}$ and $\oplus$ denotes modulo-2 addition). One might also think about more general \emph{global relabellings} that depend on both inputs (for instance maps of the form $X\mapsto X\oplus\alpha A\oplus\beta B\oplus\gamma$ with $\alpha, \beta, \gamma \in \{0,1\}$), but these do not preserve the non-signaling set in general so will not be considered here. The only global relabelling we consider is exchange of the two parties, which corresponds to transposing the distribution in the matrix notation of Equation~\eqref{eq: dist} and always preserves non-signaling.

More specifically, the non-signaling set is restricted only by the conditional independences $X\indep B|A$ and $Y\indep A|B$ (no-signaling constraints)\footnote{The independence $A\indep B$ (measurement independence) also follow from the graph, but this is not relevant when we are considering the conditional distributions $P_{XY|AB}$.}, which follow from the d-separation relations in $\cG_B$ (Figure~\ref{fig: Bell}), and naturally the positivity and normalisation conditions since these are valid probability distributions. These restrictions can be written out as linear inequality/equality constraints, and characterise the $\mathcal{H}$ representation of the non-signaling polytope. The corresponding $\mathcal{V}$-representation is characterised by the extremal non-signaling distributions. On the other hand, the local set is the set of all observed distributions $P_{XY|AB}$ that can arise when $\Lambda$ is classical corresponds to the set of correlations that admit a local hidden variable model, i.e., the set of distributions that can be expressed in the form 
 \begin{equation}
 \label{eq: loccaus}
 P_{XY|AB}=\int_{\Lambda}\mathrm{d}\Lambda\, P_{\Lambda}\, P_{A}\, P_{B}\, P_{X|A\Lambda}\, P_{Y|B\Lambda}.
 \end{equation}
Here, $\Lambda$ can be any (possibly continuous) random variable. In the rest of the thesis, we will refer to such correlations either as \emph{local} or as \emph{classical} and denote the set of all such distributions $\mathscr{L}$. We also use $\mathscr{L}^{(i_A,i_B,o_A,o_B)}$ to denote the local set in the $(i_A,i_B,o_A,o_B)$ scenario. 

The local set forms a convex polytope, which can be specified in terms of a finite set of Bell inequalities, each a necessary condition for classicality of the correlation. Along with the positivity and normalisation constraints, these define the $\mathcal{H}$-representation of $\mathscr{L}^{(i_A,i_B,o_A,o_B)}$. The $\mathcal{V}$-representation of the local polytope is defined by the set of all \emph{local deterministic distributions} i.e., distributions $P_{XY|AB}$ that are deterministic over $X$ and $Y$ for every value of $A$ and $B$. There are $o_A^{i_A}\cdot o_B^{i_B}$ distinct local deterministic distributions in a $(i_A,i_B,o_A,o_B)$ scenario. In the matrix notation~\eqref{eq: dist}, each block corresponds to a fixed value for $A$ and $B$. The local deterministic distributions expressed in this notation have exactly one non-zero entry ($=1$) in each block in accordance with the no-signaling condition required by the causal structure. The set of all such distributions can be easily enumerated. Hence, the $\mathcal{V}$-representation of the local polytope (local deterministic distributions) and the $\mathcal{H}$-representation of the non-signaling polytope (no-signaling, positivity and normalisation constraints) can be easily found for a given scenario. The $\mathcal{H}$-representation of the local polytope (Bell inequalities) and the $\mathcal{V}$-representation of the non-signaling polytope (extremal non-signaling distributions) can in principle be found through facet and vertex enumeration respectively. However, this becomes computationally intractable for larger cardinalities of the observed variables. In the following, we describe the structure of these sets in more detail for the $(2,2,2,2)$ and $(2,2,3,3)$  bipartite Bell scenarios, which are of particular interest in this thesis.

\subsection{The bipartite Bell scenario with 2 inputs and 2 outputs per party}
\label{ssec: 2222prob}

In the $(2,2,2,2)$ scenario, there are eight extremal Bell inequalities (facets of the local polytope), which are equivalent under local relabellings to the following inequality~\cite{CHSH69, Cirelson93}
\begin{equation}
\begin{split}
\label{eq: CHSH}
I_{\chsh}:=&P(X=Y|A=0,B=0)+P(X=Y|A=0,B=1)+P(X=Y|A=1,B=0)\\&+P(X\neq Y|A=1,B=1)\leq 3
\end{split}
\end{equation}
 We denote these by $I_{\chsh}^k$ for
$k\in [8]$, where $I_{\chsh}^1:=I_{\chsh}$ and $[n]$ stands for the
set $\{1,2,....,n\}$ where $n$ is a positive integer. This provides
the facet description of the $(2,2,2,2)$ local polytope.  One can also express $I_{\chsh}$ in matrix form using
\begin{equation}\label{eq:mat1}
    M_{\chsh}=\begin{array}{|cc|cc|}\hline 1&0&1&0\\0&1&0&1\\\hline1&0&0&1\\0&1&1&0\\\hline\end{array}\,,
\end{equation}
so that the Bell inequality can be written $\tr\left(M_{\chsh}^T P\right)\leq3$, where $P$ is the distrubution expressed in the matrix form~\eqref{eq: dist} and $T$ denotes the transpose.

In the vertex picture, the $(2,2,2,2)$ local polytope has 16 local deterministic vertices and the $(2,2,2,2)$ non-signaling polytope shares the vertices of the local polytope and has eight more: the Popescu-Rohrlich (PR) box and seven distinct local relabellings~\cite{Cirelson93,Popescu1994}. The PR box distribution (discussed in Section~\ref{sec: GPT}) satisfies $X\oplus Y=A.B$ and takes the following form in the current notation
\begin{equation}
\label{eq: PR}
    P_{\PR}=
\begin{array}{ |c|c|} 
 \hline
 \frac{1}{2} \quad 0 & \frac{1}{2} \quad 0\\ 
 0 \quad \vphantom{\frac{1}{f}}\frac{1}{2} & 0 \quad \vphantom{\frac{1}{f}}\frac{1}{2}\\
 \hline
 \frac{1}{2} \quad 0 & 0 \quad \frac{1}{2}\\ 
 0 \quad \vphantom{\frac{1}{f}}\frac{1}{2} & \vphantom{\frac{1}{f}}\frac{1}{2} \quad 0\\ 
 \hline
\end{array}\,.
\end{equation}
We denote the eight extremal non-signaling vertices equivalent under local relabellings to $P_{\PR}$ by $P_{\PR}^k$, $k\in [8]$ where $P_{\PR}^1:=P_{\PR}$. Note that the 8 CHSH inequalities $\{I_{\chsh}^k\}$ are in one-to-one correspondence with these 8 extremal non-signaling points i.e., each $P_{\PR}^k$ violates exactly one CHSH inequality and each CHSH inequality is violated by exactly one $P_{\PR}^k$.

\subsection{The bipartite Bell scenario with 2 inputs and 3 outputs per party}
\label{ssec: 2233rev}

In the $(2,2,3,3)$ Bell scenario, there are two classes of Bell inequalities that completely characterize the local polytope: the CHSH inequalities and the $I_{2233}$ inequalities~\cite{Kaszlikowski2002, CGLMP02}. A representative example of the latter is
%\begin{widetext}
\begin{equation}
\label{eq: CGLMP}
\begin{split}
I_{2233}:=\ &\big[P(X=Y|A=0,B=1)+P(X=Y-1|A=1,B=1)+P(X=Y|A=1,B=0)\\&+P(X=Y|A=0,B=0)\big]-\big[P(X=Y-1|X=0,B=1)+P(X=Y|A=1,B=1)\\&+P(X=Y-1|A=1,B=0)+P(X=Y+1|A=0,B=0)\big]\leq 2\,,
\end{split}
\end{equation}
%\end{widetext}
where all the random variables take values in $\{0,1,2\}$ and all additions and subtractions of the random variables are modulo 3.  CHSH-type inequalities for the $(2,2,3,3)$ scenario can be obtained from those of the $(2,2,2,2)$ scenario (Equation~\eqref{eq:mat1}) through a procedure known as ``lifting''. The inequalities of the larger scenario are known as the \emph{lifted CHSH inequalities} \cite{Pironio2005}. Evaluating the value of a lifted CHSH inequality attained by a given distribution in the $(2,2,3,3)$ scenario is equivalent to first coarse-graining the distribution by combining two outcomes into a single outcome, and then evaluating the corresponding CHSH inequality of the $(2,2,2,2)$ scenario. For instance, the lifted CHSH inequality corresponding to $M_{\chsh}$~\eqref{eq:mat1} and the coarse-graining of always mapping outcomes 1 and 2 to 1 is given in matrix form in the following equation, along with a representative inequality of the $I_{2233}$ type.
\begin{equation}\label{eq:mat2}
    M_{\chsh}^{(2,2,3,3)}=\begin{array}{|ccc|ccc|}\hline 1&0&0&1&0&0\\0&1&1&0&1&1\\0&1&1&0&1&1\\\hline1&0&0&0&1&1\\0&1&1&1&0&0\\0&1&1&1&0&0\\\hline\end{array}\quad\text{and}\quad
    M_{I_{2233}}=\begin{array}{|ccc|ccc|}\hline 1&0&-1&1&-1&0\\-1&1&0&0&1&-1\\0&-1&1&-1&0&1\\\hline1&-1&0&-1&1&0\\0&1&-1&0&-1&1\\-1&0&1&1&0&-1\\\hline\end{array}\,.
\end{equation}

The $(2,2,3,3)$ local polytope has a total of 1116 facets, $36$ of which correspond to positivity constraints, $648$ to CHSH facets (these are equivalent to first coarse-graining two of the outputs into one (for each party and each input) and then applying one of the eight $(2,2,2,2)$ CHSH inequalities), and the remaining $432$ are $I_{2233}$-type~\cite{CollinsGisin04} (we label these $I_{2233}^i$ for $i\in\{1,2,\ldots,432\}$ with $I_{2233}^1=I_{2233}$).

The facets of the non-signaling polytope correspond to positivity constraints, since the no-signaling constraints being equalities, only reduce the dimension of the polytope. Converting this facet description to the vertex description (e.g., using the {\sc Porta} software~\cite{porta}) one can obtain all the vertices of the $(2,2,3,3)$ non-signaling polytope.  This comprises $81$ local deterministic vertices, $648$ PR-box type vertices and $432$ extremal non-signaling vertices (for each of the $I_{2233}$ inequalities there is one of the latter that gives maximal violation).  We call these new vertices the \emph{$I_{2233}$-vertices}.  The specific vertex that maximally violates~\eqref{eq: CGLMP} is
\begin{equation}
\label{eq: NL3}
P_{\NL}:=
    \begin{array}{ |c|c|} 
 \hline
 \frac{1}{3} \quad 0 \quad 0 & \frac{1}{3} \quad 0 \quad 0\\ 
 0 \quad \frac{1}{3} \quad 0 & 0 \quad \frac{1}{3} \quad 0\\
 0 \quad 0 \quad \vphantom{\frac{1}{f}}\frac{1}{3} & 0 \quad 0 \quad \vphantom{\frac{1}{f}}\frac{1}{3}\\
 \hline
 \frac{1}{3} \quad 0 \quad 0 & 0 \quad \frac{1}{3} \quad 0\\ 
 0 \quad \frac{1}{3} \quad 0 & 0 \quad 0 \quad \frac{1}{3}\\
 0 \quad 0 \quad \vphantom{\frac{1}{f}}\frac{1}{3} & \vphantom{\frac{1}{f}}\frac{1}{3} \quad 0 \quad 0\\
 \hline
\end{array}
\end{equation}

The $432$ $I_{2233}$ vertices of the $(2,2,3,3)$ non-signaling polytope are related to each other through local relabellings\footnote{In general, equivalent points of the non-signaling polytope may be related by local relabellings or exchange of the two parties (which is a global operation). In the $(2,2,3,3)$ scenario there are $2\times(2\times 6^2)^2=10368$ such operations, twice the number of local relabellings. To count these, note that for each party there are $2$ ways to permute the inputs, and $6$ ways to permute the outputs for each of the $2$ inputs. All 432 extremal points of the $I_{2233}$ type (those which maximally violate a $I_{2233}$ inequality) can be generated using only local relabellings of $P_{\NL}$, and similarly all 648 extremal points of the CHSH type can be generated through local relabellings of $P_{\PR}$ embedded in the $(2,2,3,3)$ scenario (by adding zero probabilities to the third outcome).}. The $I_{2233}$ inequalities are a special case of the CGLMP inequalities which correspond to facets of the local polytope $\mathscr{L}^{(2,2,d,d)}$ ($d\geq 3$) \cite{CGLMP02, Masanes2002}. For the $d>3$ case however, there may in general be other new classes of inequalities apart from the CHSH and CGLMP inequalities.

\section{Certifying non-classicality in entropy space}
\label{sec:entvec}

\subsection{Motivation for the entropic analysis of causal structures}
In the bipartite Bell causal structure (Figure~\ref{fig: Bell}), that we discussed above, the set of all joint conditional distributions $P_{XY|AB}$ over the observed nodes $X$,$Y$,$A$,$B$ that can arise when $\Lambda$ is classical is relatively well understood. For fixed input and output sizes, it forms a convex polytope and hence membership can be checked using a linear program (although the size of the linear program scales exponentially with the number of inputs and the problem is NP-complete \cite{Pitowski89}). Because of this, the complete set of Bell inequalities characterizing these polytopes are unknown for $|X|$, $|Y|>3$ or $|A|$, $|B|>5$ \cite{Masanes2002, Bancal2010, Cope2019}. In causal structures with more unobserved common causes (such as the Triangle causal structure of Figure~\ref{fig: Triangle}), the set of compatible correlations is not always convex, and hence not well understood. The inflation technique proposed in \cite{wolfe2016} has been used to derive some necessary constraints (or Bell inequalities) on the classical sets of correlations in such causal structures. This technique in principle certify whether
or not a given distribution belongs to the classical set $\mathscr{P}(\cG^C)$ of a causal structure $\cG$ \cite{Navascues2017}, and involves constructing a new causal structure, known as the inflation of the original causal structure such that Bell inequalities for the inflated causal structure also hold for the original causal structure. However, the
method does not tell us how to construct a suitable inflation of the causal structure in order to achieve this, or how large this inflation needs to be. Thus, in general, using the inflation technique becomes intractable in practice. 

One approach to overcoming the difficulties of non-convex correlation sets is to analyse the problem in entropy space \cite{Yeung97}. This has proven to be useful in a number of cases (see, e.g., \cite{FritzChaves2013, Chaves13}, or \cite{Weilenmann2017} for a detailed review), since the problem is convex in entropy space and the entropic inequalities characterising the relevant sets are independent of the number of measurement outcomes. These advantages have motivated the entropic analysis of causal structures and the entropy vector method is employed for this purpose. In causal structures with observed parentless nodes (such as the Bell scenario), this method can be supplemented with the post-selection technique \cite{BraunsteinCaves88} that can allow for the detection of non-classicality that was previously undetectable in entropy space. In the following, we outline the basics of the entropy vector method, with and without post-selection. For a more thorough overview of entropic techniques for analysing causal structures, we refer the reader to \cite{Weilenmann2017}.

\subsection{The entropy vector method}

The results of this thesis are related to certifying the gap between the classical and quantum sets of correlations in entropy space. For this, one must find necessary conditions satisfied by the set of classical entropies over the observed nodes, and look for quantum violations of these conditions. Hence the entropy vector method in this case only involves classical entropies. In the following, we describe the entropy vector method for the Shannon entropy $H()$, but the same concepts will be later applied to other measures such as the Tsallis or R\'enyi entropies.

\begin{definition}[Entropy vector]
Given a joint
distribution $P_{X_1,\ldots,X_n}\in \mathscr{P}_n$ over $n$ random variables
$X_1,X_2,\ldots,X_n$, the \emph{entropy vector} of $P_{X_1,\ldots,X_n}$ is defined as a vector
with $2^n-1$ components, each of which correspond to the entropy of an
element of the powerset of $\{X_1,X_2,\ldots,X_n\}$ (excluding the empty
set), i.e., the entropy vector can be expressed as
$$\big(H(X_1),...,H(X_n),H(X_1X_2),H(X_1X_3)...,H(X_1X_2...X_n)\big).$$
The entropy vector of a distribution $P\in \mathscr{P}_n$ is denoted as $\mathbf{H}(P)\in \mathbb{R}^{2^n-1}$.
\end{definition}

\begin{definition}[Entropy cone]
\label{definition: entcone}
 Let $\Gamma_n^*$ be set of all vectors in $\mathbb{R}^{2^n-1}$ that are entropy vectors of a probability distribution $P\in \mathscr{P}_n$, i.e., $\Gamma_n^*=\{v\in \mathbb{R}^{2^n-1}: \exists P_{X_1,\ldots,X_n} \text{s.t. } v=\mathbf{H}(P_{X_1,\ldots,X_n})\}$. Its closure $\overline{\Gamma_n^*}$ is known as the \emph{entropy cone} and it includes vectors $v$ for which there exists a sequence $P_k \in \mathscr{P}_n$ such that $\mathbf{H}(P_k)$ tends to $v$ as $k \rightarrow \infty$.
\end{definition}

 Note that the conditional entropies and mutual
informations can be encoded in the entropy vector description through the chain rule~\eqref{eq: chainrule} which gives the relations $H(X|Y)=H(XY)-H(Y)$,
$I(X:Y)=H(X)+H(Y)-H(XY)$ and $I(X:Y|Z)=H(XZ)+H(YZ)-H(XYZ)-H(Z)$. The \emph{entropy cone} $\overline{\Gamma_n^*}$ is known
to be a convex set for any $n$ \cite{Zhang1997ANC}, but it is difficult to characterise and various approximations to it have been considered \cite{Weilenmann2017}.

An outer approximation to the entropy cone, known as the \emph{Shannon cone} is obtained by noting that the information-theoretic properties of entropies (discussed in Section~\ref{ssec: ShanEnt}) imply constraints on valid entropy vectors. These
include non-negativity of the entropies, monotonicity~\eqref{eq: monotonicity} (i.e.,
$H(R)\leq H(RS)$) and submodularity~\eqref{eq: SSA} (also known as
strong-subadditivity; $H(RT)+H(ST)\geq H(RST)+H(T)$). As noted in Section~\ref{ssec: ShanEnt}, in the Shannon case, monotonicity and
submodularity are equivalent to the non-negativity of the conditional
entropy $H(S|R)$ and the conditional mutual information $I(R:S|T)$
respectively and hold for any three disjoint subsets $R$, $S$ and $T$
of $\{X_1,\ldots,X_n\}$. 

\begin{definition}[Shannon constraints and the Shannon cone]
\label{definition: Shancone}
The set of linear constraints comprising of non-negativity, monotonicity and submodularity are together
known as the \emph{Shannon constraints} and the set of vectors
$u \in \mathbb{R}^{2^n-1}$ obeying all the Shannon constraints form
the convex cone known as the \emph{Shannon cone}, $\Gamma_n$.
\end{definition}
  Following standard practice, we will include non-negativity implicitly, such that every entropy vector $v$ belongs to the space $\mathbb{R}_{\geq 0}^{2^n-1}$. Excluding non-negativity, there are a total of $n+n(n-1)2^{n-3}$ independent Shannon
constraints for $n$ variables \cite{Yeung97}. By definition, the
Shannon cone is an outer approximation to $\overline{\Gamma_n^*}$
i.e., $\overline{\Gamma_n^*} \subseteq \Gamma_n$.\footnote{For
  $n\leq3$, the cones coincide, but for $n\geq 4$ they do not
  \cite{Zhang1997ANC}.} Hence all entropy vectors derived from a
probability distribution $P\in \mathscr{P}_n$ obey the Shannon
constraints but not all vectors $u \in \mathbb{R}^{2^n-1}$ obeying the
Shannon constraints are such that $\mathbf{H}(P)=u$
for some joint distribution $P\in \mathscr{P}_n$. 
\paragraph{Entropic causal constraints: } The constraints on the entropy vectors mentioned so far are independent of the causal structure. We know from Section~\ref{ssec: classCM} that a causal structure imposes further constraints on the entropies of the random variables associated with its nodes. In the case of Shannon entropies and a classical causal structure $\cG^C$ over $n$ nodes, all the constraints implied by the causal structure on the entropies can be derived from the following $n$ constraints (of Equation~\eqref{eq: shancausmain}, reproduced here for convenience), one for each node $X_i$
\begin{equation}
\label{eq: entCI}
    I(X_i:X_i^{\nuparrow}|X_i^{\downarrow_1})=0
\end{equation}
These encode the conditional independence of each node from its non-descendants given its parents. Then the set of all entropy vectors $v\in \Gamma^*_n$ that also satisfy the $n$ entropic causal constraints~\eqref{eq: entCI} for a classical causal structure $\cG^C$ are denoted as $\Gamma^*_n(\cG^C)$. Similarly, the set of all entropy vectors in the outer approximation, $v\in \Gamma_n$ that satisfy the causal constraints for $\cG^C$ are denoted as $\Gamma_n(\cG^C)$. Then $\Gamma^*_n\subseteq \Gamma_n$ by construction. Further, it can be shown (Lemma 9 of \cite{Weilenmann2017}) that $\Gamma^*_n(\cG^C)$ is indeed the set of all achievable entropy vectors in $\cG^C$ i.e., 
\begin{equation}
    \Gamma^*_n(\cG^*)=\{v\in \mathbb{R}_{\geq 0}^{2^n-1}|\exists P=\Pi_{i=1}^nP_{X_i|X_i^{\downarrow_1}}\in \mathscr{P}_n \text{ s.t. } \mathbf{H}(P)=v\},
\end{equation}
and that its closure $\overline{\Gamma^*}_n(\cG^C)$ is convex. Note that it need not be finitely generated. However, its outer approximation $\Gamma_n(\cG^C)$ is by construction defined through a finite set of inequalities i.e., the $n+n(n-1)2^{n-3}$ Shannon constraints and the $n$ causal constrains~\eqref{eq: entCI}. 

Given a causal structure $\cG$, we wish to characterise the set of entropy vectors over the observed nodes that can arise in the classical causal structure $\cG^C$. Say the causal structure $\cG$ has $n$ nodes $\{X_1,...,X_n\}$ of which, without loss of generality, the first $k$ are observed and the remaining $n-k$ are unobserved. Then the marginalisation is performed by projecting the entropy cone in $\mathbb{R}^{2^n-1}$ (over all the nodes) to its marginal entropy cone in $\mathbb{R}^{2^k-1}$ (over the observed nodes only). The finitely generated outer approximation, $\Gamma_n(\cG^C)$ is often considered for this purpose since the cone $\overline{\Gamma}^*_n(\cG^C)$ is difficult to characterise. The projection of the cone $\Gamma_n(\cG^C)$ into the subspace of the $k$ observed nodes defines the marginal cone $\Gamma_k(\cG^C)$, and can be obtained through Fourier-Motzkin Elimination \cite{Williams1986} (Section~\ref{ssec: FME}). This provides a $\mathcal{H}$-representation of the marginal cone $\Gamma_k(\cG^C)$, and every marginal entropy vector over the observed nodes arising from a compatible distribution in the classical causal structure $\cG^C$ necessarily belongs to this cone. 
Since non-classical causal
structures do not satisfy the initial assumption of the existence of
the joint distribution/entropies, they may give rise to correlations
that do not satisfy the marginal constraints on the observed nodes
obtained through this procedure. A violation of any of the inequalities defining $\Gamma_k(\cG^C)$ by the entropy vector $v=\mathbf{H}(P_Q)$ of a distribution $P_Q$ compatible with the quantum causal structure $\cG^Q$, certifies the non-classicality of $P_Q$.

%For line-like causal structures (of which the bipartite Bell causal structure of Figure~\ref{fig: Bell} is an instance), the classical and quantum Shannon entropy cones coincide and Shannon entropic inequalities cannot certify the non-classicality of these causal structures even though they support non-classical correlations~\cite{Weilenmann16}. Further, in other scenarios such as the Triangle which is also known to support non-classical correlations~\cite{Fritz2012}, known Shannon entropic inequalities such as those of~\cite{Chaves2014, Weilenmann2018} have no known quantum violations. The main question of the current work is whether using Tsallis entropies can provide tighter, quantum violatable entropic inequalities and avoid these limitations.

\subsection{Post-selected causal structures}\label{ssec: PS}

\begin{figure}[t!]
\centering
	\subfloat[]{\includegraphics[scale=1.0]{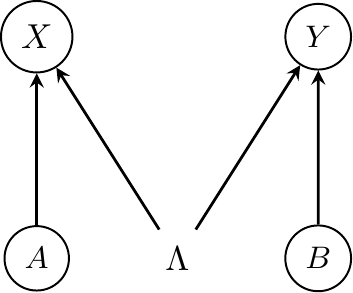}\label{fig: Bell2}}\qquad\qquad\qquad\qquad\qquad\qquad
\subfloat[]{\includegraphics[scale=1.0]{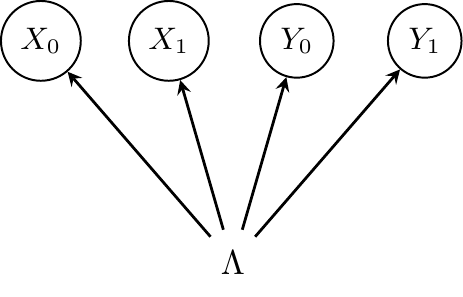}\label{fig: BellPS}}
\caption[The bipartite Bell causal structure and its post-selected version]{\textbf{The bipartite Bell causal structure and its post-selected version: }\textbf{(a)} The bipartite Bell causal structure. The nodes $A$ and
  $B$ represent the random variables corresponding to independently
  chosen inputs, while $X$ and $Y$ represent the random
  variables corresponding to the outputs. $\Lambda$ is an unobserved
  node representing the common cause of $X$ and $Y$. \textbf{(b)} The
  post-selected bipartite Bell causal structure for the case of binary inputs. The observed
  nodes $X_a$ represent the outputs when the input is $a\in\{0,1\}$
  and likewise for $Y_b$. Note that $X_0$ and $X_1$ are never
  simultaneously observed and likewise $Y_0$ and $Y_1$.}
\end{figure}
 In causal structures with one or more observed parentless nodes, supplementing the entropy vector method with an additional technique can allow for tighter characterisations of the marginal scenario. This technique involves defining, for each given causal structure with at least one parentless node, a new causal structure called the post-selected causal structure. The general technique for doing this can be found in~\cite{Weilenmann2017} (for example). Here, we will only consider the bipartite Bell causal structure (Figure~~\ref{fig: Bell}, and reproduced in Figure~\ref{fig: Bell2} for convenience) with two inputs per party and the post-selected version thereof. 
 
 The post-selected causal structure is obtained by removing the parentless observed nodes $A$ and $B$ in the original causal structure~\ref{fig: Bell2} and replacing the descendants $X$ and $Y$ with two copies of each i.e., $X_{A=0}$, $X_{A=1}$, $Y_{B=0}$, $Y_{B=0}$ such that the original causal relations are preserved and there is no mixing between the copies (this is shown in Figure~\ref{fig: BellPS}).  It makes sense to do this in the classical case because classical information can be copied, so we can simultaneously consider the outcome $X$ given $A=0$ and that given $A=1$.  By contrast, in the quantum case the values of $A$ correspond to different measurements that are used to generate $X$, and the associated variables $X_{A=0}$ and $X_{A=1}$ may not co-exist. It hence does not make sense to consider a joint distribution over $X_{A=0}$ and $X_{A=1}$ in this case.  We therefore only consider the subsets of the observed variables that co-exist
\begin{equation}
\label{eq: coexisting}
 \mathcal{S}:=\{X_0,\: X_1,\: Y_0,\: Y_1,\: X_0Y_0,\: X_0Y_1,\: X_1Y_0,\: X_1Y_1\} ,  
\end{equation}
where we use the short form $X_0Y_0$ for the set $\{X_0,Y_0\}$ etc.
Any non-trivial inequalities derived for the co-existing sets in the classical case can admit quantum or GPT violations.

\subsection{Entropy vectors and post-selection}
\label{ssec: entineq}
In \cite{BraunsteinCaves88}, Braunstein and Caves derived a set of
constraints on the post-selected causal structure of Figure~\ref{fig:
  BellPS} and showed that these constraints can be violated by
quantum correlations. In fact, it is now known that in the absence of post-selection, the sets of entropies over the observed nodes in the classical and quantum causal structures, $\cG_B^C$ and $\cG_B^Q$ coincide and therefore, the non-classicality of $\cG_B$ cannot be detected entropically \cite{Weilenmann16}. To discuss post-selected entropic inequalities, we first introduce the notion
of \emph{entropic classicality}. For every distribution $P_{XY|AB}$ in the Bell causal structure (Figure~\ref{fig: Bell2}), we can associate an entropy vector $v\in\mathbb{R}^{8}$ in the post-selected causal structure (Figure~\ref{fig: BellPS}) whose components are the entropies of each element of the set $\mathcal{S}$ (Equation~\eqref{eq: coexisting}) distributed according to $P_{X_aY_b}:=P_{XY|A=a,B=b}$. Let $\mathbf{H}$ be
the map that takes the observed distribution to
its corresponding entropy vector in the post-selected causal structure. 

\begin{definition}[Entropic classicality]
\label{definition: entclass}
An entropy vector $v\in\mathbb{R}^8$ is \emph{classical} with respect to the bipartite Bell causal structure (Figure~\ref{fig: Bell2}) if there exists a classical distribution $P_{XY|AB}\in\mathscr{L}$ such that $\mathbf{H}(P_{XY|AB})=v$, where $\mathscr{L}$ is the local or classical polytope (Equation~\eqref{eq: loccaus}). Further, a distribution $P_{XY|AB}$ is \emph{entropically classical} if there exists a classical distribution with the same entropy vector, i.e., if there exists a classical entropy vector $v$ such that $\mathbf{H}(P_{XY|AB})=v$.
\end{definition}
 
The set of all classical entropy vectors forms a convex cone. We now review how the Braunstein Caves (BC) Inequalities are derived for the case when the observed parentless nodes $A$ and $B$ are binary. In this case, the post-selected causal structure~\ref{fig: BellPS} imposes no additional constraints on the distribution (or entropies) of the observed nodes $X_0$, $X_1$, $Y_0$ and $Y_1$ because they share a common parent and thus any joint distribution over $X_0$, $X_1$, $Y_0$ and $Y_1$ can be realised in the causal structure~\ref{fig: BellPS}. By contrast, any correlations in the original causal structure~\ref{fig: Bell2} must obey the no-signaling constraints over the observed nodes $A$, $B$, $X$ and $Y$ since $A$ does not influence $Y$ and $B$ does not influence $X$ in this causal structure. The inequalities derived by Braunstein and Caves follow by applying monotonicity~\eqref{eq: monotonicity}, strong subadditivity in unconditional form~\eqref{eq: SSA} and the chain rule~\eqref{eq: chainrule} to the entropies of the observed variables $\{X_0,X_1,Y_0,Y_1\}$ in the post-selected causal structure. The derived relations hold for the classical causal structure (and not necessarily for the quantum and GPT cases) because only in the classical case does it make sense to consider a joint distribution over these four variables that in the quantum and GPT cases do not co-exist (c.f.\ Section~\ref{ssec: PS}). The BC inequalities are entropic Bell inequalities i.e., they hold for every classical entropy vector in the post-selected causal structure~\ref{fig: BellPS}.  There are four BC inequalities
%\begin{widetext}
\begin{equation}
\label{eq: BCineqs}
    \begin{split}
        I_{\BC}^1:=H(X_0Y_0)+H(X_1)+H(Y_1)-H(X_0Y_1)-H(X_1Y_0)-H(X_1Y_1)\leq 0\\
        I_{\BC}^2:=H(X_0Y_1)+H(X_1)+H(Y_0)-H(X_0Y_0)-H(X_1Y_0)-H(X_1Y_1)\leq 0\\
        I_{\BC}^3:=H(X_1Y_0)+H(X_0)+H(Y_1)-H(X_0Y_0)-H(X_0Y_1)-H(X_1Y_1)\leq 0\\
        I_{\BC}^4:=H(X_1Y_1)+H(X_0)+H(Y_0)-H(X_0Y_0)-H(X_0Y_1)-H(X_1Y_0)\leq 0\\
    \end{split}
\end{equation}
%\end{widetext}
It has been shown in~\cite{Fritz13} that these four
inequalities are complete in the following sense (the lemma below is
implied by Corollary~V.3 in~\cite{Fritz13}).
\begin{lemma}\label{lem:BCcomplete}
  A distribution in the post-selected Bell scenario with binary $A$ and $B$ is entropically classical if and only if it satisfies the four BC inequalities~\eqref{eq: BCineqs}.
\end{lemma}
%It turns out that in the $(2,2,2,2)$ Bell scenario, non-classical distributions that do not violate the BC inequalities can be made to do so with some additional post-processing, as shown in~\cite{Chaves13}. We review this result below before analysing the same question in the $(2,2,3,3)$ scenario.
Note that the BC inequalities~\eqref{eq: BCineqs} and this completeness result are independent of the cardinality of $X$ and $Y$. However, a crucial point is that entropic classicality does not imply classicality of the associated distribution. For example, the entropy vector $\mathbf{H}(P_{PR})$ associated with the maximally non-classical distribution $P_{PR}$ (Equation~\eqref{eq: PR}) satisfies all the BC inequalities is hence entropically classical. Since entropic inequalities are non-linear in the probabilities, it is nevertheless possible to have a convex combination of two entropically classical distributions that is entropically non-classical. This allows for the non-classicality of a distribution such as $P_{PR}$ to be ``activated'' in entropy space by mixing with a classical distribution. Hence a natural question that arises is whether post-selected entropic inequalities can identify all non-classical distributions in this manner, after a suitable and reasonable post-processing operation (such as mixing with a classical distribution). This is known to be the case for $(d,d,2,2)$ Bell scenarios with $d\geq 2$ \cite{Chaves2012}, but has been an open question for the non-binary outcome case. In Chapter~\ref{chapter: mixingpaper} (which is based on \cite{Vilasini2020}), we analyse this question for the 2 input, $d$ outcome Bell scenarios or $(2,2,d,d)$ scenarios with $d\geq 3$, and our results suggest that the answer is negative in this case for both Shannon and Tsallis entropic inequalities. The Triangle causal structure (Figure~\ref{fig: Triangle}) on the other hand cannot be analysed using post-selection since it has no observed parentless nodes. We analyse this causal structure using Tsallis entropies, in the absence of post-selection in Chapter~\ref{chapter: Tsallispaper} (based on \cite{Vilasini2019}). We leave further details of the entropic technique and of these results to Chapters~\ref{chapter: Tsallispaper} and~\ref{chapter: mixingpaper}.

\chapter{Entropic analysis of causal structures without post-selection}
\label{chapter: Tsallispaper}

\lettrine[nindent=0em, slope=-.5em,lines=2]{A}{s} discussed in  Section~\ref{sec:entvec}, the main motivation for the entropic technique for analysing causal structures stems from their ability to overcome certain problems that arise in probability space, such as the non-convexity of correlation sets and the drastic scaling of the number of Bell inequalities in the cardinality of the variables.
The entropic method for analysing causal structures has proven to be useful in a number of cases (see, e.g.,~\cite{Fritz13,
  Chaves13}, or~\cite{Weilenmann2017} for a detailed review). However, it was shown
in~\cite{Weilenmann16} that the entropy vector method with Shannon
entropies cannot detect the classical-quantum gap for line-like causal
structures, in the absence of post-selection. This includes causal structures such as the Bell scenario which are well known to support non-classical correlations. Although new Shannon
entropic inequalities have been derived using this method (e.g., in the Triangle causal structure), no quantum
violation of these have been found even when non-classical correlations are known to
exist in these causal structures~\cite{Weilenmann2018, Chaves2014}. Due to these limitations of
Shannon entropies, it is natural to ask whether other entropic
quantities could do better. 

This chapter is predominantly based on work carried out with Roger Colbeck \cite{Vilasini2019}, while Appendix~\ref{appendix: Renyiresults} reports observations from a side-project with Mirjam Weilenmann. The main question we address here is whether generalised entropic inequalities can overcome these limitations of Shannon entropies in the absence of post-selection. In particular we employ Tsallis entropies and analyse the Bell~\ref{fig: Bell} and Triangle causal structures~\ref{fig: Triangle}. A motivation for considering Tsallis
entropies for the task is that they are a family with an additional
(real) parameter.  The set of entropies for all possible values of
this parameter conveys more information about the underlying
probability distribution than a single member of the family and hence
the ability to vary a parameter may give advantages for analysing causal structures.  Tsallis entropies appear to be a good candidate since they satisfy monotonicity, strong subadditivity (in both conditional and unconditional forms)
and the chain rule which are desirable properties for their use in the
entropy vector method.\footnote{Other examples of more
  general entropy measures such as the R\'enyi entropy~\cite{Renyi1961} do
  not satisfy one or more of these properties, making it more difficult to get entropic constraints on them using the entropy vector method (c.f. Table~\ref{tab:ent_prop}).} Tsallis entropies
have been considered in the context of causal structures
before~\cite{Wajs15} where they were shown to give an advantage over
Shannon entropy in detecting the non-classicality of certain states in
the post-selected Bell scenario~\ref{fig: BellPS}\footnote{Note that non-classicality cannot be
  detected Shannon entropically in the Bell causal structure (Figure~\ref{fig:
    Bell}) without post-selection \cite{Weilenmann16}.}. In this chapter we analyse causal structures  in the absence of post-selection and consider post-selection in Chapter~\ref{chapter: mixingpaper}. Note however that the post-selection technique (c.f. Section~\ref{ssec: PS}) can only be applied to causal structures with at least one observed parentless node i.e., it can be applied to the Bell, but not to the Triangle causal structure.

\section{Summary of contributions: in words and in poetry} \label{sec:introduction} 

The entropy vector method requires as input the constraints implied by a causal structure on the entropies of its nodes. Finding that these constraints in for Tsallis entropies cannot take the same form as the Shannon ones (Equation~\eqref{eq: entCI}), we derive the constraints
on the classical Tsallis entropies that are implied by a given causal
structure in Section~\ref{sec: tsalcaus}. As an additional result, we generalise these entropic causal constraints to quantum Tsallis entropies for certain cases, in Appendix~\ref{appendix: quantum}. In Section
\ref{ssec: tsalentvec}, we use these causal constraints in the entropy vector
method with Tsallis entropies but find that the computational
procedure becomes too time consuming even for simple causal structures
such as the bipartite Bell scenario. Despite this limitation, we
derive new Tsallis entropic inequalities for the Triangle causal
structure in Section~\ref{sec: newineq}, using known Shannon entropic
inequalities of~\cite{Chaves2014} and our Tsallis constraints of
Section~\ref{sec: tsalcaus}. In Section~\ref{sec: causaldiscussion},
we discuss the reasons for the computational difficulty of this
method, the drawbacks of using Tsallis entropies for analysing causal
structure and identify potential future directions stemming from our work. In Appendix~\ref{appendix: LPA}, we provide details of the Mathematica package {\sc LPAssumptions} \cite{LPAssumptions} that we developed and used for obtaining some of the computational results of this, and the next chapter. Additionally, in Appendix~\ref{appendix: Renyiresults}, we summarise the results obtained together with Mirjam Weilenmann where we also found limitations of using R\'enyi entropies for certifying non-classicality in causal structures.
Below is a poetic summary of our paper \cite{Vilasini2019} on which the current chapter is based, following which we present the results (in a more rigorous, non-poetic fashion).

\textit{About underlying physics we want to tell,\\
Using observed correlations, as did Bell\\
For a \hyperref[fig: Bell]{scenario} that was blessed,\\
In \hyperref[sec: Probspace]{probability space}, with convex correlation sets.\\
Generally though,\\
That may not be so.}

\textit{So we move to \hyperref[sec:entvec]{entropy space},\\
Putting convexity back in place,\\
But we can not be too smug,\\
For the Shannon cones aren't too snug.}

\textit{So here we try our luck with Tsallis entropies,\\
Hoping they give us tighter inequalities.\\
On the way we find some \hyperref[sec: tsalcaus]{constraints} that are new,\\
That follow from \hyperref[{corollary: mainconstraint}]{classical causal structures}, and some \hyperref[appendix: quantum]{quantum ones too}}

\textit{Then \hyperref[ssec: tsalentvec]{we use them} in the method of entropy vectors.\\
Alas, that seems too much for our little computers.\\
But we have a cheeky way to put our constraints to use,\\
Using known Shannon results, \hyperref[eq: Tsineq]{new Tsallis inequalities} we produce.}

\textit{Though we found nice math results along the way,\\
Our Tsallis inequalities also \hyperref[ssec: qviol]{seem hard to violate},\\
Computationally costly to find tighter inequalities,\\
A feature or a bug of the entropic technique?}

%{\raggedleft \textit{-Vilasini Venkatesh (2019)}\par}

\section{Causal constraints and Tsallis entropy vectors}
\label{sec: tsalcaus}

In Section \ref{ssec: tsalprop}, we discussed some of the general properties of Tsallis entropy that hold irrespective of the underlying causal structure over the variables. These include non-negativity, monotonicity~\eqref{eq: monotonicity}, and strong subadditivity~\eqref{eq: SSA} which characterise the set of \emph{Shannon constraints} (c.f. Definition~\ref{definition: Shancone}). We summarise these properties here for convenience. For any joint distribution
over the random variables involved the following properties hold. 
\begin{enumerate}
    \item \label{prop: pseudo}\textbf{Pseudo-additivity \cite{Curado1991}:} For two independent random
      variables $X$ and $Y$ i.e., $P_{XY}=P_XP_Y$, and for all $q$, the Tsallis entropies satisfy
    \begin{equation}
        \label{eq: pseudoadd2}
        S_q(XY)=S_q(X)+S_q(Y)+(1-q)S_q(X)S_q(Y).
    \end{equation}
    Note that in the Shannon case ($q=1$), we recover additivity for
    independent random variables.
  \item \label{prop: upper} \textbf{Upper bound \cite{Furuichirelentropy}:} For $q\geq 0$ we have
    $S_q(X)\leq\ln_qd_X$.  For $q>0$ equality is achieved if and only
    if $P_X(x)=1/d_X$ for all $x$ (i.e., if the distribution on $X$ is
    uniform).
    \item \textbf{Monotonicity \cite{Daroczy1970}:} For all $q$,
    \begin{equation}
        \label{eq: mono}
        S_q(X)\leq S_q(XY).
    \end{equation}
    \item \textbf{Strong subadditivity \cite{Furuichi04}:} For $q\geq1$,
    \begin{equation}
    \label{eq: subadd}
           S_q(XYZ)+S_q(Z)\leq S_q(XZ)+S_q(YZ).
    \end{equation}
   
    \item \textbf{Chain rule~\cite{Furuichi04}:} For all $q$,
    \begin{equation}
        \label{eq: chain2}
        S_q(X_1,X_2,\ldots,X_n|Y)=\sum_{i=1}^n S_q(X_i|X_{i-1},\ldots,X_1,Y).
    \end{equation}
\end{enumerate}
   The chain rules $S_q(XY)=S_q(X)+S_q(Y|X)$ and
   $S_q(XY|Z)=S_q(X|Z)+S_q(Y|XZ)$ emerge as particular cases and allow
   the Tsallis mutual informations (Equation~\eqref{eq: tsalmi1}) to be
   written as follows.\footnote{Using the chain rule, the monotonicity and strong subadditivity relations (Equations~\eqref{eq: mono} and~\eqref{eq: subadd}) are equivalent to the non-negativity of the unconditional and conditional Tsallis mutual
informations. For $q<1$, strong subadditivity does not hold in
general~\cite{Furuichi04}, hence we often restrict to the case $q\geq
1$ in what follows.} Note that, due to the chain rules, the conditional  version of the properties are also satisfied. 
    \begin{equation}
\label{eq: tsalmi2}
    \begin{split}
        I_q(X:Y)&=S_q(X)+S_q(Y)-S_q(XY),\\
        I_q(X:Y|Z)&=S_q(XZ)+S_q(YZ)-S_q(Z)-S_q(XYZ).
    \end{split}
\end{equation}

The causal structure imposes the causal Markov constraints on the joint probability distribution as well as entropic causal constraints over the variables involved (Section~\ref{ssec: classCM}). We have seen in Section~\ref{sec:entvec} that the entropy vector method requires as an input, the complete set of entropic causal constraints that encode all the conditional independences of the causal structure. In the case of Shannon entropies, for an $n$ node causal structure, the $n$ Shannon entropic causal constraints of Equation~\eqref{eq: entCI} imply all the conditional independences that follow from d-separation for that causal structure. To analyse causal structures using Tsallis entropies, we require analogous Tsallis entropic causal constraints. 

A first observation is that Tsallis entropy vectors do not in general satisfy the causal constraints (Equation~\eqref{eq: entCI}) satisfied by their Shannon counterparts. For a concrete counterexample, consider the simple, three variable causal structure where $Z$ is a common cause of $X$ and $Y$, and where there are no other causal relations. In terms of Shannon entropies, the only causal constraint in this case is $I(X:Y|Z)=0$. Taking $X,Y$ and $Z$ to be binary variables with possible values $0$ and $1$, the distribution $P(xyz)=1/4$ $\forall x\in X, y\in Y$ if $z=0$ and $P(xyz)=0$ otherwise, satisfies $P(xy|z)=P(x|z)P(y|z)$ $\forall x\in X, y\in Y$ and $z\in Z$ but has a $q=2$ Tsallis conditional mutual information of $I_2(X:Y|Z)=\frac{1}{4}$. Hence when using Tsallis entropies (and conditional Tsallis entropy as defined in Section~\ref{ssec: tsalprop}), the causal constraint cannot be simply encoded by $I_q(X:Y|Z)=0$ for $q>1$.

Given this observation, it is natural to ask whether there are constraints for Tsallis entropies implied by the causal Markov condition (Equation~\eqref{eq: markov}). We answer this question with the following Theorems. Note that we will be using the notation $d_X$ to denote the cardinality/alphabet size $|X|$ of a random variable $X$ in order the make the notation consistent with the quantum case (Section~\ref{appendix: quantum}) where $d_X$ will denote the dimension of the Hilbert space associated with the subsystem $X$. Then for classical states, $d_X$ coincides with the cardinality of the corresponding random variable, which justifies this notation.

\begin{theorem}
\label{theorem: mibound}
If a joint probability distribution $P_{XY}$ over random variables $X$ and $Y$ with alphabet sizes $d_X$ and $d_Y$ factorises as $P_{XY}=P_XP_Y$, then for all $q\in[0,\infty)$, the Tsallis mutual information $I_q(X:Y)$ is upper bounded by
\begin{equation*}
    I_q(X:Y)\leq f(q,d_X,d_Y)\,,
\end{equation*}
where the function $f(q,d_X,d_Y)$ is given by
\begin{equation*}
    f(q,d_X,d_Y)=\frac{1}{(q-1)}\left(1-\frac{1}{d_X^{q-1}}\right)\left(1-\frac{1}{d_Y^{q-1}}\right)=(q-1)\ln_qd_X\ln_qd_Y.
\end{equation*}
For $q\in(0,\infty)\setminus\{1\}$, the bound is saturated if and
only if $P_{XY}$ is the uniform distribution over $X$ and $Y$.
\end{theorem}
\begin{proof}
The proof follows from the pseudo-additivity of Tsallis entropies
(Property~\ref{prop: pseudo}) and the upper bound (Property~\ref{prop:
  upper}).  Using these, for all $q\geq 0$ and for all separable
distributions $P_{XY}=P_XP_Y$, we have
\begin{equation}
    \label{eq: mibound}
    I_q(X:Y)=S_q(X)+S_q(Y)-S_q(XY)=(q-1)S_q(X)S_q(Y)\leq \frac{\left(1-\frac{1}{d_X^{q-1}}\right)\left(1-\frac{1}{d_Y^{q-1}}\right)}{q-1}=f(q,d_X,d_Y)\,.
\end{equation}
Whenever $q\in(0,\infty)\setminus\{1\}$, the bound is saturated if
and only if $P_{XY}$ is uniform over $X$ and $Y$ since,
for these values of $q$, $S_q(X)$ and $S_q(Y)$ both attain their
maximum values if and only if this is the case.
\end{proof}

\begin{theorem}
\label{theorem: causal}
If a joint probability distribution $P_{XYZ}$ satisfies the conditional independence $P_{XY|Z}=P_{X|Z}P_{Y|Z}$, then for all $q\geq 1$ the Tsallis conditional mutual information $I_q(X:Y|Z)$ is upper bounded by
\begin{equation*}
    I_{q}(X:Y|Z)\leq f(q,d_X,d_Y)\,.
\end{equation*}
 For $q>1$, the bound is saturated only by distributions in which for
 some fixed value $k$ the joint probabilities are given by
 $P(xyz)=\begin{cases}\frac{1}{d_Xd_Y} \quad \text{if}\quad z=k\\ 0
   \quad \qquad \text{otherwise}\end{cases}$ for all $x$, $y$ and
 $z$\footnote{These distributions have deterministic $Z$ and there is
   one such distribution for each value that $Z$ can take.}.
\end{theorem}
\begin{proof}
Writing out $I_q(X:Y|Z)$ in terms of probabilities we have
\begin{align*}
    I_q(X:Y|Z)&=\frac{1}{q-1}\big[\sum_{xyz}P^q(xyz)+\sum_z P^q(z) -\sum_{xz} P^q(xz)-\sum_{yz}P^q(yz)\big]\\
    &=\frac{1}{q-1}\sum_zP^q(z)\big[\sum_{xy}P^q(xy|z)+1 -\sum_{x} P^q(x|z)-\sum_{y}P^q(y|z)\big]\\
    &=\sum_z P^q(z)I_q(X:Y)_{P_{XY|Z=z}}.
\end{align*}
Using this and Theorem \ref{theorem: mibound}, we can bound $I_q(X:Y|Z)$ as 
\begin{align*}
\max\limits_{\substack{P_{XYZ}=P_ZP_{X|Z}P_{Y|Z}}} I_q(X:Y|Z) &=\max\limits_{\substack{P_{XYZ}=P_ZP_{X|Z}P_{Y|Z}}}\sum_z P(z)^qI_q(X:Y)_{P_{XY|Z=z}}\\
&\leq \max\limits_{P_Z}\sum_z P(z)^q \max\limits_{P_{X|Z=z}P_{Y|Z=z}}I_q(X:Y)_{P_{XY|Z=z}}\\
&=\max\limits_{P_Z}\sum_zP(z)^qf(q,d_X,d_Y)=f(q,d_X,d_Y)\,.
\end{align*}
The last step holds because for all $q>1$, $\sum_zP(z)^q$ is
maximized by deterministic distributions over $Z$ with a maximum value
of $1$ i.e., only distributions $P_{XYZ}$ that are deterministic over
$Z$ saturate the upper bound of $f(q,d_X,d_Y)$.  This completes the proof.
\end{proof}

Two corollaries of Theorem~\ref{theorem: causal} naturally follow.
\begin{corollary}
\label{corollary: causal} Let $X$, $Y$ and $Z$ be random variables
with fixed alphabet sizes.  Then for all $q\geq1$ we have
\begin{equation*}
   \max\limits_{\substack{P_{XYZ}\\ P_{XY|Z}=P_{X|Z}P_{Y|Z}}} I_q(X:Y|Z) = \max\limits_{\substack{P_{XY}\\ P_{XY}=P_XP_Y}} I_q(X:Y)\,.
\end{equation*}
Furthermore, for $q>1$, the maximum on the left hand side is achieved only by distributions in which for some fixed value $k$ the joint probabilities are given by $P(xyz)=\begin{cases}\frac{1}{d_Xd_Y} \quad \text{if} \quad z=k\\ 0
  \quad \qquad \text{otherwise}\end{cases}$, while the maximum on the right hand side occurs if and
only if $P_{XY}$ is the uniform distribution.
\end{corollary}

The significance of these new relations for causal structures is then
given by the following corollary.

\begin{corollary}
\label{corollary: mainconstraint}
Let $P_{X_1\ldots X_n}$ be a distribution compatible with the classical
causal structure $\cG^\rC$ and $X$, $Y$ and $Z$ be disjoint subsets of
$\{X_1,\ldots,X_n\}$ such that $X$ and $Y$ are d-separated given $Z$.
Then for all $q\geq1$ we have
\begin{equation*}
    I_q(X:Y|Z)\leq f(q,d_X,d_Y)\,,
\end{equation*}
where $d_X$ is the product of $d_{X_i}$ for all $X_i\in X$, and
likewise for $d_Y$.
\end{corollary}

\begin{remark}
The results of this section can be generalised to the quantum case under certain assumptions i.e., as constraints on quantum Tsallis entropies implied by certain quantum causal structures (see Appendix~\ref{appendix: quantum} for details). Note that only constraints on the classical Tsallis entropy vectors derived in this section are required to detect the classical-quantum gap. Hence, Appendix~\ref{appendix: quantum} is not pertinent to the main results of this chapter but can be seen as additional results regarding the properties of quantum Tsallis entropies.
\end{remark}

\subsection{Number of independent Tsallis entropic causal constraints}
\label{ssec: numconst}
We saw previously that in the Shannon case ($q=1$), the $n$ conditions~\eqref{eq: entCI}
of the form $I(X_i:X_i^\nuparrow|X_i^{\downarrow_1})=0$
($i=1,\ldots,n$) imply all the independence relations that follow from
the causal structure. In the Tsallis case however, the $n$ conditions
of the form
$I_q(X_i:X_i^{\nuparrow}|X_i^{\downarrow_1})\leq
f(q,d_{X_i},d_{X_i^{\nuparrow}})$ (Corollary~\ref{corollary: mainconstraint}) do not do the same.  In the
bipartite Bell and Triangle causal structures, in the case where the dimension (cardinality) of each
individual node is taken to be $d$, we find that there are 53 and 126 distinct Tsallis entropic
inequalities that are implied by the respective causal structures. These are in one-to-one correspondence with the d-separation relations in the
corresponding DAGs, and there are no redundancies in these constraints i.e., all of them are required for characterising the conditional independences in the DAGs.  In more detail, we used linear
programming to show that each implication of d-separation yields a
non-trivial entropic causal constraint for all $q>1$ and $d>2$ for the
bipartite Bell and Triangle causal structures. By comparison, in these causal structures five and six independent Shannon entropic constraints imply all the others.  As an illustration of the
difference, in the Shannon case, $I(A:BC)=0$ implies $I(A:B)=I(A:C)=0$, whereas
the analogous implication does not hold in the Tsallis case in
general: although $I_q(A:BC)\leq f(q,d_A,d_{BC})$ implies
$I_q(A:B)\leq f(q,d_A,d_{BC})$, it is not the case that
$I_q(A:BC)\leq f(q,d_A,d_{BC})$ implies
$I_q(A:B)\leq f(q,d_A,d_B)$.\footnote{For an explicit counterexample,
  consider
  $P_{ABC}=\{\frac{3}{10},0,\frac{2}{10},0,\frac{1}{10},\frac{1}{10},\frac{2}{10},\frac{1}{10}\}$
  over binary $A$, $B$ and $C$ for which $I_2(A:BC)=9/25<3/8=f(2,2,4)$
  but $I_2(A:B)=13/50>1/4=f(2,2,2)$.}

The number of distinct conditional independences (and hence the number
of independent Tsallis constraints that follow from d-separation) in a
DAG depends on the specific graph, however for any DAG $\cG_n$ with
$n$ nodes, the number of such constraints can be upper bounded by that
of the $n$-node DAG where all $n$ nodes are independent i.e., the $n$
node DAG with no edges. The number of conditions in this DAG can be
thought of as the number of ways of partitioning $n$ objects into four
disjoint subsets\footnote{The four subsets correspond
to the three arguments of the conditional mutual information and a set
of `leftovers'.} such that the first two are non-empty and where the
ordering of the first two does not matter. Therefore, there are at most
$\frac{1}{2}(4^n-2\times3^n+2^n)$ such conditions.

\subsection{Using Tsallis entropies in the entropy vector method}
\label{ssec: tsalentvec}

We used the causal constraints of Corollary \ref{corollary:
  mainconstraint} in the entropy vector method (Section~\ref{sec:entvec}) with the aim of
deriving new quantum-violatable entropic inequalities for the Triangle
causal structure (Figure \ref{fig: Triangle}). To do so, we started
with the variables $A,B,C,X,Y,Z$ of the Triangle causal structure, the
Shannon constraints and causal constraints satisfied by the Tsallis
entropy vectors over these variables (Corollary~\ref{corollary:
  mainconstraint}) and used a Fourier-Motzkin (FM) elimination (Section~\ref{ssec: FME})
algorithm (from the {\sc porta} software \cite{porta}) to
eliminate the Tsallis entropy components involving the unobserved
variables $A,B,C$ and obtain the constraints on the observed
nodes $X,Y,Z$. The Tsallis entropy vector for the six nodes has $2^6-1=63$ components. The
required marginal scenario with the observed nodes $X,Y,Z$ has Tsallis
entropy vectors with $2^3-1=7$ components and in this case, the Fourier-Motzkin algorithm has to run $56$ iterations, each of which eliminates
one variable.

Starting with the full set of 126 Tsallis entropic causal constraints
for the Triangle causal structure as well as the 246 independent
Shannon constraints, the Fourier-Motzkin elimination algorithm did not
finish within several days on a standard desktop PC and the number of
intermediate inequalities generated grew to about 90,000 after 11 steps. Because of this we instead tried starting with a subset
comprising 15 of the 126 Tsallis entropic causal
constraints\footnote{These included the 6 that follow from ``each node
  $N_i$ is conditionally independent of its descendants given its
  parents'' (denoted as $N_i\perp^d N_i^{\nuparrow}|N_i^{\downarrow_1}$)
  and 9 more chosen arbitrarily from the total of 126 independent
  Tsallis constraints we found for the Triangle. The 6 former
  constraints for the Triangle (Figure~\ref{fig: Triangle}) are
  $A\perp^d CXB$, $B\perp^d CYA$, $C\perp^d BZA$, $X\perp^d YAZ|CB$,
  $Y\perp^d XBZ|AC$ and $Z\perp^d YCX|AB$. An example of 9 more
  constraints for which the procedure did not work are $X\perp^d Y|CB$,
  $X\perp^d A|CB$, $X\perp^d Z|CB$, $Y\perp^d X|AC$, $Y\perp^d B|AC$,
  $Y\perp^d Z|AC$, $Z\perp^d Y|AC$, $Z\perp^d C|AB$ and $Z\perp^d X|AB$.  We
  also tried some other choices and number of constraints but this did
  not lead to any improvement.}  i.e., 261 constraints on 63
dimensional vectors. We considered the case of $q=2$ and where the six
random variables are all binary. Again, in this case the algorithm did
not finish after several days. We also tried starting with fewer
causal constraints (for example, the six constraints analogous to the
Shannon case) as well as using a modified code, optimised to deal with
redundancies better but both of these attempts made no significant
difference to this outcome.

Such a rapid increase of the number of inequalities in each step is a
known problem with Fourier-Motzkin elimination where an elimination
step over $n$ inequalities can result in up to $n^2/4$ inequalities in
the output and running $d$ successive elimination steps can yield a
double exponential complexity of $4(n/4)^{2^d}$
\cite{Williams1986}. This rate of increase can be kept under control
when the resulting set of inequalities has many redundancies. This
happens in the Shannon case where the causal constraints are simple
equalities and the system of 246 Shannon constraints plus 6 Shannon
entropic causal constraints reduces to a system of just 91 independent
inequalities before the FM elimination. In the Tsallis case, no
reduction of the system of inequalities is possible in general due to
the nature of the causal constraints. The fact that the Tsallis
entropic causal constraints are inequality constraints rather than
equalities also contributes to the computational difficulty since each
independent equality constraint in effect reduces the dimension of the
problem by~1.

We also tried the same procedure on the bipartite Bell causal
structure (Figure \ref{fig: Bell}), again for $q=2$ and binary
variables. Starting with the full set of 53 causal constraints,
again resulted in the program running for over a week without nearing
the end, and a similar result was obtained when starting only with
8--10 causal constraints. While starting with fewer causal constraints
such as the 5 conditional independence constraints (one for each node)
resulted in a terminating program, no non-trivial entropic
inequalities were obtained (i.e., we only obtained those relating to the Shannon constraints or following directly from d-separation).\footnote{For example, we were able to obtain
$I_2(A:BY)\leq \frac{7}{16}$ and $I_2(B:AX)\leq \frac{7}{16}$, while,
in the case of binary variables and $q=2$, the independences in the
DAG together with Theorem~\ref{theorem: mibound} imply
$I_2(A:BY)\leq \frac{6}{16}$ and $I_2(B:AX)\leq \frac{6}{16}$, which
are the Tsallis entropic equivalents of the two no-signaling
constraints.}

\section{New Tsallis entropic inequalities for the Triangle causal structure}\label{sec: newineq}
Despite the limitations encountered in applying the entropy vector
method to Tsallis entropies (Section~\ref{ssec: tsalentvec}), here we
find new Tsallis entropic inequalities for the Triangle causal
structure for all $q\geq1$ by using known inequalities for the Shannon entropy~\cite{Chaves2014} and the causal constraints derived in Section~\ref{sec: tsalcaus}. Using the entropy vector method for Shannon entropies, the following three classes of entropic inequalities were obtained for the Triangle causal structure (Figure~\ref{fig: Triangle}) in~\cite{Chaves2014}.\footnote{Note that a tighter  entropic characterization was found in~\cite{Weilenmann2018} based on non-Shannon inequalities, and that the techniques introduced here could also be applied to these.} Including all permutations of $X$, $Y$ and $Z$, these yield 7 inequalities.

\begin{subequations}
\begin{equation}
\label{eq: Shineq1}
    -H(X)-H(Y)-H(Z)+H(XY)+H(XZ)\geq 0,
\end{equation}
\begin{equation}
\label{eq: Shineq2}
    -5H(X)-5H(Y)-5H(Z)+4H(XY)+4H(XZ)+4H(YZ)-2H(XYZ)\geq 0,
\end{equation}
\begin{equation}
\label{eq: Shineq3}
    -3H(X)-3H(Y)-3H(Z)+2H(XY)+2H(XZ)+3H(YZ)-H(XYZ)\geq 0.
\end{equation}
\end{subequations}

By replacing the Shannon entropy $H()$ with the Tsallis entropy $S_q()$
on the left hand side of these inequalities and minimizing the
resultant expression over our outer approximation to the classical
Tsallis entropy cone for the Triangle causal structure, one can obtain
valid Tsallis entropic inequalities for this causal structure. More
precisely, the outer approximation to the classical Tsallis entropy
cone for the Triangle is characterised by the $6+6(6-1)2^{6-3}=246$
independent Shannon constraints (monotonicity and strong subadditivity
constraints) and the $126$ causal constraints (one for each
conditional independence implied by the causal structure). To perform
this minimization we used {\sc
  LPAssumptions}~\cite{LPAssumptions} (Appendix~\ref{appendix: LPA}), a Mathematica package that we developed for solving linear programs involving unspecified variables (other than those being optimised over), by implementing the simplex method (Section~\ref{ssec:simplexalgo}).  In our case, we assumed that the dimensions of all the unobserved nodes
($A$,$B$ and $C$) are equal to $d_u$ and those of all the observed
nodes ($X$, $Y$ and $Z$) is $d_o$, and so the unspecified variables are $q\geq 1$,
$d_u\geq 2$ and $d_o\geq 2$. We obtained the following 
Tsallis entropic inequalities for the Triangle.
\begin{subequations}
\label{eq: Tsineq}
\begin{equation}
\label{eq: Tsineq1}
    -S_q(X)-S_q(Y)-S_q(Z)+S_q(XY)+S_q(XZ)\geq B_1(q,d_o,d_u), 
\end{equation}

\begin{align}
\begin{split}
\label{eq: Tsineq2}
   -5S_q(X)-5S_q(Y)-5S_q(Z)+4S_q(XY)+4S_q(XZ)+4S_q(YZ)-2S_q(XYZ)\\
    \geq B_2(q,d_o,d_u):=\max\big(B_{21}(q,d_o,d_u),B_{22}(q,d_o,d_u)\big), 
\end{split}
\end{align}

\begin{equation}
\label{eq: Tsineq3}
    -3S_q(X)-3S_q(Y)-3S_q(Z)+2S_q(XY)+2S_q(XZ)+3S_q(YZ)-S_q(XYZ)\geq B_3(q,d_o,d_u), 
\end{equation}
\end{subequations}
where,
\begin{subequations}
\begin{equation}
\label{eq: Bndineq1}
    B_1(q,d_o,d_u)=-\frac{1}{q-1}\Bigg(1-d_o^{1-q}\Bigg)\Bigg(2-d_o^{1-q}-d_u^{1-q}\Bigg),
\end{equation}

\begin{align}
\begin{split}
\label{eq: Bndineq2}
    B_{21}(q,d_o,d_u)&=-\frac{1}{q-1}\Bigg(11+d_u^{3-3q}+6d_o^{2-2q}+3d_o^{1-q}d_u^{1-q}-6d_u^{1-q}-15d_o^{1-q}\Bigg),\\
    B_{22}(q,d_o,d_u)&=-\frac{1}{q-1}\Bigg(10+d_o^{1-q}d_u^{3-3q}+5d_o^{2-2q}+2d_o^{1-q}d_u^{1-q}-5d_u^{1-q}-13d_o^{1-q}\Bigg),
\end{split}
\end{align}

\begin{equation}
\label{eq: Bndineq3}
    B_3(q,d_o,d_u)=-\frac{1}{q-1}\Bigg(6+d_o^{1-q}d_u^{2-2q}+3d_o^{2-2q}+d_o^{1-q}d_u^{1-q}-3d_u^{1-q}-8d_o^{1-q}\Bigg).
\end{equation}
\end{subequations}
Note that
$\lim_{q\rightarrow 1}B_1=\lim_{q\rightarrow 1}B_2=\lim_{q\rightarrow
  1}B_3=0$ $\forall d_u,d_o \geq 2$, recovering the original
inequalities for Shannon entropies (Equations~(\ref{eq:
  Shineq1})--(\ref{eq: Shineq3})) as a special case.

In~\cite{Rosset2017}, an upper bound on the dimensions of classical unobserved systems needed to reproduce a set of observed correlations is derived in terms of the dimensions of the observed systems. In the case of the Triangle causal structure with
$d_X=d_Y=d_Z=d_o$ and $d_A=d_B=d_C=d_u$ as considered here, the result
of~\cite{Rosset2017} implies that all classical correlations $P_{XYZ}$
can be reproduced by using hidden systems of dimension at most $d_o^3-d_o$. Since the dimension
of the unobserved systems is unknown, it makes sense to take the
minimum of the derived bounds over all $d_u$ between $2$ and $d_o^3-d_o$. By
taking their derivative, one can verify that for $q>1$ each of the
functions $B_1$, $B_{21}$, $B_{22}$ and $B_3$ is monotonically
decreasing in $d_o$ and $d_u$, and hence that the minimum is obtained
for $d_u=d_o^3-d_o$ for any given $d_o\geq
2$. It follows that for all $q>1$ and $d_o\geq2$ relations of the same
form as Equations~\eqref{eq: Tsineq1}--\eqref{eq: Tsineq3} hold, with
the quantities on the right hand sides replaced by
\begin{subequations}
\begin{align}
\begin{split}
\label{eq: B2ndineq1}
   &B_1^{*}(q,d_o)=B_1(q,d_o,d_o^3-d_o)\\
  & =\frac{-1}{q-1}\Bigg(2+d_o^{2-2q}-3d_o^{1-q}+d_o(-d_o + d_o^3)^{-q}-d_o^3 (-d_o + d_o^3)^{-q}-d_o^{2 - q} (-d_o + d_o^3)^{-q}+d_o^{4 - q} (-d_o + d_o^3)^{-q}\Bigg)
  \end{split}
\end{align}

\begin{align}
\begin{split}
\label{eq: B2ndineq2}
    &B_{21}^{*}(q,d_o)=B_{21}(q,d_o,d_o^3-d_o)\\
    &=\frac{-1}{q-1}\Bigg(11 + 6 d_o^{2 - 2 q} - 
 15 d_o^{1 - q} + (-d_o + d_o^3)^{3 - 3 q} - 
 6 (-d_o + d_o^3)^{1 - q} + 
 3 d_o^{1 - q} (-d_o + d_o^3)^{1 - q}\Bigg),\\
   & B_{22}^{*}(q,d_o)=B_{22}(q,d_o,d_o^3-d_o)\\
    &=\frac{-1}{q-1}\Bigg(10 + 5 d_o^{2 - 2 q} - 13 d_o^{1 - q} + 
 d_o^{1 - q} (-d_o + d_o^3)^{3 - 3 q} - 
 5 (-d_o + d_o^3)^{1 - q} + 
 2 d_o^{1 - q} (-d_o + d_o^3)^{1 - q}\Bigg),
\end{split}
\end{align}

\begin{align}
\begin{split}
\label{eq: B2ndineq3}
    &B_3^{*}(q,d_o)=B_3(q,d_o,d_o^3-d_o)\\
    &=\frac{-1}{q-1}\Bigg(6 + 3 d_o^{2 - 2 q} - 8 d_o^{1 - q} + 
 d_o^{1 - q} (-d_o + d_o^3)^{2 - 2 q}- 
 3 (-d_o + d_o^3)^{1 - q} + 
 d_o^{1 - q} (-d_o + d_o^3)^{1 - q}\Bigg)\,.
 \end{split}
\end{align}
\end{subequations}
A quantum violation of any of these bounds would imply that no
unobserved classical systems of arbitrary dimension could
reproduce those quantum correlations.

\begin{remark}
\label{remark: boundlimit}
Because they are monotonically decreasing, the bounds for $d_u=d_o^3-d_o$ are not as tight as the $d_u$-dependent bounds for general $q>1$. Nevertheless, as $q\to 1$, all the bounds $B^*(q,d_o)$ tend to 0, reproducing the known result of~\cite{Fritz13} for the Shannon case.
\end{remark}

\begin{remark}
In some cases it may be interesting to show quantum violations of these inequalities for low values of $d_u$, hence ruling out classical explanations with hidden systems of low dimensions, while possibly leaving open the case of arbitrary classical explanations. This would be interesting if it could be established that using hidden quantum systems allows for much lower dimensions than for hidden classical systems, for example.
\end{remark}

\subsection{Looking for quantum violations}\label{ssec: qviol}
It is known that the Triangle causal structure (Figure~\ref{fig:
  Triangle}) admits non-classical correlations such as Fritz's
distribution~\cite{Fritz2012}. The idea behind this distribution is to
embed the CHSH game in the Triangle causal structure such that
non-locality for the Triangle follows from the non-locality of the
CHSH game. To do so, $C$ is replaced by the sharing of a maximally
entangled pair of qubits, and $A$ and $B$ are taken to be uniformly
random classical bits. The observed variables $X$, $Y$ and $Z$ in
Figure~\ref{fig: Triangle} are taken to be pairs of the form
$X:=(\tilde{X},B)$, $Y:=(\tilde{Y},A)$ and $Z:=(A,B)$, where $\tilde{X}$
and $\tilde{Y}$ are generated by measurements on the halves of the
entangled pair with $B$ and $A$ used to choose the settings such that
the joint distribution $P_{\tilde{X}\tilde{Y}|BA}$ maximally violates
a CHSH inequality.  By a similar post-processing of other non-local
distributions in the bipartite Bell causal structure (Figure~\ref{fig:
  Bell}) such as the Mermin-Peres magic square game~\cite{Mermin1990,
  Peres1990} and chained Bell inequalities~\cite{BraunsteinCaves88},
one can obtain other non-local distributions in the Triangle that
cannot be reproduced using classical systems.  We explore whether any
of these violate any of our new inequalities.

Since the values of $B_i(q,d_o,d_u)$ are monotonically decreasing in
$d_o$ and $d_u$, if a distribution realisable in a quantum causal
structure does not violate the bounds~\eqref{eq: Tsineq1}--\eqref{eq:
  Tsineq3} for all $q\geq1$ and some fixed values of $d_o$ and $d_u$,
then no violations are possible for $d_o'>d_o$, $d_u'>d_u$. We
therefore take the smallest possible values of $d_o$ and $d_u$ when
showing that a particular distribution cannot violate any of the bounds.
  
For Fritz's distribution~\cite{Fritz2012}, $C$ is a two-qubit maximally
entangled state, $A$ and $B$ are binary random variables while $X$, $Y$ and $Z$ are random variables of dimension 4, i.e., the actual observed dimensions are
$(d_X,d_Y,d_Z)=(4,4,4)$ in this case. Here we see that taking
$d_o=4$ and the smallest possible $d_u$ which is $d_u=2$, the left
hand sides of Equations~\eqref{eq: Tsineq1}--\eqref{eq: Tsineq3}
evaluated for Fritz's distribution do not violate the corresponding bounds $B_i(q,d_o=4,d_u=2)$ for any
$q\geq 1$. This means that it is not possible to detect any quantum
advantage of this distribution (even over the case where the
unobserved systems are classical bits) using this method, and
automatically implies that it cannot violate the bounds
$B_i(q,d_o=4,d_u)$ for $d_u\geq 2$.

We also considered the chained Bell and magic square correlations
embedded in the Triangle causal structure analogously to the case
discussed above.  For each of these, we define $d^i$ to be the
smallest value of $d_o$ for which the bound $B_i(q,d_o=d^i,d_u=2)$
cannot be violated for any $q>1$. The values of $d^i$ are given in
Table~\ref{table: violations} for the different cases of the chained
Bell correlations and the magic square. Since the values of $d^i$ are
always lower than the smallest of the observed dimensions in the
problem, and due to the monotonicity of the bounds it follows that
none of these quantum distributions violate any of our inequalities
when the observed dimension is set to $d_o^{\min}$.

\begin{table}[t]
\begin{center}
\resizebox{\textwidth}{!}{%
\begin{tabular}{ ccccc } 
\hline
\hline
\quad & \multicolumn{3}{c}{$d^i$} & smallest\\
\cline{2-4}
Scenario & Ineq.~\eqref{eq: Tsineq1} ($i=1$)\ \ \ \  & Ineq.~\eqref{eq: Tsineq2} ($i=2$)\ \ \ \  & Ineq.~\eqref{eq: Tsineq3} ($i=3$)\ & \ \ observed dim. ($d_o^{\min}$)\\
 \hline
$N=2$ & 2 & 2 & 2 & 4 \\ 
$N=3$ & 3 & 2 & 3 & 6 \\ 
$N=4$ & 4 & 2 & 4 & 8 \\ 
$N=5$ & 5 & 2 & 5 & 10 \\ 
$N=6$ & 6 & 2 & 6 & 12 \\ 
$N=7$ & 7 & 2 & 7 & 14 \\ 
$N=8$ & 8 & 2 & 8 & 16 \\ 
$N=9$ & 9 & 3 & 9 & 18 \\ 
$N=10$ & 10 & 3 & 10 & 20 \\ 
Magic Sq. & 4 & 2 & 4 & 9 \\ 
 \hline
 \hline
\end{tabular}}
\end{center}
\caption[Looking for violations of our Tsallis entropic inequalities]{\textbf{Values of $d^i$ for the chained Bell and magic square
    correlations embedded in the Triangle causal structure.} The
  values of $N$ correspond to the number of inputs per party in the
  chained Bell inequality, which always has two outputs per party (the
  $N=2$ case corresponds to Fritz's
  distribution~\cite{Fritz2012}). When embedded in the Triangle, the
  number of outcomes of the observed nodes are
  $(d_X,d_Y,d_Z)=(2N,2N,N^2)$.  The last column of the table gives the
  minimum of the observed node dimensions $(d_X,d_Y,d_Z)$ for each
  $N$, which is simply $2N$. For the magic square, the dimensions
  $(d_X,d_Y,d_Z)$ are $(12,12,9)$. In all cases, the minimum value of
  $d^i$ such that the Inequalities~(\ref{eq: Tsineq1})--(\ref{eq:
    Tsineq3}) with bounds $B_i(q,d_o=d^i,d_u=2)$ are not violated for
  any $q\geq 1$ is less than the minimum observed dimension
  $d_o^{\min}$, and hence no violations of~\eqref{eq:
    Tsineq1}--\eqref{eq: Tsineq3} could be found for the relevant case
  with $d_o=d^{\min}_o$.}
\label{table: violations}
\end{table}

We further checked for violations of Inequalities~(\ref{eq:
  Tsineq1})--(\ref{eq: Tsineq3}) by sampling random quantum states for
the systems $A$, $B$ and $C$ and random quantum measurements whose
outcomes would correspond to the classical variables $X$, $Y$ and
$Z$. The value of $q$ was also sampled randomly between $1$ and
$100$. We considered the cases where the shared systems were pairs of
qubits with 4 outcome measurements ($d_X=d_Y=d_Z=4$) and qutrits with
9 outcome measurements ($d_X=d_Y=d_Z=9$) but were unable to find
violations of any of the inequalities even for the bounds with the
$d_o=4, d_u=2$ (two qubit case) and $d_o=9, d_u=2$ (two qutrit case),
i.e., the bounds obtained when the unobserved systems are classical
bits.

\begin{remark}
In the derivation of Inequalities~(\ref{eq: Tsineq1})--(\ref{eq:
  Tsineq3}), we set the dimensions of the observed nodes
$X$, $Y$ and $Z$ to all be equal and those of the unobserved
nodes $A$, $B$ and $C$ to also all be equal. One could in principle
repeat the same procedure taking different dimensions for all 6
variables but we found the computational procedure too
demanding. However, Table~\ref{table: violations} shows that even when we consider the bounds $B_i(q,d_o,d_u)$ with $d_o$ and $d_u$ much smaller than the actual dimensions, known non-local distributions in the Triangle considered in Table~\ref{table: violations} do not violate the corresponding Inequalities~(\ref{eq: Tsineq1})--(\ref{eq:
  Tsineq3}) for any $q\geq 1$. Since the bounds are monotonically decreasing in $d_u$ and $d_o$, even if we obtained the general bounds for arbitrary dimensions of $X$, $Y$, $Z$, $A$, $B$ and $C$, they would be strictly weaker than $B_i(q,d^i,d_u=2)$ $\forall i\in \{1,2,3\}, q\geq 1$ and can certainly not be violated by these distributions.
\end{remark}

\section{Discussion}\label{sec: causaldiscussion}
We have investigated the use of Tsallis entropies within the entropy
vector method to causal structures, showing how causal constraints
imply bounds on the Tsallis entropies of the variables
involved. Although Tsallis entropies for $q\geq1$ possess many
properties that aid their use in the entropy vector method, the nature
of the causal constraints makes the problem significantly more
computationally challenging than in the case of Shannon entropy.  This
meant that we were unable to complete the desired computations in the
former case, even for some of the simplest causal structures.
Nevertheless, we were able to derive new classical causal constraints
expressed in terms of Tsallis entropy by analogy with known Shannon
constraints, but were unable to find cases where these
were violated, even using quantum distributions that are known not to
be classically realisable.  This mirrors an analogous result for
Shannon entropies~\cite{Weilenmann2018}.

%Tsallis entropies are known to give improvements~\cite{Wajs15} in cases that involve post-selection.  While post-selection cannot be used for general causal structures (including the Triangle), it would be interesting to understand whether using Tsallis entropy helps in other cases for which post-selection is applicable.

One could also investigate whether other entropic quantities could be
used in a similar way, and in Appendix~\ref{appendix: Renyiresults} we have considered the use of R\'enyi entropies. The R\'enyi entropies of order $\alpha$ do not
satisfy strong subadditivity for $\alpha \neq 0,1$, while the R\'enyi
as well as the min and max entropies fail to obey the chain rules for
conditional entropies.  Thus, use of these in the entropy vector method, would
require an entropy vector with components for all possible conditional
entropies as well as unconditional ones, considerably increasing the
dimensionality of the problem, which we would expect to make the
computations harder. In some cases, not having a chain rule
  has not been prohibitive~\cite{WeilenmannGPT}, but our results of Appendix~\ref{appendix: Renyiresults} reveal significant limitations of R\'enyi entropies for analysing causal structure.
  
Further, one could consider using algorithms other than Fourier-Motzkin
elimination to obtain non-trivial Tsallis entropic constraints
over observed nodes starting from the Tsallis cone over all the nodes
(see e.g.,~\cite{Gl_le_2018}). These could in principle yield
solutions even in cases where FM elimination becomes
intractable. However, we found that the FM elimination procedure
became intractable even when starting out with only a small subset of
the Tsallis entropic causal constraints for a simple causal structure
such as the Bell one. %\footnote{As mentioned previously, with very few (less than 10) initial causal constraints, the FM elimination completed but yielded no non-trivial inequalities.} 
This suggests that the difficulty is not only with the number of constraints, but also with their nature (in particular, that they are not equalities and depend non-trivially on the dimensions). Consequently, we bypassed FM elimination and used an alternative technique to obtain new Tsallis entropic inequalities for the Triangle causal structure (Section~\ref{sec: newineq}). %This method turned out to be feasible for the case where $d_X=d_Y=d_Z:=d_o$ and  $d_A=d_B=d_C:=d_u$ but proved to be computationally intractable for arbitrary dimensions/cardinalities.

It is also worth noting that the following alternative definition of the Tsallis
conditional entropy was proposed in~\cite{ABE2001157}.
%\begin{equation}
%    \tilde{S_q}(X|Y=y):=\frac{1}{1-q}\left(\sum_{x}p_{x|y}^q-1\right)=\frac{1}{1-q}\left(\frac{\sum_x p_{xy}^q}{p_y^q}-1\right),
%\end{equation}
\begin{equation}
    \tilde{S_q}(X|Y)=\frac{1}{1-q}\frac{\sum_yP(y)^qS_q(X|Y=y)}{\sum_y P(y)^q}=\frac{1}{1-q}\left(\frac{\sum_{x,y}P(xy)^q}{\sum_y P(y)^q}-1\right).
\end{equation}
Using this definition, Tsallis entropies would satisfy the same causal
constraints as the Shannon entropy (Equation~\eqref{eq:
  shancausmain}). However, the conditional entropies defined this way
do not satisfy the chain rules of Equation~\eqref{eq: chainrule} but
instead obey a non-linear chain rule,
$S_q(XY)=S_q(X)+S_q(Y|X)+(1-q)S_q(X)S_q(Y|X)$~\cite{ABE2001157}. This
would again mean that conditional entropies would need to be included
in the entropy vector.  Furthermore, since Fourier-Motzkin elimination
only works for linear constraints, an alternative algorithm would be
required to use this chain rule in conjunction with the entropy vector
method.

That the inequalities for Tsallis entropy derived in this work depend
on the dimensions of the systems involved could be used to certify
that particular observed correlations in a classical causal structure
require a certain minimal dimension of unobserved systems to be
realisable. To show this would require showing that
classically-realisable correlations violate one of the
inequalities for some $d_u$.  Such bounds would then complement
the upper bounds of~\cite{Rosset2017}.  However, in some cases we know our bounds are not tight enough to do this.  As a simple example, within the Triangle causal structure we tried taking $X=(X_B,X_C)$, $Y=(Y_A,Y_C)$ and $Z=(Z_A,Z_B)$ with $X_B=Z_B$, $X_C=Y_C$ and $Y_A=Z_A$ where each are uniformly distributed with cardinality $D$, for $D\in\{3,\ldots,10\}$.  In this case it is clear that the correlations cannot be achieved with classical unobserved systems with $d_u=2$. Taking the bound with $d_u=2$ and $d_o=D^2$ no violations of~\eqref{eq: Tsineq1}--\eqref{eq: Tsineq3} were seen by plotting the graphs for $q\in[1,20]$, for the range of $D$ above. Hence, our bounds are too loose to certify lower bounds on $d_u$ in this case.

While our analysis highlights significant drawbacks of using Tsallis
entropies for analysing causal structures, it does not rule out the
possibility of Tsallis entropies being able to detect the
classical-quantum gap\footnote{Proving that Tsallis entropies are
  unable to do this would also be difficult.  For instance, the proof
  of~\cite{Weilenmann16} that Shannon entropies are unable to detect
  the gap in line-like causal structures involves first characterising
  the marginal polytope through Fourier-Motzkin elimination, which
  itself proved to be computationally infeasible with Tsallis
  entropies even for the simplest line-like causal structure, the
  bipartite Bell scenario.} in these causal structures, or others.  To overcome the difficulties we encountered we would either need increased computational power, and/or the development of new, alternative techniques for analysing causal structures (with or without entropies).

%............. Appendix ..............

\section{Appendix}

\subsection{Quantum generalisations of Theorems \ref{theorem: mibound} and \ref{theorem: causal}}\label{appendix: quantum}

The main results of this Chapter only required the Tsallis entropic causal constraints of Theorems \ref{theorem: mibound} and \ref{theorem: causal}, which were derived for classical causal structures.  Here, we present additional results that generalise these theorems to certain quantum causal structures. For these, we recap some of the properties of quantum Tsallis entropies encountered in Section~\ref{ssec: tsalprop}.
%In the following, for a (finite dimensional) Hilbert space $\mathscr{H}$, we use $\mathcal{L}(\mathscr{H})$ to represent the set of linear operators on $\mathscr{H}$, $\mathscr{P}(\mathscr{H})$ to represent the set of positive (semi-definite) operators on $\mathscr{H}$, and $\mathscr{S}(\mathscr{H})$ to denote the set of density operators on $\mathscr{H}$ (positive and trace 1).

Tsallis entropies as defined for classical random variables in
Section~\ref{ssec: tsalprop} are easily generalised to the quantum case
by replacing the probability distribution by a density matrix
\cite{Hu2006}. For a quantum system described by the density matrix
$\rho \in \mathscr{S}(\mathscr{H})$ on the Hilbert space $\mathscr{H}$
and $q>0$, the quantum Tsallis entropy is defined by
\begin{equation}
 S_q(\rho)=
 \begin{cases}
   -\Tr \rho^q\ln_q \rho, & q\neq 1.\\
    H(\rho), & q=1.
  \end{cases}
\end{equation}
where $H(\rho)= -\Tr \rho \ln \rho$ is the von Neumann entropy of
$\rho$ and $\ln_q (x)=\frac{x^{1-q}-1}{1-q}$ as in Section~\ref{ssec:
  tsalprop}.\footnote{Analogously to the classical case we keep it
  implicit that if $\rho$ has any 0 eigenvalues these do not
  contribute to the trace.}

Given a density operator $\rho_{AB}\in\mathscr{S}(\cH_A\ot\cH_B)$, the
conditional quantum Tsallis entropy of $A$ given $B$ can then be
defined by $S_q(A|B)_\rho=S_q(AB)-S_q(B)$, the mutual information
between $A$ and $B$ by $I_q(A:B)_\rho=S_q(A)+S_q(B)-S_q(AB)$, and for
$\rho_{ABC}\in\mathscr{S}(\cH_A\ot\cH_B\ot\cH_C)$ the conditional Tsallis
information between $A$ and $B$ given $C$ is defined by
$I_q(A:B|C)_\rho=S_q(A|C)+S_q(B|C)-S_q(AB|C)$.  In this section we use
$d_S$ to represent the dimensions of the Hilbert space
$\mathscr{H}_S$.

The following properties of quantum Tsallis entropies will be useful for what follows.
\begin{enumerate}
    \item \textbf{Pseudo-additivity \cite{Tsallis1988}:} \label{prop:Qpseudo}If $\rho_{AB}=\rho_A\ot\rho_B$,
      then 
    \begin{equation}
        \label{eq:Qpseudoadd}
        S_q(AB)=S_q(A)+S_q(B)+(1-q)S_q(A)S_q(B)\,.
    \end{equation}
  \item \label{prop:Qupper} \textbf{Upper bound \cite{Audenaert2007}:} For all $q>0$, we have
    $S_q(A)\leq\ln_qd_A$ and equality is achieved if and only
    if $\rho_A=\id_A/d_A$.
    \item \textbf{Subadditivity \cite{Audenaert2007}:} For any density matrix $\rho_{AB}$ with marginals $\rho_A$ and $\rho_B$, the following holds for all $q\geq 1$,
    \begin{equation}
        \label{eq:Qsubadd}
        S_q(AB)\leq S_q(A)+S_q(B)\,.
    \end{equation}
    
\end{enumerate}

Using these we can generalize Theorem~\ref{theorem: mibound} to the
quantum case. This corresponds to the causal structure with two
independent quantum nodes and no edges in between them.

\begin{theorem}
\label{theorem: qmibound}
For all bipartite density operators in product form, i.e., $\rho_{AB}=\rho_A\otimes \rho_B$ with
$\rho_A\in\mathscr{S}(\mathscr{H}_A)$ and $\rho_B\in\mathscr{S}(\mathscr{H}_B)$, the quantum Tsallis mutual information $I_q(A:B)_{\rho}$ is upper bounded as follows for all $q>0$
\begin{equation*}
    I_q(A:B)_{\rho}\leq f(q,d_A,d_B)\,,
\end{equation*}
where the function $f(q,d_A,d_B)$ is given by
\begin{equation*}
    f(q,d_A,d_B)=\frac{1}{(q-1)}\left(1-\frac{1}{d_A^{q-1}}\right)\left(1-\frac{1}{d_B^{q-1}}\right)=(q-1)\ln_qd_A\ln_qd_B\,.
\end{equation*}
The bound is saturated if and only if
$\rho_{AB}=\frac{\mathds{1}_A}{d_A}\otimes\frac{\mathds{1}_B}{d_B}$.
\end{theorem}
\begin{proof}
  The proof goes through in the same way as the proof of
  Theorem~\ref{theorem: mibound} for the classical case
  (Properties~\ref{prop:Qpseudo} and~\ref{prop:Qupper} are analogous
  to those needed in the classical proof).
\end{proof}

Next, we generalise Theorem~\ref{theorem: causal} and
Corollaries~\ref{corollary: causal} and~\ref{corollary:
  mainconstraint}. This would correspond to the causal constraints on
quantum Tsallis entropies implied by the common cause causal structure
with $C$ being a complete common cause of $A$ and $B$ (which share no
causal relations among themselves). Here, one must be careful in
precisely defining the conditional mutual information and interpreting
it physically. For example, if the common case $C$ were quantum and
the nodes $A$ and $B$ were classical outcomes of measurements on $C$,
then $A$, $B$ and $C$ do not coexist and there is no joint state
$\rho_{ABC}$ in such a case. This is a significant difference in
quantum causal modelling compared to the classical case, and there
have been several proposals for how do deal with it \cite{Leifer2013,
  Costa2016, Allen2017, Pienaar2019}. In the following we
consider two cases:
\begin{enumerate}
\item When $C$ is classical, all 3 systems coexist and $\rho_{ABC}$
  can be described by a classical-quantum state (See
  Theorem~\ref{theorem: qcausal1}).
\item When $C$ is quantum, one approach is to view $\rho_{ABC}$ not as
  the joint state of the 3 systems but as being related to the
  Choi-Jamiolkowski representations of the quantum channels from $C$
  to $A$ and $B$ (See Section~\ref{ssec: noncoexisting}) as done in
  \cite{Allen2017}.
\end{enumerate}

The following Lemma proven in~\cite{Kim2016} is required for our
generalization of Theorem~\ref{theorem: causal} in the first case.
\begin{lemma}[\cite{Kim2016}, Lemma~1]
\label{lemma: condtsal}
Let $\mathscr{H}_A$ and $\mathscr{H}_Z$ be two Hilbert spaces and $\{\ket{z}\}_z$ be an orthonormal basis of $\mathscr{H}_Z$. Let $\rho_{AZ}$ be classical on $\mathscr{H}_Z$ with respect to this basis i.e., 
$$\rho_{AZ}=\sum_z P(z) \rho_A^{(z)} \otimes \ket{z}\bra{z},$$
where $\sum_zP(z)=1$ and $\rho_A^{(z)} \in S(\mathscr{H}_A)\ \forall z$. Then for all $q>0$, $$S_q(AZ)_{\rho}=\sum_zP(z)^qS_q(\rho_A^{(z)})+S_q(Z),$$
where $S_q(Z)$ is the classical Tsallis entropy of the variable $Z$ distributed according to $P_Z$.
\end{lemma}
Note that the above Lemma immediately implies that
\begin{equation}\label{eq:cond}
  S_q(A|Z)_{\rho}=\sum_zP(z)^qS_q(\rho_A^{(z)})\,.
\end{equation}

\begin{theorem}
\label{theorem: qcausal1}
Let $\rho_{ABC}=\sum_c P(c)\rho_{AB}^{(c)}\otimes \proj{c}$, where $\rho_{AB}^{(c)}=\rho_A^{(c)}\otimes \rho_B^{(c)}$ $\forall c$, then, for all $q\geq1$,
$$I_q(A:B|C)_{\rho_{ABC}}\leq f(q,d_A,d_B)\,.$$

For $q>1$ the bound is saturated if and only if $\rho_{ABC}=\frac{\mathds{1}_A}{d_A}\otimes \frac{\mathds{1}_B}{d_B}\otimes \ket{c}\bra{c}_C$.
\end{theorem}
\begin{proof}
Using~\eqref{eq:cond} we have,
\begin{align*}
      I_q(A:B|C)_{\rho_{ABC}}
    &=S_q(A|C)_{\rho}+S_q(B|C)_{\rho}-S_q(AB|C)_{\rho}\\
    &=\sum_cP(c)^q[S_q(\rho_A^{(c)})+S_q(\rho_B^{(c)})-S_q(\rho_{AB}^{(c)})]\\
    &=\sum_cP(c)^qI_q(A:B)_{\rho_{AB}^{(c)}}\,.
\end{align*}
The rest of the proof is analogous to Theorem~\ref{theorem: causal},
where using the above, Theorem~\ref{theorem: qmibound} and defining
the set
$\mathcal{R}= \{\rho_{ABC} \in
\mathscr{H}_A\otimes\mathscr{H}_B\otimes\mathscr{H}_C:
\rho_{ABC}=\sum_cP(c)\rho_A^{(c)}\otimes\rho_B^{(c)}\otimes\ket{c}\bra{c}\}$
we have,

\begin{align*}
\max\limits_{\mathcal{R}} I_q(A:B|C)_{\rho} &=\max\limits_{\mathcal{R}}\sum_c P(c)^qI_q(A:B)_{\rho_{AB}^{(c)}}\\
&\leq \max\limits_{\{P(c)\}_c}\sum_c P(c)^q(c) \max\limits_{\{\rho_A^{(c)}\}_c,\{\rho_B^{(c)}\}_c}I_q(A:B)_{\rho_{AB}^{(c)}}\\
&=f(q,d_A,d_B)\,,
\end{align*}
where the last step follows because for all $q\geq1$, $\sum_cP(c)^q$ is
maximized by deterministic distributions over $C$ with a maximum value
of $1$\footnote{For $q>1$ such deterministic distributions are the
  only way to obtain the bound.} and $I_q(A:B)_{\rho_{AB}^{(c)}}$ for
product states is maximised by the maximally mixed state over $A$ and
$B$ for all $c$ (Theorem~\ref{theorem: qmibound}). Thus, for $q>1$,
the bound is saturated if and only if
$\rho_{ABC}=\frac{\mathds{1}_A}{d_A}\otimes
\frac{\mathds{1}_B}{d_B}\otimes \ket{c}\bra{c}_C$ for some value $c$
of $C$.
\end{proof}

\subsubsection{A generalisation: when systems do not coexist}
\label{ssec: noncoexisting}
There is a fundamental problem with naively generalising classical conditional independences such as $P_{XY|Z}=P_{X|Z}P_{Y|Z}$ to the quantum case by replacing joint distributions by density matrices: it is not clear what is meant by a conditional quantum state e.g., $\rho_{A|C}$ since it is not clear what it means to condition on a quantum system, specially when the (joint state of the) system under consideration and the one being conditioned upon do not coexist. There are a number of approaches for tackling this problem, from describing quantum states in space and time on an equal footing \cite{Horsman2016} to quantum analogues of Bayesian inference \cite{Leifer2013} and causal modelling \cite{Costa2016, Allen2017, Pienaar2019}. In the following, we will focus on one such approach that is motivated by the framework of \cite{Allen2017}. Central to this approach is the Choi-Jamio\l{}kowski isomorphism \cite{Jamiolkowski1972, Choi1975} from which one can define conditional quantum states.

\begin{definition}[Choi state]
Let $\ket{\gamma}=\sum_i\ket{i}_R\ket{i}_{R^*}\in \mathscr{H}_R\otimes\mathscr{H}_{R^*}$, where $\mathscr{H}_{R^*}$ is the dual space to $\mathscr{H}_R$ and $\{\ket{i}_R\}_i$, $\{\ket{i}_{R^*}\}_i$ are orthonormal bases of $\mathscr{H}_R$ and $\mathscr{H}_{R^*}$ respectively. Given a channel $\mathscr{E}_{R|S}:\mathscr{S}(\mathscr{H}_R)\to\mathscr{S}(\mathscr{H}_S)$, the \emph{Choi state of
  the channel} is defined by
$$\rho_{S|R}=(\mathscr{E}_{R|S}\otimes\mathcal{I})(\ket{\gamma}\bra{\gamma})=\sum_{ij}\mathscr{E}(\ket{i}\bra{j}_R)\otimes\ket{i}\bra{j}_{R^*}\,.$$
Thus, $\rho_{S|R}\in\mathscr{P}(\mathscr{H}_S\otimes\mathscr{H}_{R^*})$.
\end{definition}

\begin{figure}[t]
    \centering
\includegraphics[scale=1.0]{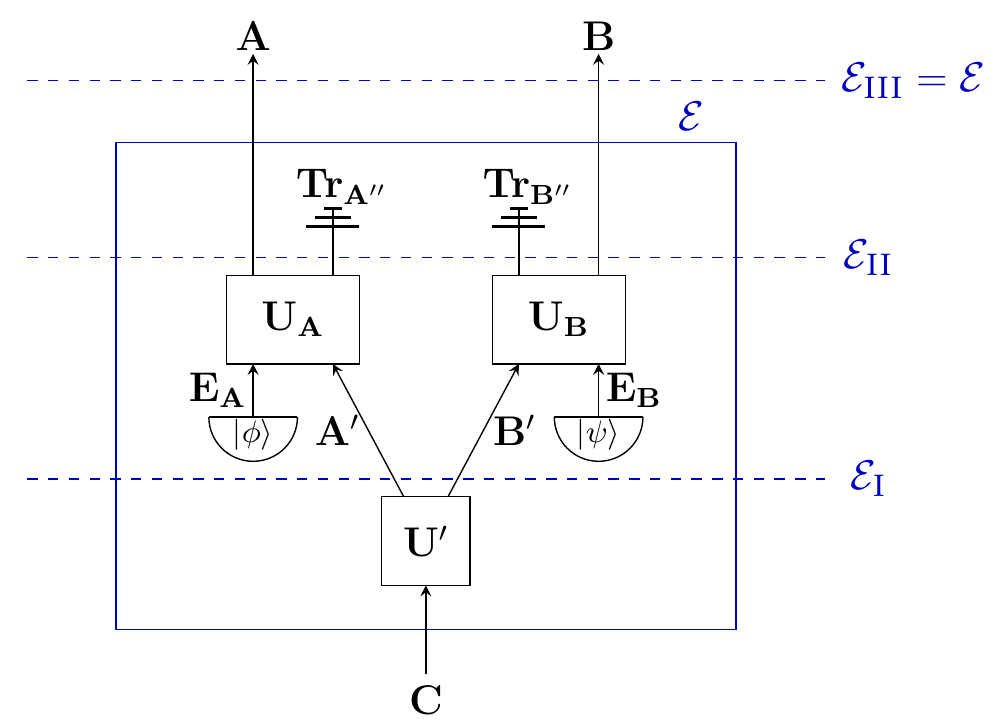}
    \caption[Circuit decomposition of quantum common cause channel]{\textbf{A circuit decomposition of the channel $\mathbf{\mathscr{E}:\mathscr{S}(\mathscr{H}_C)\rightarrow \mathscr{S}(\mathscr{H}_A\otimes\mathscr{H}_B)}$ when $\mathbf{C}$ is a \emph{complete common cause} of $\mathbf{A}$ and $\mathbf{B}$:} If the map $\mathscr{E}$ from the system $C$ to the systems $A$ and $B$ can be decomposed as shown here, then $C$ is a complete common cause of $A$ and $B$ (\cite{Allen2017}). We build up our result step by step considering the channels given by $\mathscr{E}_{\textsc{i}}$ (unitary), $\mathscr{E}_{\textsc{ii}}$ (unitary followed by local isometries) and $\mathscr{E}_{\textsc{iii}}=\mathscr{E}$.}
    \label{fig: qchannel}
\end{figure}

Now, if a quantum system $C$ evolves through a unitary channel
$\mathscr{E}_I(\cdot)=U'(\cdot)U'^{\dagger}$ to two systems $A'$ and
$B'$ where
$U':\mathscr{H}_C\rightarrow \mathscr{H}_{A'}\otimes\mathscr{H}_{B'}$,
it is reasonable to call the system $C$ a quantum common cause of the
systems $A'$ and $B'$. Further, this would still be reasonable if one
were to then perform local completely positive trace preserving (CPTP)
maps on the $A'$ and $B'$ systems. By the Stinespring dilation
theorem, these local CPTP maps can be seen as local isometries
followed by partial traces, and the local isometries can be seen as
the introduction of an ancilla in a pure state followed by a joint
unitary on the system and ancilla. This is illustrated in
Figure~\ref{fig: qchannel} and is compatible with the definition of
quantum common causes presented in \cite{Allen2017}.  In other words,
a system $C$ can be said to be a complete (quantum) common cause of
systems $A$ and $B$ if the corresponding
channel
$\mathscr{E}:\mathscr{S}(\mathscr{H}_C)\rightarrow
\mathscr{S}(\mathscr{H}_A\otimes\mathscr{H}_B)$ can be decomposed as
in Figure~\ref{fig: qchannel} for some choice of unitaries $U'$,
$U_A$, $U_B$ and pure states $\ket{\phi}_{E_A}$,
$\ket{\psi}_{E_B}$. Note that a more general set of channels fit the
definition of quantum common cause in Ref.~\cite{Allen2017} than we
use here; whether the theorems here extend to this case we leave as an open
question.

%Quantum channel figure

In \cite{Allen2017} it is shown that whenever a system $C$ is a
complete common cause of systems $A$ and $B$ then the Shannon conditional mutual information evaluated
on the state $\tau_{ABC^*}=\frac{1}{d_A}\rho_{AB|C}$ satisfies
$I(A:B|C^*)_\tau=0$ where $\rho_{AB|C}$ is the Choi state of the channel
from $C$ to $A$ and $B$. We generalise this result to Tsallis
entropies for $q\geq 1$ for certain types of channels. We present the
result in three cases, each with increasing levels of generality. These are explained in Figure~\ref{fig: qchannel} and
correspond to the cases where the map from the complete common cause
$C$ to its children $A$ and $B$ is ({\sc i}) unitary ($\mathscr{E}_{\textsc{i}}=U'$); ({\sc ii}) unitary followed by local isometries ($\mathscr{E}_{\textsc{ii}}$); ({\sc iii}) Unitary followed by local isometries followed by partial traces on local systems ($\mathscr{E}_{\textsc{iii}}=\mathscr{E}$). 

\begin{lemma}\label{lemma: qstageI}
Let $\mathscr{E}_{\textsc{i}}: \mathscr{S}(\mathscr{H}_C)\rightarrow \mathscr{S}(\mathscr{H}_{A'}\otimes\mathscr{H}_{B'})$ be a unitary quantum channel i.e., $$\mathscr{E}_{\textsc{i}}(\cdot)= U'(\cdot)U'^{\dagger},$$
where $U': \mathscr{H}_C\rightarrow \mathscr{H}_{A'}\otimes\mathscr{H}_{B'}$ is an arbitrary unitary operator. If $\rho_{A'B'|C}$ is the corresponding Choi state, then the Tsallis conditional mutual information evaluated on the state $\tau_{A'B'C^*}=\frac{1}{d_C}\rho_{A'B'|C}\in \mathscr{S}(\mathscr{H}_{A'}\otimes\mathscr{H}_{B'}\otimes\mathscr{H}_{C^*})$ satisfies
  $$I_q(A':B'|C^*)_\tau= f(q,d_{A'},d_{B'}) \qquad \forall q>0.$$
\end{lemma}
\begin{proof}
 The conditional mutual information $I_q(A':B'|C^*)_\tau$ can be written as
 \begin{equation}
 \label{eq: qCMI-I}
     I_q(A':B'|C^*)_\tau=\frac{1}{q-1}\big(\Tr_{A'B'C^*}\tau_{A'B'C^*}^q+\Tr_{C^*}\tau_{C^*}^q-\Tr_{A'C^*}\tau_{A'C^*}^q-\Tr_{B'C^*}\tau_{B'C^*}^q\big).
 \end{equation}
 We will now evaluate every term in the above expression for the case where the channel that maps the $C$ system to the $A'$ and $B'$ systems is unitary. In this case, $\tau_{A'B'C^*}$ is a pure state and can be written as $\tau_{A'B'C^*}=\ket{\tau}\bra{\tau}_{A'B'C^*}$ where
 \begin{equation}
 \label{eq: vectau-I}
     \ket{\tau}_{A'B'C^*}=\frac{1}{\sqrt{d_C}}\sum_i U'\ket{i}_C\otimes \ket{i}_{C^*}.
 \end{equation}
 This means that
 $\Tr_{A'B'C^*}\tau_{A'B'C^*}^q=\Tr_{A'B'C^*}\tau_{A'B'C^*}$ $\forall
 q>0$. Since $\tau_{A'B'C^*}$ is a valid quantum state, it must be a trace one operator and we have
 \begin{equation}
  \label{eq: qterm1-I}
     \Tr_{A'B'C^*}\tau_{A'B'C^*}^q=1 \quad \forall q>0
 \end{equation}
 Further, we have $\tau_{C^*}=\Tr_{A'B'}\tau_{A'B'C^*}=\frac{\mathds{1}_{C^*}}{d_C}$ and hence
 \begin{equation}
  \label{eq: qterm2-I}
     \Tr_{C^*}\tau_{C^*}^q=\frac{1}{d_C^{q-1}}=\frac{1}{d_{A'}^{q-1}d_{B'}^{q-1}}.
 \end{equation}
 The second step follows from the fact that $U': \mathscr{H}_C\rightarrow \mathscr{H}_{A'}\otimes\mathscr{H}_{B'}$ is unitary so $d_C=d_{A'}d_{B'}$. 
 
 Now, the marginals over $A'$ and $B'$ are
 $\tau_{A'}=\Tr_{B'C^*}\tau_{A'B'C^*}=\frac{\mathds{1}_{A'}}{d_{A'}}$
 and
 $\tau_{B'}=\Tr_{A'C^*}\tau_{A'B'C^*}=\frac{\mathds{1}_{B'}}{d_{B'}}$. By the Schmidt decomposition of $\tau_{A'B'C^*}$, the non-zero
 eigenvalues of $\tau_{A'}$ are the same as those of  $\tau_{B'C^*}$.
 Since the Tsallis entropy depends only on the non-zero eigenvalues,
 $S_q(A')=S_q(B'C^*)$ and hence
 \begin{equation}
  \label{eq: qterm3-I}
     \Tr_{B'C^*}\tau_{B'C^*}^q=d_{A'}\Bigg(\frac{1}{d_{A'}^q}\Bigg)=\frac{1}{d_{A'}^{q-1}}\,.
 \end{equation}
 By the same argument it follows that
  \begin{equation}
   \label{eq: qterm4-I}
     \Tr_{A'C^*}\tau_{A'C^*}^q=d_{B'}\Bigg(\frac{1}{d_{B'}^q}\Bigg)=\frac{1}{d_{B'}^{q-1}}\,.
 \end{equation}
 Combining Equations~\eqref{eq: qCMI-I}-\eqref{eq: qterm4-I}, we have
 \begin{align}
     I_q(A':B'|C^*)_\tau=\frac{1}{q-1}\Bigg(1+\frac{1}{d_{A'}^{q-1}d_{B'}^{q-1}}-\frac{1}{d_{A'}^{q-1}}-\frac{1}{d_{B'}^{q-1}}\Bigg)
     =f(q,d_{A'},d_{B'}) \quad \forall q>0\,.
 \end{align}
\end{proof}

\begin{lemma}\label{lemma: qstageII}
Let $\mathscr{E}_{\textsc{ii}}: \mathscr{S}(\mathscr{H}_C)\rightarrow \mathscr{S}(\mathscr{H}_{\tilde{A}}\otimes\mathscr{H}_{\tilde{B}})$ be a quantum channel of the form $$\mathscr{E}_{\textsc{ii}}(\cdot)=(U_A\otimes U_B)\big[\ket{\phi}\bra{\phi}_{E_A}\otimes U'(\cdot)U'^{\dagger}\otimes \ket{\psi}\bra{\psi}_{E_B}\big](U_A\otimes U_B)^{\dagger},$$
where $U': \mathscr{H}_C\rightarrow \mathscr{H}_{A'}\otimes\mathscr{H}_{B'}$, $U_A:\mathscr{H}_{E_A}\otimes\mathscr{H}_{A'}\rightarrow \mathscr{H}_{\tilde{A}}$ and $U_B:\mathscr{H}_{B'}\otimes\mathscr{H}_{E_B}\rightarrow \mathscr{H}_{\tilde{B}}$ are arbitrary unitaries and $\ket{\phi}_{E_A}$ and $\ket{\psi}_{E_B}$ are arbitrary pure states. If $\rho_{\tilde{A}\tilde{B}|C}$ is the corresponding Choi state, then the Tsallis conditional mutual information evaluated on the state $\tau_{\tilde{A}\tilde{B}C^*}=\frac{1}{d_C}\rho_{\tilde{A}\tilde{B}|C}\in \mathscr{S}(\mathscr{H}_{\tilde{A}}\otimes\mathscr{H}_{\tilde{B}}\otimes\mathscr{H}_{C^*})$ satisfies
  $$I_q(\tilde{A}:\tilde{B}|C^*)_\tau= f(q,d_{A'},d_{B'}) \qquad \forall q>0.$$
\end{lemma}
\begin{proof}
  Note that the map $\mathscr{E}_{\textsc{ii}}$ is the unitary map
  $\mathscr{E}_{\textsc{i}}(\cdot)=U'(\cdot)U'^{\dagger}$ followed by
  local isometries $V_A$ and $V_B$ on the $A'$ and $B'$ systems
  respectively. Since the expression for the conditional mutual
  information $I_q(\tilde{A}:\tilde{B}|C^*)_\tau$ can be written in
  terms of entropies, which are functions of the eigenvalues of the
  relevant reduced density operators, and since the eigenvalues are
  unchanged by local isometries, this conditional mutual information
  is invariant under local isometries.  The rest of the proof is
  identical to that of Lemma~\ref{lemma: qstageI} resulting in
  \begin{equation}
      I_q(\tilde{A}:\tilde{B}|C^*)_\tau=  I_q(A':B'|C^*)_\tau=f(q,d_{A'},d_{B'}) \qquad \forall q>0.
  \end{equation}
\end{proof}

For the last case where $\mathscr{E}_{\textsc{iii}}(\cdot)=\Tr_{A''B''}\Big[(U_A\otimes U_B)\big[\ket{\phi}\bra{\phi}_{E_A}\otimes U'(\cdot)U'^{\dagger}\otimes \ket{\psi}\bra{\psi}_{E_B}\big](U_A\otimes U_B)^{\dagger}\Big]$, one could intuitively argue that tracing out systems could not increase the mutual information and one would expect that
\begin{equation}
\label{eq: wish}
   I_q(AA'':BB''|C^*)_\tau\geq I_q(A:B|C^*)_\tau.
\end{equation}  
Since
$I_q(AA'':BB''|C^*)_\tau=I_q(A:B|C^*)_\tau+I_q(AA'':B''|BC^*)_\tau+I_q(A'':B|AC^*)_\tau$,
Equation~\eqref{eq: wish} would follow from strong subadditivity used
twice i.e., $I_q(AA'':B''|BC^*)_\tau\geq 0$ and
$I_q(A'':B|AC^*)_\tau\geq 0$. However, it is known that strong
subadditivity does not hold in general for Tsallis entropies for
$q>1$~\cite{Petz2014}. Ref.~\cite{Petz2014} also provides a
sufficiency condition for strong subadditivity to hold for Tsallis
entropies. In the following Lemma, we provide another, simple
sufficiency condition that also helps bound the Tsallis mutual information $I_q(AA'':B|C)_{\tau}$ (or $I_q(A:BB''|C)_{\tau}$) corresponding to the map $\mathscr{E}_{\textsc{iii}}$ where only one of $A''$ or $B''$ is traced out but not both.

\begin{lemma}[Sufficiency condition for strong subadditivity of Tsallis entropies]
\label{lemma: sufficient}
If $\rho_{ABC}$ is a pure quantum state, then for all $q\geq 1$ we
have $I_q(A:B|C)_\rho\geq 0$.
\end{lemma}
\begin{proof}
  We have
  \begin{align*}
          I_q(A:B|C)=S_q(AC)+S_q(BC)-S_q(ABC)-S_q(C).
  \end{align*}
  Since $\rho_{ABC}$ is pure we have $S_q(ABC)=0$ $\forall q> 0$
  and (from the Schmidt decomposition argument mentioned earlier)
  $S_q(AC)=S_q(B)$, $S_q(BC)=S_q(A)$ and $S_q(C)=S_q(AB)$. Thus,
   \begin{align*}
          I_q(A:B|C)=S_q(A)+S_q(B)-S_q(AB)=I_q(A:B)\geq 0,
  \end{align*}
  which follows from subadditivity of quantum Tsallis entropies for $q\geq 1$ \cite{Audenaert2007}. In other words, for pure $\rho_{ABC}$, strong subadditivity of Tsallis entropies is equivalent to their subadditivity which holds whenever $q\geq 1$.
\end{proof}

\begin{corollary}
\label{corollary: qstageIII}
Let $\mathscr{E}^1_{\textsc{iii}}: \mathscr{S}(\mathscr{H}_C)\rightarrow \mathscr{S}(\mathscr{H}_{\tilde{A}}\otimes\mathscr{H}_{B})$ be a quantum channel of the form $$\mathscr{E}^1_{\textsc{iii}}(\cdot)=\Tr_{B''}\Big[(U_A\otimes U_B)\big[\ket{\phi}\bra{\phi}_{E_A}\otimes U'(\cdot)U'^{\dagger}\otimes \ket{\psi}\bra{\psi}_{E_B}\big](U_A\otimes U_B)^{\dagger}\Big],$$
where $U': \mathscr{H}_C\rightarrow \mathscr{H}_{A'}\otimes\mathscr{H}_{B'}$, $U_A:\mathscr{H}_{E_A}\otimes\mathscr{H}_{A'}\rightarrow\mathscr{H}_{\tilde{A}}\cong\mathscr{H}_{A}\otimes\mathscr{H}_{A''}$ and $U_B:\mathscr{H}_{B'}\otimes\mathscr{H}_{E_B}\rightarrow\mathscr{H}_{\tilde{B}}\cong\mathscr{H}_{B}\otimes\mathscr{H}_{B''}$ are arbitrary unitaries and $\ket{\phi}_{E_A}$ and $\ket{\psi}_{E_B}$ are arbitrary pure states. If $\rho_{\tilde{A}B|C}$ is the corresponding Choi state, then the Tsallis conditional mutual information evaluated on the state $\tau_{\tilde{A}BC^*}=\frac{1}{d_C}\rho_{\tilde{A}B|C}\in \mathscr{S}(\mathscr{H}_{\tilde{A}}\otimes\mathscr{H}_{B}\otimes\mathscr{H}_{C^*})$ satisfies
  $$I_q(\tilde{A}:B|C^*):=I_q(AA'':B|C^*)_\tau\leq f(q,d_{A'},d_{B'}) \qquad \forall q\geq 1.$$
\end{corollary}
\begin{proof}
  Since $I_q(AA'':BB''|C^*)_{\tau}=I_q(AA'':B|C^*)_{\tau}+I_q(AA'':B''|BC^*)_{\tau}$, the purity of $\tau_{\tilde{A}\tilde{B}C^*}=\tau_{AA''BB''C^*}$ and Lemma~\ref{lemma: sufficient} imply that $$I_q(AA'':BB''|C^*)_{\tau}\geq I_q(AA'':B|C^*)_{\tau}, \forall q\geq 1,$$ or (equivalently) in more concise notation,
  $$I_q(\tilde{A}:\tilde{B}|C^*)_{\tau}\geq I_q(\tilde{A}:B|C^*)_{\tau} \quad \forall q\geq 1.$$ Finally, using Lemma~\ref{lemma: qstageII} we obtain the required result.
\end{proof}

Now, for Equation~\eqref{eq: wish} to hold, we do not necessarily need strong subadditivity. Even if $I_q(A'':B|AC)_{\tau}\geq 0$ does not hold, Equation~\eqref{eq: wish} would still hold if $I_q(AA'':B''|BC)_{\tau}+I_q(A'':B|AC)_{\tau}\geq 0$. This motivates the following conjecture.

\begin{conjecture}\label{conjecture: qstageIII}
Let $\mathscr{E}_{\textsc{iii}}: \mathscr{S}(\mathscr{H}_C)\rightarrow \mathscr{S}(\mathscr{H}_{A}\otimes\mathscr{H}_{B})$ be a quantum channel of the form $$\mathscr{E}_{\textsc{iii}}(\cdot)=\Tr_{A''B''}\Big[(U_A\otimes U_B)\big[\ket{\phi}\bra{\phi}_{E_A}\otimes U'(\cdot)U'^{\dagger}\otimes \ket{\psi}\bra{\psi}_{E_B}\big](U_A\otimes U_B)^{\dagger}\Big],$$
where $U': \mathscr{H}_C\rightarrow \mathscr{H}_{A'}\otimes\mathscr{H}_{B'}$, $U_A:\mathscr{H}_{E_A}\otimes\mathscr{H}_{A'}\rightarrow \mathscr{H}_{A}\otimes\mathscr{H}_{A''}$ and $U_B:\mathscr{H}_{B'}\otimes\mathscr{H}_{E_B}\rightarrow \mathscr{H}_{B}\otimes\mathscr{H}_{B''}$ are arbitrary unitaries and $\ket{\phi}_{E_A}$ and $\ket{\psi}_{E_B}$ are arbitrary pure states. If $\rho_{AB|C}$ is the corresponding Choi state, then the Tsallis conditional mutual information evaluated on the state $\tau_{ABC^*}=\frac{1}{d_C}\rho_{AB|C}\in \mathscr{S}(\mathscr{H}_{A}\otimes\mathscr{H}_{B}\otimes\mathscr{H}_{C^*})$ satisfies
  $$I_q(A:B|C^*)_\tau\leq f(q,d_{A'},d_{B'}) \qquad \forall q\geq 1.$$
\end{conjecture}

Notice that in Corollary~\ref{corollary: qstageIII} and
Conjecture~\ref{conjecture: qstageIII}, the bounds are functions of
$d_{A'}$ and $d_{B'}$ and not of the dimensions of the systems $A$ and
$B$ (those in the quantity on the left hand side). In the case that
$d_A\geq d_{A'}$ and $d_B\geq d_{B'}$, the fact that $f(q,d_A,d_B)$ is
a strictly increasing function of $d_A$ and $d_B$ $\forall q\geq 0$
allows us to write $I_q(\tilde{A}:B|C^*)_\tau\leq
f(q,d_{\tilde{A}},d_B)$ and $I_q(A:B|C^*)_\tau\leq f(q,d_A,d_B)$ under
the conditions of Corollary~\ref{corollary: qstageIII} and
Conjecture~\ref{conjecture: qstageIII} respectively. However, if
$d_A\leq d_{A'}$ and/or $d_B\leq d_{B'}$, the bounds
$f(q,d_{\tilde{A}},d_B)$ and $f(q,d_A,d_B)$ are tighter than the bound
$f(q,d_{A'},d_{B'})$ and so not implied. However, based on the several examples that we have checked, we further conjecture the following.

\begin{conjecture}\label{conjecture: qstageIII2}
Under the same conditions as Conjecture~\ref{conjecture: qstageIII}
  $$I_q(A:B|C^*)_\tau\leq f(q,d_A,d_B) \qquad \forall q\geq 1.$$
\end{conjecture}

Further, it is shown in \cite{Allen2017} that if $C$ is a \emph{complete common cause} of $A$ and $B$ then the corresponding Choi state, $\rho_{AB|C}$ decomposes as $\rho_{AB|C}=(\rho_{A|C}\otimes\mathds{1}_B)(\mathds{1}_A\otimes\rho_{B|C})$ or $\rho_{AB|C}=\rho_{A|C}\rho_{B|C}$ in analogy with the classical case where if a classical random variable $Z$ is a common cause of the random variables $X$ and $Y$, then the joint distribution over these variables factorises as $p_{XY|Z}=p_{X|Z}p_{Y|Z}$. Then we have that $\tau_{ABC^*}=\frac{1}{d_C}\rho_{AB|C}=\frac{1}{d_C}\rho_{A|C}\rho_{B|C}$. By further analogy with the classical results of Section~\ref{sec: tsalcaus}, one may also consider
instead a state of the form
$\hat{\sigma}_{ABCC^*}=\sigma_C\otimes\frac{1}{d_C}\rho_{A|C}\rho_{B|C}=\sigma_C\otimes\tau_{ABC^*}$,
where $\sigma_C\in\mathscr{S}(\mathscr{H}_C)$.\footnote{This is the
  analogue of the statement $P_{ABC}=P_CP_{A|C}P_{B|C}$ for
  probability distributions.}  Note that $\hat{\sigma}_{ABCC^*}$ is a
valid density operator on $\mathscr{H}_A\otimes\mathscr{H}_B\otimes\mathscr{H}_C\otimes\mathscr{H}_{C^*}$.

\begin{lemma}\label{lemma: sigma}
  The state $\hat{\sigma}_{ABCC^*}=\sigma_C\otimes\tau_{ABC^*}$ defined above
  satisfies $$I_q(A:B|CC^*)_{\hat{\sigma}}\leq f(q,d_A,d_B)\,,$$
whenever $I_q(A:B|C^*)_{\tau}\leq f(q,d_A,d_B)$ holds for the state $\tau_{ABC^*}=\frac{1}{d_A}\rho_{AB|C}$, where $\rho_{AB|C}$ represents the quantum channel from $C$ to $A$ and $B$ and $\sigma_C$ is the input quantum state to this channel.
\end{lemma}
\begin{proof}
  Since $\hat{\sigma}$ is a product state between the $C$ and $ABC^*$ subsystems, by the pseudo-additivity of quantum Tsallis entropies and
  the chain rule we have
  \begin{align*}
    I_q(A:B|CC^*)_{\hat{\sigma}}&=S_q(ACC^*)+S_q(BCC^*)-S_q(ABCC^*)-S_q(CC^*)\\                   &=S_q(AC^*)+S_q(BC^*)-S_q(ABC^*)-S_q(C^*)\\
    &\ \ \ -(q-1)S_q(C)\left(S_q(AC^*)+S_q(BC^*)-S_q(ABC^*)-S_q(C^*)\right)\\
                                &=(1-(1-q)S_q(C))I(A:B|C^*)\\
                                &=\Tr(\sigma_C^q)I(A:B|C^*)\,.
  \end{align*}

Now let $p_c$ be the distribution whose entries are the eigenvalues
of $\sigma_C$.  We have $\Tr(\sigma_C^q)=\sum_cp_c^q$.  Thus if $q>1$,
$\sum_cp_c^q\leq 1$ with equality if and only if $p_c=1$ for
some value of $c$.  It follows that
$$I_q(A:B|CC^*)_{\hat{\sigma}}\leq I_q(A:B|C^*)_\tau\,.$$
Therefore, if $I_q(A:B|C^*)_\tau\leq f(q,d_A,d_B)$, we also have
$I_q(A:B|CC^*)_{\hat{\sigma}}\leq f(q,d_A,d_B)$.
\end{proof}

\subsection{{\sc LPAssumptions}: A new linear program solver for Mathematica}
\label{appendix: LPA}
Our new Tsallis entropic inequalities for the Triangle causal structure, presented in Section~\ref{sec: newineq} were obtained using the {\sc LPAssumptions} Mathematica package that we developed for solving linear programs involving unspecified variables. The package is available as open source on Github \cite{LPAssumptions}. Here we provide a brief overview of the main functionality of this package and how it was used for obtaining the results presented earlier in this chapter. 

Linear programs are discussed in Section~\ref{sssec: LP} of the preliminaries, but we restate the standard form of a linear program (according to the convention followed in this thesis) here for completeness. A generic linear programming problem can be specified by giving a vector $c$, a matrix $M$ and a vector $b$ such that the problem corresponds to 
\begin{align*}
    \text{Minimize } & c.x\\
    \text{Subject to }& Mx\geq b\\
    &x\geq0
\end{align*}
Any linear programming problem can be written in this form and this is the same form as used by Mathematica's inbuilt {\sf LinearProgramming} function. If $c$, $M$ and $b$ are completely specified (i.e., consist of numerical values), {\sf LinearProgramming} can already be used to solve the problem.  The aim of our package \cite{LPAssumptions} is to also be able to cope with the case where there are unspecified constants in vector $c$ representing the objective function.  For example, we might wish to optimize $x_1+ax_2$, where $a$ is an unspecified constant in the range $0\leq a\leq 1$, returning the solution for all values of $a$ in the range. In our case, for the results of Section~\ref{sec: newineq}, we optimised the entropic expressions of Equations~\eqref{eq: Shineq1}-\eqref{eq: Shineq3} with the Shannon entropies $H()$ replaced by Tsallis entropies $S_q()$  over our outer approximation to the classical Tsallis entropy cone for the Triangle causal structure. This outer approximation is characterised by the Shannon constraints and our Tsallis entropic causal constraints (Corollary~\ref{corollary: mainconstraint}). The latter involve unspecified variables other than those being optimised over, namely the dimensions/cardinality of the variables and this optimisation cannot be performed using Mathematica's inbuilt {\sf LinearProgramming} function, and requires our {\sc LPAssumptions} package. 

{\sc LPAssumptions} works by using the two phase simplex algorithm~\cite{Dantzig1990} (Section~\ref{ssec:simplexalgo}). The algorithm involves visiting adjacent vertices of the feasible region (a polyhedron), starting from an initial vertex, until the vertex corresponding to the optimal solution is reached. The coefficients of the objective function and the constraints are encoded in a tableau and at each step, the next vertex to be visited is decided by checking the sign of certain elements in this tableau, which would indicate whether or not moving to that vertex would improve the value of the objective function.  In our case, these elements of the tableau can depend on the additional unspecified variables in the problem. {\sc LPAssumptions} relies on Mathematica's {\sf Simplify[expr,assump]} command to decide on on the sign of these elements, which will then tell it how to proceed through the computation. Here, {\sf expr} is an inequality that checks for the sign and {\sf assump} are the constraints on the unspecified variables that can be input by the user, along with any additional assumptions on the expressions that are made during the computation.  If Mathematica is unable to decide whether the inequality is true or false, the algorithm splits into two cases, one in which it assumes {\sf expr} is true, and the other in which it assumes it is false. It then carries on adding additional assumptions as necessary until termination (or the number of iterations is exceeded).  Problems can arise if {\sf expr} is true (or false) but Mathematica's {\sf Simplify} is unable to determine this (see the examples notebook mentioned in the next paragraph for such a case). 

The main function of the package, is {\sf LPAssumptions[c, M, b, assump, (Options)]}. It takes as input {\sf c} (a vector corresponds to the objective function), {\sf M} (the constraint matrix), {\sf b} (the constraint vector) and {\sf assump} (any initial assumptions on the unspecified variables involved). The output is a set of pairs $\{(\text{constr}_i,\text{vec}_i)\}_i$, where $\text{constr}_i$ specifies the set of constraints under which the optimum is achieved by the vector $\text{vec}_i$. With no assumptions and no unspecified variables, provided the problem is feasible and bounded, the answer given by {\sf LPAssumptions} should match that of {\sf LinearProgramming} (up to the slightly different structure of the output). Note that the current version of our package can only handle linear programs in which all the additional, unspecified variables occur in the objective function. Linear programs where all such variables occur in the constraint vector can also be solved using our program by the following trick: convert the original LP to the desired form (variables appearing only in the objective vector $c$) by taking the dual of the LP, solve the dual LP using {\sc LPAsssumptions} and use the duality theorems of Section~\ref{sssec: LP} to obtain a solution for the original LP. We had to use this trick for obtaining our results of Section~\ref{sec: newineq}.

%The default output takes the form $\{\{\text{constr}_1,\text{vec}_1\},\{\text{constr}_2,\text{vec}_2\},\ldots\}$,where $\text{constr}_i$ are the set of constraints under which the optimum is achieved by the vector $\text{vec}_i$.  The values of the objective function at the optima can then be computed using $c.\text{vec}_i$. Additional options such as the maximum number of iterations, outputting the optimal point vs the optimal value etc. 
A detailed explanation of the various functionalities and commands can be found in the user manual we have made available along with the package \cite{LPAssumptions}, and a Mathematica notebook  ({\tt LPAssumptions\_examples.nb}) with examples illustrating the use of this package is available at \definecolor{mylinkcolor}{rgb}{0,0,0}\url{http://www-users.york.ac.uk/~rc973/LPAssumptions.html}.\definecolor{mylinkcolor}{rgb}{0,0,0.7} In the examples, we also show how to compute the local weight (defined in Appendix~\ref{appendix: proofs}, see also~\cite{Zukowski99,Cope2019}) of a non-signaling probability distribution where the distribution has unspecified parameters (specifically we consider a noisy PR-box with inefficient detectors, see Section~IV~D~F of~\cite{Pirandola2019}). This package was also used in some of the results of \cite{Vilasini2020} which are presented in Chapter~\ref{chapter: mixingpaper}.

\subsection{Analysing causal structures using conditional R\'enyi entropies}
\label{appendix: Renyiresults}
Given the limitations of Shannon as well as Tsallis entropies found in \cite{Weilenmann16} and the work presented in this chapter \cite{Vilasini2019}, a natural question that arises is--- what about other generalised entropies? We already noted that other entropies like the R\'enyi, min or max entropies do not satisfy the chain rule (Table~\ref{tab:ent_prop}). The chain rule is important for the entropy vector method since it allows conditional entropies to be written in terms of unconditional ones (Equation~\eqref{eq: chain2}) and hence these need not be included as independent components of the entropy vector. In the absence of a chain rule, one would need to include the conditional entropies involving all possible subsets of nodes as additional components of the entropy vector, thereby significantly increasing the dimensionality of the problem and one would expect this to make the computations harder. This was an important motivation behind the choice of Tsallis entropies for the work presented in this chapter. However, in some cases, it has been possible to obtain non-trivial results using the entropic technique even in the absence of a chain rule \cite{WeilenmannGPT}. 

In a side project with Mirjam Weilenmann, we investigated the possibility of using R\'enyi entropies for analysing causal structures. For this, we considered the conditional R\'enyi entropies of Definition~\ref{definition: condRenyi}. Some of the properties of R\'enyi entropies and this conditional version were discussed in Section~\ref{ssec: renyi} and summarised in Table~\ref{tab:ent_prop}, these include: the non-negativity of the entropy i.e. $H_{\alpha}(X)\geq 0$, $\forall \alpha\geq 0$, additivity i.e., if $P_{XY}=P_XP_Y$, then $H_\alpha(XY)=H_\alpha(X)+H_\alpha(Y)$ $\forall \alpha\geq 0$ \cite{Renyi1961}, monotonicity i.e., $H_\alpha(X)\leq H_\alpha (XY)$ $\forall \alpha \in (0,1)\cup (1,\infty)$ \cite{Linden2013} and the conditional form of strong subadditivity, also known as data processing $H_\alpha(X|Y)\geq H_\alpha(X|YZ)$ $\forall \alpha\geq 0$ \cite{Iwamoto2013}. There are further properties of these entropies which are already known or can be easily derived, as we show below. Importantly, one of these (Property~\ref{prop: Renyichain}) is a generalised chain rule for the conditional R\'enyi entropies.

\begin{itemize}
    \item \textbf{Property 1:} (Non-negativity of conditional entropy) $H_\alpha(X|Y)\geq 0$ $\forall \alpha\geq 1$.
    \begin{proof}
    Since $\sum_xP(x|y)^\alpha\leq \sum_x P(x|y)=1$ $\forall y\in Y$ and $ \alpha\geq 1$, we have that $\sum_yP(y)\sum_xP(x|y)^\alpha$ $\leq \sum_yP(y)=1$ and hence $\log(\sum_yP(y)\sum_xP(x|y)^\alpha)\leq 0$ $\forall \alpha\geq 1$. Thus $H_\alpha(X|Y)\geq 0$ $\forall \alpha\geq 1$. 
    \end{proof}
    \item\textbf{Property 2:} (Independence) If $P_{XY}=P_XP_Y$, then $H_\alpha(X|Y)=H_\alpha(X)$ $\forall \alpha\geq 0$.
    
    Note: This is a special case of Property~6 and will also be part of that in the implementation.
    
    \begin{proof}
    This directly follows from using $P(xy)=P(x)P(y)$ $\forall x\in X, y\in Y$ in the definition of the conditional R\'enyi entropy (Equation~\eqref{eq: renyi2}),
    \begin{equation*}
        H_\alpha(X|Y)=\frac{1}{1-\alpha} \log (\sum_yP(y)\sum_xP(x)^\alpha)=\frac{1}{1-\alpha} \log (\sum_xP(x)^\alpha)=H_\alpha(X).
    \end{equation*}
    
    \end{proof}
    
   \item\label{prop: Renyichain} \textbf{Property 3:} (Generalised chain rule) $\forall \alpha \in (0,1) \cup (1,\infty)$, $\forall P_{XYZ}$,
    \begin{equation*}
        H_\alpha(X|YZ)\geq H_\alpha(XZ|Y)-\log d_Z
    \end{equation*} 
    
    This is derived for quantum R\'enyi entropies in \cite{Muller-Lennert2013} and follows as a special case for our classical definition.
    
   \item \textbf{Property 4:} (Conditional independence or entropic causal constraint) If $P_{XY|Z}=P_{X|Z}P_{Y|Z}$, then $H_\alpha(X|YZ)=H_\alpha(X|Z)$ or equivalently, $I_{\alpha}(X:Y|Z)=0$ $\forall \alpha \geq 0$. 
   \begin{proof}
   Using the conditional independence $P_{XY|Z}=P_{X|Z}P_{Y|Z}$ $\Rightarrow$ $P(xyz)=P(x|z)P(y|z)P(z)$ $=P(x|z)P(yz)$ $\forall x\in X, y\in Y, z\in Z$ in the definition of the conditional R\'enyi entropy (Equation~\eqref{eq: renyi2}) we have
\begin{align}
    \begin{split}
        H_\alpha(X|YZ)
        &=\frac{1}{1-\alpha}\log(\sum_{xyz} P(xyz)^\alpha P(yz)^{1-\alpha})\\
        &= \frac{1}{1-\alpha}\log(\sum_{xyz} P(x|z)^\alpha P(yz)^\alpha P(yz)^{1-\alpha})\\
        &= \frac{1}{1-\alpha}\log(\sum_{xz} P(x|z)^\alpha \sum_y P(yz))\\
         &= \frac{1}{1-\alpha}\log(\sum_{xz} P(x|z)^\alpha P(z))=H_\alpha(X|Z).
    \end{split}
\end{align}
By the definition of the conditional mutual information, $I_{\alpha}(X:Y|Z)=H(X|YZ)-H(Y|Z)$, it follows that $I_{\alpha}(X:Y|Z)=0$ in this case.
   \end{proof}
   
   \item \textbf{Property 5:} (Monotonicity under conditioning) $H_\alpha(X|Z)\leq H_\alpha(XY|Z)$ $\forall \alpha \geq 0$. 
    \begin{proof}
  Noting that $\sum_y P(xy|z)^\alpha\leq \big(\sum_y P(xy|z)\big)^\alpha=P(x|z)^\alpha$ $\forall \alpha \geq 0$, we have
  \begin{align}
      \begin{split}
          H_\alpha(XY|Z)&=\frac{1}{1-\alpha}\log(\sum_z P(z)\sum_{x}\sum_yP(xy|z)^\alpha)\\
          &\geq \frac{1}{1-\alpha}\log(\sum_z P(z)\sum_{x}P(x|z)^\alpha)
          =H_\alpha(X|Z).
      \end{split}
  \end{align}
   \end{proof}
   
\end{itemize}

Using these constraints to define the outer approximation to the classical R\'enyi entropy cone for the bipartite Bell causal structure (Figure~\ref{fig: Bell2}), we applied the entropy vector method for this case. The range of $\alpha$ for which all these constraints hold is $\alpha\geq 1$, and the following applies to this range. The program terminated and no non-trivial entropic inequalities were obtained, indicating that this outer approximation to the classical marginal R\'enyi entropy cone coincides with the quantum one for this causal structure. This is the same outcome as the Shannon case ($\alpha=1$) found in \cite{Weilenmann16}, while for the Tsallis case analysed in the main part of this chapter, we found the problem to be computationally too costly and were hence unable to find any non-trivial Tsallis entropic inequalities.
In the R\'enyi case, the program ran for several days (5-7 days on a regular desktop PC), probably owing to the dimension-dependency of the generalised chain rule (Property~\ref{prop: Renyichain}), suggesting that this method would be computationally costly for larger causal structures such as the Triangle (Figure~\ref{fig: Triangle}). 

In the presence of post-selection, we saw that the Shannon entropic Braunstein-Caves inequalities~\eqref{eq: BCineqs} can be derived for the post-selected Bell causal structure of Figure~\ref{fig: BellPS}. These are derived by using the monotonicity~\eqref{eq: monotonicity} and strong-subadditivity~\eqref{eq: SSA} (unconditional form) on the entropies of the set $\{X_0,X_1,Y_0,Y_1\}$. In the R\'enyi case however, strong subadditivity does not hold in the unconditional form, and such inequalities cannot be derived analogously. Implementing the entropy vector method for R\'enyi entropies in the post-selected Bell causal structure, it was found by Mirjam that no non-trivial inequalities are obtained. We believe that this failure to detect non-classicality using R\'enyi entropies is because the generalised chain rule~\ref{prop: Renyichain} is not a tight enough constraint, and as a consequence, the initial outer approximation to classical entropy cone is not a good one.

\chapter{Entropic analysis of causal structures with post-selection}
\label{chapter: mixingpaper}

\lettrine[nindent=0em, slope=-.5em,lines=2]{T}{he} results of Chapter~\ref{chapter: Tsallispaper} along with those of \cite{Weilenmann16} have revealed important drawbacks of the entropic technique for analysing causal structures in the absence of post-selection, using Shannon, Tsallis and well as R\'enyi entropies. In the present chapter, we analyse the post-selected Bell causal structure (Figure~\ref{fig: BellPS}) using entropies. We have seen that post-selection allows for the derivation of non-trivial entropic inequalities~\eqref{eq: BCineqs} in causal structures such as the Bell scenario that do not support a classical-quantum gap in the absence of post-selection \cite{Weilenmann16}. It is then natural to ask whether the non-classicality of a distribution can always be detected through post-selected entropic inequalities such as those of Equation~\eqref{eq: BCineqs}. For the $(d,d,2,2)$ Bell scenarios with $d\geq 2$, this is known to be the case~\cite{Chaves13} in the following sense.  For every non-classical distribution in the $(d,d,2,2)$ Bell scenario, there is a transformation that does not make any non-classical distribution classical, and such that the resulting distribution violates one of the BC entropic inequalities. The main purpose in this chapter is to investigate whether a similar result holds for non-binary outcomes. To do so, we need to specify a class of post-processing operations. The most general operations that we could consider are the \emph{non-classicality non-generating (NCNG)} operations, i.e., those that do not map any classical distribution to a non-classical one.  An interesting subset of these is the class of post-processings achievable through \emph{local operations and shared randomness (LOSR)}, which are physical in the sense that two separated parties with shared randomness could perform them.\footnote{In general NCNG operations used on the correlations prior to evaluating an entropic inequality need not be physical in this sense.} Because of the difficulty of dealing with arbitrary NCNG operations, for the majority of our analysis we consider LOSR supplemented with the additional (NCNG) operation where the parties are exchanged (and convex combinations). We use LOSR+E to refer to this supplemented set.

We study the $(2,2,3,3)$ Bell scenario with LOSR+E post-processing operations, to see whether when applied to any non-classical distribution the result violates an entropic Bell inequality. We investigate this using both Shannon and Tsallis entropies. Our motivation for considering Tsallis entropies is that they are known to provide an advantage over the Shannon entropy in detecting non-classicality in the absence of post-processing~\cite{Wajs15} in the sense that there are non-classical distributions that violate Tsallis entropic inequalities but not the analogous Shannon-entropic inequality. In the $(2,2,2,2)$ case however, due to the result of \cite{Chaves13}, this advantage is less apparent when post-processings are considered. It is unclear whether or not this is also the case for the $(2,2,3,3)$ scenario, and hence we consider Tsallis entropies in this work. The present chapter is based on the paper \cite{Vilasini2020} which is joint work with Roger Colbeck.

\section{Summary of contributions: in words and in poetry}

The result of~\cite{Chaves13} that Shannon inequalities are always sufficient for detecting distributions that violate the CHSH inequality in the $(2,2,2,2)$ scenario readily extends to the lifted-CHSH inequalities of the $(2,2,3,3)$ scenario (c.f.\ Corollary~\ref{coroll:chaves}). We summarise this result in Section~\ref{ssec: 2222}. Due to Corollary~\ref{coroll:chaves}, we are particularly interesting in the region containing non-classical distributions that satisfy all the CHSH-type inequalities in the $(2,2,3,3)$ scenario. In Section~\ref{sec: 2233prob}, we find the vertex description of the convex polytope containing non-signaling distributions in the $(2,2,3,3)$ scenario that satisfy all the lifted-CHSH inequalities. We show that every non-classical distribution in this polytope violates exactly one $I_{2233}$-type inequality (Proposition~\ref{proposition: probspace}) and identify non-classical distributions in this region that violate entropic inequalities. Further, in Section~\ref{sec: 2233ent}, we first consider a subset of the post-processing operations $LOSR+E$, that correspond to mixing with a classical distribution, and identify a family of non-classical distributions in $\Pi_{CHSH}^{\{2,2,3,3\}}$ for which we conjecture that the non-classicality cannot be detected through arbitrary Shannon entropic inequalities and a class of Tsallis entropic inequalities. 
In Section~\ref{ssec: CGrelab}, we show analytically that if these conjectures hold, then they also extend also the whole of $LOSR+E$. This suggests that post-selected entropic inequalities, both of Shannon and Tsallis types, are in general insufficient for detecting non-classicality in the bipartite Bell causal structure.
 Following the trend set in the previous chapter, a poetic summary of this chapter is provided before delving into the technical results.

\textit{\hyperref[chapter: Tsallispaper]{Recall}, we found it hard to certify,\\
That a causal structure must be described,\\
By systems that obey non-classical physics,\\
Based on the entropies of the observed statistics.}

\textit{But for some causal structures, we realise,\\
That an additional technique does materialise.\\
\hyperref[ssec: PS]{Post-selection} provides a trick,\\
To detect non-classicality that would otherwise escape our grip.}

\textit{But does it allow us to identify,\\
All the non-classical correlations that can arise?\\
In some cases\footnote{Particularly, this refers to\\
Bell scenarios with output cardinality two,\\
Here we consider,\\
Cardinalities of three or larger.}, there's indeed a technique \cite{Chaves13},\\
But does it generalise? An answer we seek.}

\textit{We consider a set of operations $O$ you can do,\\
On correlations before their entropies are put to test,\\
For a subset $M$ of $O$, our computations give a \hyperref[appendix: evidence]{clue}\\
That the final entropies fail to reveal non-classicality, even at their best.}

\textit{If our computations for $M$ are true,\\
We \hyperref[corollary: final]{prove} that they continue to be true for all of $O$,\\
Along the way, we get some related \hyperref[sec: 2233prob]{results} too,\\
Through a minor geometric detour.}

\textit{So it seems that post-selected entropic inequalities,\\
Be it of Shannon or the Tsallis kind,\\
Are insufficient for detecting non-classicality,\\
Even if we post-process our data before we try.\\}

%{\raggedleft \textit{-Vilasini Venkatesh (2020)}\par}

\section{Existing result for the \texorpdfstring{$(2,2,2,2)$}{(2,2,2,2)} Bell scenario in entropy space}\label{ssec: 2222}
 
The current section summarises the relevant results of~\cite{Chaves13}
regarding the sufficiency of entropic inequalities in the $(2,2,2,2)$
scenario. As previously mentioned, it is possible for a non-classical
distribution to have the same entropy vector as a classical one and hence to
be entropically classical. For example, the maximally non-classical
distribution in probability space, $P_{\PR}$ (Equation~\eqref{eq: PR})
is entropically classical since it has the same entropy vector as the
classical distribution
\begin{equation}
    \label{eq: Pc}
     P_{\Cl}=
\begin{array}{ |c|c|} 
 \hline
 \frac{1}{2} \quad 0 & \frac{1}{2} \quad 0\\ 
 0 \quad \vphantom{\frac{1}{f}}\frac{1}{2} & 0 \quad \vphantom{\frac{1}{f}}\frac{1}{2}\\
 \hline
 \frac{1}{2} \quad 0 & \frac{1}{2} \quad 0\\ 
 0 \quad \vphantom{\frac{1}{f}}\frac{1}{2} & 0 \quad \vphantom{\frac{1}{f}}\frac{1}{2}\\
 \hline
\end{array}
\end{equation}
and hence cannot violate any of the BC inequalities\footnote{$P_{\PR}$ and $P_{\Cl}$
  are related by a permutation of the entries in the bottom right
  $2\times 2$ block and entropies are invariant under such
  permutations.}. However, the distribution
$\frac{1}{2}P_{\PR}+\frac{1}{2}P_{\Cl}$ maximally violates $I_{\BC}^4\leq0$
attaining a value of $\ln 2$. That convex mixtures of non-violating distributions can lead to a violation is due to the
fact that entropic inequalities are non-linear in the underlying
probabilities (in contrast to the facet Bell inequalities in
probability space).

In~\cite{Chaves13}, it was shown that such a procedure is possible for every non-classical distribution in the $(d,d,2,2)$ Bell scenario with $d\geq 2$ i.e., for every such distribution, there exists an LOSR transformation such that the resultant distribution violates a Shannon entropic BC inequality~(\ref{eq: BCineqs}).  Thus, non-classicality can be detected in this scenario by processing the observed correlations (in a way that cannot generate non-classicality) before using a BC entropic inequality on the result.  In this sense the BC entropic inequalities provide a necessary and sufficient test for non-classicality in these scenarios.

In more detail, for the $(2,2,2,2)$ case this works as follows.  First, one defines a special class of distributions, \emph{isotropic distributions}, as follows for some $k\in \{1,2,...,8\}:=[8]$ and $\epsilon \in [0,1]$.
\begin{equation}
   P_{\iso}^k=\epsilon P_{\PR}^k+(1-\epsilon)P_{\noise},
\end{equation}
where $P_{\noise}$ is white noise i.e., the distribution with all entries equal to $1/4$. In the $(2,2,2,2)$ Bell scenario the isotropic distribution $P_{\iso}^k$ is non-classical if and only if $\epsilon>1/2$. The LOSR transformation used in~\cite{Chaves13} involves first transforming the observed distribution into an isotropic distribution through a local depolarisation procedure that cannot generate non-classicality. Second, it is shown that for any non-classical isotropic distribution i.e., a $P_{\iso}^k$ with $\epsilon>1/2$, there exists a classical distribution $P_{\Cl}^k$ such that the distribution $P_{\E,v}^k=vP_{\iso}^k+(1-v)P_{\Cl}^k$ violates one of the BC entropic inequalities for sufficiently small $v>0$. In particular, the value of $I_{\BC}^k$ for $P_{\E,v}^k$ can be expanded for small $v$ as
\begin{equation}
\label{eq: BCexpand2}
    I_{\BC}^k \approx \frac{v}{\ln 4}[f(\epsilon)-(4\epsilon-2)\ln v],
\end{equation}
where $f(\epsilon)$ is a function of $\epsilon$, independent of $v$
(see~\cite{Chaves13} for details). Thus for any $\epsilon>1/2$, the
corresponding isotropic distributions are non-classical and taking $v$
arbitrarily small can make $I_{\BC}^k$ positive which is a violation
of the entropic inequality. We summarise the main result
of~\cite{Chaves13} for $(2,2,2,2)$ Bell scenarios in the following
Theorem (which is implicit in~\cite{Chaves13}).
\begin{theorem}
\label{theorem: chaves}
For every non-classical distribution, $P_{XY|AB}$ in the $(2,2,2,2)$ Bell scenario, there exists an LOSR transformation $\mathcal{T}$, such that $\mathcal{T}(P_{XY|AB})$ violates one of the BC entropic inequalities~\eqref{eq: BCineqs}.
\end{theorem}

One of the aims of the present chapter is to study whether this result
extends to the case where the number of outcomes per party is more
than two. In general, the $(2,2,d,d)$ Bell polytope for $d>2$ has new,
distinct classes of Bell inequalities and extremal non-signaling
vertices other than the CHSH inequalities and the PR boxes. We analyse this problem for the $d=3$ case, for which the following results pertaining to the $(2,2,3,3)$ scenario in probability space will be helpful.

\section{New results for the \texorpdfstring{$(2,2,3,3)$}{(2,2,3,3)} Bell scenario in probability space}\label{sec: 2233prob}
In this section we compute the vertex description of the \emph{CHSH-classical polytope}, $\Pi_{\chsh}^{(2,2,3,3)}$, i.e., the polytope whose facets are the 648 CHSH inequalities and the positivity constraints of the $(2,2,3,3)$ Bell scenario.  As previously mentioned, this will be the main region of interest in the remainder of this work since the non-classicality of distributions not belonging to this region can always be certified using Shannon-entropic inequalities (Corollary~\ref{coroll:chaves}). 
% Non-classical distributions in this polytope can thus only violate $I_{2233}$ inequalities. Here, we completely characterise this polytope by finding all its extremal points.
The following result allows us to significantly speed up the vertex enumeration problem.

\begin{proposition}
\label{proposition: probspace}
Every non-classical distribution in $\Pi_{\chsh}^{(2,2,3,3)}$ violates only one $I_{2233}$ inequality. 
\end{proposition}
\begin{proof}
 Let $i\in\{2,3,\ldots,432\}$ and consider the linear program that maximises the value of $\epsilon \geq 0$ subject to there existing a non-signaling distribution that
 \begin{itemize}
     \item violates $I_{2233}^1\leq2$ and $I_{2233}^i\leq2$ by at least $\epsilon$, i.e., $I_{2233}^1-\epsilon \geq 2$ and $I_{2233}^i-\epsilon \geq 2$;
     \item satisfies all CHSH-type inequalities.
 \end{itemize}
 We run over $i\in\{2,3,\ldots,432\}$ and check that in all cases either the output of this linear program is $\epsilon=0$ (meaning that the two $I_{2233}$ inequalities can be jointly saturated but not violated) or that the program is infeasible (the two $I_{2233}$ inequalities cannot even be jointly saturated). By symmetry it follows that no pair of $I_{2233}$ inequalities can be simultaneously violated when all the CHSH-type inequalities are satisfied.
\end{proof}

Note that in the $(2,2,3,3)$ scenario there exist extremal non-signaling distributions that violate multiple Bell inequalities. For example, the distributions $P_{\NL}$ (Equation~\eqref{eq: NL3}) and $P_{\NL}^*$ (Equation~\eqref{eq: NL*}) violate $I_{2233}\leq2$ (c.f. Equation~\eqref{eq: CGLMP}) although only $P_{\NL}$ violates it maximally. By symmetry, $P_{\NL}$ also violates another $I_{2233}$ inequality. %, namely the one maximally violated by $P_{\NL}^*$.
In addition, $P_{\NL}$ violates the CHSH-type inequality whose evaluation is equivalent to applying the output coarse-graining $0\mapsto0$, $1\mapsto1$ and $2\mapsto1$ for each party and then evaluating~\eqref{eq: CHSH}. This is in contrast to the $(2,2,2,2)$ scenario where there is a one-to-one correspondence between the extremal non-signaling vertices and the CHSH inequalities in the sense that each such vertex violates exactly one CHSH inequality.\footnote{Note that this correspondence breaks down in the $(2,2,3,3)$ scenario where it is possible for a CHSH-type vertex to violate multiple CHSH-type inequalities (these correspond to the same 2-outcome CHSH inequality after coarse-graining).}

Due to the symmetries, all the vertices of $\Pi_{\chsh}^{(2,2,3,3)}$ can be enumerated by first finding all the vertices for which the $I_{2233}$ inequality of Equation~\eqref{eq: CGLMP} is saturated or violated i.e., $I_{2233}\geq 2$, and taking the orbit of these vertices under local relabellings and exchange of the parties. The vertex enumeration for this case yields 47 extremal points of which 30 are the local deterministic points that saturate $I_{2233}\geq2$ and 17 are non-classical points that violate only this inequality. These are listed in Table~\ref{tab:vertices}. By applying all symmetries and removing duplicate vertices we find that $\Pi_{\chsh}^{(2,2,3,3)}$ has 7425 vertices (including the 81 local deterministic vertices).

\section{New results for the \texorpdfstring{$(2,2,3,3)$}{(2,2,3,3)} Bell scenario in entropy space}
\label{sec: 2233ent}

We now investigate whether entropic inequalities are necessary and sufficient for non-classicality in the $(2,2,3,3)$ Bell scenario. In $(d,d,2,2)$ scenarios with $d\geq 2$, only $2d$ Shannon entropic inequalities are required for the Shannon entropic characterisation of the scenario~\cite{Chaves2012, Fritz13}; in the $(2,2,2,2)$, these are the four inequalities of~\eqref{eq: BCineqs}. It may at first seem surprising that these can always be used to decide whether a distribution is classical because the number of extremal Bell inequalities grows very rapidly in $d$ in the $(d,d,2,2)$ scenario~\cite{Cope2019}, and deciding whether a distribution is classical is NP-complete~\cite{Avis2004}. The reduction in the number of inequalities in entropy space is compensated by the need to identify a suitable post-processing operation (of which there are uncountably many possibilities) in order to detect violations. 

The first observation is a corollary of Theorem~\ref{theorem: chaves}.
\begin{corollary}\label{coroll:chaves}
Let $P_{XY|AB}$ be a distribution in the $(2,2,3,3)$ Bell scenario that violates at least one CHSH-type inequality. Then there exists an LOSR transformation $\mathcal{T}$, such that $\mathcal{T}(P_{XY|AB})$ violates one of the BC entropic inequalities~\eqref{eq: BCineqs}.
\end{corollary}
\begin{proof}
For each CHSH-type inequality in the $(2,2,3,3)$ scenario, there exists a coarse-graining in which two of the outcomes are mapped to one (for each party and each input) such that for any initial distribution in the $(2,2,3,3)$ scenario that violates the CHSH-type inequality the coarse-grained distribution violates one of the CHSH-inequalities in the $(2,2,2,2)$ scenario. Hence, for the given $P_{XY|AB}$, after applying the corresponding coarse-graining for the violated CHSH-type inequality, followed by the LOSR operation from Theorem~\ref{theorem: chaves} we violate one of the BC entropic inequalities.
\end{proof}
This corollary means that we can limit our analysis to $\Pi^{(2,2,3,3)}_{\chsh}$, the polytope in which all the CHSH inequalities are satisfied, and, in particular, the non-classical region of this. This is the region in which one of the $I_{2233}$ inequalities is violated.

In going from the $(2,2,2,2)$ to $(2,2,3,3)$ scenario, a new class of inequalities (the $I_{2233}$ inequalities) become relevant in probability space but the entropic characterisation remains unchanged, since entropic inequalities do not depend on the number of measurement outcomes.  It is natural to ask whether all non-classical distributions in the $(2,2,3,3)$ scenario that satisfy all the CHSH inequalities cannot be certified entropically. However, this is not the case as shown by the following proposition.

\begin{proposition}
\label{proposition: countereg}
The polytope $\Pi_{\chsh}^{(2,2,3,3)}$ is not entropically classical.
\end{proposition}
\begin{proof}
  Consider the distribution
\begin{equation}
\label{eq: countereg}
P_e:=\frac{1}{50}\, 
    \begin{array}{ |ccc|ccc|} 
 \hline
 21 & 0 & 0 & 21 & 0 & 0\\ 
 0 & 2 & 0 & 1 & 1 & 0\\
11 & 0 & \vphantom{\frac{1}{f}}16 & 0 & 1 & \vphantom{\frac{1}{f}}26\\
 \hline
 31 & 0 & 0 & 20 & 1 & 10\\ 
 1 & 1 & 0 & 1 & 0 & 1\\
 0 & 1 & \vphantom{\frac{1}{f}}16 & \vphantom{\frac{1}{f}}1 & 1 & 15\\
 \hline
\end{array}\,.
\end{equation}
This is formed by mixing the non-local extremal point number~8 of $\Pi_{\chsh}^{(2,2,3,3)}$ (see Table~\ref{tab:vertices}) with the three local deterministic points 18, 26 and 47 with respective weights $1/10$, $3/10$, $1/5$ and $2/5$, and hence is in $\Pi_{\chsh}^{(2,2,3,3)}$.  It achieves a $I_{\BC}^4$ value of $0.0199733$, in violation of $I_{\BC}^4\leq0$, so is not entropically classical.
\end{proof}

\begin{figure}[t!]
\centering
    \includegraphics[scale=.3]{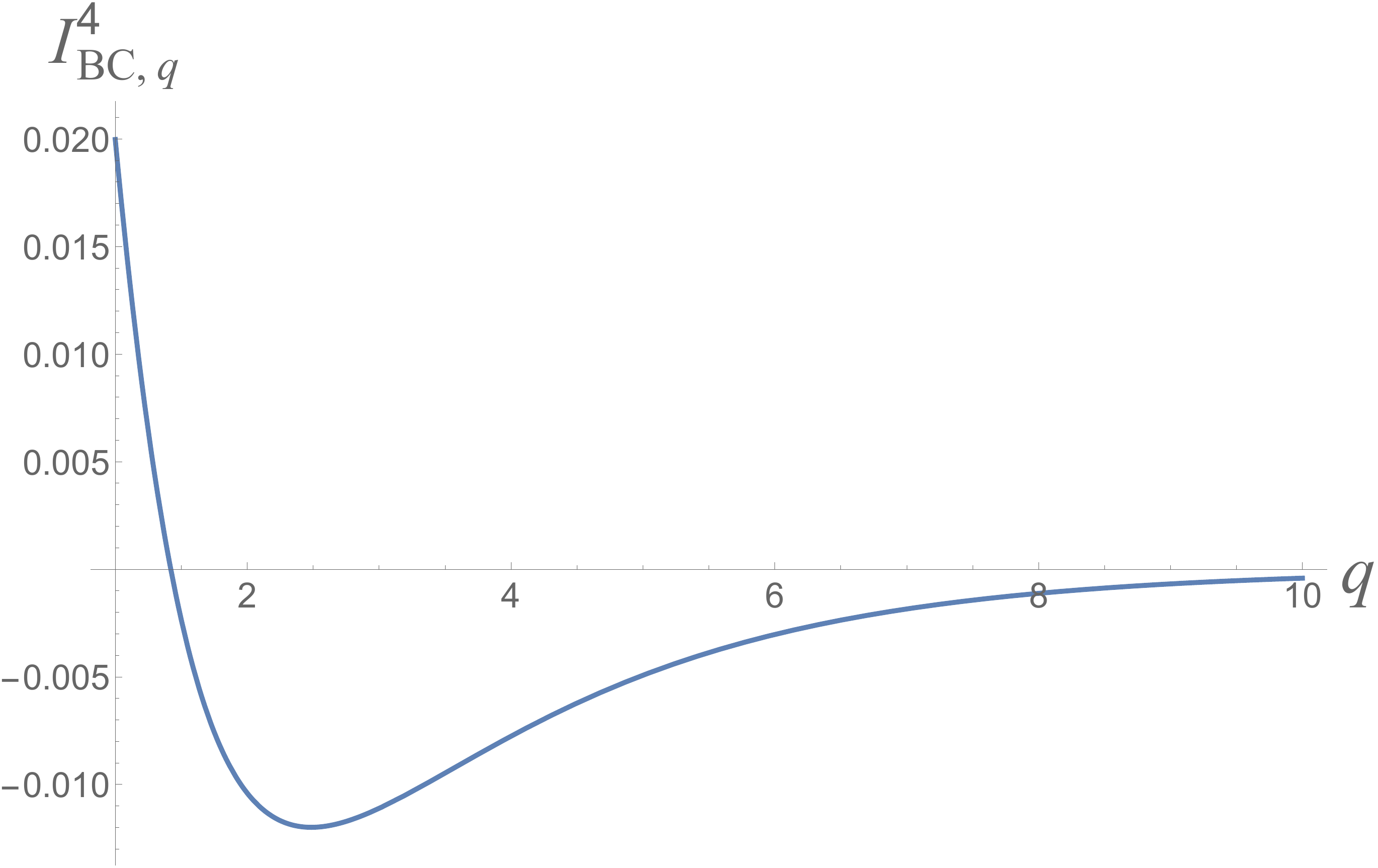}

    \caption[Plot of the Tsallis entropic BC expression as a function of the Tsallis parameter $q$ for the distribution of Equation~\eqref{eq: countereg}.]{\textbf{Plot of the $I_{\BC,q}^4$ value as a function of the Tsallis parameter $q$ for the distribution $p_e$ (Equation~\eqref{eq: countereg}).} %This non-classical distribution does not violate any of the CHSH inequalities, but it violates the $I_{2233}$ Inequality~\eqref{eq: CGLMP}.
      As seen from the plot, the distribution violates the BC inequality $I_{\BC,q}^4\leq 0$ for $q$ values between $1$ (Shannon case) and just under $1.5$ and the violation is maximum in the Shannon case, indicating that it is preferable to use Shannon rather than Tsallis entropies with $q>1$ in this case.}
    \label{fig: countereg_dist}
  \end{figure}

We remark that by mixing with more local deterministic distributions and varying the weights, larger violations of $I_{\BC}^4\leq0$ can be found; the distribution $P_e$ used in the previous proposition was chosen for its relative simplicity. Interestingly, we find that the Shannon entropic BC inequalities appear 
to give the largest violation among the Tsallis entropic inequalities 
for $q\geq 1$ when applied to $P_e$. This can be seen in Figure~\ref{fig: countereg_dist}.

In light of Proposition~\ref{proposition: countereg}, it is natural to ask whether the non-classicality of all distributions in $\Pi_{\chsh}^{(2,2,3,3)}$ can be detected through entropic inequalities. We find numerical evidence that suggests the contrary, i.e., that there are non-classical distributions in $\Pi_{\chsh}^{(2,2,3,3)}$ whose non-classicality cannot be detected through entropic inequalities using a general class of post-processing operations, and hence that these entropic inequalities are not sufficient for detecting non-classicality in the $(2,2,3,3)$ scenario. Before presenting these results, we briefly overview the post-processing operations considered in this work.

\paragraph{Post-processing operations }
In this chapter we study whether entropic inequalities can always detect non-classicality in the $(2,2,3,3)$ Bell scenario. In order to do so we could in principle consider applying any NCNG operation to the distribution prior to evaluating the entropic inequality. However, due to the difficulty in dealing with arbitrary NCNG operations, we consider the subset of these corresponding to LOSR+E instead.  In~\cite{Wolfe2019} it was shown that all LOSR operations can be generated by convex combinations of local deterministic operations.  These can be thought of in the following way.  Each party first does a deterministic function on their input, uses the result as the input to their device, then does a deterministic function on their input and the output of their device to form the final output.  All such operations correspond to local relabellings and local coarse-grainings. Note that deterministic classical distributions can be formed as a special case of coarse-graining (a local deterministic distribution is formed when each party coarse-grains all of their outputs to one output for each of their inputs). For the distributions we consider for our main conjectures, it turns out that all the coarse-grainings give rise to local distributions (c.f.\ Proposition~\ref{proposition: CG}), so, by considering mixing with deterministic classical distributions, local relabelling and exchange of parties we can cover all LOSR+E operations. We hence start by separately considering mixing with classical distributions, and then consider relabelling and exchange of parties.

\subsection{Mixing with classical distributions}
\label{ssec: locmix}

Analogously to the $(2,2,2,2)$ case, we can define a family of distributions $P_{\iso, \epsilon}^{(2,2,3,3)}=\epsilon P_{\NL}+(1-\epsilon)P_{\noise}^{(2,2,3,3)}$, where $P_{\noise}^{(2,2,3,3)}$ is the uniform distribution with all entries equal to $1/9$ and $\epsilon\in[0,1]$. This class of distributions is isotropic in the sense that the marginal distributions are uniform for each input of each party. In order to show the insufficiency of entropic inequalities, one needs to identify at least one non-classical distribution whose non-classicality cannot be detected through entropic inequalities. We will discuss this for the class $P_{\iso, \epsilon}^{(2,2,3,3)}$ and only consider distributions of this form in the rest of the chapter. Further, without loss of generality, we consider only the BC inequality $I_{\BC}^4\leq 0$ in what follows (by symmetry all the arguments will also hold for isotropic distributions corresponding to relabelled versions of $P_{\NL}$ and the corresponding BC inequalities).

\subsubsection{Using Shannon entropy}
\label{sssec: 2233shan}
In the entropic picture of the $(2,2,3,3)$ scenario, the 4 BC Inequalities~\eqref{eq: BCineqs} still hold (these are valid independently of the cardinality of the random variables). Again, analogously to the $(2,2,2,2)$ case, the maximally non-local distribution, $P_{\NL}$ has the same entropy vector as the classical distribution
\begin{equation}
    \label{eq: Pc2233}
    P_{\Cl}^{(2,2,3,3)}=
    \begin{array}{ |c|c|} 
 \hline
 \frac{1}{3} \quad 0 \quad 0 & \frac{1}{3} \quad 0 \quad 0\\ 
 0 \quad \frac{1}{3} \quad 0 & 0 \quad \frac{1}{3} \quad 0\\
 0 \quad 0 \quad \vphantom{\frac{1}{f}}\frac{1}{3} & 0 \quad 0 \quad \vphantom{\frac{1}{f}}\frac{1}{3}\\
 \hline
 \frac{1}{3} \quad 0 \quad 0 & \frac{1}{3} \quad 0 \quad 0\\ 
 0 \quad \frac{1}{3} \quad 0 & 0 \quad \frac{1}{3} \quad 0\\
 0 \quad 0 \quad \vphantom{\frac{1}{f}}\frac{1}{3} & 0 \quad 0 \quad \vphantom{\frac{1}{f}}\frac{1}{3}\\
 \hline
\end{array}
\end{equation}
(amongst others).
The distribution $P_{\NL}$ is hence entropically classical. However, in contrast to the $(2,2,2,2)$ case, we have evidence suggesting that there are values of $\epsilon$ for which $P_{\iso, \epsilon}^{(2,2,3,3)}$ is non-classical, but such that the mixture $vP_{\iso, \epsilon}^{(2,2,3,3)}+(1-v)P_{\Loc}$ is entropically classical for all classical distributions $P_{\Loc}$ and all $v\in [0,1]$, i.e., there exist non-classical distributions in the $(2,2,3,3)$ scenario for which mixing with classical distributions never gives rise to a non-classical entropy vector.

We begin by considering mixing $P_{\iso, \epsilon}^{(2,2,3,3)}$ with $P_{\Cl}^{(2,2,3,3)}$ in analogy with the treatment of the $(2,2,2,2)$ case. Although we have not fully proven this, from our numerics, this mixing appears to be optimal in the sense that when it does not allow for entropic violations, no other mixing can either. This allows us to identify a range of $\epsilon$ for which the mixture $P_{\iso, \epsilon}^{(2,2,3,3)}$ is non-classical, yet appears to remain entropically classical even when mixed with arbitrary classical distributions.  We begin with two propositions whose proofs can be found in Appendix~\ref{appendix: proofs}.

\begin{restatable}{proposition}{Propnonloc} \label{prop:nonloc} $P_{\iso, \epsilon}^{(2,2,3,3)}$ is non-classical if and only if $\epsilon>1/2$. Further, for $\epsilon\leq4/7$, $P_{\iso, \epsilon}^{(2,2,3,3)}$ satisfies all the CHSH-type inequalities, while for $\epsilon>4/7$ it violates at least one CHSH-type inequality.
\end{restatable}
By analogy with the $(2,2,2,2)$ case, we consider the violation of
$I_{\BC}^4$ attainable by mixing $P_{\iso, \epsilon}^{(2,2,3,3)}$ with
$P_{\Cl}^{(2,2,3,3)}$.  We find that for $\epsilon \in (1/2,4/7]$,
$P_{\iso, \epsilon}^{(2,2,3,3)}$ is non-classical but does not violate any
of the BC inequalities. As shown in the above proposition, these distributions are in the CHSH-classical polytope $\Pi_{\chsh}^{(2,2,3,3)}$ and hence lie in our region of interest.

\begin{restatable}{proposition}{PropBCshan}
\label{proposition: BCshan2233}
For $\epsilon \leq 4/7$,  $P_{\E,\epsilon,v}^{(2,2,3,3)}=vP_{\iso,
  \epsilon}^{(2,2,3,3)}+(1-v)P_{\Cl}^{(2,2,3,3)}$ does not violate any
of the BC entropic inequalities~\eqref{eq: BCineqs} for any $v \in [0,1]$. However, for all $\epsilon > 4/7$, there exists a $v\in [0,1]$ such that the entropic inequality $I_{\BC}^4\leq 0$ is violated by $P_{\E,\epsilon,v}^{(2,2,3,3)}$.  
\end{restatable}
The second part of this proposition already follows from Corollary~\ref{coroll:chaves} and Proposition~\ref{prop:nonloc}.

\begin{corollary}\label{cor:entcl}
For $\epsilon \leq 4/7$,  $P_{\E,\epsilon,v}^{(2,2,3,3)}=vP_{\iso,
  \epsilon}^{(2,2,3,3)}+(1-v)P_{\Cl}^{(2,2,3,3)}$ is entropically
classical for all $v\in[0,1]$.
\end{corollary}
\begin{proof}
This follows from Proposition~\ref{proposition: BCshan2233}
and Lemma~\ref{lem:BCcomplete}.
\end{proof}

While Proposition~\ref{proposition: BCshan2233} shows that the proof strategy of~\cite{Chaves13} does not directly generalize to all non-classical distributions in the $(2,2,3,3)$ case, it does not rule out the possibility that there may exist other mixings with classical distributions that could transform $P_{\iso, \epsilon}^{(2,2,3,3)}$ for $\epsilon \in (1/2,4/7]$ into a distribution that violates one of the BC inequalities. To investigate this, we can consider the polytope formed by mixing $P_{\iso, \epsilon}^{(2,2,3,3)}$ with classical distributions for some $\epsilon \leq 4/7$, i.e., the polytope $\Conv(\{P_{\iso, \epsilon}^{(2,2,3,3)}\}\bigcup\{P^{\Loc,k}\}_k)$, where $\{P^{\Loc,k}\}_k$ denotes the set of all (81) local deterministic vertices of the $(2,2,3,3)$ Bell-local polytope, $\{P_{\mathrm{iso},\epsilon}^{(2,2,3,3)}\}$ is a set with a single element and $\Conv()$ denotes the convex hull. We considered several values of $\epsilon \leq 4/7$ and numerically optimised the entropic expression $I_{\BC}^4$ over the non-classical region of this polytope\footnote{To restrict to the non-classical region, it is sufficient to mix with a subset of these 81 locals--- see Appendix~\ref{appendix: evidence} for more detail.} for each\footnote{Note however that it is enough that these results hold for one value of $\epsilon \in (1/2,4/7]$ in order to conclude that entropic inequalities are not sufficient for detecting non-classicality in this scenario.} and were unable to find violations. The optimization involves a non-linear objective function with linear constraints.  Hence, it is possible that the numerical approach missed the global optimum. Nevertheless, this is evidence for the following conjecture and is presented in more detail in Appendix~\ref{appendix: evidence}.  Proposition~\ref{proposition: BCshan2233} along with Figures~\ref{fig: 2233shantsal} and evidence in Appendix~\ref{appendix: evidence} also suggest this conjecture.

\begin{conjecture}
\label{conjecture: 2233shan}
Let $\epsilon\leq4/7$. For all mixtures of the distribution $P_{\iso,\epsilon}^{(2,2,3,3)}$ with classical distributions in the $(2,2,3,3)$ Bell scenario, the resulting distribution is entropically classical, i.e., all distributions in $\Conv(\{P_{\iso, \epsilon}^{(2,2,3,3)}\}\bigcup\{p^{\Loc,k}\}_k\})$ are entropically classical.
\end{conjecture}

The interesting cases of Conjecture~\ref{conjecture: 2233shan} are for non-classical distributions (i.e., for $\epsilon>1/2$), and the most relevant of these are those that can be achieved in quantum theory. The next remark addresses this case.
\begin{sloppypar}
\begin{remark}
  There exist non-classical quantum distributions that lie in the polytope $\Conv(\{P_{\iso, \epsilon=4/7}^{(2,2,3,3)}\}\bigcup\{P^{\Loc,k}\}_k\})$ and in this case, our results suggest that the non-classicality of the corresponding distributions cannot be detected through entropic inequalities. Let $P_{\QM}$ be the quantum distribution from~\cite[Equation~(14) with $d=3$]{CGLMP02} with Bob's inputs relabelled. This violates $I_{\BC}^4\leq 0$ through mixing with $P_{\Cl}^{(2,2,3,3)}$. Consider then mixing $P_{\QM}$ with uniform noise $P_{\noise}^{(2,2,3,3)}$ to obtain $P_{\mix}(u):=uP_{\QM}+(1-u)P_{\noise}^{(2,2,3,3)}$ ($u\in [0,1]$). We found that for some values of $u$ (e.g., $u=7/10$), $P_{\mix}(u)$ is non-classical. %and belongs to the polytope $\Conv(P_{\iso, \epsilon=4/7}^{(2,2,3,3)}\bigcup\{P^{\Loc,k}\}_k\})$. %% This is clear by definition of P_mix
  Further, $P_{\mix}(u)$ is quantum achievable since it can be obtained from the density operator $u\ket{\psi'}\bra{\psi'}+(1-u)\frac{\mathbb{I}}{9}$ (where $\ket{\psi'}$ is the two qutrit state producing $P_{\QM}$) and the same quantum measurements that produce $P_{\QM}$ from $\ket{\psi'}$.
\end{remark}
\end{sloppypar}
\subsubsection{Using Tsallis entropies}
\label{sssec: 2233tsal}

Given the results (Proposition~\ref{proposition: BCshan2233} and Conjecture~\ref{conjecture: 2233shan}) of the previous section for Shannon entropic inequalities, a natural question is whether other entropic measures can provide an advantage over the Shannon entropy in detecting non-classicality.  Here, we look at Tsallis entropies and find that similar results hold in this case as well, suggesting that Tsallis entropies also do not allow us to completely solve the problem.

The properties of monotonicity, strong-subadditivity and the chain rule are sufficient to derive the BC inequalities, which hence also hold for Tsallis entropy when $q\geq 1$. Other generalized entropies such as R\'enyi or min/max entropies do not satisfy one or more of these properties in general and it is not clear whether the analogues of~\eqref{eq: BCineqs} hold for these. In the R\'enyi case, we saw in Appendix~\ref{appendix: Renyiresults} that no analogous inequalities were obtained. Hence we focus on the case of Tsallis entropies in the remainder of this chapter, where for all $q\geq 1$ we have
%\begin{widetext}
\begin{equation}
\label{eq: BCineqsTs}
    \begin{split}
I_{\BC,q}^1=S_q(X_0Y_0)+S_q(X_1)+S_q(Y_1)-S_q(X_0Y_1)-S_q(X_1Y_0)-S_q(X_1Y_1)\leq 0\\
        I_{\BC,q}^2=S_q(X_0Y_1)+S_q(X_1)+S_q(Y_0)-S_q(X_0Y_0)-S_q(X_1Y_0)-S_q(X_1Y_1)\leq 0\\
        I_{\BC,q}^3=S_q(X_1Y_0)+S_q(X_0)+S_q(Y_1)-S_q(X_0Y_0)-S_q(X_0Y_1)-S_q(X_1Y_1)\leq 0\\
        I_{\BC,q}^4=S_q(X_1Y_1)+S_q(X_0)+S_q(Y_0)-S_q(X_0Y_0)-S_q(X_0Y_1)-S_q(X_1Y_0)\leq 0\\
    \end{split}
\end{equation}
%\end{widetext}
and we refer to these as the Tsallis entropic BC inequalities. Entropic classicality in Tsallis entropy space can be defined analogously to Definition~\ref{definition: entclass}, in terms of Tsallis entropy vectors over the set of variables $\mathcal{S}$ (Equation~\eqref{eq: coexisting}). We say that a distribution is \emph{$q$-entropically classical} if its entropy vector written in terms of the Tsallis entropy of order $q$ is achievable using a classical distribution.  In the case of the Shannon entropy, we used the fact (Lemma~\ref{lem:BCcomplete}) that the BC Inequalities~\eqref{eq: BCineqs} are known to be necessary and sufficient for entropic classicality for 2-input Bell scenarios \cite{Fritz13}. However, it is not clear if the result of \cite{Fritz13} generalises to Tsallis entropies for $q>1$. Thus our results in the Tsallis case are weaker than those for Shannon, being stated only for the BC inequalities.  We leave the generalization to arbitrary Tsallis entropic inequalities as an open problem.

\begin{restatable}{proposition}{PropBCtsal}
\label{proposition: BCtsal2233}
For $\epsilon \leq 4/7$,  $P_{\E,\epsilon,v}^{(2,2,3,3)}=vP_{\iso,
  \epsilon}^{(2,2,3,3)}+(1-v)P_{\Cl}^{(2,2,3,3)}$ does not violate
any of the Tsallis BC inequalities~\eqref{eq: BCineqsTs} for any $v \in
[0,1]$ and $q>1$. However, for $\epsilon > 4/7$ and every $q>1$, there always exists a $v\in [0,1]$ such that the entropic inequality $I_{\BC,q}^4\leq 0$ is violated by $P_{\E,\epsilon,v}^{(2,2,3,3)}$.  
\end{restatable}

We refer the reader to Appendix~\ref{appendix: proofs} for a proof of
this Proposition. To investigate the extension to other mixings, we tried the same computational procedure (see Appendix~\ref{appendix: evidence}) as in the Shannon case. We found no violation of the Tsallis entropic BC inequalities for any mixings of $P_{\iso, \epsilon}^{(2,2,3,3)}$ with classical distributions, for several values of $q>1$ and $\epsilon \in (1/2,4/7]$, leading to the following conjecture, which is similar to Conjecture~\ref{conjecture: 2233shan}.
\begin{conjecture}
\label{conjecture: 2233tsal}
Let $\epsilon\leq4/7$. For all mixtures of the distribution $P_{\iso, \epsilon}^{(2,2,3,3)}$ with classical distributions in the $(2,2,3,3)$ Bell scenario, the resulting distribution does not violate any of the Tsallis entropic BC inequalities for any $q>1$, i.e., all distributions in $\Conv(\{P_{\iso, \epsilon}^{(2,2,3,3)}\}\bigcup\{P^{\Loc,k}\}_k)$ satisfy the Tsallis entropic BC Inequalities~\eqref{eq: BCineqsTs} for all $q>1$.
\end{conjecture}

Figure~\ref{fig: 2233shantsal1}, shows the values of $\epsilon$ and $v$ for which $I_{\BC,q}^4$ (for $q=1, 2 , 8$) evaluated with $P_{\E,\epsilon,v}^{(2,2,3,3)}$ is positive, which is also suggestive of this conjecture.

\begin{remark}
  Any impossibility result for the $(2,2,3,3)$ scenario also holds in the $(2,2,d,d)$ case for $d>3$ because the former is always embedded in the latter i.e., every distribution in the $(2,2,3,3)$ scenario has a corresponding distribution in all the $(2,2,d,d)$ scenarios with $d>3$ which can be obtained by assigning a zero probability to the additional outcomes. Further, the entropic Inequalities~\eqref{eq: BCineqs} remain the same for all these scenarios as they do not depend on the cardinality of the random variables involved. Thus the existence non-classical distributions for the $d=3$ case whose non-classicality cannot be detected by entropic inequalities implies the same result for all $d>3$.
\end{remark}

\subsection{Beyond classical mixings}
\label{ssec: CGrelab}

So far, we only considered mixing with classical distributions to obtain entropic violations and gave evidence that this does not work for some non-classical distributions in the $(2,2,3,3)$ scenario. This motivates us to study whether using arbitrary LOSR+E operations allows us to detect this non-classicality through entropic violations. We show in this section that if Conjectures~\ref{conjecture: 2233shan} and~\ref{conjecture: 2233tsal} hold then they also hold for all LOSR+E operations. First consider the following example.

The maximum possible violation of the BC inequalities in the $(2,2,2,2)$ case is $I_{\BC}^4=\ln 2$~\cite{Chaves13}. This is derived by considering only Shannon inequalities within the coexisting sets, and the bound that the maximum entropy of a binary variable is $\ln 2$.  An analogous proof holds in the $(2,2,3,3)$ case, except that the bound is then $\ln 3$. In the former case we have $P_{\E,\epsilon=1,v=1/2}=\frac{1}{2}P_{\PR}+\frac{1}{2}P_{\Cl}$, which maximally violates $I_{\BC}^4\leq0$, while in the latter case, one such distribution is formed by $(P_{\NL}+P_{\NL}^*+P_{\Cl}^{(2,2,3,3)})/3$, where $P_{\NL}^*$ is another extremal non-local distribution:
\begin{equation}
    \label{eq: NL*}
    P_{\NL}^*=
    \begin{array}{ |c|c|} 
 \hline
 \frac{1}{3} \quad 0 \quad 0 & \frac{1}{3} \quad 0 \quad 0\\ 
 0 \quad \frac{1}{3} \quad 0 & 0 \quad \frac{1}{3} \quad 0\\
 0 \quad 0 \quad \vphantom{\frac{1}{f}}\frac{1}{3} & 0 \quad 0 \quad \vphantom{\frac{1}{f}}\frac{1}{3}\\
 \hline
 \frac{1}{3} \quad 0 \quad 0 & 0 \quad 0 \quad \frac{1}{3}\\ 
 0 \quad \frac{1}{3} \quad 0 & \frac{1}{3} \quad 0 \quad 0\\
 0 \quad 0 \quad \vphantom{\frac{1}{f}}\frac{1}{3} & 0 \quad \vphantom{\frac{1}{f}}\frac{1}{3} \quad 0\\
 \hline
\end{array}\,.
\end{equation}

\begin{figure}[t!]
\centering
\subfloat[]
{
    \includegraphics[scale=.5]{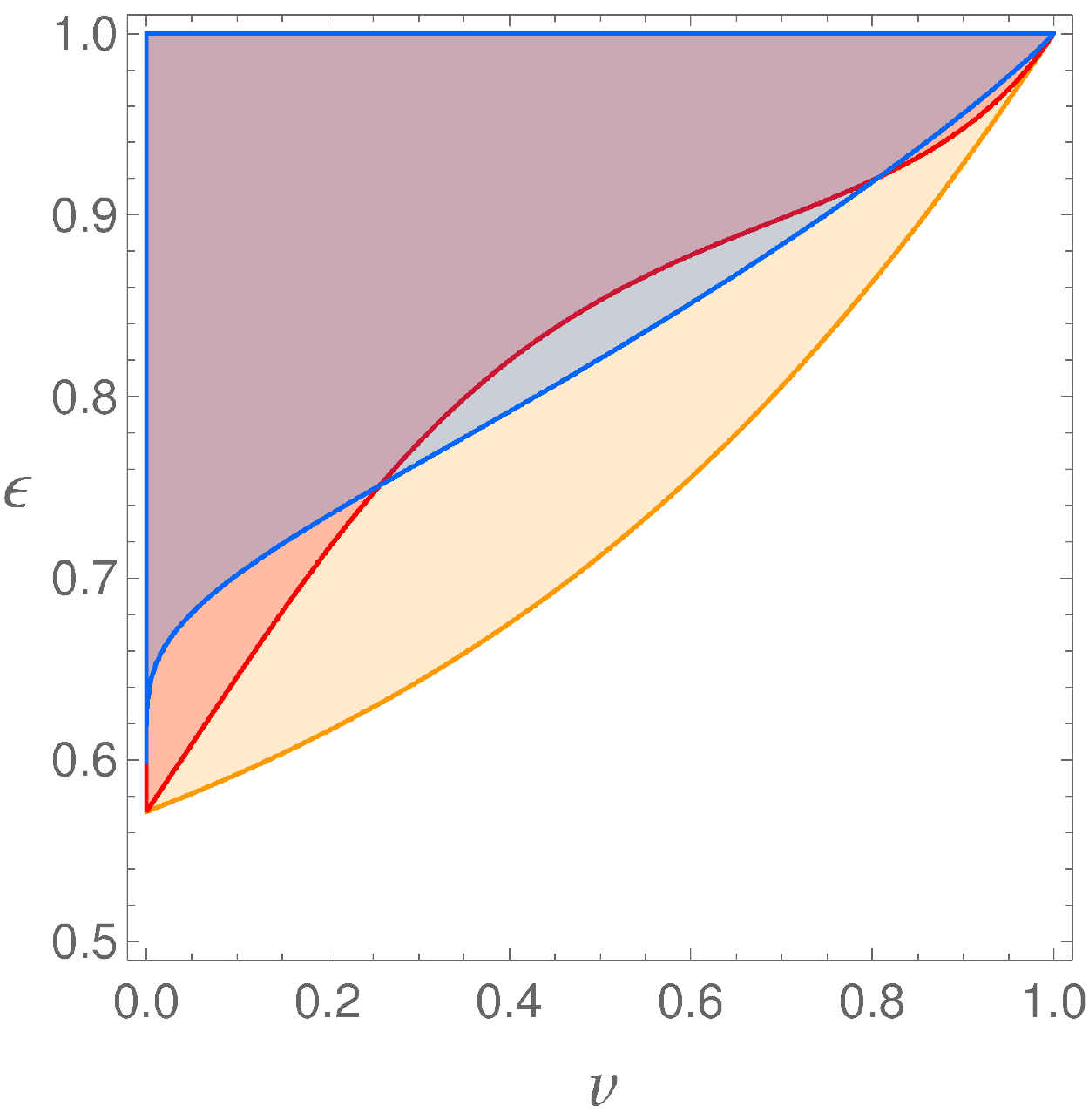}
    \label{fig: 2233shantsal1}
}\qquad
\subfloat[]
{
    \includegraphics[scale=.5]{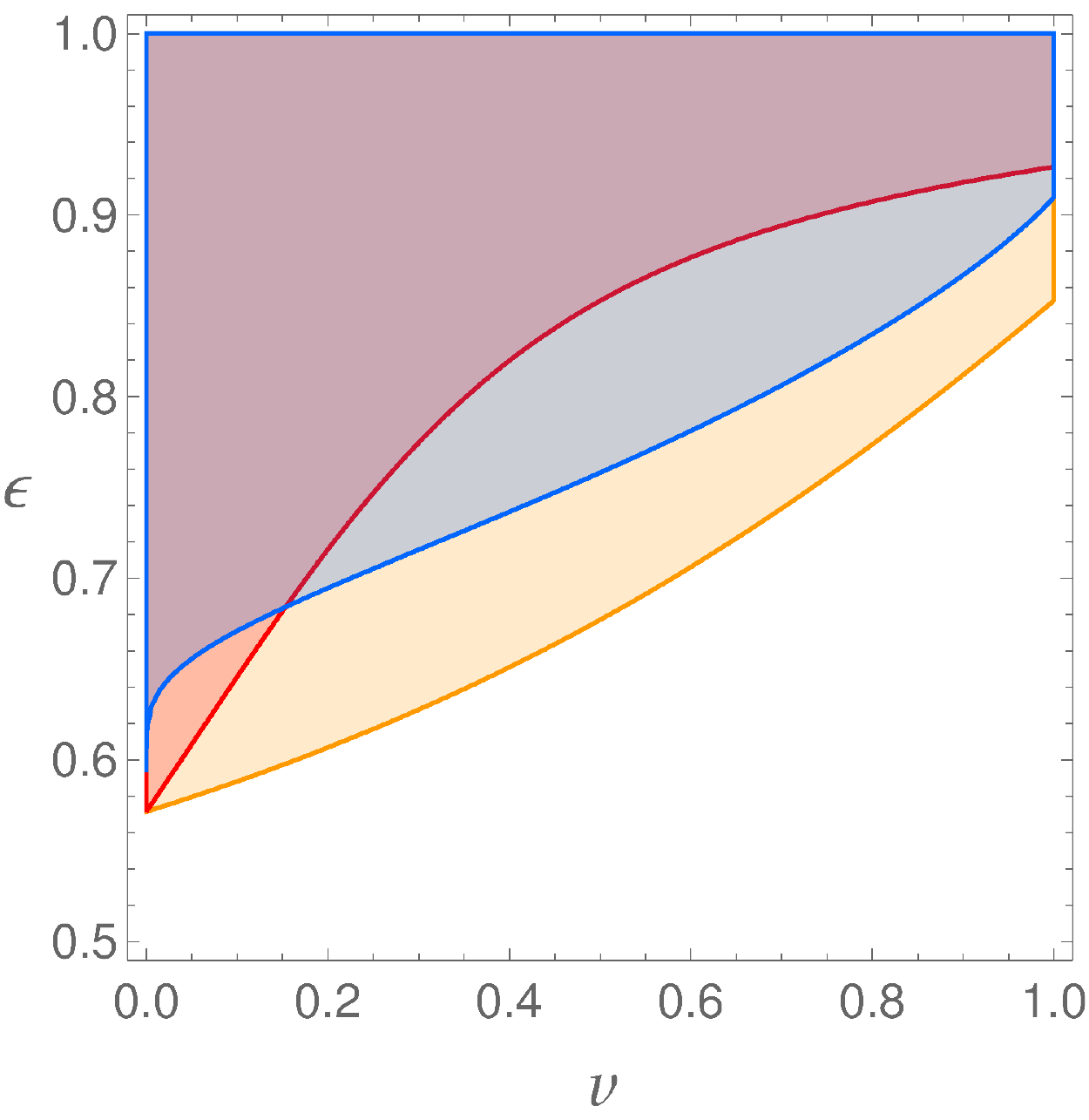}
    \label{fig: 2233shantsal2}
}
 \caption[Regions in the $v-\epsilon$ plane where Tsallis entropic BC inequalities can be violated.]{The regions in the $v-\epsilon$ plane where the Shannon entropic inequality $I_{\BC,1}^4:=I_{\BC}^4 \leq 0$ (blue), the Tsallis entropic inequality, $I_{\BC,q}^4 \leq 0$ for $q=2$ (orange) and $q=8$ (red) are violated by the distributions \text{(a)} $P_{\E,\epsilon,v}^{(2,2,3,3)}=vP_{\iso,\epsilon}^{(2,2,3,3)}+(1-v)P_{\Cl}^{(2,2,3,3)}$ \textbf{(b)} $\tilde{P}_{\E,\epsilon,v}^{(2,2,3,3)}=v\tilde{P}_{\iso,\epsilon}^{(2,2,3,3)}+(1-v)P_{\Cl}^{(2,2,3,3)}$. Here $P_{\iso,\epsilon}^{(2,2,3,3)}=\epsilon P_{\NL}+(1-\epsilon)P_{\noise}^{(2,2,3,3)}$ and $\tilde{P}_{\iso,\epsilon}^{(2,2,3,3)}=\epsilon (1/2P_{\NL}+1/2P_{\NL}^*)+(1-\epsilon)P_{\noise}^{(2,2,3,3)}$. For both (a) and (b) $I_{\BC,q}^4\leq 0$ is not violated when $\epsilon \leq 4/7 \approx 0.5714$ but for $\epsilon > 4/7$, there is a violation of this inequality for a larger range of $v$ values in the latter case, and also for a larger range in the $q=2$ case as compared to the other two cases.}
    \label{fig: 2233shantsal}
  \end{figure}

Since the equal mixture $(P_{\NL}+P_{\Cl}^{(2,2,3,3)})/2$ violates
$I_{\BC}^4\leq0$ non-maximally, one may be motivated to use the
non-local distribution $\tilde{P}_{\NL}=(P_{\NL}+P_{\NL}^*)/2$ in place
of $P_{\NL}$ in the
definition of $P_{\iso,\epsilon}^{(2,2,3,3)}$, i.e., to take
\begin{equation*}
    \tilde{P}_{\iso,\epsilon}^{(2,2,3,3)}=\epsilon \tilde{P}_{\NL}+(1-\epsilon)P_{\Cl}^{(2,2,3,3)}.
\end{equation*}
One could then consider whether for $\epsilon \in (1/2,4/7]$, $\tilde{P}_{\E,\epsilon,v}^{(2,2,3,3)}=v\tilde{P}_{\iso,\epsilon}^{(2,2,3,3)}+(1-v)P_{\Cl}^{(2,2,3,3)}$ violates $I_{\BC}^4\leq0$. Interestingly, while $\tilde{P}_{\E,\epsilon,v}^{(2,2,3,3)}$ violates $I_{\BC}^4\leq0$ for a larger range of $v$ values whenever $\epsilon> 4/7$, it does not give any violation (for any value of $v$) when $\epsilon \leq 4/7$, and Propositions~\ref{prop:nonloc} and \ref{proposition: BCshan2233} also hold if $\tilde{P}_{\iso,\epsilon}^{(2,2,3,3)}$ replaces $P_{\iso,\epsilon}^{(2,2,3,3)}$ (see Figure~\ref{fig: 2233shantsal2} for an illustration). The corresponding results also hold for the Tsallis case with $q>1$, i.e., Proposition~\ref{proposition: BCtsal2233} also holds with $\tilde{P}_{\iso,\epsilon}^{(2,2,3,3)}$ replacing $P_{\iso,\epsilon}^{(2,2,3,3)}$ (see also Figure~\ref{fig: 2233shantsal}). These suggest that mixing with relabellings in addition to mixing with classical distributions may also not help to violate entropic inequalities when $\epsilon \leq 4/7$.\bigskip

In the remainder of this section we consider the full set of LOSR+E operations. We first note that all input coarse-grainings of $P_{\iso,\epsilon}^{(2,2,3,3)}$ result in local distributions (there are no Bell inequalities if either of the parties have only one input). Similarly, considering output coarse-grainings, whenever three outcomes are mapped to one the resulting distribution is always classical because there are no Bell inequalities if one party always makes a fixed outcome for one of their inputs. We henceforth only consider coarse-grainings that take two outcomes to one. We can choose two of the three outcomes to combine into one for each party and each local input. For the four input choices $\{A=0,A=1,B=0,B=1\}$, there are 81, 108, 54 and 12 distinct coarse-grainings of this type when the outcomes of either 4, 3, 2 or 1 input choices are coarse-grained. Thus there are a total of $255$ coarse-grainings that remain.

If we apply all such coarse-grainings to $P_{\iso, \epsilon}^{(2,2,3,3)}$, this generates $255$ possible distributions that we denote $\{P^{\CG,i}_\epsilon\}_i$, $i\in[255]$. There are also $432$ distinct local relabellings of $P_{\iso,\epsilon}^{(2,2,3,3)}$, which we denote by $\{P^{\RL,j}_\epsilon\}_j$, $j\in[432]$ (this set includes $P_{\iso,\epsilon}^{(2,2,3,3)}=P^{\RL,1}_\epsilon$). Due to symmetries of $P_{\iso,\epsilon}^{(2,2,3,3)}$, it turns out that exchanging parties can be achieved through local relabellings for these distributions, so we do not need to separately consider the exchange in our results pertaining to $P_{\iso,\epsilon}^{(2,2,3,3)}$.  The set of all distributions that can be achieved through a convex mixture of $P_{\iso,\epsilon}^{(2,2,3,3)}$ with its coarse-grainings, relabellings and classical distributions is a convex polytope $\Pi_{\epsilon}$ for each $\epsilon$ and is the convex hull of these $255+432+81=768$ points, i.e.,
$$\Pi_{\epsilon}:=\Conv\left(\{P^{\CG,i}_\epsilon\}_i\bigcup\{P^{\RL,j}_\epsilon\}_j\bigcup\{P^{\Loc,k}\}_k\}\right)\,.$$
We present the results for the remaining coarse-grainings and relabellings separately below. Firstly, we show that the coarse-grainings $\{P^{\CG,i}_\epsilon\}_i$ of $P_{\iso, \epsilon}^{(2,2,3,3)}$ are classical if and only if $\epsilon\leq 4/7$.

\begin{restatable}{proposition}{PropCGs}
\label{proposition: CG}
The distribution $P^{\CG,i}_\epsilon$ is classical for all $i$ if and
only if $\epsilon \leq 4/7$.
\end{restatable}
This is intuitive because $P_{\iso,\epsilon}^{(2,2,3,3)}$ satisfies all the CHSH-type inequalities if and only if $\epsilon\leq4/7$. Since coarse-grainings cannot generate non-classicality and correspond to reducing the number of outcomes, and since $I_{2233}$ requires three outcomes, after coarse-graining the only relevant thing is whether there is a CHSH-violation.  A full proof is given in Appendix~\ref{appendix: proofs}.

Proposition~\ref{proposition: CG} implies that $\Pi_{\epsilon}=\Conv(\{P^{\RL,j}_\epsilon\}_j\bigcup\{P^{\Loc,k}\}_k)$ $\forall \epsilon \leq 4/7$, and that it is not necessary to consider coarse-grainings for such values of $\epsilon$. Our next results are that if Conjectures~\ref{conjecture: 2233shan} and~\ref{conjecture: 2233tsal} hold for $P_{\iso, \epsilon}^{(2,2,3,3)}$ for $\epsilon \leq 4/7$, then they continue to hold even when we consider arbitrary convex combinations with classical distributions and local relabellings of $P_{\iso, \epsilon}^{(2,2,3,3)}$.

\begin{restatable}{proposition}{ProprelabsA}
\label{proposition: relab1}
Let $\epsilon\leq4/7$. If Conjecture~\ref{conjecture: 2233shan} holds, then every distribution in $\Pi_{\epsilon}$ is Shannon entropically classical. 
\end{restatable}

\begin{restatable}{proposition}{ProprelabsB}
\label{proposition: relab2}
Let $\epsilon\leq4/7$. If Conjecture~\ref{conjecture: 2233tsal} holds, then every distribution in $\Pi_{\epsilon}$ satisfies the Tsallis entropic BC Inequalities~\eqref{eq: BCineqsTs} $\forall q>1$. 
\end{restatable}
These are proven in Appendix~\ref{appendix: proofs} and give the
following corollary.

\begin{corollary}\label{corollary: final}
  Let $\epsilon\leq4/7$. If Conjectures~\ref{conjecture: 2233shan} and~\ref{conjecture: 2233tsal} hold, then for any operation $\mathcal{O}$ in LOSR+E, $\mathcal{O}(P_{\iso, \epsilon}^{(2,2,3,3)})$ does not violate a Shannon or Tsallis ($q>1$) entropic BC inequality.
 \end{corollary}
  
\section{Discussion}
\label{sec: disc&conc}
We have provided evidence that there are distributions in the $(2,2,d,d)$ ($d\geq 3$) scenarios for which arbitrary LOSR+E operations do not enable detection of non-classicality with any Shannon entropic inequalities or Tsallis entropic BC inequalities. This is in contrast to the $(2,2,2,2)$ case~\cite{Chaves13}, where non-classicality can always be certified using the Shannon BC inequalities and LOSR post-processings. The region where the BC inequalities can fail to detect non-classicality would contain  non-classical distributions that satisfy all CHSH-type Bell inequalities. We found all the vertices that characterize this region for the $d=3$ case and identified distributions in it that violate BC entropic inequalities. Thus, the set of all non-classical distributions in the $(2,2,3,3)$ scenario that cannot be certified through entropic inequalities under LOSR+E post-processings is not characterized by the CHSH-type inequalities.

Although we considered LOSR+E operations, a natural next question is to what extent the results can be generalized to more general NCNG operations. In particular, there could be a non-linear NCNG map that allows the entropic BC inequalities to detect a wider range of non-classical distributions. An interesting open question would be to see whether for any non-classical distribution of the form $P_{\iso,\epsilon}^{(2,2,3,3)}$ with $1/2<\epsilon\leq 4/7$ (conjectured to be entropically classical with respect to LOSR+E), one of these more general operations would allow its non-classicality to be detected entropically. 

In~\cite{Wajs15} it was shown that (without post-processing) Tsallis entropic inequalities can detect non-classicality undetectable by Shannon entropic inequalities in the $(2,2,2,2)$ and $(2,2,3,3)$ Bell scenarios. In the presence of LOSR+E operations, we did not find any advantage of Tsallis entropies over the Shannon entropy in the $(2,2,3,3)$ Bell scenario. In fact, for some non-classical distributions such as that of Equation~\eqref{eq: countereg}, the Shannon entropic inequalities appear to give the largest violations among Tsallis entropies with $q\geq1$\footnote{Which corresponds to those for which the BC inequalities can be derived in the classical case}. On the other hand, for the family of distributions $P_{\iso,\epsilon}^{(2,2,3,3)}$, our results suggest that the range of $\epsilon$ for which post-processing via mixing with classical distributions enables non-classicality detection is the same for the Shannon as well as Tsallis entropic BC inequalities for $q>1$. However, when entropic detection of non-classicality is possible, using Tsallis entropy can make it easier to do this detection in the sense that there is a wider range of mixings that achieve this (see Figure~\ref{fig: 2233shantsal}).\footnote{As a specific example, consider the distribution $P_{\E,\epsilon=0.7,v=0.4}^{(2,2,3,3)}$. Figure~\ref{fig: 2233shantsal1} indicates that this distribution violates the Tsallis entropic inequality $I_{\BC,q=2}^4 \leq 0$ but does not violate the Shannon entropic inequality $I_{\BC}^4 \leq 0$. However, we can always further mix $P_{\E,\epsilon=0.7,v=0.4}^{(2,2,3,3)}$ with the classical distribution $P_{\Cl}^{(2,2,3,3)}$ to obtain $0.05P_{\E,\epsilon=0.7,v=0.4}^{(2,2,3,3)}+0.95P_{\Cl}^{(2,2,3,3)}=0.02P_{\iso, \epsilon}^{(2,2,3,3)}+0.98P_{\Cl}^{(2,2,3,3)}$ which violates the Shannon entropic inequality $I_{\BC}^4\leq0$. This is also in agreement with the results of \cite{Wajs15}, since when mixing is not considered, we also find examples where it is advantageous to use Tsallis entropy in the $(2,2,3,3)$ scenario.}

%The derivation of BC inequalities in both the Shannon~\eqref{eq: BCineqs} and Tsallis cases~\eqref{eq: BCineqsTs} is based on applying the strong subadditivity in the unconditional form, which does not hold for other entropy measures such as the R\'enyi, min and max entropies (c.f. Table~\ref{tab:ent_prop}). These inequalities are complete (in the Shannon case) as stated by Lemma~\ref{lem:BCcomplete} which means that if one were to apply the entropy vector method to the post-selected Bell causal structure with binary inputs, these would be the only non-trivial Shannon entropic inequalities obtained. In contrast, as mentioned in Appendix~\ref{appendix: Renyiresults} of Chapter~\ref{chapter: Tsallispaper}, no non-trivial R\'enyi entropic inequalities for the post-selected Bell causal structure was obtained even through the entropy vector method.

In conclusion, while the entropic approach for detecting non-classicality is useful in a number of scenarios, it is known to have disadvantages in others.  In particular, in the absence of post-selection we are not aware of any cases where entropic inequalities can be violated~\cite{Weilenmann16, Weilenmann2018, Vilasini2019}. We found that the entropic approach also suffers drawbacks in the presence of post-selection as it may fail to detect non-classicality under a natural class of post-processing operations, both in the case of Shannon and Tsallis entropies.  However, this method remains of use since in many cases non-classicality can be detected using it.

\section{Appendix}
\subsection{\texorpdfstring{$\mathcal{V}$}{V}-representation of the polytope \texorpdfstring{$\Pi_{\chsh}^{(2,2,3,3)}$}{the CHSH-classical region}}
\label{appendix: vertices}
Table~\ref{tab:vertices} enumerates the 47 extremal points of the polytope $\Pi_{\chsh}^{(2,2,3,3)}$ that saturate or violate the inequality ${I_{2233}\leq2}$~\eqref{eq: CGLMP} while satisfying all the CHSH-type inequalities in the $(2,2,3,3)$ Bell scenario. The first 17 of these are non-classical while the remaining 30 are local deterministic vertices. Due to Proposition~\ref{proposition: probspace} and the symmetries of the scenario, the remaining vertices of $\Pi_{\chsh}^{(2,2,3,3)}$ can be generated by taking the orbit of these vertices under local relabellings and exchange of parties. In Table~\ref{tab:vertices}, each extremal point is given by a single 36 dimensional vector which corresponds to writing the point in the notation explained in Section~\ref{sec: Probspace} (a $6\times 6$ matrix) and ``flattening'' it by writing one row after another in order.

\subsection{Evidence for Conjectures~\ref{conjecture: 2233shan} and \ref{conjecture: 2233tsal}}
\label{appendix: evidence}
In order to check for violations of the Shannon and Tsallis entropic inequalities $I_{\BC}^4\leq 0$ and $I_{\BC,q}^4\leq 0$ that could be obtained by mixing $P_{\iso, \epsilon}^{(2,2,3,3)}$ (Equation~\eqref{eq: NL3}) with arbitrary classical distributions, we maximized the left hand sides of these inequalities over the polytope $\Conv(\{P_{\iso, \epsilon}^{(2,2,3,3)}\}\bigcup\{P^{\Loc,k}\}_k)$ for some $\epsilon$ values in $(1/2,4/7]$ (such as $\epsilon=4/7,5/9$) numerically using {\sc Mathematica}. Note that this polytope contains the local polytope where by definition, entropic inequalities cannot be violated. Thus we can simplify the optimization and increase its reliability by only optimizing over the non-classical part of the polytope $\Conv(\{P_{\iso, \epsilon}^{(2,2,3,3)}\}\bigcup\{P^{\Loc,k}\}_k)$. We find this region as follows. For $1/2<\epsilon\leq 4/7$, we know from Proposition~\ref{prop:nonloc} that $P_{\iso, \epsilon}^{(2,2,3,3)}$ is non-classical but does not violate any of the CHSH inequalities, while it violates $I_{2233}\leq2$~\eqref{eq: CGLMP}. By Proposition~\ref{proposition: probspace}, this is the only Bell inequality that $P_{\iso, \epsilon}^{(2,2,3,3)}$ violates for this range of $\epsilon$. Thus, the non-classical part of the polytope $\Conv(\{P_{\iso,\epsilon}^{(2,2,3,3)}\}\bigcup\{P^{\Loc,k}\}_k)$ is the convex hull of $P_{\iso, \epsilon}^{(2,2,3,3)}$ and all the local deterministic points that satisfy $I_{2233}=2$. These are the 30 local deterministic points of Table~\ref{tab:vertices}. Hence we only need to optimise over convex combinations of $P_{\iso, \epsilon}^{(2,2,3,3)}$ with these 30 points and not all 81 local deterministic points, which reduces the size of the optimization (number of variables) and increases the chances of it being effective in detecting entropic violations if there are any.

\afterpage{\FloatBarrier}
{\setlength{\tabcolsep}{1em}
\begin{table}[t!]
    \centering
    \resizebox{\textwidth}{!}{%
    \begin{tabular}{|c|c|}
    \hline
    Number & Vertex\\
    \hline
       1  &  $\frac{1}{6}$(1, 1, 0, 1, 0, 1, 0, 1, 1, 1, 1, 0, 1, 0, 1, 0, 1, 1, 1, 0, 1, 0, 2, 0, 1, 1, 0, 0, 0, 2, 0, 1, 1, 2, 0, 0) \\
       2  &  $\frac{1}{6}$(1, 1, 0, 1, 0, 1, 0, 1, 1, 1, 1, 0,  1, 0, 1, 0, 1, 1, 2, 0, 0, 0, 1, 1, 0, 2, 0, 1, 0, 1, 0, 0, 2, 1, 1, 0)\\
       3  &  $\frac{1}{6}$(1, 1, 0, 2, 0, 0, 0, 1, 1, 0, 2, 0, 1, 0, 1, 0, 0, 2, 1, 0, 1, 0, 1, 1, 1, 1, 0, 1, 0, 1, 0, 1, 1, 1, 1, 0)\\
       4 &  $\frac{1}{6}$(2, 0, 0, 1, 0, 1, 0, 2, 0, 1, 1, 0, 0, 0, 2, 0, 1, 1, 1, 0, 1, 0, 1, 1, 1, 1, 0, 1, 0, 1, 0, 1, 1, 1, 1, 0)\\
       5  &  $\frac{1}{5}$(1, 0, 0, 1, 0, 0, 0, 1, 1, 1, 1, 0, 1, 0, 1, 0, 0, 2, 1, 0, 1, 0, 1, 1, 1, 1, 0, 1, 0, 1, 0, 0, 1, 1, 0, 0)\\
       6  &  $\frac{1}{5}$(1, 0, 0, 1, 0, 0, 0, 1, 1, 1, 1, 0, 1, 0, 1, 0, 1, 1, 1, 0, 0, 0, 1, 0, 1, 1, 0, 1, 0, 1, 0, 0, 2, 1, 1, 0)\\
       7  &  $\frac{1}{5}$(1, 0, 0, 1, 0, 0, 0, 1, 1, 1, 1, 0, 1, 0, 1, 0, 1, 1, 1, 0, 1, 0, 2, 0, 1, 1, 0, 1, 0, 1, 0, 0, 1, 1, 0, 0)\\
       8  &  $\frac{1}{5}$(1, 0, 0, 1, 0, 0, 0, 2, 0, 1, 1, 0, 1, 0, 1, 0, 1, 1, 1, 0, 0, 0, 1, 0, 1, 1, 0, 1, 0, 1, 0, 1, 1, 1, 1, 0)\\
       9  &  $\frac{1}{5}$(1, 1, 0, 1, 0, 1, 0, 1, 0, 0, 1, 0, 0, 0, 2, 0, 1, 1, 1, 0, 1, 0, 1, 1, 0, 1, 0, 0, 0, 1, 0, 1, 1, 1, 1, 0)\\
       10  &  $\frac{1}{5}$(1, 1, 0, 1, 0, 1, 0, 1, 0, 0, 1, 0, 1, 0, 1, 0, 1, 1, 1, 0, 0, 0, 1, 0, 1, 1, 0, 0, 0, 2, 0, 1, 1, 1, 1, 0)\\
       11 &  $\frac{1}{5}$(1, 1, 0, 1, 0, 1, 0, 1, 0, 0, 1, 0, 1, 0, 1, 0, 1, 1, 2, 0, 0, 0, 1, 1, 0, 1, 0, 0, 0, 1, 0, 1, 1, 1, 1, 0)\\
       12 &  $\frac{1}{5}$(1, 1, 0, 1, 0, 1, 0, 1, 1, 0, 2, 0, 0, 0, 1, 0, 0, 1, 1, 0, 1, 0, 1, 1, 0, 1, 0, 0, 0, 1, 0, 1, 1, 1, 1, 0)\\
       13 &  $\frac{1}{5}$(1, 1, 0, 1, 0, 1, 0, 1, 1, 1, 1, 0, 0, 0, 1, 0, 0, 1, 1, 0, 1, 0, 1, 1, 0, 1, 0, 0, 0,1, 0, 1, 1, 2, 0, 0)\\
       14 &  $\frac{1}{5}$(1, 1, 0, 1, 0, 1, 0, 1, 1, 1, 1, 0, 0, 0, 1, 0, 0, 1, 1, 0, 1, 0, 1, 1, 0, 2, 0, 1, 0, 1, 0, 0, 1, 1, 0, 0)\\
       15 &  $\frac{1}{5}$(1, 1, 0, 2, 0, 0, 0, 1, 0, 0, 1, 0, 1, 0, 1, 0, 1, 1, 1, 0, 0, 0, 1, 0, 1, 1, 0, 1, 0, 1, 0, 1, 1, 1, 1, 0)\\
       16 &  $\frac{1}{5}$(2, 0, 0, 1, 0, 1, 0, 1, 1, 1, 1, 0, 0, 0, 1, 0, 0, 1, 1, 0, 1, 0, 1, 1, 1, 1, 0, 1, 0, 1, 0, 0, 1, 1, 0, 0)\\
       17 &  $\frac{1}{9}$(2, 1, 0, 2, 0, 1, 0, 2, 1, 1, 2, 0, 1, 0, 2, 0, 1, 2, 2, 0, 1, 0, 2, 1, 1, 2, 0, 1, 0, 2, 0, 1, 2, 2, 1, 0)\\
       18 &  (0, 0, 0, 0, 0, 0, 0, 0, 0, 0, 0, 0, 0, 0, 1, 0, 0, 1, 0, 0, 0, 0, 0, 0, 0, 0, 0, 0, 0, 0, 0, 0, 1, 0, 0, 1)\\
       19 &  (0, 0, 0, 0, 0, 0, 0, 0, 0, 0, 0, 0, 0, 0, 1, 0, 0, 1, 0, 0, 0, 0, 0, 0, 0, 0, 1, 0, 0, 1, 0, 0, 0, 0, 0, 0)\\
       20 &  (0, 0, 0, 0, 0, 0, 0, 0, 0, 0, 0, 0, 0, 0, 1, 0, 0, 1, 0, 0, 1, 0, 0, 1, 0, 0, 0, 0, 0, 0, 0, 0, 0, 0, 0, 0)\\
       21 &  (0, 0, 0, 0, 0, 0, 0, 0, 0, 0, 0, 0, 0, 0, 1, 0, 1, 0, 0, 0, 0, 0, 0, 0, 0, 0, 0, 0, 0, 0, 0, 0, 1, 0, 1, 0)\\
       22 &  (0, 0, 0, 0, 0, 0, 0, 0, 0, 0, 0, 0, 0, 0, 1, 0, 1, 0, 0, 0, 1, 0, 1, 0, 0, 0, 0, 0, 0, 0, 0, 0, 0, 0, 0, 0)\\
       23 &  (0, 0, 0, 0, 0, 0, 0, 0, 0, 0, 0, 0, 0, 0, 1, 1, 0, 0, 0, 0, 0, 0, 0, 0, 0, 0, 0, 0, 0, 0, 0, 0, 1, 1, 0, 0)\\
       24 &  (0, 0, 0, 0, 0, 0, 0, 0, 0, 0, 0, 0, 0, 1, 0, 0, 0, 1, 0, 0, 0, 0, 0, 0, 0, 1, 0, 0, 0, 1, 0, 0, 0, 0, 0, 0)\\
       25 &  (0, 0, 0, 0, 0, 0, 0, 0, 0, 0, 0, 0, 1, 0, 0, 0, 0, 1, 0, 0, 0, 0, 0, 0, 1, 0, 0, 0, 0, 1, 0, 0, 0, 0, 0, 0)\\
       26 &  (0, 0, 0, 0, 0, 0, 0, 0, 0, 0, 0, 0, 1, 0, 0, 0, 0, 1, 1, 0, 0, 0, 0, 1, 0, 0, 0, 0, 0, 0, 0, 0, 0, 0, 0, 0)\\
       27 &  (0, 0, 0, 0, 0, 0, 0, 0, 0, 0, 0, 0, 1, 0, 0, 0, 1, 0, 1, 0, 0, 0, 1, 0, 0, 0, 0, 0, 0, 0, 0, 0, 0, 0, 0, 0)\\
       28 &  (0, 0, 0, 0, 0, 0, 0, 0, 1, 0, 1, 0, 0, 0, 0, 0, 0, 0, 0, 0, 0, 0, 0, 0, 0, 0, 0, 0, 0, 0, 0, 0, 1, 0, 1, 0)\\
       29 &  (0, 0, 0, 0, 0, 0, 0, 0, 1, 0, 1, 0, 0, 0, 0, 0, 0, 0, 0, 0, 1, 0, 1, 0, 0, 0, 0, 0, 0, 0, 0, 0, 0, 0, 0, 0)\\
       30 &  (0, 0, 0, 0, 0, 0, 0, 0, 1, 1, 0, 0, 0, 0, 0, 0, 0, 0, 0, 0, 0, 0, 0, 0, 0, 0, 0, 0, 0, 0, 0, 0, 1, 1, 0, 0)\\
       31 &  (0, 0, 0, 0, 0, 0, 0, 1, 0, 0, 0, 1, 0, 0, 0, 0, 0, 0, 0, 0, 0, 0, 0, 0, 0, 1, 0, 0, 0, 1, 0, 0, 0, 0, 0, 0)\\
       32 &  (0, 0, 0, 0, 0, 0, 0, 1, 0, 0, 1, 0, 0, 0, 0, 0, 0, 0, 0, 0, 0, 0, 0, 0, 0, 0, 0, 0, 0, 0, 0, 1, 0, 0, 1, 0)\\
       33 &  (0, 0, 0, 0, 0, 0, 0, 1, 0, 0, 1, 0, 0, 0, 0, 0, 0, 0, 0, 0, 0, 0, 0, 0, 0, 1, 0, 0, 1, 0, 0, 0, 0, 0, 0, 0)\\
       34 &  (0, 0, 0, 0, 0, 0, 0, 1, 0, 0, 1, 0, 0, 0, 0, 0, 0, 0, 0, 1, 0, 0, 1, 0, 0, 0, 0, 0, 0, 0, 0, 0, 0, 0, 0, 0)\\
       35 &  (0, 0, 0, 0, 0, 0, 0, 1, 0, 1, 0, 0, 0, 0, 0, 0, 0, 0, 0, 0, 0, 0, 0, 0, 0, 0, 0, 0, 0, 0, 0, 1, 0, 1, 0, 0)\\
       36 &  (0, 0, 0, 0, 0, 0, 0, 1, 0, 1, 0, 0, 0, 0, 0, 0, 0, 0, 0, 0, 0, 0, 0, 0, 0, 1, 0, 1, 0, 0, 0, 0, 0, 0, 0, 0)\\
       37 &  (0, 0, 0, 0, 0, 0, 1, 0, 0, 0, 1, 0, 0, 0, 0, 0, 0, 0, 1, 0, 0, 0, 1, 0, 0, 0, 0, 0, 0, 0, 0, 0, 0, 0, 0, 0)\\
       38 &  (0, 0, 1, 1, 0, 0, 0, 0, 0, 0, 0, 0, 0, 0, 0, 0, 0, 0, 0, 0, 0, 0, 0, 0, 0, 0, 0, 0, 0, 0, 0, 0, 1, 1, 0, 0)\\
       39 &  (0, 1, 0, 0, 0, 1, 0, 0, 0, 0, 0, 0, 0, 0, 0, 0, 0, 0, 0, 0, 0, 0, 0, 0, 0, 1, 0, 0, 0, 1, 0, 0, 0, 0, 0, 0)\\
       40 &  (0, 1, 0, 1, 0, 0, 0, 0, 0, 0, 0, 0, 0, 0, 0, 0, 0, 0, 0, 0, 0, 0, 0, 0, 0, 0, 0, 0, 0, 0, 0, 1, 0, 1, 0, 0)\\
       41 &  (0, 1, 0, 1, 0, 0, 0, 0, 0, 0, 0, 0, 0, 0, 0, 0, 0, 0, 0, 0, 0, 0, 0, 0, 0, 1, 0, 1, 0, 0, 0, 0, 0, 0, 0, 0)\\
       42 &  (1, 0, 0, 0, 0, 1, 0, 0, 0, 0, 0, 0, 0, 0, 0, 0, 0, 0, 0, 0, 0, 0, 0, 0, 1, 0, 0, 0, 0, 1, 0, 0, 0, 0, 0, 0)\\
       43 &  (1, 0, 0, 0, 0, 1, 0, 0, 0, 0, 0, 0, 0, 0, 0, 0, 0, 0, 1, 0, 0, 0, 0, 1, 0, 0, 0, 0, 0, 0, 0, 0, 0, 0, 0, 0)\\
       44 &  (1, 0, 0, 0, 1, 0, 0, 0, 0, 0, 0, 0, 0, 0, 0, 0, 0, 0, 1, 0, 0, 0, 1, 0, 0, 0, 0, 0, 0, 0, 0, 0, 0, 0, 0, 0)\\
       45 &  (1, 0, 0, 1, 0, 0, 0, 0, 0, 0, 0, 0, 0, 0, 0, 0, 0, 0, 0, 0, 0, 0, 0, 0, 0, 0, 0, 0, 0, 0, 1, 0, 0, 1, 0, 0)\\
       46 &  (1, 0, 0, 1, 0, 0, 0, 0, 0, 0, 0, 0, 0, 0, 0, 0, 0, 0, 0, 0, 0, 0, 0, 0, 1, 0, 0, 1, 0, 0, 0, 0, 0, 0, 0, 0)\\
       47 &  (1, 0, 0, 1, 0, 0, 0, 0, 0, 0, 0, 0, 0, 0, 0, 0, 0, 0, 1, 0, 0, 1, 0, 0, 0, 0, 0, 0, 0, 0, 0, 0, 0, 0, 0, 0)\\
       \hline
    \end{tabular}}
    \caption[Vertex representation of the polytope $\Pi_{\chsh}^{(2,2,3,3)}$.]{\textbf{The vertices of $\Pi_{\chsh}^{(2,2,3,3)}$ that saturate or violate the $I_{2233}$ Inequality~\eqref{eq: CGLMP}.} All the vertices of the polytope can be obtained from the vertices listed here through local relabellings or exchange of parties.}
    \label{tab:vertices}
\end{table}}

Performing the optimization as outlined above, we found the maximum value to always be non-positive for both Shannon case and the Tsallis case with $q=1.1,2,3,10,50$. We obtained similar results when taking other values of $\epsilon\leq 4/7$ in the distribution $P_{\iso, \epsilon}^{(2,2,3,3)}$ and also when considering the inequalities $I_{\BC,q}^i$ for $i\in \{1,2,3\}$. This suggests that no point in the polytope $\Conv(\{P_{\iso, \epsilon}^{(2,2,3,3)}\}\bigcup\{P^{\Loc,k}\}_k)$ violates any of the (Shannon or Tsallis entropic) BC inequalities for $\epsilon\leq 4/7$. For $\epsilon>4/7$, some mixing of $P_{\iso, \epsilon}^{(2,2,3,3)}$ with $P_{\Cl}^{(2,2,3,3)}$ gives a distribution that violates $I_{\BC,q}^4\leq0$ $\forall q\geq 1$ (c.f.\ Proposition~\ref{proposition: BCshan2233}). Note that in the Shannon ($q=1$) case, the range of values of the mixing parameter $v$ for which a violation can be found becomes arbitrarily small as $\epsilon$ approaches $4/7$ from above (see Figure~\ref{fig: 2233shantsal}). This limits the effectiveness of numerical tests for $q$ close to $1$.  For instance, in the Shannon case our program was not able to detect violations of $I^4_{\BC}\leq 0$ for $\epsilon<4.2/7$ (even though our analytic argument shows that these are present), while it was for $\epsilon\geq 4.2/7$. Similarly, for the $q=2$ Tsallis case, violations of $I^4_{\BC,2}\leq 0$ could be found for $\epsilon\geq 4.00001/7$, but not below. The reason for this difference is in line with what one might expect by comparing the plots in Figure~\ref{fig: 2233shantsal}, where for $\epsilon>4/7$ the range of values of $v$ for which a violation is possible is larger in the $q=2$ case. This highlights an advantage of using Tsallis entropy and gives us further confidence that for $1/2\leq\epsilon<4/7$ there is no violation.

However, because of the form of our objective function, the optimisation methods available do not guarantee to find the global maximum. Thus, our findings only constitute evidence for the conjectures and are not conclusive. In general, finding global optima for non-linear, non-convex/concave functions is an open question. A potential avenue for proving these conjectures is using DC (difference of convex) programming~\cite{Horst1999} since our objective function being a linear combination of entropies is a difference of convex functions.

\subsection{Proofs} \label{appendix: proofs}
For the proofs we need the concept of the local weight of a
non-signaling distribution~\cite{Zukowski99, Cope2019}
\begin{definition}
The \emph{local weight} of a non-signaling distribution $P_{XY|AB}$ is
the largest $\alpha\in[0,1]$ such that we can write
$$P_{XY|AB}=\alpha Q^{\Loc}_{XY|AB}+(1-\alpha)Q^{\NL}_{XY|AB}\,,$$
where $Q^{\Loc}_{XY|AB}$ is an arbitrary local distribution and
$Q^{\NL}_{XY|AB}$ is an arbitrary non-signaling distribution.  We
denote the local weight by $l(P_{XY|AB})$.
\end{definition}
The local weight of a distribution can be found by linear programming.

\Propnonloc*

\begin{proof}
The distribution $P_{\iso, \epsilon}^{(2,2,3,3)}=\epsilon P_{\NL}+(1-\epsilon)P_{\noise}^{(2,2,3,3)}$ can be written as follows.
\begin{equation}
\label{eq: pisoAB}
   P_{\iso, \epsilon}^{(2,2,3,3)}=
     \begin{tabular}{ |c|c|} 
 \hline
 $A$ \quad $B$ \quad $B$ & $A$ \quad $B$ \quad $B$\\ 
 $B$ \quad $A$ \quad $B$ & $B$ \quad $A$ \quad $B$\\
 $B$ \quad $B$ \quad $A$ & $B$ \quad $B$ \quad $A$\\
 \hline
$A$ \quad $B$ \quad $B$ & $B$ \quad $A$ \quad $B$\\ 
 $B$ \quad $A$ \quad $B$ & $B$ \quad $B$ \quad $A$\\
 $B$ \quad $B$ \quad $A$ & $A$ \quad $B$ \quad $B$\\
 \hline
\end{tabular}
\end{equation}
where $A=(2\epsilon+1)/9$ and $B=(1-\epsilon)/9$. We used the {\sc LPAssumptions} linear program solver~\cite{LPAssumptions} (see Appendix~\ref{appendix: LPA} of Chapter~\ref{chapter: Tsallispaper} for details) to find the local weight of $P_{\iso, \epsilon}^{(2,2,3,3)}$, as a function of $\epsilon$ to be
\begin{equation*}
    l(P_{\iso, \epsilon}^{(2,2,3,3)})=\begin{cases}
    1  &0\leq \epsilon \leq \frac{1}{2}\\
    2(1-\epsilon)  &\frac{1}{2}<\epsilon \leq 1
    \end{cases}
\end{equation*}
which establishes the first part of the claim.

The second part can be confirmed by computing the value of each CHSH-type quantity for the distribution $P_{\iso,\epsilon}^{(2,2,3,3)}$ and determining that each has a saturating $\epsilon$ of at most $4/7$.
\end{proof}

\PropBCshan*

\begin{proof}

Consider the function $f:(0,1)\times(0,1)\to\mathbb{R}$ given by
\begin{align*}
   f(\epsilon,v):=&3(3-2(1-\epsilon)v)\ln[3-2(1-\epsilon)v]+5(1-\epsilon)v\ln[(1-\epsilon)v]-(1+2\epsilon)v\ln[(1+2\epsilon)v]\\&-(3-(2+\epsilon)v)\ln[3-(2+\epsilon)v]-3\ln
9\, 
\end{align*}

where we implicitly extend the domain to $[0,1]\times[0,1]$ by taking
the relevant limit. The Shannon entropic expression $I_{\BC}^4(\epsilon,v)$ evaluated for the distribution $P_{\E,\epsilon,v}^{(2,2,3,3)}=vP_{\iso, \epsilon}^{(2,2,3,3)}+(1-v)P_{\Cl}^{(2,2,3,3)}$ (seen as a function of $\epsilon$ and $v$) is then given as $$I_{\BC}^4=\frac{1}{3\ln[2]}f(\epsilon,v)$$
Thus all the following arguments for $f(\epsilon,v)$ also hold for $I_{\BC}^4$. 
\par
  We first use that for $c>0$ and $a\in\mathbb{R}$ for sufficiently
  small $v$ we have
  $$\ln[c+av]=\ln[c]+\frac{av}{c}+O(v^2)\,.$$
Using this we can expand $f(\epsilon,v)$ for small $v$ as
\begin{equation}\label{eq:sm}
  f(\epsilon,v)=(4-7\epsilon)v\ln[v]-(4-7\epsilon)v(1+\ln[3])+v(5(1-\epsilon)\ln[1-\epsilon]-(1+2\epsilon)\ln[1+2\epsilon])+O(v^2)\,.
\end{equation}
Thus, since $\lim_{v\to0}v\ln[v]=0$ we have $\lim_{v\to0}f[\epsilon,v]=0$.

We also have
\begin{equation}
\label{eq:deriv}    
\begin{split}
  \frac{\partial}{\partial v}f(\epsilon,v)&=(4-7\epsilon)\ln\left[v\right]+5(1-\epsilon)\ln[1-\epsilon]-(1+2\epsilon)\ln[1+2\epsilon]\\
 & -6(1-\epsilon)\ln[3-2(1-\epsilon)v]+(2+\epsilon)\ln[3-(2+\epsilon)v]\,.
\end{split}
\end{equation}

Note that $5(1-\epsilon)\ln[1-\epsilon]\leq 0$,
$-(1+2\epsilon)\ln[1+2\epsilon]\leq 0$ $\forall \epsilon \in
[0,1]$. Further, since $3-2(1-\epsilon)v\geq 3-(2+\epsilon)v$ and both
terms are positive, $6(1-\epsilon)\geq(2+\epsilon)$ $\forall \epsilon
<4/7$, and using the fact that $\ln[]$ is an increasing function, we have $-6(1-\epsilon)\ln[3-2(1-\epsilon)v]+(2+\epsilon)\ln[3-(2+\epsilon)v]\leq 0$ $\forall \epsilon \in [0,4/7]$, $v\in [0,1]$. This in turn implies that
\begin{align}
  \frac{\partial}{\partial v}f(\epsilon,v)\leq(4-7\epsilon)\ln\left[v\right]\qquad \forall\ 0\leq \epsilon \leq 4/7, 0\leq v \leq 1 \nonumber
\end{align}
Hence we can conclude that for $\epsilon\leq4/7$,
$ \frac{\partial}{\partial v}f(\epsilon,v)<0$ for all $v\in [0,1]$. Thus, $f(\epsilon,v)$ is zero at $v=0$ and, for $\epsilon\leq4/7$, decreases
with $v$, implying that $f(\epsilon,v)\leq 0$ $\forall \epsilon\in
[0,4/7]$, $v\in [0,1]$. Note that $P_{\E,\epsilon,v}^{(2,2,3,3)}=vp_{\iso,
  \epsilon}^{(2,2,3,3)}+(1-v)p_C^{(2,2,3,3)}$ does not violate any of
the analogous inequalities  $I_{\BC}^i\leq 0$ 
for any $i\in \{1,2,3\}$, $\epsilon,v  \in [0,1]$.
This is because for this distribution, we always have
$H(X_0Y_0)=H(X_0Y_1)=H(X_1Y_0)$, $H(X_0)=H(X_1)=H(Y_0)=H(Y_1)$. Thus
all three inequalities $I_{\BC}^1\leq 0$, $I_{\BC}^2\leq 0$ and $I_{\BC}^3\leq 0$ reduce to
$H(X_1)+H(Y_1)-H(X_1Y_0)-H(X_1Y_1)\leq 0$, which is always satisfied since $H(X_1)\leq H(X_1Y_0)$ and $H(Y_1)\leq H(X_1Y_1)$ by the monotonicity of Shannon entropy.

Further, using the expression for the derivative of $f(\epsilon,v)$
with respect to $v$ in Equation~\eqref{eq:deriv}, we find that for
$\epsilon>4/7$,
$\lim_{v\to0} \frac{\partial}{\partial v}f(\epsilon,v)=\infty$. Thus,
since $f(\epsilon,v)=0$ for $v=0$, sufficiently close to $v=0$ there
exists a $v$ such that $f(\epsilon,v)>0$. This proves the claim.
\end{proof}

\PropBCtsal*
\begin{proof}
The Tsallis entropic expression $I_{\BC,q}^4(\epsilon,v)$ evaluated for the distribution $P_{\E,\epsilon,v}^{(2,2,3,3)}=vP_{\iso, \epsilon}^{(2,2,3,3)}+(1-v)P_{\Cl}^{(2,2,3,3)}$ (seen as a function of $q$, $\epsilon$ and $v$) is given as
\begin{align*}
    I_{\BC,q}^4&=\frac{1}{q-1}\Bigg[9\Bigg(\frac{3-2(1-\epsilon)v}{9}\Bigg)^q+15\Bigg(\frac{(1-\epsilon)v}{9}\Bigg)^q-\frac{6}{3^q}-3\Bigg(\frac{3-(2+\epsilon)v}{9}\Bigg)^q-3\Bigg(\frac{(1+2\epsilon)v}{9}\Bigg)^q\Bigg]\\&=:\frac{g(q,\epsilon,v)}{q-1}
\end{align*}
For $q>1$, the following arguments for $g(q,\epsilon,v)$ also hold for
$I_{\BC,q}^4$. Note that
\begin{align}
\begin{split}
\label{eq: derivTsal}
  \frac{\partial}{\partial v}g(q,\epsilon,v)&=\frac{q}{3^{2q-1}}\Big[-6(1-\epsilon)\Big(3-2(1-\epsilon)v\Big)^{q-1}+5(1-\epsilon)\Big((1-\epsilon)v\Big)^{q-1}\\&+(2+\epsilon)\Big(3-(2+\epsilon)v\Big)^{q-1}-(1+2\epsilon)\Big((1+2\epsilon)v\Big)^{q-1}\Big] \,.
  \end{split}
\end{align}

Then, since
\begin{align*}
  &6(1-\epsilon)\Big(3-2(1-\epsilon)v\Big)^{q-1}\geq (2+\epsilon)\Big(3-(2+\epsilon)v\Big)^{q-1}\qquad \text{and}\\ 
  &(1+2\epsilon)\Big((1+2\epsilon)v\Big)^{q-1}\geq 5(1-\epsilon)\Big((1-\epsilon)v\Big)^{q-1}
\end{align*}
hold for all $\epsilon\leq4/7$, $v\in[0,1]$ and $q>1$, we have 
$$\frac{\partial}{\partial v}g(q,\epsilon,v)\leq 0 \qquad \forall \epsilon\leq 4/7,\ v\in [0,1],\ q> 1.$$
Since $g(q,\epsilon,v=0)=0$, this implies that
$g(q,\epsilon,v)\leq 0$ $\forall \epsilon\leq 4/7, v\in [0,1], q>
1$. Hence, for $\epsilon\leq 4/7$ we cannot violate $I_{\BC,q}^4\leq0$
for any $v\in[0,1]$, $q>1$.

Analogously to the Shannon case, $P_{\E,\epsilon,v}^{(2,2,3,3)}=vP_{\iso,
  \epsilon}^{(2,2,3,3)}+(1-v)P_{\Cl}^{(2,2,3,3)}$ also does not
violate any of the inequalities $I_{\BC,q}^1\leq0$, $I_{\BC,q}^2\leq0$
or $I_{\BC,q}^3\leq0$ for any $\epsilon,v\in[0,1]$ and $q>1$ by the
same argument as in Proposition~\ref{proposition: BCshan2233}.

Further, Equation~\eqref{eq: derivTsal} implies $\lim_{v\to0}\frac{\partial}{\partial v}g(q,\epsilon,v)=\frac{q}{3^q}(7\epsilon-4)$ which is always positive for $\epsilon>4/7$. Since $g(q,\epsilon,v)=0$ for $v=0$, this allows us to conclude that for $\epsilon>4/7$, there
  exists a $v$ sufficiently close to $v=0$ such that $g(q,\epsilon,v)>0$. This establishes the claim.
\end{proof}

\PropCGs*
\begin{proof}
For the ``if'' part of the proof, we calculated all the 255
coarse-grainings of $P_{\iso, \epsilon=4/7}^{(2,2,3,3)}$ and used a
linear programming algorithm to find that all of these are local
(their local weight equals $1$). Since decreasing $\epsilon$ in $P_{\iso,
  \epsilon}^{(2,2,3,3)}$ cannot increase the violation of any (probability space) Bell inequality and neither can coarse-grainings, it follows that if $\epsilon\leq 4/7$, all coarse-grainings of $P_{\iso, \epsilon}^{(2,2,3,3)}$ are classical.

For the ``only if'' part we need to show that if all coarse-grainings of $P_{\iso, \epsilon}^{(2,2,3,3)}$ are classical then $\epsilon\leq 4/7$ or equivalently, if $\epsilon>4/7$, there exists at least one coarse-graining of $P_{\iso, \epsilon}^{(2,2,3,3)}$ that is non-classical. Consider the coarse-graining that involves combining the second output with the first for all 4 input choices. For $P_{\iso, \epsilon}^{(2,2,3,3)}$ as in Equation~\eqref{eq: pisoAB}, this coarse-graining gives
\begin{equation}
\label{eq: pisoCG}
\setlength{\tabcolsep}{0.7em}
    P_{\CG, \epsilon}^{(2,2,3,3)}=
     \begin{tabular}{ |ccc|ccc|} 
 \hline
 $2(A+B)$ & 0 & $2B$ & $2(A+B)$ & 0 & $2B$\\ 
 0 & 0 & 0 & 0 & 0 & 0\\
 $2B$ & 0 & $A$ & $2B$ & 0 & $A$\\
 \hline
 $2(A+B)$ & 0 & $2B$ & $3B+A$ & 0 & $A+B$\\ 
 0 & 0 & 0 & 0 & 0 & 0\\
 $2B$ & 0 & $A$ & $A+B$ & 0 & $B$\\
 \hline
\end{tabular}\,,
\end{equation}
where $A=(2\epsilon+1)/9$ and $B=(1-\epsilon)/9$.  The $I_{2233}$ value or left hand side of Equation~\eqref{eq: CGLMP} for this distribution is $9A-3B$. For this to be classical, we require that $9A-3B\leq 2$ which gives $\epsilon\leq 4/7$. Again using the {\sc LPAssumptions} linear program solver~\cite{LPAssumptions} we found the local weight of this distribution as a function of $\epsilon$, which gives the following.
\begin{equation*}
    l(P_{\CG, \epsilon}^{(2,2,3,3)})=\begin{cases}
    1,  &0\leq \epsilon \leq \frac{4}{7}.\\
    \frac{1}{9}(17-14\epsilon),  &\frac{4}{7}<\epsilon \leq 1.
    \end{cases}
\end{equation*}
In other words, if $\epsilon>4/7$, then the coarse-graining $P_{\CG,\epsilon}^{(2,2,3,3)}$ of $P_{\iso,\epsilon}^{(2,2,3,3)}$ violates the $I_{2233}$ Inequality~(\ref{eq: CGLMP}) and is hence non-classical. This concludes the proof.
\end{proof}

We prove the following two propositions together as they only differ in one step.
\ProprelabsA*
\ProprelabsB*
\begin{proof}
  If Conjectures~\ref{conjecture: 2233shan} and~\ref{conjecture: 2233tsal} hold, then for any $\epsilon\leq4/7$, $I_{\BC,q}^i\leq 0$ holds $\forall q\geq 1$, $\forall i \in \{1,2,3,4\}$ and for all distributions in the polytope $\Conv(\{P_{\iso, \epsilon}^{(2,2,3,3)}\}\bigcup\{P^{\Loc,k}\}_k\})$.\footnote{Note that $q=1$ covers the Shannon case.} We want to show that this implies the same for the larger polytope that comprises the convex hull of not just $P_{\iso, \epsilon}^{(2,2,3,3)}$ and local deterministic distributions $\{P^{\Loc,k}\}_k$, but also all the local relabellings of $P_{\iso, \epsilon}^{(2,2,3,3)}$, i.e., the polytope $\Pi_{\epsilon}=\Conv(\{P^{\RL,j}_\epsilon\}_j\bigcup\{P^{\Loc,k}\}_k)$. While we considered $P_{\iso, \epsilon}^{(2,2,3,3)}$ in Propositions~\ref{proposition: BCshan2233} and \ref{proposition: BCtsal2233} and Conjectures~\ref{conjecture: 2233shan} and~\ref{conjecture: 2233tsal}, due to symmetries, these also apply to every relabelling of $P_{\iso, \epsilon}^{(2,2,3,3)}$, i.e., $I_{\BC,q}^i\leq 0$ ($\forall i\in \{1,2,3,4\}$) throughout every polytope in the set $\{\Conv(\{P^{\RL,j}_\epsilon\}\bigcup\{P^{\Loc,k}\}_k)\}_j$ (where $j$ runs over all the local relabellings of $P_{\iso, \epsilon}^{(2,2,3,3)}$)\footnote{A note on notation: $\{P^{\RL,j}_\epsilon\}$ is a set comprising a single element, while $\{P^{\RL,j}_\epsilon\}_j$ is a set whose elements are the distributions for every $j$.}. This is because for every input-output relabelling of $P_{\iso, \epsilon}^{(2,2,3,3)}$, we can correspondingly relabel the inequality expression $I_{\BC,q}^4$ (for $q\geq 1$) and the same arguments as Propositions~\ref{proposition: BCshan2233} and \ref{proposition: BCtsal2233} again hold, and similarly Conjectures~\ref{conjecture: 2233shan} and~\ref{conjecture: 2233tsal} also follow for this case.\footnote{Note that output relabellings don't change the entropic expression but input relabellings (4 in number) can give either one of $I_{\BC,q}^i\leq 0$ for $i\in \{1,2,3,4\}$. Thus for an output relabelling of $P_{\iso, \epsilon}^{(2,2,3,3)}$, one can continue using $I_{\BC,q}^4$ in Propositions~\ref{proposition: BCshan2233} and \ref{proposition: BCtsal2233} and the following Conjectures while for input relabellings, one simply needs to relabel the inequalities accordingly and run the same arguments.} From this argument, it follows that if Conjectures~\ref{conjecture: 2233shan} and~\ref{conjecture: 2233tsal} hold, then $I_{\BC,q}^i\leq 0$ $\forall q \geq 1$, $\forall i \in \{1,2,3,4\}$ everywhere in the union of the polytopes, i.e., everywhere in $\bigcup_j\Conv(\{P^{\RL,j}_\epsilon\}\bigcup\{P^{\Loc,k}\}_k)$.

  To conclude the proof it remains to show that $\bigcup_j\Conv(\{P^{\RL,j}_\epsilon\}\bigcup\{P^{\Loc,k}\}_k)=\Pi_\epsilon$ $\forall \epsilon \leq 4/7$. This is established below. Then, Proposition~\ref{proposition: relab2} automatically follows while Proposition~\ref{proposition: relab1} follows from this and Lemma~\ref{lem:BCcomplete}.
\end{proof}

\begin{proposition}
\label{proposition: R1}
$P^j_{\mix,\epsilon}:= \frac{1}{2}P_{\iso, \epsilon}^{(2,2,3,3)}+\frac{1}{2}P^{\RL,j}_\epsilon$ is local $\forall j\neq 1$ if and only if $\epsilon \leq 4/7$.
\end{proposition}
\begin{proof}
For the ``if'' part of the proof, we used a linear program to confirm that $P^j_{\mix,\epsilon=4/7}$ is local\footnote{$j=1$ is excluded since $P^1_{\mix,\epsilon}=P_{\iso, \epsilon}^{(2,2,3,3)}$ which is non-classical for $\epsilon > 1/2$.} $\forall j\neq 1$. Since reducing $\epsilon$ in $P^j_{\mix,\epsilon}$ cannot decrease the local weight, this also holds for $\epsilon<4/7$.

The ``only if'' part of the proof is equivalent to showing that $\forall \epsilon > 4/7$, $\exists j$ such that $P^j_{\mix,\epsilon}$ is non-classical. Consider the particular local relabelling that corresponds to Alice swapping the outputs ``1'' and ``2'' only when her input is $A=1$. Let the distribution obtained by applying this relabelling to $P_{\iso, \epsilon}^{(2,2,3,3)}$ be $P^{\RL}_\epsilon$. Then $P_{\mix,\epsilon}=\frac{1}{2}P_{\iso, \epsilon}^{(2,2,3,3)}+\frac{1}{2}P^{\RL}_\epsilon$. More explicitly, 
\begin{equation}
    P^{\RL}_\epsilon=
     \begin{array}{|c|c|} 
 \hline
 A \quad B \quad B & A \quad B \quad B\\ 
 B \quad A \quad B & B \quad A \quad B\\
 B \quad B \quad A & B \quad B \quad A\\
 \hline
A \quad B \quad B & B \quad A \quad B\\ 
 B \quad B \quad A & A \quad B \quad B\\
  B \quad A \quad B & B \quad B \quad A\\
 \hline
\end{array}\qquad \text{and} \qquad p_{\mix,\epsilon}=
     \begin{array}{ |c|c|} 
 \hline
 A \quad B \quad B & A \quad B \quad B\\ 
 B \quad A \quad B & B \quad A \quad B\\
 B \quad B \quad A & B \quad B \quad A\\
 \hline
A \quad B \quad B & B \quad A \quad B\\ 
 B \quad * \quad * & * \quad B \quad *\\
 B \quad * \quad * & * \quad B \quad *\\
 \hline
\end{array}\quad,
\end{equation}
where $*=\frac{A+B}{2}$, $A=(2\epsilon+1)/9$ and $B=(1-\epsilon)/9$. Now consider the Bell inequality $\Tr(M^TP)\geq 1$ where $P$ is an arbitrary distribution in the $(2,2,3,3)$ scenario and $M$ is the following matrix.
\begin{equation}
    M:=
     \begin{array}{ |c|c|} 
 \hline
 0 \quad 1 \quad 1 & 0 \quad 1 \quad 1\\ 
 1 \quad 0 \quad 1 & 1 \quad 0 \quad 1\\
 1 \quad 1 \quad 1 & 0 \quad 0 \quad 0\\
 \hline
0 \quad 1 \quad 0 & 1 \quad 0 \quad 0\\ 
 1 \quad 0 \quad 0 & 0 \quad 1 \quad 0\\
 1 \quad 0 \quad 0 & 0 \quad 1 \quad 0\\
 \hline
\end{array}
\end{equation}
Then, the condition for $P_{\mix,\epsilon}$ to be non-classical with respect to this Bell inequality i.e., $\Tr(M^TP_{\mix,\epsilon})<1$ gives us $\epsilon>4/7$. Since $P^{\RL}_\epsilon$ is a local relabelling of $P_{\iso, \epsilon}^{(2,2,3,3)}$, there exists a $j$ such that $P^{\RL}_\epsilon=P^{\RL,j}_\epsilon$ and hence $P_{\mix,\epsilon}=P^j_{\mix,\epsilon}$. Thus we have shown that whenever $\epsilon>4/7$, $\exists j$ such that $P^j_{\mix,\epsilon}$ is non-classical and hence $P^j_{\mix,\epsilon}$ is local for all $j$ implies that $\epsilon \leq 4/7$ which concludes the proof.
\end{proof}

By symmetry, there is an analogue of Proposition~\ref{proposition: R1} with $P_{\iso, \epsilon}^{(2,2,3,3)}$ replaced by $P^{\RL,i}_\epsilon$ for any $i$, so we have the following corollary.
\begin{corollary}
\label{corollary: R1}
$\tilde{P}^{i,j}_{\mix,\epsilon}:= \frac{1}{2}P^{\RL,i}_\epsilon+\frac{1}{2}P^{\RL,j}_\epsilon$ is local $\forall j\neq i$ if and only if $\epsilon \leq 4/7$.
\end{corollary}

\begin{theorem}[Bemporad et.\ al 2001 \cite{Bemporad2001}]
\label{theorem: bemporad}
Let $\mathcal{P}$ and $\mathcal{Q}$ be polytopes with vertex sets $V$ and $W$ respectively i.e., $\mathcal{P}=\Conv(V)$ and $\mathcal{Q}=\Conv(W)$. Then $\mathcal{P}\bigcup\mathcal{Q}$ is convex if and only if the line-segment $[v,w]$ is contained in $\mathcal{P}\bigcup\mathcal{Q}$ $\forall v\in V$ and $w \in W$.
\end{theorem}

Let $\cP_j=\Conv(\{P^{\RL,j}_\epsilon\}\bigcup\{P^{\Loc,k}\}_k\})$ and $\mathbb{P}$ be the set of polytopes $\mathbb{P}:=\{\cP_j\}_j$. We use the above theorem to prove the final result that establishes Propositions~\ref{proposition: relab1} and~\ref{proposition: relab2}.
\begin{lemma}
Let $V_j$ be the vertex set of the polytope $\cP_j \in \mathbb{P}$ and $V:=\bigcup_j V_j$. Then, $\bigcup_{\cP_j\in \mathbb{P}}\cP_j=\Conv(\bigcup_i V_j)=\Conv(V)=\Pi_\epsilon$ $\forall \epsilon \leq 4/7$.
\end{lemma}
\begin{proof}
  By Corollary~\ref{corollary: R1}, for $i\neq j$ we have that $\frac{1}{2}\left(P^{\RL,i}_\epsilon+P^{\RL,j}_\epsilon\right) \in \mathcal{L}^{(2,2,3,3)}=\cP_i\bigcap\cP_j$ $\forall \epsilon \leq 4/7$. This implies that $\alpha P^{\RL,i}_\epsilon+(1-\alpha)P^{\RL,j}_\epsilon\in \cP_i\bigcup\cP_j$ $\forall \epsilon \leq 4/7$, $\alpha\in [0,1]$, i.e., the line segment $[P^{\RL,i}_\epsilon,P^{\RL,j}_\epsilon]$ is completely contained in the union of the corresponding polytopes $\cP_i\bigcup\cP_j$. Note that all other line segments $[v_i,v_j]$ with $v_i \in V_i$ and $v_j \in V_j$ are contained in $\cP_i\bigcup\cP_j$ by construction since at least one of $v_i$ or $v_j$ would be a local-deterministic vertex. Therefore, by Theorem~\ref{theorem: bemporad}, $\cP_i\bigcup\cP_j$ is convex $\forall i,j$ and $\epsilon \leq 4/7$. We can then apply Proposition~\ref{proposition: R1} and Theorem~\ref{theorem: bemporad} to the convex polytopes $\cP_i\bigcup\cP_j$ and $\cP_k$ and show that $\cP_i\bigcup\cP_j\bigcup\cP_k$ is convex $\forall i,j,k$ and $\epsilon \leq 4/7$. Proceeding in this way, we conclude that $\bigcup_{\cP_i\in \mathbb{P}}\cP_i$ is convex $\forall \epsilon \leq 4/7$, and hence $\bigcup_{\cP_j\in \mathbb{P}} \cP_j=\Conv(\bigcup_i V_j)=\Conv(V)=\Pi_\epsilon$.
\end{proof}

\chapter[Cyclic and fine-tuned causal models and compatibility with relativistic\texorpdfstring{\\}{lb} principles]{Cyclic and fine-tuned causal models and compatibility with relativistic principles}
\label{chapter: jammingpaper}

\lettrine[nindent=0em, slope=-.5em,lines=2]{T}{he} notion of causation is closely tied with that of space-time, but how are the two related? Causation can be modelled operationally, without any reference to space-time, based only on observed correlations and interventions on systems \cite{Pearl2009}. It can be represented through directed graphs that are in principle independent from the partial order imposed by space-time. That the arrows of this directed graph are compatible with the light cone structure of space-time is an empirical observation that supports the notion of relativistic causality. For example, if an external intervention on a variable $A$ produces an observable change in another variable $B$, then one would say that $A$ affects $B$ and call $A$ a cause of $B$. Assigning space-time locations to the variables and requiring the effect $B$ to always be in the future light cone of the cause $A$ makes this causal relationship compatible with the partial order of Minkowski space-time. In general, a variable may jointly affect a set of variables without affecting individual variables in the set. Such scenarios correspond to causal models where the correlations are \emph{fine-tuned} to hide certain causal influences. For certain embeddings of such causal models in space-time (such as the \emph{jamming} scenario considered in \cite{Grunhaus1996, Horodecki2019}), this can lead to superluminal influence without superluminal signalling, or closed timelike curves that do not lead to an observable signalling to the past, some of these are strictly post-quantum features that do not arise in standard quantum theory. For a deeper understanding of causality in quantum and post-quantum theories, it is thus important to disentangle the two notions of causation-- the operational one and the relativistic one, and to characterise the relationship(s) between them in these general scenarios. A rigorous framework for causally modelling such scenarios and formalising their relationship with a space-time structure is not available. Such an analysis can deepen our understanding of the principles of causality other than broad relativistic principles (such as no superluminal signalling) that apply in the quantum world. This chapter is based on joint work with Roger Colbeck that we hope to turn into a paper soon.

%More specifically, our goal here is to develop a causal modelling approach to analyse such fine-tuned and cyclic causal scenarios with respect to an underlying space-time structure. We wish to understand under what conditions such causal models can/cannot lead to violations of relativistic principles and also to gain some perspective on whether scenarios that don't lead to such violations make sense physically. \cite{Horodecki2019} claim to derive necessary and sufficient conditions for satisfying relativistic causality in multipartite Bell scenarios. However an overarching concern regarding these claims is that they are derived from definitions (such as ``relativistic causality'' or ``no causal loops'') that are not based on a mathematical rigorous framework, but only stated in words. Such definitions are particularly tricky in the presence of fine-tuning (which as we will see, is necessitated by the examples considered in \cite{Horodecki2019}) where a distinction needs to be made between causal loops and influences that are operationally detectable vs those that are not. This can be characterised through a formal framework  for cyclic and fine-tuned causal models in non-classical theories. 

\section{Summary of contributions: in words and in poetry}
We develop an operational framework for analysing cyclic and fine-tuned causal models in non-classical theories, and their compatibility with a space-time structure. This in particular includes the jamming scenarios envisaged in \cite{Grunhaus1996, Horodecki2019} where a party superluminally influences correlations between two other parties. We present the framework in two parts, the first concerns causal models (Section~\ref{sec: causmod}) and the second characterises the embedding of these causal models in a space-time structure (Section~\ref{sec:spacetime}). In Sections~\ref{sec: nosig} and~\ref{sec: noloops}, we derive necessary and sufficient conditions for the causal model to be compatibly embedded in a space-time structure and to rule out operationally detectable causal loops.
Following this, in Section~\ref{sec: jamming}, we analyse the jamming scenario \cite{Grunhaus1996, Horodecki2019} using causal modelling. We prove in Theorem~\ref{theorem: jamming} that if the shared joint system is classical and accessible to the parties, then the jamming scenario can indeed lead to superluminal signalling, contrary to what is claimed in \cite{Grunhaus1996, Horodecki2019}. In Section~\ref{sec: problems}, we analyse the results of \cite{Horodecki2019} regarding such scenarios, providing counter examples to some of their claims, and discussing how our results provide possible resolutions. Keeping in theme with the many cycles and paradoxes that the reader will encounter in this thesis, we present them with a mildly self-referential poem before looping back into the technicalities.
%A main task we wish to complete before submitting this work to the arXiv or a journal is to better formalise Pearl's do-calculus in the general scenarios considered here-- non-classical causal models involving directed cycles and fine-tuned correlations. This is required for defining interventions in our framework, which in turn provides a criterion for operationally identifying causal influences, and hence causal loops. Currently, we are able to define the interventions, derive Pearl's rules of do-calclulus under certain minimal assumptions and calculate all the pre and post-intervention distributions for certain examples. Ideally, we would like our framework to provide a general procedure for calculating these distributions in a general class of cyclic causal structures involving non-classical nodes and fine-tuned correlations. We note that the full do-calculus in this manner has been worked out for classical cyclic causal models which are faithful, but not for the quantum/post-quantum and unfaithful cases.

\textit{What is causation?\\
Can we always discover its presence?\\
If it is not mere correlation,\\ 
Then what is its essence?}

\textit{What is it that orders the cause \\
Before that what we dub the effect?\\
What brings order to the chaos\\
Telling us what we can or cannot \href{definition: affects}{affect}?}

\textit{Time\footnote{What is time, \\
does it really ``exist’’?\\
That is a question for another time,\\
Another poem, and I must resist.} gives us a direction ``forward’’.\\
To pose questions without already knowing the answer.\\
To drive causation through its invisible arrow,\\
A rule that seems doomed to be followed.}

\textit{Causality and space-time,\\
At first, we disconnect.\\
By actions and observable changes, \\
We \href{sec: causmod}{model cause} and effect.}

\textit{With space-time out of the way, we abstract away,\\
Allowing \href{definition: causalloops}{causal loops}, and hidden influences\\
That from our observations may evade.\\
But we’d like to avoid paradoxes\\
That put our grandfathers’ lives at stake.\\
Requiring a joint probability distribution\\
Keeps these inconsistencies at bay.}

\textit{Bringing one foot back to reality,\\
We embed the causal model in \href{sec:spacetime}{space-time}\\
Derive conditions for its \href{theorem:poset}{compatibility}.\\
Align the actions and changes\\
With the past and future,\\
But this doesn’t suffice to exorcise \\
All the loops we set afoot earlier.}

\textit{Some lurk, beyond the observable realm.\\
Finely tuned parameters that don’t reveal them.\\
We can only rule out loops that manifest\\
In observations that we can operationally test.\\
So we \href{sec: problems}{critically analyse} a previous claim \cite{Horodecki2019}\\
That specifies how all the loops can be tamed.}

\textit{A \href{sec: discJamming}{plethora of questions} begging for answers,\\
Some to be addressed before publishing.\\
Therefore, much like the work presented here,\\
This poem too needs some fine-tuning.}

% {\raggedleft \textit{-Vilasini Venkatesh (2020)}\par}

\section{Motivation for analysing cyclic and fine-tuned causal models}
\label{sec:motiveg}
A ubiquitous assumption made in majority of the literature on causal modelling is that of \emph{faithfulness}, which we briefly discussed in Section~\ref{ssec: classCM} (in particular, Example~\ref{example: fine-tune1}). This corresponds to the requirement that all the (conditional) independences observed in the correlations arising from a causal structure must be a consequence of certain graph separation properties (i.e., d-separation relations, Definition~\ref{definition:dsep}) of that causal structure. In \emph{unfaithful} causal models, the causal parameters are fine-tuned such that some causal influences exactly cancel out to yield additional independences beyond what is implied by d-separation. Fine-tuning creates problems for causal inference since it leads to correlations that do not faithfully reflect the causal relations and it is considered an undesirable property of causal models and often avoided in the literature, also on the grounds that fine-tuned causal models constitute a set of measure-zero. In fact quantum correlations in the Bell causal structure~\ref{fig: Bell2} can be explained using classical causal models if we allow for additional, fine-tuned causal influences (e.g., from $A$ to $X$, see Figure~\ref{fig: Qcycle}) \cite{Wood2015}, but a faithful explanation using quantum causal models is often preferred.

However, there are a number of examples, as we will see below, which necessitate a fine-tuned explanation irrespective of whether the causal structure is classical and non-classical. These include certain everyday scenarios, cryptographic protocols as well as those arising in certain post-quantum theory. Moreover, another common assumption in the causality literature is that the causal structure is acyclic, this is justifiably so since our observations suggest that causal influences only propagate in one direction. Nevertheless, allowing fine-tuned causal influences makes possible cyclic causal structures that are compatible with minimal notions of relativistic causality, such as the impossibility of signalling superluminally at the observed level. Cyclic causal models have also been employed in the classical literature for describing systems with feedback loops \cite{Pearl2013, Forre2017}. 

\subsection{Friedman's thermostat and one-time pad}

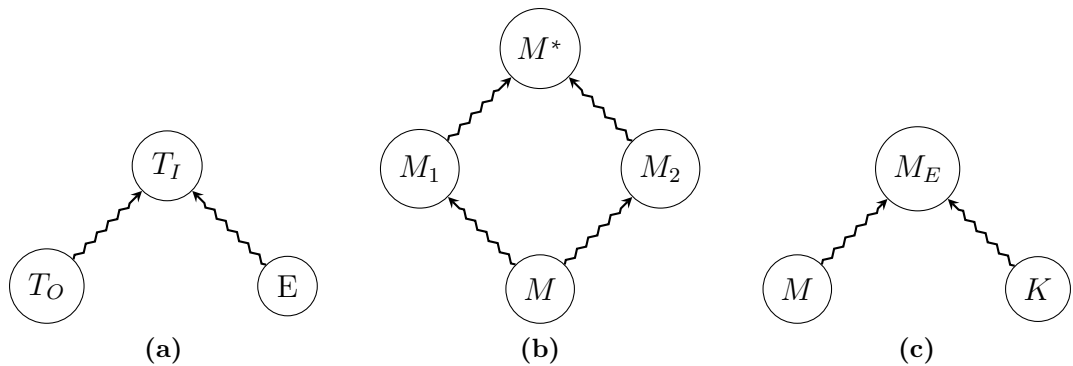
\begin{figure}[t!]
 \centering\subfloat[\label{fig: thermostat}]{\begin{tikzpicture}[scale=0.8] \node[shape=circle,draw=black] (I) at (0,0) {$T_I$};
    \node[shape=circle,draw=black] (O) at (-2,-2) {$T_O$};
    \node[shape=circle,draw=black] (E) at (2,-2) {E};  \draw [decorate, decoration={zigzag, segment length=+6pt, amplitude=+.95pt,post length=+4pt}, arrows={-stealth}, thick] (O) -- (I);

     \draw[decorate, decoration={zigzag, segment length=+6pt, amplitude=+.95pt,post length=+4pt}, arrows={-stealth}, thick] (E) -- (I); %\draw [decorate, decoration={zigzag, segment length=+6pt, amplitude=+.95pt,post length=+4pt}, arrows={-stealth}, thick] (I.south east) to [out=260,in=170]  (E.north west);
     \end{tikzpicture}}\qquad
 \subfloat[\label{fig: traitor}]{\begin{tikzpicture}[scale=0.8]
 \node[shape=circle,draw=black] (M) at (0,-4) {$M$};\node[shape=circle,draw=black] (Mp) at (0,0) {$M^*$};
    \node[shape=circle,draw=black] (M1) at (-2,-2) {$M_1$};
    \node[shape=circle,draw=black] (M2) at (2,-2) {$M_2$};  \draw [decorate, decoration={zigzag, segment length=+6pt, amplitude=+.95pt,post length=+4pt}, arrows={-stealth}, thick] (M1) --(Mp); 
     \draw [decorate, decoration={zigzag, segment length=+6pt, amplitude=+.95pt,post length=+4pt}, arrows={-stealth}, thick] (M2) -- (Mp);\draw[decorate, decoration={zigzag, segment length=+6pt, amplitude=+.95pt,post length=+4pt}, arrows={-stealth}, thick] (M) -- (M1); \draw[decorate, decoration={zigzag, segment length=+6pt, amplitude=+.95pt,post length=+4pt}, arrows={-stealth}, thick] (M) -- (M2);\end{tikzpicture}}\qquad \subfloat[\label{fig: otp}]{\begin{tikzpicture}[scale=0.8] \node[shape=circle,draw=black] (E) at (0,0) {$M_E$};
    \node[shape=circle,draw=black] (M) at (-2,-2) {$M$};
    \node[shape=circle,draw=black] (K) at (2,-2) {$K$};  \draw[decorate, decoration={zigzag, segment length=+6pt, amplitude=+.95pt,post length=+4pt}, arrows={-stealth}, thick] (M) -- (E);
     \draw[decorate, decoration={zigzag, segment length=+6pt, amplitude=+.95pt,post length=+4pt}, arrows={-stealth}, thick] (K) -- (E);\end{tikzpicture}}
    \caption[Motivating examples for cyclic and fine-tuned causal models]{\textbf{Causal structures for the motivating examples described in the main text: } \textbf{(a)} Friedman's thermostat \textbf{(b)} Traitorous Lieutenant \textbf{(c)} One-time pad. In this chapter, we use squiggly arrows $\longrsquigarrow$ to denote causal influence, as this will later be classified into solid and dashed arrows. Note that there may be additional causal influences, for example there can be a direct influence of the outside temperature $T_O$ and/or the inside temperature $T_I$ on the energy consumption $E$ in (a), the latter would make it a cyclic causal model. Further, in examples like (b), we will later see that an additional common cause between $M_1$ and $M_2$ will be required to fully explain the correlations (c.f. Figure~\ref{fig: jamming}).}
    \label{fig: motiv_eg}
\end{figure}

Consider a house with an ideal thermostat. Such a thermostat would maintain a constant inside temperature $T_I$ throughout the year by adjusting the energy consumption $E$ in accordance with the outside temperature $T_O$. An individual who does not know how a thermostat works might conclude that $T_O$ and $E$ which are correlated have a causal relationship between each other while the indoor temperature $T_I$ is causally independent of everything else. However, an engineer who is more well-versed with the workings of a thermostat knows that both $T_O$ and $E$ exert a causal influence on $T_I$, and that these influences must perfectly cancel each other out for the thermostat to function ideally. The causal model in this case is fine-tuned since the independence of $T_I$ from $T_O$ and $E$ does not correspond to a d-separation relation in the causal structure (Figure~\ref{fig: thermostat}). This thermostat analogy which is attributed to Milton Friedman \cite{Friedman2003}, can be extended to a number of other scenarios such as the effect of fiscal and monetary policies on economic growth \cite{Rowe2009}, or physical systems where several forces exactly balance out. 

In cryptographic settings, examples that necessitate fine-tuning include the one-time pad and the ``traitorous lieutenant problem'' \cite{Brul2017}. Consider a general who wishes to relay an important secret message $M$ to an ally and has two lieutenants available as messengers, but one of them is a traitor who might leak the message to enemies. Consider for simplicity that $M$ is a single bit. The general could then adopt the following strategy: Depending on $M=0$ or $M=1$, two uniformly distributed bits $M_1$ and $M_2$ are generated such that $M_1=M_2$ or $M_1\neq M_2$. $M_1$ is given to the first and $M_2$ to the second lieutenant to relay to the ally. Then the ally would receive $M_1$ and $M_2$ and can simply use modulo-2 addition $\oplus$ to obtain $M^*$ which would be identical to the original message $M^*=M=M_1\oplus M_2$ (Figure~\ref{fig: traitor}). More importantly, the individual messages $M_1$ and $M_2$ contain no information about $M$ and hence neither lieutenant has any information about the secret message. A similar protocol underlies the one-time pad where a message $M$ is encrypted using a secret key $K$ (both binary for this example) to produce an encrypted message $M_E=M\oplus K$ which can be sent through a public channel as it will carry no information about the original message $M$ if the key $K$ is uniformly distributed and is kept private. Only a receiver of $M_E$ who knows the key $K$ can decrypt the message $M=M_E\oplus K$ (Figure~\ref{fig: otp}). Hence fine-tuning of causal influences i.e., causation in the absence of correlation, is key to the working of such cryptographic protocols. 

Further, cyclic causal models have been analysed in the classical literature \cite{Pearl2013, Forre2017} for the purpose of describing complex systems involving feedback loops. Note that the cyclic dependences here do not correspond to closed time-like curves since the variables under question are considered over a period of time--- e.g., demand at time $t_1$ influences the price at time $t_2>t_1$, which in turn influences the demand at time $t_3>t_2$. Therefore in order to characterise genuine closed time-like curves one must consider not only the pattern of causal influences, but also how the relevant variables are embedded in a space-time structure.

\subsection{Jamming non-local correlations}
\begin{figure}[t!]
\centering
	\subfloat[]{	\centering
	\begin{tikzpicture}[line width=0.20mm, scale=0.9, transform shape]
		\draw[arrows={-stealth}] (-1,0)--(-1,1); \node[align=center] at (-1,1.2){$t$}; \draw[arrows={-stealth}] (-1,0)--(0,0); \node[align=center] at (0.2,0){$x$}; 
%		\filldraw[black] (0,2) circle (0.5pt) node[anchor=east] {$t$}; \filldraw[black] (0,3) circle (0.5pt) node[anchor=east] {$t+1$}; \filldraw[black] (2,0) circle (0.5pt) node[anchor=north] {$x_A$};
%		\filldraw[black] (4,0) circle (0.5pt) node[anchor=north] {$x_B$}; \filldraw[black] (6,0) circle (0.5pt) node[anchor=north] {$x_C$}; 
		\filldraw[black] (2,2) circle (1.5pt) node[anchor=north] {$A$}; \filldraw[black] (2,3) circle (1.5pt) node[anchor=north] {$X$}; 
		\filldraw[black] (4,2) circle (1.5pt) node[anchor=north] {$B$}; \filldraw[black] (4,3) circle (1.5pt) node[anchor=north] {$Y$}; 
		\filldraw[black] (6,2) circle (1.5pt) node[anchor=north] {$C$}; \filldraw[black] (6,3) circle (1.5pt) node[anchor=north] {$Z$}; \filldraw[black] (4,1) circle (1.5pt) node[anchor=north] {$\Lambda$};
		\filldraw[black] (4,5) circle (1.5pt); \node[align=center] at (4,4.5) {$C_{XZ}$};
		\draw (2,3)--(5,6); \draw (2,3)--(0.2,4.8); \draw (4,2)--(6,4); \draw (4,2)--(2,4); \draw (6,3)--(7.8,4.8); \draw (6,3)--(3,6);
		\draw (2,3)--(4.8,0.2); \draw (2,3)--(0.2,1.2); \draw (6,3)--(7.8,1.2); \draw (6,3)--(3.2,0.2);
		\draw [fill=blue, fill opacity=0.3, draw=none] (4,5)--(3,6)--(5,6)--cycle;
	\end{tikzpicture}}\qquad\quad
\subfloat[]{	\centering
	\begin{tikzpicture}[line width=0.20mm, scale=1.3]
		\node [draw, circle, name=c1] at (0,0) {\small{$A$}}; \node [draw, circle, name=c2] at (1.5,0) {\small{$B$}}; 
		\node [draw, circle, name=c3] at (3,0) {\small{$C$}}; 
		
		\node [draw, circle, name=c4] at (0,1.5) {$X$}; \node [draw, circle, name=c5] at (1.5,1.5) {\small{$Y$}};
		\node [draw, circle, name=c6] at (3,1.5) {\small{$Z$}};
		\node [draw, circle, name=c7, draw opacity=0] at (1.5,3) {$\Lambda$};
		
		\draw [thick, arrows={-stealth}] (c1) -- (c4); \draw [thick, arrows={-stealth}] (c2) -- (c5); 
		\draw [thick, arrows={-stealth}] (c3) -- (c6); 
		\draw [thick, arrows={-stealth}] (c7) -- (c4); 	\draw [thick, arrows={-stealth}] (c7) -- (c5); 	\draw [thick, arrows={-stealth}] (c7) -- (c6);
	\end{tikzpicture}}

	\caption[Space-time diagram and causal structure for a tripartite Bell experiment]{\textbf{A tripartite Bell experiment:} Three parties Alice, Bob and Charlie share a tripartite system $\Lambda$, they measure their subsystem using the freely chosen measurement settings $A$, $B$ and $C$, producing the outcomes $X$, $Y$ and $Z$ respectively. \textbf{(a)} Space-time configuration for the jamming scenario \cite{Grunhaus1996, Horodecki2019}: The measurement events of the three parties are space-like separated such that the future of Bob's input $B$ contains the entire joint future (in blue) of Alice and Charlie's outputs $X$ and $Z$. Here, $B$ is allowed to signal to $X$ and $Z$ jointly (which can only be verified in the blue region) but not individually. In \cite{Horodecki2019}, a new variable $C_{XZ}$ is introduced, located at the earliest point in the joint future of $X$ and $Z$ and representing the correlations between $X$ and $Z$. \textbf{(b)} Causal structure for the usual tripartite Bell experiment. Note that in order to explain jamming correlations, we must either include additional causal arrows from $B$ to $X$ or $B$ to $Z$ or both (Proposition~\ref{prop:finetune}), or introduce a new node $C_{XZ}$ with an incoming arrow from $B$ \cite{Horodecki2019}. We adopt the former approach in the rest of this chapter.}
	\label{fig: trins}
\end{figure}
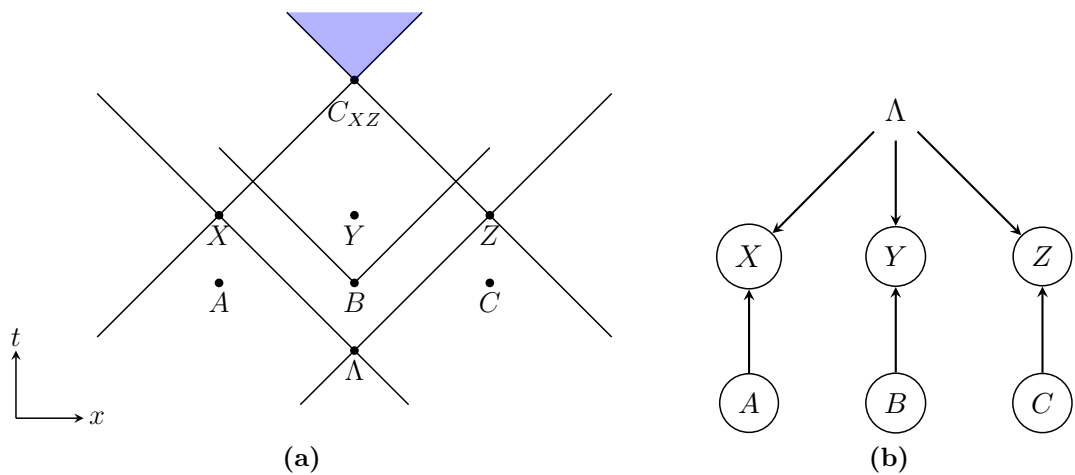

Another example that involves fine-tuning, even though it has not been motivated or discussed in this context, is that of jamming non-local correlations introduced in \cite{Grunhaus1996}. Consider three, space-like separated parties, Alice, Bob and Charlie sharing a tripartite system $\Lambda$ which they measure using measurement settings $A$, $B$ and $C$, producing outcomes $X$, $Y$ and $Z$ respectively. Suppose that their space-time locations are such that Bob's future light cone entirely contains the joint future of Alice and Charlie, as shown in Figure~\ref{fig: trins}. The standard no-signaling conditions forbid the input of each party from being correlated with the outputs of any subset of the remaining parties, in particular, the joint distribution $P_{XYZ|ABC}$ satisfies $P_{XZ|ABC}=P_{XZ|AC}$. In \cite{Grunhaus1996} it is argued that a violation of this requirement does not lead to superluminal signalling in the space-time configuration of Figure~\ref{fig: trins}, as long as $X$ and $Z$ are individually independent from $B$ i.e., $P_{X|ABC}=P_{X|A}$ and $P_{Z|ABC}=P_{Z|C}$. This is because any influence that $B$ exerts jointly on $X$ and $Z$ can only be checked when $X$ and $Z$ are brought together to evaluate the correlations $P_{XZ|ABC}$, which is only possible in their joint future, that is by construction contained in the future of $B$. This scenario allows Bob to \emph{jam} the correlations between Alice and Charlie non-locally.

In \cite{Horodecki2019}, where jamming is further analysed, the causal structure for such an experiment is represented by introducing a new random variable $C_{XZ}$ associated with the set $\{X,Z\}$ that encodes the correlations between its elements. Then $B$ is seen as a cause of $C_{XZ}$ but not as a cause of either $X$ or $Z$. In general scenarios, this representation would require adding a new variable for every non-empty subset of the observed nodes, which can become intractable.\footnote{In general, this representation would include up to $2^n-1$ observed variables whenever the original set of observed variables has $n$ elements.} Further, this representation does not always correspond to what is physically going on--- for instance, in the example of the traitorous lieutenant, this would introduce a new variable $C_{M_1M_2}$ that is observably influenced by the general's original message $M$, while $M$ would no longer be seen as a cause of $M_1$ or $M_2$. However, we know that we physically generated $M_1$ and $M_2$ using\footnote{And possibly some additional information to explain the distribution over the individual variables. As we will see later in Figure~\ref{fig: jamming}, a common cause $\Lambda$ between $M_1$ and $M_2$ would also be required in such examples.} $M$, hence it is indeed a cause of at least one of them.  Therefore, we aim to develop a new approach to causally modelling in a general class of fine-tuned and cyclic scenarios, using only the original variables/systems. The following proposition illustrates that the jamming scenario considered in \cite{Grunhaus1996, Horodecki2019} necessarily corresponds to a fine-tuned causal model over the original variables. Here, jamming is considered in the context of multipartite Bell scenarios where the jamming variable is a freely chosen input of one of the parties. In the causal model approach adopted here, we will take free choice of a variable to correspond to the exogeneity of that variable in the causal structure.\footnote{Free choice is often of a variable $B$ is often defined through the condition that $B$ can only be correlated with variables in its future \cite{CR2013}. We will discuss the relation between this notion of free choice and ours, namely taking $B$ to be exogenous, later in the chapter (Section~\ref{sec: problems}).}

\begin{proposition}
\label{prop:finetune}
Consider a tripartite Bell experiment where three parties Alice, Bob and Charlie share a system $\Lambda$ which they measure using the setting choices $A$, $B$ and $C$, producing the measurement outcomes $X$, $Y$ and $Z$ respectively. Let $\cG$ be any causal structure with only $\{A,B,C,X,Y,Z\}$ as the observed nodes where $A$, $B$ and $C$ are exogenous. Then any joint distribution $P_{XYZ|ABC}$ corresponding to the jamming correlations of \cite{Grunhaus1996, Horodecki2019} defines a fine-tuned causal model over $\cG$, irrespective of the nature (classical, quantum or GPT) of $\Lambda$. 
\end{proposition}
\begin{proof}
Jamming allows Bob's input $B$ to be correlated jointly with $X$ and $Z$ but not individually with $X$ or $Z$. Hence jamming correlations in the tripartite Bell experiment of \cite{Grunhaus1996, Horodecki2019} are characterised by the conditions $B\indep X$ and $B\indep Z$ while $B\not\indep \{X,Z\}$. Since $B$ is exogenous (i.e., has no incoming arrows), the only way to explain the correlation between $B$ and $\{X,Z\}$ is through an outgoing arrow or a directed path from $B$ to the set $\{X,Z\}$ i.e., either an arrow from $B$ to $X$, or from $B$ to $Z$ or both.\footnote{If this were not the case, $B$ would be d-separated from $\{X,Z\}$ and therefore cannot be correlated with it.} Since we require both independences $B\indep X$ and $B\indep Z$ to hold, at least one of these will not be a consequence of d-separation and hence the causal model must be fine-tuned in order to produce these correlations in the causal structure $\cG$.
\end{proof}

The simplest example of a jamming where $B=X\oplus Z$ and all variables are binary uniformly distributed (the remaining variables are irrelevant here), and we will revisit this example several times in this chapter. These are in fact the same correlations as the traitorous lieutenant example.  However in the jamming case, the three variables involved are pairwise space-like separated and since $B$ is exogenous, this corresponds to a situation where $B$ superluminally influences the correlations between $X$ and $Z$. 

%The idea is to distinguish between operationally detectable vs undetectable causal influences between variables by formalising Pearl's do-calculus in our framework, and representing such influences using different types of causal arrows. This would also allow us to provide a formal, mathematical definition of a causal loop that allows observable signalling to the past, which is what would lead to violations of relativistic causality as opposed to operationally undetectable causal loops. This is lacking in previous works such as \cite{Horodecki2019} that consider such scenarios (see Section~\ref{sec: problems} for comments on \cite{Horodecki2019}), which have not made this important distinction.

\section{The framework, Part 1: Causal models}
\label{sec: causmod}  
\subsection{Cyclic and fine-tuned causal models}
In Section~\ref{sec: causalstr} we have reviewed the standard literature on acyclic and faithful causal models, both in the classical as well as non-classical cases. Here, following the motivation set out in the previous sections, we wish to relax the assumptions of acyclicity and faithfulness and extend these methods to cyclic and fine-tuned causal structures. While quantum cyclic causal models have been previously studied \cite{Barrett2020}, these have only been analysed in the faithful case and are based on the split-node causal modelling approach of \cite{Allen2017} (this approach is briefly discussed in Section~\ref{ssec: caus_other}). This approach is not equivalent to the standard causal modelling approach such as \cite{Henson2014} in the cyclic case, for example the former forbids faithful 2 node cyclic causal structures \cite{Barrett2020} but the latter does not, and the former admits a Markov factorisation (analogous to Equation~\eqref{eq: markov}) while the latter does not in general (as we will see later in the chapter). To the best of our knowledge, there is no prior framework for causally modelling cyclic and unfaithful causal structures in the presence of quantum and post-quantum latent nodes, the lack of a Markov factorisation posing a particular difficulty. Here, we develop the bare bones of such a framework that will suffice for our main results. We note that there may be other, inequivalent ways to do the same.\footnote{Based on a different condition for compatibility of a distribution with a causal structure, for example.} We will define causal models in terms of minimal conditions that they must satisfy at the level of the observed nodes which are classical. 
%The framework is defined operationally (following Pearl's framework \cite{Pearl2009}), in terms observed correlations and interventions on the observed systems which are classical RVs (e.g., classical inputs/outputs of experiments). However, our analysis is by no means restricted to classical causal models, the framework allows for non-classical unobserved systems, and is defined in terms of concepts such as d-separation that also apply to non-classical causal models \cite{Henson2014} (Theorem~\ref{theorem: gen_dsep}). 
\par
\paragraph{Causal structure:} Causal structures will be represented using directed graphs, of which the directed acyclic graphs of Section~\ref{sec: causalstr} are special cases. Edges in these graphs will be denoted using $\longrsquigarrow$ (unlike $\longrightarrow$ in previous chapters), as it will be useful to later classify these edges into solid $\longrightarrow$ and dashed $\xdashrightarrow{}$ ones based on certain operational conditions for detecting causation. These causal structures can have observed as well as unobserved nodes, where the former are classical random variables and the latter can be classical or non-classical systems. We will then assume the existence of such a causal structure (though it may be unknown) and use the following definition of \emph{cause} that directly arises from this assumption.

\begin{definition}[Cause]
\label{def: cause}
Given a causal structure represented by a directed graph $\cG$, possibly containing observed as well as unobserved nodes, we say that a node $N_i$ is a \emph{cause} of another node $N_j$ if and only if there is a directed path $N_i\longrsquigarrow...\longrsquigarrow N_j$ from $N_i$ to $N_j$ in $\cG$.
\end{definition}

\paragraph{Observed distribution: }In classical acyclic causal models, the causal Markov condition~\eqref{eq: markov} is used for defining the compatibility of the observed distribution with the causal structure \cite{Pearl2009}. In the non-classical case, the generalised Markov condition of \cite{Henson2014} provides a compatibility condition (Section~\ref{sssec: gen_causalstr}). However, in cyclic causal models, demanding such a factorisation will be too restrictive even in the classical case. For example, consider the simplest cyclic causal structure, the 2-cycle where $X\longrsquigarrow Y$ and $Y\longrsquigarrow X$, with $X$ and $Y$ observed and $X=Y$. The Markov condition would imply that $P_{XY}=P_{X|Y}P_{Y|X}$. Since $X=Y$, the right hand side is a product of deterministic distributions, which forces $P_{XY}$ to also be deterministic in order to be a valid distribution. Therefore, we instead use a weaker compatibility condition in terms of d-separation  between observed nodes (Definition~\ref{definition:dsep}), a concept that can also be applied to non-classical causal structures (Theorem~\ref{theorem: gen_dsep}), and define compatibility of the observed distribution with a cyclic causal structure as follows within our framework.

\begin{definition}[Compatibility of observed distribution with a causal structure] 
\label{definition: compatdist}
Let $\{X_1,...,X_n\}$ be a set of random variables denoting the observed nodes of a directed graph $\mathcal{G}$, and $P_{X_1,...,X_n}$ be a joint probability distribution over them. Then $P$ is said to be \emph{compatible} with $\mathcal{G}$ if for all disjoint subsets $X$, $Y$ and $Z$ of $\{X_1,...,X_n\}$,
\begin{equation*}
    X\perp^d Y|Z \quad\Rightarrow\quad X\indep Y|Z \quad \text{ i.e., $P_{XY|Z}=P_{X|Z}P_{Y|Z}$,}
\end{equation*}
for the marginal distribution, $P_{XYZ}$ of $P_{X_1,...,X_n}$.\footnote{Note that we only need to consider d-separation between observed sets of variables in this definition, however the paths being considered may involve unobserved nodes. For example, if the observed variables $X$ and $Y$ have an unobserved common cause $\Lambda$, then $X$ and $Y$ are not d-separated by the empty set since there is an unblocked path between $X$ and $Y$ through the unobserved common cause, and naturally we don't expect $X$ and $Y$ to be independent in this case.}
\end{definition}

Definition~\ref{definition: compatdist} is essentially the \emph{soundness} of d-separation (c.f. Theorem~\ref{theorem: dsepcomplete}) and is satisfied by classical as well as non-classical causal models in the acyclic case \cite{Pearl2009, Henson2014}. It is sometimes referred to as the \emph{global directed Markov condition} \cite{Richardson1996} and holds in several classical cyclic causal models \cite{Pearl2013, Forre2017} and well as non-classical acyclic causal models \cite{Henson2014}. In Appendix~\ref{appendix: doMechanisms}, we provide an example of a quantum cyclic causal model where this holds. However there also exist cyclic causal models producing observed distributions that do not satisfy Definition~\ref{definition: compatdist}, we discuss this further in the Appendix as well. There, we also present further motivation for the compatibility condition of Definition~\ref{definition: compatdist} in terms of the properties of the underlying causal mechanisms (e.g., functional dependences in the classical case, completely positive maps in the quantum case) and outline possible methods for identifying when this condition might hold for non-classical cyclic causal models. Even in the classical case, several inequivalent definitions of compatibility are possible (which become equivalent in the acyclic case) and \cite{Forre2017} presents a detailed analysis of these conditions and the relationships between them. Such an analysis for the non-classical case is beyond the scope of the present thesis. For the rest of this chapter, we will only consider causal models that satisfy the compatibility condition~\ref{definition: compatdist}.
\par
We will work with the following minimal definition of a causal model in this chapter which is in terms of the graph and observed distribution only. Further details such as the functional relationships between classical variables, choice of quantum states/transformations, or generalised tests (Section~\ref{sssec: gen_causalstr}) can also be included in the full specification of the causal model. These constitute the \emph{causal mechanisms} of the model. Developing a complete and formal specification of these mechanisms and deriving the conditions for their compatibility with cyclic and fine-tuned causal models is a tricky problem, we outline possible ideas for this in Appendix~\ref{appendix: doMechanisms} and leave the full problem for future work. 

\begin{definition}[Causal model]
\label{def: causalmodel}
A causal model over a set of observed random variables $\{X_1,...,X_n\}$ consists of a directed graph $\mathcal{G}$ over them (possibly involving classical/quantum/GPT unobserved systems) and a joint distribution $P_{X_1,...,X_n}$ that is compatible with this graph according to Definition~\ref{definition: compatdist}. 
\end{definition}
 Definition~\ref{definition: compatdist} allows for fine-tuned distributions to be compatible with the causal structure since it only requires that d-separation implies conditional independence and not the converse.  Fine-tuned causal models may in general have an arbitrary number of additional conditional independences that are not implied by the d-separation relations in the corresponding causal graph. The following lemma shows that some additional conditional independences that are not directly implied by d-separation can be derived using d-separation and other independences that may be known.

\begin{restatable}{lemma}{CI}
\label{lemma: CI} 
Let $S_1$, $S_2$ and $S_3$ be three disjoint sets of RVs such that $S_1\indep S_2|S_3$. If $S$ is a set of RVs that is d-separated from these sets in a directed graph $\mathcal{G}$ containing $S_1$, $S_2$, $S_3$ and $S$ are nodes i.e. $S\perp^d S_i$ $\forall i\in\{1,2,3\}$, then any distribution $P$ that is compatible with $\mathcal{G}$ also satisfies the following conditional independences,
$$(S_1\cup S)\indep S_2|S_3,\quad S_1\indep(S_2\cup S)|S_3 \text{ and} \quad S_1\indep S_2|(S_3\cup S).$$
\end{restatable}
A proof can be found in Appendix~\ref{appendix: ProofsJamming}.
Note that this lemma is trivial in the case of faithful causal models. This is because, the independence $S_1\indep S_2|S_3$ implies the d-separation $S_1\perp^d S_2|S_3$ in this case and combined with $S\perp^d S_i$, this would immediately imply the d-separations $(S_1\cup S)\perp^d S_2|S_3$,$S_1\perp^d(S_2\cup S)|S_3$ and $S_1\perp^d S_2|(S_3\cup S)$, which in turn imply the corresponding independences. This property is not so evident in the case of fine-tuned causal models but as we have shown, it nevertheless holds here. Specific examples of this property for fine-tuned causal models are discussed in Section~\ref{ssec: examples}.

\subsection{Interventions and affects relations}
\label{ssec:affects}
%The compatibility condition~\ref{definition: compatdist} characterises the correlations that we can possibly observe in a given causal structure. However, when the causal structure is unknown, correlations alone do not suffice for inferring the underlying underlying causal structure since correlations are symmetric while causal relationships are directional. As we have seen in Section~\ref{sec: causalstr}, a notion of external interventions is required in order to distinguish correlation and causation, which is provided by Pearl's do-calculus \cite{Pearl2009}. An intervention corresponds to operationally probing the existence of a causal explanation-- if forcing a variable $X$ to take a certain value (which over-writes other causal mechanisms that may be influencing $X$) we observe a change in the distribution over another variable $Y$, other things remaining unchanged, we would conclude that $X$ is a cause of $Y$. 
We adopt the characterisation of interventions outlined in Section~\ref{ssec: qinterventions}, following the augmented graph approach \cite{Pearl2009} presented in Section~\ref{sssec: interventions}. The main intuition behind interventions in the acyclic case carries forth to the cyclic case as well, but we will not be allowed to use a Markov factorisation condition such as Equation~\eqref{eq: markov} for classical cyclic causal structures or its non-classical analogue of \cite{Henson2014} since this does not holds for cyclic causal structures in general. We summarise these concepts here for completeness. Given a generalised, cyclic causal structure $\cG$, an intervention on an observed node $X$ of $\cG$ is described as follows. 
\begin{itemize}
    \item An intervention variable $I_X$ taking values in the set $\{idle, \{\text{do}(x)\}_{x\in X}\}$ is introduced where $I_X=idle$ corresponds to the original pre-intervention scenario and $I_X=\text{do}(x)$ sets the variable $X$ to a particular value $x\in X$, cutting off its dependence on all other parents in $\cG$. 
    \item Two new causal structures are introduced to keep track of what happens during such an intervention, the \emph{augmented graph} $\cG_{I_X}$ and the \emph{do-graph} $\cG_{\text{do}(X)}$. The former is obtained from the original graph $\cG$ simply by adding $I_X$ as a node an including an edge $I_X\longrsquigarrow X$ (with the rest remaining unchanged). The latter corresponds to the case when a non-trivial intervention is performed i.e., when $I_X\neq idle$. Hence, $\cG_{\text{do}(X)}$ is obtained from $\cG_{I_X}$ by cutting off all incoming arrows to $X$, except the one from $I_X$, and $I_X$ only takes values in $\{\text{do}(x)\}_{x\in X}$ in $\cG_{\text{do}(X)}$. 
    \item The causal mechanisms involved in the causal model are updated accordingly i.e., they remain unchanged when $I_X=idle$ and when $I_X= \text{do}(x)$, all causal mechanisms, except that of $X$ remain unchanged while $X$ is deterministically set to the value $x$ (irrespective of the states of the subsystems corresponding to incoming edges to $X$ in $\cG$).\footnote{In general, the distribution over $I_X$ may be arbitrary, but given a specific value of $I_X$ in the post-intervention scenario, such as $I_X=\text{do}(x)$, $X$ is completely determined.} The do-conditional $P(Y=y|\text{do}(x))$ encodes the correlations between a variable $Y$ and the intervened variable $X$ (or equivalently $I_X$, since $X$ and $I_X$ are perfectly correlated) in the post-intervention graph i.e., $P(Y=y|\text{do}(x)):=P_{\cG_{\text{do}(X)}}(Y=y|I_X=\text{do}(x))$ (c.f. Equation~\eqref{eq: do}), where $P_{\cG_{\text{do}(X)}}$ is compatible with the do-graph $\cG_{\text{do}(X)}$. We will use the graph as a subscript to the distribution where it is useful to explicitly indicate the causal structure that the distribution is compatible with.
\end{itemize}

This method naturally extends to simultaneous interventions on subsets of nodes (as described in Section~\ref{sssec: interventions}). Only interventions on the observed nodes (which are classical) need to be considered since unobserved nodes are by construction experimentally inaccessible.\footnote{Manipulation of non-classical systems such as quantum states, which are by construction unobserved, can be modelled by introducing observed classical variables that specify the choice of preparation/transformation or measurement acting on the system.}

%Note that in the classical acyclic case, the Markov condition was crucial for deriving the transformation~\eqref{eq: do_transf} between the pre and post-intervention distributions. We noted the difficulties of deriving an analogous transformation in the non-classical case in Section~\ref{ssec: qinterventions}, in particular that this is known to be impossible in general for quantum causal structures even in the acyclic case as shown in \cite{Pienaar2020}. 
At the level of the causal mechanisms (if these are also given), the causal mechanisms of $\cG_{do(X)}$ can be obtained from those of $\cG$ simply by updating the causal mechanisms for each node $X_i$ in $X$ as $X_i=x_i$ iff $I_{X_i}=do(x_i)$ (while leaving the causal mechanisms for all other nodes unchanged) i.e., $P_{\cG_{do(X)}}(X)$ is fully specified by $P_{\cG_{do(X)}}(I_X)$ which can be chosen arbitrarily for the exogenous set $I_X$. Physically, the post-intervention distribution (or the do-conditional) corresponds to additional empirical data that is collected in an experiment, that can, in general be different from the experiment generating the original, pre-intervention data. For example, when the original experiment involves passive observation of correlations between the smoking tendencies and presence of cancer in a group of individuals, an intervention model may involve forcing certain individuals to take up smoking and then studying their chances of developing cancer. In repeated trials, the proportion of individuals who are passively observed and those that are actively intervened upon may be chosen as desired. The latter type of experiments may not necessarily be ethical but are nevertheless a physical possibility. 
In certain cases, it may be possible to fully deduce the post-intervention statistics counterfactually from the pre-intervention data alone (e.g., using a relation such as~\eqref{eq: do_transf}), and the former experiment need not be actually performed, sparing us some ethical dilemmas.\footnote{As discussed in Section~\ref{sec: causalstr}, whenever this is possible, the causal model is called \emph{identifiable} and in Example~\ref{example:nonidentifiable} we visited an example of a simple non-indentifiable scenario involving a classical latent common cause.} In some cases however, this may not be possible (c.f. Example~\ref{example:nonidentifiable}). 
%A complete characterisation of the post-intervention distribution based on causal mechanisms for general cyclic and non-classical causal structures is a subject of ongoing work, which we hope to resolve to a better extent before publishing this work.

Therefore a complete specification of the post-intervention distribution in terms of the pre-intervention may not always be possible. However, the compatibility condition of Definition~\ref{definition: compatdist} along with the definition Equations~\eqref{eq: intervention} and \eqref{eq: do} allow us to derive further useful relationships between these distributions, in particular the three rules of Pearl's do-calculus \cite{Pearl95, Pearl2009}. These rules have been originally derived for classical causal models satisfying the causal Markov property~\eqref{eq: markov} which does not hold in the general scenarios considered here. However, we note that the derivation of these rules don't require the Markov property but only the weaker d-separation condition of Definition~\ref{definition: compatdist} along with the defining Equations~\eqref{eq: intervention} and \eqref{eq: do}. This is captured in the following theorem and we present a proof of the same in Appendix~\ref{appendix: ProofsJamming} for completeness, even though this is predominantly based on the original proof of \cite{Pearl95}. In the following, we will use $\cG_{\overline{X}}$ to denote the graph obtained by deleting all incoming edges to $X$ in a graph $\cG$, where $X$ is some subset of the observed nodes. Similarly, $\cG_{\underline{X}}$ denotes the graph obtained by deleting all outgoing edges from a subset $X$ of nodes in a graph $\cG$.

\begin{restatable}{theorem}{DoRules}
\label{theorem:dorules}
Given a causal model over a set $S$ of observed nodes, associated causal graph $\cG$ and a distribution $P_S$ compatible with $\cG$ according to Definition~\ref{definition: compatdist}, the following 3 rules of do-calculus \cite{Pearl2009} hold for interventions on this causal model. 
\begin{itemize}
    \item \textbf{Rule 1: Ignoring observations} 
    \begin{equation}
    \label{eq: rule1}
        P(y|\text{do}(x),z,w)=P(y|\text{do}(x),w) \qquad \text{if } (Y\perp^d Z|X, W)_{\cG_{\overline{X}}}
    \end{equation}
     \item \textbf{Rule 2: Action/observation exchange} 
    \begin{equation}
     \label{eq: rule2}
        P(y|\text{do}(x),\text{do}(z),w)=P(y|\text{do}(x),z,w) \qquad \text{if } (Y\perp^d Z|X, W)_{\cG_{\overline{X},\underline{Z}}}
    \end{equation}
    \item \textbf{Rule 3: Ignoring actions/interventions} 
     \begin{equation}
      \label{eq: rule3}
        P(y|\text{do}(x),\text{do}(z),w)=P(y|\text{do}(x),w) \qquad \text{if } (Y\perp^d Z|X, W)_{\cG_{\overline{X},\overline{Z(W)}}},
    \end{equation}
\end{itemize}
where $X$, $Y$, $Z$ and $W$ are disjoint subsets of the observed nodes, $Z(W)$ denotes the set of nodes in $Z$ which are not ancestors of $W$ and the above statements hold for all values $w$, $x$, $y$ and $z$ of the variables $W$, $X$, $Y$ and $Z$.
\end{restatable}

While the observed distribution in the post-intervention causal model may not be completely specified by the pre-intervention observed distribution alone, considering the underlying causal mechanisms e.g., the states, transformations, measurements involved in the original causal model should allow for the complete specification of the post-intervention distribution. To the best of our knowledge, this problem has not been studied in non-classical and cyclic causal models, we discuss this point in further detail in Appendix~\ref{appendix: doMechanisms}, providing examples of non-classical cyclic causal models where the post-intervention distribution can be calculated from the causal mechanisms. The full solution to this problem will not be relevant to the results of this thesis. Using these concepts, we now define the \emph{affects relation} that is central to the results of this chapter.

\begin{definition}[Affects relation]
\label{definition: affects}
Consider a causal model over a set of random variables $S$ with causal graph $\mathcal{G}$ and a joint distribution $P$ compatible with it. For $X$, $Y\subseteq S$, if there exists a value $x$ of $X$ such that
\begin{equation*}
   P(Y|\text{do}(x))\neq P(Y),
\end{equation*}
then we say that $X$ \emph{affects} $Y$.
\end{definition}

The distribution on the left hand side of the above equation is compatible with the post-intervention graph $\cG_{\text{do}(X)}$ (as given by Equation~\eqref{eq: do}) while that on the right hand side is compatible with the original graph $\cG$ i.e., $P(Y)$ is short for $P_{\cG}(Y)$. Operationally, $X$ \emph{affects} $Y$ is equivalent to saying that $X$ \emph{signals to} $Y$. With this definition, we are ready to state two useful corollaries of Theorem~\ref{theorem:dorules}.

\begin{corollary}
\label{corollary:exogenous}
If $X$ is a subset of observed exogenous nodes of a causal graph $\cG$, then for any subset $Y$ of nodes disjoint to $X$ the do-conditional and the regular conditional with respect to $X$ coincide i.e.,
$$P(y|\text{do}(x))=P(y|x)\quad \forall x,y$$
In other words, for any subset $X$ of the observed exogenous nodes, correlation between $X$ and a disjoint set of observed nodes $Y$ in $\cG$ guarantees that $X$ affects $Y$. 
\end{corollary}
\begin{proof}
Since $X$ consists only of exogenous nodes, it can only be d-connected to other nodes through outgoing arrows. Then in the graph $\cG_{\underline{X}}$ (where all outgoing arrows from $X$ are cut off), $X$ becomes d-separated from all other nodes. This d-separation, $(Y\perp^d X)_{\cG_{\underline{X}}}$ implies, by Rule 2 of Theorem~\ref{theorem:dorules} that $P(y|do(x))=P(y|x) \quad \forall x,y$. Further if $X$ and $Y$ are correlated in $\cG$, i.e., $\exists x, y$ such that $P(y|x)\neq P(y)$, the equation previously established along with Definition~\ref{definition: affects} imply that $X$ affects $Y$.
\end{proof}

\begin{corollary}
\label{corollary:dsep-affects}
If $X$ and $Y$ are two disjoint subsets of the observed nodes such that $(X\perp^dY)_{\cG_{do(X)}}$, then $X$ \emph{does not affect} $Y$ and $P_{\cG_{do(X)}}(Y)=P_{\cG}(Y)$.
\end{corollary}
\begin{proof}
The d-separation $(X\perp^d Y)_{\cG_{do(X)}}$ trivially implies the d-separation $(X\perp^d Y)_{\cG_{\overline{X}}}$ since $\cG_{do(X)}$ and $\cG_{\overline{X}}$ only differ by the inclusion of the intervention nodes $I_{X_i}$ and the corresponding edges $I_{X_i}\longrightarrow X_i$ for each $X_i\in X$. Then by Rule 3 of Theorem~\ref{theorem:dorules} we have 
$P(y|\text{do}(x))=P(y)$ $\forall x,y$,
which by Definition~\ref{definition: affects} stands for $X$ does not affect $Y$. Further, the d-separation implies the independence $(X\indep Y)_{\cG_{do(X)}}$ i.e., $P_{\cG_{do(X)}}(y|x)=P_{\cG_{do(X)}}(y)$ $\forall x,y$ where the left hand side equals the do-conditional $P(y|\text{do}(x))$ by definition. Along with the result that $X$ does not affect $Y$, this yields the required equation $P_{\cG_{do(X)}}(y)=P_{\cG}(y)$ $\forall y$.
\end{proof}

Note that $X$ \emph{affects} $Y$ implies that there must be a directed path from $X$ to $Y$ in $\mathcal{G}$ (which is equivalent to $X$ is a cause of $Y$, c.f. Definition~\ref{def: cause}). This follows from the contrapositive statement of Corollary~\ref{corollary:dsep-affects}--- $X$ affects $Y$ implies that $X$ and $Y$ are not d-separated in $\cG_{do(X)}$ and since this graph has no incoming arrows to $X$ (except those from the intervention nodes in $I_X$), the only way for $X$ and $Y$ to be d-connected in $\cG_{do(X)}$ is through a directed path from $X$ to $Y$. 
However, the converse is not true. A directed path from $X$ to $Y$ in $\cG$ does not imply that $X$ affects $Y$ in the presence of fine-tuning (as illustrated by Example~\ref{example: jamming}), even though it does imply d-connection between $X$ and $Y$ in $\cG_{do(X)}$. This motivates the following classification of the causal arrows $\longrsquigarrow$ between observed nodes. The arrows  $\longrsquigarrow$ emanating from/pointing to an unobserved node cannot be operationally probed and hence need not be classified.
%Note that $X$ \emph{affects} $Y$ implies that there must be a directed path from $X$ to $Y$ in $\mathcal{G}$ (i.e., $X$ is a cause of $Y$, c.f. Definition~\ref{def: cause}) which can be shown as follows by establishing the equivalent contrapositive statement. Suppose that there are no directed paths from $X$ to $Y$ in $\cG$, then we would have $(X\perp^d Y)_{\cG_{\text{do}(X)}}$ and consequently $(X\perp^d Y)_{\cG_{\overline{X}}}$, since $\cG_{\text{do}(X)}$ and $\cG_{\overline{X}}$ only differ by the inclusion of the intervention nodes $I_{X_i}$ and the corresponding edges $I_{X_i}\longrightarrow X_i$ for each $X_i\in X$, while both graphs have no other incoming arrows to $X$. Then by Rule 3 of Theorem~\ref{theorem:dorules} we have $P(y|\text{do}(x))=P(y)$ $\forall y,x$ which, by Definition~\ref{definition: affects} stands for $X$ does not affect $Y$. However, the converse is not true. A directed path $X\longrsquigarrow... \longrsquigarrow Y$ or even a direct causal relationship $X\longrsquigarrow Y$does not imply an affects relation $X$ affects $Y$, in the presence of fine-tuning (as illustrated in the examples of Section~\ref{ssec: examples}). This motivates the following classification of the causal arrows $\longrsquigarrow$ between observed nodes. The arrows  $\longrsquigarrow$ emanating from/pointing to an unobserved node cannot be operationally probed and hence need not be classified. 

\begin{definition}[Solid and dashed arrows]
\label{definition: solidasharrows}
Given a causal graph $\mathcal{G}$, if two observed nodes $X$ and $Y$ in $\mathcal{G}$ sharing a directed edge $X\longrsquigarrow Y$ are such that $X$ affects $Y$, then the causal arrow $\longrsquigarrow$ between those nodes is called a \emph{solid arrow}, denoted $X\longrightarrow Y$. Further, all arrows $\longrsquigarrow$ between observed nodes in $\mathcal{G}$ that are \emph{not} solid arrows are called \emph{dashed arrows}, denoted $\xdashrightarrow{}$. In other words, $X\xdashrightarrow{} Y$ for any two RVs $X$ and $Y$ in $\cG$ implies that the $X$ does not affect $Y$.
\end{definition}

\begin{remark}[Exogenous nodes]
Note that if $X$ is an exogenous node that is a direct cause of another node $Y$ in a causal graph $\cG$  i.e. $X\longrsquigarrow Y$, and $X$ and $Y$ are correlated in the corresponding causal model, then by Corollary~\ref{corollary:exogenous} and Definition~\ref{definition: solidasharrows} this would imply that the arrow from $X$ to $Y$ must be a solid one. Applying this to the graphs $\cG_{I_X}$ and $\cG_{do(X)}$, where $I_X$ is exogenous and correlated with $X$ by construction (Equation~\eqref{eq: intervention}), we can conclude that  the arrow from every intervention variable to the corresponding intervened variable must be a solid arrow, i.e., $I_X\longrightarrow X$.
\end{remark}

Finally, another noteworthy implication is encapsulated in the following lemma.
 \begin{lemma}
 \label{lemma: correl-affects}
 Given a causal graph $\cG$ and two disjoint subsets $X$ and $Y$ of observed nodes therein,
 $$(X\not\indep Y)_{\cG_{do(X)}}\Rightarrow X \text{ affects } Y.$$
 \end{lemma}
 \begin{proof}
 Suppose that $X$ does not affect $Y$. By Definition~\ref{definition: affects}, this implies that $P(y|\text{do}(x))=P(y)$ $\forall x,y$. Further suppose also that $(X\not\indep Y)_{\cG_{do(X)}}$. This means that there exist two distinct values $x$ and $x'$ of $X$ and some value $y$ of $Y$ such that $P(y|\text{do}(x))\neq P(y|\text{do}(x'))$, which contradicts $P(y|\text{do}(x))=P(y)$ $\forall x,y$. Therefore $(X\not\indep Y)_{\cG_{do(X)}}$ must imply $X \text{ affects } Y$.
 \end{proof}

\subsection{Causal loops}
We distinguish between two types of causal loops that can arise in our framework. The first will be called an \emph{affects causal loop} which is based on the affects relation of Definition~\ref{definition: affects} that corresponds to physically detectable causal relationships. The second will be called a \emph{functional causal loop} and corresponds to a loop at the level of the causal mechanisms (which are functional dependences in the classical case). These function as causal loops at the underlying level even though they may not be detectable at the observable level.

 %\begin{definition}[Extended causal model]
% \label{definition: ext}
% Given a causal model over a set $S$ of RVs, the extended causal model is a causal model over a set $S'\supset S$ which contains arbitrary copies of elements in $S$.
% \end{definition}

\begin{definition}[Causal loops]
\label{definition: causalloops}
A causal model associated with a causal structure $\cG$ over a set $S$ of RVs is said to have an \emph{affects causal loop} if there exist two distinct random variables $X$ and $Y$ in $S$ such that $X$ affects $Y$ \emph{and} $Y$ affects $X$. Hence every affects causal loop corresponds to a directed cycle in $\cG$. Any directed cycle in $\cG$ that is not an affects causal loop is called a \emph{functional causal loop}.
\end{definition}

Note that the \emph{affects causal loop} is defined in terms of single RVs $X$ and $Y$ and not sets of RVs. Replacing $X$ and $Y$ by sets of observed variables $S_1$ and $S_2$ in this definition does not imply the existence of a directed cycle. For example, consider a causal structure $\cG$ with 4 nodes $A$, $B$, $C$ and $D$, all of which are observed such that the only edges in $\cG$ are the solid arrows $A\longrightarrow B$ and $C\longrightarrow D$, with $A$ affects $B$ and $C$ affects $D$. Then, if $S_1=\{A,D\}$ and $S_2=\{B,C\}$ we have $S_1$ affects $S_2$ and $S_2$ affects $S_1$ even though $\cG$ is clearly acyclic. On the other hand, it is also possible to have a cyclic causal model that has no affects causal loops but admits cyclic affects relations in terms of subsets of variables (Example~\ref{sssec: egcycle}). Such loops fall into the category of functional causal loops, even though it might be possible to operationally detect their existence through simultaneous interventions on multiple nodes and additional conditions. We will discuss these further detail later in the chapter. In particular, we will see that such causal loops can be embedded in space-time without leading to superluminal signalling (Figure~\ref{fig: egloop}).

\begin{remark}
A causal modelling approach allows us to define causal loops without a reference to space-time. This would correspond to a closed time-like curve when the variables $X$ and $Y$ in the loop are each associated with single, but mutually distinct space-time points. Cyclic causal models have also been used to describe variables such as demand and supply that are studied over a period of time. These cyclic scenarios do not correspond to closed time-like curves but to physical feedback mechanisms. 
%Before we discuss the embedding and compatibility of our causal models within a space-time structure, it will be illustrative to present the relationships between the concepts introduced so far, since these differ from the standard case of acyclic and faithful causal models. We will also describe several useful examples along the way. 
\end{remark}
\subsection{Relationships between concepts}
\begin{figure}[t!]
    \centering
    \begin{tikzcd}[arrows=Rightarrow, row sep=1.7cm, column sep=1.7cm]
\fbox{$(X\not\indep Y)_{\mathcal{G}_{do(X)}}$}\arrow[shift left=1.5ex]{d}{Lem.\ref{lemma: correl-affects}}\arrow[shift left=1.5ex]{r}{Def.\ref{definition: compatdist}}&\fbox{$(X\not\perp^d Y)_{\mathcal{G}_{do(X)}}$}\arrow[red,degil]{dl}{\textcolor{red}{\textbf{3}}}\arrow[red,degil]{l}{\textcolor{red}{\textbf{1}}}\arrow[shift left=1.5ex]{r}&\mlnode{$\exists$ a directed path \\from $X$ to $Y$ in $\mathcal{G}$\\i.e., $X$ is a cause of $Y$}\arrow{l}\arrow[red,degil,shift left=1.5ex]{d}{\textcolor{red}{\textbf{4}}}\\
   \fbox{$X$ affects $Y$} \arrow[shift left=1.5ex]{ur}{Cor.\ref{corollary:dsep-affects}}\arrow[red,degil,shift left=1.5ex]{rr}{\textcolor{red}{\textbf{5}}}\arrow[red,degil]{d}{\textcolor{red}{\textbf{7}}}\arrow[red,degil]{u}{\textcolor{red}{\textbf{2}}} &    &\mlnode{$X\longrsquigarrow Y$ in $\cG$\\i.e., $X$ is a direct cause of $Y$}\arrow[red,degil]{ddll}{\textcolor{red}{\textbf{13}}}\arrow{u}{Def.\ref{def: cause}} \arrow[ red,degil]{ll}{\textcolor{red}{\textbf{6}}} {}\arrow[red,degil]{d}{\textcolor{red}{\textbf{9}}}\arrow[red,degil, shift left=1.5ex]{dll}{\textcolor{red}{\textbf{8}}} \\
      \fbox{$X\longrightarrow Y$ in $\mathcal{G}$}\arrow{urr}{Def.\ref{definition: solidasharrows}}  \arrow[shift left=1.5ex]{u}{Def.\ref{definition: solidasharrows}}\arrow[red,degil]{d}{\textcolor{red}{\textbf{11}}} &  &  \fbox{$X\xdashrightarrow{} Y$ in $\mathcal{G}$}\arrow[shift left=1.5ex]{dl}{Def.\ref{definition: solidasharrows}}\arrow[shift left=1.5ex]{u}{Def.\ref{definition: solidasharrows}}\arrow[red,degil]{d}{\textcolor{red}{\textbf{16}}}\\
      \fbox{$(X\not\indep Y)_{\mathcal{G}}$}\arrow[red,degil, shift left=1.5ex]{uurr}{\textcolor{red}{\textbf{12}}}\arrow[red,degil, shift left=1.5ex]{u}{\textcolor{red}{\textbf{10}}}&\fbox{$X$ does not affect $Y$}\arrow[red,degil]{ur}{\textcolor{red}{\textbf{14}}}\arrow[red,degil]{r}{\textcolor{red}{\textbf{17}}} &\fbox{$(X\indep Y)_{\mathcal{G}}$}\arrow[red,degil, shift left=1.5ex]{u}{\textcolor{red}{\textbf{15}}}\arrow[red,degil, shift left=1.5ex]{l}{\textcolor{red}{\textbf{18}}}
\end{tikzcd}
  \caption[Relationships between concepts associated with causal models]{\textbf{Relationships between concepts relating to causal models: }  The black arrows denote implications while red (crossed out) arrows denote non-implications. The numbers label the counter-examples corresponding to each non-implication, which are explained in the main text. The equivalence between ``$\exists$ a directed path from $X$ to $Y$ in $\cG$'' and $(X\not\perp^d Y)_{\cG_{do(X)}}$ is explained in the paragraph following Corollary~\ref{corollary:dsep-affects}. $X\longrightarrow Y$ and $X\xdashrightarrow{} Y$ imply $X\longrsquigarrow Y$ since solid and dashed arrows are simply special instances of the more general, squiggly arrow by Definition~\ref{definition: solidasharrows}.
 }
    \label{fig: Ch6_Fig1}
\end{figure}
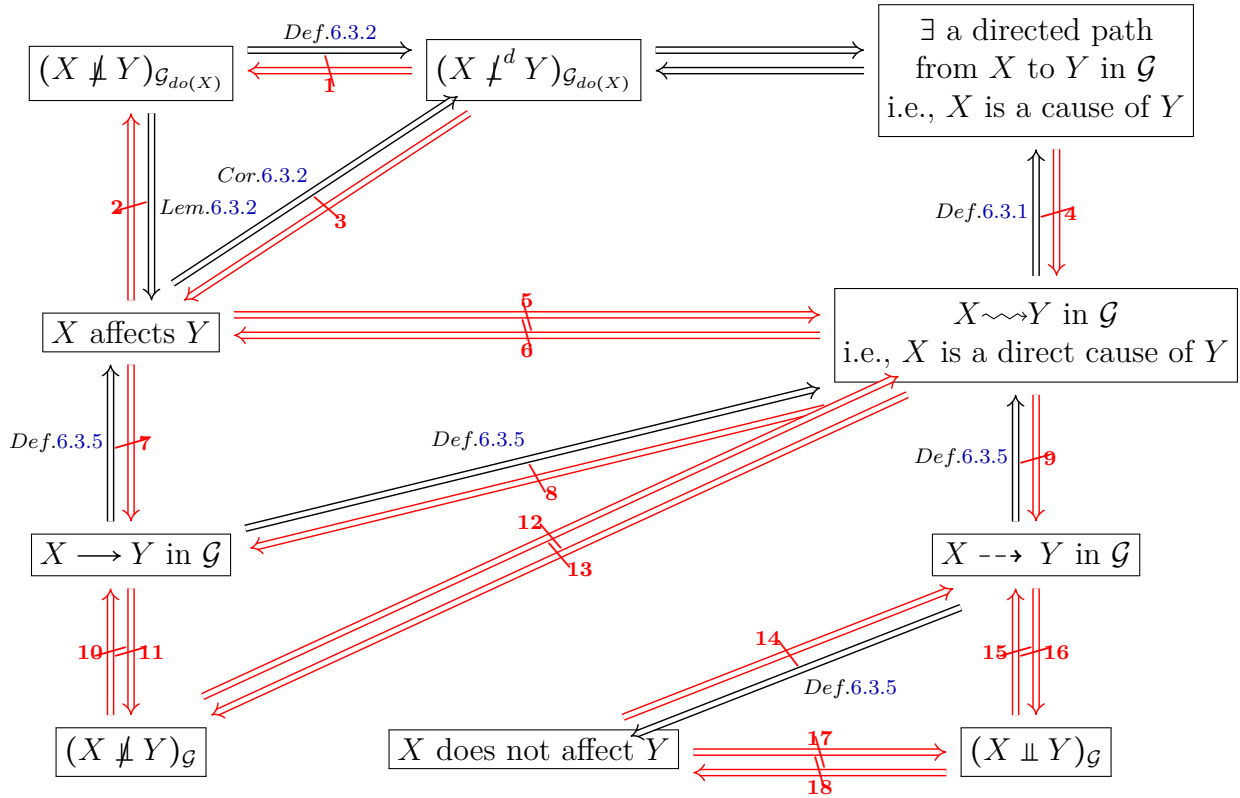

Due to the presence of fine-tuning and the introduction of the 2 types of causal arrows (solid and dashed), a number of concepts that are equivalent in faithful causal models are not equivalent for the causal models described in our framework. It will hence be illustrative to present some of relationships between the concepts relating to our causal models, before discussing the relation to space-time structure. These are illustrated in Figure~\ref{fig: Ch6_Fig1}. The reason for every implication is explained in the figure and its caption, and for every implication that fails, we provide a counter-example below. There are 14 implications in Figure~\ref{fig: Ch6_Fig1} that do not hold. Some of these can be explained by the same counter-example or are immediately evident from the definitions. Therefore we first group these 14 cases based on the corresponding counter-example/argument needed for explaining them, in the end we will only need a few distinct counter-examples to cover all these cases. Note that if we restrict to faithful and/or acyclic causal models, not all of these non-implications would hold. For instance, in the case of faithful and acyclic causal models commonly considered in the literature, non-implications 1, 2, 3, 4, 5, 9 and 12 will become implications. 

\begin{enumerate}
 \item \textbf{Non-implication 1:} In unfaithful causal models, $X$ and $Y$ can be independent even when they are d-connected, as we have seen in the examples of Figure~\ref{fig: motiv_eg}. 
  \item \textbf{Non-implications 2, 11, 18:} These are covered by Example~\ref{example: eg3}.
 %  \item \textbf{Non-implication 3:} If we have $X\xdashrightarrow{} Y$ in $\cG_{do(X)}$, then $(X\not\perp^d Y)_{\cG_{do(X)}}$ yet $X$ does not affect $Y$ (by Definition~\ref{definition: solidasharrows} of dashed arrows).
    \item \textbf{Non-implications 3, 6, 8, 13:} These are covered by Example~\ref{example: jamming}.
     \item \textbf{Non-implications 4, 5:} $X$ is a cause of $Y$ does not imply that it is a direct cause of $Y$, it can be an indirect cause. Further $X$ can affect $Y$ even when it is an indirect cause, for example $X\longrightarrow Z\longrightarrow Y$.
    \item \textbf{Non-implication 7:} This is covered by Example~\ref{example: eg2}.
    \item \textbf{Non-implication 9:} It is evident that ``$X$ is a direct cause of $Y$'' does not imply $X\xdashrightarrow{} Y$, since it can also be a cause through a solid arrow.
    \item \textbf{Non-implications 10, 12:} These are just a consequence of the fact that correlation does not imply causation. Correlation between $X$ and $Y$ can arise when they share a common cause, without being a cause (direct or indirect) of each other.
     \item \textbf{Non-implications 14, 17:} In a simple common cause scenario, i.e., $Z\longrightarrow X$ and $Z\longrightarrow Y$ with $X=Y=Z$, $X$ does not affect $Y$ however $X$ is correlated with $Y$ and there is no dashed arrow from $X$ to $Y$.
    \item \textbf{Non-implication 15:} It is evident that independence of $X$ and $Y$ does not imply that there is a dashed arrow between them, they can also be d-separated.
     \item \textbf{Non-implication 16:} We can consider scenarios where two variables connected by a dashed arrow are also connected by an alternate, correlating path such as a common cause, and hence the dashed arrow does not by itself rule out the possibility of the variables being correlated.
\end{enumerate}

\begin{figure}[t!]
    \centering
 \begin{tikzpicture}
  \node[shape=circle,draw=black] (S) at (0,0) {$S$};  \node[shape=circle,draw=black] (E) at (-2,2) {$E$};  \node[shape=circle,draw=black] (H) at (2,2) {$H$}; 
  \draw [thick, arrows=-stealth] (S) -- (E);   \draw [thick, arrows=-stealth, dashed] (S) -- (H);   \draw [thick, arrows=-stealth] (E) -- (H);
 \end{tikzpicture}
    \caption[Operationally detectable causation without correlation]{\textbf{Affects relation does not imply correlation: } This is a causal structure for Example~\ref{example: eg3} which demonstrates a scenario where $E$ affects $H$ even though $P_{EH}=P_EP_H$, i.e., solid arrows can also be fine-tuned and the ability to detect causation through an active intervention) does not imply that we will see correlation upon passive observation.}
    \label{fig:eg3}
\end{figure}
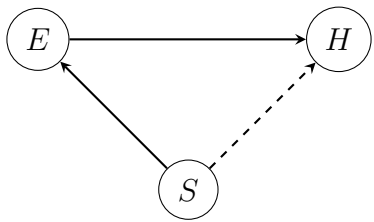

 \begin{example}
 \label{example: eg3}
Consider the causal structure of Figure~\ref{fig:eg3}. Let the three variables $S$, $E$ ad $H$ be binary and correlated as $H=S\oplus E$ and $S=E$. These relations imply that $H=0$ deterministically while $S=E$. Now, when we intervene on $E$, we can choose its value independently of $S$ and whenever we choose $E\neq S$, we will see that $H=1$ occurs with non-zero probability. In other words, there exists a value $e$ of $E$ such that $P(H=1|do(e))\neq P(H=1)=0$ i.e., $E$ affects $H$. As $E$ is a direct cause of $H$ in $\cG$, this further implies that the causal arrow from $E$ to $H$ is a solid one, even though $E$ and $H$ are independent in both the pre and post-intervention causal models i.e., $(E \indep H)_{\cG}$ and $(E \indep H)_{\cG_{do(X)}}$ both hold. The former since $H$ is deterministic in the original causal model, irrespective of the value of $E$ and the latter due since $H$ is uniform in the post-intervention model, again irrespective of the value of $E$, due to the fine-tuned nature of the correlations. Therefore the existence of an affects relation between two sets of observed variables does not imply correlation between them either in the pre or the post intervention causal model. Further, $S$ does not affect $H$ since the exogeneity of $S$ implies that $P_{\cG{do(S)}}(H|S)=P(H|S)$ (Corollary~\ref{corollary:exogenous}), and the independence of $S$ and $H$ in $\cG$ gives  $P(H|S)=P(H)$.
 \end{example}

\begin{figure}[t]
    \centering
  \subfloat[\label{fig: jamming}]{ \begin{tikzpicture}[scale=0.8]
      \node[shape=circle,draw=black] (B) at (0,0.5) {B};
    \node[shape=circle,draw=black] (A) at (-2,2) {A};
    \node[shape=circle,draw=black] (C) at (2,2) {C};
      \node[shape=circle,draw=none] (L) at (0,-1.5) {$\Lambda$};
     \path [thick, dashed, arrows=-stealth] (B) edge (A);
     \path [thick, dashed, arrows=-stealth] (B) edge (C);
       \draw[decorate, decoration={zigzag, segment length=+6pt, amplitude=+.95pt,post length=+4pt}, arrows={-stealth}, thick] (L) -- (A);  \draw[decorate, decoration={zigzag, segment length=+6pt, amplitude=+.95pt,post length=+4pt}, arrows={-stealth}, thick] (L) -- (C);
      \end{tikzpicture}}\qquad\qquad\qquad\qquad\qquad\qquad\subfloat[\label{fig:example1}]{ \begin{tikzpicture}[scale=0.8]
      \node[shape=circle,draw=black] (B) at (0,0.5) {B};
    \node[shape=circle,draw=black] (A) at (-2,2) {A};
    \node[shape=circle,draw=black] (C) at (2,2) {C};
     \node[shape=circle,draw=black] (D) at (0,3.5) {D};
      \node[shape=circle,draw=none] (L) at (0,-1.5) {$\Lambda$};
     \path [thick, dashed, arrows=-stealth] (B) edge (A);
     \path [thick, dashed, arrows=-stealth] (B) edge (C);
      \path [thick, dashed, arrows=-stealth] (A) edge (D); \path [thick, dashed, arrows=-stealth] (C) edge (D);
       \draw[decorate, decoration={zigzag, segment length=+6pt, amplitude=+.95pt,post length=+4pt}, arrows={-stealth}, thick] (L) -- (A);  \draw[decorate, decoration={zigzag, segment length=+6pt, amplitude=+.95pt,post length=+4pt}, arrows={-stealth}, thick] (L) -- (C);
    \end{tikzpicture}}
    \caption[Some fine-tuned causal structures]{\textbf{Some fine-tuned causal structures: } \textbf{(a)} The jamming causal structure of Example~\ref{example: jamming}. Note that the common cause $\Lambda$ is essential to this example,  because without $\Lambda$, $A$ and $C$ would be d-separated given $B$ which would imply the conditional independence $P_{AC|B}=P_{A|B}P_{C|B}$. The dashed arrows would imply the independence of $A$ and $B$ as well as $C$ and $B$ and hence the observed distribution would factorise as $P_{ABC}=P_AP_BP_C$. Then no pairs of disjoint subsets of $\{A,B,C\}$ would affect each other contrary to the original example. \text{(b)} Causal structure for Example~\ref{example: eg2} where $B$ affects $D$ even though there is no solid arrow path from $B$ to $D$. }
\end{figure}
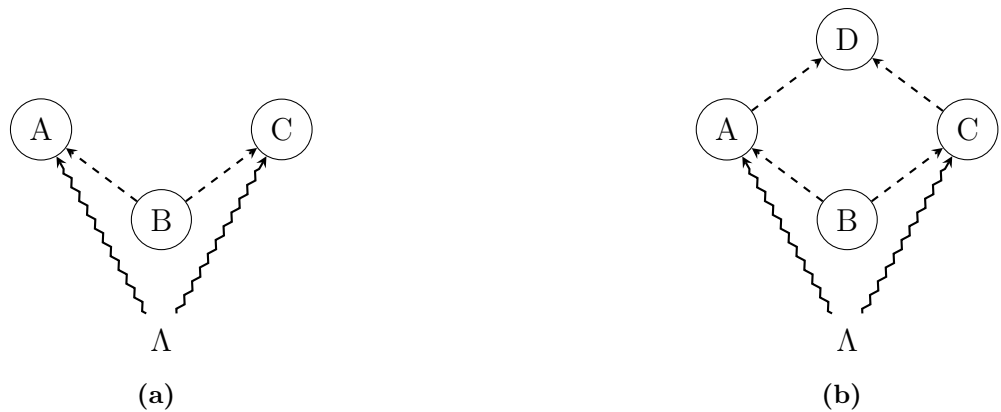

\begin{example}[Jamming]
\label{example: jamming}
Consider the causal structure of Figure~\ref{fig: jamming} where $B\xdashrightarrow{} A$, $B\xdashrightarrow{} C$ and the RVs $A$ and $C$ share an unobserved common cause $\Lambda$. By Definition~\ref{definition: solidasharrows} of the dashed arrows, we have $B$ does not affect $A$ and $B$ does not affect $C$. Suppose that $B$ affects the set $\{A,C\}$. When $A$, $B$ and $C$ are binary, a probability distribution compatible with this situation is one where all 3 RVs are uniformly distributed and correlated as $B=A\oplus C$, where $\oplus$ stands for modulo-2 addition. Then, $A$ and $C$ individually carry no information about $B$ but $A$ and $C$ jointly determine the exact value of $B$. In this case, $B$ is a cause of $A$ and of $C$ but due to fine-tuning, $B$ and $A$ as well as $B$ and $C$ are uncorrelated and there are no pairwise affects relations such that the causal influence of $B$ on $A$ (or $B$ on $C$) can only be detected when all 3 variables are jointly accessed. The common cause is crucial to this example as explained in Figure~\ref{fig: jamming}, and the causal structure compatible with the distribution and affects relations of this example is not unique. An alternative causal structure would be one where one of the dashed arrows $B\xdashrightarrow{} A$ or $B\xdashrightarrow{} C$ is dropped, as we will see in Section~\ref{fig: jammingproof}.
\end{example}
This example by itself makes no reference to space-time or the tripartite Bell scenario. However, for the particular embedding of the variables $A$, $B$ and $C$ in space-time where they are pairwise space-like separated and taken to correspond to the output of Alice, input of Bob and output of Charlie respectively, this becomes a special case of the tripartite jamming scenario of \cite{Grunhaus1996, Horodecki2019} (Figure~\ref{fig: trins}).\footnote{Barring the slight change of notation: In Figure~\ref{fig: trins} $A$ and $C$ correspond to the inputs of Alice and Charlie while $X$ and $Z$ correspond to the outputs that are jammed by $B$. We don't make a distinction between inputs and outputs in general since we will also consider situations where the jamming variable is not exogenous for example.} In the rest of the chapter, such examples, where an RV has dashed arrows to a set of RVs will be referred to as instances of ``jamming'' in accordance with the terminology of \cite{Grunhaus1996}, irrespective of the space-time configuration. We will further discuss the relation of such causal models to space-time structure later in the chapter, and will also revisit Example~\ref{example: jamming} several times in this process.

\begin{example}
\label{example: eg2}
Consider a causal model over observed variables $\{A,B,C,D\}$ associated with the causal graph $\mathcal{G}$ given in Figure~\ref{fig:example1}. Here, there are no pairs of variables sharing an edge such that one of them affects the other. A correlation compatible with this graph is obtained by taking $B=A\oplus C=D$ where all variables are binary and uniformly distributed. Here, $B$ affects $D$ even though there are no solid arrow paths from $B$ to $D$. 
\end{example}

\subsection{Identifying conditional independences and affects relations: Examples}
\label{ssec: examples}
Here we provide examples that better illustrate some of the definitions and rules of the framework laid out so far. In particular, how one can deduce the conditional independences and affects relations in a given causal model. We briefly summarise these below.
\begin{itemize}
    \item \emph{Conditional independences:} Given a causal graph $\cG$ with the set of observed nodes $S$, some of the conditional independences satisfied by the joint distribution $P_S$ can be identified using Definition~\ref{definition: compatdist} i.e., by listing all the conditional independences implied by d-separation relations in $\cG$. Further independences may be found if there are dashed arrows emanating from exogenous nodes, since $X\xdashrightarrow{} Y$ implies $X$ does not affect $Y$ (by Definition~\ref{definition: solidasharrows}) which implies $X\indep Y$ if $X$ is exogenous (c.f. Corollary~\ref{corollary:exogenous}). Lemma~\ref{lemma: CI} can also be used to list further independences not directly implied by d-separation in $\cG$. There may still be more conditional independences in $P_S$ that cannot be listed using the methods mentioned above, for instance, in Example~\ref{example: jamming}, we also have $A\indep C$ but this neither follows from d-separation or the dashed arrow structure. Since we allow for fine-tuning, there could be arbitrarily many independences in $P$, but those mentioned above are sufficient for compatibility with the causal model. 
    %Note however that certain independences are not allowed, for example we cannot have $A\indep C|B$ in the causal structure~\ref{fig: jamming}, for if this were the case $B$ would not be correlated with $A$, $C$ or $\{A,C\}$ and we would draw a different causal structure where there are no arrows (solid or dashed) between $B$ and $\{A,C\}$. 
    \item \emph{Affects relations:} Some of the affects relations can be identified from the pre and post-intervention causal models using the fact that correlation between two disjoint sets $X$ and $Y$ of observed nodes in the post-intervention graph $\cG_{\text{do}(X)}$ implies that $X$ affects $Y$ (Lemma~\ref{lemma: correl-affects}). Note however that an independence $(X\indep Y)_{\cG_{\text{do}(X)}}$ in the post-intervention model need not imply that $X$ does not affect $Y$ (i.e., a non-affects relation) unless $X$ is exogenous (c.f. non-implication 2 of Figure~\ref{fig: Ch6_Fig1} and Corollary~\ref{corollary:exogenous}), but the d-separation $(X\perp^d Y)_{\cG_{\text{do}(X)}}$ in the post-intervention model does imply $X$ does not affect $Y$ (Corollary~\ref{corollary:dsep-affects}). Therefore one can check for non-independences and d-separations in the post-intervention causal model to identify affects and non-affects relations. Again, due to fine-tuning, this identification may not be exhaustive.

    %We have seen that a non-trivial intervention on a subset $S$ of the observed nodes corresponds to a new causal graph $\cG_{\text{do}(S)}$ in which all the nodes contained in $S$ are exogenous. Then the condition that $S$ affects another subset $R$ of the observed nodes (i.e., there exists a values $s$ of $S$ such that $P(R|\text{do}(s))\neq P(S)$), corresponds to the condition that $S$ and $R$ are \emph{not} independent in the post-intervention causal structure $\cG_{\text{do}(S)}$. Hence one can apply the procedure of identifying independences explained in the previous point to the post-intervention causal structure, to identify affects relations. Again, due to fine-tuning, this identification may not be exhaustive.
\end{itemize}
In case some or all of the causal mechanisms are also given in addition to the observed distributions, it may be possible to identify further independences and affects relations in the model.

\subsubsection{Jamming (Figure~\ref{fig: jamming})}
\label{sssec: egjamming}
In the jamming causal structure $\mathcal{G}^{jam}$ of Figure~\ref{fig: jamming} and Example~\ref{example: jamming}, Definition~\ref{definition: compatdist} does not impose any conditional independences on the observed distribution $P_{ABC}$ since $\Lambda$ is unobserved.\footnote {If $\Lambda$ in Figure~\ref{fig: jamming} were observed, $A$ and $C$ would be d-separated given $\{B,\Lambda\}$ and we would have the conditional independence $P_{AC|B\Lambda}=P_{A|B\Lambda}P_{C|B\Lambda}$.} However, from Definition~\ref{definition: solidasharrows} of dashed arrows we know that $B$ affects neither $A$ nor $C$ individually and we are given that $B$ affects $\{A,C\}$. Using the exogeneity of $B$ (c.f. Corollary~\ref{corollary:exogenous}), this implies the independences $A\indep B$ and $C \indep B$  and the non-independence $B\not \indep \{A,C\}$ in $\cG^{jam}$. Now, consider an intervention on $A$. The post-intervention causal structure $\mathcal{G}^{jam}_{\text{do}(A)}$ only has the edges $B\xdashrightarrow{} C$ and $\Lambda \longrsquigarrow C$ (along with $I_A\longrightarrow A$ of course). The d-separation $(A\perp^d C)_{\mathcal{G}^{jam}_{\text{do}(A)}}$ implies the independence $(A\indep C)_{\mathcal{G}^{jam}_{\text{do}(A)}}$ and also that $A$ does not affect $C$. Similarly, we can derive $C$ does not affect $A$, $A$ does not affect $\{B,C\}$,  $C$ does not affect $\{A,B\}$ and $\{A,C\}$ does not affect $B$. Further, using Lemma~\ref{lemma: CI} and the exogeneity of $B$, we can derive $\{A,B\}$ does not affect $C$ as follows. In the causal structure $\cG^{jam}_{\text{do}(\{A,B\})}$, $A$ is d-separated from $B$ and $C$, while $B$ and $C$ are independent of each other due to the exogeneity of $B$ and the dashed arrow connecting them. Using the lemma, this gives $(\{A,B\}\indep C)_{\cG^{jam}_{\text{do}(\{A,B\})}}$ which can be explicitly written as $P_{\cG^{jam}_{\text{do}(\{A,B\})}}(c|a,b)=P_{\cG^{jam}_{\text{do}(\{A,B\})}}(c)$ $\forall a,b,c$. The left hand side equals the do-conditional $P(c|\text{do}(a),\text{do}(b))$ by definition and the right hand side can be simplified in the following two steps. Firstly as $P_{\cG^{jam}_{\text{do}(\{A,B\})}}(c)=P_{\cG^{jam}_{\text{do}(A)}}(c)$ noting that $\cG^{jam}_{\text{do}(\{A,B\})}$ and $\cG^{jam}_{\text{do}(A)}$ are effectively the same graph due to the exogeneity of $B$. Then the d-separation $(A\perp^d C)_{\cG^{jam}_{\text{do}(A)}}$ implies the independence $P_{\cG^{jam}_{\text{do}(A)}}(c|a)=P_{\cG^{jam}_{\text{do}(A)}}(c)$ $\forall a,c$, which along with $A$ does not affect $C$ (as noted earlier) gives $P_{\cG^{jam}_{\text{do}(A)}}(c)=P_\cG(c):=P(c)$ $\forall c$. Together, this gives $P_{\cG^{jam}_{\text{do}(\{A,B\})}}(c|a,b)=P(c)$ $\forall a,b,c$ i.e., $\{A.B\}$ does not affect $C$. Similarly, one can obtain $\{B,C\}$ does not affect $A$.

\begin{figure}[t]
 \centering\subfloat[\label{fig: collider}]{\begin{tikzpicture}[scale=0.8] \node[shape=circle,draw=black] (B) at (0,0) {B};
    \node[shape=circle,draw=black] (A) at (-2,-2) {A};
    \node[shape=circle,draw=black] (C) at (2,-2) {C};  \path [thick, dashed, arrows=-stealth] (A) edge (B);
     \path [thick, dashed, arrows=-stealth] (C) edge (B);\end{tikzpicture}}\qquad\qquad
 \subfloat[\label{fig: eqcyclemain}]{\begin{tikzpicture}[scale=0.8]
      \node[shape=circle,draw=black] (B) at (0,0) {B};
    \node[shape=circle,draw=black] (A) at (-2,-2) {A};
    \node[shape=circle,draw=black] (C) at (2,-2) {C};  \node[shape=circle,draw=none] (L) at (0,-4) {$\Lambda$};
     \path [thick, dashed, arrows=-stealth] (A) edge (B);
     \path [thick, dashed, arrows=-stealth] (C) edge[bend left=20] (B); 
     %\path [thick, dashed, arrows=-stealth] (B) edge[bend right=20] (A);
     \path [thick, dashed, arrows=-stealth] (B) edge[bend left=20] (C); 
     \draw[decorate, decoration={zigzag, segment length=+6pt, amplitude=+.95pt,post length=+4pt}, arrows={-stealth}, thick] (L) -- (A);  \draw[decorate, decoration={zigzag, segment length=+6pt, amplitude=+.95pt,post length=+4pt}, arrows={-stealth}, thick] (L) -- (C);
      \end{tikzpicture}}\qquad\qquad
  \subfloat[\label{fig: causalloop}]{\begin{tikzpicture}[line width=0.20mm, scale=0.8]
		\node [draw, circle, name=c1, draw opacity=0] at (0,0) {$A$}; \node [draw, circle, name=c2, draw opacity=0] at (-2,2) {$B$};
		\node [draw, circle, name=c4, draw opacity=0] at (2,2) {$D$};
		\node [draw, circle, name=c3, draw opacity=0] at (0,4) {$C$};
		\draw [thick, arrows={-stealth}] (c1) -- (c2); \draw [thick, arrows={-stealth}] (c2) -- (c3); 
		\draw [thick, arrows={-stealth}] (c3) -- (c4); 
		\draw [thick, arrows={-stealth}] (c4) -- (c1);
    \end{tikzpicture}}
    \caption[Some fine-tuned and/or cyclic causal structure]{\textbf{Some fine-tuned and/or cyclic causal structures: } \textbf{(a)} A fine-tuned collider \textbf{(b)} A functional causal loop \textbf{(c)} An affects causal loop}
    \label{fig: examples}
\end{figure}
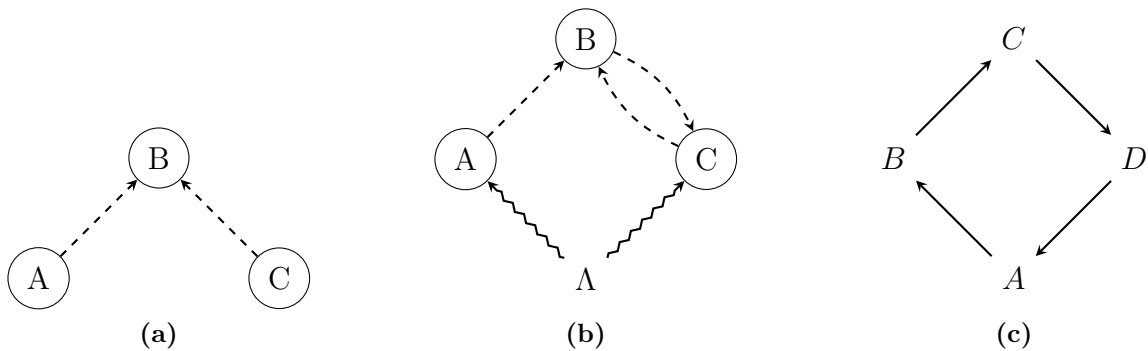

\subsubsection{Fine-tuned collider (Figure~\ref{fig: collider})}
\label{sssec: egcollider}
In the causal structure of Figure~\ref{fig: collider}, the independence $A\indep C$ follows from Definition~\ref{definition: compatdist}, while $A\indep B$ and $C\indep B$ follow from the dashed arrow structure. These are the same independences as the case in the previous example of jamming with unobserved $\Lambda$ (where $A\indep C$ was an additional independence in the jamming example but follows from d-separation in this case). Thus the distribution $P_{ABC}$ from Example~\ref{example: jamming} is compatible with both the jamming (Figure~\ref{fig: jamming}) as well the fine-tuned collider (Figure~\ref{fig: collider}) causal structures.\footnote{Note that this is essentially the one-time pad example from earlier.} However interventions on the two causal structures yield different results. We have $\{A,C\}$ affects $B$ for the fine-tuned collider (since $\{A,C\}$ consists of exogenous nodes and is correlated with $B$) but not for the jamming case. We also have $A$ affects $\{B,C\}$ and $C$ affects $\{A,B\}$ for the fine-tuned collider even though $A$ and $C$ do not individually affect $B$ due to the dashed arrow structure. This follows from the exogeneity of $A$ and $C$ and the joint correlations $A=B\oplus C$. Further, $\{A,B\}$ does not affect $C$ since these sets become d-separated upon intervention on $\{A,B\}$ and by a similar reasoning, $\{B,C\}$ does not affect $A$ and $B$ does not affect $\{A,C\}$ (in contrast with the jamming case where $B$ affects $\{A,C\}$).

\subsubsection{A functional causal loop (Figure~\ref{fig: eqcyclemain})}
\label{sssec: egcycle}
\begin{figure}[t!]
 \centering
\begin{tikzpicture}[scale=0.6]
   	\filldraw[black] (2,2) circle (1.5pt) node[anchor=north] {$\mathcal{A}$}; 
		\filldraw[black] (5,5) circle (1.5pt) node[anchor=north] {$\mathcal{B}$}; 
		\filldraw[black] (8,2) circle (1.5pt) node[anchor=north] {$\mathcal{C}$}; 
		\draw (2,2)--(7,7); \draw (2,2)--(0.2,3.8);\draw (8,2)--(9.8,3.8); \draw (8,2)--(3,7); 
		\draw [fill=blue, fill opacity=0.3, draw=none] (5,5)--(3,7)--(7,7)--cycle;
   \end{tikzpicture}
    \caption[Compatible space-time embedding of a cyclic causal model]{\textbf{A space-time embdedding for the cyclic causal model of Figure~\ref{fig: eqcyclemain}: } In the causal model, we have $B$ affects $\{A,C\}$ and $\{A,C\}$ affects $B$. In this above space-time embedding, the future of the space-time random variable $\cB$ exactly coincides with the joint future of $\cA$ and $\cC$ (blue region) i.e., the causal model is compatible with the space-time (Definition~\ref{definition: compatposet}) for this special embedding and does not lead to superluminal signalling even though it contains functional causal loops. Note that for affects causal loops, the only compatible embedding in space-time is the one where all the loop variables are assigned the same space-time location (c.f. Corollary~\ref{corollary: loops}).}
    \label{fig: egloop}
\end{figure}
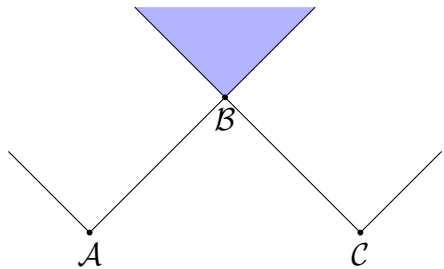

Consider the cyclic causal structure $\cG^{fl}$ (``fl'' stands for functional loop) of Figure~\ref{fig: eqcyclemain} along with the following classical causal mechanisms where all 4 variables are taken to be binary: $A=\Lambda$, $B=A\oplus C$, $C=B\oplus \Lambda$, where the exogenous variable $\Lambda$ is uniformly distributed. One can check that the distribution $P_{ABC}$ obtained through these mechanisms would be the same as that of the jamming as well as the fine-tuned collider examples above, but the affects relations differ. Firstly, in the causal model of $\cG^{fl}_{\text{do}(A)}$, $\Lambda$ is no longer a parent of $A$, but using the remaining causal mechanisms $B=A\oplus C$ and $C=B\oplus \Lambda$ (which remain the same), we can still obtain $A=\Lambda$. Therefore the intervention on $A$ does not change the observed distribution and $A$ and $B$ continue to be independent in $\cG^{fl}$ as well as $\cG^{fl}_{\text{do}(A)}$, and in both graphs the marginal distributions over $A$, $B$ and $C$ are uniform, which gives $A$ does not affect $B$. On the other hand, $B$ does not affect $A$ can be established simply from the d-separation $(B\perp^d A)_{\cG^{fl}_{\text{do}(B)}}$. In the causal model of $\cG^{fl}_{\text{do}(C)}$, neither $B$ nor $\Lambda$ are parents of $C$ but the remaining mechanisms $A=\Lambda$ and $B=A\oplus C$ give $C=B\oplus \Lambda$. Again, the observed distribution here is the same as the pre-intervention distribution, which gives $C$ does not affect $B$. By a similar argument, $B$ does not affect $C$ can also be established. Further, we have both $B$ affects $\{A,C\}$ (as in the jamming case) \emph{and} $\{A,C\}$ affects $B$ (as in the fine-tuned collider) since $P(b|\text{do}(a),\text{do}(c))$ and $P(a,c|\text{do}(b))$ are deterministic while $P(b)$ and $P(a,c)$ are uniform. We also have $A$ affects $\{B,C\}$ and $C$ affects $\{A,B\}$ as in the fine-tuned collider, which can be verified using the causal mechanisms given.\footnote{Note that in the absence of the causal mechanisms, many of the affects/non-affects relations may not be identifiable. For example, to deduce that $\{A,B\}$ does not affect $C$ in the jamming case, we used Lemma~\ref{lemma: CI} along with the fact that $B$ was exogenous in $\cG^{jam}$. But the same argument cannot be applied here since $B$ is not exogenous in $\cG^{fl}$.} As in the jamming case, we also get $\{A,B\}$ does not affect $C$ and $\{B,C\}$ does not affect $A$. Since we have no pair of variables in $\{A,B,C\}$ that affect each other, this does not correspond to an affects causal loop (Definition~\ref{definition: causalloops}). Furthermore,  even though we have directed cycles and the cyclic affects relations $B$ affects $\{A,C\}$ and $\{A,C\}$ affects $B$, the variables $A$, $B$ and $C$ can be embedded in Minkowski space-time such that this causal model does not lead to signalling superluminally, as we will see later (Figure~\ref{fig: egloop}).

\subsubsection{An affects causal loop (Figure~\ref{fig: causalloop})}
\label{sssec: egloops}
Consider the causal structure of Figure~\ref{fig: causalloop}. Applying Definition~\ref{definition: solidasharrows} of solid arrows, to this causal structure we have $A$ affects $B$, $B$ affects $C$, $C$ affects $D$ and $D$ affects $A$, which forms an affects causal loop. The conditional independences that follow from d-separation are $A\indep C|\{B,D\}$ and $B\indep D|\{A,C\}$ and any joint distribution satisfying these would be compatible with the causal structure. To further illustrate the kind of causal loops allowed in this framework, consider the pairwise correlations $A=B$, $B=C$, $C=D$ and $D\neq A$. Since this system of equations has no solutions, there exists no joint distribution $P_{ABCD}$ from which the pairwise marginals producing these correlations can be obtained. Such examples correspond to grandfather type paradoxes and cannot be modelled in frameworks that demand the existence of a valid joint probability distribution over all variables involved in a causal loop. On the other hand, examples of solid arrow directed cycles where the functional dependences of the loop variables admit solutions such as $A=B=C=D$ (with any probability) or the examples considered in \cite{Pearl2013} for other cyclic causal structures can be modelled in our framework. Additionally, there can also be affects causal loops that do not involve any solid arrows, for example through a concatenation of structures such as that of Figure~\ref{fig:example1}. We discuss causal loops in more detail in Appendix~\ref{appendix: doMechanisms}, also in the case of quantum causal structures.

\section{The framework, Part 2: Space-time}
\label{sec:spacetime}
\subsection{Space-time structure}
 We model space-time simply by a partially ordered set $\mathcal{T}$ without assuming any further structure/symmetries. A particular example of $\mathcal{T}$ is Minkowski space-time, where the partial order corresponds to the light-cone structure and the elements of $\mathcal{T}$ can be seen space-time coordinates in some frame of reference. Our results will only depend on the order relations of $\mathcal{T}$ and not on the representation of its particular elements. To make operational statements about $\mathcal{T}$, we must embed physical systems into it. In our case, we can only do so for the observed systems in the causal model which are random variables. We embed them in this space-time by assigning an element of $\mathcal{T}$ to specify the space-time location of each variable, and refer to such variables as \emph{ordered random variables} or ORVs. Here, the order of an ORV corresponds to that of the space-time $\mathcal{T}$ (and not of the causal model).\footnote{This can be seen as an abstract version of space-time random variables.}

\begin{definition}[Ordered random variable (ORV)]
\label{def: ORV}
Each ORV $\mathcal{X}$ is defined by a pair $\mathcal{X}:=(R(\mathcal{X}),O(\mathcal{X}))$ where $R(\mathcal{X})$ (equivalently denoted by the corresponding, non-calligraphic letter $X$) is a random variable and $O(\mathcal{X})\in \mathcal{T}$ specifies the location of $\mathcal{X}$ with respect to a partially ordered set $\mathcal{T}$.
\end{definition}

\emph{Notation:} We use $\prec$, $\succ$ and $\nprec\nsucc$ to denote the order relations for a given partially ordered set $\mathcal{T}$, where for $\alpha$, $\beta\in \mathcal{T}$, $\alpha \nprec\nsucc \beta$ corresponds to $\alpha$ and $\beta$ being unordered with respect to $\mathcal{T}$. This is not to be confused with $\alpha=\beta$ which corresponds to the two elements being equal. These relations carry forth in an obvious way to ORVs and we say for example that 2 ORVs $\mathcal{X}$ and $\mathcal{Y}$ are ordered as $\mathcal{X}\prec \mathcal{Y}$ iff $O(\mathcal{X})\prec O(\mathcal{Y})$. We will also use $\mathcal{X}=\mathcal{Y}$ as short hand for $X=Y$ and $O(\mathcal{X})=O(\mathcal{Y})$.

\begin{definition}
\label{definition: future}
The \emph{future} of an ORV is the set 
\begin{equation*}
    \mathcal{F}(\mathcal{X}):=\{\alpha \in \mathcal{T}: \alpha\succ O(\mathcal{X})\}.
\end{equation*}
 Then, we say that an ORV $\mathcal{Y}$ lies in the future of an ORV $\mathcal{X}$ iff $O(\mathcal{Y})\in \mathcal{F}(\mathcal{X})$. In a slight abuse of notation, we will simply write this as $\mathcal{Y}\in \mathcal{F}(\mathcal{X})$, which is equivalent to $\mathcal{X}\prec \mathcal{Y}$.
\end{definition}

 \begin{definition}
 \label{definition: incfuture}
 The \emph{inclusive future} of an ORV is the set 
 \begin{equation*}
    \overline{\mathcal{F}}(\mathcal{X}):=\{\alpha \in \mathcal{T}: \alpha\succeq O(\mathcal{X})\}.
\end{equation*}
 \end{definition}

Note that $\mathcal{X}\in \overline{\mathcal{F}}(\mathcal{X})$ but $\mathcal{X}\notin \mathcal{F}(\mathcal{X})$, hence the name ``inclusive'' future. Further, any probabilities written in terms of ORVs should be understood as being probability distributions over the corresponding random variables.

\subsection{Compatibility of a causal model with the embedding space-time}
\label{ssec: compat}

We have encountered two types of order relations, one which is encoded in the arrows $\longrsquigarrow$ of the causal structure and the other specified by the order relation $\prec$ of the space-time $\mathcal{T}$. The former, which can include directed cycles corresponds to a pre-order while the latter is by construction a partial order. We now describe the compatibility condition will tell us when the order provided by a causal model is compatible with the order provided by the space-time $\mathcal{T}$ at the operational level, here the classification of the arrows $\longrsquigarrow$ into solid and dashed arrows will play a role. In an operational theory, we must be able to consider scenarios where a random variable may be copied, distributed or acted upon. For ordered random variables, we need to consider a set of locations in the embedding partial order at which copies of RVs can be accessed for further information processing. To this effect, we first propose the following definitions before defining compatibility.

\begin{definition}[Copy of a RV]
\label{definition:copy}
A random variable $X'$ is called a copy of a random variable $X$ if $X=X'$ and $X$ affects $X'$. Hence every RV $X$ is a copy of itself, we call this the \emph{trivial copy} of $X$.
\end{definition}

For ordered random variables $\mathcal{X}$ and $\mathcal{X'}$, we will simply say $\mathcal{X}'$ is a copy of $\mathcal{X}$ whenever the corresponding RV $X':=R(\mathcal{X}')$ is a copy of the RV $X:=R(\mathcal{X})$.
\begin{definition}[Accessible region of an ORV]
\label{definition:accreg}
With each ordered random variable $\mathcal{X}:=(R(\mathcal{X}),O(\mathcal{X}))$ ordered with respect to some partially ordered set $\mathcal{T}$, we can associate a subset $\mathcal{R}_{\mathcal{X}}\subseteq \mathcal{T}$ called its \emph{accessible region}, such that any copy $\mathcal{X}'$ of $\mathcal{X}$ must belong to this set i.e., $O(\mathcal{X}')\in \mathcal{R}_{\mathcal{X}}$ for all copies $\mathcal{X}'$ of $\mathcal{X}$, and $\mathcal{R}_{\mathcal{X}}$ is the smallest subset of $\mathcal{T}$ that has this property. We say that $\mathcal{X}$ is accessible in $R(\mathcal{X})$.
\end{definition}
The accessible region has the property that $\mathcal{R}(\mathcal{X}')\subseteq \mathcal{R}(\mathcal{X})$ for all copies $\mathcal{X}'$ of $\mathcal{X}$ (since the copy of a copy is a copy), and $\mathcal{X}\in \mathcal{R}_{\mathcal{X}}$ for all ORVs $\mathcal{X}$.
\begin{remark}
\label{remark: subsets}
Definition~\ref{definition:accreg} allows us to naturally associate accessible regions to sets of ORVs in terms of the accessible regions of their individual members. The accessible region of a set $\mathcal{S}=\{\mathcal{X}_1,...,\mathcal{X}_k\}$ of ORVs is the smallest subset of $\mathcal{T}$ within which all the ORVs in the set can be jointly accessed i.e., $\mathcal{R}_{\mathcal{S}}=\bigcap\limits_{\mathcal{X}_i\in\mathcal{S}}\mathcal{R}(\mathcal{X}_i)$. Each set $\cS$ of ORVs can be thought of as being comprised of one copy of each of its members, and its accessible region is the region of space-time where a copy of each ORV in the set can be accessed.
\end{remark}

\begin{definition}[Embedding]
\label{definition: embedding}
An embedding of a set of RVs $S$ in a partially ordered set $\mathcal{T}$ produces a corresponding set of ORVs $\mathcal{S}$ by assigning a location $O(\mathcal{X})\in \mathcal{T}$, and an accessible region $\mathcal{R}_{\mathcal{X}}\subset \mathcal{T}$ to each RV $X$, which defines the associated ORV $\mathcal{X}=(X,O(\mathcal{X}))$.  
\end{definition}
We now define what it means for a causal model to be compatible with the embedding partial order $\mathcal{T}$ which captures the idea of no superluminal signalling with respect to the space-time $\mathcal{T}$. Intuitively, the definition corresponds to the condition that ``it is possible to signal \emph{everywhere} in the future and nowhere else''. This is not the same as ``it is possible to  signal only to the future (but not necessarily to every point in the future)''. We will see that this subtle difference plays a role in our results of the following sections (c.f. Remark~\ref{remark: subtlediff}).

\begin{definition}[Compatibility of a causal model with the embedding partial order ($\mathbf{compat}$)]
\label{definition: compatposet}
Let $\mathcal{S}$ be a set of ORVs ordered with respect to the partially ordered set $\mathcal{T}$. Then a causal model over $\mathcal{S}$ is said to be \emph{compatible} with the partial order $\mathcal{T}$ if the following conditions are satisfied:
\begin{enumerate}
    \item For all $\mathcal{X}\in \mathcal{S}$, $\mathcal{R}_{\mathcal{X}}=\overline{\mathcal{F}}(\mathcal{X})$, and
     \item For all subsets $\mathcal{S}_1, \mathcal{S}_2 \subseteq \mathcal{S}$, such that no proper subset of $\mathcal{S}_1$ affects $\mathcal{S}_2$, $$\mathcal{S}_1 \text{ affects } \mathcal{S}_2\Rightarrow\mathcal{R}_{\mathcal{S}_2}\subseteq\mathcal{R}_{\mathcal{S}_1}$$.
\end{enumerate}
\end{definition}
\begin{remark}
 In the second statement above, the condition that no proper subset of $\mathcal{S}_1$ affects $\mathcal{S}_2$ is required because whenever a set $\mathcal{S}$ of ORVs affects another set $\mathcal{S}_2$, any $\mathcal{S}_1\supset \cS$ will also trivially affect $\mathcal{S}_2$. 
 In such situations, one is only interested in the non-trivial affects relation $\cS$ affects $\cS_2$ (given that no further proper subsets of $\cS$ affect $\cS_2$) and not $\cS_1$ affects $\cS_2$, since $\cS_1\supset \cS$ could be arbitrarily large in this case.\footnote{In the case of single variables, $\cX$ affects $\cY$ implies that $\cY$ must be in the future of $\cX$ in order to be compatible with the space-time. We don't require $\cY$ to be in the future of every superset of $\cX$, which would trivially affect it.}
 
% Here, $\cS_1$ has a proper subset $\cS$ that affects $\cS_2$ we will only require the accessible region of $\cS_2$ to be contained in that of $\cS$ and not in In such situations, we only require $\mathcal{Y}$ to be in the future of $\mathcal{X}$ (the smallest subset that affects $\mathcal{Y}$ such that no proper subset of it does), and not in the joint future of any other subset containing $\mathcal{X}$ that would trivially affect it. 
 %Therefore, if we had a proper subset $\mathcal{S}'_1$ of $\mathcal{S}_1$ that affects $\mathcal{S}_2$ (without any proper subset of $\mathcal{S}'_1$ affecting $\mathcal{S}_2$), then we would require condition 2. of $\mathbf{compat}$ to hold for $\mathcal{S}'_1$ and $\mathcal{S}_2$ instead of $\mathcal{S}_1$ and $\mathcal{S}_2$. 
\end{remark}

%\begin{remark}
%$\mathbf{compat}$ and $\mathbf{compat^*}$ are clearly not equivalent. However if there exists a poset $\mathcal{T}$ with which a causal model is compatible according to $\mathbf{compat}$, then there also exists a poset $\mathcal{T}'$ with which the causal model is compatible according to $\mathbf{compat^*}$ (and vice-versa). The important point is that $\mathcal{T}$ and $\mathcal{T'}$ need not be the same. Going from $\mathbf{compat}$ with $\mathcal{T}$ to $\mathbf{compat^*}$ with $\mathcal{T}'$ would simply involve shrinking the future to coincide with the accessible region if the accessible region happens to be a strict subset (and not changing the future otherwise).
%\end{remark}

%\subsection{Examples}

\section{Necessary and sufficient conditions}
\label{sec: necsuff}
In this section, we derive necessary and sufficient conditions for a causal model to be compatible with a partial order $\mathcal{T}$ (Section~\ref{sec: nosig}) and for it to have no affects causal loops (Section~\ref{sec: noloops}). In \cite{Horodecki2019}, a set of necessary and sufficient conditions for no causal loops in the bipartite and tripartite Bell scenarios have been proposed. The results of this section identify implicit assumptions in these claims \cite{Horodecki2019} and provide necessary and sufficient conditions for compatibility with the space-time and for no affects causal loops for arbitrary causal structures. The other type of causal loops, functional causal loops cannot in general be detected operationally and therefore cannot be ruled out. We then discuss the relationships between the various conditions derived in our results in Section~\ref{ssec: STrelations} and consider \cite{Horodecki2019} in further detail in Section~\ref{sec: problems}.  
%however their definition of ``no causal loops'' has been stated only in words, and not in a precise manner which makes the claims particularly problematic. We discuss in Section~\ref{sec: problems}, while providing concrete counter-examples to their claims stemming from the observation that their argument cannot rule out fine-tuned, functional causal loops. Further, even if we take ``no causal loops'' to mean ``no affects causal loops'', it appears that the bipartite and relaxed tripartite no signalling conditions of \cite{Horodecki2019} are neither necessary nor sufficient for no causal loops in the bipartite and tripartite Bell scenarios respectively. 

\subsection{Conditions for compatibility with the embedding space-time}
\label{sec: nosig}

\begin{restatable}{theorem}{CompatST}[Necessary and sufficient conditions for compatibility with $\mathcal{T}$]
\label{theorem:poset}
    Given a causal model over a set of ORVs $\mathcal{S}$, the following condition ($\mathbf{cond}$) is necessary for the causal model to be compatible with the embedding partial order $\mathcal{T}$ according to $\mathbf{compat}$ (Definition~\ref{definition: compatposet}).
\begin{align}
\begin{split}
\label{eq: cond}
\forall \text{subsets }\mathcal{S}_1, \mathcal{S}_2 \subseteq \mathcal{S}\text{, such that no proper subset of } \mathcal{S}_1 \text{ affects } \mathcal{S}_2, \\
\bigcap\limits_{s_2\in \mathcal{S}_2}\overline{\mathcal{F}}(s_2) \nsubseteq \bigcap\limits_{s_1\in \mathcal{S}_1}\overline{\mathcal{F}}(s_1) \quad\Rightarrow\quad \mathcal{S}_1 \text{ does not affect } \mathcal{S}_2.
\end{split}
\end{align}
While $\mathbf{cond}$ alone is not sufficient for $\mathbf{compat}$, along with the following additional requirement it provides a sufficient condition for $\mathbf{compat}$
\begin{equation}
\label{eq: assump}
     \forall \mathcal{X}\in \mathcal{S}, \quad \overline{\mathcal{F}}(\mathcal{X})\subseteq\mathcal{R}_{\mathcal{X}}
\end{equation}
\end{restatable}
A proof of this theorem can be found in Appendix~\ref{appendix: ProofsJamming}.

\subsection{Conditions for no affects causal loops}
\label{sec: noloops}
\begin{restatable}{theorem}{NoLoops}[Necessary and sufficient condition for no affects causal loops]
\label{theorem: loops1}
 A necessary condition for a causal model over a set $S$ of RVs to have \text{no affects causal loops} is that there exists an embedding of $S$ in Minkowski space-time $\mathcal{T}$ such that the corresponding ORVs $\mathcal{S}$ satisfies the condition $\mathbf{cond}$ of Equation~\eqref{eq: cond}.
$\mathbf{cond}$ along with the additional assumption that any 2 distinct ORVs $\mathcal{X}$ and $\mathcal{Y}$ such that one affects the other, cannot share the same location in $\mathcal{T}$, are sufficient for having no affects causal loops in the causal model.
\end{restatable}

A proof of the above theorem can be found in Appendix~\ref{appendix: ProofsJamming}. Note that a subset of the above conditions, would already be sufficient for no affects causal loops, since these causal loops are defined in terms of a chain of affects relations involving single variables (and not sets of variables). Suppose that for every pair of affects relations in a causal model, $X$ affects $Y$ where $X$ and $Y$ are single variables, we can ensure that the corresponding ordered random variables satisfy $\cY\in \cF(\cX)$, this would already be sufficient to conclude that the causal model has no affects loops. However this is not sufficient for compatibility with the space-time since we may have $X$ affects $\{W,Z\}$ for some variables $W$ and $Z$ in the model that are not individually affected by $X$ such that $\overline{\mathcal{F}}(\mathcal{W})\bigcap \overline{\mathcal{F}}(\cZ)\not\subseteq \overline{\mathcal{F}}(\cX)$. Based on this, a third class of causal loops can be defined, those that can be operationally detected only when we consider affects relations of the form $S_1$ affects $S_2$ where at least one of $S_1$ and $S_2$ is a set of cardinality greater than one. An example would be the cyclic causal structure of Figure~\ref{fig: eqcyclemain} where $\{A,C\}$ affects $B$ and $B$ affects $\{A,C\}$. With our current definition~\ref{definition: causalloops}, this gets classified as a functional causal loop and not an affects loop. This does not, however, affect the results presented here and the further classification and characterisation of such loops may be left for future work. 

Keeping this in mind and applying Theorem~\ref{theorem: loops1} to the bipartite and tripartite Bell experiments respectively, where the parties are space-like separated, we have Corollaries~\ref{corollary: bipart} and \ref{corollary: tripart}.  %Note that Theorem~\ref{theorem: loops1} does not refer to the particular space-time configuration but these corollaries do, which will allow us to make important comparisons with the results of \cite{Horodecki2019} for these Bell scenarios. 

\begin{corollary}[Bipartite Bell scenario]
\label{corollary: bipart}
Consider a causal model over the observed variables $\{A,B,X,Y\}$ where $A$ and $B$ are exogenous, and $X$ and $Y$ share an unobserved common cause $\Lambda$. Suppose that $\cT$ corresponds to Minkowski space-time. Let the observed ORVs defined with respect to $\mathcal{T}$ be such that $\mathcal{A} \nprec\nsucc \mathcal{B}$, $\mathcal{A} \nprec\nsucc \mathcal{Y}$, $\mathcal{B} \nprec\nsucc \mathcal{X}$, $\mathcal{X} \nprec\nsucc \mathcal{Y}$, $\mathcal{A} \prec \mathcal{X}$ and $\mathcal{B} \prec \mathcal{Y}$ and the accessible region of every ORV coincides with its inclusive future. In this scenario, a set of sufficient conditions for the causal model to have no affects causal loops are
\begin{enumerate}
    \item $\mathcal{A}$ does not affect any subset of ORVs not containing $\mathcal{X}$.
      \item $\mathcal{B}$ does not affect any subset of ORVs not containing $\mathcal{Y}$.
      \item $\mathcal{X}$ and $\mathcal{Y}$ do not affect any other ORVs.
\end{enumerate}
1. and 2. are equivalent to the bipartite no signalling conditions (NS2) given the exogeneity of $A$ and $B$.
\begin{align}
\label{eq: bipartiteNS}
    \begin{split}
         P_{X|A}(x|a)&:=\sum_yP_{X,Y|A,B}(x,y|a,b)=\sum_y P_{X,Y|A,B}(x,y|a,b')\quad \forall x,a,b,b',\\
        P_{Y|B}(y|b)&:= \sum_xP_{X,Y|A,B}(x,y|a,b)=\sum_x P_{X,Y|A,B}(x,y|a',b)\quad \forall y,a,a',b\\
    \end{split}
\end{align}
Hence NS2 along with 3. are sufficient for no affects causal loops in this scenario.
\end{corollary}

\begin{corollary}[Tripartite Bell scenario]
\label{corollary: tripart}
Consider a causal model over the observed variables $\{A,B,C,X,Y,Z\}$ where $A$, $B$ and $C$ are exogenous, and $X$, $Y$ and $Z$ share an unobserved common cause $\Lambda$. Suppose that $\cT$ corresponds to Minkowski space-time. Let the observed ORVs defined with respect to $\mathcal{T}$ be such that $\mathcal{A} \prec \mathcal{X}$,  $\mathcal{B} \prec \mathcal{Y}$, $\mathcal{C} \prec \mathcal{Z}$, $\overline{\mathcal{F}}(\mathcal{X})\bigcap\overline{\mathcal{F}}(\mathcal{Z})\subseteq \overline{\mathcal{F}}(\mathcal{B})$, all other pairs of observed ORVs are unordered with respect to $\mathcal{T}$ and the accessible region of every ORV coincides with its inclusive future. In this scenario, a set of sufficient conditions for the causal model to have no affects causal loops are
\begin{enumerate}
    \item $\mathcal{A}$ does not affect any subset of ORVs not containing $\mathcal{X}$.
      \item $\mathcal{B}$ does not affect any subset of ORVs containing neither $\mathcal{Y}$, nor the set $\{\mathcal{X},\mathcal{Z}\}$.
          \item $\mathcal{C}$ does not affect any subset of ORVs not containing $\mathcal{Z}$.
      \item $\mathcal{X}$, $\mathcal{Y}$ and $\mathcal{Z}$ do not affect any other ORVs.
\end{enumerate}
Given the exogeneity of $A$, $B$ and $C$, 1., 2. and 3. are equivalent to the relaxed tripartite no-signaling conditions (NS3') of \cite{Horodecki2019}.
\begin{align}
\label{eq: tripartiteNS}
    \begin{split}
         P_{X,Y|A,B}(x,y|a,b)&:=\sum_zP_{X,Y,Z|A,B,C}(x,y,z|a,b,c)=\sum_z P_{X,Y,Z|A,B,C}(x,y,z|a,b,c')\quad \forall x,y,a,b,c,c'\\
          P_{Y,Z|B,C}(y,z|b,c)&:=\sum_xP_{X,Y,Z|A,B,C}(x,y,z|a,b,c)=\sum_x P_{X,Y,Z|A,B,C}(x,y,z|a',b,c)\quad \forall y,z,a,a',b,c\\
         P_{X|A}(x|a)&:=\sum_{y,z} P_{X,Y,Z|A,B,C}(x,y,z|a,b,c)=\sum_{y,z} P_{X,Y,Z|A,B,C}(x,y,z|a,b',c')\quad \forall x,a,b,b',c,c'\\
         P_{Z|C}(z|c)&:=\sum_{x,y} P_{X,Y,Z|A,B,C}(x,y,z|a,b,c)=\sum_{x,y} P_{X,Y,Z|A,B,C}(x,y,z|a',b',c)\quad \forall z,a,a',b,b',c\\
    \end{split}
\end{align}
Hence NS3' along with 4. are sufficient for no affects causal loops in this scenario.
\end{corollary}

An important point to note here is that the conditions are only sufficient and not necessary. This is because the Bell experiments considered in \cite{Horodecki2019} correspond to particular embeddings of the causal model where the measurement events are space-like separated, while our Theorem~\ref{theorem: loops1} only requires the existence of a suitable embedding which need not be this particular one. More explicitly, a violation of the no-signaling condition~\eqref{eq: bipartiteNS} in a Bell experiment is not necessarily a problem when the parties are not space-like separated, and if they are, they would need to be able to signal superluminally in more than one frame to create a causal loop from this ability to signal superluminally. But superluminal signalling in one frame need not imply the same for other frames. Hence the no-signaling conditions~\eqref{eq: bipartiteNS} and \eqref{eq: tripartiteNS} are are not necessary for ruling out causal loops in the bipartite and tripartite Bell scenarios, contrary to the claim of \cite{Horodecki2019}. We further discuss \cite{Horodecki2019} in the context of our framework in Section~\ref{sec: problems} providing explicit counter examples to these claims. Furthermore, note that these corollaries only rule out affects causal loops and not all causal loops, they also do not guarantee compatibility with the space-time. For example, a causal model whereby $\{A,X\}$ affects $Y$ and $\{B,Y\}$ affects $X$ while neither $A$ nor $X$ affect $Y$ and neither $B$ nor $Y$ affect $X$, would create a directed cycle (a functional causal loop) between $X$ and $Y$ and also lead to signalling outside the future for the space-time configuration of Corollary~\ref{corollary: bipart}. This is not ruled out by the conditions of the above corollaries, but can nevertheless be obtained from Theorems~\ref{theorem:poset} and~\ref{theorem: loops1}. We now present another noteworthy corollary that can be derived from Theorems~\ref{theorem:poset} and \ref{theorem: loops1}, a proof of which can be found in Appendix~\ref{appendix: ProofsJamming}.

\begin{restatable}{corollary}{looposet}[No affects causal loops and compatibility with space-time]
\label{corollary: loops}
A necessary condition for a causal model over a set $S$ of RVs to have no affects causal loops is that  $\exists$ an embedding of $S$ in Minkowski space-time $\mathcal{T}$ such that the causal model is compatible (Definition~\ref{definition: compatposet}) with $\mathcal{T}$.
This condition along with the additional assumption that any 2 distinct ORVs $\mathcal{X}$ and $\mathcal{Y}$ such that one affects the other, cannot share the same location in $\mathcal{T}$, are sufficient for having no affects causal loops in the causal model.
\end{restatable}

\begin{remark}
\label{remark: subtlediff}
We have seen two distinct necessary and sufficient conditions for no causal loops: a) there exists an embedding in $\mathcal{T}$ such that $\mathbf{cond}$ (Equation~\eqref{eq: cond}) holds (Theorem~\ref{theorem: loops1}), and b) there exists an embedding in $\mathcal{T}$ such that the causal model is compatible with $\mathcal{T}$ (Corollary~\ref{corollary: loops}). These two conditions namely $\mathbf{cond}$ and compatibility with $\mathcal{T}$ are not equivalent, which follows from Theorem~\ref{theorem:poset} due to the additional assumption required in one direction of the proof. The difference has to do with whether or not the accessible region coincides with the future for each ORV or is only a subset of it. The embedding for which a) holds may in general be different from the embedding for which b) holds, since an embedding also involves a specification of the accessible region (Definition~\ref{definition: embedding}). In the latter case, the accessible region must coincide with the future by definition of compatibility (Definition~\ref{definition: compatposet}), while in the former case, it need not. In the case that only a) holds, we can have a jamming scenario (Figure~\ref{fig: jamming}) where the jammed variables can be jointly accessed only within a strict subset $\mathcal{R}_{\cA\cC}$ of their joint future $\overline{\mathcal{F}}(\cA)\cap\overline{\mathcal{F}}(\cC)$. Then it is enough if this subset is contained in the future of the jamming variable $B$ and not the entire joint future. Note that modifying Definition~\ref{definition: compatposet} of compatibility to require $\mathcal{R}_{\cX}\subseteq\overline{\cF}(\cX)$ instead of $\mathcal{R}_{\cX}=\overline{\cF}(\cX)$, would not alter the fact that Theorem~\ref{theorem:poset} would require an additional assumption along with $\mathbf{cond}$ for one of the directions, it would only alter the direction for which this is required.
\end{remark}

%\begin{remark}[Free choice and exogeneity]
%\label{remark: freechoice}
%Free choice of a variable $B$ is typically defined with respect to a space-time structure, as the condition that the space-time random variable $\cB$ must be uncorrelated with any variables that are not in its future \cite{CR2013}. This definition of free choice has to be modified to consider the jamming scenario where $\cB$ can be correlated with a set of variables $\{\cX,\cZ\}$ that can only be jointly accessed in $\cB$'s future, even if the individual variables lie outside its future. \cite{Horodecki2019} proposes a definition of free choice for multi-partite Bell scenarios that allows for jamming. This definition however, does not easily generalise beyond the Bell scenarios. A natural way to define free choice in terms of the causal model and without reference to space-time, would be to demand independence of a free variable with all its non-descendants in the causal structure.....

%In this work, we have modelled free choice of a variable by its exogeneity in the causal model, without any reference to space-time. This ensures that the variable is not 
%\end{remark}
\subsection{Relationships between the conditions}
\label{ssec: STrelations}

\begin{figure}[t!]
    \centering
\begin{tikzcd}[arrows=Rightarrow, row sep=2cm, column sep=-2cm]
&\fbox{\href{eq: cond}{\textbf{cond}} $\land$ $\Big(\mathcal{X}$ \href{definition: affects}{affects} $\mathcal{Y}$ $\Rightarrow$ $O(\mathcal{X})\neq O(\mathcal{Y})\Big)$}\arrow{dl}{Thm.~\ref{theorem: loops1}}\arrow[red,degil,shift left=1.5ex]{dr}{\textcolor{red}{\textbf{II}}}&\\
   \fbox{No \href{definition: causalloops}{affects causal loops}}\arrow[red,degil,shift left=1.5ex]{ur}{\textcolor{red}{\textbf{I}}}\arrow[inner sep=-3pt, "Cor.~\ref{corollary: loops}" ']{rr}\arrow{d}{Thm.~\ref{theorem: loops1}}  &    &\fbox{$\exists$ a \href{definition: compatposet}{compatible embedding  in space-time}}\arrow[red,degil]{ul}{\textcolor{red}{\textbf{III}}}\arrow[red, degil, shift right=1.5ex, outer sep=2pt, "\textcolor{red}{\textbf{IV}}" ']{ll}\arrow{d}{Thm.~\ref{theorem:poset}}\arrow[shift left=1.5ex]{dll}{Thm.~\ref{theorem:poset}}\\
      \fbox{\href{eq: cond}{\textbf{cond}}}\arrow[red,degil,shift left=1.5ex]{u}{\textcolor{red}{\textbf{V}}}\arrow[red, degil]{urr}{\textcolor{red}{\textbf{VI}}}\arrow{dr}{\substack{Thm.~\ref{theorem:poset} \\ (\href{proof:poset}{proof})}}&  &  \fbox{\href{eq: cond}{\textbf{cond}} $\land$ $\Big(\overline{\mathcal{F}}(\mathcal{X})\subseteq\mathcal{R}_{\mathcal{X}}$ $\forall \mathcal{X}\in \mathcal{S}\Big)$}\arrow[shift left=1.5ex]{u}\arrow[shift left=1.5ex]{dl}{\substack{Thm.~\ref{theorem:poset} \\ (\href{proof:poset}{proof})}}\\
     & \fbox{$\mathcal{R}_{\mathcal{X}}\subseteq \overline{\mathcal{F}}(\mathcal{X}),\quad \forall \mathcal{X}\in \mathcal{S}$}\arrow[red,degil,shift left=1.5ex]{ul}{\textcolor{red}{\textbf{VII}}}\arrow[red,degil]{ur}{\textcolor{red}{\textbf{VIII}}}&
\end{tikzcd}
    \caption[Relationships between concepts associated with space-time structure]{\textbf{Illustrative summary of results pertaining to embeddings of causal models in a space-time.}}
    \label{fig: Ch6_Fig2}
\end{figure}
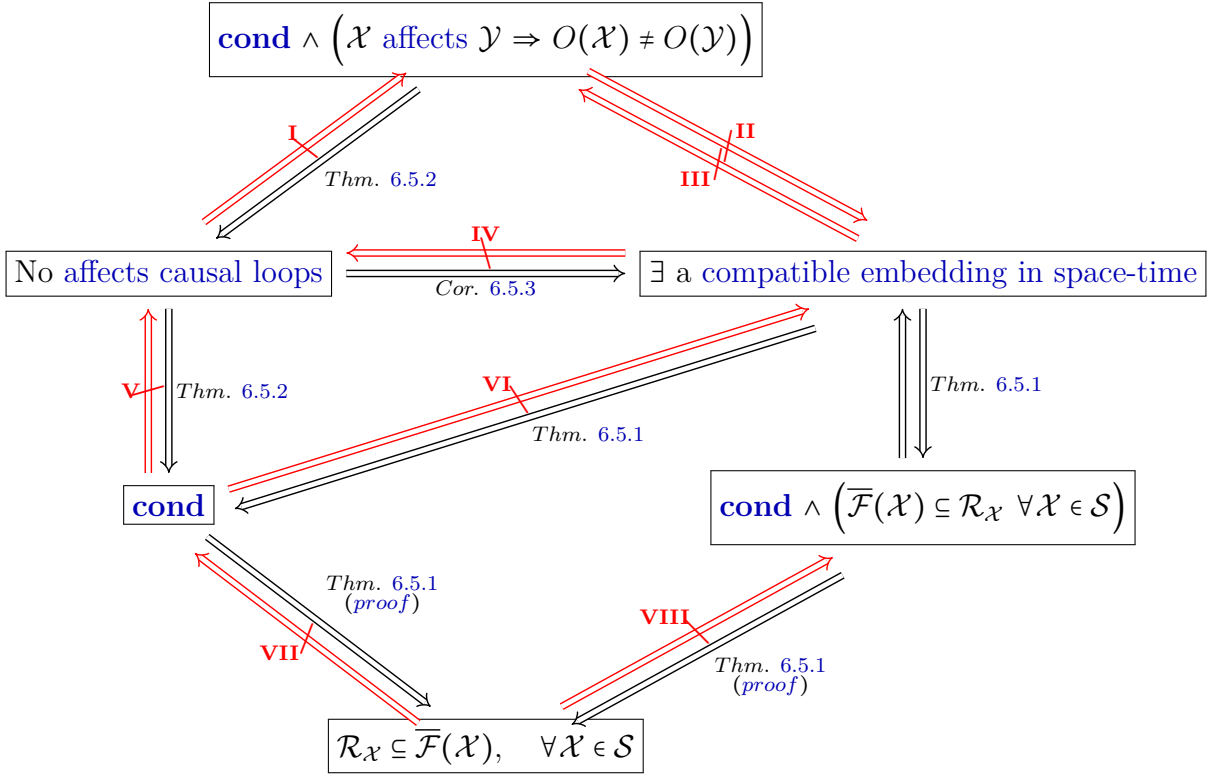

The results of Sections~\ref{sec: nosig} and~\ref{sec: noloops}, along with related (non-)implications are summarised in Figure~\ref{fig: Ch6_Fig2}. All the implications are covered by the results of the previous sections as explained in the Figure. Here, we describe counter examples to establish the 7 non-implications of the Figure, which can be grouped as follows.
\begin{enumerate}
 \item \textbf{Non-implications I, III: } While no affects causal loops implies \textbf{cond} by Theorem~\ref{theorem:poset}, it does not imply that $O(\cX)\neq O(\cY)$ whenever $\cX$ affects $\cY$. This is because we can have an acyclic causal model with non-trivial affects relations that is embedded trivially in space-time, i.e., where all the variables are embedded at the same space-time location. This explains non-implication I. Further, this trivial embedding of any causal model is clearly compatible with the space-time, which explains non-implication III.
    \item \textbf{Non-implications IV, V:} These are immediately established by embedding the 2-cycle affects loop (i.e., the two variable causal model with $X$ affects $Y$ affects $X$) in space-time such that $X$ and $Y$ are assigned the same space-time location $O(\mathcal{X})=O(\mathcal{Y})$. Such  causal loops between variables located at the same space-time point do not violate the compatibility condition~\ref{definition: compatposet} or lead to superluminal signalling, but they are rather uninteresting. %The idealisation of taking variables with an affects relation (such as the input and output of a local measurement) at the same space-time point is often made (including \cite{Horodecki2019}), but one needs to be careful about this idealisation in such scenarios.
    \item \textbf{Non-implications II, VI: } These are covered by the example provided in Remark~\ref{remark: subtlediff}: compatibility requires that the accessible region coincides with the future while \textbf{cond} only implies that the accessible region is a subset of the future.
    \item \textbf{Non-implications VII, VIII: } This is because the accessible region (Definition~\ref{definition:accreg}) of an ordered random variable $\cX$ is defined only in terms of copies of $\cX$. Therefore we can have an embedding for which $\cX$ affects another ORV $\cY$ which is not a copy of itself and does not lie in its future, even if the accessible region of each variable is contained in the variable's future. Such an embedding would violate \textbf{cond}.
\end{enumerate}

\section{Theories with superluminal influence without superluminal signalling}
\label{sec: jamming}

\subsection{Jamming and superluminal signalling}
\label{sec: jammingproof}

The jamming scenario analysed in the literature \cite{Grunhaus1996, Horodecki2019} has been considered only in the context of multipartite Bell scenarios, in particular the tripartite case, as discussed in Section~\ref{sec:motiveg}. There, the inputs of all the parties are considered to be freely chosen and the parties share a joint system $\Lambda$ that provides the initial correlations for the Bell scenario. Additionally, jamming allows the input of one party to jointly signal to the outputs of a set of other parties, without signalling to them individually. In the causal modelling paradigm, we have formalised freely chosen variables as parentless nodes in the causal structure, the notion of signalling through the affects relation, and have modelled $\Lambda$ as a common cause between outputs of all the parties. In the following, we will consider a particular class of jamming scenarios arising in the tripartite Bell experiment of \cite{Grunhaus1996, Horodecki2019}, those where Bob's input jams the outputs of Alice and Charlie, while the other inputs and outputs do not feature in the correlations. In this section, we simply denote these variables as $B$, $A$ and $C$ respectively.\footnote{Even though the naming convention is slightly different from the full tripartite Bell scenario of Figure~\ref{fig: trins}} In Figure~\ref{fig: jamming}, we have explained that a common cause $\Lambda$ between $A$ and $C$ is necessary for producing jamming correlations in a causal structure where the input $B$ is parentless and the outputs $A$ and $C$ have no outgoing arrows. In the remainder of this section, we explore the possibility of superluminal signalling using such a causal model.\footnote{When causally modelling Bell scenarios, it is standard to take input variables to be exogenous and output variables to have no outgoing arrows, and be related by a common cause, as in Figures~\ref{fig: Bell2} and~\ref{fig: trins}. However, one can explain the correlations and affects relations of the Bell scenario using alternative causal models whereby the inputs have incoming fine-tuned arrows, and/or outputs have outgoing fine-tuned arrows. These can also be analysed in our framework, but we will not consider them here.}

It would be illustrative to sketch a simple protocol that can be used to obtain the jamming correlations of Example~\ref{example: jamming}, that can lead to superluminal signalling in the space-time embedding considered in \cite{Grunhaus1996, Horodecki2019}. This protocol captures the main intuition behind Theorem~\ref{theorem: jamming} that will follow, and is illustrated in Figure~\ref{fig: jammingproof}. It also illustrates that Figure~\ref{fig: jamming} is not the unique causal structure that is compatible with the correlations and affects relations of Example~\ref{example: jamming}.

\paragraph{A protocol for jamming: }
 In Example~\ref{example: jamming}, suppose that $\Lambda$ is classical and all four variables, including $\Lambda$ are binary and uniformly distributed. The output $A$ is generated from $\Lambda$ by flipping its value whenever $B=1$ and outputting $A=\Lambda$ otherwise i.e., $A=\Lambda\oplus B$. $C$ is set to always be equal to $\Lambda$ irrespective of $B$. This immediately gives the required correlation $B=A\oplus C$ and reproduces the affects relations of the example. Even though $B$ does not affect $A$, it was physically sent to $A$ and even though $B$ jointly affects $A$ and $C$, it was not sent to $C$. This corresponds to the causal structure of Figure~\ref{fig: jamming} but with the arrow $B\xdashrightarrow{} C$ removed, as shown in Figure~\ref{fig: jammingproof}. Equivalently, the same correlations and affects relations can be generated by setting $A=\Lambda$ and $C=\Lambda\oplus B$ which would correspond to the causal structure of Figure~\ref{fig: jamming} but with the arrow $B\xdashrightarrow{} A$ removed instead. In this example, without additional information, we cannot distinguish between the three causal structures where $B$ has a dashed arrow to $A$, to $C$ or to both. All 3 causal structures (also illustrated in Figure~\ref{fig: jammingproof2}) are compatible with the correlations and the affects relations of the example. 
 
Note that $A$ and $\Lambda$ together are enough to determine $B$. Therefore, in a space-time configuration such as that of \cite{Grunhaus1996, Horodecki2019} where the space-time random variables $\cA$, $\cB$ and $\cC$ are pairwise space-like separated, and $\Uplambda$ is in the common past of $\cA$ and $\cC$, information about the freely chosen variable $\cB$ can already be available at the space-time location of $\cA$. This is outside the future of $\cB$, and can lead to superluminal signalling from the space-time location of $\cB$ to that of $\cA$. More generally, note that $\{B,\Lambda\}$ affects $A$ but neither $B$ nor $\Lambda$ affect $A$ (as denoted by the dashed arrows in Figure~\ref{fig: jammingproof}). By Theorem~\ref{theorem:poset}, $\cA$ must be in the joint future of $\cB$ and $\Uplambda$ for compatibility with the space-time. However, the space-like separation between $\cA$ and $\cB$ leads to a violation of this condition and hence to superluminal signalling, irrespective of the space-time location assigned to $\Lambda$. The following theorem (a proof of which can be found in Appendix~\ref{appendix: ProofsJamming}) establishes the possibility of superluminal signalling using an observed $\Lambda$ in a general class of jamming scenarios.

\afterpage{\FloatBarrier}
\begin{figure}[t!]
 \centering
  \subfloat[\label{fig:jammingsptime}]{\begin{tikzpicture}[scale=0.6]
   	\filldraw[black] (2,2) circle (1.5pt) node[anchor=north] {$\mathcal{A}$}; 
		\filldraw[black] (5,2) circle (1.5pt) node[anchor=north] {$\mathcal{B}$}; 
		\filldraw[black] (8,2) circle (1.5pt) node[anchor=north] {$\mathcal{C}$}; 	\filldraw[black] (5,-1.3) circle (1.5pt) node[anchor=north] {$\Uplambda$};
		\draw (2,2)--(6,6); \draw (2,2)--(0.2,3.8);\draw (8,2)--(9.8,3.8); \draw (8,2)--(4,6); 	\draw (5,2)--(9,6); \draw (5,2)--(1,6); 	\draw (2,2)--(6,-2); \draw (2,2)--(0.2,0.2); \draw (8,2)--(9.8,0.2); \draw (8,2)--(4,-2);
		\draw [fill=blue, fill opacity=0.3, draw=none] (5,5)--(4,6)--(6,6)--cycle; \draw [fill=red, fill opacity=0.3, draw=none] (5,-1)--(4,-2)--(6,-2)--cycle;
   \end{tikzpicture}} \qquad\subfloat[\label{fig: jammingobs}]{\begin{tikzpicture}[scale=0.8]
      \node[shape=circle,draw=black] (B) at (0,0.5) {B};
    \node[shape=circle,draw=black] (A) at (-2,2) {A};
    \node[shape=circle,draw=black] (C) at (2,2) {C};
      \node[shape=circle,draw=black] (L) at (0,-1.5) {$\Lambda$};
     \path [thick, dashed, arrows=-stealth] (B) edge (A);
     %\path [thick, dashed, arrows=-stealth] (B) edge (C);
      \path [thick, dashed, arrows=-stealth] (L) edge (A);
     \path [thick, arrows=-stealth] (L) edge (C);
      \end{tikzpicture}}\qquad\subfloat[\label{fig: jammingtable}]{\begin{tikzpicture}
 \node[align=center] at (0,0) {\setlength{\extrarowheight}{4pt}\setlength{\tabcolsep}{0.4em}\begin{tabular}{ | c | c | c | c | c| }
    \hline
   A  & B & C & $\Lambda$ & $P_{ABC}$ \\\hline
    0 & 0 & 0 & 0 & 1/4\\ 
    1 & 0 & 1 & 1 & 1/4\\
    0 & 1 & 1 & 0 & 1/4\\ 
    1 & 1 & 0 & 1 & 1/4\\
    \hline
  \end{tabular}};
 \end{tikzpicture}}
    \caption[A protocol that can lead to superluminal signalling in the jamming scenario]{\textbf{A simple protocol that can lead to superluminal signalling in the jamming scenario of \cite{Grunhaus1996, Horodecki2019}:} \textbf{(a)} Space-time embedding of the variables $A$, $B$, $C$ and $\Lambda$ suggested in \cite{Grunhaus1996, Horodecki2019}, where the joint future of the space-time variables $\cA$ and $\cC$ (blue region) is contained in the future of $\cB$. Here, $B$ corresponds to the input of a party, Bob while $A$ and $C$ are the outputs of Alice and Charlie respectively. In this configuration, \cite{Grunhaus1996, Horodecki2019} claim that there will be no superluminal signalling since $\cA$ and $\cC$ are jointly accessible only in the future of $\cB$ and $B=A\oplus C$ can not be computed from $A$ or $C$ individually. However, if $\Lambda$ is observed, then our protocol explained in the main text allows for superluminal signalling in this space-time configuration, contrary to this claim. In the protocol, $C=\Lambda$ and $B=A\oplus \Lambda$ freely chosen variable $B$ can already be determined at the location of $\cA$ which is outside the future of $\cB$ and hence can lead to superluminal signalling. \textbf{(b)}  The causal structure corresponding to the protocol. Note that this causal structure is missing the dashed arrow from $B$ to $C$ that is present in the causal structure of Figure~\ref{fig: jamming} even though both explain the same correlations and affects relations. \textbf{(c)} The correlations obtained from the protocol.}
    \label{fig: jammingproof}
\end{figure}
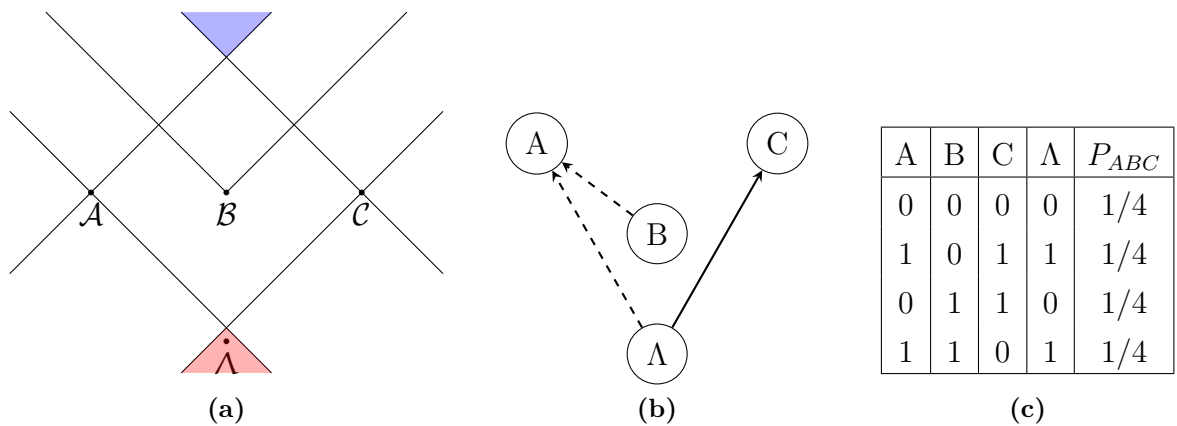
\begin{restatable}{theorem}{Jamming}
\label{theorem: jamming}
Consider a causal model over four nodes $A$, $B$, $C$ and $\Lambda$, all of which are observed. Let $B$ be exogenous, $A$ and $C$ have no outgoing arrows and $\Lambda$ be a common cause of $A$ and $C$. Further, suppose that $B$ affects the set $\{A,C\}$ without affecting its individual elements i.e., $B$ jams the correlations between $A$ and $C$. This causal model is not compatible with Minkowski space-time $\mathcal{T}$ i.e., leads to superluminal signalling when the variables $A$, $B$ and $C$ are embedded in $\mathcal{T}$ as proposed in \cite{Grunhaus1996, Horodecki2019}, irrespective of the space-time location of $\Lambda$. The proposed embedding is: $\mathcal{A} \nprec\nsucc \mathcal{B}$, $\mathcal{B} \nprec\nsucc \mathcal{C}$, $\mathcal{A} \nprec\nsucc \mathcal{C}$, $\overline{\mathcal{F}}(\mathcal{A})\bigcap\overline{\mathcal{F}}(\mathcal{C})\subseteq \overline{\mathcal{F}}(\mathcal{B})$, and the accessible region of each observed ORV coincides with its inclusive future. 
\end{restatable}

\begin{remark}
The classical protocol described in this section can also be extended to jamming scenarios where Bob jams non-classical correlations between Alice and Charlie. For example, let Alice and Charlie share a bipartite state, measure it using the inputs $A$ and $C$ and obtain the outcomes $X$ and $Z$, and let $B$ be an input of Bob that may be sent to one of them. Suppose, depending on $B$, we would like Alice and Charlie to share either correlations arising from measurements on the Bell state $\ket{\psi_0}=\frac{1}{\sqrt{2}}(\ket{00}+\ket{11})$ or those arising from the same measurements on a different Bell state,  $\ket{\psi_1}=\frac{1}{\sqrt{2}}(\ket{01}+\ket{10})$. For this, Alice and Charlie simply need to share the first Bell state but one of the parties, say Alice must perform a controlled NOT on her local subsystem controlled on the value of $B$, prior to the measurement. Then the parties would measure $\ket{\psi_{0}}$ whenever $B=0$ and $\ket{\psi_{1}}$ whenever $B=1$ as required. Similarly, $B$ can be used to decide whether Alice and Charlie share one relabelling of a PR box (Equation~\eqref{eq: PR}) or another.\footnote{Similarly, one can devise protocols to realise other types of jamming correlations by allowing the underlying mechanisms (like the choice of local operation prior to measurement) to depend on $B$ in a way that this does not reflect in the local statistics. Further one can also consider jamming correlations where Bob's output $Y$ is non-trivially involved in the process, which we do not analyse here as it is not directly relevant to the results of this thesis.} Note however that in a tripartite Bell scenario where all the parties are non-communicating, these correlations cannot be generated quantum mechanically or in boxworld, and therefore define a more general class of post-quantum theories \cite{Horodecki2019, Salazar2020}.
\end{remark}

\subsection{Comments on Horodecki and Ramanathan, Nat. Comms. 2019}
\label{sec: problems}

Here, we address the claim of \cite{Horodecki2019} (Propositions 2 and 3) that the no-signaling conditions NS2 (Equation~\eqref{eq: bipartiteNS}) and NS3' (Equation~\eqref{eq: tripartiteNS}) are necessary and sufficient for no causal loops in the bipartite and tripartite Bell scenarios respectively. The definition of no causal loops used by \cite{Horodecki2019} in the proof of this statement is the following, which we quote verbatim below.
\begin{quote}
    No causal loops occur, where a causal loop is a sequence of events, in which one event is among the causes of another event, which in turn is among the causes of the first event.
\end{quote}
Here ``events'' correspond to measurement events, i.e., an input and output pair of space-time random variables both of which are assigned the same space-time location, and ``cause'' does not seem to be defined explicitly. A rigorous formalisation of these definitions is offered by the present framework--- events correspond to ordered (or space-time) random variables (Definition~\ref{def: ORV}), causation and signalling (which are non-equivalent) are defined in terms of a causal model (Definition~\ref{def: cause} and \ref{definition: affects}) and different types of causal loops have been distinguished (Definition~\ref{definition: causalloops}). Based on this, it appears that neither direction of these claims hold. In the following we provide concrete counter-examples to illustrate our points, and explain how our results provide possible resolutions. Since the definitions of \cite{Horodecki2019} appear to be relatively ambiguous, we note that our framework may not be the only possible way to formalise them, but we are not aware of any other.

\bigskip
\paragraph{Definition of causal loops:} As we have seen, fine-tuning allows for two distinct types of causal loops (Definition~\ref{definition: causalloops}), the affects loops which can be detected at the observed level and the functional loops that may in principle be operationally undetectable (\emph{functional causal loops}). An example of the latter in the bipartite Bell scenario is provided in Figure~\ref{fig: counteregloop}, where the observed distribution $P_{XYAB}$ over the parties' inputs $A$ and $B$, and outputs $X$ and $Y$ exhibits no correlations i.e., $P_{XYAB}=P_XP_YP_AP_B$, and there are no affects relations i.e., none of the causal influences can be operationally detected, when $\Lambda$ is unobserved. Nevertheless, the causal structure involves a (functional) causal loop, which would qualify as a causal loop also according to the definition of \cite{Horodecki2019}. Clearly we could never rule out such causal loops based on any operational conditions, especially using conditions that only restrict the correlations (such as NS2 and NS3'). We can only rule out causal loops of the affects type as done in Theorem~\ref{theorem: loops1}, where have taken into account correlations as well as an operational notion of causation defined through active interventions. For this reason, even if we take ``no causal loops'' of \cite{Horodecki2019} to mean ``no affects causal loops'', the no signalling conditions $(NS2)$~\eqref{eq: bipartiteNS} and $(NS3')$~\eqref{eq: tripartiteNS} are neither necessary nor sufficient, which we discuss below with explicit examples.
\afterpage{\FloatBarrier}
\begin{figure}[t!]
    \centering
\subfloat[]{	\begin{tikzpicture}[line width=0.20mm, scale=1.3]
		\node [draw, circle, name=c1] at (0,0) {\small{$A$}}; \node [draw, circle, name=c2] at (2.2,0) {\small{$B$}}; 
		\node [draw, circle, name=c3] at (0,1.735) {$X$}; \node [draw, circle, name=c4] at (2.2,1.735) {\small{$Y$}};
		\node [draw, circle, name=c5, draw opacity=0] at (1.1,0) {$\Lambda$};
		\draw [thick, dashed, arrows={-stealth}] (c5) -- (c3); 
		\draw [thick, dashed, arrows={-stealth}] (c5) -- (c4); 
		  \path [thick, dashed, arrows=-stealth] (c3) edge[bend right=20] (c4);
		  \path [thick, dashed, arrows=-stealth] (c4) edge[bend right=20] (c3);
	\end{tikzpicture}  \label{fig: counteregloop}}\qquad\qquad\subfloat[]{	\begin{tikzpicture}[line width=0.20mm, scale=1.3]
		\node [draw, circle, name=c1] at (0,0) {\small{$A$}}; \node [draw, circle, name=c2] at (2.2,0) {\small{$B$}}; 
		\node [draw, circle, name=c3] at (0,1.735) {$X$}; \node [draw, circle, name=c4] at (2.2,1.735) {\small{$Y$}};
		\node [draw, circle, name=c5, draw opacity=0] at (1.1,0) {$\Lambda$};
		\draw [thick, arrows={-stealth}] (c1) -- (c3); \draw [thick, arrows={-stealth}] (c2) -- (c4); 
		\draw [thick, arrows={-stealth}] (c5) -- (c3); 
		\draw [thick,  arrows={-stealth}] (c5) -- (c4); 
		\draw [thick,  arrows={-stealth}] (c1) -- (c4); 
	\draw [thick,  arrows={-stealth}] (c1) -- (c3); 
		 
	\end{tikzpicture}  \label{fig: counteregnec}}\qquad\qquad\subfloat[]{		\begin{tikzpicture}[line width=0.20mm, scale=1.3]
		\node [draw, circle, name=c1] at (0,0) {\small{$A$}}; \node [draw, circle, name=c2] at (2.2,0) {\small{$B$}}; 
		\node [draw, circle, name=c3] at (0,1.735) {$X$}; \node [draw, circle, name=c4] at (2.2,1.735) {\small{$Y$}};
		\node [draw, circle, name=c5, draw opacity=0] at (1.1,0) {$\Lambda$};
		\draw [thick, dashed, arrows={-stealth}] (c5) -- (c3); 
		\draw [thick, arrows={-stealth}] (c5) -- (c4); 
		\draw [thick, dashed, arrows={-stealth}] (c3) -- (c4); 
		  	 \path [thick, dashed, arrows=-stealth] (c1) edge[bend right=20] (c3);
		  	 \path [thick, dashed, arrows=-stealth] (c3) edge[bend right=20] (c1);
		  	 %\draw [thick, dashed, arrows={-stealth}] (c2) -- (c4); 
	\end{tikzpicture}  \label{fig: counteregmmt}}
    \caption[Counter-examples to the claims of Ref. \cite{Horodecki2019}]{\textbf{Counter-examples to the claims of \cite{Horodecki2019}: } All variables are assumed to be binary in these examples. \textbf{(a)} \emph{Causal loops do not imply signalling: } Let the dependence of $X$ and $Y$ on their parents be $X=Y\oplus \Lambda$ and $Y=X\oplus \Lambda$, with $X$ and $Y$ being uniformly distributed.\footnote{In this case, both these dependences correspond to the same relation.} This, along with the d-separation condition (Definition~\ref{definition: compatdist}) yields $P_{XYAB}=P_XP_YP_AP_B$, and the no-signaling conditions~\eqref{eq: bipartiteNS} trivially hold. Nevertheless, we have a causal loop between $X$ and $Y$ that cannot be operationally detected when $\Lambda$ is unobserved. \textbf{(b)} \emph{Signalling does not imply causal loops: } When $A$ and $Y$ are embedded in space-time such that they are space-like separated, the causal model is not compatible (Definition~\ref{definition: compatposet}) with the space-time and leads to superluminal signalling. However, it has no causal loops. \textbf{(c)} \emph{Instantaneous measurements are problematic: } Let all the variables be uniformly distributed and related as $X=A\oplus Y$ ($X$ jams $A$ and $Y$), $A\oplus \Lambda=X$ ($A$ and $\Lambda$ collide into $X$). This gives $Y=\Lambda$ (hence the solid arrow). If $A$ and $X$ are assigned the same space-time location by idealising the local measurement to be instantaneous, then no superluminal signalling ensues despite the causal loop. Note that even if we had a solid arrow 2-cycle with $X$ affects $A$ affects $X$, this need not pose a problem (either to the free choice definition used in \cite{Horodecki2019} or to superluminal signalling) when they are embedded at the same location.}
\end{figure}

\bigskip
\paragraph{Sufficient conditions for no causal loops:} The no-signaling conditions $NS2$~\eqref{eq: bipartiteNS} and $NS3'$~\eqref{eq: tripartiteNS} are not sufficient for no causal loops (of either type) in the corresponding Bell scenarios, and there are a number of reasons for this. Firstly, in Theorem~\ref{theorem: jamming} we have shown that the jamming scenario in the space-time configuration of \cite{Horodecki2019} can lead to superluminal signalling if the common cause $\Lambda$ is an observed variable. The scenarios for which we have proven this result are a special case of the tripartite Bell experiment where the inputs of Alice and Charlie, and the output of Bob are ignored, and they satisfy the tripartite no-signaling conditions $NS3'$ of \cite{Horodecki2019}. This shows that  $NS3'$ alone are not sufficient for ruling out superluminal signalling and subsequent causal loops that can arise, we have also seen that there exists an explicit protocol (Figure~\ref{fig: jammingproof}) that achieves superluminal signalling when $\Lambda$ is observed.
Furthermore, including this assumption still does not suffice to rule out causal loops, even in the bipartite Bell experiment, as shown in Corollaries~\ref{corollary: bipart} and~\ref{corollary: tripart}. To rule out affects causal loops, we also require that the output variables do not affect any other variables. For an example of an affects causal loop that is not ruled out by the no-signaling conditions $NS2$ of the bipartite Bell scenario, simply consider a 2-cycle between the outputs $X$ and $Y$ such that $X \longrightarrow Y$ and $Y\longrightarrow X$ with $X=Y$. Since the inputs $A$ and $B$ play no role here, no-signaling is trivially satisfied despite the causal loop that allows $X$ to affect $Y$ and $Y$ in turn to affect $X$. In addition, the inclusion of these extra assumptions still falls short for ruling out all causal loops and superluminal signalling. For the latter, as mentioned in Section~\ref{sec: noloops}, we require further assumptions involving affects relations such as $\{A,X\}$ does not affect $Y$. We note that a previous work by Baumeler, Degorre and Wolf \cite{Baumeler2018} also illustrates that non-classical, non-signaling correlations in the bipartite Bell scenario admit an alternative explanation using causal loops, pointing to insufficiency of the no-signaling conditions for ruling out causal loops. %In our opinion, the problem arises because correlation and causation have not been clearly distinguished in \cite{Horodecki2019}, the no-signaling conditions only relate to the former and do not suffice to rule out signalling/causal loops which relate to the latter.

%For example, in the bipartite Bell scenario, the no-signaling conditions only forbid causal influence from Alice's input to Bob's output and vice versa, but they do not forbid a causal loop between the two outputs $X$ and $Y$ which are correlated, or the output of a party from retrocausally affecting their input. The additional implicit assumption is that $X$ and $Y$ do not affect any other variables, which would imply that the correlations between $X$ and $Y$ can only be explained by a common cause, ruling out causal loops between $X$ and $Y$ or between the input and output of each party. This assumption that the outputs of all parties do not affect any other variables (in this case, equivalent to having no outgoing arrows in the causal model), is implicit in \cite{Horodecki2019}, but not stated in their claims. Without this assumption the no-signaling conditions NS2 and NS3' are clearly not sufficient for no causal loops in the bipartite and tripartite Bell scenarios as claimed in \cite{Horodecki2019}.
\bigskip
\paragraph{Necessary conditions for no causal loops:} We also noted in Section~\ref{sec: jammingproof} that the no-signaling conditions are not necessary for no causal loops in the bipartite and tripartite Bell scenarios. Consider the causal structure of Figure~\ref{fig: counteregnec} where we have the usual bipartite Bell scenario, and in addition that Alice's input $A$ affects Bob's output $Y$. This violates the no-signaling conditions. The particular embedding of this causal model in Minkowski space-time $\cT$ where the space-time random variables $\cA$ and $\cY$ are space-like separated is not compatible with the space-time and leads to superluminal signalling. However, the causal structure is clearly acyclic and hence the causal model has no causal loops (in this case both of the affects or the functional kind). Further, there clearly exists a compatible embedding of the causal model in $\cT$, this would be one where $\cY$ is taken to be in the future of $\cA$ (in accordance with Corollary~\ref{corollary: loops}).
The fact that superluminal signalling does not imply causal loops has been pointed out in \cite{Horodecki2019} itself. If superluminal signalling were allowed only in one special reference frame, this cannot be used to create causal loops.\footnote{Though we have not explicitly considered reference frames in this work, they could in principle be taken into account by introducing a partially ordered set for every choice of reference frame, such that the space-time is characterised by a collection of partially ordered sets. And compatibility of a causal model with the space-time (Definition~\ref{definition: compatposet}) would correspond to compatibility (or no signalling outside the future) with all these partially ordered sets, such that signalling outside the light cone even in one frame would be a violation of compatibility.} The conclusion that the no-signaling conditions are necessary for no causal loops appears to be at odds with this observation, and the proof of \cite{Horodecki2019} seems to be implicitly assuming that superluminal signalling in one frame would enable such signalling in all frames.  
%This problem is addressed in our framework by noting that causation cannot be inferred only using conditions on the correlations (such as the no-signaling conditions), a notion of interventions is also required. Formalising this idea using the causal modelling approach for cyclic and fine-tuned scenarios has allowed us to derive necessary and sufficient conditions for no affects causal loops in arbitrary scenarios as done in Theorem~\ref{theorem: loops1}, and for compatibility with Minkowski space-time (or any space-time modelled by a partial order $\mathcal{T}$) as done in Theorem~\ref{theorem:poset}. 

%\afterpage{\FloatBarrier}

\bigskip
\paragraph{Instantaneous measurements:} As an idealisation, measurements are considered to be instantaneous in \cite{Horodecki2019}, such that the input and output variables of each party are embedded at the same space-time location. While this is an idealisation that is often made, in the present case it must be treated with caution, since causal loops (even of the affects type) between variables at the same space-time location do not necessarily lead to superluminal signalling (c.f. Theorem~\ref{theorem: loops1}, Figure~\ref{fig: counteregmmt}). In the bipartite Bell scenario, when Alice's input and output are at the same space-time location, this allows for an affects causal loop $A\longrightarrow X$ and $X\longrightarrow A$, while still satisfying free choice in the sense of: a free variable is not correlated with variables outside its (inclusive) future.
This is yet another reason for the insufficiency for the no-signaling conditions for ruling out causal loops.
From a physical standpoint, it would be natural to expect that in a causal theory, non-trivial operations cannot be performed instantaneously and some frameworks for causality, such as \cite{Portmann2017} explicitly forbid this. Note that the sufficiency proof of Theorem~\ref{theorem: loops1} explicitly forbids such instantaneous embeddings.
\bigskip

\paragraph{Causal structure vs space-time structure:} It is important to point out that the directed graphs presented here differ from those of \cite{Horodecki2019} in what they represent. In \cite{Horodecki2019}, arrows between space-time variables (which are always represented as solid) $\cX\longrightarrow \cY$ represent the space-time structure i.e., that $\cY$ is in the future of $\cX$. This does not necessarily mean that $\cX$ is a cause of $\cY$. In this thesis, arrows are used to represent the causal structures such that $\cX\longrightarrow \cY$ stands for $\cX$ is a cause of $\cY$ (Definition~\ref{def: cause}). If the causal model is compatible with the space-time, this would imply that $\cY$ is indeed in the future of $\cX$, but apriory this need not be the case in our framework. \cite{Horodecki2019} assumes that for any two space-time random variables $\cX$ is a cause of $\cY$ implies that $\cY$ is in the future of $\cX$. Again the definition of ``cause'' is important here, since this need not hold for fine-tuned causes. For example if $X$ is a fine-tuned cause of $Y$ (i.e., $X\xdashrightarrow{} Y$), $\cY$ being in the future of $\cX$ is not necessary for ruling out superluminal signalling in such cases (Theorem~\ref{theorem:poset}). Formalising this notion in terms of the affects relation (introduced in Definition~\ref{definition: affects}) we have  $\cX$ affects $\cY$ implies that $\cY$ is in the (inclusive) future of $\cX$. Note that this follows from our compatibility condition (Definition~\ref{definition: compatposet}). 
\bigskip
\paragraph{Free choice:} Free choice of inputs in the Bell scenario is crucial to the arguments of \cite{Horodecki2019} since without this, outputs of parties can be correlated with inputs of space-like separated parties through common causes, hence violating the no-signaling conditions without leading to causal loops. The commonly used definition of free choice proposed in \cite{CR2013} needs to be modified when considering jamming correlations in multipartite Bell scenarios as noted in \cite{Horodecki2019}. The former requires each party's input to only be correlated with variables in its future while the latter allows Bob's input in the tripartite Bell experiment to be correlated with Alice and Charlie's outputs jointly even though the each output is space-like separated to the input variable. Accordingly, the definition will change depending on the scenario considered, and does not generalise easily. Here we have modelled free choice of a random variable by taking it to be exogenous in the causal graph, which generalises readily to arbitrary causal structures. Though admittedly, this is not the only way to model free choice in causal structures.%\footnote{A weaker definition of free choice as opposed to exogenity would be to demand that a free variable is not affected by any subset of its non-descendants. This would allow non-exogenous variables to be considered as free (akin to superdeterminism), but would be operationally indistinguishable from the exogenous case and therefore would not alter any of the predictions made using the latter.}   
  Modelling free choice is not a major concern in this work since our main results, Theorems~\ref{theorem:poset} and~\ref{theorem: loops1} do not assume any form of free choice and apply to arbitrary causal models.

\begin{remark}
It can be shown that defining free choice through exogeneity would recover the free choice definitions of \cite{CR2013, Horodecki2019} in the Bell scenarios when the causal model is embedded compatibly in the space-time (using the results of Section~\ref{sec: noloops}). However, the converse is not true since the latter are defined in terms of correlations and the lack of correlation does not imply a lack of causation, specially when we allow for fine-tuning. Therefore the definitions are not equivalent. Other inequivalent definitions may also be possible. We leave for future work, a deeper exploration of the relationship(s) between the different notions of free choice that can arise in the presence of fine-tuning.
\end{remark}
%Furthermore, in the presence of fine-tuned correlations, it might not suffice to define free-choice only in terms of correlations as done in \cite{Horodecki2019}. As we have seen in the example of Friedman's thermostat (Section~\ref{sec:motiveg}) and Figure~\ref{fig:eg3}, it is possible for a variable (such as the inside temperature $T_I$) to be uncorrelated with all other variables in a causal model and nevertheless be affected by some of them. Such variables would certainly not be considered as ``freely chosen'' since their causes can be operationally detected through interventions (and not by observing the correlations alone). Modelling free choice through exogeneity rules out not only correlation but also causation, thereby avoiding this issue.

\section{Discussion}
\label{sec: discJamming}
%The work presented in this chapter is part of an ongoing project with plenty of scope for future work in different directions. Some of these questions are of immediate interest while the rest of broader and more long-term interest. The latter can reveal connections between our framework and some of the other approaches to causality discussed in Section~\ref{ssec: caus_other}. 

\paragraph{D-separation and affects relations: } The \emph{affects relation} (Definition~\ref{definition: affects}), based on the notion of interventions which is crucial for distinguishing between correlation and causation. In acyclic causal structures \cite{Pearl2009, Henson2014} and in classical cyclic causal structures \cite{Forre2017}, existing frameworks completely describe how the post-intervention distribution can be calculated, from the observed distribution and/or the underlying causal mechanisms. In non-classical cyclic causal structures, such a characterisation is not available. In Section~\ref{sec: causmod}, we have used the d-separation condition (Definition~\ref{definition: compatdist}) on the observed distribution to obtain a partial characterisation which suffices for the current purpose, but this does not fully specify the post-intervention distribution. In Appendix~\ref{appendix: doMechanisms}, we outline a possible method for doing so, given the underlying causal mechanisms. We note that this method may not always recover the d-separation condition~\ref{definition: compatdist}. In the classical cyclic case, it has been shown that the d-separation condition is recovered whenever all the variables are discrete and the causal mechanisms of the model satisfy a certain property known as \emph{ancestrally unique solvability} (discussed in Appendix~\ref{appendix: doMechanisms}). All the examples presented in the main chapter are either acyclic or cyclic and satisfy these properties. Therefore our main results are not impacted by making d-separation a defining condition. Defining the framework without this assumption, using the causal mechanisms as primitives would only generalise it, and could potentially be of independent interest as a general framework for causal modelling. Further, another observation made in Appendix~\ref{appendix: doMechanisms} is that the presence of causal loops could allow us to distinguish between a faithful, non-classical explanations vs unfaithful classical explanations (e.g., using non-local hidden variables) of quantum correlations, which cannot be operationally distinguished otherwise. Formalising this intuition would provide another interesting line of investigation stemming from this work.

\paragraph{Causal loops and paradoxes: } A full formalisation of our causal model framework, in a manner described in the above paragraph would allow us to make a precise comparison with the two main types of closed timelike curves or CTCs that have been proposed in the literature namely, Deutsch's CTCs (DCTCs) \cite{Deutsch1991} and post-selected CTCs (PCTCs) \cite{Bennett2005, Svetlichny2011, Lloyd2011a, Lloyd2011}. These CTCs have provided insights into complexity classes in information theory, known to have different computing powers \cite{Aaronson2008, Lloyd2011} and to provide different resolutions to the grandfather and unsolved theorem paradoxes \cite{Lloyd2011}. In our framework, grandfather type paradoxes are forbidden by the assumption that a valid joint probability distribution observed variables indeed exists. An example of a paradoxical scenario is a 2-cycle between $X$ and $Y$ where the influence $X\longrightarrow Y$ defines the functional dependence $Y=X$ and $Y\longrightarrow X$ gives the dependence $X\neq Y$. These equations are mutually inconsistent and there is no joint distribution $P_{XY}$ compatible with these dependences.\footnote{This is similar in structure to the Liar's paradox, e.g., ``this statement is false'' or more generally a liar cycle. Interestingly, this will crop up again in our discussion of multi-agent paradoxes and contextuality in Chapter~\ref{chapter: PRdoxespaper}. The striking similarities between such paradoxes in philosophy, time-travel, quantum contextuality and multi-agent scenarios have also been pointed out in \cite{Dourdent2020}.} This consistency condition appears to be similar in spirit to that of PCTCs where the paradoxical scenario is never compatible with the post-selection i.e., occurs with zero probability. The unproved theorem paradox on the other hand, can depend on how the framework is formalised. For example, in classical cyclic causal models, an assumption regarding the unique solvability of the underlying functional dependences is often made. In particular, this could be seen as the requirement that any information involved in a loop (such as the unproved theorem) must be fully determined by the mechanisms of the causal model thereby eliminating the paradox of a proof that ``came from nowehere''. 

\paragraph{Causal inference in the presence of fine-tuning: } The \emph{no fine-tuning} assumption, in some form or the other is required by all causal discovery algorithms, which are algorithms that infer an unknown causal structure from observed correlations, interventions and other additional information that may be available. Relaxations to the assumption have been considered where certain forms of fine-tuning have been allowed \cite{Zhalama2017}. In the general case where arbitrary amount of fine-tuning is allowed, it can be fundamentally impossible to infer the causal structure since we have an arbitrary number of fine-tuned causal arrows that produce absolutely no observable effects. Take for instance the example of Figure~\ref{fig: counteregloop}, where there is a fine-tuned causal loop even though no correlations can be seen between the observed variables even when we intervene on them. Therefore, to address the problem of causal inference in our framework i.e., in the presence of cycles, fine-tuning and non-classical systems, we would need to make certain assumptions on the kind of fine-tuning that is allowed. One reasonable assumption could be that whenever there is a fine-tuned arrow between two variables $X\xdashrightarrow{} Y$ that cannot be detected when considering those variables alone, the causal influence must become detectable when we consider additional variables along with $X$ and $Y$, as is the case in the jamming and collider examples of Section~\ref{ssec: examples}. This is another interesting direction for future work.

\paragraph{Jamming theories and physical principles: } An important motivation for analysing post-quantum theories and jamming type scenarios constrained by relativistic principles is to gain deeper insights into the physical principles that single out quantum theory. One property of jamming theories that becomes suggested by our results is that they would involve some form of fine-tuning as well as variables and influences that must remain fundamentally inaccessible to observation. Even though jamming correlations in the space-like separated configuration of \cite{Grunhaus1996, Horodecki2019} are post-quantum phenomena, they are similar in spirit to non-local hidden variable explanations of quantum theory such as the De broglie-Bohm pilot wave theory \cite{Bohm1952} which also involve fine-tuned influences. Here the pilot wave which is not operationally accessible leads to the operational predictions of quantum theory, but can involve non-local or superluminal influences at the ontological level. An interesting future direction would be to formalise the concept of jamming theories in a similar manner to generalised probabilistic theories and consider other physical properties thereof. One difficulty in doing this is that the post-quantum jamming scenario requires a particular space-time configuration, while GPTs can be formalised independently of a space-time structure. Disentangling the notions of causality and space-time through operational considerations, as done here, could be seen as first step towards such a more general characterisation of jamming theories.

\paragraph{Indefinite causal orders: } Going beyond the notion of a fixed (but possibly unknown) causal structure, causal orders that may be fundamentally indefinite has been considered \cite{Oreshkov2012}. Here processes are said to be \emph{causally separable} if they admit an explanation in terms of a classical mixture of fixed causal order processes, and \emph{causally non-separable} otherwise. Even though our framework assumes a fixed causal and space-time structure, we note the following point regarding causally separable processes. Consider a scenario where the causal order $X\longrightarrow Y$ or $Y\longrightarrow X$ between two classical random variables is determined probabilistically by a third, binary variable $\Lambda$. Let the first causal order correspond to the dependence $Y=X$ and the second correspond to the dependence $X=Y\oplus 1$ for binary $X$ and $Y$. This can be modelled in the causal structure of Figure~\ref{fig: counteregloop} with the dashed arrows replaced with solid arrows, considering only the variables $X$, $Y$ and $\Lambda$. The causal mechanisms would be $X=\Lambda. (Y\oplus 1) \oplus (\Lambda\oplus 1). E_X$ and  $Y=(\Lambda\oplus 1). X \oplus \Lambda. E_Y$, where $E_X$ and $E_Y$ are mutually independent, uniformly distributed binary variables. Conditioned on $\Lambda=0$, $X$ affects $Y$ but $Y$ does not affect $X$ i.e., $P(X|\text{do}(Y=y),\Lambda=0)=P(X|\Lambda=0)$ $\forall y$ and similarly $P(Y|\text{do}(X=x),\Lambda=1)=P(Y|\Lambda=1)$ $\forall x$. However, when $\Lambda$ is not given, it can be checked that $X$ affects $Y$ \emph{and} $Y$ affects $X$ which is an affects causal loop.\footnote{Note that the contradictory conditions $X=Y$ and $Y=X\oplus 1$ are not a problem here since they never have to be satisfied simultaneously.} At first sight this might seem counter-intuitive to the idea that classical mixtures of causal orders should be implementable in a lab by flipping a coin to decide the order of two operations. Such a physical implementation corresponds to a situation where one operation (say, that which generates $X$) is performed at a time $t_1$ and the other (which generates $Y$) at time $t_2>t_1$ when $\Lambda=0$ while the second operation is performed at $t_1$ and first at $t_2$ when $\Lambda=1$ i.e., it is a directed acyclic graph over 4 variables, with $X_{t_1}\longrightarrow Y_{t_2}$, $Y_{t_1}\longrightarrow X_{t_2}$. This suggests the importance of the space-time embeddings in understanding the physicality of such processes. It would be interesting to analyse the quantum counterpart of this example, namely the quantum switch \cite{Chiribella2013} (which implements a quantum controlled superposition of the orders) in a similar manner, through a suitable generalisation of our framework. Further, indefinite causal orders that violate so-called
\emph{causal inequalities} \cite{Oreshkov2012} have also been proposed in frameworks that do not assume a fixed background space-time structure (as we have assumed in the present work). However such frameworks appear to disallow certain types of causal loops that are allowed by ours, and it would also be interesting to consider whether genuine\footnote{When no assumptions are made, causal inequalities, like Bell inequalities can be trivially violated. Identifying a genuine violation hence depends on the assumptions under which the causal inequality is derived, and these are not yet completely formalised for general scenarios.} causal inequality violations can be modelled in our framework.

\paragraph{Indefinite space-time locations: } Physically, it is possible to implement scenarios where a quantum system is superposed not only between spatial locations, but also in time. These spatio-temporal superpositions are required in the physical implementation of the quantum switch mentioned above \cite{Procopio2015, Rubino2017, Portmann2017}. It would therefore be of interest to generalise our framework to also consider unobserved systems that may be space-time delocalised. The framework currently uses the standard quantum formalism and the Born rule which only apply to quantum states and composite systems defined at a single instant of time. Generalisations to the standard formalism that define quantum states over space and time have been proposed and a particularly relevant one for this purpose would be the causal box formalism of \cite{Portmann2017} which models quantum information processing mechanisms that act on such systems. In an ongoing collaboration with members of ETH Zurich\footnote{Vilasini,  V.,  del  Rio,  L. and Renner, R. \emph{Causality in definite and indefinite space-times.} In preparation (2020). \definecolor{mylinkcolor}{rgb}{0,0,0}  \url{https://wdi.centralesupelec.fr/users/valiron/qplmfps/papers/qs01t3.pdf}\definecolor{mylinkcolor}{rgb}{0,0,0.7} }, we explore the relations between the causal box formalism and indefinite causality, outlining a way to recover a probability rule analogous to the Born rule in certain cases. This suggests one possible way to generalise the results of the current chapter to situations where the systems are delocalised in space and time.

\section{Appendix}
\subsection{Do-conditionals from causal mechanisms}
\label{appendix: doMechanisms}

In Section~\ref{sec: causmod} we have outlined how interventions and do-conditionals (i.e., the post intervention distribution) are defined in our framework, and Theorem~\ref{theorem:dorules} provides some conditions under which the post and pre intervention distributions can be related. Ideally though, one would expect that it should be possible to fully specify the post-intervention distribution if we are given all the underlying causal mechanisms involved in the causal structure. For example, in the classical case, the structural equations of the causal model \cite{Pearl2009} provide these causal mechasisms. Here for each node $X$ in the causal structure, the dependence of $X$ on its parents par$(X)$ corresponds to a stochastic map, which can be written in terms of a deterministic function $X=f_X(\text{par}(X),E_X)$ by including an additional exogenous and unobserved error variable $E_X$ for each node $X$. This is called a \emph{structural equation}. If the structural equations for all the nodes are known, then the complete post-intervention distribution can be calculated. This has been shown to be the case for classical cyclic causal models in \cite{Forre2017}. An intervention $\text{do}(x)$ on $X$, corresponds to updating the structural equation for $X$ to $X=x$ while keeping the remaining structural equations the same. Another important result for the classical case derived in \cite{Forre2017} is that the d-separation property or the \emph{global directed Markov condition} of Definition~\ref{definition: compatdist} is recovered whenever all the random variables are discrete and the structural equations of the causal model satisfy a property known as \emph{ancestrally unique solvability} (anSEP). Roughly, this property demands that the structural equation for each node must admit a unique solution given the values of the node's ancestors. We need not define this concept formally for our purposes here.

Our goal would be to extend these ideas to quantum and post-quantum cyclic causal structures, where the causal mechanisms involve measurements and transformations on non-classical systems, which cannot be expressed using deterministic structural equations. In the non-classical case, it is unclear what conditions allow for the d-separation condition to be recovered. Even to make this question precise in the non-classical case, one would need to specify the analog of structural equations for such causal models which is an open problem. Here, we present a possible method for achieving this by explaining it using the following example and sketching how it might generalise to a larger class of causal models. %Formalising this approach to construct a complete framework for non-classical cyclic and fine-tuned causal models is a subject of our ongoing and planned future work, which is also of independent interest apart from its relation to the work presented here. 

\begin{figure}[t!]
    \centering
\subfloat[]{\begin{tikzpicture}[scale=0.8]
      \node[shape=circle,draw=black] (A) at (0,0.5) {A};
    \node[shape=circle,draw=black] (X) at (-2,2) {X};
    \node[shape=circle,draw=black] (Y) at (2,2) {Y};
     \node[shape=circle,draw=black] (B) at (0,3.5) {B};
      \node[shape=circle,draw=none] (L) at (0,-1.5) {$\Lambda$};
     \draw[decorate, decoration={zigzag, segment length=+6pt, amplitude=+.95pt,post length=+4pt}, arrows={-stealth}, thick] (A) -- (X);
     \draw[decorate, decoration={zigzag, segment length=+6pt, amplitude=+.95pt,post length=+4pt}, arrows={-stealth}, thick](X) -- (B);
      \draw[decorate, decoration={zigzag, segment length=+6pt, amplitude=+.95pt,post length=+4pt}, arrows={-stealth}, thick] (B) -- (Y); \draw[decorate, decoration={zigzag, segment length=+6pt, amplitude=+.95pt,post length=+4pt}, arrows={-stealth}, thick](Y) -- (A);
       \draw[decorate, decoration={zigzag, segment length=+6pt, amplitude=+.95pt,post length=+4pt}, arrows={-stealth}, thick] (L) -- (X);  \draw[decorate, decoration={zigzag, segment length=+6pt, amplitude=+.95pt,post length=+4pt}, arrows={-stealth}, thick] (L) -- (Y);
    \end{tikzpicture}\label{fig: Qcycle1}}\qquad\subfloat[]{\begin{tikzpicture}[scale=0.8]
      \node[shape=circle,draw=black] (A) at (0,0.5) {A};
         \node[shape=circle,draw=black] (As) at (-2,-1.5) {$A^*$};
    \node[shape=circle,draw=black] (X) at (-2,2) {X};
    \node[shape=circle,draw=black] (Y) at (2,2) {Y};
     \node[shape=circle,draw=black] (B) at (0,3.5) {B};    \node[shape=circle,draw=black] (Bs) at (2,-1.5) {$B^*$};
      \node[shape=circle,draw=none] (L) at (0,-1.5) {$\Lambda$};
     \draw[decorate, decoration={zigzag, segment length=+6pt, amplitude=+.95pt,post length=+4pt}, arrows={-stealth}, thick] (As) -- (X);
     \draw[decorate, decoration={zigzag, segment length=+6pt, amplitude=+.95pt,post length=+4pt}, arrows={-stealth}, thick] (X) -- (B);
      \draw[decorate, decoration={zigzag, segment length=+6pt, amplitude=+.95pt,post length=+4pt}, arrows={-stealth}, thick] (Bs) -- (Y); \draw[decorate, decoration={zigzag, segment length=+6pt, amplitude=+.95pt,post length=+4pt}, arrows={-stealth}, thick](Y) -- (A);
       \draw[decorate, decoration={zigzag, segment length=+6pt, amplitude=+.95pt,post length=+4pt}, arrows={-stealth}, thick] (L) -- (X);  \draw[decorate, decoration={zigzag, segment length=+6pt, amplitude=+.95pt,post length=+4pt}, arrows={-stealth}, thick] (L) -- (Y);
    \end{tikzpicture}\label{fig: Qcycle2}}\qquad\subfloat[]{\begin{tikzpicture}[scale=0.8]
         \node[shape=circle,draw=black] (A) at (-2,3.5) {A};
    \node[shape=circle,draw=black] (X) at (-2,1) {X};
    \node[shape=circle,draw=black] (Y) at (2,1) {Y};
  \node[shape=circle,draw=black] (B) at (2,3.5) {B};
      \node[shape=circle,draw=none] (L) at (0,-1.5) {$\Lambda$};
     \draw[decorate, decoration={zigzag, segment length=+6pt, amplitude=+.95pt,post length=+4pt}, arrows={-stealth}, thick] (A) -- (X); \draw[decorate, decoration={zigzag, segment length=+6pt, amplitude=+.95pt,post length=+4pt}, arrows={-stealth}, thick](B) -- (Y);
       \draw[decorate, decoration={zigzag, segment length=+6pt, amplitude=+.95pt,post length=+4pt}, arrows={-stealth}, thick] (L) -- (X);  \draw[decorate, decoration={zigzag, segment length=+6pt, amplitude=+.95pt,post length=+4pt}, arrows={-stealth}, thick] (L) -- (Y);
       \draw[thick, dashed, -stealth] (A)--(Y); \draw[thick, dashed, -stealth] (B)--(X);
    \end{tikzpicture}\label{fig: Qcycle3}}\qquad\qquad\subfloat[]{\begin{tikzpicture}
 \node[align=center] at (0,0) {\setlength{\extrarowheight}{12pt}\setlength{\tabcolsep}{0.7em}\begin{tabular}{ | c | c | c | c | c| c | c | }
    \hline
   X  & Y & A & B & Measurements, outcomes & $P^{QM}_{XYAB}$ & $P_{XYAB}$\\\hline
    0 & 0 & 0 & 0 & $\sigma_Z\otimes \sigma_Z$, $(0,0)$ &$\frac{1}{8}$&$\frac{1}{3}$\\ 
    0 & 0 & 1 & 0 & $\sigma_X\otimes \sigma_Z$, $(+,1)$&$\frac{1}{16}$&$\frac{1}{6}$\\
    1 & 0 & 0 & 0 & $\sigma_Z\otimes \sigma_X$, $(1,+)$&$\frac{1}{16}$&$\frac{1}{6}$\\ 
    1 & 1 & 1 & 1 & $\sigma_X\otimes \sigma_X$, $(-,-)$&$\frac{1}{8}$&$\frac{1}{3}$\\[4pt]
    \hline
  \end{tabular}\label{fig: Qcycle4}};
 \end{tikzpicture}}
    \caption[A cyclic quantum causal model]{\textbf{A cyclic quantum causal model: } \textbf{(a)} A cyclic variation of the bipartite Bell causal structure (Figure~\ref{fig: Bell2}). \textbf{(b)} A method to calculate the observed distribution of (a) when $\Lambda$ is non-classical involves this intermediate causal structure. This is obtained from (a) by copying the nodes $A$ and $B$ and removing the directed cycle as shown. This gives an acyclic causal structure for which the distribution $P_{XYABA^*B^*}$ can be calculated using known methods. Then, post-selecting on $A=A^*$ and $B=B^*$ gives the distribution $P_{XYAB}$ for the original cyclic causal structure of (a). \textbf{(c)} A classical causal, fine-tuned structure that can generate, all non-classical correlations of the bipartite Bell causal structure. Creating a causal loop in this case by adding the arrows $X\longrsquigarrow B$ and $Y\longrsquigarrow A$ does not lead to the same predictions as (a), which corresponds to adding these arrows to the original Bell causal structure. This method explained in the main text. \textbf{(d)} The table provides the observed distribution for Example~\ref{example: qloop} calculated using the proposed method. The only values of $A$, $B$, $X$ and $Y$ that are compatible with the loop conditions $A=Y$ and $B=X$ are those listed here, and the fifth column lists the measurements and outcomes that these values correspond to, according to Example~\ref{example: qloop}. $P_{XYAB}^{QM}$ denotes the probabilities of the measurements and outcomes listed in the fifth column calculated using the Born rule. These values are sub-normalised, and upon renormalisation, the observed distribution $P_{XYAB}$ for the cyclic causal structure (a) is obtained. Note that the d-separation condition~\ref{definition: compatdist} is satisfied in this case.}
\label{fig: Qcycle}
\end{figure}
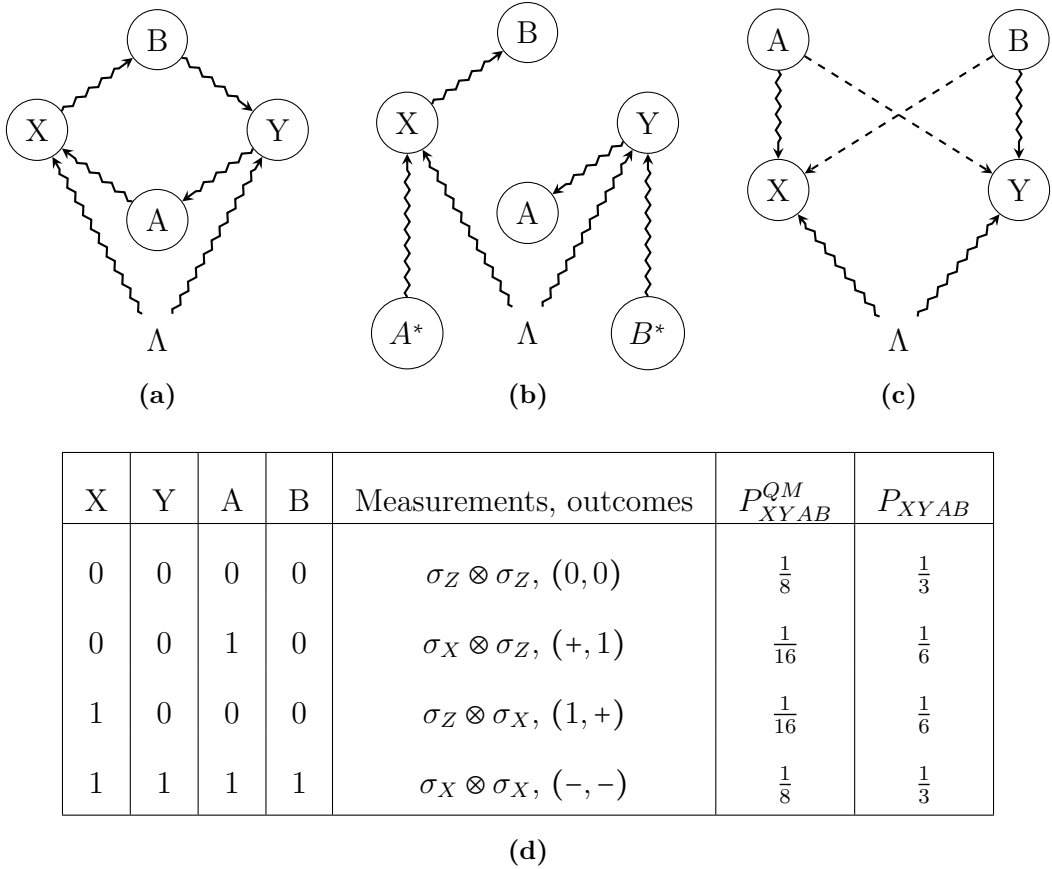

\begin{example}[A quantum cyclic causal model]
\label{example: qloop}
Consider the cyclic variation of the bipartite Bell causal structure illustrated in Figure~\ref{fig: Qcycle1}. Let the common cause $\Lambda$ correspond to the Bell state $\ket{\psi_{\Lambda}}=\frac{1}{\sqrt{2}}(\ket{00}+\ket{11})$. Suppose that $A$ and $B$ are the settings of local measurements on the two subsystems such that when these variables take the value $0$, it denotes a $\sigma_Z$ or computational basis ($\{\ket{0},\ket{1}\}$) measurement on the associated subsystem, and the value $1$ denotes a $\sigma_X$ or Hadamard basis ($\{\ket{+},\ket{-}\}$)  measurement. $X$ and $Y$ are the binary outcomes of these measurements where 0 and 1 for the Hadamard basis measurements denote the outcomes $+$ and $-$ respectively. The additional constraints coming from the causal loop are that $B=X$ and $A=Y$. This specifies all the causal mechanisms, how do we calculate the observed distribution $P_{XYAB}$?
\end{example}

\paragraph{A method based on post-selection: } One method is to first calculate the observed correlations for the specified state and measurements in the original Bell scenario (Figure~\ref{fig: Bell2}), and then post-select on the observations that obey the loop conditions $B=X$ and $A=Y$. More formally, this corresponds to transforming the original cyclic causal structure of Figure~\ref{fig: Qcycle1} to the acyclic causal structure of Figure~\ref{fig: Qcycle2} by cutting off the edges $A\longrsquigarrow X$ and $B\longrsquigarrow Y$ and replacing them with the edges $A^*\longrsquigarrow X$ and $B^*\longrsquigarrow Y$ by introducing two exogenous nodes $A^*$ and $B^*$. Then the inputs $A^*$ and $B^*$ and outputs $X$ and $Y$ along with the shared system $\Lambda$ define a Bell scenario, while the variables $A=Y$ and $B=X$ can simply be seen as local post-processings of the outcomes. We can then calculate the observed probabilities for this acyclic causal structure using the Born rule, and post-select on $A^*=A$ and $B^*=B$, which effectively achieves the post-selection $A=Y$ and $B=X$ in the original Bell scenario (Figure~\ref{fig: Bell2}). These probabilities will not be normalised and one has to renormalise them to obtain the observed distribution $P_{XYAB}$. This is calculated in Figure~\ref{fig: Qcycle} and can be used to find all the affects relations. An intervention on $A$ would cut off the arrow from $Y$ to $A$. We can then deduce that $A$ does not affect $X$ as follows. Since $A$ is effectively exogenous in the post-intervention causal structure and $\Lambda$ is a Bell state, $A$ will be uncorrelated with $X$ here i.e., $P_{\cG_{\text{do}(A)}}(x|a)=P_{\cG_{\text{do}(A)}}(x)$ $\forall a,x$ and both equal the uniform distribution. From the pre-intervention distribution calculated in the last column of Figure~\ref{fig: Qcycle4}, we have that $P(X=x)$ is also uniform which gives $P_{\cG_{\text{do}(A)}}(x|a)=P(x)$ $\forall a,x$. Similarly $B$ does not affect $Y$ can also be established. Further, $\{A,B\}$ affects $\{X,Y\}$ because a joint intervention on $A$ and $B$ takes us back to the original Bell scenario in which these sets are correlated (c.f. Lemma~\ref{lemma: correl-affects}). Similarly, we can also establish that $X$ affects $B$ and $Y$ affects $A$ using the loop conditions $A=Y$ and $B=X$ which immediately yields $\{X,Y\}$ affects $\{A,B\}$. In this example, $\cG_{\text{do}(A,B)}$ corresponds to a quantum causal structure (the Bell scenario) while $\cG_{\text{do}(X,Y)}$ is a simple classical causal structure for the causal structure $\cG$ of Figure~\ref{fig: Qcycle1} and the (observed) arrows $\longrsquigarrow$ of the figure can be classified into dashed and solid arrows as: $A\xdashrightarrow{} X$, $B\xdashrightarrow{} Y$, $X\longrightarrow B$ and $Y\longrightarrow A$. Note that this example corresponds to a functional causal loop as per Definition~\ref{definition: causalloops}, and the observed distribution satisfies the d-separation condition of Definition~\ref{definition: compatdist}.\footnote{The observed d-separations here are $A\perp^d B|\{X,Y\}$ and $X\perp^d Y|\{A,B\}$ and the observed distribution $P_{XYAB}$ satisfies the conditional independences $A\indep B|\{X,Y\}$ and $X\indep Y|\{A,B\}$.} The post-intervention distribution is fully specified here because, all interventions (except the one on $\Lambda$ alone) are associated with acyclic post-intervention graphs and for interventions on the exogenous $\Lambda$, the post and pre-intervention distributions coincide.

 \paragraph{Applying the method to fine-tuned explanations of non-classical correlations: } It is known that certain non-classical correlations arising in the bipartite Bell causal structure cannot be obtained in the same causal structure if the common cause $\Lambda$ was classical. However, these correlations can be easily generated in the classical, fine-tuned causal structure of Figure~\ref{fig: Qcycle3}, which differs from the original causal structure by the inclusion of fine-tuned causal influences from each party's input to the other party's output. We now explain how this is achieved and then apply the post-selection method explained above to create a causal loop in Figure~\ref{fig: Qcycle3} by adding $X\longrightarrow B$ and $Y\longrightarrow A$. This will demonstrate that, even though the same non-classical correlations and affects relations can be obtained in the original Bell causal structure and its fine-tuned classical counterpart~\ref{fig: Qcycle3}, the two causal structures may behave differently in the presence of causal loops.

 First consider the PR box (Equation~\eqref{eq: PR}), which is one of the maximally non-classical correlations of the Bell causal structure. It is defined by the condition $X\oplus Y=A.B$ where all the variables are binary. This is easily generated in the classical causal structure of Figure~\ref{fig: Qcycle3} by the structural equations $\Lambda=E$, $Y=E$ and $X=E\oplus A.B$ (where $E$ is binary and uniformly distributed). Other non-classical correlations can be obtained by adding some ``noise'' to this PR box example. Let $\Lambda=(E,F)$ correspond to two variables $E$ and $F$ both binary, and the former distributed uniformly. Then the structural equations $Y=E$ and $X=E\oplus F \oplus A.B$ for different distributions over the exogenous variable $F$ correspond to the PR box mixed with different levels of noise.\footnote{Note that the model can be symmetrised by including an additional, uniformly distributed binary variable $G$ in the description of $\Lambda=(E,F,G)$ and using the structural equations $ X=E\oplus (G\oplus 1)(A.B\oplus F)$ and $Y=E\oplus G(A.B\oplus F)$.}
  \begin{align}
 \label{eq: Bellfinetune}
     \begin{split}
         X&=A.B\oplus E\oplus F,\\
          Y&=E.
     \end{split}
 \end{align}
Therefore, the causal mechanisms that allow us to produce non-classical correlations $P_{XYAB}$ in the acyclic causal structure~\ref{fig: Qcycle3} are the functional dependences~\eqref{eq: Bellfinetune} along with a specification of the distributions over the exogenous variables $E$ and $F$ that constitute $\Lambda$. $E$ is uniform while $F$ can vary depending on the correlation to be generated. We now construct the causal loop by including the additional arrows $X\longrsquigarrow B$ and $Y\longrsquigarrow A$ and by effectively post-selecting on the loop condition $A=Y$ and $B=X$. These, along with the causal mechanisms~\eqref{eq: Bellfinetune} of the acyclic case define the mechanisms for the cyclic causal structure.  
We will now see that these causal mechanisms are incompatible with each other. We have $Y=E$, $X=E\oplus F\oplus A.B$, $A=Y$ and $B=X$, which gives $X=E.X\oplus E\oplus F$ and $Y=E$. Therefore for $(E,F)=(0,0)$, we have $(X,Y)=(0,0)$ and for $(E,F)=(0,1)$ we have $(X,Y)=(1,0)$. However for $(E,F)=(1,0)$ we get $X=X\oplus 1$ which does not have a solution. For $(E,F)=(1,1)$ we get $X=X$ which is not a unique solution. Therefore if we demand unique solvability, we must require $E=0$ deterministically which contradicts the initial assumption that $E$ is uniform. Even if we don't require uniqueness, we can not have $(E,F)=(1,0)$ and forbidding this would make $E$ and $F$ correlated and non-uniform. 

Therefore, in the classical, fine-tuned explanation of the Bell correlations, adding the loop is not consistent with the causal mechanisms that generate the non-classical correlations in the absence of the loop--- in particular, they are in conflict with the preparation of the exogenous variable $\Lambda$. If we have a consistent loop, then intervention on $A$ and $B$ will no longer recover the original non-classical correlations. This is in contrast to the faithful case analysed in Figure~\ref{fig: Qcycle} (and explained previously in the text), when $\text{do}(A,B)$ gives back the non-classical correlations of the Bell scenario. This suggests that certain (non-local) hidden variable explanations for quantum correlations (in a Bell experiment) can in principle be distinguished from the explanation provided by standard quantum mechanics in the presence of causal loops. We have only shown this for a particular set of functions or causal mechanisms for generating the former and it would be interesting to consider if this generalises, in particular to causal mechanisms provided by Bohmian mechanisms \cite{Bohm1952}, a non-local hidden variable theory.

%Since this method reduces the problem to a classical one, one can impose ancestrally unique solvability to recover the d-separation property \cite{Forre2017} which is the compatibility condition (Definition~\ref{definition: compatdist}) used in our framework. We have sketched two different methods for calculating the observed and post intervention distributions from non-classical causal mechanisms in cyclic causal structures.The formalisation and generalisation of one or both of these methods is left for future work.

\afterpage{\FloatBarrier}

\paragraph{Generalising to other causal structures: } The idea behind the post-selection method employed for Example~\ref{example: qloop} above can in principle be generalised to other non-classical, cyclic causal structures where every directed cycle includes at least one edge $W\longrsquigarrow Z$ connecting classical nodes $W$ and $Z$. The intuition is that cutting off such an edge in every direct cycle and replacing it with an edge $W^*\longrsquigarrow Z$, by introducing an additional, exogenous variable $W^*$ would result in a directed acyclic graph (DAG). One can then apply the generalised causal model framework of \cite{Henson2014} (reviewed in Section~\ref{sssec: gen_causalstr}) to obtain the observed distribution in this DAG and then post-select on $W=W^*$ for all the edges that were cut off. Then a way to recover the d-separation condition (using the result of \cite{Forre2017}) would be to check whether there exists a classical causal model for the same cyclic causal structure that produces identical observed correlations and satisfies the anSEP property. Note that this classical causal model need not necessarily yield the same post-intervention distributions. In the example of Figure~\ref{fig: Qcycle1}, an intervention on $A$ and $B$ gives the Bell scenario, which as we know produces non-classical correlations that cannot be obtained in the corresponding classical causal model \cite{Wood2015}. Finally, it would be interesting to compare this method with the framework of post-selected closed time-like curves \cite{Lloyd2011}. % and analyse the possibility of generalising the post-selection method suitably to cycles containing only non-classical systems. 

\begin{remark}
We note that assuming the d-separation condition of Definition~\ref{definition: compatdist} as a primitive property of the framework rules out certain cyclic causal structures from being described in our current framework. In the classical case, these are precisely those cyclic causal models that do not satisfy anSEP or those involving continuous random variables (due to the result of \cite{Forre2017}). A good example of such a causal model is that of \cite{Neal2000}, and \cite{Forre2017} proposes a generalisation of d-separation called $\sigma-$separation through which they derive a \emph{generalised global directed Markov} condition that applies to classical causal models involving continuous variables and/or do not satisfy anSEP. This reduces to d-separation in the acyclic case. Therefore, one option would be to replace d-separation with $\sigma$-separation in Definition~\ref{definition: compatdist} to generalise our framework for cyclic causal models. 
\end{remark}

\subsection{Proofs}
\label{appendix: ProofsJamming}

\CI*

\begin{proof}
The conditional independendence $S_1\indep S_2|S_3$ stands for $P_{S_1S_2|S_3}=P_{S_1|S_3}P_{S_2|S_3}$, which implies
\begin{equation}
\label{eq: lemma11}
    P_{S_1|S_2S_3}=P_{S_1|S_3}.
\end{equation}
The 3 d-separation relations $S\perp^d S_i$ for $i\in \{1,2,3\}$ imply that $S$ is d-separated from every subset of the union $S_1\cup S_2\cup S_3$. This implies the following independences by Definition~\ref{definition: compatdist} of compatibility of the distribution $P$ with the causal model represented by $\mathcal{G}$,
\begin{equation}
    \label{eq: lemma12}
    P_{S|S'}=P_S\quad \forall S'\subseteq S_1\cup S_2 \cup S_3.
\end{equation}
Now consider the conditional distribution $P_{S_2|SS_1S_3}$. We have,
\begin{align}
\label{eq: lemma13}
    \begin{split}
       P_{S_2|SS_1S_3}&=\frac{P_{S_2SS_1S_3}}{P_{SS_1S_3}} \\
       &=\frac{P_{S_3}P_{S_2|S_3}P_{S_1|S_2S_3}P_{S|S_1S_2S_3}}{P_{SS_1S_3}}\\
       &=\frac{P_{S_3}P_{S_2|S_3}P_{S_1|S_3}P_{S}}{P_{S}P_{S_1S_3}}\\
       &=P_{S_2|S_3},
    \end{split}
\end{align}
where we have used Equations~\eqref{eq: lemma11} and~\eqref{eq: lemma12} in the third line, noting that $P_{S|S_1S_3}=P_S\Rightarrow P_{SS_1S_3}=P_{S}P_{S_1S_3}$. Equation~\eqref{eq: lemma13} is equivalent to $P_{SS_1S_2|S_3}=P_{SS_1|S_3}P_{S_2|S_3}$ which denotes the conditional independence $(S\cup S_1)\indep S_2|S_3$. The conditional independence $S_1\indep (S\cup S_2)|S_3$ can be derived analogously due to the symmetry between $S_1$ and $S_2$. 

Finally, we have
\begin{align}
        P_{S_2|SS_3}=\frac{P_{S_2SS_3}}{P_{SS_3}}=\frac{P_{S_3}P_{S_2|S_3}P_{S|S_2S_3}}{P_{S}P_{S_3}}=P_{S_2|S_3},
\end{align}
and similarly $P_{S_1|SS_3}=P_{S_1|S_3}$. Together with Equation~\eqref{eq: lemma13}, this implies $P_{S_2|SS_1S_3}=P_{S_2|SS_3}$. This is equivalent to $P_{S_1S_2|SS_3}=P_{S_1|SS_3}P_{S_2|SS_3}$ which denotes the final conditional independence $S_1\indep S_2|(S\cup S_3)$.
\end{proof}

\DoRules*

\begin{proof}
\textbf{Rule 1:} We first note that the graph $\cG_{\text{do}(X)}$ differs from $\cG_{\overline{X}}$ only by the inclusion of the additional nodes $I_{X_i}$ and corresponding edge $I_{X_i}\longrsquigarrow X_i$ for each $X_i\in X$. Therefore, the d-separation condition $(Y\perp^d Z|X, W)_{\cG_{\overline{X}}}$  for the latter implies the same condition $(Y\perp^d Z|X, W)_{\cG_{\text{do}(X)}}$ for the former. 
Using the d-separation condition of Definition~\ref{definition: compatdist} this implies the conditional independence of $Y$ and $Z$ given $\{X,W\}$ for the distribution $P_{\cG_{\text{do}(X)}}$ compatible with $\cG_{\text{do}(X)}$. Using the definition of the do-conditional (Equation~\eqref{eq: do}),  $P_{\cG_{\text{do}(X)}}(y,z|x,w):=P(y,z|\text{do}(x),w)=P(y|\text{do}(x),w)P(z|\text{do}(x),w)$. This conditional independence automatically gives the required Equation~\eqref{eq: rule1}.

\textbf{Rule 2:}
$\cG_{\overline{X},\underline{Z}}$ is the graph where all incoming arrows to $X$ and outgoing arrows from $Z$ are removed in $\cG$. Hence, the d-separation condition $(Y\perp^d Z|X, W)_{\cG_{\overline{X},\underline{Z}}}$ implies that the only paths between $Y$ and $Z$ in the graph $\cG_{\overline{X}}$ that are not blocked by $X$ and $W$ are paths involving an outgoing arrow from $Z$. These are precisely the paths that get removed in going from $\cG_{\overline{X}}$ to $\cG_{\overline{X},\underline{Z}}$, resulting in the required d-separation there. 
The same statement holds for the graph $\cG_{do(X)}$ (by the argument used in the proof of Rule 1), and also for the graph $\cG_{do(X), I_Z}$ which corresponds to adding the nodes $I_{Z_i}$ and edges $I_{Z_i}\longrsquigarrow Z_i$ to $\cG_{do(X)}$ for each $Z_i\in Z$. The latter holds true since the addition of the $I_{Z_i}$ nodes and $I_{Z_i}\longrsquigarrow Z_i$ edges cannot create any additional paths between $Z$ and $Y$ that are left unblocked by $X$ and $W$. This implies that the only paths between $Y$ and the set $I_Z:=\{I_{Z_i}\}_i$ not blocked by $X$ and $W$ in $\cG_{do(X), I_Z}$ are paths from $I_Z$, going through $Z$ and involving an outgoing arrow from $Z$ i.e., paths involving the subgraph $I_Z\longrsquigarrow Z \longrsquigarrow...$. All these paths would get blocked when conditioning additionally on $Z$. This gives $(Y\perp^d I_Z|X, W,Z)_{\cG_{do(X), I_Z}}$, which through the compatibility condition (Definition~\ref{definition: compatdist}) implies the conditional independence $(Y\indep I_Z|X, W,Z)_{\cG_{do(X), I_Z}}$, equivalently expressed as
\begin{equation}
    \label{eq:rule2proof}
    P(y|\text{do}(x),w,z,I_Z=idle)=P(y|\text{do}(x),w,z,I_Z=\text{do}(z)) \quad \forall y,x,w,z.
\end{equation}

By definition (see Section~\ref{ssec:affects}) we have $P(y|\text{do}(x),w,z,I_Z=idle)=P(y|\text{do}(x),w,z)$ and $(y|\text{do}(x),w,z,I_Z=\text{do}(z))=P(y|\text{do}(x),w,\text{do}(z))$ $\forall y,x,w,z$. Along with Equation~\eqref{eq:rule2proof}, this gives the required Equation~\ref{eq: rule2}. In other words, once $X$, $W$ and $Z$ are given, $Y$ does not depend on whether the given value $z$ of $Z$ was obtained through an intervention ($I_Z=do(z)$) or passive observation (i.e., where $I_{Z_i}=idle$ for all $i$, which is the causal model where no interventions are made on elements of $Z$).

\begin{sloppypar}\textbf{Rule 3:} Consider the graph $\cG_{\text{do}(X),I_Z}$ which is the post-intervention graph with respect to the nodes $X$ augmented with $I_{Z_i}\longrsquigarrow Z_i$ for all $Z_i \in Z$. In this graph, suppose we had the d-separation relation $(Y\perp^d I_Z|X,W)_{\cG_{\text{do}(X),I_Z}}$. By Definition~\ref{definition: compatdist}, this would result in the conditional independence $(Y\indep I_Z|X,W)_{\cG_{\text{do}(X),I_Z}}$ which can be expressed as
$$P(y|w,\text{do}(x),I_Z=\text{idle})=P(y|w,\text{do}(x),I_Z=\text{do}(z))\quad \forall y,w,x,z$$
Note that by definition of the augmented and post-intervention causal models, we have $P(y|w,\text{do}(x),I_Z=\text{idle})=P(y|w,\text{do}(x))$  and $P(y|w,\text{do}(x),I_Z=\text{do}(z))=P(y|w,\text{do}(x),\text{do}(z))$ $\forall y,w,x,z$ (c.f. Equations~\eqref{eq: intervention} and \eqref{eq: do}), and consequently $P(y|w,\text{do}(x),\text{do}(z))=P(y|w,\text{do}(x))$ $\forall y,w,x,z$, which is the required Equation~\eqref{eq: rule3}. Therefore, showing that the d-separation condition $(Y\perp^d Z|X,W)_{\cG_{\overline{X},\overline{Z(W)}}}$ implies the d-separation relation $(Y\perp^d I_Z|X,W)_{\cG_{\text{do}(X),I_Z}}$ would complete the proof. This is shown by contradiction. Suppose that $(Y\perp^d Z|X,W)_{\cG_{\overline{X},\overline{Z(W)}}}$ and $(Y\not\perp^d I_Z|X,W)_{\cG_{\text{do}(X),I_Z}}$. Then there must exist a path from a member $I_{Z_i}$ of $I_Z$ to a member $Y_j$ of $Y$ in $\cG_{\text{do}(X),I_Z}$
that is unblocked by $X$ and $W$. There are two possibilities for such a path: either it contains the subgraph $I_{Z_i}\longrsquigarrow Z_i \longrsquigarrow ... Y_j$ or the subgraph $I_{Z_i}\longrsquigarrow Z_i \longlsquigarrow ... Y_j$. Denoting these possibilities as cases 1 and 2 respectively, let $\mathscr{P}$ be the shortest such path. We will show that a contraction arises in each case.

 \emph{Case 1:} Consider the first case where $\mathscr{P}$ contains the subgraph $I_{Z_i}\longrsquigarrow Z_i \longrsquigarrow ... Y_j$. In this case, $(Y\not\perp^d I_Z|X,W)_{\cG_{\text{do}(X),I_Z}}$ (which we have assumed) implies $(Y\not\perp^d Z_i|X,W)_{\cG_{\text{do}(X)}}$. Along with the assumption that $(Y\perp^d Z_i|X,W)_{\cG_{\overline{X},\overline{Z(W)}}}$, this implies that there exists a path from $Z_i$ to $Y$ in $\cG_{\text{do}(X)}$ unblocked by $X$ and $W$ that passes through some member $Z_k$ of $Z(W)$, which would blocked when the incoming arrows to $Z_k$ are removed. This leads to the following subcases where the path from $Z_i$ to $Y_j$ in $\cG_{\text{do}(X)}$ contains the following subgraphs:
    \begin{itemize}
        \item \emph{Case 1a:} $Z_i\longrsquigarrow...\longrsquigarrow Z_k\longlsquigarrow... Y_j$ or
        \item \emph{Case 1b:}$Z_i\longrsquigarrow...\longlsquigarrow Z_k\longlsquigarrow... Y_j$ or
         \item \emph{Case 1c:} $Z_i\longrsquigarrow...\longrsquigarrow Z_k\longrsquigarrow... Y_j$
    \end{itemize}

Each of these cases are not possible due to the following reasons. In \emph{Case 1a}, some descendant of $Z_k$ must be in $W$ for the path to be unblocked in $\cG_{\text{do}(X)}$ but by definition, $Z(W)$ (which contains $Z_k$) is the set of all nodes in $Z$ that don't have descendants in $W$. In \emph{Case 1b}, the path between $Z_i$ and $Z_k$ must contain a collider. For this path to be unblocked by $X$ and $W$ in $\cG_{\text{do}(X)}$ the collider node must have a descendant in $W$ but the other requirement that this path must be blocked in $\cG_{\overline{X}, \overline{Z(W)}}$ implies that the same collider node must be a member of $Z(W)$ which by definition does not have any descendants in $W$, yielding a contradiction. In \emph{Case 1c}, there is either a directed path from $Z_k\in Z$ to $Y_j\in Y$ in $\cG_{\text{do}(X)}$ or a collider in the path between $Z_k$ and $Y_j$. The latter is ruled out by the same argument used in Case 1b. If there is a directed path from $Z_k$ to $Y_j$, then there is a directed path from $I_{Z_k}$ to $Y_j$ in $\cG_{\text{do}(X),I_Z}$ i.e., there is a path from a member of $Z$ to $Y$ that is unblocked by $X$ and $W$ in $\cG_{\text{do}(X),I_Z}$ and that is shorter than the shortest path $\mathscr{P}$, which is not possible.

Finally, consider \emph{Case 2} where the path $\mathscr{P}$ contains the subgraph $I_{Z_i}\longrsquigarrow Z_i \longlsquigarrow ... Y_j$. The initial assumption that $(Y\not\perp^d I_Z|X,W)_{\cG_{\text{do}(X),I_Z}}$ implies that the collider node $Z_i$ must have descendants in the conditioning set $W$ i.e., $Z_i\not\in Z(W)$. However, in this case we will violate the assumption that $(Y\perp^d Z|X,W)_{\cG_{\overline{X},\overline{Z(W)}}}$. On the other hand, to satisfy this d-separation, we would require $Z_i\in Z(W)$ but this would violate $(Y\not\perp^d I_Z|X,W)_{\cG_{\text{do}(X),I_Z}}$. 
Hence we have shown that $(Y\perp^d Z|X,W)_{\cG_{\overline{X},\overline{Z(W)}}}$ and $(Y\not\perp^d I_Z|X,W)_{\cG_{\text{do}(X),I_Z}}$ can never be simultaneously satisfied and hence that $(Y\perp^d Z|X,W)_{\cG_{\overline{X},\overline{Z(W)}}}$ implies $(Y\not\perp^d I_Z|X,W)_{\cG_{\text{do}(X),I_Z}}$ which in turn implies the required Equation~\eqref{eq: rule3}.
\end{sloppypar}
\end{proof}

\CompatST*

\begin{proof}
\label{proof:poset}
\emph{Necessity of condition:} For any two subsets $\mathcal{S}_1, \mathcal{S}_2\subseteq \mathcal{S}$ such that no proper subset of $\mathcal{S}_1$ affects $\mathcal{S}_2$, $\mathbf{compat}$ implies that $\mathcal{R}_{\mathcal{S}_2}\subseteq \mathcal{R}_{\mathcal{S}_1}$ whenever $\mathcal{S}_1$ affects $\mathcal{S}_2$.
 Further, by $\mathbf{compat}$, $\mathcal{R}_{\mathcal{X}}=\overline{\mathcal{F}}(\mathcal{X})$ for all ORVs $\mathcal{X}$. Using these along with the fact that for any subset $\mathcal{S}_i\subseteq\mathcal{S}$, $\mathcal{R}_{\mathcal{S}_i}=\bigcap\limits_{s_i\in\mathcal{S}_i}\mathcal{R}_{s_i}$ (see Remark~\ref{remark: subsets}), we have $\mathcal{R}_{\mathcal{S}_i}=\bigcap\limits_{s_i\in\mathcal{S}_i}\overline{\mathcal{F}}(s_i)$ for all subsets $\mathcal{S}_i$ whenever $\mathbf{compat}$ holds. Hence we have the following whenever $\mathbf{compat}$ holds, which is equivalent to Equation~\eqref{eq: cond}
   \begin{equation*}
\mathcal{S}_1 \text{ affects } \mathcal{S}_2 \Rightarrow \bigcap\limits_{s_2\in \mathcal{S}_2}\overline{\mathcal{F}}(s_2) \subseteq \bigcap\limits_{s_1\in \mathcal{S}_1}\overline{\mathcal{F}}(s_1). \end{equation*}
\newline
\par
\emph{Sufficiency of condition + additional assumption:}
 Applying the condition $\mathbf{cond}$ of Equation~\eqref{eq: cond} to the case where $\mathcal{Y}=\{\mathcal{X}\}$ (i.e., $\mathcal{Y}$ is any arbitrary copy of $\mathcal{X}$), we have $\mathcal{X}$ affects $\mathcal{Y}$ (by Definition~\ref{definition:copy} of a copy). This gives $\overline{\mathcal{F}}(\mathcal{X}')\subseteq \overline{\mathcal{F}}(\mathcal{X})$ and hence $\mathcal{X}'\in\overline{\mathcal{F}}(\mathcal{X})$ $\forall$ copies $\mathcal{X}'$ of $\mathcal{X}$. Since by Definition~\ref{definition:accreg}, the accessible region of $\mathcal{X}$ is the smallest subset of $\mathcal{T}$ containing all possible copies of $\mathcal{X}$, we must have $\mathcal{R}_{\mathcal{X}}\subseteq\overline{\mathcal{F}}(\mathcal{X})$. Hence $\mathbf{cond}$ by itself is not sufficient for $\mathbf{compat}$ as it does not require the accessible region to coincide with the future, only that it is a subset of the future. Along with the additional assumption that $\mathcal{R}_{\mathcal{X}}\supseteq\overline{\mathcal{F}}(\mathcal{X})$ $\forall \mathcal{X}\in \mathcal{S}$, this gives
    \begin{equation}
        \label{eq: suff1}
        \mathcal{R}_{\mathcal{X}}=\overline{\mathcal{F}}(\mathcal{X}), \quad \forall \mathcal{X}\in \mathcal{S}.
    \end{equation}
    \item Again, the accessible region of any subset $\mathcal{S}_i\subseteq\mathcal{S}$ is $\mathcal{R}_{\mathcal{S}_i}=\bigcap\limits_{s_i\in\mathcal{S}_i}\mathcal{R}_{s_i}$ (Remark~\ref{remark: subsets}). Equation~\eqref{eq: suff1} then implies that
    \begin{equation}
    \label{eq: suff2}
        \mathcal{R}_{\mathcal{S}_i}=\bigcap\limits_{s_i\in\mathcal{S}_i}\overline{\mathcal{F}}(s_i) \qquad \forall \mathcal{S}_i\subseteq \mathcal{S}
    \end{equation}
Hence we have that the condition of Equation~\eqref{eq: cond} along with the additional requirement of Equation~\eqref{eq: assump} imply Equation~\eqref{eq: suff1} and~\eqref{eq: suff2}, where Equation~\eqref{eq: suff1} is the first condition for $\mathbf{compat}$ (Definition~\ref{definition: compatposet}).  Further, Equation~\eqref{eq: suff2} along with the condition Equation~\eqref{eq: cond} then imply the second condition for $\mathbf{compat}$ which is
    \begin{equation}
     \label{eq: suff3}
      \forall \text{subsets }\mathcal{S}_1, \mathcal{S}_2 \subseteq \mathcal{S} \text{, such that no proper subset of } \mathcal{S}_1 \text{ affects } \mathcal{S}_2\text{, } \mathcal{S}_1 \text{ affects } \mathcal{S}_2 \Rightarrow \mathcal{R}_{\mathcal{S}_2}\subseteq\mathcal{R}_{\mathcal{S}_1}.
    \end{equation}
\end{proof}

\NoLoops*

\begin{proof}
\emph{Necessity of condition:} 
Consider a causal model over a set $S$ of RVs that has no affects causal loops. Taking $\mathcal{T}$ to be Minkowski space-time, embed the RVs in $\mathcal{T}$ such that whenever an RV $X$ affects another RV $Y$, then the corresponding ORVs satisfy $\mathcal{Y}\in \mathcal{F}(\mathcal{X})$. Since $X$ affects $Y$ implies that $Y$ does not affect $X$ (by the assumption of no affects causal loops), this is always possible for any pair of RVs. For embedding the remaining RVs, consider the following conditions for two sets $S_1$ and $S_2$ of the RVs,
\begin{enumerate}
   \item No proper subset of $S_1$ affects $S_2$,
  \item No proper subset of $S_2$ is affected by $S_1$,
   \item No proper subset of $S_2$ affects $S_1$ and,
  \item No proper subset of $S_1$ is affected by $S_2$,
\end{enumerate}
Then embed the elements of $S_1$ and $S_2$ in $\mathcal{T}$ as follows
\begin{itemize}
    \item If conditions 1. and 2. hold and $S_1$ affects $S_2$: $\bigcap\limits_{s_2\in \mathcal{S}_2}\overline{\mathcal{F}}(s_2) \subseteq \bigcap\limits_{s_1\in \mathcal{S}_1}\overline{\mathcal{F}}(s_1)$.
    \end{itemize}
    This automatically gives (since the sets are arbitrary):
    \begin{itemize}
   \item If conditions 3. and 4. hold and $S_2$ affects $S_1$: $\bigcap\limits_{s_2\in \mathcal{S}_2}\overline{\mathcal{F}}(s_2) \supseteq \bigcap\limits_{s_1\in \mathcal{S}_1}\overline{\mathcal{F}}(s_1)$,
      \item If conditions 1.-4. hold, $S_1$ affects $S_2$ and $S_2$ affects $S_1$: $\bigcap\limits_{s_2\in \mathcal{S}_2}\overline{\mathcal{F}}(s_2) = \bigcap\limits_{s_1\in \mathcal{S}_1}\overline{\mathcal{F}}(s_1)$.
\end{itemize}
Such an embedding is always possible because the joint inclusive future of some set $T$ of points in Minkowski space-time can be seen as the inclusive future of a single point $L_T$, which would be the earliest point that is in the future of all points in $T$, and the embedding imposes an order on the points $L_T$ corresponding to a set. For example, for every pair of ORV sets $\cS_1$ and $\cS_2$ satisfying 1. and 2. and $\cS_1$ affects $\cS_2$, the embedding requires $L_{\cS_1}\preceq L_{\cS_2}$. For the third case where 1.-4. are satisfied and the sets affect each other, we would have $L_{\cS_1}=L_{\cS_2}$. Note that this is possible even when the RVs are assigned distinct locations in $\cT$, since only the joint futures of sets need to coincide (this was the case in Figure~\ref{fig: eqcyclemain} where $B$ and $\{A,C\}$ affect each other). In the case that $S_1$ and $S_2$ consist of single elements, this embedding clearly agrees with $X$ affects $Y$ implies $\mathcal{Y}\in \mathcal{F}(\mathcal{X})$, since no causal loops forbids the third case. In particular, for all subsets $S_1$ and $S_2$ satisfying the conditions 1. and 2., we have
\begin{equation}
\label{eq: setaffects}
   \mathcal{S}_1 \text{ affects } \mathcal{S}_2 \Rightarrow \bigcap\limits_{s_2\in \mathcal{S}_2}\overline{\mathcal{F}}(s_2) \subseteq \bigcap\limits_{s_1\in \mathcal{S}_1}\overline{\mathcal{F}}(s_1) 
\end{equation}
Now, suppose $S_1$ and $S_2$ are two sets of variables that satisfy 1. and $S_1$ affects $S_2$. Then, either they also satisfy 2. or they do not satisfy 2., in which case there exists a proper subset $S_2'\subset S_2$ such that $S_1$ affects $S_2'$. Without loss of generality, we can assume that $S_1$ and $S_2'$ satisfy both 1. and 2. (if not, we can repeat the argument for this case by taking a proper subset of $S_2'$ that will satisfy these conditions). In other words $S_1$ and $S_2'\subseteq S_2$ satisfy 1. and 2. and also the affects relation $S_1$ affects $S_2'$. Therefore, for the embedding described above, we have $\bigcap\limits_{s'_2\in \mathcal{S}'_2}\overline{\mathcal{F}}(s'_2) \subseteq \bigcap\limits_{s_1\in \mathcal{S}_1}\overline{\mathcal{F}}(s_1) $ due to Equation~\eqref{eq: setaffects}. Further, $S_2\supseteq S'_2$ implies that $\bigcap\limits_{s_2\in \mathcal{S}_2}\overline{\mathcal{F}}(s_2) \subseteq \bigcap\limits_{s'_2\in \mathcal{S}'_2}\overline{\mathcal{F}}(s'_2)$. Therefore for any $S_1$ and $S_2$, which satisfy 1. and  $S_1$ affects $S_2$, we have $\bigcap\limits_{s_2\in \mathcal{S}_2}\overline{\mathcal{F}}(s_2) \subseteq \bigcap\limits_{s_1\in \mathcal{S}_1}\overline{\mathcal{F}}(s_1)$. In other words, for the embedding described above, we have the following which is equivalent to $\mathbf{cond}$ (Equation~\eqref{eq: cond}).
   \begin{equation*}
   \begin{split}
\forall \text{subsets } \mathcal{S}_1, \mathcal{S}_2 \subseteq \mathcal{S}\text{, such that no proper subset of } \mathcal{S}_1 \text{ affects } \mathcal{S}_2, \\
\mathcal{S}_1 \text{ affects } \mathcal{S}_2 \Rightarrow \bigcap\limits_{s_2\in \mathcal{S}_2}\overline{\mathcal{F}}(s_2) \subseteq \bigcap\limits_{s_1\in \mathcal{S}_1}\overline{\mathcal{F}}(s_1). 
\end{split}
\end{equation*}
\par

For all other RVs (if any remain), the embedding can be arbitrary. For example, these may be superfluous RVs that are neither affected nor affect any other sets of RVs in the model. Hence we have shown that if a causal model has no causal loops, there exists an embedding in Minkowski space-time $\mathcal{T}$ such that $\mathbf{cond}$ is satisfied.
\par
\emph{Sufficiency of condition+additional assumption:} $\mathbf{cond}$ (Equation~\eqref{eq: cond}) implies in particular that when the sets correspond to single elements $\mathcal{X}$ and $\mathcal{Y}$, $\mathcal{X}$ affects $\mathcal{Y}$ implies that $\overline{\mathcal{F}}(\mathcal{Y})\subseteq\overline{\mathcal{F}}(\mathcal{X})$ (i.e., $\mathcal{Y}\in \overline{\mathcal{F}}(\mathcal{X})$). Hence we can have a causal loop $\mathcal{X}$ affects $\mathcal{Y}$ and $\mathcal{Y}$ affects $\mathcal{X}$ only if $\mathcal{X}$ and $\mathcal{Y}$ share the exact same space-time location i.e., $O(\mathcal{X})=O(\mathcal{Y})$. The additional assumption forbids this and hence forbids causal loops.
\end{proof}

\looposet*
\begin{proof}
\emph{Necessity of condition:} The proof is identical to that of Theorem~\ref{theorem: loops1}, with the additional requirement that $\overline{\mathcal{F}}(\mathcal{X})=\mathcal{R}_{\mathcal{X}}$. Note that this is a particular choice of embedding since an embedding involves an assignment of locations in $\mathcal{T}$ to each RV along with an assignment of accessible regions (see Definition~\ref{definition: embedding}). In Theorem~\ref{theorem:poset}, we proved the sufficiency of $\mathbf{cond}$ (Equation~\eqref{eq: cond}) along with the additional assumption $\overline{\mathcal{F}}(\mathcal{X})\subseteq\mathcal{R}_{\mathcal{X}}$ for compatibility of a causal model with the embedding $\cT$. $\mathbf{cond}$ implies that $\overline{\mathcal{F}}(\mathcal{X})\supseteq\mathcal{R}_{\mathcal{X}}$ which along with $\overline{\mathcal{F}}(\mathcal{X})\subseteq\mathcal{R}_{\mathcal{X}}$ implies $\overline{\mathcal{F}}(\mathcal{X})=\mathcal{R}_{\mathcal{X}}$, and this was required for the proof. Here, we already have $\overline{\mathcal{F}}(\mathcal{X})=\mathcal{R}_{\mathcal{X}}$ by our choice of embedding and Theorem~\ref{theorem: loops1} shows that no causal loops imply the existence of an embedding for which $\mathbf{cond}$ holds. Hence with the current choice of embedding, this also implies compatibility of the causal model with the embedding partial order $\mathcal{T}$.
\par\emph{Sufficiency of condition+additional assumption:} The proof is identical to the that of the sufficiency part of Theorem~\ref{theorem: loops1}.
\end{proof}

\Jamming*

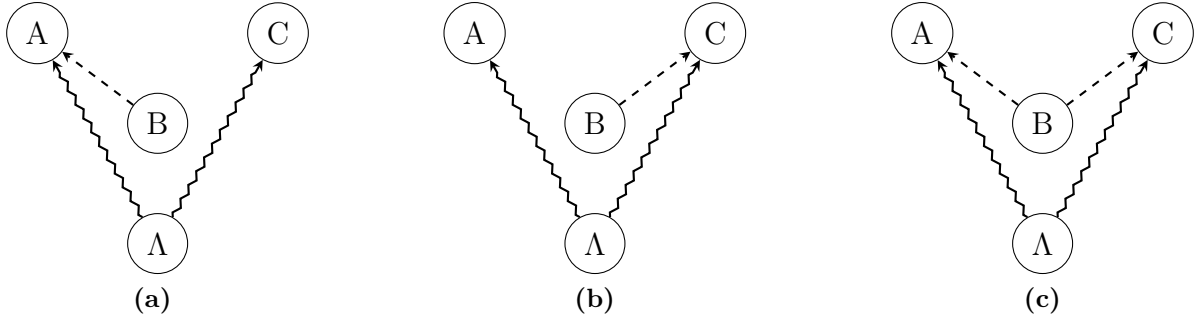
\begin{figure}[t]
    \centering
  \subfloat[]{ \begin{tikzpicture}[scale=0.8]
      \node[shape=circle,draw=black] (B) at (0,0.5) {B};
    \node[shape=circle,draw=black] (A) at (-2,2) {A};
    \node[shape=circle,draw=black] (C) at (2,2) {C};
      \node[shape=circle,draw=black] (L) at (0,-1.5) {$\Lambda$};
     \path [thick, dashed, arrows=-stealth] (B) edge (A);
     %\path [thick, dashed, arrows=-stealth] (B) edge (C);
       \draw[decorate, decoration={zigzag, segment length=+6pt, amplitude=+.95pt,post length=+4pt}, arrows={-stealth}, thick] (L) -- (A);  \draw[decorate, decoration={zigzag, segment length=+6pt, amplitude=+.95pt,post length=+4pt}, arrows={-stealth}, thick] (L) -- (C);
      \end{tikzpicture}}\qquad\qquad\subfloat[]{ \begin{tikzpicture}[scale=0.8]
      \node[shape=circle,draw=black] (B) at (0,0.5) {B};
    \node[shape=circle,draw=black] (A) at (-2,2) {A};
    \node[shape=circle,draw=black] (C) at (2,2) {C};
      \node[shape=circle,draw=black] (L) at (0,-1.5) {$\Lambda$};
     %\path [thick, dashed, arrows=-stealth] (B) edge (A);
     \path [thick, dashed, arrows=-stealth] (B) edge (C);
       \draw[decorate, decoration={zigzag, segment length=+6pt, amplitude=+.95pt,post length=+4pt}, arrows={-stealth}, thick] (L) -- (A);  \draw[decorate, decoration={zigzag, segment length=+6pt, amplitude=+.95pt,post length=+4pt}, arrows={-stealth}, thick] (L) -- (C);
      \end{tikzpicture}
 }\qquad\qquad\subfloat[]{ \begin{tikzpicture}[scale=0.8]
      \node[shape=circle,draw=black] (B) at (0,0.5) {B};
    \node[shape=circle,draw=black] (A) at (-2,2) {A};
    \node[shape=circle,draw=black] (C) at (2,2) {C};
      \node[shape=circle,draw=black] (L) at (0,-1.5) {$\Lambda$};
     \path [thick, dashed, arrows=-stealth] (B) edge (A);
     \path [thick, dashed, arrows=-stealth] (B) edge (C);
       \draw[decorate, decoration={zigzag, segment length=+6pt, amplitude=+.95pt,post length=+4pt}, arrows={-stealth}, thick] (L) -- (A);  \draw[decorate, decoration={zigzag, segment length=+6pt, amplitude=+.95pt,post length=+4pt}, arrows={-stealth}, thick] (L) -- (C);
      \end{tikzpicture}
 }
    \caption[Possible causal structures for Theorem~\ref{theorem: jamming}.]{\textbf{Possible causal structures for Theorem~\ref{theorem: jamming}.}}
    \label{fig: jammingproof2}
\end{figure}

\begin{proof}
$B$ affects $\{A,C\}$ implies the existence of a directed path from $B$ to at least one of $A$ and $C$. Given that $B$ is exogenous, $A$ and $C$ have no outgoing arrows and share a common cause $\Lambda$, this yields the 3 possible causal structures of Figure~\ref{fig: jammingproof2}, where the arrows from $\Lambda$ may be solid of dashed. In all these causal structures, we have the d-separation relations $A\perp^d C|\{B,\Lambda\}$ and $B\perp^d \Lambda$. By Definition~\ref{definition: compatdist}, these imply the conditional independences
\begin{equation}
\label{eq: jamproof1}
    A\indep C|\{B,\Lambda\} \quad \text{and}\quad B\indep \Lambda.
\end{equation}
Further, the given affects relations along with the exogeneity of $B$ imply that
\begin{equation}
\label{eq: jamproof2}
    B\not\indep \{A,C\}\quad \text{and}\quad B\indep A \quad \text{and}\quad B\indep C.
\end{equation}
We now show that in all the causal structures, irrespective of the classification of the outgoing arrows from $\Lambda$ and the space-time location that it is assigned, the causal model is not compatible with Minkowski space-time with respect to the embedding given in the theorem statement, and hence leads to superluminal signalling. For this, our first step is to show, using Equations~\eqref{eq: jamproof1} and~\eqref{eq: jamproof2} that $B$ affects at least one of $\{A,\Lambda\}$ and $\{C,\Lambda\}$. Suppose that $B$ affects neither of the sets, then by its exogeneity, this is equivalent to 
\begin{equation}
\label{eq: jamproof3}
 B\indep \{A,\Lambda\}\quad \text{and}\quad B\indep \{C,\Lambda\}   
\end{equation}
Then we have
\begin{equation}
\label{eq: jamproof4}
    \begin{split}
        P_{A|CB\Lambda}&=P_{A|B\Lambda}=\frac{P_{AB\Lambda}}{P_{B\Lambda}}=\frac{P_{A\Lambda|B}P_B}{P_{B\Lambda}}\\
        &=\frac{P_{A\Lambda}P_B}{P_{B}P_{\Lambda}}=P_{A|\Lambda},
    \end{split}
\end{equation}
    Where we have used $A\indep C|\{B,\Lambda\}$ in the first line and $B\indep \Lambda$ and  $B\indep \{A,\Lambda\}$ in the second line. Further, $B\indep \{C,\Lambda\}$ implies that $P_{BC\Lambda}=P_BP_{C\Lambda}$. Using this in Equation~\eqref{eq: jamproof4}, 
    \begin{equation}
\label{eq: jamproof5}
        P_{ABC\Lambda}=P_{A|\Lambda}P_{BC\Lambda}=P_BP_{A|\Lambda}P_{C\Lambda}
    \end{equation}
    
   Summing over $B$ and $\Lambda$, we obtain $P_{AC}=\sum_{\Lambda}P_{A|\Lambda}P_{C\Lambda}$. Using this in Equation~\eqref{eq: jamproof5} gives
    
    \begin{equation}
P_{ABC}=P_B\sum_{\Lambda}P_{A|\Lambda}P_{C\Lambda}=P_BP_{AC}.
    \end{equation}
    The last line is equivalent to $B\indep \{A,C\}$, which contradicts Equation~\eqref{eq: jamproof2}. Therefore, given the conditions in the theorem statement, Equation~\eqref{eq: jamproof3} cannot hold i.e., $B$ is correlated with at least one of $\{A,\Lambda\}$ and $\{C,\Lambda\}$. Using the exogeneity of $B$, this implies that $B$ affects at least one of these sets. For compatibility with the space-time, this affects relation implies the following constraint on the space-time embedding where the accessible region of each ORV is taken to coincide with its inclusive future.
    \begin{equation}
    \label{eq: jamproof6}
        \overline{\cF}(\cA)\cap\overline{\cF}(\Uplambda)\subseteq \overline{\cF}(\cB) \quad \lor\quad \overline{\cF}(\cC)\cap\overline{\cF}(\Uplambda)\subseteq \overline{\cF}(\cB).
    \end{equation}
 Now, note that Equation~\eqref{eq: jamproof2} implies that
    \begin{equation}
        \label{eq: jamproof7}
        A\not\indep C|B
    \end{equation}
This is because if we had the contrary, i.e., $A\indep C|B$, then along with $B\indep A$ this gives $P_{A|BC}=P_{A|B}=P_A$. Using this along with $B\indep C$ gives $P_{ABC}=P_{A|BC}P_{BC}=P_AP_BP_C$. This in turn implies $B\indep \{A,C\}$, which contradicts the first condition of Equation~\eqref{eq: jamproof2}. 

Next, we show that $\{B,\Lambda\}$ affects $A$ and $\{B,\Lambda\}$ affects $C$, again by contradiction. Suppose that $\{B,\Lambda\}$ does not affect $A$, this is equivalent to $\{B,\Lambda\}\indep A$ (i.e., $P_{A|B\Lambda}=P_A$) since $B$ and $\Lambda$ are exogenous. Along with Equation~\eqref{eq: jamproof1}, this implies that $P_{A|BC\Lambda}=P_{A|B\Lambda}=P_A$. Then $P_{ABC}=\sum_{\Lambda}P_AP_{BC\Lambda}=P_AP_{BC}$, which contradicts Equation~\eqref{eq: jamproof2}. Therefore we must have $\{B, \Lambda\}$ affects $A$. Similarly, one can show that $\{B, \Lambda\}$ affects $C$ must also hold under the assumptions given in the theorem statement. Now, we have four cases depending on whether or not $\Lambda$ affects $A$ or $C$, which we enumerate below.
\begin{itemize}
    \item \textit{Case 1: $\Lambda$ affects neither $A$ nor $C$} In this case, both the outgoing arrows from $\Lambda$ would be dashed (by Definition~\ref{definition: solidasharrows}). In this case, any compatible embedding of the causal model in space-time must satisfy (c.f. Theorem~\ref{theorem:poset})
    \begin{equation}
        \cA\in \overline{\cF}(\cB)\cap \overline{\cF}(\Uplambda) \quad\land\quad \cC\in \overline{\cF}(\cB)\cap \overline{\cF}(\Uplambda).
    \end{equation}
     \item \textit{Case 2: $\Lambda$ affects $A$ but not $C$} In this case we have $\Lambda\longrightarrow A$ and $\Lambda \xdashrightarrow{} C$, and a necessary condition for compatibly embedding this causal model in space-time is
         \begin{equation}
        \cA\in \overline{\cF}(\Uplambda) \quad\land\quad \cC\in \overline{\cF}(\cB)\cap \overline{\cF}(\Uplambda).
    \end{equation}
      \item \textit{Case 3: $\Lambda$ affects $C$ but not $A$} In this case we have $\Lambda\longrightarrow C$ and $\Lambda \xdashrightarrow{} A$ and compatibly with the space-time requires that
         \begin{equation}
        \cA\in \overline{\cF}(\cB)\cap\overline{\cF}(\Uplambda) \quad\land\quad \cC\in \overline{\cF}(\Uplambda).
    \end{equation}
       \item \textit{Case 4: $\Lambda$ affects $A$ as well as $C$} Here we have $\Lambda\longrightarrow A$ and $\Lambda\longrightarrow C$ and compatibility with the space-time necessitates
         \begin{equation}
         \label{eq: jamproof8}
        \cA\in \overline{\cF}(\Uplambda) \quad\land\quad \cC\in \overline{\cF}(\Uplambda).
    \end{equation}
\end{itemize}
In Cases 1-3, we can immediately see that at least one of $\cA$ and $\cC$ must be in the future of $\cB$ to restrict the causal model from signalling superluminally. In Case 4, this follows from Equations~\eqref{eq: jamproof6} and~\eqref{eq: jamproof7}--- since $\cA$ and $\cC$ must be in the future of $\Uplambda$, the joint futures of each of these ORVs and $\Uplambda$ coincides with the future of the ORV itself. Therefore, the space-time compatibility condition is always violated  by the embedding of \cite{Grunhaus1996, Horodecki2019} where $\cA$, $\cB$ and $\cC$ are pairwise space-like separated, and for this embedding, the jamming causal model considered here, with observed $\Lambda$ leads to superluminal signalling.
\end{proof}

\part{Multi-agent paradoxes}
\captionsetup[figure]{margin=1.3cm,font=small,labelfont={bf},name={Figure},labelsep=colon}
%\captionsetup[table]{margin=1.3cm,font=small,labelfont={bf},name={Table},labelsep=colon}
\begin{center}
    \thispagestyle{empty}
    \vspace*{\fill}
   \large{\textit{ The ‘paradox’ is only a conflict between reality and your feeling of what reality ‘ought to be’.
}}\\\vspace{3mm}
\raggedleft\large{\textit{  - Richard Feynman }}
    \vspace*{\fill}
\end{center}

\chapter{Multi-agent paradoxes in quantum theory}
\label{chapter: FR}

\lettrine[nindent=0em, slope=-.5em,lines=2]{P}{rocessing} empirically acquired data and making inferences about the world form a crucial part of the scientific method. For a consistent description of the world, these inferences should be based on a sound system of logic--- simple reasoning principles applicable to general situations, on which there is common agreement. For example, we would like to make inferences such as ``if I know that $a$ holds, and I know that $a$ implies $b$, then I know that $b$ holds'' independently of the nature of $a$ and $b$. When considering scenarios with several rational agents, inferences may involve reasoning about each other's knowledge. In such cases, we often use logical primitives such as, ``If I know that she knows $a$, and I know that she arrived at $a$ using a set of rules that we commonly agreed upon, then I know $a$''. Examples include games like poker, complex auctions, cryptographic scenarios, and \href{https://en.wikipedia.org/wiki/Hat_puzzle}{logical hat puzzles}, where we must process complex statements of the sort ``I know that she knows that he does not know $a$'' based on some logical primitives, and the common knowledge of the agents. A system of logic that provides these simple and intuitive rules for multi-agent reasoning is modal logic, which we outlined in Section~\ref{sec: logic}. 

On the other hand, when agents (such as ourselves) describe the world through physical theories, we would like to be able to model the agents also as physical systems of the theory, in order to develop a more complete understanding of the physical world. In particular, we would like our theory to model at least those parts of the agent that are responsible for storing and processing empirical data, such as their memory. When that theory is quantum mechanics, it turns out that these two desiderata (applying standard rules of logic to reason about each other's knowledge, and modelling agents' memories as physical systems) are incompatible with agents' empirical observations. This incompatibility (or ``paradox'') was first formalised by Frauchiger and Renner, in a thought experiment where agents who can measure each others memories (modelled as quantum systems) and reason about shared and individual knowledge may reach conclusions that contradict their own observations \cite{Frauchiger2018}. 

The FR paradox is originally presented in terms of a prepare and measure scenario where parties exchange quantum states and thereby generate quantum correlations between each other. However it can be equivalently described by an entanglement-based scenario where the parties do not communicate but pre-share suitable quantum correlations, as illustrated  in Figure~\ref{fig:circuits}. %Further, the original version of the paradox is not directly formulated in terms of logic, but \cite{Nurgalieva2018} provides an equivalent reformulation of the original prepare and measure experiment in the framework of epistemic modal logic. 
Here, we present the entanglement version of the paradox (Section~\ref{sec: FRorig}) as this will be the most convenient form for generalising the paradox to post-quantum settings as we will do in Chapter~\ref{chapter: PRdoxespaper}. The Frauchiger-Renner thought experiment has fuelled an enormous volume of discussion and debate in the scientific community that is still ongoing. In Section~\ref{sec: FRdiscuss}, we aim to provide a brief overview of this debate and an outlook on this matter, which will also serve to motivate the results of the next chapter where we generalise the analysis beyond quantum theory.

\section{An entanglement version of the Frauchiger-Renner thought experiment}
\label{sec: FRorig}
\subsection{The assumptions}
The main result of Frauchiger and Renner establishes the impossibility of a physical theory $\mathcal{T}$ to simultaneously satisfy three assumptions, denoted by $\mathbf{Q}$, $\mathbf{C}$ and $\mathbf{S}$. In \cite{Nurgalieva2018}, it was pointed out that the FR argument involved an additional implicit assumption, $\mathbf{U}$. We present these four assumptions below as stated in \cite{Frauchiger2018} and \cite{Nurgalieva2018}, before we proceed to demonstrate their incompatibility with the entanglement version of FR's argument.

The first assumption $\mathbf{Q}$ pertains to the validity of certain predictions of quantum mechanics. For FR's argument, the full framework of quantum theory need not be assumed, but only requires a weaker version of the Born rule that is applicable for making deterministic predictions.
\begin{assumption}[$\mathbf{Q}$]
A theory $\mathcal{T}$ that satisfies $\mathbf{Q}$ allows any agent Alice to reason as follows. Suppose Alice is in the vicinity of a system $S$ associated with the Hilbert space $\mathscr{H}_S$, and she knows that ``The system $S$ is in a state $\ket{\psi}_S\in \mathscr{H}_S$ at time $t_0$''. Furthermore, suppose that Alice also knows that`` the value $x$ is obtained by a measurement of $S$ w.r.t. the family $\{\pi_x^{t_0}\}_{x\in X}$ of Heisenberg operators relative to time $t_0$, and the measurement is completed at time $t$''. If $\bra{\psi}\pi_{\xi}^{t_0}\ket{\psi}=1$ for some $\xi\in X$, then Alice can conclude that ``I am certain at time $t_0$ that $x=\xi$ at time $t$''.
\end{assumption}

The second assumption $\mathbf{C}$, originally referred to as ``self-consistency'', pertains to how agents reason about each other's knowledge.
\begin{assumption}[$\mathbf{C}$]
A theory $\mathcal{T}$ that satisfies $\mathbf{C}$ allows any agent Alice to reason as follows. If Alice has established that ``I am certain that an agent Bob, upon reasoning using theory $\mathcal{T}$, is certain that $x =\xi$ at time $t$''  then Alice can conclude ``I am certain that $x=\xi$ at time $t$.”
\end{assumption}

The third assumption, $\mathbf{S}$ (originally referred to as ``single-world'') pertains to the intuition that an agent experiences a single outcome when they perform or witness a measurement.
\begin{assumption}[$\mathbf{S}$]
A theory $\mathcal{T}$ that satisfies $\mathbf{S}$ disallows any agent Alice from making both the statements ``I am certain at time $t_0$ that $x=\xi$ at time $t$.'' and ``I am certain at time $t_0$ that $x\neq\xi$ at time $t$'', where $x$ is a value that can be observed at time $t > t_0$.
\end{assumption}

The fourth assumption $\mathbf{U}$, requires a bit more explanation. Suppose an agent Alice, has a system $S$ prepared in the state $\ket{\psi}_S=\frac{1}{\sqrt{2}}(\ket{0}+\ket{1})$. Alice measures $S$ in the $Z$ basis,  $\{\ket{0},\ket{1}\}$, and records the outcome $x$ of the measurement in her memory $A$ (which was initialised to the state $\ket{0}_A$. From Alice's perspective, $x$ is a uniformly distributed classical bit. Suppose also that (rather idealistically) that this measurement happens in Alice's lab which is a closed system consisting of the subsystems $S$ and $A$ that does not leak any information to the environment. Now, consider how an outside agent Bob who models $S$ and $A$ both as quantum systems  would describe the measurement process. Since Alice's lab is a closed system, Bob knows nothing about Alice's measurement outcome $x$ and would describe the evolution of systems $A$ and $S$ in the lab through a unitary map $U_{AS}$. Further, due to the perfect correlations between the measurement outcome and the memory state that records it, this unitary would be a coherent controlled NOT with the $S$ as control and $A$ as target. That is, Bob describes Alice's measurement through the evolution,
\begin{equation}
\label{eq: unitary}
   \frac{1}{\sqrt{2}}\Big(\ket{0}_S+\ket{1}_S\Big)\otimes\ket{0}_A\quad \xrightarrow{U_{AS}}\quad  \frac{1}{\sqrt{2}}\Big(\ket{0}_S\ket{0}_A+\ket{1}_S\ket{1}_A\Big).
\end{equation}
Hence, Bob would see Alice's memory $A$ to be entangled with her measured system $S$. The FR thought experiment assumes that outside agents such as Bob, who model an inside agent such as Alice as a quantum system would model measurements performed by the latter in through such reversible evolutions. This was noted in \cite{Nurgalieva2018}, and is encapsulated in the assumption $\mathbf{U}$ proposed by therein.
\begin{assumption}[$\mathbf{U}$]
A theory $\mathcal{T}$ that satisfies $\mathbf{U}$ allows any agent Bob to model measurements performed by any other agent Alice as reversible evolutions in Alice’s lab — for example, a unitary evolution $U_{AS}$ of the joint state of Alice’s memory $A$ and the system $S$ measured by her.
\end{assumption}

\begin{figure}[t!]
\centering
   \subfloat[Inside perspective]{
        \centering
       \includegraphics[width=0.4\textwidth]{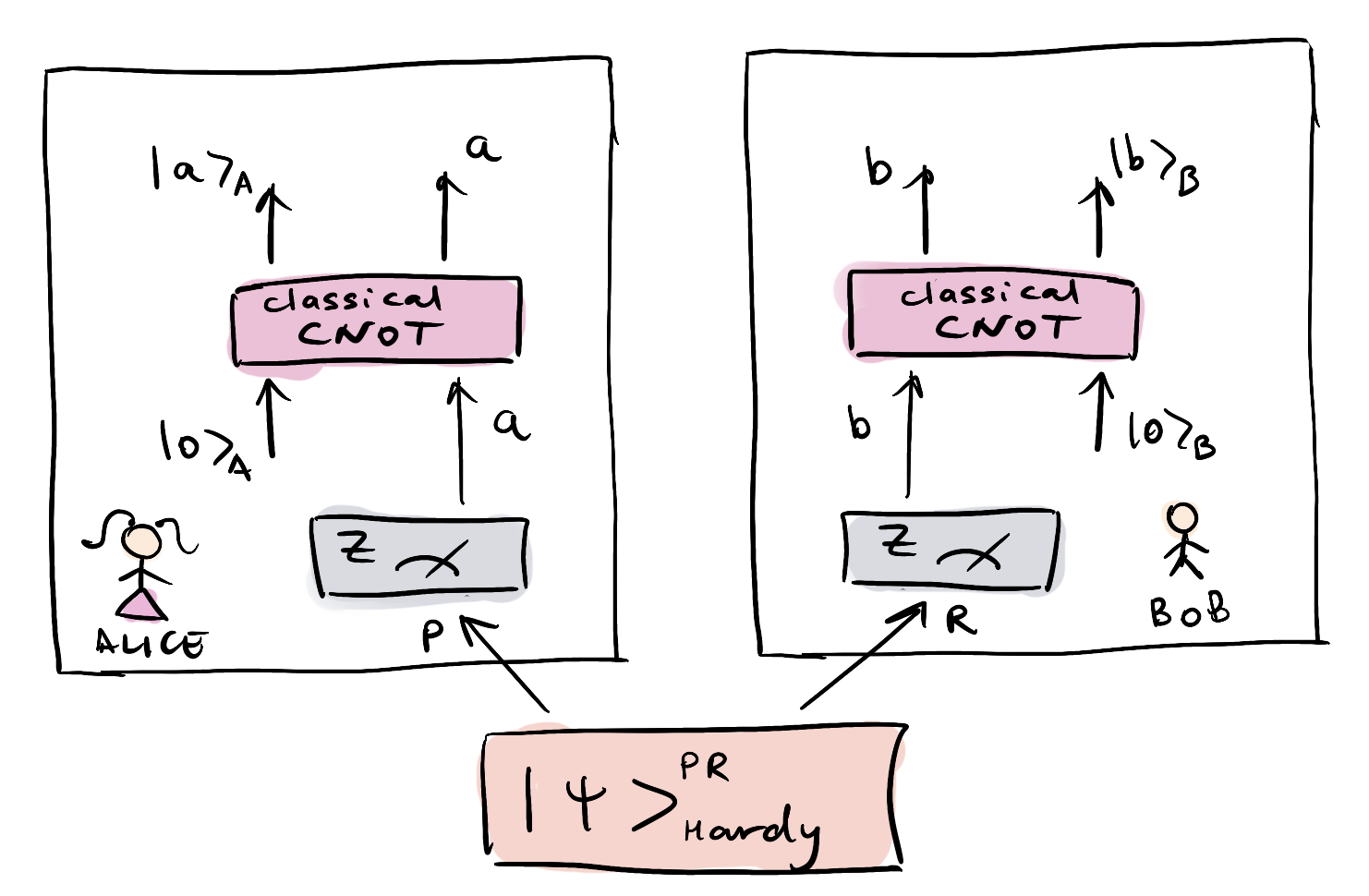}
        \label{fig:inside_persp}
}
  \subfloat[Outside perspective]{
    \centering
        \includegraphics[width=0.5\textwidth]{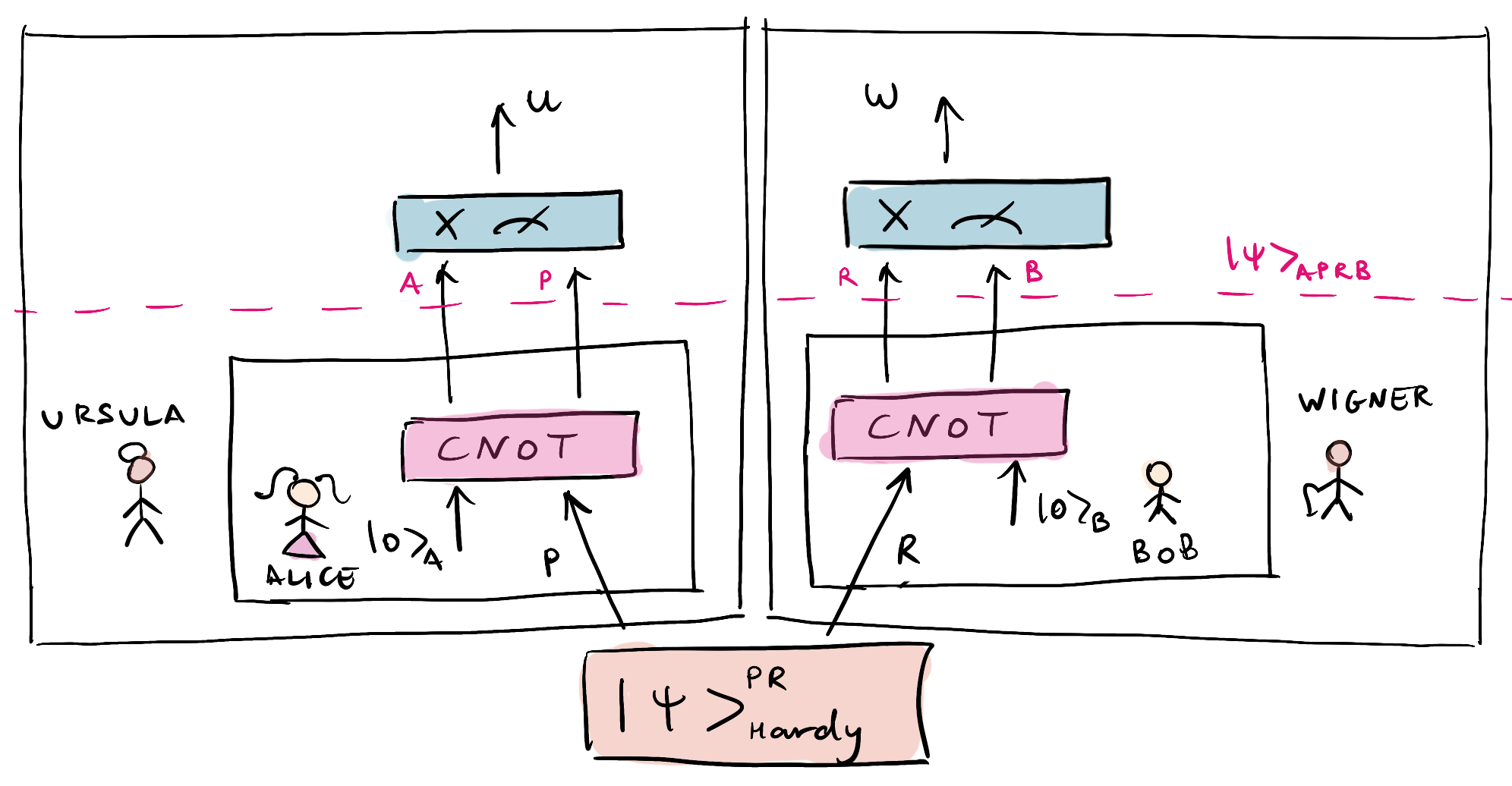}
        \label{fig:outside_persp}
}
\caption[Entanglement based version of the Frauchiger-Renner thought experiment]{{\bf An entanglement-based version of the Frauchiger-Renner setting \cite{Frauchiger2018} from different perspectives.} Alice and Bob (inside agents) share a Hardy state $\ket{\Psi}_{PR}=(\ket{00}+\ket{10}+\ket{11})/\sqrt{3}$, measure each their qubit ($P$ and $R$ respectively) and update their memories $A$ and $B$ accordingly. Their labs are contained inside the labs of the outside observers Ursula and Wigner, who can measure the systems $AP$ and $RB$ respectively. The paradox arises when one tries to combine the inside and outside perspectives of quantum measurements on an entangled system into a single perspective.
{\bf (a)} From their viewpoints, Alice and Bob measure their halves of $\ket{\Psi}_{PR}$ in the $Z$ basis $\{\ket{0},\ket{1}\}$ to obtain the outcomes $a$ and $b$. They then perform a classical CNOT (i.e., classical copy) to copy their classical outcome into their memories $A$ and $B$ both initialised to $\ket{0}$.
{\bf (b)} Ursula and Wigner perceive Alice and Bob's memory updates as implementing quantum CNOTs on $A$ controlled by $P$ and $B$ controlled by $R$ respectively. The resultant joint state is $\ket{\Psi}_{APRB}=(\ket{0000}+\ket{1100}+\ket{1111})/\sqrt{3}$. Hence, they see quantum correlations between the systems and memories of the inside agents. Later, they measure the joint systems $AP$ and $RB$ in the  ``$X$ basis'' $\{\ket{ok}=(\ket{00}-\ket{11})/\sqrt{2},\ket{fail}=(\ket{00}+\ket{11})/\sqrt{2}\}$ to obtain the outcomes $u$ and $w$ respectively. 
If they obtain $u=w=ok$, the agents can reason about each others' knowledge to arrive at the paradoxical chain of statements $u=w=ok\Rightarrow b=1 \Rightarrow a=1 \Rightarrow w=fail$. In Chapter~\ref{chapter: PRdoxespaper}, we extend this scenario to box world where Alice and Bob share a PR box instead of the Hardy state and find a suitable memory update operation and measurements for the parties such that a stronger version of the paradox is recovered, independently of the outcomes obtained.\footnotemark}
\label{fig:circuits}
\end{figure}

\subsection{The paradox: incompatibility of the assumptions}
One may have already spotted a tension between the assumptions $\mathbf{S}$ and $\mathbf{U}$ which describe the ``inside'' and ``outside'' perspectives of a quantum measurement i.e., the subjective experience of a single classical outcome for Alice, vs the entangled superposition state of Alice and her system as seen by Bob. This is precisely the tension that Wigner originally illustrated in his thought experiment. In fact, Wigner's thought experiment essentially corresponds to the two agent scenario described in the previous section (paragraph preceding assumption $\mathbf{U}$), where the inside observer Alice plays the role of ``Wigner's friend'' and the outside observer Bob who describes Alice as a quantum system plays the role of Wigner. By including the assumptions $\mathbf{Q}$ and $\mathbf{C}$ which introduce the element of reasoning about the knowledge of other agents, the FR experiment elevates this apparent tension to an explicit mathematical contradiction between the four assumptions. The following entanglement version of the thought experiment is entirely equivalent to the original prepare and measure version, but simplifies the description. 
\footnotetext{This figure is taken from our paper \cite{Vilasini_PRdoxes} which is joint work with Nuriya Nurgalieva and L\'idia del Rio, and the credits for this pretty illustration goes to Nuriya.}
\paragraph{Experimental setup.} The experiment involves four agents, whom we call Alice, Bob, Ursula and Wigner\footnote{In accordance with the naming convention of \cite{Nurgalieva2018, Vilasini_PRdoxes}.}. The experiment starts with Alice and Bob sharing the following bipartite state
\begin{equation}
    \label{eq: Hardy1}
    \ket{\psi}_{PR}=\frac{1}{\sqrt{3}}\Big(\ket{0}_P\ket{0}_R+\ket{1}_P\ket{0}_R+\ket{1}_P\ket{1}_R\Big),
\end{equation}
where $P$ and $R$ denote the subsystems held by Alice and Bob respectively (for reasons that will become more apparent in the next chapter). This state is known as a \emph{Hardy state} due to its particular relevance in Hardy's paradox \cite{Hardy1992, Hardy1993}. Let $A$ and $B$ denote the subsystems that correspond to Alice's and Bob's memory respectively, where they store the outcome of any measurement that they may perform.  Suppose that Alice's lab is located inside the lab of another agent, Ursula who can  perform joint measurements on Alice's system $P$ and her memory $A$. Similarly, let Bob's lab be located inside Wigner's lab, such that Wigner can perform joint measurements on Bob's system $R$ and his memory $B$. Alice's and Bob's labs are isolated such that no information about their measurement outcomes leaks out. 
\paragraph{The protocol: } The protocol is the following:
\begin{itemize}
    \item[\bf t=1] Alice measures her half ($P$) of joint state $\ket{\psi}_{PR}$, in the $\{\ket{0},\ket{1}\}$ (i.e., $Z$ or computational) basis, and stores the outcome $a$ in her memory $A$. 
    \item[\bf t=2] Bob measures his half ($R$) of joint state $\ket{\psi}_{PR}$, also in the $Z$ basis, and stores the outcome $b$ in his memory $B$. By the assumption $\mathbf{U}$, the final joint state of Alice and Bob's systems and memories after their measurements at times $t=1$ and $t=2$, as described from the outside would be
    \begin{equation}
    \label{eq: Hardy2}
    \ket{\psi}_{PARB}=\frac{1}{\sqrt{3}}\Big(\ket{0}_P\ket{0}_A\ket{0}_R\ket{0}_B+\ket{1}_P\ket{1}_A\ket{0}_R\ket{0}_B+\ket{1}_P\ket{1}_A\ket{1}_R\ket{1}_B\Big)
\end{equation}
    
    \item[\bf t=3] Ursula jointly measures the systems $P$ and $A$ in Alice's lab, in the $\{\ket{ok},\ket{fail}\}$ basis defined below and obtains the outcome $u$.
    \begin{equation}
        \begin{split}
 \ket{ok}_{PA}&:=\frac{1}{\sqrt{2}}\Big(\ket{0}_P\ket{0}_A-\ket{1}_P\ket{1}_A\Big)\\
          \ket{fail}_{PA}&:=\frac{1}{\sqrt{2}}\Big(\ket{0}_P\ket{0}_A+\ket{1}_P\ket{1}_A\Big)
        \end{split}
    \end{equation}
    \item[\bf t=4] Wigner jointly measures the systems $R$ and $B$ in Bob's lab, also in the $\{\ket{ok},\ket{fail}\}$ basis and obtains the outcome $w$.
        \begin{equation}
        \begin{split}
 \ket{ok}_{RB}&:=\frac{1}{\sqrt{2}}\Big(\ket{0}_R\ket{0}_B-\ket{1}_R\ket{1}_B\Big)\\
          \ket{fail}_{RB}&:=\frac{1}{\sqrt{2}}\Big(\ket{0}_R\ket{0}_B+\ket{1})R\ket{1}_B\Big)
        \end{split}
    \end{equation}
     \item[\bf t=5] Ursula and Wigner compare the outcomes of their measurements. If they were both ``ok'', they halt the experiment. Otherwise, they reset the timer and all systems to the initial conditions, and repeat the experiment. Note that there is a non-zero probability for both Ursula and Wigner to obtain the outcome ``ok'' when measuring the subsystems $PA$ and $RB$ in the state $\ket{\psi}_{PARB}$ of Equation~\eqref{eq: Hardy2} i.e., \begin{equation}
     \label{eq: okok}
         P(u=w=``ok''|\psi_{PARB})=|\Big(\bra{ok}_{PA}\otimes\bra{ok}_{RB}\Big)\cdot\ket{\psi}_{PARB}|^2=\frac{1}{12}.
     \end{equation}
\end{itemize}
The measurement basis for each agent mentioned above are agreed on beforehand and the agents do not communicate once the experiment begins, except for the communication between Ursula and Wigner at $t=5$. We now discuss how the agents reason about each other's knowledge to arrive at the contradiction, given that the experiment halted i.e., post-selecting on  the condition that $$u=w=ok$$. 

\paragraph{Agents' reasoning: } 
\begin{enumerate}
    \item \textit{Ursula reasons about Bob: } Rewriting the joint state $\ket{\psi}_{PARB}$ as follows
    \begin{equation}
    \begin{split}
      \ket{\psi}_{PARB}&=\sqrt{\frac{2}{3}}\ket{fail}_{PA}\ket{0}_R\ket{0}_B+\frac{1}{\sqrt{3}}\ket{1}_P\ket{1}_A\ket{1}_R\ket{1}_B,\\
      &=\sqrt{\frac{2}{3}}\ket{fail}_{PA}\ket{0}_R\ket{0}_B+\frac{1}{\sqrt{6}}\ket{fail}_{PA}\ket{1}_R\ket{1}_B-\frac{1}{\sqrt{6}}\ket{ok}_{PA}\ket{1}_R\ket{1}_B,
      \end{split}
    \end{equation}
    Ursula at $t=3$ reasons (using $\mathbf{Q}$) on obtaining the outcome $u=ok$ ,that she knows with certainty that Bob must have obtained the outcome $b=1$ at $t=2$. This is because Ursula knows that her (normalised) post-measurement state on obtaining the outcome $u=ok$ would be $\ket{ok}_{PA}\ket{1}_R\ket{1}_B$, and there is unit probability of Bob obtaining $b=1$ for his measurement on $B$ given this outcome. Thus Ursula (at $t=3$) knows that $$u=ok\Rightarrow b=1.$$
    \item \textit{Bob reasons about Alice: } From the original form of the joint state~\eqref{eq: Hardy2}, Bob reasons using $\mathbf{Q}$, on obtaining the outcome $b=1$ at time $t=2$ that he knows with certainty that Alice would have obtained the outcome $a=1$ at $t=1$ i.e., Bob (at $t=2$) knows that $$b=1\Rightarrow a=1.$$
     \item \textit{Alice reasons about Wigner: } Rewriting the joint state as 
         \begin{equation}
      \ket{\psi}_{PARB}=\frac{1}{\sqrt{3}}\ket{0}_P\ket{0}_A\ket{0}_R\ket{0}_B+\sqrt{\frac{2}{3}}\ket{1}_P\ket{1}_A\ket{fail}_{RB},
    \end{equation}
    Alice, on obtaining the outcome $a=1$ at $t=1$, uses $\mathbf{Q}$ to reason about what Wigner would observe at $t=4$ when he measures Bob's lab. Similar to previous steps, she would conclude with certainty that Wigner observes $w=fail$ i.e., Alice (at $t=1$) knows that $$a=1\Rightarrow w=fail.$$
\end{enumerate}
Note that in each of the steps above, $\mathbf{S}$ is implicitly used. These statements reflect the knowledge and reasoning of different agents, and are combined using the assumption $\mathbf{C}$ to give the following paradoxical chain of statements when $u=w=ok$ is observed at $t=4$.
\begin{equation}
\label{eq: FRchain}
    u=w=ok\Rightarrow b=1\Rightarrow a=1\Rightarrow w=fail
\end{equation}
In other words, whenever the experiment halts with  $u = w = ok$, the agents can make deterministic statements about each other’s reasoning and measurement outcomes, and thereby conclude that Alice had predicted $w = fail$ with certainty. This is in conflict with the assumption $\mathbf{S}$ which forbids $w$ from taking both the values with certainty. 

In the modal logic language, the assumption $\mathbf{C}$ can be seen as encoding the distribution axiom (Axiom~\ref{axiom:distribution}) i.e., ``If an agent knows a statement $\phi$ and also that $\phi\Rightarrow \psi$, then they know $\psi$'', along with the concept of trust (Definition~\ref{def:trust}) i.e., an agent $A$ trusts another agent $B$ if and only if ``$A$ knows that $B$ knows $\phi$ implies that $A$ knows $\phi$''. In \cite{Nurgalieva2018}, it is shown that the FR paradox can be reformulated as: the Kripke structure satisfying the knowledge axioms of modal logic described in Section~\ref{sec: logic} is incompatible with agents who apply quantum theory to model each other and reason using that logical system. They show that this is the case even if the truth axiom (Axiom \ref{axiom:truth}) is replaced with the weaker notion of a trust structure (Definition~\ref{def:trust}). We refer the reader to \cite{Nurgalieva2018} for a detailed analysis of this paradox in terms of the axioms of modal logic and the trust relations at different times. These ideas will also become clearer in Chapter~\ref{chapter: PRdoxespaper} where we derive an analogous paradox in box world, a generalised probabilistic theory (see Section~\ref{sec: GPT} for an overview of GPTs) by generalising the quantum theory dependent assumptions of FR to the general setting, in the language of modal logic. We now proceed to discuss some of the implications of FR's result and its relation to other no-go theorems.

\begin{remark}
It is important to note here, also as stressed in the original paper \cite{Frauchiger2018} that an ``agent'' need not necessarily correspond to a conscious observer in such thought experiments. An agent simply refers to any entity capable of performing measurements on other systems and applying simple deductive principles to process empirical information. For example, a small quantum computer would in principle fulfil the desiderata for agency in such settings. Hence, FR-type thought experiments could possibly be implementable in the near future, with progress in scalable quantum computing, even though they cannot be implemented with current technology.
\end{remark}

\section{Discussion}
\label{sec: FRdiscuss}

\subsection{Intuition and relation to Hardy's paradox}
The intuition behind the FR paradox described above comes from Bell's theorem, combined with the unitary description of the measurement from the outside perspective (Equation~\eqref{eq: unitary}). The entanglement of the shared Hardy state $\ket{\psi}_{PR}$ implies the inability to assign simultaneous values to the outcomes corresponding to different measurement settings \cite{Hardy1993}, the FR experiment corresponds to a situation where all the choices of measurements are actually implemented in a single run of the experiment since the unitarity of the measurement map~\eqref{eq: unitary} allows the outside agents to reverse the effect of the inside agents' measurement before performing their measurement. This also means that the final joint state $\ket{\psi}_{PARB}$ that Ursula and Wigner measure contains the initial correlations of $\ket{\psi}_{PR}$ as evident from Equations~\eqref{eq: Hardy1} and~\eqref{eq: Hardy2}. This leads to the paradoxical situation that the outcomes of all four parties cannot admit simultaneous value assignments. FR's paradox can be viewed as a manifestation of Hardy's proof of Bell's theorem \cite{Hardy1993} which is not based on the violation of Bell inequalities, but on deriving a logical contradiction between the prediction of quantum mechanics for certain entangled states (such as~\eqref{eq: Hardy1}) and the assumption of local hidden variables for the outcomes of possible measurements. Hardy derives such a contradiction for almost all bipartite entangled states in \cite{Hardy1993}.\footnote{This class excludes the maximally entangled Bell states which are too symmetric to derive a logical contradiction akin to the chain~\eqref{eq: FRchain}.} It can be shown that in the 4 agent setting of FR (Figure~\ref{fig:circuits}), replacing the shared state $\ket{\psi}_{PR}$ by any one state from the class of Hardy states, and modifying the definition of the $\{\ket{ok},\ket{fail}\}$ basis accordingly (also as proposed in \cite{Hardy1993}), one can elevate Hardy's argument for the outcomes of potentially unperformed measurements in the regular bipartite Bell scenario to the outcomes of the four measurements (one per agent) actually performed by in an FR-type setup. Hence the FR paradox is closely related to Hardy's paradox \cite{Hardy1992}, as was also pointed out in \cite{Pusey2018}. We leave the formal derivation of this and a more general characterisation of the states and measurements required for deriving FR-type paradoxes in $N$-agent experiments to future work. 

\subsection{Relation to other no-go theorems}
There are other no-go theorems in a similar vein as FR, that are based on an extension of Wigner's thought experiment. All these theorems illustrate the incompatibility between an objective assignment of values to the measurement outcomes of all agents in scenarios where quantum predictions are extended to the level of agents. The most notable of these is Brukner's no-go theorem \cite{Brukner2018} which states that the following four assumptions are incompatible: 1. quantum theory is universally valid (also when applied to observers), 2. locality i.e., the setting of one party doesn't affect the outcome of the other, 3. freedom of choice 4. the ability to jointly assign truth values to observed outcomes of all the observers. Brukner derives the CHSH inequalities from these assumptions 2.-4. and the well known violation of these Bell inequalities in quantum theory demonstrates the aforementioned incompatibility of the assumptions 1.-4.  Assumption 1 roughly corresponds to the assumptions $\mathbf{Q}$ and $\mathbf{U}$ of FR, with the difference being that the former would need to assume full validity of the Born rule rather than the weaker, deterministic version given by $\mathbf{Q}$. Assumption 4 is similar in motivation as the assumption $\mathbf{C}$ of FR, with the difference being that $\mathbf{C}$ relates to combining deterministic statements made by agents to assign values to all their outcomes. In this sense, $\mathbf{C}$ is a stronger assumption than 4. which does not require the value assignments to be deterministic. Assumptions 2 and 3 are different from the FR setting, where the measurement choice for each party is fixed. One could say that FR's no-go theorem is to Brukner's what Hardy's is to Bell's. The former, in both cases are based on logical arguments, that can be made without reference to inequalities but require certain probabilities to be deterministic i.e., have 0 or 1 values. For example, in the FR case, we require $P(b=0|u=ok)=P(a=0|b=1)=P(w=ok|a=1)=0$ in order to make deterministic logical inferences, which we cannot have for measurements on the maximally entangled Bell states (as per Hardy's argument \cite{Hardy1993}). However the latter could still be certified as incompatible with Brukner's assumptions since it violates CHSH inequalities. 

\cite{Bong2020} proposed a modified version of Brukner's theorem, noting a limitation of the latter. They pointed out that the Bell inequalities derived through assumptions 2.-4., can also be derived using the assumptions 3. (free choice) and Kochen-Spekker non-contextuality \cite{Kochen1968}, without reference to the objectivity of the agents' observations, which is supposed to be the subject of such theorems. They show that Bell inequality violations are in general, not sufficient to demonstrate a contradiction between quantum theory and the existence of observer independent or objective facts about the world, and derive a new set of inequalities from weaker assumptions that are indeed sufficient for this purpose. The difference is that, Brukner's version includes hypothetical measurements that may not have been actually performed in the considered experimental run, while \cite{Bong2020} only considers measurement outcomes that were actually observed by the agents, and tests for the objectivity (i.e., possibility of assigning universal truth values) of the observed outcomes. The exact connections between FR's paradox and these no-go theorems is not known. But these examples suggest that contextuality would be a necessary but not sufficient condition to derive FR-type paradoxes. We discuss the relation to contextuality in Chapter~\ref{chapter: PRdoxespaper}, Section~\ref{ssec: contextuality} and leave a characterisation of the set of all contextual scenarios leading to FR-type paradoxes to future work. Such a characterisation would require a general definition of an FR-type logical paradox which is currently lacking, and to the best of our knowledge the only examples of such logical multi-agent paradoxes are the original FR paradox \cite{Frauchiger2018} and our box world version \cite{Vilasini_PRdoxes} (Chapter~\ref{chapter: PRdoxespaper}). Generalising the study of such paradoxes to arbitrary theories as attempted in \cite{Vilasini_PRdoxes} (Chapter~\ref{chapter: PRdoxespaper}) provides a starting point for addressing such questions.

\subsection{Interpretations of quantum theory}
When treating agents as physical systems, the divide between the ``observer'' and the ``observed'' becomes subjective--- we always put ourselves in the former category, but may be put in the latter category by another observer. Wigner \cite{Wigner1961} was among the first to point out through his famous thought experiment, that this leads to conceptual issues in the quantum case, and this problem is sometimes colloquially referred to as the \emph{shifty split}. FR's extension of the Wigner's friend scenario formalises this conceptual argument as a no-go theorem, thereby providing a concrete footing for the longstanding debates regarding the measurement problem and in turn, interpretations of quantum theory, which represent the different stances that one can take on the measurement problem. We begin this discussion by noting the importance of differentiating between objections against the FR result vs objections against specific assumptions used in deriving the result. The result itself, demonstrates an incompatibility between a certain set of assumptions and subsumes the latter type of objections--- it leaves open the choice of which assumption(s) among $\mathbf{Q}$, $\mathbf{C}$, $\mathbf{S}$ and $\mathbf{U}$ must be dropped to resolve the paradox, it is agnostic to the choice itself. This allows for a classification of the different interpretations of quantum theory based on the choice they make, and also indicates that any interpretation that satisfies all four assumptions should be ruled out. It appears that the main interpretation(s) under threat of being ruled out are those presented in standard textbooks which albeit ambiguous, seem to be trying to have the cake and eat it too \cite{Pusey2018}. Possible resolutions to the apparent paradox involve dropping one or more of the assumptions.

\paragraph{Dropping $\mathbf{Q}$ and/or $\mathbf{U}$: } Objections against FR such as ``you cannot put agents in superpositions'', or ``the outside description of the measurement~\eqref{eq: unitary} should lead to a mixed state and not a pure, entangled state'' do not refute the validity of the theorem but that of the assumptions $\mathbf{U}$ and/or $\mathbf{Q}$ about extending quantum theory to agents. This is one possible resolution allowed by the theorem.\footnote{Similar misinterpretations can occur with Bell's theorem. For example, if we interpret Bell’s theorem as telling us that there is fundamental uncertainty in quantum theory, that cannot be explained away by conditioning on classical information. A realist would argue that this is not true, since we can have fine-tuned, classical explanations i.e., non-local HV theories. Stating Bell’s theorem as a no-go result, this argument is subsumed: you can either drop ``realism'' or ``locality'', you have a choice, the theorem doesn't tell you which one to pick.} Further, in \cite{Sudbery2017}, Anthony Sudbery analyses Bohmian mechanics \cite{DeBroglie1927, Bohm1952, Durr2009} in the light of FR's result, arguing that the interpretation satisfies all assumptions, except possibly $\mathbf{U}$ regarding the particular evolution of the labs. This would imply that Bohmian mechanics does not agree with the predictions of quantum theory at all scales, as is commonly believed. Veronika Baumann and Stefan Wolf \cite{Baumann2018} have provided an interesting analysis of the thought experiment in the relative state formalism \cite{Everett1957, Wheeler1957}, considering different ways of describing the evolution of the labs and showing that these can lead to different predictions. They also compare other interpretations in this regard and show that the relative state formalism admitting unitary and universal quantum description deviates from the standard Born rule. 

\paragraph{Dropping $\mathbf{C}$: } Certain versions of the Copenhagen interpretation \cite{Heisenberg1935, Bohr1949} also turn out to problematic for similar reasons as the textbook approach, and Matthew Leifer argues (for example, in this \href{https://www.youtube.com/watch?v=MaRjP5H0vR4}{talk}) that only a \emph{perspectival} version of the interpretation, which gives up the assumption $\mathbf{C}$ would survive. QBism \cite{Fuchs2013} and relational quantum mechanics are also perspectival in this sense and propose a similar resolution--- measurement results are real only from the subjective perspective of specific agents, and not objective properties of the world.\footnote{However, as pointed out in \cite{Nurgalieva2018, Frauchiger2018}, it may be difficult to fully analyse such multi-agent experiments in QBism which tends to limit its focus on the experience and actions of single agents.} Further, a theory that satisfies all assumptions except $\mathbf{C}$, would call for a revision of our understanding of causal and logical inference. The latter, since dropping $\mathbf{C}$ implies a break down of standard rules of classical modal logic \cite{Nurgalieva2018}. The former, due to the following reason. As we have seen in Part I of this thesis, standard approaches to causal inference depict causal structures as having certain observed and certain unobserved nodes. The former are classical variables corresponding to the observed measurement statistics. In theories where agents as modelled as quantum systems according to $\mathbf{Q}$ and $\mathbf{U}$, whether measurement outcomes are seen as classical variables or as entangled quantum subsystems is subjective, which challenges existing models for causal inference (both in classical and quantum theory). This suggests that causal structure itself may be subjective in such scenarios. It would be interesting to study causal and logical inference in such scenarios and understand how these differ from situations where agents are modelled as classical systems.

\paragraph{Dropping $\mathbf{S}$: } This appears to be an intuitive assumption but a violation of this is not necessarily in conflict with the quantum formalism. It is argued that interpretations such as many-worlds (and its numerous variations) \cite{DeWitt1970, Deutsch1985} as well as the relative state formalism mentioned earlier, do not satisfy this assumption \cite{Frauchiger2018}. In the former, measurement is seen as leading to a branching into different possible worlds such that every possible outcome of the measurement occurs in a corresponding world. Whether many-worlds type interpretations satisfy the other assumptions, may depend on the particular variation being considered, and how the branching is defined therein.

\paragraph{Other assumptions: } Other implicit assumptions of the FR experiment have been pointed out \cite{Sudbery2019, Nurgalieva2018}. Some of these assumptions relate to idealizations such as the ability to perfectly prepare pure states and perform perfect projective measurements. While it is true that a relaxation of such assumptions may no longer result in a deterministic, logical paradox, these would still result in a conceptual problem regarding the incompatibility of universal quantum theory and objectivity of measurement results, akin to those pointed out by \cite{Brukner2018, Bong2020}. Further, one can argue that most interpretations would allow for such idealizations to be made, which are common place in theoretical studies. It would be interesting to analyse whether some of the other implicit assumptions in the result pointed out in \cite{Sudbery2019} are no longer needed in the entanglement version of the experiment presented here. This discussion by no means does full justice to the massive volume of discussions and literature generated by the FR result. We refer the reader to the original paper \cite{Frauchiger2018} as well as \cite{Nurgalieva2018} for more in-depth discussions on interpretations.

\subsection{Outlook}
In conclusion, we draw yet another analogy with Bell's theorem \cite{Bell}. The theorem has resulted in significant conceptual and practical advancement by ruling out certain classical descriptions of the world (i.e., local hidden variables) due to their incompatibility with our observations, which agree with quantum theory. It has also faced several objections and criticism over the years, and generated several alternate derivations/versions \cite{CHSH69, Clauser1974, Hardy1993, GHZ1989, Wood2015} which reveal different facets of quantum theory that deviate from our understanding of the classical world. In a similar manner, theorems such as those of Frauchiger and Renner have deep foundational implications for the nature of objectivity, agents' experience, logical reasoning and causal inference. Critically examining the assumptions behind such theorems and characterising general scenarios where such incompatibilities can arise are likely to reveal important connections between these fundamental concepts and the role that quantum theory plays in them. Finally, as previously noted, FR-type thought experiments could possibly be implemented in the near future depending on the advancements in scalable quantum computing, and don't necessarily require the agents to be conscious observers \cite{Frauchiger2018}. This would in principle allow for the interpretations of quantum theory to be subject to experimental tests. The caveat being that a clear cut answer would require a loophole-free implementation. This would be significantly more difficult, as we have seen in the case of Bell's theorem where the first experimental test \cite{Freedman1972} was realised within a decade but the first loophole free tests \cite{Giustina2015, Hensen2015, Shalm2015} were only realised more than 5 decades after the original theorem \cite{Bell} was proposed.

\chapter{Multi-agent paradoxes beyond quantum theory}
\label{chapter: PRdoxespaper}

%At the core, the Frauchiger-Renner argument relies on entanglement and unitarity of quantum theory-- the former gives rise to a shared state that is incompatible with a local hidden variable model, and the latter allows the agents to implement all 4 measurements (2 per side) in one run of the experiment by unitarily reversing the effect of the first measurement before performing the second. 
\lettrine[nindent=0em, slope=-.5em,lines=2]{T}{he} Frauchiger-Renner thought experiment \cite{Frauchiger2018} reveals an incompatibility between extending quantum theory to reasoning agents and certain simple rules of logical deduction \cite{Nurgalieva2018}, which we have discussed in Chapter~\ref{chapter: FR}. Our goal here is to understand whether this incompatibility is a peculiar feature of quantum theory, or whether modelling reasoning agents using other physical theories can also lead to such contradictions. Previous works have analysed such multi-agent paradoxes only in quantum theory, and a theory-independent analysis could possibly reveal more fundamental aspects these examples. For example, is the reversibility (in the quantum case, unitarity) of the global measurement transformation a necessary property for such paradoxes or the entanglement of the shared state (in the quantum case, the Hardy state), or both? Generalising this study beyond the quantum setting would not only help us identify the features of physical theories responsible for such paradoxes, but may also shed some light on the structure of logic in non-classical theories. Here, we investigate this question within the landscape of generalized probabilistic theories \cite{Hardy01, Barrett07}. This chapter is based on our paper \cite{Vilasini_PRdoxes}, which is coauthored by Nuriya Nurgalieva and L\'idia del Rio.

\section{Summary of contributions: in words and in poetry}
%The assumptions involved in the result of Frauchiger and Renner rely on quantum theory, and the first step towards analysing these paradoxes in arbitrary physical theories is to formulate these assumptions in theory-independent manner such that they can be instantiated in a theory of choice. 
In Section~\ref{sec:conditions}, we generalize the Frauchiger-Renner conditions using the laguage of modal logic, so that they can be applied to any physical theory. 
In particular, in Section~\ref{ssec: mmtsupdate}, we introduce a way to describe an agent's measurement from the perspective of other agents in generalised probabilistic theories (GPT). Finally, in Section~\ref{sec:paradox} we derive a logical inconsistency akin to one found in \cite{Frauchiger2018}, using a setup where agents share a PR box, a maximally non-local resource in box world (a particular GPT). The paradox found is stronger than the quantum one, in the sense that it does not rely on post-selection.\footnote{The joint state and the probability distributions of the original Frauchiger-Renner paradox are akin to those of Hardy's paradox~\cite{Hardy1993}. For a comparison of Hardy's paradox and PR box and why the latter allows for a contradiction without post-selection, see \cite{Abramsky15}. An entanglement version of the Frauchiger-Renner experiment and it's relation to our extension is explained in Figure \ref{fig:circuits}.}. Another important point brought to light by our version of the paradox is that the reversibility of the measurement map (akin to the assumption $\mathbf{U}$ in the quantum case) is not necessary for deriving such paradoxes in general. In Section~\ref{sec:discussion}, we provide a detailed discussion which includes a comparison of our results with the quantum case (Section~\ref{ssec: discuss_quantum}), its implications for measurements in box world (Section~\ref{ssec: discuss_mmts}), relationships between multi-agent paradoxes and contextuality (Section~\ref{ssec: contextuality}), and the plentiful scope for future work. Based on the knowledge of the summary sections of previous chapters, the reader might logically reason and thereby consider possible a world $s$ where the present chapter also contains a poetic summary. This is indeed that world.

\textit{We talk, so we reason.\\
We reason about what we know,\\
We reason about what others know\\
And if we \hyperref[def:trust]{trust} them, make their knowledge our own.}

\textit{We learn, so we store.\\
We store our knowledge in a part of our memory.\\
We model that memory by a physical theory.\\
But if that theory is quantum, this leads to an inconsistency \cite{Frauchiger2018}. }

\textit{We learn, so we wonder.\\
Which theories lead to such apparent inconsistensies,\\
Between reasoning agents and their memories modelled ``physically''?\\
Proposing a \hyperref[def:memory_update]{mathematical map} for update of memories,\\
An example we find in box world, a GPT,\\
where using a PR box, agents find a \hyperref[sec:paradox]{stronger paradox}.}

\textit{We answer questions, so we question more.\\
What properties of theories lead to these paradoxes galore?\\
\hyperref[ssec: contextuality]{Must be contextuality}, which forms of it, we are not yet sure\\
That, in future work we will explore!}

%{\raggedleft \textit{-Vilasini Venkatesh (2019)}\par}

\section{Generalized reasoning, memories and measurements}
\label{sec:conditions}

Here we generalize the assumptions underlying the Frauchiger-Renner result to general physical theories. The conditions can be instantiated by each specific theory.  This includes but is not limited to theories framed in the approach of generalized probabilistic theories (GPTs) \cite{Hardy01}. 
In some theories, like quantum mechanics and box world (a GPT), we will find these four conditions to be incompatible, by finding a direct contradiction in examples like the Frauchiger-Renner experiment or the PR-box experiment described in Section~\ref{sec:paradox}. In other theories (like classical mechanics and Spekkens' toy theory \cite{Spekkens07}) these four conditions may be compatible.  A complete characterization of theories where one can find these paradoxes is the subject of future work.

\subsection{Reasoning about knowledge}

%\begin{figure}[t]
%\centering
%\subfloat[An agent using deduction, applying the distribution axiom of modal logic.]{
 %   \vspace{1cm}
 %   \includegraphics[scale=0.1]{Ch7_Fig2a.png}
%    \label{fig:distribution}}
%\subfloat[An agent $A_i$ trusts another agent $A_j$, denoted by $A_j \leadsto A_i$, if they take all of $A_j$'s knowledge to be true for $A_i$ as well.]{
%    \includegraphics[scale=0.115]{Ch7_Fig2b.png}
%    \label{fig:trust}
%}
%\caption[Illustration of reasoning agents using the modal logic]{\textbf{Agents use logic to reason.} A desiderata for useful physical theories is that agents be allowed to make deductions and transfer knowledge from one another, given a trust relation (Definition \ref{def:reasoning}). For a short review of the modal logic framework and axioms, see Section~\ref{sec: logic}.}
%\end{figure}

This condition is theory-independent. It tells us that rational agents can reason about each other's knowledge in the usual way. This is formalized by a weaker version of \emph{epistemic modal logic}, which we explain in the following. For the full derivation of the form used here see \cite{Nurgalieva2018}, and for an overview of the modal logic framework, see Section~\ref{sec: logic}.

We start with a simple example.
The goal of modal logic is to allow us to operate with chained statements like ``Alice knows that Bob knows that Eve doesn't know the secret key $k$, and Alice further knows that $k=1$,''  which can be expressed as $$K_A\ [(K_B\ \neg K_E\  k ) \ \wedge \ k=1],$$ where the operators $K_i$ stand for ``agent $i$ knows.''
If in addition Alice trusts Bob to be a rational, reliable agent, she can deduce from the statement ``I know that Bob knows that Eve doesn't know the key'' that ``I know that Eve doesn't know the key'', and forget about the source of information  (Bob). This is expressed as   $$K_A ( K_B\  \neg K_E \ k) \implies K_A \ \neg K_E \ \ k.$$
We should also allow Alice to make deductions of the type ``since Eve does not know the secret key, and one would  need to know the key in order to recover the encrypted message $m$, I conclude that Eve cannot know the secret message,'' which can be encoded as $$K_A [(\neg K_E\  k) \wedge (K_i\ m \implies K_i\ k,\ \forall \ i)] \implies K_A \neg K_E\  m.$$ 
Generalizing from this example, this gives us the following structure. 

\begin{definition}[Reasoning agents]
\label{def:reasoning}
An experimental setup with multiple agents $A_1, \dots A_N$ can be described by knowledge operators $K_1, \dots K_N$ and statements $\phi \in \Phi$, such that $K_i \phi$ denotes ``agent $A_i$ knows $\phi$.'' It should allow agents to make \emph{deductions}\footnote{Note that this is the distribution axiom, (Axiom~\ref{axiom:distribution}).}, that is
 $$K_i [\phi \wedge (\phi \implies \psi) ] \implies K_i\ \psi.$$
 
Furthermore, each experimental setup defines a \emph{trust relation} between agents (Definition~\ref{def:trust}): we say that an agent $A_i$ trusts another agent $A_j$ (and denote it by $A_j \leadsto A_j$) iff  for all statements $\phi$, we have $$ K_i (K_j \ \phi) \implies K_i\ \phi.$$
\end{definition}

%For the purposes of following the example of Section~\ref{sec:paradox}, this informal definition suffices. The full formal version of the axioms of modal logic used here can be found in Section~\ref{sec: logic}.

\begin{remark}[One human $\neq$ one agent]
Note that in general `one human $\neq$ one agent.' For example, consider a setting where we know that Alice's memory will be tampered with at time $\tau$ (much like the original Frauchiger-Renner experiment, or the sleeping beauty paradox \cite{Elga2000}). We can define  two different agents $A_{t<\tau}$ and $A_{t>\tau}$ to represent Alice before and after the tampering--- and then for example Bob could trust pre-tampering (but not post-tampering) Alice, $A_{t<\tau} \leadsto B$.
\end{remark}

\begin{remark}[Complexity cost of reasoning]
Note that in general, even the most rational physical agents may be limited by bounded processing power and memory capacity, and will not be able to chain an indefinite number of deductions within sensible time scales. That is, these axioms for reasoning are an idealization of absolutely rational agents with unbounded processing power (see~\cite{Aaronson2017} for an overview of this and related issues). If we would like modal logic to apply to realistic, physical agents, we might account for a cost (in time, or in memory) of each logical deduction, and require it to stay below a given threshold, much like a resource theory for complexity. However, in the examples of this chapter, agents only need to make a handful of logical deductions, and these complexity concerns do not play a significant role. 
\end{remark}

\subsection{Physical theories as common knowledge}

This condition is to be instantiated by each physical theory, and is the way that we incorporate the physical theory into the reasoning framework used by agents in a given setting.
If all agents use the same  theory to model the operational experiment (like quantum mechanics, special relativity, classical statistical physics, or box world), this is included in the \emph{common knowledge} shared by the agents.  
For example, in the case of quantum theory, we have that ``everyone knows that the probability of obtaining outcome $\ket x$ when measuring a state $\ket\psi$ is given by $|\braket x\psi|^2$, and everyone knows that everyone knows this, and so on.''

%\begin{figure}[t!]
%    \centering
%    \includegraphics[scale=0.12]{Ch7_Fig3.png}
 %   \caption[A theory as common knowledge possessed by agents]{\textbf{Common knowledge.} Here, a shared physical theory $\mathbb T$ is common knowledge: all agents know that all agents know that ... (and so on) ... that theory $\mathbb T$ holds.}
  %  \label{fig:theory}
%\end{figure}

\begin{definition}[Common knowledge]
We model a physical theory shared by all agents $\{A_i\}_i$ in a given setting as a set $\mathbbm T$ of statements that are common knowledge shared by all agents, i.e.
$$ \phi \in \mathbb T \iff  (\{K_i\}_i)^n \ \phi, \quad \forall\ n\in \mathbb N,$$
where $(\{K_i\}_i)^n$ is the set of all possible sequences of $n$ operators picked from $\{K_i\}_i$. For example, $(K_1 \ K_5 \ K_1 \ K_2) \in (\{K_i\}_i)^4$ and stands for ``agent $A_1$ knows that agent $A_5$ knows that agent $A_1$ knows that agent $A_2$ knows.'' 
\end{definition} 

Note that the set $\mathbb T$ of common knowledge may include statements about the settings of the experiment, as well as complex derivations.\footnote{One can also alternatively model a physical theory as a subset $\mathbb T_P$ of the set $\mathbb T$ of common knowledge, $\mathbb T_P\subseteq\mathbb T$, in the case when details of experimental setup are not relevant to the theoretical formalism.} 
To find our paradoxical contradiction, we may only need a very weak version of a full physical theory: for example Frauchiger and Renner only require a possibilistic version of the Born rule, which tells us whether an outcome will be observed with certainty \cite{Frauchiger2018}. This will also be the case in box world.

\subsection{Agents as physical systems}

In operational experiments, a reasoning agent can make statements about systems that she studies; consequently, the theory used by the agent must be able to produce a description or a model of such a system, namely, in terms of a set of states. For example, in quantum theory a two-state quantum system with a ground state $\ket{0}$ and an excited state $\ket{1}$ (\textit{qubit}) can be fully described by a set of states $\{\ket{\psi}\}$ in the Hilbert space $\mathbb{C}^2$, where $\ket{\psi}=\alpha\ket{0}+\beta\ket{1}$ with $\alpha,\beta\in\mathbb{C}$ and $|\alpha|^2+|\beta|^2=1$. Other examples of theories and respective descriptions of states of systems include: GPTs, where e.g.~a generalised bit (\textit{gbit}) is a system completely characterized by two binary measurements which can be performed on it \cite{Barrett07} (a review of GPTs can be found in Section \ref{sec: GPT}); algebraic quantum mechanics, with states defined as linear functionals $\rho:A\to\mathbb{C}$, where $A$ is a $C*$-algebra \cite{vonNeumann1955}; or resource theories with some state space $\Omega$, and epistemically defined subsystems \cite{delRio2015, Kraemer2018}.

\begin{definition}[Systems]
\label{def:systems}
A ``physical system'' (or simply ``system'') is anything that can be an object of a physical study\footnote{We strive to be as general as possible and do not suppose or impose any structure on systems and connections between them; in particular, we don't make any assumptions about how composite systems are formally described in terms of their parts.}. A system can be characterized, according to the theory $\mathbb{T}$, by a set of possible states $\mathscr{P}_S$. In addition, a system is associated with a set of allowed operations, $\mathcal{O}_S: \mathscr{P}_S \mapsto \mathscr{P}_S$ on these states.
\end{definition}

\begin{definition}[Parallel composition]
\label{def: parallel}
For any two systems $S_1$ and $S_2$, the union of the two defines a new system $S_1\cup S_2$ or simply $S_1S_2$. The operator $\parallel$ denotes parallel composition of states and operations such that $P_{S_1}\parallel P_{S_2} \in \mathscr{P}_{S_1S_2}$ whenever  $P_{S_1}\in \mathscr{P}_{S_1}$ and $P_{S_2}\in \mathscr{P}_{S_2}$ and similarly, $O_{S_1}\parallel O_{S_2} \in \mathcal{O}_{S_1S_2}$ whenever  $O_{S_1}\in \mathcal{O}_{S_1}$ and $O_{S_2}\in \mathcal{O}_{S_2}$. In other words, the state $P_{S_1}\parallel P_{S_2}$ of $S_1S_2$ can be prepared by simply preparing the states $P_{S_1}$ and $P_{S_2}$ of the individual systems $S_1$ and $S_2$ and the operation $O_{S_1}\parallel O_{S_2}$ can be implemented by locally performing the operations $O_{S_1}$ and $O_{S_2}$ on the individual systems.
\end{definition}
We assume no further structure to this operator. Note also that we do not assume that a given composite system can be split into/described in terms of its parts even though combining individual systems in this manner allows us to define certain states of composite systems\footnote{In fact, in box world, we can consider operations on two initial systems that transform it into a new, larger system that can no longer be seen as being made up of 2 smaller systems. We call this ``supergluing", see Section~\ref{sec:discussion} for a discussion.}. Now we introduce agents into the picture. 
\begin{definition}[Agents]
\label{def:agents}
A physical setting may be associated with a set $\mathcal A$ of agents. 
An agent $A_i \in\mathcal{A}$ is described by a knowledge operator $K_i\in\mathcal{K_\mathcal{A}}$ and a physical system $M_i\in\mathcal{M_\mathcal{A}}$, which we call a ``memory.'' Each agent may study other systems according to the theory $\mathbb T$. An agent's memory $M_i$ records the results and the consequences of the studies conducted by $A_i$. The memory may be itself an object of a study by other agents.
\end{definition}

\subsection{Measurements and memory update}
\label{ssec: mmtsupdate}
Here we consider measurements both from the perspective of an agent who performs them, and that of another agent who is modeling the first agent's memory. In an experiment involving measurements, each agent has the subjective experience of only observing one outcome (independently of how others may model her memory), and we can see this as the definition of a measurement: if there is no subjective experience of observing a single outcome, we don't call it a measurement. We can express this experience as statements such as $\phi_0 =$ ``The outcome was 0, and the system is now in state $\ket0$.'' We explain this further after the formal definition.

\begin{definition}[Measurements]
\label{def:measurements}
A measurement is a type of study that can be conducted by an agent $A_i$ on a system $S$, the essential result of which is the obtained ``outcome'' $x\in\mathcal{X}_S$. If \emph{witnessed} by another agent $A_j$ (who knows that $A_i$ performed the measurement but does not know the outcome), the measurement is characterized by a set of propositions $\{\phi_x\}\in\Phi$, where $\phi_x$ corresponds to the outcome $x$, satisfying:
\begin{itemize}
\item $K_j(K_i(\exists \ x\in\mathcal{X}_S: K_i\ \phi_x))$,
\item $K_j \ K_i\ \phi_x \implies K_j \ K_i\ \neg (\phi_y), \quad \forall \ y \neq x$.
\end{itemize}
\end{definition}
The first condition tells us that $A_j$ knows that from $A_i$'s perspective, she must have observed one outcome $x\in X$, and $A_i$ would have used this knowledge to derive all the relevant conclusions, as expressed by the proposition $\phi_x$. For example, if the measurement represents a perfect $Z$ measurement of a qubit, $\phi_0$ may include statements like ``the qubit is now in state $\ket0$; before the measurement it was not in state $\ket1$; if I measure it again in the same way, I will obtain outcome 0'' and so on. Note that this condition does not imply that the measurement outcome stored in $A_i$'s memory is classical for $A_j$. In fact, in the quantum case $A_j$ may see $A_i$'s memory as a quantum system entangled with the system that $A_i$ measured. Despite this, $A_j$ knows that from $A_i$'s perspective, this outcome appears to be classical, which is what the first condition captures.
The second condition implements $A_i$'s experience of observing a single outcome, and the fact that the outside agent $A_j$ knows that this is the case from $A_i$'s perspective. If $A_i$ observes $x$, they conclude that the conclusions $\phi_y$ that they would have derived had they observed a different outcome $y$ are not valid and $A_j$ knows that $A_i$ would do so. In the previous example, they would know that it does not hold $\phi_1=$ ``the qubit is now in state $\ket1$; before the measurement it was not in state $\ket0$; if I measure it again I will see outcome 1.'' This condition also ensures that the conclusions $\{\phi_x\}_x$ are mutually incompatible, i.e.\ that the measurement is tightly characterized. 

A measurement of another agent's memory is also an example of a valid measurement. In other words, agent $A_j$ can choose  $A_i$'s lab, consisting of $A_i$'s memory and another system $S$ (which $A_i$ studies), as an object of her study. Thus, any agent's memory can be modelled by the other agents as a physical system undergoing an evolution that correlates it with  the measured system. In quantum theory, this corresponds to the unitary evolution 
\begin{align}
    \left( \sum_{x=0}^{N-1} p_x\ \ket x_{\text{system}}\right) \otimes \ket{0}_{\text{memory}} \to  \sum_{x=0}^{N-1} p_x \underbrace{\ket x_{\text{system}} \otimes \ket{x}_{\text{memory}}}_{=: \ \ket{\tilde x}_{SM}}.
    \label{eq:entangling_measurement}
\end{align}
The key aspect here is that the set of states of the joint system of observed system and memory, 
$\mathscr{P}_{SM} = \operatorname{span} \{  \ket x_{\text{system}} \otimes \ket{x}_{\text{memory}}  \}_{x=0}^{N-1} $ is post-measurement isomorphic to the set of states $\mathscr{P}_S$ system alone. That is, for every transformation $\epsilon_S$ that you could apply  to the system before the measurement, there is a corresponding transformation $\epsilon_{SM}$ acting on the $\mathscr{P}_{SM}$ that is operationally identical. By this we mean that an outside observer would not be able to tell if they are operating with $\epsilon_S$ on a single system before the measurement, or with  $\epsilon_{SM}$ on system and memory after the measurement. In particular, if $\epsilon_S$ is itself another measurement on $S$ within a probabilistic theory, it should yield the same statistics as post-measurement $\epsilon_{SM}$. 
For a quantum example that helps clarify these notions, consider $S$ to be a qubit initially in an arbitrary state $\alpha \ket0_S + \beta \ket1_S$. An agent Alice measures $S$ in the $Z$ basis and stores the outcome in her memory $A$.  While she has a subjective experience of seeing only one possible outcome, an outside observer Bob could model the joint evolution of $S$ and $A$ as
$$ \left( \alpha \ket0_S + \beta \ket1_S\right) \otimes \ket0_A \ \to \   \alpha \ket0_S \ket0_A + \beta \ket1_S \ket1_A. $$
Suppose now that (before Alice's measurement) Bob was interested in performing an $X$ measurement on $S$. This would have been a measurement with projectors $\{ \proj +_S, \proj -_S\}$, where $ \ket \pm_S = \frac1{\sqrt2} (\ket0_S \pm \ket1_S)$. 
However, he arrived too late: Alice has already performed her $Z$ measurement on $S$. If now Bob simply measured $X$ on $S$, he would obtain uniform statistics, which would be uncorrelated with the initial state of $S$. So what can he do? It may not be very friendly, but he can measure $S$ and Alice's memory $A$ jointly, by projecting onto
\begin{align*}
    \ket+_{SA} &= \frac1{\sqrt2} (\ket0_S \ket0_A + \ket1_S \ket1_A) \\ \ket-_{SA} &= \frac1{\sqrt2} (\ket0_S \ket0_A - \ket1_S \ket1_A),
\end{align*}
which yields the same statistics of Bob's originally planned measurement on $S$, had Alice not measured it first. In fact, this is precisely the $\{\ket{ok},\ket{fail}\}$ basis in which the outside observers, Ursula and Wigner measure, in the Frauchiger-Renner case (c.f. Section~\ref{sec: FRorig}).
Ideally, this equivalence should also hold in the more general case where the observed system may have been previously correlated with some other reference system: such correlations would be preserved in the measurement process, as modelled by the ``outside'' observer Bob. 

There are many options to formalize this notion that ``every way that an outside observer could have manipulated the system before the measurement, he may now manipulate a subspace of `system and observer's memory,' with the same results.'' 
A possible simplification to restrict our options is to take subsystems and the tensor product structure  as primitives of the theory, which may apply to GPTs \cite{Barrett07} but not for general physical theories (like field theories; for a discussion see \cite{Kraemer2018}). Here, we will for now restrict ourselves to this case, and leave a more general formulation of this condition as future work. We also restrict ourselves to information-preserving measurements (excluding for now those where some information may have leaked to an  environment external to Alice's memory), which are sufficient to derive the contradiction.

\begin{figure}[t!]
   \centering
\hspace{0.1cm}\subfloat[][{ \bf  Alice's perspective.}  The measurement in Z basis performed by Alice, who writes the classical result down to her memory $A$.]{\qquad\qquad\includegraphics[scale=1.1]{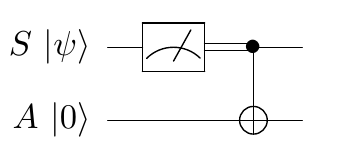}\qquad\qquad\quad \label{fig:view_of_alice}}\hspace{-0.4cm}\subfloat[][{ \bf  Bob's perspective on Alice performing a measurement.} The memory update of Alice, after she measures the system $S$ in Z basis, as seen from the point of view of the outside observer, corresponding to the memory update $U$.]{\qquad\qquad\qquad\qquad\includegraphics[scale=1.1]{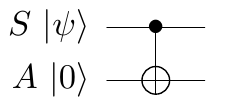}\quad\qquad\qquad\quad \label{fig:view_on_alice}}\\
\subfloat[][{ \bf  Bob's perspective.} Bob performs a measurement in the  $X$ basis of a system $S$.]{\qquad\includegraphics[scale=1.1]{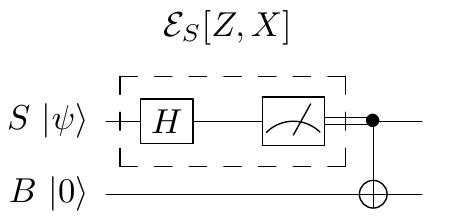}\qquad\qquad \label{fig:operation_s}}\hspace{-0.6cm}
   \subfloat[][{ \bf  Bob's perspective on performing a measurement after Alice.} Bob performing a measurement in $\tilde X$ basis of systems $S$ and $A$, after Alice's memory update $U$.]{\qquad\quad\includegraphics[scale=1.1]{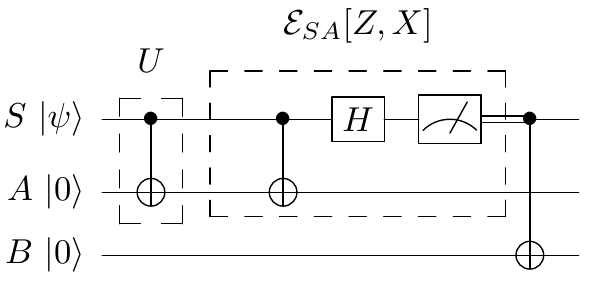}\label{fig:operation_sa}\qquad}
   \caption[Measurement and memory update in quantum theory from different perspectives]{{ \bf  The measurement and memory update in quantum theory from different perspectives.} 
From Alice's point of view, the measurement of the system $S$ either in Z basis yields a classical result, which she records to her memory $A$, performing a classical CNOT (Figure \ref{fig:view_of_alice}). For an outside observer, Bob who is not aware of Alice's measurement result, Alice's memory is entangled with the system and the CNOT is a quantum entangling operation which corresponds to the memory update $U$ (Figure \ref{fig:view_on_alice}). Further, there is no classical measurement outcome from Bob's perspective even though he knows that Alice would perceive one. If Bob had access to the system $S$ prior to the measurement by $A$, and wanted to measure it in X basis ($\{\ket{+}_S,\ket{-}_S\}$), he would have to perform an operation $\mathscr{E}_S [Z,X]$ (and then copy the classical result into his memory $B$) (Figure \ref{fig:operation_s}). If the system $S$ was initially in a state $\ket{\psi} = \ket{+}_S$, then a proposition which would correspond to this operation is $\phi[\mathscr{E}_S [Z,X](\ket{\psi}_S)] = ``s=+"$. However, if the measurement in Z is already performed by $A$ and the result is written to her memory, the whole process described by Bob as a memory update $u$, and in order to comply his initial wish to measure $S$ only, he can perform an operation $\mathscr{E}_{SA}[Z,X]$ on $S$ and $A$ together instead, which is a measurement in $\{\ket{+}_{SA},\ket{-}_{SA}\}$ basis (Figure \ref{fig:operation_sa}). A proposition which this operation yields is $\phi[\mathscr{E}_{SM_i}\circ u(\ket{\phi}_S\otimes \ket{0}_A)] = ``sa=+"$ (as $\ket{\xi}_{SA} = \ket{+}_{SA}$), which naturally follows from $``s=+"$, given the structure of the memory update $u$.\footnotemark\vspace{5mm}}
\label{fig:circuits_views}
\end{figure}
\footnotetext{This figure is taken from our paper \cite{Vilasini_PRdoxes} and was made by coauthors.}

\bigskip
\begin{definition}[Information-preserving memory update]
\label{def:memory_update}
Let $\mathscr{P}_S$ be a set of states of a system $S$ that is being studied by an agent $A_i$ with a memory $M_i$, and $\mathscr{P}_{SM_i}$ be a set of states of the joint system $SM_i$. If for a given initial state $Q_{M_i}^{in}\in \mathscr{P}_{M_i}$ of the memory, there exists a corresponding map $U^Q: \mathscr{P}_{SM_i}\rightarrow \mathscr{P}_{SM_i}$ ($\in \mathcal{O}_{SM_i}$) that satisfies the following conditions $(1)$ and $(2)$, then $U^Q$ is called an \emph{information-preserving memory update}. 

\begin{enumerate}
    \item Local operations on $S$ before the memory update can be simulated by joint operations on $S$ and $M_i$ after the update. That is, for all $P_S\in \mathscr{P}_S, \ O_S \in \mathcal{O}_S, \ A_j\in\mathcal{A},  \ \phi$, there exists an operation $O_{SM_i} \in \mathcal{O}_{SM_i}$ such that
\begin{gather*}
 K_j \  \phi[O_S(P_s)] \Rightarrow K_j \ \phi[O_{SM_i}\circ U^Q(P_S\parallel Q_{M_i}^{in})],
\end{gather*}
where $\phi[\dots]$ are arbitrary statements that  depend on the argument.

\item The memory update does not factorize into local operations. That is, there exist no operations $O'_S\in \mathcal{O}_S$ and $O'_{M_i}\in \mathcal{O}_{M_i}$ such that $$U^Q=O'_S\parallel O'_{M_i}$$

\end{enumerate}

\end{definition}

Condition $(1)$ was explained in previous paragraphs. Condition $(2)$ is required because the trivial map which entails doing nothing to the system and memory (i.e., the identity) satisfies  Condition $(1)$ even though such an operation should certainly not be regarded as a memory update. Condition $(2)$ requires that $U^Q$ does not factorise into local operations over $S$ and $M_i$ is required in order to rule out such trivial operations that cannot be taken to represent a memory update.
 See Figure~\ref{fig:circuits_views} for an example of $U^Q$ in the quantum case where it is a reversible unitary operation and the initial state of the memory, $Q^{in}_{M_i}$ is $\ket{0}_{M_i}$. In general, the memory update map $U^Q$ need not be reversible; for example, in box world it is an irreversible transformation, as we will see later. 
\par
 Note that it is enough to consider the memory update map $U^Q$ corresponding to a particular choice of initial state $Q^{in}_{M_i}$ since the map $U^{Q'}$ corresponding to any other state $Q'^{in}_{M_i}\in \mathscr{P}_{M_i}$ can be obtained by first locally transforming the memory state into $Q^{in}_{M_i}$ and then applying $U^Q$. Thus without loss of generality, we will consider only specific initial states in the chapter and drop the label $Q$ on this map, simply calling it $U$. For example, in the quantum case, it is enough to consider the memory update with the memory initialised to the state, $\ket{0}_{M_i}$.
 \par
The characterization of measurements introduced in this section is rather minimal. In physical theories like classical and quantum mechanics, measurements have other natural properties that we do not require here. Two striking examples are ``after her measurement, Alice's memory becomes correlated with the system measured in such a way that, for any subsequent operation that Bob could perform on the system, there is an equivalent operation he may perform on her memory'' and ``the correlations are such that there exists a joint operation on the system and Alice's memory that would allow Bob to conclude which measurement Alice performed.'' While these properties hold in the familiar classical and quantum worlds, we do not know of other physical theories where measurements can satisfy them, and they require Bob to be able to act independently on the system and on Alice's memory, which may not always be possible. For example, we will see that in box world, these two subsystems become \emph{superglued} after Alice's measurement, and that Bob only has access to them as a whole and not as individual components.\footnote{Thus the state-space $\mathscr{P}_{SM_i}$ can also contain such ``super-glued states".}  As such, we will not require these properties out of measurements, for now. We revisit this discussion in Section~\ref{sec:discussion}.

\section{Agents, memory and measurement in box world}

In Section~\ref{sec: GPT}, we reviewed Barrett's framework \cite{Barrett07} for generalised probabilistic theories, in particular describing states, transformations and measurements in box world, a particular GPT. We will now instantiate our general conditions for agents, memories and measurements (definitions \cref{def:measurements,def:agents,def:memory_update}) in box world. As there is no physical theory for the dynamics behind box world, there is plenty of freedom in the choice of implementation. In principle each such choice could represent a different physical theory leading to the same black-box behaviour in the limit of a single agent with an implicit memory. This is analogous to the way in which different versions of quantum theory (Bohmian mechanics, collapse theories, unitary quantum mechanics with von Neumann measurements) result in the same  effective theory in that limit.

\begin{definition}[Agents in box world]
\label{def:agentsbox}
Let $\mathbb T$ be the theory that describes box world, according to~\cite{Barrett07} (see also Section~\ref{sec: GPT}). 
As per definition~\ref{def:agents}, an agent $A_i \in\mathcal{A}$ is described by a knowledge operator $K_i\in\mathcal{K_\mathcal{A}}$ and a physical memory $M_i\in\mathcal{M_\mathcal{A}}$. 

We will focus on the case where the memory consists of bit or gbits. Each agent may study other systems according to the theory $\mathbb T$. An agent's memory records the results and the consequences of the studies conducted by them, and may be an object of a study by other agents.
\end{definition}

It is worth mentioning that boxes do not correspond to physical systems, but to input/output functions that can only be evaluated once. As such, the post-measurement state of a physical system is described by a whole new box. The notion of an individual system itself, as we will see, may be unstable under measurements--- some measurements \emph{glue} the system to the observer's memory, in a way that makes individual access to the original system impossible.

%\begin{figure}[t!]
%    \centering
 %   \includegraphics[width=0.5\textwidth]{Ch7_Fig7.pdf}
  %  \caption[Inside perspective of a measurement in GPTs]{\textbf{Measurement: observer's perspective.} An agent Alice measures a system with measurement setting $X$, and obtains outcome $a$ with a given probability. In the language of GPTs this corresponds to running the box that encodes the measurement statistics. Alice may then the measurement data (input and output) to memory. If this is a classical memory, like a notebook, the procedure corresponds to preparing a new box (to be run later by herself), which outputs the pair $(X,a)$ deterministically.}
   % \label{fig:measurement_Alice}
%\end{figure}

\paragraph{Measurement: observer's perspective.} From the point of view of the observer who is measuring (say Alice), making a measurement on a system corresponds simply to running the box whose state vector encodes the measurement statistics. 
Alice may then commit the result of her measurement to a physical memory, like a notebook where she writes `I measured observable $X$ and obtained outcome $a$.' To be useful, this should be a memory that may be consulted later, i.e.\ it could receive an input $Y=$`start: open and read the memory', and output the pair $(X,a)$. In the language of GPTs, this means that Alice, from her own perspective, prepares a new box with a trivial input $Y=$`start' and two outputs $(X',a')$, with the behaviour $\vec P(X',a'|Y) = \delta_{X, X'} \delta_{a,a'}$, which depends on her observations. She may later run this box (look at her notebook) and recover the measurement data. The exact dimension of the box will depend on how Alice perceives and models her own memory; for example it could consist of two bits, or two gbits, or, if we think that before the measurement she stored the information about the choice of observable elsewhere, it could be a single bit or gbit encoding only the outcome. We leave this open for now, as we do not want to constrain the theory too much at this stage. 

\begin{figure}[t!]
    \centering
\subfloat[Generally, in GPTs with some notion of subsystems, Ursula can think of the physical system measured by Alice, and Alice's memory pre-measurement as two boxes, which Ursula could in principle run if Alice chose not to measure (left).  From Ursula's perspective, Alice's measurement corresponds to a transformation of the two initial boxes that results in a new, post-measurement box available to Ursula, whose behaviour will depend on the concrete physical theory.]{
    \qquad\includegraphics[width=0.36\textwidth]{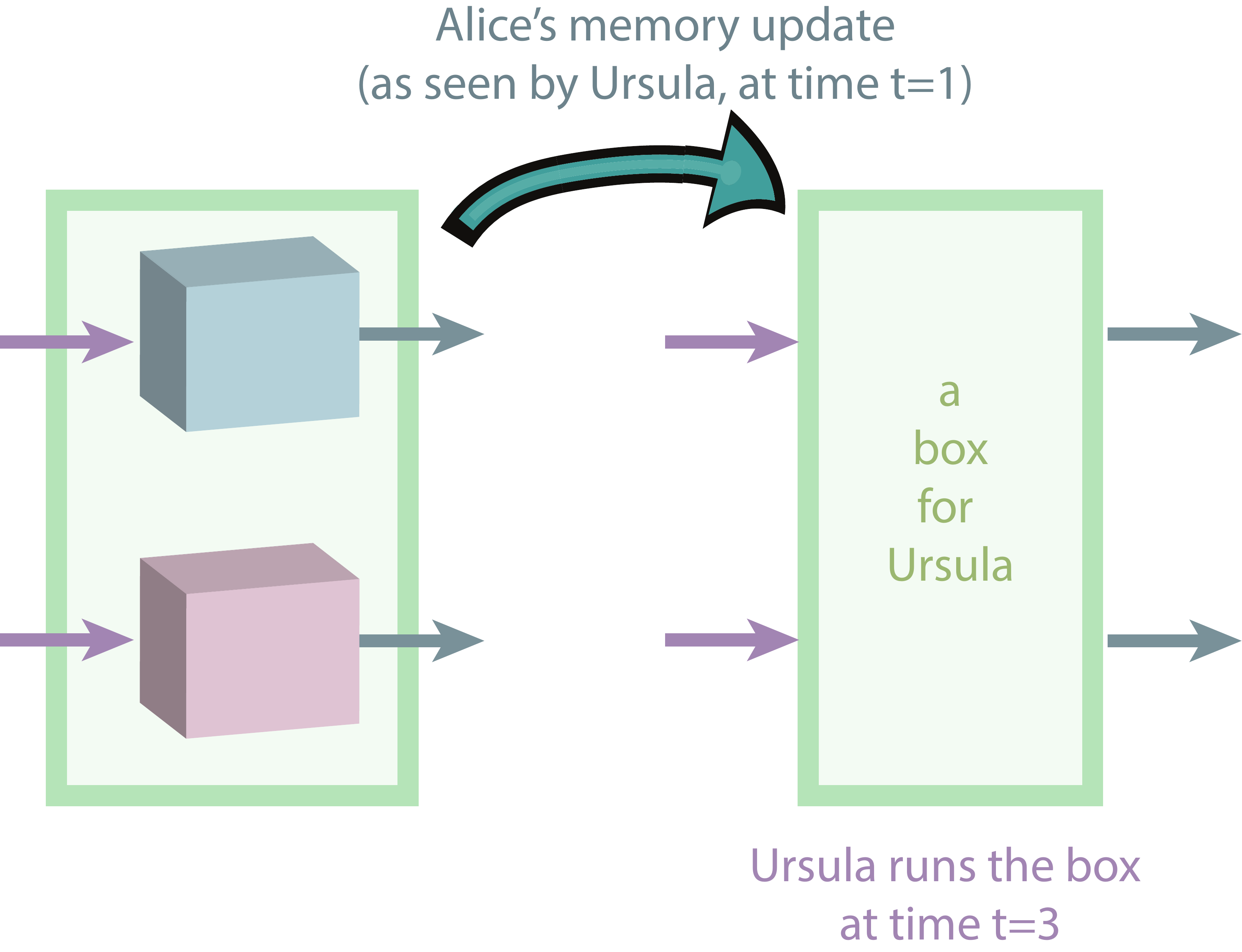}\qquad
    \label{fig:update_general}
}\subfloat[In  box world, suppose the two initial systems correspond to gbits, and Alice's memory is initialized as shown. If we want to preserve the global system dimensions, then the rules for allowed transformations (Definition~\ref{def:memory_update}) limit the statistics of Ursula's final box to be of the form shown in the right (Appendix~\ref{appendix:proofs}). The asterisks represent arbitrary values, which will depend on the choice of implementation of Alice's measurement. ]{
 \qquad\includegraphics[width=0.36\textwidth]{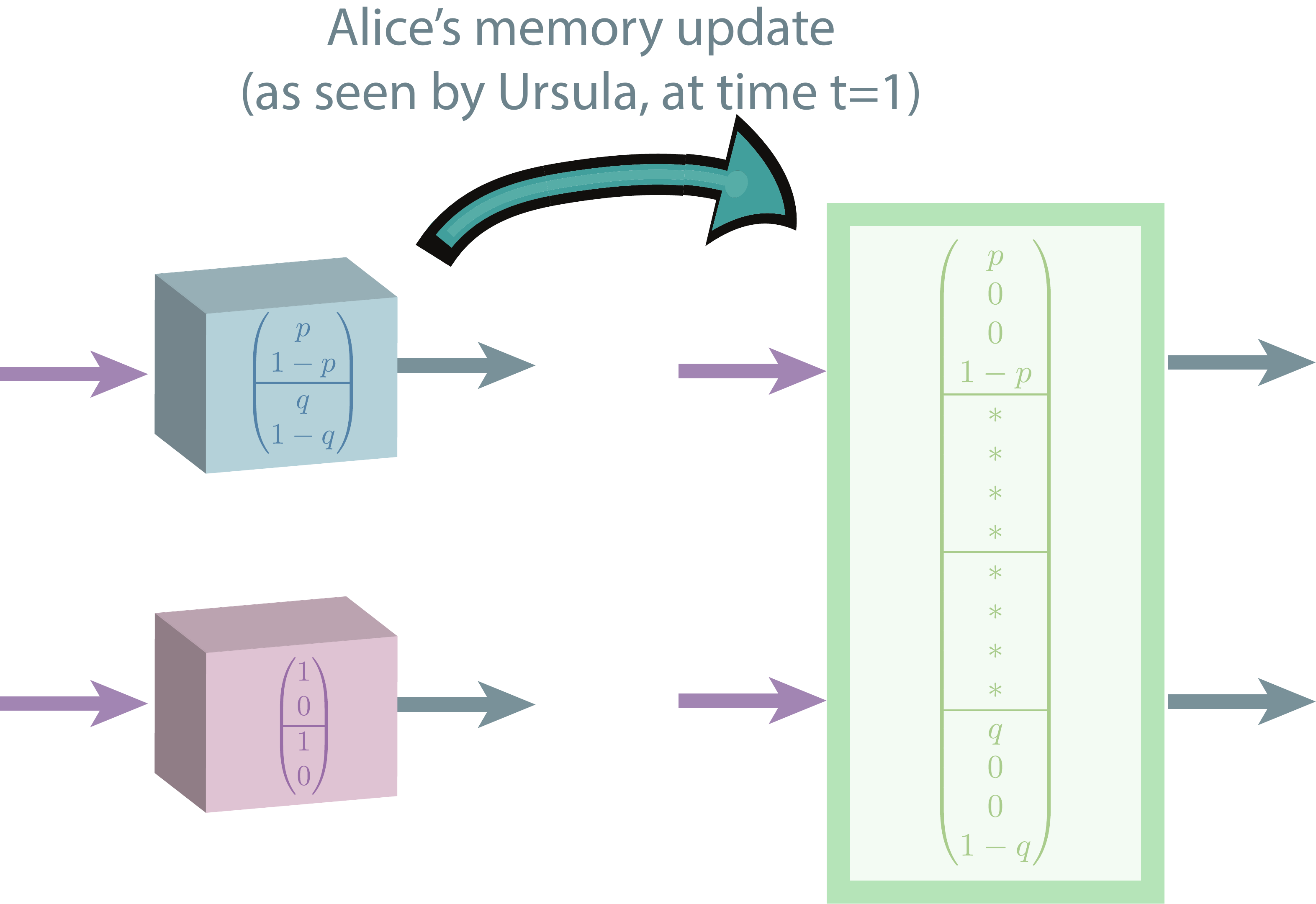}\qquad
    \label{fig:update_conditions}
}
\caption[Outside perspective of a measurement and memory update in GPTs]{{\bf Memory update after a measurement: an outsider's perspective.} Here Alice makes a measurement of a system  (blue, top) at time $t=1$ and stores her outcome in her memory (pink, bottom). The question is how an outsider, Ursula, models Alice's measurement. Note that this transformation need not be reversible, in fact this is true of the box world case.\footnotemark}
    \label{fig:update_basics}
\end{figure}

\paragraph{Measurements: inferences.}\footnotetext{This figure is taken from our paper \cite{Vilasini_PRdoxes} and was made by L\'idia del Rio.}
To see the kind of inferences and conclusions that an agent can take from a measurement in box world, it's convenient to look at the example where Alice and Bob share a PR box (Figure~\ref{fig:GTP_states}). Suppose that Alice measured her half of the box with input $X=1$ and obtained outcome $a=0$. From the PR correlations, $XY=a\oplus b$, she can conclude that if Bob measures $Y=0$, he must obtain $b=0$, and if he measures $Y=1$, he must obtain $b=1$. This is independent of whether Bob's measurement happens before or after Alice (or even space-like separated).
She could reach similar deterministic conclusions for her other choice of measurement and possible outcomes. 
In the language of Definition~\ref{def:measurements}, we have 
\begin{align*}
    \phi_{X=0, a=0} &= ``[Y=0 \implies b=0] \wedge [Y=1 \implies b=0]", \\
    \phi_{X=0, a=1} &= ``[Y=0 \implies b=1] \wedge [Y=1 \implies b=1]", \\
    \phi_{X=1, a=0} &= ``[Y=0 \implies b=0] \wedge [Y=1 \implies b=1]", \\
    \phi_{X=1, a=1} &= ``[Y=0 \implies b=1] \wedge [Y=1 \implies b=0]".
\end{align*}

\paragraph{Measurement: memory update from an outsider's perspective.}
Next we need to model how an outside agent, Ursula, models Alice's measurement, in the case where Alice does not communicate her outcome to Ursula.\footnote{Naming convention: as we will see in Section~\ref{sec:paradox}, the proposed experiment feature two ``internal'' agents, Alice and Bob, who will in turn be measured by two ``external'' agents, Ursula and Wigner. In the example of Section~\ref{sec:conditions}, the internal agent was Alice and the external Bob, so that their different pronouns could help keep track of whose memory we were referring to.} 
Suppose that all agents share a time reference frame, and Alice makes her measurement at time $t=1$. From Ursula's perspective, in the most general case, this will correspond to Alice preparing a new box, with some number of inputs and outputs, which Ursula can later run (Figure~\ref{fig:update_general}). The exact form of this box will depend on the underlying physical theory for measurements: in the quantum case it corresponds to a box with the measurement statistics of a state that's entangled between the system measured and Alice's memory, as we saw. In classical mechanics, it will correspond to perfect classical correlations between those two subsystems. In the other extreme, we could imagine a theory of very destructive measurements, where after Alice's measurement, the physical system she had measured  would vanish. From Ursula's perspective, this could be modelled by a box with a void associated distribution. Now suppose that we would like to have a physical theory where the dimension of systems is preserved by measurements: for example, if the system that Alice measures is instantiated by a box with binary input and output (e.g.~a gbit, or half of a PR-box), and Alice's memory, where she stores the outcome of the measurement is also represented as a gbit, then we would want the post-measurement box accessible to Ursula to have in total two binary inputs and two binary outputs (or more generally, four possible inputs and four possible outputs). Note that this is not a required condition for a theory to be \emph{physical} per se--- it is just a familiar rule of thumb that gives some persistent meaning to the notion of subsystems and dimensions. In such a theory that supports box world correlations, we find that the allowed statistics of Ursula's box must satisfy the conditions of Figure~\ref{fig:update_conditions} (proof in Appendix~\ref{appendix:proofs}). These conditions still leave us some wiggle room for possible different implementations. 

\begin{figure}[t!]
    \centering
 \subfloat[Alice measures a system (inner blue box) and stores the outcome in her memory (inner pink box). From Ursula's perspective,  this appears as though Alice simply wired the outputs of the two boxes with a controlled-{\sc{not}} gate, so that  the measurement output is copied to the output of the memory. When Ursula later runs the outer green box, she provides two inputs, which go through this circuit, resulting in two identical outputs.]{\qquad\qquad
    \includegraphics[width=0.3\textwidth]{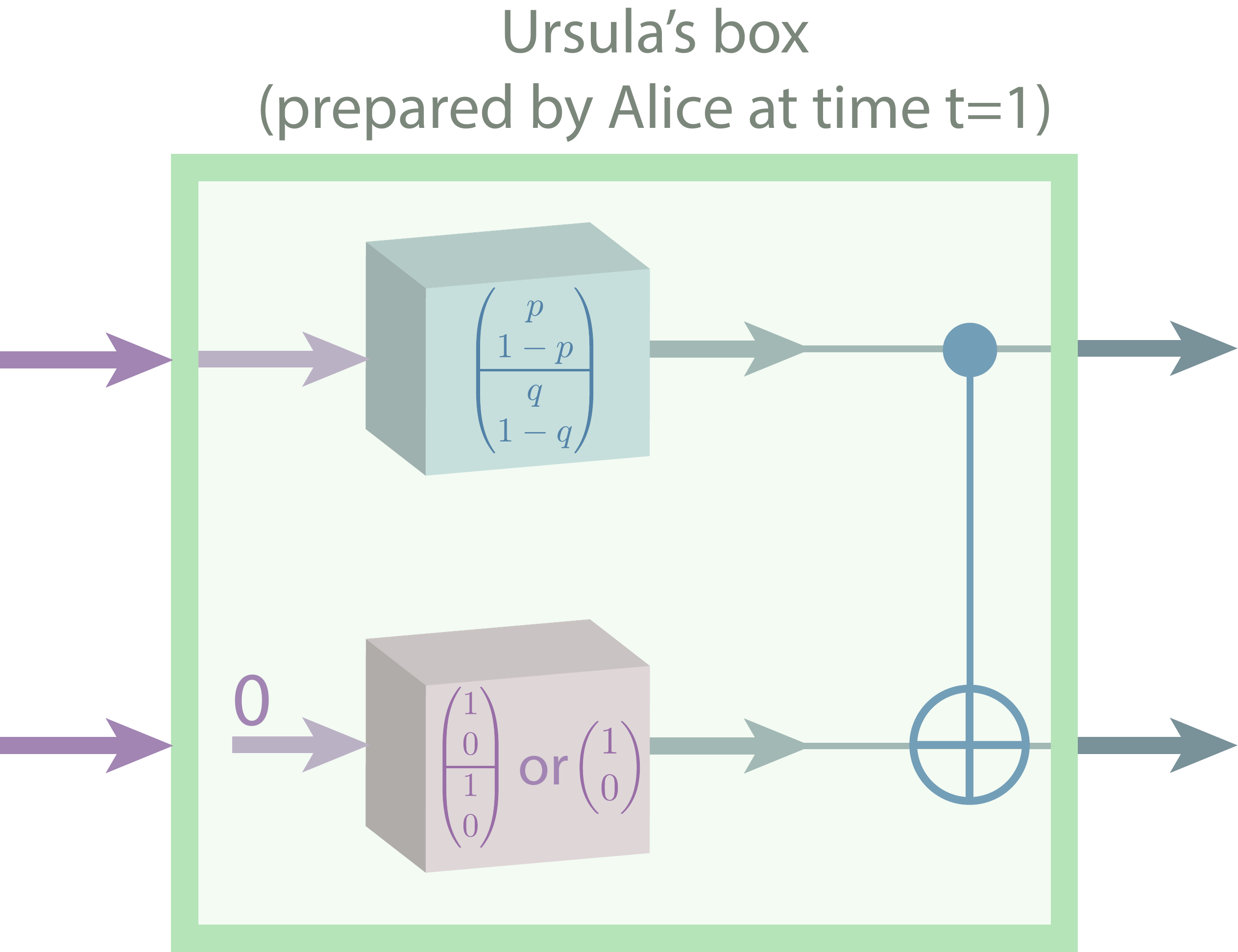}\qquad\qquad
    \label{fig:update_copy}
}\subfloat[In an information-preserving model, for every measurement $X$ on the initial state of the system measured by Alice, there exists a measurement $\tilde X$ that Ursula can do jointly on the system and Alice's memory that reproduces the same statistics. This is achieved, for example, by Ursula fixing her second input to 0, and simply discarding the second (trivial) output as shown.]{
   \qquad \includegraphics[width=0.37\textwidth]{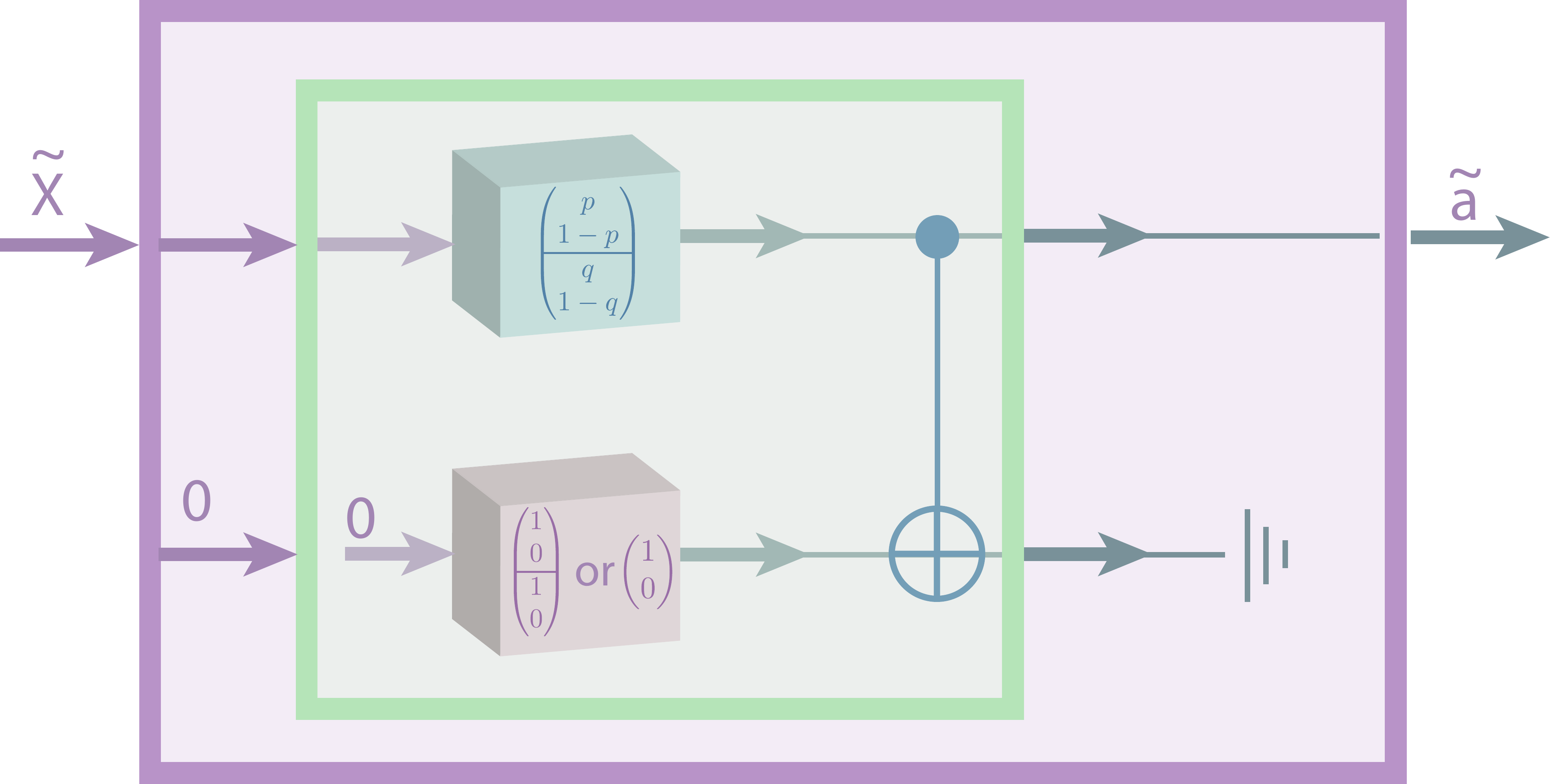}\qquad
    \label{fig:update_ursula}
}
\caption[Information-preserving memory update in box world]{\textbf{Information-preserving memory update.} This (trivial) physical implementation of Alice's measurement in box world satisfies the conditions of Figure~\ref{fig:update_conditions} and is information-preserving (Definition~\ref{def:memory_update}). The crucial detail is that Ursula is not allowed to open her box (in green) and access the circuitry inside. Note that is Ursula can perform arbitrary measurements on the joint system then the pre and post measurement states give different statistics, the former is a product state while the latter isn't. Further,  there are other possibilities for modelling measurements--- this is the simplest one that still allows us to derive the paradox. Details and proofs in Appendix~\ref{appendix:proofs}.\footnotemark}
    \label{fig:update_implementation}
\end{figure}

\paragraph{Measurements: information-preserving memory update.}
In order to find a multi-agent paradox, we will need a model of memory update that is information-preserving, in the sense of Definition~\ref{def:memory_update}.  This does not imply that Alice's transformation (as seen by Ursula) be reversible. Firstly, it is known that all reversible dynamics in box world is trivial and cannot correlate product states \cite{Gross2010}, hence the memory update map (Definition~\ref{def:memory_update}) in box world cannot be reversible. Further, we find that in general, the memory update (from the outside perspective) can \emph{glue} two subsystems such that Ursula will only be able to address them as a whole (since separating them could lead to a violation of no-signaling). But the relevant fact is that Ursula can apply some post-processing in order to obtain a new box with the same behaviour as the pre-measurement system that Alice observed. \footnotetext{This figure is taken from our paper \cite{Vilasini_PRdoxes} and was made by L\'idia del Rio.} In Figure~\ref{fig:update_implementation} we give an example of a model that satisfies these conditions, in addition to the conditions of Figure~\ref{fig:update_conditions}. It is a minimal implementation among many possible, which already allows us to derive such a paradox beyond quantum theory. We further discuss some of the limitations and alternatives to this model in Section~\ref{sec:discussion}. 
What is important here (and proven in Appendix~\ref{appendix:proofs}) is that this model generalizes to the case where Alice measures half of a bipartite state, like a PR box. That is, suppose that Alice and Bob share a PR box. Imagine that at time $t=1$ Alice makes her measurement $X$, obtaining (from her perspective) an outcome $a$, and that Bob makes his measurement $Y$ at time $t=2$, obtaining outcome $b$. As usual, if Alice and Bob were to communicate at this point, they would find that $XY=a\oplus b$, and indeed the propositions $\phi_{X,a}$ and $\phi_{Y,b}$ that represent their subjective measurement experience would hold. But now suppose that Alice and Bob do not get the chance to communicate and compare their input and outputs; instead, at time $t=3$, an observer Ursula, who models Alice's measurement as in Figure~\ref{fig:update_copy}, runs the box corresponding to Alice's half of the PR box and Alice's memory, and applies the post-processing of Figure~\ref{fig:update_ursula}. Ursula's input is $\tilde X$ and her output is $\tilde a$. Then the claim is that $\tilde XY=\tilde a\oplus b$: that is, Ursula and Bob effectively share a PR box. This is proven in Appendix~\ref{appendix:proofs}. 
We now have all the ingredients needed to find a multi-agent epistemic paradox in box world. 

 \section{Finding the paradox}
\label{sec:paradox}

We now describe a scenario in box world where reasoning, physical agents reach a logical paradox.

\paragraph{Experimental setup.} The proposed thought experiment is similar in spirit to the one proposed by Frauchiger and Renner~\cite{Frauchiger2018}, with the PR box playing the role of the Hardy state (c.f. Figure~\ref{fig:circuits}). Alice and Bob share a PR box (the corresponding box world state is given in Figure~\ref{fig:PRbox}); they each will measure their half of the PR box and store the outcomes in their local memories. Let Alice's lab be located inside the lab of another agent, Ursula's lab such that Ursula can now perform joint measurements on Alice's system (her half of the PR box) and memory, as seen in the previous section. Similarly, let Bob's lab be located inside Wigner's lab, such that Wigner can perform joint measurements on Bob's system and memory. We assume that Alice's and Bob's labs are isolated such that no information about their measurement outcomes leaks out. 
The protocol is the following:
\begin{itemize}
    \item[\bf t=1] Alice measures her half of the PR box, with measurement setting $X$, and stores the outcome $a$ in her memory $A$. 
    \item[\bf t=2] Bob measures his half of the PR box, with measurement setting $Y$, and stores the outcome $b$ in his memory $B$. 
    \item[\bf t=3] Ursula measures the box corresponding to Alice's lab (as in Figure~\ref{fig:update_ursula}), with measurement setting $\tilde X = X\oplus 1$, obtaining outcome $\tilde a$. 
    \item[\bf t=4] Wigner measures the box corresponding to Bob's lab, with measurement setting $\tilde Y = Y\oplus 1$, obtaining outcome $\tilde b$. 
\end{itemize}
Agents can agree on their measurement settings beforehand, but should not communicate once the experiment begins. The trust relation, which specifies which agents consider each other to be rational agents (as opposed to mere physical systems), is 
\begin{align*}
    A_{t=1,2} &\leftrightsquigarrow B_{t=1,2} \\ B_{t=2,3} &\leftrightsquigarrow U_{t=3} \\ U_{t=3,4} &\leftrightsquigarrow W_{t=4}\\  W_{t=4} &\leftrightsquigarrow A_{t=1}.
\end{align*}
Note that the time information is important here as it tells us that the trust relations hold for the pre-tampering versions of the ``inside'' agents, Alice and Bob (who are measured by the ``outside'' agents Ursula and Wigner). The common knowledge $\mathbbm{T}$ shared by all four agents includes the PR box correlations, the way the external agents model Alice and Bob's measurements, and the trust structure above.  

\paragraph{Reasoning.} Now the agents can reason about the events in other agents' labs.
We take here the example where the measurement settings are $X=Y=0, \tilde X = \tilde Y=1$, and where Wigner obtained the outcome $\tilde b=0$; the reasoning is analogous for the remaining cases. 
\begin{enumerate}
    \item \emph{Wigner reasons about Ursula's outcome.} At time $t=4$, Wigner knows that, by virtue of their information-preserving modelling of Alice and Bob's measurements, he and Ursula effectively shared a PR box\footnoteremember{proof}{See Appendix~\ref{appendix:proofs} for a proof.}. 
    He can therefore use the PR correlations $\tilde X \tilde Y = \tilde a \oplus \tilde b$ to conclude that Ursula's output must be $1$, 
    $$ K_W (\tilde{b}=0 \implies \tilde a=1).$$
    
    \item \emph{Wigner reasons about Ursula's reasoning.} Now Wigner thinks about what Ursula may have concluded regarding Bob's outcome. He knows that at time $t=3$, Ursula and Bob effectively shared a PR box\footnoterecall{proof}, satisfying $\tilde X Y = \tilde a \oplus b$, and that therefore Ursula must have concluded
    $$ K_W K_U (\tilde{a}=1 \implies b=1).$$
    
    \item \emph{Wigner reasons about Ursula's reasoning about Bob's reasoning.} Next, Wigner wonders ``What could Ursula, at time $t=3$, conclude about Bob's reasoning at time $t=2$?" Well, Wigner knows that she knows that Bob knew that at time $t=2$ he effectively shared a PR box with Alice, satisfying $XY=a\oplus b$, and therefore concludes
     $$ K_W K_U K_B (b=1 \implies a=1).$$
     
    \item \emph{Wigner reasons about Ursula's reasoning about Bob's reasoning about Alice's reasoning.} Now Wigner thinks about Alice's perspective at time $t=1$, through the lenses of Bob (at time $t=2$) and Ursula ($t=3$). Back then, Alice knew that she obtained some outcome $a$, and that Wigner would model Bob's measurement in an information-preserving way, such that Alice (at time $t=1$) and Wigner (of time $t=4$) share an effective PR box\footnoterecall{proof}, satisfying $X \tilde Y = a \oplus \tilde b$, which results, in particular, in
    $$ K_W K_U K_B K_A (a=1 \implies \tilde b=1).$$
    
    \item \emph{Wigner applies trust relations.} In order to combine the statements obtained above, we need to apply the trust relations described above, starting from the inside of each proposition, for example, 
    \begin{align*}
         K_W K_U K_B K_A &(a=1 \implies \tilde b=1) \\
         \implies K_W K_U K_B  &(a=1 \implies \tilde b=1) \qquad \qquad (\because\quad A \leadsto B)\\
         \implies K_W K_U  &(a=1 \implies \tilde b=1) \qquad\qquad (\because\quad B \leadsto U)\\
         \implies K_W &(a=1 \implies \tilde b=1) \qquad\qquad (\because\quad U \leadsto W)
    \end{align*}
    and similarly for the other statements (where $A \leadsto B$ denotes $B$ trusts $A$, c.f. Definition~\ref{def:trust}), so that we obtain 
    \begin{align*}
         & K_W  \big[ (\tilde{b}=0 \implies \tilde a=1) \wedge (\tilde{a}=1 \implies b=1) \wedge  (b=1 \implies a=1) \wedge (a=1 \implies \tilde b=1) \big] \\
    \implies &K_W (\tilde{b}=0 \implies \tilde b =1).
    \end{align*}
        
\end{enumerate}
We could have equally taken the point of view of any other observer, and from any particular outcome or choice of measurement, and through similar reasoning chains reached the following contradictions, 
\begin{align*}
    &K_A [ ({a}=0 \implies a =1) \wedge ({a}=1 \implies  a =0) ],\\
    &K_B [ (b=0 \implies b =1) \wedge (b=1 \implies b =0) ],\\
    &K_U [ (\tilde{a}=0 \implies \tilde a =1) \wedge (\tilde{a}=1 \implies \tilde a =0) ],\\
    &K_W [ (\tilde{b}=0 \implies \tilde b =1) \wedge (\tilde{b}=1 \implies \tilde b =0) ].
\end{align*}

\section{Discussion}
\label{sec:discussion}

We have generalized the conditions of the Frauchiger-Renner theorem and made them applicable to arbitrary physical theories, including the framework of \emph{generalized probability theories}. We then applied these conditions to the GPT of box world and found an experimental setting that leads to a multi-agent logical paradox.

\subsection{Comparison with the quantum thought experiment}
\label{ssec: discuss_quantum}
We showed that box world agents reasoning about each others' knowledge can come to a deterministic contradiction, which is stronger than the original paradox, as it can be reached without post-selection, from the point of view of every agent and for any measurement outcome obtained by them.

\paragraph{Post-selection.}
In contrast to the original Frauchiger-Renner experiment of \cite{Frauchiger2018}, no post-selection was required to arrive at this contradictory chain of statements as, in fact, all the implications above are symmetric, for example
$$\tilde{a}=0 \iff b=0 \iff a=0 \iff \tilde{b}=0 \iff \tilde{a}=1.$$
As a result, one can arrive at a similar (symmetric) paradoxical chain of statements irrespective of the choice of agent and outcome for the first statement. In other words, irrespective of the outcomes observed by every agent, each agent will arrive at a contradiction when they try to reason about the outcomes of other agents. This is because, as shown in \cite{Abramsky15}, the PR box exhibits strong contextuality and no global assignments of outcome values for all four measurements exists for any choice of local assignments. In contrast, the original paradox of \cite{Frauchiger2018} admits the same distribution as that of Hardy's paradox \cite{Hardy1993}. It is shown in \cite{Abramsky15} that this distribution is an example of logical contextuality where for a particular choice of local assignments (the ones that are post-selected on in the original Frauchiger-Renner experiment), a global assignment of values compatible with the support of the distribution fails to exist, but this is not true for all local assignments. This makes the paradox even stronger in box world, since it can be found without post-selection and by any of the agents, for any outcome that they observe. In particular, the paradox would already arise in a single run of the experiment. For a simple method to enumerate all possible contradictory statements that the agents may make, see the analysis of the PR box presented in \cite{Abramsky15}.

%\paragraph{Communication \emph{vs} prepare-and-measure.}
%One might note that a distinction between our proposal and  the original Frauchiger-Renner experiment is that there is no communication between Alice and Bob in our PR box version. However, the original quantum  scenario can be replaced by a protocol where Alice and Bob receive an appropriately prepared quantum state and perform measurements on it without communicating to each other, and the original paradox would still hold in such a case.
\paragraph{Reversibility of the memory update map} 
As mentioned previously, the memory update map $U$ in the quantum case is quantum CNOT gate which is a unitary and hence reversible. In box world however, this map cannot be reversible since it is known that all reversible maps in box world map product states to product states \cite{Gross2010} and hence no reversible $U$ in box world could satisfy Definition~\ref{def:memory_update} of an information-preserving memory update. The map we propose here for box world is clearly irreversible as it leads to correlations between the initially uncorrelated system and memory.

\subsection{Physical measurements in box world}
\label{ssec: discuss_mmts}

Since we lack a physical theory to explain how measurements and transformations are instantiated for generalised non-signaling boxes, and only have access to their input/output behaviour, all allowed transformations consist of pre- and post-processing. In the quantum case, we have in addition to a description of possible input-output correlations, a mathematical framework for the underlying states producing those correlations, the theory of von Neumann measurements and transformations as CPTP maps. In box world, introduction of dynamical features (for example, a memory update algorithm)  is less intuitive and requires additional constructions. In the following, we outline the main limitations we found.

\paragraph{Systems vs boxes.} In quantum theory, a system corresponds to a physical substrate that can be acted on more than once. For example, Alice could measure a spin first in the $Z$ basis and then in $X$ basis (obviously with different results than if she had measured first $X$ and then $Z$). The predictions for each subsequent measurement are represented by a different box in the GPT formalism, such that each box encodes the current state of the system in terms of the measurement statistics of a tomographically complete set of measurements. After each measurement, the corresponding box disappears, but quantum mechanics gives us a rule to compute the post-measurement state of the underlying system, which in turn specifies the box for future measurements. On the other hand, the default theory for box world lacks the notion of underlying physical systems and a definite rule to compute the post-measurement vector state of something that has been measured once.  Indeed, Equations~\ref{eq: op1}-\ref{eq: op3} tell us that post-measurement states is only partially specified: for instance, if  the measurement performed was fiducial, we know that the block corresponding to that measurement in the post-measurement state would have a ``1'' corresponding to the outcome obtained and ``0'' for all other outcomes in the block. However, we still have freedom in defining the entries in the remaining blocks.
Our model proposes a possible physical mechanism for updating boxes (which could be read as updating the state of the underlying system), but so far only for the case where we compare the perspectives of different agents, and we leave it open whether Alice has a subjective update rule that would allow her to make subsequent measurements on the same physical system.

\paragraph{Verifying a measurement.} In our simple model, the external observer Ursula has no way to know which measurement Alice performed, or whether she measured anything at all--- the connection between Alice's and Ursula's views is postulated rather than derived from a physical theory. Indeed,  Alice could have simply wired the boxes as in Figure \ref{fig:update_copy} without actually performing the measurement, and Ursula will not know the difference: she obtains the same joint state of Alice's memory and the system she measured.
In contrast, consider the case of quantum mechanics with standard von Neumann measurements. There, Alice's memory gets entangled with the system, and the post-measurement state depends on the basis in which Alice measured her system. For example, if Alice's qubit $S$ starts off in the normalised pure state $\ket{\psi}=\alpha\ket{0}_S+\beta\ket{1}_S$ and her memory $M$ initialised to $\ket{0}_M$, the initial state of her system and memory from Ursula's perspective is $\ket{\Psi}^{in}_{SM}= [\alpha\ket{0}_S+\beta\ket{1}_S]\otimes \ket{0}_M=[(\frac{\alpha+\beta}{\sqrt{2}})\ket{+}_S+(\frac{\alpha-\beta}{\sqrt{2}})\ket{-}_S)]\otimes \ket{0}_M$
If Alice measures the system in the $Z$ basis, the post-measurement state from Ursula's perspective is $ \ket{\Psi}^{out,Z}_{SM}= \alpha\ket{0}_S\ket{0}_M+\beta\ket{1}_S\ket{1}_M$, which is an entangled state. If instead, Alice measured in the Hadamard (X) basis, the post-measurement state would be $ \ket{\Psi}^{out,X}_{SM}
   =(\frac{\alpha+\beta}{\sqrt{2}})\ket{+}_S\ket{0}_M+(\frac{\alpha-\beta}{\sqrt{2}})\ket{-}_S\ket{1}_M$. Clearly the measurement statistics of $\ket{\Psi}^{in}_{SM}$, $ \ket{\Psi}^{out,Z}_{SM}$ and  $\ket{\Psi}^{out,X}_{SM}$ are different and Ursula can thus (in principle, with some probability) tell whether or not Alice performed a measurement and which measurement was performed by her. 
In the absence of a physical theory backing box world, we can still lift this degenerancy between the three situations (Alice didn't measure, she measured $X=0$, or she measured $X=1$) by adding another classical system to the circuitry of~\ref{fig:update_copy}: for example, a trit that stores what Alice did, and which Ursula could consult independently of the glued box of system and Alice's memory. However, we'd still have a postulated connection between what's stored in this trit and what Alice actually did, and not one that is physically motivated.

\paragraph{\emph{Supergluing} of non-signaling boxes.}
For the memory update circuit (from Ursula's perspective) of Figure \ref{fig:update_copy}, and the initial state of Equation~\eqref{eq: PSMin}, the final state would be (following the notation of Section~\ref{sec: GPT}) $\vec{P}_{fin}^{SM}=(p\quad 0\quad 0\quad 1-p|p\quad 0\quad 0\quad 1-p|q\quad 0\quad 0\quad 1-q|q\quad 0\quad 0\quad 1-q)^T_{SM}$. Note that while the reduced final state of $S$ does not depend on the input $X'$ to $M$, the reduced final state on Alice's memory $M$, $\vec{P}_{fin}^{M}$ clearly depends on the input $X$ of the system $S$ if $p\neq q$. If $X=0$, $\vec{P}^M_{fin}=(p\quad 1-p|p\quad 1-p)^T$ and if $X=1$, $\vec{P}^M_{fin}=(q\quad 1-q|q\quad 1-q)^T$, i.e., the systems $S$ and $M$ are \emph{signalling}. This is expected since there is clearly a transfer of information from $S$ to $M$ during the measurement as seen in Figure~\ref{fig: bipartitemeas}. However, this means that the state $\vec{P}_{fin}^{SM}$ is not a valid box world state of 2 systems $S$ and $M$ but a valid state of a single system $SM$ i.e., after Alice performs her wiring/measurement, it is not possible to physically separate Alice's system $S$ from her memory $M$ from Ursula's perspective. For if this were possible, there would be a violation of the no-signaling principle and the notion of relativistic causality. In quantum theory, on the other hand it is always possible to perform separate measurements on Alice's system and her memory even after she measures. We call this feature \emph{supergluing} of post-measurement boxes, where it is no longer possible for Ursula to separately measure $S$ or $M$, but she can only jointly measure $SM$ as though it were a single system. Note that this is only the case for $p\neq q$ and in our example with the PR-box (Section \ref{sec:paradox}), $p=q=1/2$ and $\vec{P}_{fin}^{SM}$ remains a valid bipartite non-signaling state in this particular, fine-tuned case of the PR box and there is no supergluing in the particular example described in Section~\ref{sec:paradox}.

\paragraph{A glass half full.}
The above-mentioned  features of the memory update in box world are certainly not desirable, and not what one would expect to find in a physical theory with meaningful notions of subsystems. An optimistic way to look at these limitations  is to see them as providing us with further intuition for why PR boxes have not yet been found in nature. One of the main contributions of this chapter is the finding that despite these peculiar features of box world and the fact that it has no entangling bipartite joint measurements (a crucial step in the original quantum paradox), a consistent outside perspective of the memory update exists such that with our generalised assumptions, a multi-agent paradox can be recovered. This indicates that the reversibility of dynamics akin to quantum unitarity is not necessary to derive this kind of paradox. We suspect that contextuality, and the existence of a suitable information-preserving memory update (Definition~\ref{def:memory_update}) are necessary conditions for deriving such paradoxes. We discuss the relation to contextuality further in Section~\ref{ssec: contextuality}, and leave a detailed analysis to future work.

\paragraph{Other models for physical measurements.}
Ours is not the first attempt at coming up with a (partial) physical theory that reproduces the statistics of box world. Here we review the  approach of Skrzypczyk et al.\ in~\cite{Skrzypczyk2008}. There the authors consider a variation of box world that has a reduced set of physical states (which the authors call \emph{genuine}), which consists of the PR box and all the deterministic local boxes. The wealth of box world state vectors (i.e.\ the non-signaling polytope, or what we could call  epistemic states) is recovered by allowing classical processing of inputs and outputs via classical wirings, as well as convex combinations thereof. In contrast, box world takes all convex combinations of maximally non-signaling boxes (of which the PR box is an example) to be genuine physical states; this becomes relevant as we require the allowed physical operations to map such states to each other. For the restricted state space of~\cite{Skrzypczyk2008}, the set of allowed operations is larger than in box world, particularly for multipartite settings. For example, there we are allowed maps that implement the equivalent of entanglement swapping: if Bob shares a PR box with Alice, and another with Charlie, there is an allowed map that he can apply on his two halves which leaves Alice and Charlie sharing a PR box, with some probability. It would be interesting to try to model memory update in this modified theory, to see if (1) there is a more natural implementation of measurements within the extended set of operations, and (2) whether this theory allows for multi-agent paradoxes.

\subsection{Characterization of general theories}

While we have shown that a consistency paradox, similar to the one arising in the Frauchiger-Renner setup, can also be adapted for the box world in terms of GPTs, it still remains unclear how to characterize all possible theories where it is possible to find a setup leading to a contradiction. It seems that contextuality is a key property of such theories, this is discussed in more detail in Section~\ref{ssec: contextuality}. Another central ingredient seems to be information-preserving models for physical measurements such as our memory update of Definition~\ref{def:memory_update}, which allow us to replace counter-factuals with actual measurements, performed in sequence by different agents.

\paragraph{Beyond standard composition of systems.} 
Additionally, it is still an open problem to find an operational way to state the outside view of measurements (and a memory update operation), for theories without a prior notion of subsystems and a tensor rule for composing them. This will allow us to search for multi-agent logical paradoxes in field theories, for example. One possible direction is to use  notions of effective and subjective locality, as outlined for example in~\cite{Kraemer2018}.

\subsection{Relation to contextuality}
\label{ssec: contextuality}
Multi-agent logical paradoxes  involve chains (or possibly more general structures) of statements that cannot be simultaneously true in a consistent manner. Contextuality, on the other hand, can often be expressed in terms of the inability to consistently assign definite outcome values to a set of measurements \cite{Kochen1967,Spekkens2005}. 

The examples of Frauchiger-Renner in quantum theory and the the present one in box world, both arise in contextual theories. Given this, and the strong links between these, Hardy's paradox \cite{Hardy1992, Hardy1993} and logical contextuality \cite{Abramsky15} (also pointed out in Section~\ref{sec: FRdiscuss})--- our hypothesis is that a contextual physical theory, when applied to systems that are themselves reasoning agents, can give rise to logical multi-agent paradoxes. The fact that such theories may allow a very different description of a measurement process from the points of views of an agent performing the measurement vs an outside agent (who analyses this agent and her system together) also has an important role to play in these paradoxes. In the quantum case this is closely linked to the measurement problem, the problem of reconciling unitary dynamics (outside view) and non-unitary ``collapse" (inside view).  The existence of a connection between multi-agent paradoxes and contextuality is hard to miss, but it is the nature of this connection that is unknown i.e., are all proofs of multi-agent logical paradoxes proofs of contextuality, or vice-versa? Based on our results, and the analysis of the Frauchiger-Renner experiment presented in Section~\ref{sec: FRdiscuss}, the hunch is that contextuality is necessary, but not sufficient for deriving such paradoxes: specific forms of contextuality, combined with suitable conditions on the memory update might be required. These questions will be formally addressed in future work. Nevertheless, in the following, we provide an overview of further connections and some more specific open questions in this direction. 

\paragraph{Liar cycles.} In~\cite{Abramsky15} relations between logical paradoxes and quantum contextuality are explored; in particular, the authors point out a direct connection between contextuality and a type of classic semantic paradoxes called \emph{Liar cycles} \cite{Cook2004}. A Liar cycle of length $N$ is a chain of statements of the  form:
\begin{align}
    \phi_1 = ``\phi_2 \ \text{is true}'', \phi_2 = ``\phi_3 \ \text{is true}'', \dots, \phi_{N-1} = ``\phi_N \ \text{is true}'', \phi_N = ``\phi_1 \ \text{is false}''.
\end{align}
It can be shown that the patterns of reasoning which are used in finding a contradiction in the chain of statements above are similar to the reasoning we make use of in FR-type arguments, and can also be connected to the cases of PR box (which corresponds to a Liar cycle of length 4) and Hardy's paradox. This further suggests that multi-agent paradoxes are closely linked to the notion of contextuality. 

\paragraph{Relation to logical pre-post selection paradoxes.} In \cite{Pusey2015}, it has been shown that every proof of a logical pre-post selection paradox is a proof of contextuality. The exact connection between FR-type paradoxes and logical pre-post selection paradoxes is not known and this would be an interesting avenue to explore which would also provide insights into the relationship between FR paradoxes and contextuality.

\section{Appendix}

\subsection{Proofs}
\label{appendix:proofs}

\subsubsection{Memory update for a single lab}

In this section, we describe how a box world agent would measure a system and store the result in a memory. From the perspective of an outside observer (who does not know the outcome of the agent's measurement), we describe the initial and final states of the system and memory before and after the measurement as well as the transformation that implements this memory update in box world. In the quantum case, any initial state of the system $S$ is mapped to an isomorphic joint state of the system $S$ and memory $M$ (see Equation~\eqref{eq:entangling_measurement}) and hence the memory update map that maps the former to the latter (an isometry in this case\footnote{An isometry since it introduces an initial pure state on $M$, followed by a joint unitary on $SM$.}) satisfies Definition~\ref{def:memory_update} of an information-preserving memory update. We will now characterise the analogous memory update map in box world and show that it also satisfies Definition~\ref{def:memory_update}.

\begin{theorem}
\label{theorem: boxmemory}
In box world, there exists a valid transformation $u$ that maps every arbitrary, normalized state $\vec{P}^S_{in}$ of the system $S$ to  an isomorphic final state  $\vec{P}_{fin}^{SM}$ of the system $S$ and memory $M$ and hence constitutes an information-preserving memory update (Definition~\ref{def:memory_update}).
\end{theorem}
\begin{proof}
To simplify the argument, we will describe the proof for the case where $S$ and $M$ are gbits. For higher dimensional systems, a similar argument holds, this will be explained at the end of the proof. 

We start with the system in an arbitrary, normalized gbit state $\vec{P}^S_{in}=(p\quad 1-p|q\quad 1-q)^T$ (where the subscript T denotes transpose and $p,q \in [0,1]$) and the memory initialised to one of the 4 pure states\footnote{It does not matter which pure state the memory is initialized in, a similar argument applies in all cases.}, say $\vec{P}^M_{in}=\vec{P}_1=(1\quad 0|1\quad 0)^T$. Then the joint initial state, $ \vec{P}_{in}^{SM}=(p\quad 1-p|q\quad 1-q)^T_S\otimes (1\quad 0|1\quad 0)^T_M$ of the system and memory can be written as follows, where $P_{in}(a=i,a'=j|X=k,X'=l)$ denotes the probability of obtaining the outcomes $a=i$ and $a'=j$ when performing the fiducial measurements $X=k$ and $X'=l$ on the system and memory respectively, in the initial state $\vec{P}_{in}^{SM}$.
\begin{equation}
\label{eq: PSMin}
    \vec{P}_{in}^{SM}=\left(
\begin{array}{c}
P_{in}(a=0,a'=0|X=0,X'=0)\\
P_{in}(a=0,a'=1|X=0,X'=0)\\
P_{in}(a=1,a'=0|X=0,X'=0)\\
P_{in}(a=1,a'=1|X=0,X'=0)\\
\hline
P_{in}(a=0,a'=0|X=0,X'=1)\\
P_{in}(a=0,a'=1|X=0,X'=1)\\
P_{in}(a=1,a'=0|X=0,X'=1)\\
P_{in}(a=1,a'=1|X=0,X'=1)\\
 \hline 
P_{in}(a=0,a'=0|X=1,X'=0)\\
P_{in}(a=0,a'=1|X=1,X'=0)\\
P_{in}(a=1,a'=0|X=1,X'=0)\\
P_{in}(a=1,a'=1|X=1,X'=0)\\
 \hline 
P_{in}(a=0,a'=0|X=1,X'=1)\\
P_{in}(a=0,a'=1|X=1,X'=1)\\
P_{in}(a=1,a'=0|X=1,X'=1)\\
P_{in}(a=1,a'=1|X=1,X'=1)\\
\end{array}
\right)_{SM}=
\left(
\begin{array}{c}
p\\
0\\
1-p\\
0\\
\hline
p\\
0\\
1-p\\
0\\
 \hline 
q\\
0\\
1-q\\
0\\
 \hline 
q\\
0\\
1-q\\
0\\
\end{array}
\right)_{SM}
\end{equation}

The rest of the proof proceeds as follows: we first describe a final state $\vec{P}_{fin}^{SM}$ of the system and memory and a corresponding memory update map $U$ that satisfy Definition~\ref{def:memory_update} of a generalized information-preserving memory update. Then, we show that this map can be seen an allowed box world transformation which completes the proof. 

If an agent performs a measurement on the system, the state of the memory must be updated depending on the outcome and the final state of the system and memory after the measurement must hence be a correlated (i.e., a non-product) state. Although the full state space of the 2 gbit system $SM$ is characterised by the 4 fiducial measurements $(X,X')\in\{(0,0),(0,1),(1,0),(1,1)\}$, Definition~\ref{def:memory_update} allows us to restrict possible final states to a useful subspace of this state space that contain correlated states of a certain form. The definition requires that for every map $\mathscr{E}_S$ on the system before measurement, there exists a corresponding map $\mathscr{E}_{SM}$ on the system and memory after the measurement that is operationally identical. Thus it suffices if the joint final state $\vec{P}_{fin}^{SM}$ belongs to a subspace of the 2 gbit state space for which only 2 of the 4 fiducial measurements are relevant for characterising the state, namely any 2 fiducial measurements on $\vec{P^{SM}_{fin}}$ that are isomorphic to the 2 fiducial measurements on $\vec{P}^S_{in}$.
Note that by definition of fiducial measurements, the outcome probabilities of any measurement can be found given the outcome probabilities of all the fiducial measurements and without loss of generality, we will only consider the case where the agents perform fiducial measurements on their systems.

A natural isomorphism between fiducial measurements on $\vec{P}_{in}^S$ and those on $\vec{P}_{fin}^{SM}$ to consider here (in analogy with the quantum case) is: $X=i \Leftrightarrow (X,X')=(i,i) \quad, \forall i\in \{0,1\}$ i.e., only consider the cases where the fiducial measurements performed on $S$ and $M$ are the same.
Now, in order for the states to be isomorphic or operationally equivalent, one requires that performing the fiducial measurements $(X,X')=(i,i)$ on $\vec{P}_{fin}^{SM}$ should give the same outcome statistics as measuring $X=0$ on $\vec{P}_{in}^S$. This can be satisfied through an identical isomorphism on the outcomes: $a=i \Leftrightarrow (a,a')=(i,i) \quad, \forall i\in \{0,1\}$. Then the final state of the system and memory, $\vec{P}^{SM}_{fin}$ will be of the form
\begin{equation}
\label{eq: PSMfin}
    \vec{P}_{fin}^{SM}=\left(
\begin{array}{c}
P_{fin}(a=0,a'=0|X=0,X'=0)\\
P_{fin}(a=0,a'=1|X=0,X'=0)\\
P_{fin}(a=1,a'=0|X=0,X'=0)\\
P_{fin}(a=1,a'=1|X=0,X'=0)\\
\hline
P_{fin}(a=0,a'=0|X=0,X'=1)\\
P_{fin}(a=0,a'=1|X=0,X'=1)\\
P_{fin}(a=1,a'=0|X=0,X'=1)\\
P_{fin}(a=1,a'=1|X=0,X'=1)\\
 \hline 
P_{fin}(a=0,a'=0|X=1,X'=0)\\
P_{fin}(a=0,a'=1|X=1,X'=0)\\
P_{fin}(a=1,a'=0|X=1,X'=0)\\
P_{fin}(a=1,a'=1|X=1,X'=0)\\
 \hline 
P_{fin}(a=0,a'=0|X=1,X'=1)\\
P_{fin}(a=0,a'=1|X=1,X'=1)\\
P_{fin}(a=1,a'=0|X=1,X'=1)\\
P_{fin}(a=1,a'=1|X=1,X'=1)\\
\end{array}
\right)_{SM}=
\left(
\begin{array}{c}
p\\
0\\
0\\
1-p\\
\hline
*\\
*\\
*\\
*\\
 \hline 
*\\
*\\
*\\
*\\
 \hline 
q\\
0\\
0\\
1-q\\
\end{array}
\right)_{SM},
\end{equation}

where $*$ are arbitrary, normalised entries and where $P_{fin}(a=i,a'=j|X=k,X'=l)$ denotes the probability of obtaining the outcomes $a=i$ and $a'=j$ when performing the fiducial measurements $X=k$ and $X'=l$ on the system and memory respectively, in the final state $\vec{P}_{fin}^{SM}$. 
This final state can be compressed since the only relevant and non-zero probabilities in $\vec{P}_{fin}^{SM}$ occur when $X=X'$ and $a=a'$. We can then define new variables $\tilde{X}$ and $\tilde{a}$ such that $X=X'=i \Leftrightarrow \tilde{X}=i$ and $a=a'=j \Leftrightarrow \tilde{a}=j$ for $i,j \in \{0,1\}$ and $\vec{P}_{fin}^{SM}$ can equivalently be written as in Equation~\eqref{eq: finshort} which is clearly of the same form as $P_{in}^S$.

\begin{equation}
\label{eq: finshort}
    \vec{P}_{fin}^{SM}\equiv \left(
\begin{array}{c}
 P(\tilde{a}=0|\tilde{X}=0)\\
 P(\tilde{a}=1|\tilde{X}=0)\\
\hline
 P(\tilde{a}=0|\tilde{X}=1)\\
 P(\tilde{a}=1|\tilde{X}=1)\\
\end{array}
\right)_{SM}=\left(
\begin{array}{c}
 p\\
 1-p\\
\hline
 q\\
 1-q\\
\end{array}
\right)_{SM}
\end{equation}

Hence the initial state of the system, $\vec{P}^S_{in}=(p\quad 1-p|q\quad 1-q)^T$ (which is an arbitrary gbit state) is isomorphic to the final state of the system and memory, $\vec{P}_{fin}^{SM}$ (as evident from Equation~\eqref{eq: finshort}) with the same outcome probabilities for $X=0,1$ and $\tilde{X}=0,1$. This implies that for every transformation $\mathscr{E}_S$ on the former, there exists a transformation $\mathscr{E}_{SM}$ on the latter
such that for all outside agents $A_j$ and for all $p,q \in [0,1]$ (i.e., all possible input gbit states on the system), $ \quad K_j \phi[\mathscr{E}_S(\vec{P}^S_{in})] \Rightarrow K_j \phi[\mathscr{E}_{SM}\circ\vec{P}^{SM}_{fin}]$, where $\vec{P}^{fin}_{SM}=U(\vec{P}^{in}_{S})$. Thus any map $U$ that maps $\vec{P}_{in}^{SM}=\vec{P}^{in}_{S}\otimes P^M_{in}$ to $\vec{P}^{fin}_{SM}$ satisfies Definition~\ref{def:memory_update}. 

We now find a valid box world transformation that maps the initial state $\vec{P}_{in}^{SM}$ (Equation~\eqref{eq: PSMin}) to any final state of the form $\vec{P}_{fin}^{SM}$(Equation~\eqref{eq: PSMfin}) which would correspond to the memory update map $U$. 

Noting that all bipartite transformations in box world can be decomposed to a classical circuit of a certain form (see Appendix~\ref{ssec: boxworldoutcomes} or the original paper \cite{Barrett07} for details), In Figure~\ref{fig: memorycircuit}, we construct an explicit circuit of this form that converts $\vec{P}_{in}^{SM}$ to $\vec{P}_{fin}^{SM}$. By construction, we only need to consider the case of $X=X'$ since for $X\neq X'$, the entries of $\vec{P}_{fin}^{SM}$ can be arbitrary and are irrelevant to the argument. For $X\neq X'$, one can consider any such circuit description and it is easy to see that $\vec{P}_{in}^{SM}=(p\quad 1-p|q\quad 1-q)^T_S\otimes (1\quad 0|1\quad 0)^T_M$ is indeed transformed into $\vec{P}_{fin}^{SM}=(p\quad 0\quad 0\quad 1-p|*\quad*\quad*\quad*|*\quad*\quad*\quad*|q\quad 0\quad 0\quad 1-q)^T_{SM}$ through the map $U$ defined by these sequence of steps. For example, if the circuit description for the $X\neq X'$ case is same as that for the $X=X'$ case, then the resultant memory update map is equivalent to the circuit of Figure~\ref{fig:update_copy} which corresponds to performing a fixed measurement $X'=0$ on the initial state of $M$ and a classical CNOT on the output wire of $M$ controlled by the output wire of $S$\footnote{The output wires of boxes carry classical information after the measurement.}. The final state in that case is $(p\quad 0\quad 0\quad 1-p|p\quad 0\quad 0\quad 1-p|q\quad 0\quad 0\quad 1-q|q\quad 0\quad 0\quad 1-q)^T_{SM}$.

For higher dimensional systems $S$ with $n>2$ fiducial measurements, $X \in \{0,...,n-1\}$ and $k>2$ outcomes taking values $a\in \{0,...,k-1\}$, let $b_n$ and $b_k$ be the number of bits required to represent $n$ and $k$ in binary respectively. Then the memory $M$ would be initialized to $b_k$ copies of the pure state $\vec{P}_{in,n}^M=(1\quad 0|...|1\quad 0)^T_M$ which contains $n$ identical blocks (one for each of the $n$ fiducial measurements). One can then perform the procedure of Figure~\ref{fig: memorycircuit} ``bitwise'' combining each output bit with one pure state of $M$ and apply the same argument to obtain the result. For the specific case of the memory update transformation of Figure~\ref{fig:update_copy}, this would correspond to a bitwise CNOT on the output wires of $S$ and $M$.
\end{proof}

\subsubsection{Memory update for two labs sharing initial correlations}

So far, we have considered a single agent measuring a system in her lab. We can also consider situations where multiple agents jointly share a state and measure their local parts of the state, updating their corresponding memories. One might wonder whether the initial correlations in the shared state are preserved once the agents measure it to update their memories (clearly the local measurement probabilities remain unaltered as we saw in this section). The answer is affirmative and this is what allows us to formulate the Frauchiger-Renner paradox in box world as done in the Section~\ref{sec:paradox}, even though a coherent copy analogous to the quantum case does not exist here.

\begin{theorem}
\label{theorem: PRmemory}
Suppose that Alice and Bob share an arbitrary bipartite state $ \vec{P}^{PR}_{in}$ (which may be correlated), locally perform a fiducial measurement on their half of the state and store the outcome in their local memories $A$ and $B$. Then the final joint state $\vec{P}^{\tilde{A}\tilde{B}}_{fin}$ of the systems $\tilde{A}:=PA$ and $\tilde{B}:=RB$ as described by outside agents is isomorphic to  $ \vec{P}^{PR}_{in}$ with the systems $\tilde{A}$ and $\tilde{B}$ taking the role of the systems $P$ and $R$ i.e., local memory updates by Alice and Bob preserve any correlations initially shared between them.
\end{theorem}
\begin{proof}
In the following, we describe the proof for the case where the bipartite system shared by Alice and Bob consists of 2 gbits, however, the result easily generalises to arbitrary higher dimensional systems by the argument presented in the last paragraph of the proof of Theorem~\ref{theorem: boxmemory}. 

Let $\vec{P}^{PR}_{in}$ be an arbitrary 2 gbit state with entries $P_{in}(ab=ij|XY=kl)$ ($i,j,k,l \in \{0,1\}$), which correspond to the joint probabilities of Alice and Bob obtaining the outcomes $a=i$ and $b=j$ when measuring $X=k$ and $Y=l$ on the $P$ and $R$ subsystems when sharing that initial state. Let $X',a'\in \{0,1\}$ and $Y',b' \in \{0,1\}$ be the fiducial measurements and outcomes for the memory systems $A$ and $B$ (also gbits) respectively. We describe the measurement and memory update process for each agent separately and characterise the final state of Alice's and Bob's systems and memories after the process as would appear to outside agents who do not have access to Alice and Bob's measurement outcomes. This analysis does not depend on the order in which Alice and Bob perform the measurement as the correlations are symmetric between them, so without loss of generality, we can consider Bob's measurement first and then Alice's. 

Suppose that Bob's memory $B$ is initialised to the state $\vec{P}_{in}^{B}=\vec{P}_1^{B}=(1\quad 0|1\quad 0)^T_{B}$. Then the joint initial state of the Alice's and Bob's system and Bob's memory as described by an agent Wigner outside Bob's lab is $\vec{P}^{PRB}_{in}=\vec{P}^{PR}_{in}\otimes\vec{P}_1^{B}$. This can be expanded as follows where $P_{in}(abb'=ijk|XYY'=lmn)$ represents the probability of obtaining the binary outcomes $a=i$,$b=j$,$b'=k$ when performing the binary fiducial measurements $X=l$, $Y=m$, $Y'=n$ on the initial state $\vec{P}^{PRB}_{in}$.

\begin{equation}
\label{eq: PRBin}
    \vec{P}^{PRB}_{in}=\left(
\begin{array}{c}
 P_{in}(abb'=000|XYY'=000)\\
 P_{in}(abb'=001|XYY'=000)\\
 P_{in}(abb'=010|XYY'=000)\\
 P_{in}(abb'=011|XYY'=000)\\
 P_{in}(abb'=100|XYY'=000)\\
 P_{in}(abb'=101|XYY'=000)\\
 P_{in}(abb'=110|XYY'=000)\\
 P_{in}(abb'=111|XYY'=000)\\
\hline
\cdot\\
\cdot\\
\cdot\\
\hline
 P_{in}(abb'=000|XYY'=111)\\
 P_{in}(abb'=001|XYY'=111)\\
 P_{in}(abb'=010|XYY'=111)\\
 P_{in}(abb'=011|XYY'=111)\\
 P_{in}(abb'=100|XYY'=111)\\
 P_{in}(abb'=101|XYY'=111)\\
 P_{in}(abb'=110|XYY'=111)\\
 P_{in}(abb'=111|XYY'=111)\\
\end{array}
\right)_{PRB}=\left(
\begin{array}{c}
 P_{in}(ab=00|XY=00)\\
 0\\
 P_{in}(ab=01|XY=00)\\
  0\\
 P_{in}(ab=10|XY=00)\\
  0\\
 P_{in}(ab=11|XY=00)\\
  0\\
\hline
\cdot\\
\cdot\\
\cdot\\
 \hline P_{in}(ab=00|XY=11)\\
  0\\
 P_{in}(ab=01|XY=11)\\
  0\\
 P_{in}(ab=10|XY=11)\\
  0\\
 P_{in}(ab=11|XY=11)\\
\end{array}
\right)_{PRB} 
\end{equation}
$\vec{P}^{PRB}_{in}$ has 8 blocks $G_{XYY'}$, one for each value of $(X,Y,Y')$ and is a product state with 4 equal pairs of blocks, $G^{in}_{000}=G^{in}_{001}$,$G^{in}_{010}=G^{in}_{011}$, $G^{in}_{100}=G^{in}_{101}$, $G^{in}_{110}=G^{in}_{111}$ since both measurements on the initial state of $B$ give the same outcome.

Now, the outside observer Wigner will describe the transformation on $RB$ through 
the memory update map $U$ of Figure~\ref{fig: memorycircuit}. Let $ \vec{P}^{PRB}_{fin}$ be the final state that results by applying this map to the systems $RB$ in the initial state  $\vec{P}^{PRB}_{in}$. Any transformation on a system characterised by $n$ fiducial measurements with $k$ outcomes each can be represented by a $nk\times nk$ block matrix where each block is a $k\times k$ matrix (see \cite{Barrett07} for further details), for the system $RB$, $n=k=4$ and the memory update map $U_{RB}$ would be a $16\times 16$ block matrix of the following form where each $U_{ij}$ is a $4\times 4$ matrix.

\begin{equation*}
U_{RB}=
\left(
\begin{array}{c|c|c}
U_{11} & \cdot\quad\cdot\quad\cdot & U_{14} \\
\hline
\cdot&\quad &\cdot\\
\cdot&\quad &\cdot\\
\cdot&\quad &\cdot\\
\hline
U_{41} & \cdot\quad\cdot\quad\cdot & U_{44}
\end{array}
\right)_{RB}
\end{equation*}
Here, the first 4 rows decide the entries in the first block of the transformed matrix, the next 4, the second block and so on. Noting that the memory update transformation (Figure~\ref{fig: memorycircuit}) merely permutes elements within the relevant blocks (and does not mix elements between different blocks), the only non-zero blocks of $U_{RB}$ are the diagonal ones $U_{ii}$. Further, by the same argument as in Theorem~\ref{theorem: boxmemory}, the only relevant entries in the transformed state are when the same fiducial measurement is performed on Bob's system $R$ and memory $B$ i.e., only cases where $Y=Y'$. The remaining measurement choices maybe arbitrary for the final state (just as they are for $X\neq X'$ in Equation~\eqref{eq: PSMfin}). This means that among the 4 diagonal blocks, only 2 of them are relevant. The 4 fiducial measurements on $RB$ are $YY'=00,01,10,11$ and in that order, only the first and fourth are relevant since they correspond to $Y=Y'$. Within these relevant blocks (in this case $U_{11}$ and $U_{44}$), the operation is a CNOT on the output $b'$ controlled by the output $b$ and we have the following matrix representation of the memory update map $U$ of Figure~\ref{fig: memorycircuit}\footnote{The memory update map corresponding to the circuit of Figure~\ref{fig:update_copy} is a specific case of this map where the arbitrary blocks $*$ are also equal to $CN$}.
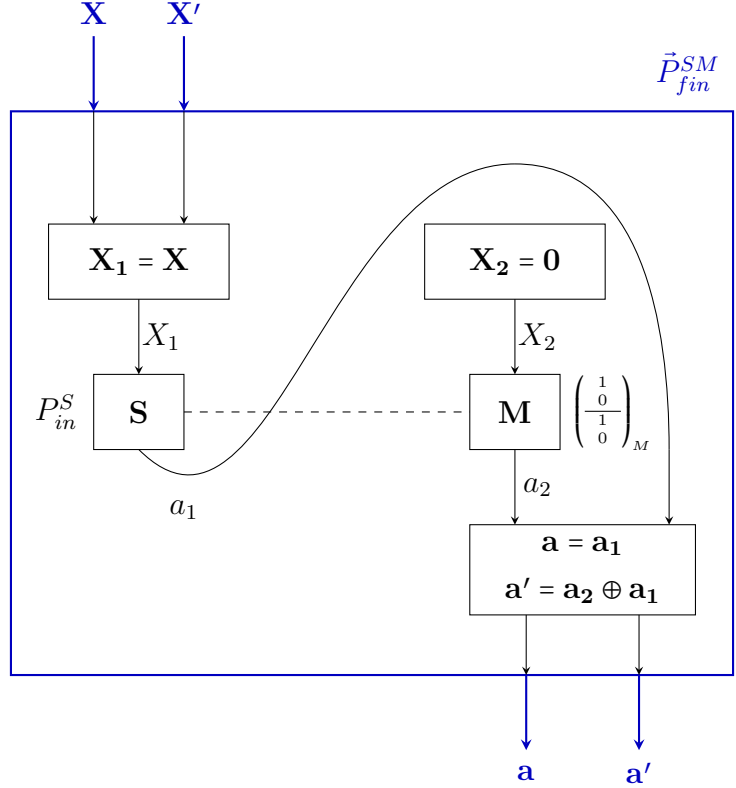
\begin{figure}[t!]
    \centering
\begin{tikzpicture}
\draw[blue!75!black,thick] (-0.5,-5) rectangle (9.1,2.5);
\draw (0,0) rectangle node[align=center]{$\mathbf{X_1=X}$} (2.4,1);
\draw (5,0) rectangle node[align=center]{$\mathbf{X_2=0}$} (7.4,1);
\draw[arrows={-stealth},blue!75!black,thick] (0.6,3.5)--(0.6,2.5); \draw[arrows={-stealth},blue!75!black,thick] (1.8,3.5)--(1.8,2.5);
\node[align=center,blue!75!black] at (0.6,3.8) {$\mathbf{X}$}; \node[align=center,blue!75!black] at (1.8,3.8) {$\mathbf{X'}$};
\draw[arrows={-stealth}] (0.6,2.5)--(0.6,1); \draw[arrows={-stealth}] (1.8,2.5)--(1.8,1);
\draw[arrows={-stealth}] (1.2,0)--(1.2,-1); \draw[arrows={-stealth}] (6.2,0)--(6.2,-1);
\node[align=center] at (1.5,-0.5) {$X_1$}; \node[align=center] at (6.5,-0.5) {$X_2$};
\draw (0.6,-2) rectangle node[align=center]{$\mathbf{S}$} (1.8,-1);
\draw (5.6,-2) rectangle node[align=center]{$\mathbf{M}$} (6.8,-1); \draw[dashed] (1.8,-1.5)--(5.6,-1.5); \draw[arrows={-stealth}] (6.2,-2)--(6.2,-3); \node[align=center] at (6.5,-2.5) {$a_2$}; \draw (5.6,-4.2) rectangle node[rectangle split,rectangle split parts=2]{$\mathbf{a=a_1}$ \nodepart{second} $\mathbf{a'=a_2\oplus a_1}$} (8.6,-3);
\node[align=center,blue!75!black] at (8.5,3) {$\vec{P}_{fin}^{SM}$};
 \draw[arrows={-stealth}] (6.35,-4.2)--(6.35,-5);  
  \draw[arrows={-stealth}] (7.85,-4.2)--(7.85,-5);  
 \draw[arrows={-stealth}] (8.25,-2)--(8.25,-3);  
\draw[arrows={-stealth},blue!75!black,thick] (6.35,-5)--(6.35,-6); \draw[arrows={-stealth},blue!75!black,thick] (7.85,-5)--(7.85,-6);
\node[align=center,blue!75!black] at (6.35,-6.3) {$\mathbf{a}$}; \node[align=center,blue!75!black] at (7.85,-6.3) {$\mathbf{a'}$}; \draw (1.2,-2) to [out=315,in=180] (6.2,1.8);
\draw (8.25,-2) to [out=90,in=0] (6.2,1.8);
\node[align=center] at (1.8,-2.8) {$a_1$};
\node[align=center] at (0.1,-1.5) {$P_{in}^S$};
\node[align=center] at (7.5,-1.5) {\tiny{$\left(\begin{array}{c}1\\0\\
\hline
1\\
0\\
\end{array}\right)_M$}};
\end{tikzpicture}
    \caption[Circuit decomposition of the memory update map in box world]{\textbf{Classical circuit decomposition of the memory update map $U$ as a box world transformation:} The blue box represents the final state of the system $S$ and memory $M$ after the memory update characterised by the fiducial measurements $X$ and $X'$ and the outcomes $a$, $a'$. Let $U$ be the memory update map that maps the initial state $\vec{P}_{in}^{SM}$ to a final state $\vec{P}_{fin}^{SM}$.  Noting that we only need to consider the case of $X=X'$ since the for $X\neq X'$, the entries of $\vec{P}_{fin}^{SM}$ can be arbitrary, the action of $\mathcal{T}$ is equivalent to the circuit shown here i.e., 1) Choose $X_1=X(=X')$ and perform this fiducial measurement on the initial state of the system $\vec{P}_{in}^S$ to obtain the outcome $a_1$. 2) Fix $X_2 =0$ (or $X_2=1$) and perform this fiducial measurement on the initial state of the memory $\vec{P}_{in}^M=(1\quad 0|1\quad 0)^T_M$ to obtain the outcome $a_2$. 3) Set $a=a_1$. 4) If $a_1=1$, set $a'=a_2$, otherwise set $a'=a_2\oplus 1$, where $\oplus$ denotes modulo 2 addition.}
    \label{fig: memorycircuit}
\end{figure}
\begin{equation}
 U_{RB}=
\left(
\begin{array}{c|c|c|c}
CN & 0 & 0 & 0 \\
\hline
0&*&0&0\\
\hline
0&0&*&0\\
\hline
0 & 0&0 & CN
\end{array}
\right)_{RB},  \qquad CN= \left(
\begin{array}{cccc}
1&0&0&0\\
0&1&0&0\\
0&0&0&1\\
0&0&1&0\\
\end{array}\right)
\end{equation}
where $0$ represents the $4\times 4$ null matrix and blocks labelled $*$ can be arbitrary. The final state $\vec{P}_{fin}^{PRB}$ as seen by Wigner is then
\begin{equation}
   \vec{P}_{fin}^{PRB}= (\mathcal{I}_P\otimes U_{RB})\vec{P}_{in}^{PRB}=(\mathcal{I}_P\otimes U_{RB})\Big[\vec{P}_{in}^{PR}\otimes \left(\begin{array}{c}
        1 \\
        0\\
        \hline
        1\\
        0\\
   \end{array}\right)_{B}\Big],
\end{equation}
where $\mathcal{I}_P$ is the identity transformation on the $P$ system. Since the $CN$ blocks are the only relevant blocks in $u_{RB}$ and each block of $\vec{P}_{in}^{PRB}$ has the same pattern of non-zero and zero entries (Equation~\eqref{eq: PRBin}), it is enough to look at the action of $\mathcal{I}_P\otimes CN$ on the first block $G^{in}_{000}$ of $\vec{P}_{in}^{PRB}$. Noting that $\mathcal{I}_P$ is a $2\times 2$ identity matrix, we have

\begin{align*}
\begin{split}
( \mathcal{I}_P\otimes CN)G^{in}_{000}&=\left(\begin{array}{cccccccc}
    1&0&0&0&0&0&0&0 \\
    0&1&0&0&0&0&0&0 \\
    0&0&0&1&0&0&0&0 \\
    0&0&1&0&0&0&0&0 \\
    0&0&0&0&1&0&0&0 \\
    0&0&0&0&0&1&0&0 \\
    0&0&0&0&0&0&0&1 \\
    0&0&0&0&0&0&1&0 \\
\end{array}\right) \left(
\begin{array}{c}
 P_{in}(ab=00|XY=00)\\
 0\\
 P_{in}(ab=01|XY=00)\\
  0\\
 P_{in}(ab=10|XY=00)\\
  0\\
 P_{in}(ab=11|XY=00)\\
  0\\
\end{array}
\right)=G^{fin}_{000}\\
&=\left(
\begin{array}{c}
 P_{in}(ab=00|XY=00)\\
 0\\
 0\\
 P_{in}(ab=01|XY=00)\\
 P_{in}(ab=10|XY=00)\\
  0\\
  0\\
 P_{in}(ab=11|XY=00)\\
\end{array}
\right)
=\left(
\begin{array}{c}
 P_{fin}(abb'=000|XYY'=000)\\
 P_{fin}(abb'=001|XYY'=000)\\
 P_{fin}(abb'=010|XYY'=000)\\
 P_{fin}(abb'=011|XYY'=000)\\
 P_{fin}(abb'=100|XYY'=000)\\
 P_{fin}(abb'=101|XYY'=000)\\
 P_{fin}(abb'=110|XYY'=000)\\
 P_{fin}(abb'=111|XYY'=000)\\
\end{array}
\right),
\end{split}
\end{align*}

where $P_{fin}(abb'=ijk|XYY'=lmn)$ represents the probability of obtaining the outcomes $a=i$, $b=j$, $b'=k$ when performing the fiducial measurements $X=l$, $Y=m$, $Y'=n$ on the final state $\vec{P}^{PRB}_{fin}$ and $G^{fin}_{000}$ is the first block of this final state. Clearly the only non-zero outcome probabilities are when $b=b'$ and this allows us to compress the final state by defining $\tilde{b}=i\Leftrightarrow b=b'=i$ for $i\in \{0,1\}$ and we have the following.
\begin{equation*}
    ( \mathcal{I}_P\otimes CN)G^{in}_{000} \equiv \left(
\begin{array}{c}
 P_{in}(ab=00|XY=00)\\
 P_{in}(ab=01|XY=00)\\
 P_{in}(ab=10|XY=00)\\
 P_{in}(ab=11|XY=00)\\
\end{array}
\right)=\left(
\begin{array}{c}
 P_{fin}(a\tilde{b}=00|XYY'=000)\\
 P_{fin}(a\tilde{b}=01|XYY'=000)\\
 P_{fin}(a\tilde{b}=10|XYY'=000)\\
 P_{fin}(a\tilde{b}=11|XYY'=000)
\end{array}
\right)=G_{00}^{in}
\end{equation*}
Here $G_{00}^{in}$ is the first block of the initial state $\vec{P}_{in}^{PR}$ and we have that the first block of the final state of $PRB$ is equivalent (up to zero entries) to the first block of the initial state over $PR$ alone or $G^{fin}_{000}=G_{00}^{in}$. Among the 8 blocks of $\vec{P}_{fin}^{PRB}$, only the 4 blocks $G^{fin}_{000}$,$G^{fin}_{011}$,$G^{fin}_{100}$ and $G^{fin}_{111}$ are the relevant ones (since $Y=Y'$ for these) and we can similarly show that $G^{fin}_{011}\equiv G^{in}_{01}$,$G^{fin}_{100}\equiv G^{in}_{10}$ and $G^{fin}_{111}\equiv G^{in}_{11}$ for the remaining 3 relevant blocks. Defining $\tilde{Y}=i \Leftrightarrow Y=Y'=i$ for $i\in\{0,1\}$, we obtain

\begin{equation}
\label{eq: PRBfin}
    \vec{P}_{fin}^{PRB}=\vec{P}_{fin}^{P\tilde{B}}\equiv\left(
\begin{array}{c}
 P_{fin}(a\tilde{b}=00|X\tilde{Y}=00)\\
 P_{fin}(a\tilde{b}=01|X\tilde{Y}=00)\\
 P_{fin}(a\tilde{b}=10|X\tilde{Y}=00)\\
 P_{fin}(a\tilde{b}=11|X\tilde{Y}=00)\\
 \hline
  P_{fin}(a\tilde{b}=00|X\tilde{Y}=01)\\
 P_{fin}(a\tilde{b}=01|X\tilde{Y}=01)\\
 P_{fin}(a\tilde{b}=10|X\tilde{Y}=01)\\
 P_{fin}(a\tilde{b}=11|X\tilde{Y}=01)\\
 \hline
  P_{fin}(a\tilde{b}=00|X\tilde{Y}=10)\\
 P_{fin}(a\tilde{b}=01|X\tilde{Y}=10)\\
 P_{fin}(a\tilde{b}=10|X\tilde{Y}=10)\\
 P_{fin}(a\tilde{b}=11|X\tilde{Y}=10)\\
 \hline
  P_{fin}(a\tilde{b}=00|X\tilde{Y}=11)\\
 P_{fin}(a\tilde{b}=01|X\tilde{Y}=11)\\
 P_{fin}(a\tilde{b}=10|X\tilde{Y}=11)\\
 P_{fin}(a\tilde{b}=11|X\tilde{Y}=11)\\
\end{array}
\right)= \left(
\begin{array}{c}
 P_{in}(ab=00|XY=00)\\
 P_{in}(ab=01|XY=00)\\
 P_{in}(ab=10|XY=00)\\
 P_{in}(ab=11|XY=00)\\
 \hline
  P_{in}(ab=00|XY=01)\\
 P_{in}(ab=01|XY=01)\\
 P_{in}(ab=10|XY=01)\\
 P_{in}(ab=11|XY=01)\\
 \hline
  P_{in}(ab=00|XY=10)\\
 P_{in}(ab=01|XY=10)\\
 P_{in}(ab=10|XY=10)\\
 P_{in}(ab=11|XY=10)\\
 \hline
  P_{in}(ab=00|XY=11)\\
 P_{in}(ab=01|XY=11)\\
 P_{in}(ab=10|XY=11)\\
 P_{in}(ab=11|XY=11)\\
\end{array}
\right)=\vec{P}_{in}^{PR}
\end{equation}
Equation~\eqref{eq: PRBfin} shows that final state $\vec{P}_{fin}^{P\tilde{B}}$ of Alice's system $P$, Bob's system $R$ and Bob's memory $B$ after Bob's local memory update is isomorphic to the initial state $\vec{P}_{in}^{PR}$ shared by Alice and Bob, having the same outcome probabilities as the latter for all the relevant measurements. Thus the initial correlations present in $\vec{P}_{in}^{PR}$ are preserved after Bob locally updates his memory according to the update procedure of Figure~\ref{fig: memorycircuit}. One can now repeat the same argument for Alice's local memory update taking $\vec{P}_{fin}^{P\tilde{B}}\otimes (1\quad 0|1 \quad 0)^T_{A}$ to be the initial state and by analogously defining $\tilde{s}=i \Leftrightarrow s=s'=i$ for $s\in \{a,X\}$,$i\in \{0,1\}$, we have the required result that the final state after both parties perform their local memory updates (as described by outside agents Ursula and Wigner) is isomorphic and operationally equivalent to the initial state shared by the parties before the memory update.
\begin{equation}
    \vec{P}_{fin}^{PARB}=\vec{P}_{fin}^{\tilde{A}\tilde{B}}\equiv \vec{P}_{in}^{PR}
\end{equation}

\end{proof}

\chapter{Conclusions and outlook}
This two part thesis proposes new techniques for analysing acyclic and cyclic causal structures as well as multi-agent logical paradoxes in non-classical theories. By developing these techniques, we have derived interesting preliminary insights and have identified a number of possible directions for future work on these aspects of quantum and post-quantum theories, which have both foundational as well as practical applications. 
%We summarise our contributions and outlook in more detail below.

The results reported in Part I pertain to the entropic analysis of causal structures (Chapters~\ref{chapter: Tsallispaper} and~\ref{chapter: mixingpaper}) and to the connections between the notions of causation and space-time (Chapter~\ref{chapter: jammingpaper}). In the former case, we have developed a systematic method for analysing causal structures using generalised entropies, providing, in particular, a new set of Tsallis entropic inequalities for the Triangle causal structure. We have also extended our analysis to post-selected causal structures and identified families of distributions whose non-classicality can be detected in Bell scenarios with non-binary outcomes. Further, the properties of classical and quantum Tsallis entropies derived in Chapter~\ref{chapter: Tsallispaper} can be potentially useful beyond the analysis of causal structures, since Tsallis entropies are widely employed in other areas of information theory as well as non-extensive statistical physics.

On the other hand, our results also reveal significant drawbacks of the entropic technique for certifying the non-classicality of causal structures, both with and without the additional post-selection method. We identified the main reasons for the computational intractability of the entropy vector method using Tsallis entropies. Although we were able to circumvent this issue to derive new Tsallis entropic inequalities for the Triangle scenario, no quantum violations of these inequalities were found even by known non-classical distributions in the scenario. We noted that R\'enyi entropies do not seem to overcome these limitations either, due to the lack of linear constraints required to construct a good initial outer approximation to the R\'enyi entropy cone. In post-selected causal structures, we provided numerical and analytical evidence suggesting that both Shannon and Tsallis entropic inequalities are in general insufficient for detecting non-classicality in Bell scenarios with binary inputs and three or more outcomes per party. %We used both Shannon and Tsallis entropic inequalities as well as a general class of post-processing operations on the observed distributions\footnote{ Namely $LOSR+E$ (local operations and shared randomness + exchange of parties).} to reach this conclusion.

Limitations for distinguishing between classical and non-classical correlations in causal structures using Shannon entropies have been previously encountered in the absence of post-selection \cite{Weilenmann16, Weilenmann2017}. By showing that similar limitations arise even for generalised entropies and in the presence of post-selection, our results shed further light on the debate regarding the usefulness of the entropic method for this purpose. It appears that the advantages of working in entropy space such as convexity and better scaling of the number of bounding inequalities come at the cost of losing valuable information about the non-classical nature of the correlations, even in post-selected causal structures. Our results illuminate some of the merits as well as limitations of the entropic technique and point to the need for developing novel methods for understanding causation in non-classical theories. 

In general, this is a difficult task--- fully characterising the bounding inequalities for the set of classical correlations in a given causal structure is known to be an NP-complete problem \cite{Pitowski89}. Therefore, increase in the computational power of current technologies and improvements to the performance of variable elimination algorithms would naturally increase the set of scenarios that can be characterised. For example, developing a variable elimination algorithm (such as Fourier-Motzkin) capable of dealing with systems of non-linear constraints would allow for tighter entropic characterisations through the inclusion of non-linear inequalities. One can also consider other algorithms for polyhedral projection that can sometimes produce results even when Fourier-Motzkin Elimination (FME) becomes intractable \cite{Gl_le_2018}. We have provided a method to bypass the FME to derive generalised entropic inequalities in cases where Shannon entropic inequalities are known. Our Tsallis inequalities thus derived depend on the cardinality of the variables in the Triangle causal structure. Hence one could, in principle, investigate dimension dependent distinctions between classical and quantum correlations using these, while also taking into account the new class of non-local distributions recently discovered in the Triangle network \cite{Renou2019, Kraft2020}. Other possibilities include employing the inflation technique \cite{wolfe2016} in combination with the entropic method, constructing more efficient algorithms (than the random search typically employed) for finding quantum violations, and considering an alternative definition of the Tsallis conditional entropies \cite{ABE2001157} that satisfy simpler causal constraints but violate the chain rule. These methods could lead to  improvements but are not guaranteed to be computationally tractable. 
%That being said, the entropic method remains useful for analysing specific cases, as we have also highlighted in the chapters. 
%For example, in the three outcome Bell scenario we considered, we have identified a class of correlations whose non-classicality can be detected entropically using an analogue of the technique that is known to be sufficient for the binary outcome case. This result can possibly be extended to larger outcome scenarios to identify other such non-trivial classes there. We also noted that a Tsallis over Shannon advantage for detecting non-classicality (found in \cite{Wajs15}) arises when post-processing operations are not considered, but seem to disappear under suitable post-processing of the distributions. 

Cryptographic and other information-processing applications have greatly benefited from the characterisation of the sets of correlations in the Bell scenario \cite{Ekert91,MayersYao,BHK, Pironio2009, CK, RogerThesis}, which are relatively well understood due to their convexity. Often, these applications also go beyond the Bell scenario and involve networks of parties sharing multiple sources of entanglement, where the set of observed correlations in probability space is no longer convex. The entanglement swapping protocol \cite{Bennett1993, Zukowski1993} is an important example, which also plays a role in the construction of quantum repeater networks \cite{Briegel1998} used in quantum communication scenarios and the quantum internet. Therefore, exploring different techniques for certifying non-classicality in and beyond the Bell causal structure, and understanding their advantages and limitations as done in this thesis is also of practical importance. At the foundational level, this reveals fundamental distinctions between the principles of causality that manifest themselves in classical and quantum regimes. The information-theoretic approach to causal structures also allow us to consider post-quantum theories and identify physical principles other than relativistic ones that single out quantum theory among this spectrum of theories. 

In Chapter~\ref{chapter: jammingpaper}, we have further explored this fundamental aspect by developing a causal modelling framework for a general class of physical theories, and formalising relationships between causality, space-time and relativistic principles (such as the inability to signal superluminally) therein. This allowed us to analyse scenarios involving superluminal influences and causal loops that nevertheless do not lead to superluminal signalling at the operational level. In particular, our framework provides an initial tool set for causally modelling post-quantum theories admitting jamming non-local correlations \cite{Grunhaus1996, Horodecki2019}, which have not been rigorously analysed before, and for identifying further physical principles underlying such theories. 

Moreover, these results also generate scope for interesting future investigations into causality and space-time structure. One promising avenue would be to extend this work to construct a more complete framework for causally modelling cyclic and fine-tuned causal structures in non-classical theories, which is likely to be of independent interest. Furthermore, this work can aid an operational analysis of closed time-like curves (CTCs) in the presence of fine-tuned causal influences, and an investigation of whether fine-tuned and non fine-tuned interpretations of quantum theory behave differently in the presence of CTCs. Yet another direction is to examine simulations of indefinite causal structures \cite{Hardy2005, Oreshkov2012, Araujo2017} using CTCs and fine-tuning, and analyse their embeddings within a fixed space-time structure. Indefinite causal structures can in theory be implemented through a quantum superposition of gravitating masses \cite{Zych2019}, they have also been claimed to be physically implemented by superposing the orders of quantum operations on a target system \cite{Procopio2015, Rubino2017}. The space-time information associated with these operations plays an important role in distinguishing between the two implementations \cite{Paunkovic2020}\footnote{As we also find in: Vilasini,  V.,  del  Rio,  L. and Renner, R. \emph{Causality in definite and indefinite space-times.} In preparation (2020). \definecolor{mylinkcolor}{rgb}{0,0,0}  \url{https://wdi.centralesupelec.fr/users/valiron/qplmfps/papers/qs01t3.pdf}\definecolor{mylinkcolor}{rgb}{0,0,0.7} }. Further, general relativity is known to admit CTC solutions \cite{VanStockum1937, Godel1949} even in the absence of quantum effects, while certain fundamental principles of quantum theory such as the no-cloning principle and linearity may no longer hold in the presence of CTCs \cite{Deutsch1991}. A deeper understanding of causality and space-time structure in non-classical theories, in the presence of causal loops would hence provide insights not only into quantum theory, but potentially also into its unification with general relativity, which has been a longstanding open problem in theoretical physics. The thesis provides a small contribution towards these larger goals.

Part II of this thesis contributes to an understanding of multi-agent logic in scenarios where reasoning agents model each other's memories as systems of a physical theory. Generalising the conditions of the Frauchiger-Renner quantum experiment to generalised probabilistic theories allowed us to derive a stronger paradox in the particular GPT of box world. Our result reveals that reversibility of the memory update akin to quantum unitarity is not necessary for deriving such multi-agent paradoxes. Rather, a logical form of contextuality along with an appropriate notion of an information-preserving memory update appear to be more fundamental for deriving a deterministic paradox. This work paves the way for a more general characterisation of Frauchiger-Renner type paradoxes, their relationships to contextuality and the structure of multi-agent logic in quantum and post-quantum theories.

Apart from the implications for logic, these multi-agent paradoxes also challenge standard approaches to causal modelling that we have discussed in the first part of this thesis. In these settings, a system, such as a measurement outcome that appears classical to one agent may be modelled as an entangled subsystem (entangled with the first agent's memory) by another agent. Therefore the distinction between observed and unobserved systems in a causal structure can become subjective, and a detailed investigation of causality in these general settings is yet to be undertaken. 

The fact that the agents do not necessarily model the measurement outcomes of other agents as classical systems implies that one cannot assume the existence of a joint distribution over all these outcomes. As we have also seen  in Part I of this thesis, no joint distribution can in general be assigned to non-coexisting systems in a non-classical causal structure. It is precisely this inability to make a global assignment of values to all measurement outcomes that results in the apparent paradox, suggesting that these outcomes cannot be seen as objective facts about the world, but only subjective to the agents who observed them. In both the quantum and box-world versions of the paradox, the states and measurements involved produce non-classical correlations in the bipartite Bell causal structure. The treatment of agents as systems of a non-classical physical theory promotes this non-classicality to the level of the outcomes observed by the agents, however it appears that not all non-classical correlations can lead to such paradoxes. A better understanding of non-classical correlations in causal structures would hence enhance our knowledge about these general settings involving complex quantum systems that are capable of logical deductions. 

With sufficient technological progress in creating and manipulating stable many-qubit superpositions, the physical implementation of multi-agent paradoxes may be a distinct possibility in the near future, since a relatively small quantum computer would fulfill the role of an agent for these purposes. There are global efforts towards achieving such macroscopic superpositions, both for the practical purposes of scalable quantum computing and for pushing the horizons of fundamental physics by probing the quantum effects of gravity. These make it imperative to develop more general theoretical frameworks for modelling multi-agent paradoxes as well as causality and space-time structure in non-classical theories.
We believe there are several deep connections between causality and multi-agent logical paradoxes in non-classical theories that are yet to be discovered, and hope that the results presented in both parts of this thesis contribute towards a more complete physical framework or theory in the future.

% -------------------------------------------------------------------
% Bibliography
% -------------------------------------------------------------------
\definecolor{mylinkcolor}{rgb}{0,0,0}
%\bibliographystyle{naturemag}
%\bibliography{refs3} 

\begin{thebibliography}{100}

\bibitem{Aaronson2017}
{\sc Aaronson, S.}
\newblock Why philosophers should care about computational complexity.
\newblock {\em arXiv:1108.1791\/} (2011).

\bibitem{Aaronson2008}
{\sc Aaronson, S., and Watrous, J.}
\newblock Closed timelike curves make quantum and classical computing
  equivalent.
\newblock {\em Proceedings of the Royal Society A: Mathematical, Physical and
  Engineering Sciences 465}, 2102 (2008), pp. 631–647.

\bibitem{ABE2001157}
{\sc Abe, S., and Rajagopal, A.}
\newblock Nonadditive conditional entropy and its significance for local
  realism.
\newblock {\em Physica A: Statistical Mechanics and its Applications 289}, 1
  (2001), pp. 157--164.

\bibitem{Abramsky15}
{\sc Abramsky, S., Barbosa, R.~S., Kishida, K., Lal, R., and Mansfield, S.}
\newblock {Contextuality, Cohomology and Paradox}.
\newblock {\em 24th EACSL Annual Conference on Computer Science Logic (CSL
  2015) 41\/} (2015), Ed. Stephan Kreutzer, pp. 211--228.

\bibitem{Acin2016}
{\sc Acin, A., and Masanes, L.}
\newblock Certified randomness in quantum physics.
\newblock {\em Nature 540}, 7632 (2016), pp. 213--219.

\bibitem{Allen2017}
{\sc Allen, J.-M.~A., Barrett, J., Horsman, D.~C., Lee, C.~M., and Spekkens,
  R.~W.}
\newblock Quantum common causes and quantum causal models.
\newblock {\em Physical Review X 7}, 3 (2017), p. 031021.

\bibitem{Araki1970}
{\sc Araki, H., and Lieb, E.~H.}
\newblock Entropy inequalities.
\newblock {\em Communications in Mathematical Physics 18}, 2 (1970), pp.
  160--170.

\bibitem{Araujo2015}
{\sc Ara{\'{u}}jo, M., Branciard, C., Costa, F., Feix, A., Giarmatzi, C., and
  Brukner, {\v{C}}.}
\newblock {Witnessing causal nonseparability}.
\newblock {\em New Journal of Physics 17}, 10 (2015), p. 102001.

\bibitem{Araujo2014b}
{\sc Ara{\'{u}}jo, M., Costa, F., and Brukner, {\v{C}}.}
\newblock {Computational Advantage from Quantum-Controlled Ordering of Gates}.
\newblock {\em Physical Review Letters 113}, 25 (2014), p. 250402.

\bibitem{Araujo2016}
{\sc Ara{\'{u}}jo, M., Feix, A., Navascu{\'{e}}s, M., and Brukner, {\v{C}}.}
\newblock {A purification postulate for quantum mechanics with indefinite
  causal order}.
\newblock {\em Quantum 1\/} (2016), p. 10.

\bibitem{Araujo2017}
{\sc Ara\'ujo, M., Gu\'erin, P.~A., and Baumeler, A.}
\newblock Quantum computation with indefinite causal structures.
\newblock {\em Physical Review A 96\/} (2017), p. 052315.

\bibitem{Arnon-Friedman2018}
{\sc Arnon-Friedman, R., Dupuis, F., Fawzi, O., Renner, R., and Vidick, T.}
\newblock Practical device-independent quantum cryptography via entropy
  accumulation.
\newblock {\em Nature Communications 9}, 1 (2018), p. 459.

\bibitem{Audenaert2007}
{\sc Audenaert, K.}
\newblock Subadditivity of q-entropies for $q>1$.
\newblock {\em Journal of Mathematical Physics 48\/} (2007), p. 083507.

\bibitem{Avis2004}
{\sc Avis, D., Imai, H., Ito, T., and Sasaki, Y.}
\newblock Deriving tight {B}ell inequalities for 2 parties with many 2-valued
  observables from facets of cut polytopes.
\newblock {\em arxiv:quant-ph/0404014\/} (2004).

\bibitem{Bancal2010}
{\sc Bancal, J.-D., Gisin, N., and Pironio, S.}
\newblock Looking for symmetric {B}ell inequalities.
\newblock {\em Journal of Physics A: Mathematical and Theoretical 43}, 38
  (2010), p. 385303.

\bibitem{Barrett07}
{\sc Barrett, J.}
\newblock Information processing in generalized probabilistic theories.
\newblock {\em Physical Review A 75\/} (2007), p. 032304.

\bibitem{BHK}
{\sc Barrett, J., Hardy, L., and Kent, A.}
\newblock No signaling and quantum key distribution.
\newblock {\em Physical Review Letters 95\/} (2005), p. 010503.

\bibitem{Barrett2020}
{\sc Barrett, J., Lorenz, R., and Oreshkov, O.}
\newblock Cyclic quantum causal models.
\newblock {\em arxiv:2002.1215731\/} (2020).

\bibitem{Baumann2018}
{\sc Baumann, V., and Wolf, S.}
\newblock On {F}ormalisms and {I}nterpretations.
\newblock {\em Quantum 2\/} (2018), p. 99.

\bibitem{Baumeler2018}
{\sc Baumeler, A., Degorre, J., and Wolf, S.}
\newblock Bell correlations and the common future.
\newblock {\em Quantum Foundations, Probability and Information\/} (2018), pp.
  255–268.

\bibitem{Baumeler2016}
{\sc Baumeler, {\"A}., and Wolf, S.}
\newblock The space of logically consistent classical processes without causal
  order.
\newblock {\em New Journal of Physics}, 1 (2016), p. 013036.

\bibitem{Bell}
{\sc Bell, J.~S.}
\newblock On the {E}instein {P}odolsky {R}osen paradox.
\newblock {\em Physics Physique Fizika 1\/} (1964), pp. 195--200.

\bibitem{Bemporad2001}
{\sc Bemporad, A., Fukuda, K., and Torrisi, F.~D.}
\newblock Convexity recognition of the union of polyhedra.
\newblock {\em Computational Geometry 18}, 3 (2001), pp. 141--154.

\bibitem{Bennett1995}
{\sc Bennett, C., Brassard, G., Crepeau, C., and Maurer, U.}
\newblock Generalized privacy amplification.
\newblock {\em IEEE Transactions on Information Theory 41\/} (1995), pp.
  1915--1923.

\bibitem{Bennett2005}
{\sc Bennett, C., and Schumacher, B.}
\newblock Talk at {QUPON} {W}ien, (2005).

\bibitem{Bennett1993}
{\sc Bennett, C.~H., Brassard, G., Cr\'epeau, C., Jozsa, R., Peres, A., and
  Wootters, W.~K.}
\newblock Teleporting an unknown quantum state via dual classical and
  {E}instein-{P}odolsky-{R}osen channels.
\newblock {\em Physical Review Letters 70\/} (1993), pp. 1895--1899.

\bibitem{Bohm1952}
{\sc Bohm, D.}
\newblock A suggested interpretation of the quantum theory in terms of
  ``hidden'' variables. {I}.
\newblock {\em Physical Review 85\/} (1952), pp. 166--179.

\bibitem{Bohr1949}
{\sc Bohr, N.}
\newblock Discussions with {E}instein on epistemological problems in atomic
  physics.
\newblock {\em Albert {E}instein: {P}hilosopher-{S}cientist, The Library of
  Living Philosophers, Ed. Schilpp, P. A.\/} (1949), pp. 200--241.

\bibitem{Boltzmann1866}
{\sc Boltzmann, L.}
\newblock \"{U}ber die mechanische bedeutung des zweiten hauptsatzes der
  wärmetheorie.
\newblock {\em Wiener Berichte 53\/} (1866), pp. 195--220.

\bibitem{Bong2020}
{\sc Bong, K.-W., Utreras-Alarc{\'o}n, A., Ghafari, F., Liang, Y.-C., Tischler,
  N., Cavalcanti, E.~G., Pryde, G.~J., and Wiseman, H.~M.}
\newblock A strong no-go theorem on the {W}igner's friend paradox.
\newblock {\em Nature Physics\/} (2020),
  https://doi.org/10.1038/s41567--020--0990--x.

\bibitem{Boyd2004}
{\sc Boyd, S., and Vandenberghe, L.}
\newblock Convex optimization.
\newblock {\em Cambridge {U}niversity {P}ress\/} (2004).

\bibitem{Branciard2016}
{\sc Branciard, C., Ara{\'{u}}jo, M., Feix, A., Costa, F., and Brukner,
  {\v{C}}.}
\newblock {The simplest causal inequalities and their violation}.
\newblock {\em New Journal of Physics 18}, 1 (2016), p. 13008.

\bibitem{Branciard2012}
{\sc Branciard, C., Rosset, D., Gisin, N., and Pironio, S.}
\newblock Bilocal versus nonbilocal correlations in entanglement-swapping
  experiments.
\newblock {\em Physical Review A 85\/} (2012), p. 032119.

\bibitem{BraunsteinCaves88}
{\sc Braunstein, S.~L., and Caves, C.~M.}
\newblock Information-theoretic {B}ell inequalities.
\newblock {\em Physical Review Letters 61\/} (1988), pp. 662--665.

\bibitem{Briegel1998}
{\sc Briegel, H.-J., D\"ur, W., Cirac, J.~I., and Zoller, P.}
\newblock Quantum repeaters: The role of imperfect local operations in quantum
  communication.
\newblock {\em Physical Review Letters 81\/} (1998), pp. 5932--5935.

\bibitem{Brukner2018}
{\sc Brukner, {\v{C}}.}
\newblock A no-go theorem for observer-independent facts.
\newblock {\em Entropy 20}, 350 (2018).

\bibitem{Brukner2004}
{\sc Brukner, {\v{C}}., Taylor, S., Cheung, S., and Vedral, V.}
\newblock Quantum entanglement in time.
\newblock {\em arXiv:quant-ph/0402127\/} (2004).

\bibitem{Brul2017}
{\sc Brul\'e, J.}
\newblock A causation coefficient and taxonomy of correlation/causation
  relationships.
\newblock {\em arxiv:1708.05069\/} (2017).

\bibitem{Brunner2014}
{\sc Brunner, N., Cavalcanti, D., Pironio, S., Scarani, V., and Wehner, S.}
\newblock Bell nonlocality.
\newblock {\em Reviews of Modern Physics 86\/} (2014), pp. 419--478.

\bibitem{Buhrman2010}
{\sc Buhrman, H., Cleve, R., Massar, S., and de~Wolf, R.}
\newblock Nonlocality and communication complexity.
\newblock {\em Reviews of Modern Physics 82\/} (2010), pp. 665--698.

\bibitem{Carnap1988}
{\sc Carnap, R.}
\newblock Meaning and necessity: a study in semantics and modal logic.
\newblock {\em University of Chicago Press\/} (1988).

\bibitem{Castro-Ruiz2020}
{\sc Castro-Ruiz, E., Giacomini, F., Belenchia, A., and Brukner, {\v{C}}.}
\newblock Quantum clocks and the temporal localisability of events in the
  presence of gravitating quantum systems.
\newblock {\em Nature Communications 11}, 1 (2020), p. 2672.

\bibitem{Cerf1997}
{\sc Cerf, N.~J., and Adami, C.}
\newblock Negative entropy and information in quantum mechanics.
\newblock {\em Physical Review Letters 79\/} (1997), pp. 5194--5197.

\bibitem{Chaves13}
{\sc Chaves, R.}
\newblock Entropic inequalities as a necessary and sufficient condition to
  noncontextuality and locality.
\newblock {\em Physical Review A 87\/} (2013), p. 022102.

\bibitem{Chaves2018}
{\sc Chaves, R., Carvacho, G., Agresti, I., Di~Giulio, V., Aolita, L.,
  Giacomini, S., and Sciarrino, F.}
\newblock Quantum violation of an instrumental test.
\newblock {\em Nature Physics 14}, 3 (2018), pp. 291--296.

\bibitem{Chaves2012}
{\sc Chaves, R., and Fritz, T.}
\newblock Entropic approach to local realism and noncontextuality.
\newblock {\em Physical Review A 85\/} (2012), pp. 032113.

\bibitem{Chaves2014}
{\sc Chaves, R., Luft, L., and Gross, D.}
\newblock Causal structures from entropic information: geometry and novel
  scenarios.
\newblock {\em New Journal of Physics 16}, 4 (2014), p. 043001.

\bibitem{Chaves2015}
{\sc Chaves, R., Majenz, C., and Gross, D.}
\newblock Information-theoretic implications of quantum causal structures.
\newblock {\em Nature Communications 6}, 1 (2015), p. 5766.

\bibitem{Chernikov1960}
{\sc Chernikov, S.~N.}
\newblock Contraction of systems of linear inequalities.
\newblock {\em Doklady Akademii Nauk SSSR 131}, 3 (1960), pp. 518--521.

\bibitem{Chernikov1965}
{\sc Chernikov, S.~N.}
\newblock The convolution of finite systems of linear inequalities.
\newblock {\em USSR Computational Mathe- matics and Mathematical Physics 5\/}
  (1965), pp. 1--24.

\bibitem{Chiribella2018}
{\sc Chiribella, G., Banik, M., Bhattacharya, S.~S., Guha, T., Alimuddin, M.,
  Roy, A., Saha, S., Agrawal, S., and Kar, G.}
\newblock Indefinite causal order enables perfect quantum communication with
  zero capacity channel.
\newblock {\em arXiv:1810.10457\/} (2018).

\bibitem{Chiribella2010}
{\sc Chiribella, G., D'Ariano, G.~M., and Perinotti, P.}
\newblock Probabilistic theories with purification.
\newblock {\em Physical Review A 81\/} (2010), p. 062348.

\bibitem{Chiribella2013}
{\sc Chiribella, G., D'Ariano, G.~M., Perinotti, P., and Valiron, B.}
\newblock {Quantum computations without definite causal structure}.
\newblock {\em Physical Review A 88}, 2 (2013), p. 022318.

\bibitem{Choi1975}
{\sc Choi, M.-D.}
\newblock Completely positive linear maps on complex matrices.
\newblock {\em Linear Algebra and its Applications 10}, 3 (1975), pp. 285--290.

\bibitem{porta}
{\sc Christof, T., and Loebel, A.}
\newblock {POlyhedron Representation Transformation Algorithm}.
\newblock {\em (PORTA)\/} (1997).

\bibitem{Clauser1974}
{\sc Clauser, J.~F., and Horne, M.~A.}
\newblock Experimental consequences of objective local theories.
\newblock {\em Physical Review D 10\/} (1974), pp. 526--535.

\bibitem{CHSH69}
{\sc Clauser, J.~F., Horne, M.~A., Shimony, A., and Holt, R.~A.}
\newblock Proposed experiment to test local hidden-variable theories.
\newblock {\em Physical Review Letters 23\/} (1969), pp. 880--884.

\bibitem{Clausius1850}
{\sc Clausius, R.}
\newblock Ueber die bewegende kraft der w{\"a}rme und die gesetze, welche sich
  daraus f{\"u}r die w{\"a}rmelehre selbst ableiten lassen.
\newblock {\em Annalen der Physik 155}, 3 (1850), pp. 368--397.

\bibitem{RogerThesis}
{\sc Colbeck, R.}
\newblock Quantum and relativistic protocols for secure multi-party
  computation.
\newblock {\em PhD Dissertation, University of Cambridge. arxiv:0911.3814\/}
  (2007).

\bibitem{CK}
{\sc Colbeck, R., and Kent, A.}
\newblock Private randomness expansion with untrusted devices.
\newblock {\em Journal of Physics A 44}, 9 (2011), p. 095305.

\bibitem{CR2013}
{\sc Colbeck, R., and Renner, R.}
\newblock A short note on the concept of free choice.
\newblock {\em arxiv:1302.4446\/} (2013).

\bibitem{LPAssumptions}
{\sc Colbeck, R., and Vilasini, V.}
\newblock L{PA}ssumptions ({M}athematica package).
\newblock https://github.com/rogercolbeck/LPAssumptions.

\bibitem{CollinsGisin04}
{\sc Collins, D., and Gisin, N.}
\newblock A relevant two qubit {B}ell inequality inequivalent to the {CHSH}
  inequality.
\newblock {\em Journal of Physics A: Mathematical and General 37}, 5 (2004), p.
  1775.

\bibitem{CGLMP02}
{\sc Collins, D., Gisin, N., Linden, N., Massar, S., and Popescu, S.}
\newblock Bell inequalities for arbitrarily high-dimensional systems.
\newblock {\em Physical Review Letters 88\/} (2002), p. 040404.

\bibitem{Cook2004}
{\sc Cook, R.~T.}
\newblock Patterns of paradox.
\newblock {\em The Journal of Symbolic Logic 69}, 3 (2004), pp. 767--774.

\bibitem{Cope2019}
{\sc Cope, T., and Colbeck, R.}
\newblock {B}ell inequalities from no-signaling distributions.
\newblock {\em Physical Review A 100\/} (2019), p. 022114.

\bibitem{Costa2016}
{\sc Costa, F., and Shrapnel, S.}
\newblock Quantum causal modelling.
\newblock {\em New Journal of Physics 18}, 6 (2016), p. 63032.

\bibitem{Curado1991}
{\sc Curado, E. M.~F., and Tsallis, C.}
\newblock Generalized statistical mechanics: connection with thermodynamics.
\newblock {\em Journal of Physics A 24\/} (1991), pp. L69--L72.

\bibitem{Dantzig1963}
{\sc Dantzig, G.~B.}
\newblock Linear programming and extensions.
\newblock {\em Princeton University Press\/} (1963).

\bibitem{Dantzig1990}
{\sc Dantzig, G.~B.}
\newblock Origins of the simplex method.
\newblock {\em A History of Scientific Computing\/} (1990), pp. 141–151.

\bibitem{Daroczy1970}
{\sc Dar{\'{o}}czy, Z.}
\newblock {Generalized information functions}.
\newblock {\em Information and Control 16}, 1 (1970), pp. 36--51.

\bibitem{DeBroglie1927}
{\sc De~Broglie, L.}
\newblock La m\'ecanique ondulatoire et la structure atomique de la mati\'ere
  et du rayonnement.
\newblock {\em Journal of Physics Radium 8}, 5 (1952), pp. 225--241.

\bibitem{delRio2015}
{\sc del Rio, L., Kr{\"a}mer, L., and Renner, R.}
\newblock Resource theories of knowledge.
\newblock {\em arXiv:1511.08818\/} (2015).

\bibitem{Deutsch1985}
{\sc Deutsch, D.}
\newblock Quantum theory as a universal physical theory.
\newblock {\em International Journal of Theoretical Physics 24}, 1 (1985), pp.
  1--41.

\bibitem{Deutsch1991}
{\sc Deutsch, D.}
\newblock Quantum mechanics near closed timelike lines.
\newblock {\em Physical Review D 44\/} (1991), pp. 3197--3217.

\bibitem{DeWitt1970}
{\sc DeWitt, B.~S.}
\newblock Quantum mechanics and reality.
\newblock {\em Physical Today 23\/} (1970), pp. 155–165.

\bibitem{Dourdent2020}
{\sc Dourdent, H.}
\newblock A quantum {G}\"odelian hunch.
\newblock {\em arXiv:2005.04274\/} (2020).

\bibitem{Durr2009}
{\sc D\"{u}rr, D., and Teufel, S.}
\newblock Bohmian mechanics: The physics and mathematics of quantum theory.
\newblock {\em Springer\/} (2009).

\bibitem{Ekert91}
{\sc Ekert, A.~K.}
\newblock Quantum cryptography based on {B}ell's theorem.
\newblock {\em Physical Review Letters 67\/} (1991), pp. 661--663.

\bibitem{Elga2000}
{\sc Elga, A.}
\newblock Self-locating belief and the sleeping beauty problem.
\newblock {\em Analysis 60}, 2 (2000), pp. 143--147.

\bibitem{Everett1957}
{\sc Everett, H.}
\newblock ``{R}elative state'' formulation of quantum mechanics.
\newblock {\em Reviews of Modern Physics 29\/} (1957), pp. 454--462.

\bibitem{Fagin2004}
{\sc Fagin, R., Halpern, J.~Y., Moses, Y., and Vardi, M.}
\newblock Reasoning about knowledge.
\newblock {\em MIT press\/} (2004).

\bibitem{Fehr2014}
{\sc Fehr, S., and Berens, S.}
\newblock On the conditional {R}{\'e}nyi entropy.
\newblock {\em IEEE Transactions on Information Theory 60\/} (2014), pp.
  6801--6810.

\bibitem{Forre2017}
{\sc Forr\'e, P., and Mooij, J.~M.}
\newblock Markov properties for graphical models with cycles and latent
  variables.
\newblock {\em arxiv:1710.08775\/} (2017).

\bibitem{Frauchiger2018}
{\sc Frauchiger, D., and Renner, R.}
\newblock Quantum theory cannot consistently describe the use of itself.
\newblock {\em Nature Communications 9}, 1 (2018), p. 3711.

\bibitem{Freedman1972}
{\sc Freedman, S.~J., and Clauser, J.~F.}
\newblock Experimental test of local hidden-variable theories.
\newblock {\em Physical Review Letters 28\/} (1972), pp. 938--941.

\bibitem{Friedman2003}
{\sc Friedman, M.}
\newblock The fed's thermostat.
\newblock {\em The {W}all {S}treet {J}ournal\/} (2003).

\bibitem{Fritz2012}
{\sc Fritz, T.}
\newblock Beyond {B}ell's theorem: correlation scenarios.
\newblock {\em New Journal of Physics 14}, 10 (2012), p. 103001.

\bibitem{FritzChaves2013}
{\sc {Fritz}, T., and {Chaves}, R.}
\newblock Entropic inequalities and marginal problems.
\newblock {\em IEEE Transactions on Information Theory 59}, 2 (2013), pp.
  803--817.

\bibitem{Fritz13}
{\sc Fritz, T., and Chaves, R.}
\newblock Entropic inequalities and marginal problems.
\newblock {\em IEEE Transactions on Information Theory 59\/} (2013), pp.
  803--817.

\bibitem{Fuchs2013}
{\sc Fuchs, C.~A., and Schack, R.}
\newblock Quantum-bayesian coherence.
\newblock {\em Reviews of Modern Physics 85\/} (2013), pp. 1693--1715.

\bibitem{Fukuda2016}
{\sc Fukuda, K.}
\newblock Polyhedral computation (lecture notes), Spring 2016.
\newblock
  \url{http://www-oldurls.inf.ethz.ch/personal/fukudak/lect/pclect/notes2016/}.

\bibitem{Furuichi04}
{\sc Furuichi, S.}
\newblock Information theoretical properties of {T}sallis entropies.
\newblock {\em Journal of Mathematical Physics 47\/} (2004).

\bibitem{Furuichirelentropy}
{\sc Furuichi, S., Yanagi, K., and Kuriyama, K.}
\newblock Fundamental properties of {T}sallis relative entropy.
\newblock {\em Journal of Mathematical Physics 45\/} (2004).

\bibitem{LogicStanford}
{\sc Garson, J.}
\newblock Modal logic.
\newblock {\em The Stanford Encyclopedia of Philosophy\/} (2016), Ed. Edward N.
  Zalta.

\bibitem{Geiger1987}
{\sc Geiger, D.}
\newblock Towards the formalization of informational dependencies.
\newblock {\em Tech. rep. 880053. UCLA Computer Science\/} (1987).

\bibitem{Gibbs1902}
{\sc Gibbs, J.~W.}
\newblock Elementary principles in statistical mechanics.
\newblock {\em Charles Scribner's Sons\/} (1902).

\bibitem{Giustina2015}
{\sc Giustina, M., Versteegh, M.~A., Wengerowsky, S., Handsteiner, J.,
  Hochrainer, A., Phelan, K., Steinlechner, F., Kofler, J., Larsson, J.-{\AA}.,
  Abell{\'{a}}n, C., Amaya, W., Pruneri, V., Mitchell, M.~W., Beyer, J.,
  Gerrits, T., Lita, A.~E., Shalm, L.~K., Nam, S.~W., Scheidl, T., Ursin, R.,
  Wittmann, B., and Zeilinger, A.}
\newblock Significant-loophole-free test of {B}ell's theorem with entangled
  photons.
\newblock {\em Physical Review Letters 115}, 25 (2015), p. 250401.

\bibitem{Gl_le_2018}
{\sc Gl{\"a}{\ss}le, T., Gross, D., and Chaves, R.}
\newblock Computational tools for solving a marginal problem with applications
  in {B}ell non-locality and causal modeling.
\newblock {\em Journal of Physics A: Mathematical and Theoretical 51}, 48
  (2018), p. 484002.

\bibitem{Godel1949}
{\sc G\"odel, K.}
\newblock An example of a new type of cosmological solutions of einstein's
  field equations of gravitation.
\newblock {\em Reviews of Modern Physics 21\/} (1949), pp. 447--450.

\bibitem{Goldblatt2006}
{\sc Goldblatt, R.}
\newblock Mathematical modal logic: A view of its evolution.
\newblock {\em Handbook of the History of Logic 7\/} (2006), pp. 1--98.

\bibitem{GHZ1989}
{\sc Greenberger~D.M., Horne~M.A., Z.~A.}
\newblock Going beyond {B}ell’s theorem.
\newblock {\em Quantum Theory and Conceptions of the Universe. Fundamental
  Theories of Physics 37\/} (1989).

\bibitem{Gross2010}
{\sc Gross, D., M\"uller, M., Colbeck, R., and Dahlsten, O. C.~O.}
\newblock All reversible dynamics in maximally nonlocal theories are trivial.
\newblock {\em Physical Review Letters 104\/} (2010), p. 080402.

\bibitem{Grotschel2012}
{\sc Gr\"otschel, M., Lovasz, L., and Shrijver, A.}
\newblock Geometric algorithms and combinatorial optimization.
\newblock {\em Springer Science and Business Media\/} (2012).

\bibitem{Grunhaus1996}
{\sc Grunhaus, J., Popescu, S., and Rohrlich, D.}
\newblock Jamming nonlocal quantum correlations.
\newblock {\em Physical Review A 53\/} (1996), pp. 3781--3784.

\bibitem{Guerin2018}
{\sc Gu{\'{e}}rin, P.~A., and Brukner, {\v{C}}.}
\newblock Observer-dependent locality of quantum events.
\newblock {\em New Journal of Physics 20}, 10 (2018), p. 103031.

\bibitem{Guerin2016}
{\sc Gu\'erin, P.~A., Feix, A., Ara\'ujo, M., and Brukner, {\v{C}}.}
\newblock Exponential communication complexity advantage from quantum
  superposition of the direction of communication.
\newblock {\em Physical Review Letters 117\/} (2016), p. 100502.

\bibitem{Hardy1992}
{\sc Hardy, L.}
\newblock Quantum mechanics, local realistic theories, and {L}orentz-invariant
  realistic theories.
\newblock {\em Physical Review Letters 68\/} (1992), pp. 2981--2984.

\bibitem{Hardy1993}
{\sc Hardy, L.}
\newblock Nonlocality for two particles without inequalities for almost all
  entangled states.
\newblock {\em Physical Review Letters 71\/} (1993), pp. 1665--1668.

\bibitem{Hardy01}
{\sc Hardy, L.}
\newblock Quantum theory from five reasonable axioms.
\newblock {\em arXiv:quant-ph/0101012\/} (2001).

\bibitem{Hardy2005}
{\sc Hardy, L.}
\newblock Probability theories with dynamic causal structure: A new framework
  for quantum gravity.
\newblock {\em arxiv:gr-qc/0509120\/} (2005).

\bibitem{Hardy2016}
{\sc Hardy, L.}
\newblock Operational general relativity: Possibilistic, probabilistic, and
  quantum.
\newblock {\em arxiv:1608.06940\/} (2016).

\bibitem{Heisenberg1935}
{\sc Heisenberg, W.}
\newblock Ist eine deterministische erg\"{a}nzung der quantenmechanik
  m\"{o}glich?
\newblock {\em Wolfgang {P}auli. {W}issenschaftlicher {B}riefwechsel mit
  {B}ohr, {E}instein, {H}eisenberg, Springer 2\/} (1935), Ed. Hermann, A. and
  von Meyenn, K. and Weisskopf, V. F. pp. 409--418.

\bibitem{Hensen2015}
{\sc Hensen, B., Bernien, H., Dr{\'{e}}au, A.~E., Reiserer, A., Kalb, N., Blok,
  M.~S., Ruitenberg, J., Vermeulen, R. F.~L., Schouten, R.~N., Abell{\'{a}}n,
  C., Amaya, W., Pruneri, V., Mitchell, M.~W., Markham, M., Twitchen, D.~J.,
  Elkouss, D., Wehner, S., Taminiau, T.~H., and Hanson, R.}
\newblock {Loophole-free Bell inequality violation using electron spins
  separated by 1.3 kilometres}.
\newblock {\em Nature 526}, 7575 (2015), pp. 682--686.

\bibitem{Henson2014}
{\sc Henson, J., Lal, R., and Pusey, M.~F.}
\newblock Theory-independent limits on correlations from generalized bayesian
  networks.
\newblock {\em New Journal of Physics 16}, 11 (2014), p. 113043.

\bibitem{Horodecki2019}
{\sc Horodecki, P., and Ramanathan, R.}
\newblock The relativistic causality versus no-signaling paradigm for
  multi-party correlations.
\newblock {\em Nature Communications 10}, 1 (2019), p. 1701.

\bibitem{Horsman2016}
{\sc Horsman, D., Heunen, C., Pusey, M.~F., Barrett, J., and Spekkens, R.~W.}
\newblock Can a quantum state over time resemble a quantum state at a single
  time?
\newblock {\em Proceedings of the Royal Society A 473\/} (2017), p. 20170395.

\bibitem{Horst1999}
{\sc Horst, R., and Thoai, N.~V.}
\newblock {DC} {P}rogramming: Overview.
\newblock {\em Journal of Optimization Theory and Applications 103}, 1 (1999),
  pp. 1--43.

\bibitem{Hu2006}
{\sc Hu, X., and Ye, Z.}
\newblock Generalized quantum entropy.
\newblock {\em Journal of Mathematical Physics 47}, 2 (2006), p. 023502.

\bibitem{Iwamoto2013}
{\sc Iwamoto, M., and Shikata, J.}
\newblock Revisiting conditional {R}\'enyi entropies and generalizing
  {S}hannon's bounds in information theoretically secure encryption?
\newblock {\em Technical report, Cryptology ePrint Archive 440\/} (2013).

\bibitem{Jamiolkowski1972}
{\sc Jamiołkowski, A.}
\newblock Linear transformations which preserve trace and positive
  semidefiniteness of operators.
\newblock {\em Reports on Mathematical Physics 3}, 4 (1972), pp. 275--278.

\bibitem{Kaszlikowski2002}
{\sc Kaszlikowski, D., Kwek, L.~C., Chen, J.-L., Zukowski, M., and Oh, C.~H.}
\newblock Clauser-horne inequality for three-state systems.
\newblock {\em Physical Review A 65\/} (2002), p. 032118.

\bibitem{Katta1983}
{\sc Katta, M.}
\newblock Linear {P}rogramming.
\newblock {\em John Wiley and Sons\/} (1983), p. 482.

\bibitem{Khachiyan2008}
{\sc Khachiyan, L., Boros, E., Borys, K., Elbassioni, K., and Gurvich, V.}
\newblock Generating all vertices of a polyhedron is hard.
\newblock {\em Discrete and Computational Geometry}, 1 (2008), pp. 174--190.

\bibitem{Kim2016}
{\sc Kim, J.~S.}
\newblock Tsallis entropy and general polygamy of multi-party quantum
  entanglement in arbitrary dimensions.
\newblock {\em Physical Review A 94\/} (2016).

\bibitem{Kochen1967}
{\sc Kochen, S., and Specker, E.}
\newblock Logical structures arising in quantum theory.
\newblock {\em in Addison, J., L. Henkin, and A. Tarski (eds.), The theory of
  models, North-Holland, Amsterdam\/} (1967), pp. 177--189.

\bibitem{Kochen1968}
{\sc Kochen, S., and Specker, E.}
\newblock The problem of hidden variables in quantum mechanics.
\newblock {\em Indiana University Mathematics Journal 17\/} (1968), pp. 59--87.

\bibitem{Konig2005}
{\sc K\"onig, R., and Renner, R.}
\newblock A de {F}inetti representation for finite symmetric quantum states.
\newblock {\em Journal of Mathematical Physics 46\/} (2005), p. 122108.

\bibitem{Kraft2020}
{\sc Kraft, T., Designolle, S., Ritz, C., Brunner, N., G\"uhne, O., and Huber,
  M.}
\newblock Quantum entanglement in the triangle network.
\newblock {\em arXiv:2002.03970\/} (2020).

\bibitem{Kraemer2018}
{\sc Kr{\"a}mer, L., and del Rio, L.}
\newblock Operational locality in global theories.
\newblock {\em Philosophical Transactions of the Royal Society of London A:
  Mathematical, Physical and Engineering Sciences 376}, 2123 (2018).

\bibitem{Kripke2007}
{\sc Kripke, S.~A.}
\newblock Semantical considerations on modal logic.
\newblock {\em Universal Logic: An Anthology\/} (2012), pp. 197--208.

\bibitem{Leifer2013}
{\sc Leifer, M.~S., and Spekkens, R.~W.}
\newblock Towards a formulation of quantum theory as a causally neutral theory
  of {B}ayesian inference.
\newblock {\em Physical Review A 88\/} (2013), p. 052130.

\bibitem{Lewis1918}
{\sc Lewis, C.~I.}
\newblock A survey of symbolic logic.
\newblock {\em University of California Press\/} (1918).

\bibitem{Lewis1959}
{\sc Lewis, C.~I., Langford, C.~H., and Lamprecht, P.}
\newblock Symbolic logic.
\newblock {\em Dover Publications New York\/} (1959).

\bibitem{Lieb1973}
{\sc Lieb, E., and Ruskai, M.}
\newblock Proof of the strong subadditivity of quantum-mechanical entropy.
\newblock {\em Journal of Mathematical Physics 14\/} (1973), pp. 1938--1941.

\bibitem{Linden2013}
{\sc Linden, N., Mosonyi, M., and Winter, A.}
\newblock The structure of r{\'e}nyi entropic inequalities.
\newblock {\em Proceedings of the Royal Society A: Mathematical, Physical and
  Engineering Sciences 469}, 2158 (2013), p. 20120737.

\bibitem{Lloyd2011a}
{\sc Lloyd, S., Maccone, L., Garcia-Patron, R., Giovannetti, V., and Shikano,
  Y.}
\newblock Quantum mechanics of time travel through post-selected teleportation.
\newblock {\em Physical Review D 84\/} (2011), p. 025007.

\bibitem{Lloyd2011}
{\sc Lloyd, S., Maccone, L., Garcia-Patron, R., Giovannetti, V., Shikano, Y.,
  Pirandola, S., Rozema, L.~A., Darabi, A., Soudagar, Y., Shalm, L.~K., and
  Steinberg, A.~M.}
\newblock Closed timelike curves via postselection: Theory and experimental
  test of consistency.
\newblock {\em Physical Review Letters 106\/} (2011), p. 040403.

\bibitem{Ludwig1964}
{\sc Ludwig, G.}
\newblock Versuch einer axiomatischen grundlegung der quantenmechanik und
  allgemeinerer physikalischer theorien.
\newblock {\em Zeitschrift f\"ur Physik 181\/} (1964), pp. 233--260.

\bibitem{Ludwig1967}
{\sc Ludwig, G.}
\newblock Attempt of an axiomatic foundation of quantum mechanics and more
  general theories {II}.
\newblock {\em Communications in Mathematical Physics 4}, 5 (1967), pp.
  331--348.

\bibitem{Ludwig1968}
{\sc Ludwig, G.}
\newblock Attempt of an axiomatic foundation of quantum mechanics and more
  general theories {III}.
\newblock {\em Communications in Mathematical Physics 9\/} (1968), pp. 1--12.

\bibitem{Martin1999}
{\sc Martin, R.~K.}
\newblock Large scale linear and integer optimization: {A} unified approach.
\newblock {\em Springer Science and Business Media\/} (1999).

\bibitem{Masanes2002}
{\sc Masanes, L.}
\newblock Tight {B}ell inequality for d-outcome measurements correlations.
\newblock {\em Quantum information and computation 3\/} (2002), p. 345.

\bibitem{Masanes2006}
{\sc Masanes, L., Acin, A., and Gisin, N.}
\newblock General properties of nonsignaling theories.
\newblock {\em Physical Review A 73\/} (2006), p. 012112.

\bibitem{MayersYao}
{\sc Mayers, D., and Yao, A.}
\newblock Quantum cryptography with imperfect apparatus.
\newblock {\em Proceedings of the 39th Annual Symposium on Foundations of
  Computer Science (FOCS-98)\/} (1998), pp. 503--509.

\bibitem{Megidish2013}
{\sc Megidish, E., Halevy, A., Shacham, T., Dvir, T., Dovrat, L., and
  Eisenberg, H.~S.}
\newblock Entanglement swapping between photons that have never coexisted.
\newblock {\em Physical Review Letters 110\/} (2013), p. 210403.

\bibitem{Mermin1990}
{\sc Mermin, N.~D.}
\newblock Simple unified form for the major no-hidden-variables theorems.
\newblock {\em Physical Review Letters 65\/} (1990), pp. 3373--3376.

\bibitem{Muller-Lennert2013}
{\sc Müller-Lennert, M., Dupuis, F., Szehr, O., Fehr, S., and Tomamichel, M.}
\newblock On quantum rényi entropies: A new generalization and some
  properties.
\newblock {\em Journal of Mathematical Physics 54}, 12 (2013), p. 122203.

\bibitem{Navascues2017}
{\sc Navascues, M., and Wolfe, E.}
\newblock The inflation technique completely solves the classical inference
  problem.
\newblock {\em arxiv:1707.06476\/} (2017).

\bibitem{Neal2000}
{\sc Neal, R.~M.}
\newblock On deducing conditional independence from d-separation in causal
  graphs with feedback (research note).
\newblock {\em Journal of Artificial Intelligence Research 12\/} (2000), pp.
  87–91.

\bibitem{NielsenChuang}
{\sc Nielsen, M.~A., and Chuang, I.~L.}
\newblock {Quantum Computation and Quantum Information: 10th Anniversary
  Edition}.
\newblock {\em Cambridge University Press\/} (2011).

\bibitem{Nurgalieva2018}
{\sc Nurgalieva, N., and del Rio, L.}
\newblock Inadequacy of modal logic in quantum settings.
\newblock {\em In Proceedings QPL 2018, EPTCS 287\/} (2019), pp. 267--297.

\bibitem{Oreshkov2019}
{\sc Oreshkov, O.}
\newblock Time-delocalized quantum subsystems and operations: on the existence
  of processes with indefinite causal structure in quantum mechanics.
\newblock {\em {Quantum} 3\/} (2019), p. 206.

\bibitem{Oreshkov2012}
{\sc Oreshkov, O., Costa, F., and Brukner, {\v{C}}.}
\newblock {Quantum correlations with no causal order}.
\newblock {\em Nature Communications 3\/} (2012), p. 1092.

\bibitem{Oreshkov2016}
{\sc Oreshkov, O., and Giarmatzi, C.}
\newblock {Causal and causally separable processes}.
\newblock {\em New Journal of Physics 18}, 9 (2016), p. 93020.

\bibitem{Paunkovic2020}
{\sc Paunkovi{\'{c}}, N., and Vojinovi{\'{c}}, M.}
\newblock Causal orders, quantum circuits and spacetime: distinguishing between
  definite and superposed causal orders.
\newblock {\em {Quantum} 4\/} (2020), p. 275.

\bibitem{Pawlowski2009}
{\sc Paw{\l}owski, M., Paterek, T., Kaszlikowski, D., Scarani, V., Winter, A.,
  and Zukowski, M.}
\newblock Information causality as a physical principle.
\newblock {\em Nature 461}, 7267 (2009), pp. 1101--1104.

\bibitem{Pearl95}
{\sc Pearl, J.}
\newblock Causal diagrams for empirical research.
\newblock {\em Biometrika 82}, 4 (1995), p. 669--688.

\bibitem{Pearl1995}
{\sc Pearl, J.}
\newblock On the testability of causal models with latent and instrumental
  variables.
\newblock {\em UAI'95: Proceedings of the Eleventh conference on Uncertainty in
  artificial intelligence\/} (1995).

\bibitem{Pearl2009}
{\sc Pearl, J.}
\newblock Causality: Models, reasoning, and inference.
\newblock {\em Second edition, Cambridge University Press\/} (2009).

\bibitem{Pearl2013}
{\sc Pearl, J., and Dechter, R.}
\newblock Identifying independencies in causal graphs with feedback.
\newblock {\em UAI'96: Proceedings of the Twelfth international conference on
  Uncertainty in artificial intelligence\/} (2013).

\bibitem{Peres1990}
{\sc Peres, A.}
\newblock Incompatible results of quantum measurements.
\newblock {\em Physics Letters A 151}, 3 (1990), pp. 107--108.

\bibitem{Petz2014}
{\sc Petz, D., and Virosztek, D.}
\newblock Some inequalities for quantum {T}sallis entropy related to the strong
  subadditivity.
\newblock {\em Mathematical Inequalities and Applications 18\/} (2014), p. 555.

\bibitem{Pienaar2019}
{\sc Pienaar, J.}
\newblock A time-reversible quantum causal model.
\newblock {\em arxiv:1902.00129\/} (2019).

\bibitem{Pienaar2020}
{\sc Pienaar, J.}
\newblock Quantum causal models via quantum bayesianism.
\newblock {\em Physical Review A 101\/} (2020), p. 012104.

\bibitem{Pienaar2015}
{\sc Pienaar, J., and Brukner, {\v{C}}.}
\newblock A graph-separation theorem for quantum causal models.
\newblock {\em New Journal of Physics 17}, 7 (2015), p. 073020.

\bibitem{Pirandola2019}
{\sc Pirandola, S., Andersen, U.~L., Banchi, L., Berta, M., Bunandar, D.,
  Colbeck, R., Englund, D., Gehring, T., Lupo, C., Ottaviani, C., Pereira, J.,
  Razavi, M., Shaari, J.~S., Tomamichel, M., Usenko, V.~C., Vallone, G.,
  Villoresi, P., and Wallden, P.}
\newblock Advances in quantum cryptography.
\newblock {\em arxiv:1906.01645\/} (2019).

\bibitem{Pironio2005}
{\sc Pironio, S.}
\newblock Lifting {B}ell inequalities.
\newblock {\em Journal of Mathematical Physics 46\/} (2005), p. 062112.

\bibitem{Pironio2009}
{\sc Pironio, S., Acín, A., Massar, S., Boyer de~la Giroday, A., Matsukevich,
  D., Maunz, P., Olmschenk, S., Hayes, D., Luo, L., A~Manning, T., and Monroe,
  C.}
\newblock Random numbers certified by {B}ell's theorem.
\newblock {\em Nature 464\/} (2010), pp. 1021--4.

\bibitem{Pitowski89}
{\sc Pitowski, I.}
\newblock Quantum probability -- quantum logic.
\newblock {\em Springer-Verlag Berlin Heidelberg 321\/} (1989).

\bibitem{Popescu1994}
{\sc Popescu, S., and Rohrlich, D.}
\newblock Quantum nonlocality as an axiom.
\newblock {\em Foundations of Physics 24}, 3 (1994), pp. 379--385.

\bibitem{Portmann2017}
{\sc Portmann, C., Matt, C., Maurer, U., Renner, R., and Tackmann, B.}
\newblock {Causal Boxes: Quantum Information-Processing Systems Closed under
  Composition}.
\newblock {\em IEEE Transactions on Information Theory 63}, 5 (2017), pp.
  3277--3305.

\bibitem{Preskill2016}
{\sc Preskill, J.}
\newblock Quantum shannon theory.
\newblock {\em arXiv:1604.07450\/} (2016).

\bibitem{Procopio2015}
{\sc Procopio, L.~M., Moqanaki, A., Ara{\'{u}}jo, M., Costa, F., {Alonso
  Calafell}, I., Dowd, E.~G., Hamel, D.~R., Rozema, L.~A., Brukner, {\v{C}}.,
  and Walther, P.}
\newblock {Experimental superposition of orders of quantum gates}.
\newblock {\em Nature Communications 6\/} (2015), p. 7913.

\bibitem{Pusey2018}
{\sc Pusey, M.~F.}
\newblock An inconsistent friend.
\newblock {\em Nature Physics 14}, 10 (2018), pp. 977--978.

\bibitem{Pusey2015}
{\sc Pusey, M.~F., and Leifer, M.~S.}
\newblock Logical pre- and post-selection paradoxes are proofs of
  contextuality.
\newblock {\em In Proceedings QPL 2015, EPTCS 195\/} (2015), pp. 295--306.

\bibitem{Reichenbach1956}
{\sc Reichenbach, H.}
\newblock The direction of time.
\newblock {\em Dover Publications\/} (1956).

\bibitem{Renner2006}
{\sc Renner, R.}
\newblock Security of quantum key distribution.
\newblock {\em PhD Dissertation, ETH Z{\"u}rich. arxiv:quant-ph/0512258\/}
  (2006).

\bibitem{Renou2019}
{\sc Renou, M.-O., B\"aumer, E., Boreiri, S., Brunner, N., Gisin, N., and
  Beigi, S.}
\newblock Genuine quantum nonlocality in the triangle network.
\newblock {\em Physical Review Letters 123\/} (2019), p. 140401.

\bibitem{Renyi1961}
{\sc R\'enyi, A.}
\newblock On measures of information and entropy.
\newblock {\em Proceedings of the 4th Berkeley Symposium on Mathematics,
  Statistics and Probability, University of California Press\/} (1961), pp.
  547--561.

\bibitem{Richardson1996}
{\sc Richardson, T.}
\newblock A discovery algorithm for directed cyclic graphs.
\newblock {\em Proceedings of the Twelfth International Conference on
  Uncertainty in Artificial Intelligence\/} (1996), pp. 454–461.

\bibitem{Ried2015}
{\sc Ried, K., Agnew, M., Vermeyden, L., Janzing, D., Spekkens, R.~W., and
  Resch, K.~J.}
\newblock A quantum advantage for inferring causal structure.
\newblock {\em Nature Physics 11}, 5 (2015), pp. 414--420.

\bibitem{Rosset2017}
{\sc Rosset, D., Gisin, N., and Wolfe, E.}
\newblock Universal bound on the cardinality of local hidden variables in
  networks.
\newblock {\em Quantum Information and Computation 18\/} (2018), pp.
  0910--0926.

\bibitem{Rowe2009}
{\sc Rowe, N.}
\newblock Why there's so little good evidence that fiscal (or monetary) policy
  works (online), (2009).
\newblock
  \url{https://worthwhile.typepad.com/worthwhile\_canadian\_initi/2009/01/why-theres-so-little-good-evidence-that-fiscal-or-monetary-policy-works.html}.

\bibitem{Rubino2017}
{\sc Rubino, G., Rozema, L.~A., Feix, A., Ara{\'{u}}jo, M., Zeuner, J.~M.,
  Procopio, L.~M., Brukner, {\v{C}}., and Walther, P.}
\newblock {Experimental verification of an indefinite causal order}.
\newblock {\em Science Advances 3}, 3 (2017).

\bibitem{Salazar2020}
{\sc Salazar, R., Kamon, M., Goyeneche, D., Horodecki, K., Saha, D.,
  Ramanathan, R., and Horodecki, P.}
\newblock A no-go theorem for device-independent security in relativistic
  causal theories.
\newblock {\em arxiv:1712.01030\/} (2020).

\bibitem{Scheines1997}
{\sc Scheines, R.}
\newblock An introduction to causal inference.
\newblock {\em In McKim and Turner (eds.)\/} (1997), pp. 185–99.

\bibitem{Schrijver1986}
{\sc Schrijver, A.}
\newblock Theory of linear and integer programming.
\newblock {\em John Wiley and Sons, Inc.\/} (1986).

\bibitem{Segal1947}
{\sc Segal, I.~E.}
\newblock Postulates for general quantum mechanics.
\newblock {\em Annals of Mathematics 48}, 4 (1947), pp. 930–948.

\bibitem{Shalm2015}
{\sc Shalm, L.~K., Meyer-Scott, E., Christensen, B.~G., Bierhorst, P., Wayne,
  M.~A., Stevens, M.~J., Gerrits, T., Glancy, S., Hamel, D.~R., Allman, M.~S.,
  Coakley, K.~J., Dyer, S.~D., Hodge, C., Lita, A.~E., Verma, V.~B., Lambrocco,
  C., Tortorici, E., Migdall, A.~L., Zhang, Y., Kumor, D.~R., Farr, W.~H.,
  Marsili, F., Shaw, M.~D., Stern, J.~A., Abell{\'{a}}n, C., Amaya, W.,
  Pruneri, V., Jennewein, T., Mitchell, M.~W., Kwiat, P.~G., Bienfang, J.~C.,
  Mirin, R.~P., Knill, E., and Nam, S.~W.}
\newblock {Strong Loophole-Free Test of Local Realism}.
\newblock {\em Physical Review Letters 115}, 25 (2015), p. 250402.

\bibitem{Shannon1948}
{\sc {Shannon}, C.~E.}
\newblock A mathematical theory of communication.
\newblock {\em The Bell System Technical Journal 27}, 3 (1948), pp. 379--423.

\bibitem{Shor1994}
{\sc {Shor}, P.~W.}
\newblock Algorithms for quantum computation: discrete logarithms and
  factoring.
\newblock {\em Proceedings 35th Annual Symposium on Foundations of Computer
  Science\/} (1994), pp. 124--134.

\bibitem{Short2006}
{\sc Short, A.~J., Popescu, S., and Gisin, N.}
\newblock Entanglement swapping for generalized nonlocal correlations.
\newblock {\em Physical Review A 73}, 1 (2006), p. 012101.

\bibitem{Skrzypczyk2008}
{\sc Skrzypczyk, P., Brunner, N., and Popescu, S.}
\newblock Emergence of quantum correlations from non-locality swapping.
\newblock {\em Physical Review Letters 102\/} (2009), p. 110402.

\bibitem{Spekkens2005}
{\sc Spekkens, R.~W.}
\newblock Contextuality for preparations, transformations, and unsharp
  measurements.
\newblock {\em Physical Review A 71}, 5 (2005), p. 052108.

\bibitem{Spekkens07}
{\sc Spekkens, R.~W.}
\newblock Evidence for the epistemic view of quantum states: A toy theory.
\newblock {\em Physical Review A 75\/} (2007), p. 032110.

\bibitem{Sudbery2017}
{\sc Sudbery, A.}
\newblock Single-world theory of the extended wigner's friend experiment.
\newblock {\em Foundations of Physics 47}, 5 (2017), pp. 658--669.

\bibitem{Sudbery2019}
{\sc Sudbery, A.}
\newblock The hidden assumptions of {F}rauchiger and {R}enner.
\newblock {\em arxiv:1905.13248\/} (2019).

\bibitem{Svetlichny2011}
{\sc Svetlichny, G.}
\newblock Time {T}ravel: {D}eutsch vs. {T}eleportation.
\newblock {\em International Journal of Theoretical Physics 50}, 12 (2011), pp.
  3903–3914.

\bibitem{Tomamichel2009}
{\sc {Tomamichel}, M., {Colbeck}, R., and {Renner}, R.}
\newblock A fully quantum asymptotic equipartition property.
\newblock {\em IEEE Transactions on Information Theory 55}, 12 (2009), pp.
  5840--5847.

\bibitem{Tsallis1988}
{\sc Tsallis, C.}
\newblock Possible generalization of {B}oltzmann-{G}ibbs statistics.
\newblock {\em Journal of Statistical Physics 52}, 1 (1988), pp. 479--487.

\bibitem{Tsirelson1993}
{\sc Tsirelson, B.}
\newblock Some results and problems on quantum {B}ell-type inequalities.
\newblock {\em Hadronic Journal Supplement\/} (1993), pp. 329–345.

\bibitem{Cirelson93}
{\sc Tsirelson, B.}
\newblock Some results and problems on quantum {B}ell-type inequalities.
\newblock {\em Hadronic Journal Supplement 8\/} (1993), pp. 329--345.

\bibitem{Himbeeck2019}
{\sc Van~Himbeeck, T., Bohr~Brask, J., Pironio, S., Ramanathan, R., Sainz,
  A.~B., and Wolfe, E.}
\newblock Quantum violations in the {I}nstrumental scenario and their relations
  to the {B}ell scenario.
\newblock {\em {Quantum} 3\/} (2019), p. 186.

\bibitem{VanStockum1937}
{\sc van Stockum, W.~J.}
\newblock The gravitational feild of a distribution of particles rotating about
  an axis of symmetry.
\newblock {\em Proceedings of the Royal Society of Edinburgh 57\/} (1937), pp.
  135--154.

\bibitem{Verma1988}
{\sc Verma, T., and Pearl, J.}
\newblock Causal networks: Semantics and expressiveness.
\newblock {\em Proceedings of the Fourth Annual Conference on Uncertainty in
  Artificial Intelligence (UAI '88)\/} (1990), pp. 69--78.

\bibitem{Vilasini2019}
{\sc Vilasini, V., and Colbeck, R.}
\newblock Analyzing causal structures using {T}sallis entropies.
\newblock {\em Physical Review A 100\/} (2019), p. 062108.

\bibitem{Vilasini2020}
{\sc Vilasini, V., and Colbeck, R.}
\newblock Limitations of entropic inequalities for detecting nonclassicality in
  the postselected {B}ell causal structure.
\newblock {\em Physical Review Research 2\/} (2020), p. 033096.

\bibitem{Vilasini_PRdoxes}
{\sc Vilasini, V., Nurgalieva, N., and del Rio, L.}
\newblock Multi-agent paradoxes beyond quantum theory.
\newblock {\em New Journal of Physics 21}, 11 (2019), p. 113028.

\bibitem{Vilasini_crypto}
{\sc Vilasini, V., Portmann, C., and del Rio, L.}
\newblock Composable security in relativistic quantum cryptography.
\newblock {\em New Journal of Physics 21}, 4 (2019), p. 043057.

\bibitem{Vitter1999}
{\sc Vitter, J.~S., Larmore, L., Leighton, T., and Raz, R.}
\newblock Exponential separation of quantum and classical communication
  complexity.
\newblock {\em Proceedings of the Thirty-First Annual ACM Symposium on Theory
  of Computing\/} (1999), pp. 358–367.

\bibitem{vonNeumann1955}
{\sc Von~Neumann, J.}
\newblock Mathematical foundations of quantum mechanics.
\newblock {\em Princeton university press\/} (1955).

\bibitem{vonNeumann1932}
{\sc von Neumann, J., Wheeler, N.~A., and Beyer, R.~T.}
\newblock Mathematical foundations of quantum mechanics.
\newblock {\em Princeton University Press\/} (1932).

\bibitem{Wajs15}
{\sc Wajs, M., Kurzynski, P., and Kaszlikowski, D.}
\newblock Information-theoretic {B}ell inequalities based on {T}sallis entropy.
\newblock {\em Physical Review A 91\/} (2015), p. 012114.

\bibitem{Weilenmann16}
{\sc Weilenmann, M., and Colbeck, R.}
\newblock Inability of the entropy vector method to certify nonclassicality in
  linelike causal structures.
\newblock {\em Physical Review A 94\/} (2016), p. 042112.

\bibitem{Weilenmann2017}
{\sc Weilenmann, M., and Colbeck, R.}
\newblock Analysing causal structures with entropy.
\newblock {\em Proceedings of the Royal Society of London A: Mathematical,
  Physical and Engineering Sciences 473}, 2207 (2017).

\bibitem{Weilenmann2018}
{\sc Weilenmann, M., and Colbeck, R.}
\newblock Non-{S}hannon inequalities in the entropy vector approach to causal
  structures.
\newblock {\em {Quantum} 2\/} (2018), p. 57.

\bibitem{WeilenmannGPT}
{\sc Weilenmann, M., and Colbeck, R.}
\newblock Analysing causal structures in generalised probabilistic theories.
\newblock {\em {Quantum} 4\/} (2020), p. 236.

\bibitem{Weilenmann2020}
{\sc Weilenmann, M., and Colbeck, R.}
\newblock Self-testing of physical theories, or, is quantum theory optimal with
  respect to some information-processing task?
\newblock {\em Physical Review Letters 125\/} (2020), p. 060406.

\bibitem{Wheeler1957}
{\sc Wheeler, J.~A.}
\newblock Assessment of {E}verett's ``relative state'' formulation of quantum
  theory.
\newblock {\em Reviews of Modern Physics 29\/} (1957), pp. 463--465.

\bibitem{Wigner1961}
{\sc Wigner, E.~P.}
\newblock Remarks on the mind-body question.
\newblock {\em The Scientist Speculates, Heineman\/} (1961), Ed. I. J. Good.

\bibitem{Williams1986}
{\sc Williams, H.~P.}
\newblock Fourier's method of linear programming and its dual.
\newblock {\em The American Mathematical Monthly 93}, 9 (1986), pp. 681--695.

\bibitem{Wolfe2019}
{\sc Wolfe, E., Schmid, D., Sainz, A.~B., Kunjwal, R., and Spekkens, R.~W.}
\newblock Quantifying {B}ell: the resource theory of nonclassicality of
  common-cause boxes.
\newblock {\em {Quantum} 4\/} (2020), p. 280.

\bibitem{wolfe2016}
{\sc Wolfe, E., Spekkens, R.~W., and Fritz, T.}
\newblock The inflation technique for causal inference with latent variables.
\newblock {\em Journal of Causal Inference 7\/} (2019).

\bibitem{Wood2015}
{\sc Wood, C.~J., and Spekkens, R.~W.}
\newblock {The lesson of causal discovery algorithms for quantum correlations:
  causal explanations of {B}ell-inequality violations require fine-tuning}.
\newblock {\em New Journal of Physics 17}, 3 (2015), p. 33002.

\bibitem{Yeung97}
{\sc Yeung, R.~W.}
\newblock A framework for linear information inequalities.
\newblock {\em IEEE Transactions on Information Theory 43}, 6 (1997), pp.
  1924--1934.

\bibitem{Zhalama2017}
{\sc Zhalama, Zhang, J., and Mayer, W.}
\newblock Weakening faithfulness: some heuristic causal discovery algorithms.
\newblock {\em International Journal of Data Science and Analytics 3}, 2
  (2017), pp. 93--104.

\bibitem{Zhang1997ANC}
{\sc Zhang, Z., and Yeung, R.~W.}
\newblock A non-{S}hannon-type conditional inequality of information
  quantities.
\newblock {\em IEEE Trans. Information Theory 43\/} (1997), pp. 1982--1986.

\bibitem{Zukowski99}
{\sc Zukowski, M., Kaszlikowski, D., Baturo, A., and Larsson, J.-{\AA}.}
\newblock Strengthening the {B}ell theorem: conditions to falsify local realism
  in an experiment.
\newblock {\em arXiv:quant-ph/9910058\/} (1999).

\bibitem{Zukowski1993}
{\sc Zukowski, M., Zeilinger, A., Horne, M.~A., and Ekert, A.~K.}
\newblock ``{E}vent-ready-detectors'' {B}ell experiment via entanglement
  swapping.
\newblock {\em Physical Review Letters 71\/} (1993), pp. 4287--4290.

\bibitem{Zych2019}
{\sc Zych, M., Costa, F., Pikovski, I., and Brukner, {\v{C}}.}
\newblock {Bell's theorem for temporal order}.
\newblock {\em Nature Communications 10}, 1 (2019), p. 3772.

\end{thebibliography}

\end{document}